\documentclass[11pt,a4paper,twoside]{book}
\usepackage[paper=a4paper,left=25mm,right=25mm,top=25mm,bottom=25mm]{geometry}
\usepackage[breaklinks=true,colorlinks=true,linkcolor=blue,urlcolor=blue,citecolor=blue]{hyperref}
\usepackage{bookmark}
\usepackage[utf8]{inputenc}
\usepackage[T1]{fontenc}
\usepackage{lmodern}
\usepackage[english]{babel}
\usepackage{comment}
\usepackage{appendix}
\usepackage{indentfirst}
\usepackage{epigraph}
\renewcommand{\textflush}{flushepinormal}
\usepackage{fancybox,framed}
\usepackage{datetime}
\newdateformat{monthyeardate}{\monthname[\THEMONTH] \THEYEAR}
\usepackage{xcolor}
\usepackage{tikz-cd}
\usepackage{cancel}
\usepackage{cite}
\usepackage{tensor}
\usepackage{lettrine}
\graphicspath{{Figures/}}
\makeatletter
\renewcommand{\@epitext}[1]{%
  \begin{minipage}{\epigraphwidth}\begin{\textflush} \hspace*{20pt}#1\\
    \ifdim\epigraphrule>\z@ \@epirule \else \vspace*{-.5\baselineskip} \fi
  \end{\textflush}\end{minipage}}
\makeatother
\usepackage[Lenny]{fncychap}
\usepackage{fancyhdr}
\usepackage{titletoc}
\usepackage{chngcntr}
\counterwithin*{section}{chapter}
\titlecontents{chapter}[0pt]{\bfseries\protect\addvspace{15pt}\titlerule\addvspace{1.5ex}}
{\contentslabel[\thecontentslabel]{0pt}\qquad}{\chaptername~}{\hfill\contentspage}[\addvspace{0.7ex}\titlerule\addvspace{1.5ex}]
\addto\captionsenglish{}
\usepackage{graphicx}
\usepackage{float}
\usepackage[font=normalsize]{subfig} 
\usepackage[export]{adjustbox}
\usepackage{dsfont}
\usepackage{mathrsfs}
\usepackage[tbtags]{amsmath}
\usepackage{amssymb,amsmath,mathtools}
\usepackage{braket}
\usepackage{amsthm} 
\usepackage[framemethod=TikZ]{mdframed}
\definecolor{mycolor}{gray}{0.93}
\newmdtheoremenv[roundcorner=5pt,backgroundcolor=mycolor,everyline=true,linewidth=1.1pt,innertopmargin=-2pt, innerbottommargin=7pt,splittopskip=17pt]{theorem}{Theorem}[chapter]
\newmdtheoremenv[roundcorner=5pt,backgroundcolor=mycolor,everyline=true,linewidth=1.1pt,innertopmargin=-2pt, innerbottommargin=7pt,splittopskip=17pt]{proposition}[theorem]{Proposition}
\newmdtheoremenv[roundcorner=5pt,backgroundcolor=mycolor,everyline=true,linewidth=1.1pt,innertopmargin=-2pt, innerbottommargin=7pt,splittopskip=17pt]{lemma}[theorem]{Lemma}
\newmdtheoremenv[roundcorner=5pt,backgroundcolor=mycolor,everyline=true,linewidth=1.1pt,innertopmargin=-2pt, innerbottommargin=7pt,splittopskip=17pt]{corollary}[theorem]{Corollary}
\newmdtheoremenv[roundcorner=5pt,backgroundcolor=mycolor,everyline=true,linewidth=1.1pt,innertopmargin=-2pt, innerbottommargin=7pt,splittopskip=17pt]{proposition2}{Proposition}
\theoremstyle{definition}
\newmdtheoremenv[roundcorner=5pt,backgroundcolor=mycolor,everyline=true,linewidth=1.1pt,innertopmargin=-2pt, innerbottommargin=7pt,splittopskip=17pt]{definition}[theorem]{Definition}
\newmdtheoremenv[roundcorner=5pt,backgroundcolor=mycolor,everyline=true,linewidth=1.1pt,innertopmargin=-2pt, innerbottommargin=7pt,splittopskip=17pt]{setup}[theorem]{Set-up}
\newtheorem{Examplep}[theorem]{Example}
\newenvironment{example}
  {\pushQED{\qed}\renewcommand{\qedsymbol}{$\blacktriangle$}
  \Examplep}{\popQED\endExamplep}
\newtheorem{Remarkp}[theorem]{Remark}
\newenvironment{remark}
  {\pushQED{\qed}\renewcommand{\qedsymbol}{$\blacktriangle$}   
  \Remarkp}{\popQED\endRemarkp}
\newtheorem{Examplesp}[theorem]{Examples}
\newenvironment{examples}
  {\pushQED{\qed}\renewcommand{\qedsymbol}{$\blacktriangle$}
  \Examplesp}{\popQED\endExamplesp}
\newtheorem{Remarksp}[theorem]{Remarks}
\newenvironment{remarks}
  {\pushQED{\qed}\renewcommand{\qedsymbol}{$\blacktriangle$}   
  \Remarksp}{\popQED\endRemarksp}
\renewcommand{\qedsymbol}{$\blacksquare$}
\definecolor{mycolor}{gray}{0.93}
\renewcommand\fbox{\fcolorbox{black}{mycolor}}
\renewcommand{\:}{\colon} 
\newcommand{\defeq}{\vcentcolon=} 
\newcommand{\verteq}{\rotatebox[origin=c]{90}{$\mkern1mu=$}}
\renewcommand{\i}{\mathrm{i}} 
\renewcommand{\d}{\mathrm{d}}
\newcommand{\I}{\mathrm{I}}
\newcommand{\bb}[1]{\mathbb{#1}}
\renewcommand{\sf}[1]{\mathsf{#1}}
\renewcommand{\cal}[1]{\mathcal{#1}}
\usepackage[hang]{footmisc}
\usepackage{lipsum}
\counterwithout{footnote}{chapter}
\let\oldbibliography\thebibliography
\renewcommand{\thebibliography}[1]{\oldbibliography{#1}
\setlength{\itemsep}{-0.1\baselineskip}}
\begin{document}
%
%
%
%
%
\begin{titlepage}

\begin{figure}[H]
\centering
\subfloat{\includegraphics[width=0.5\textwidth]{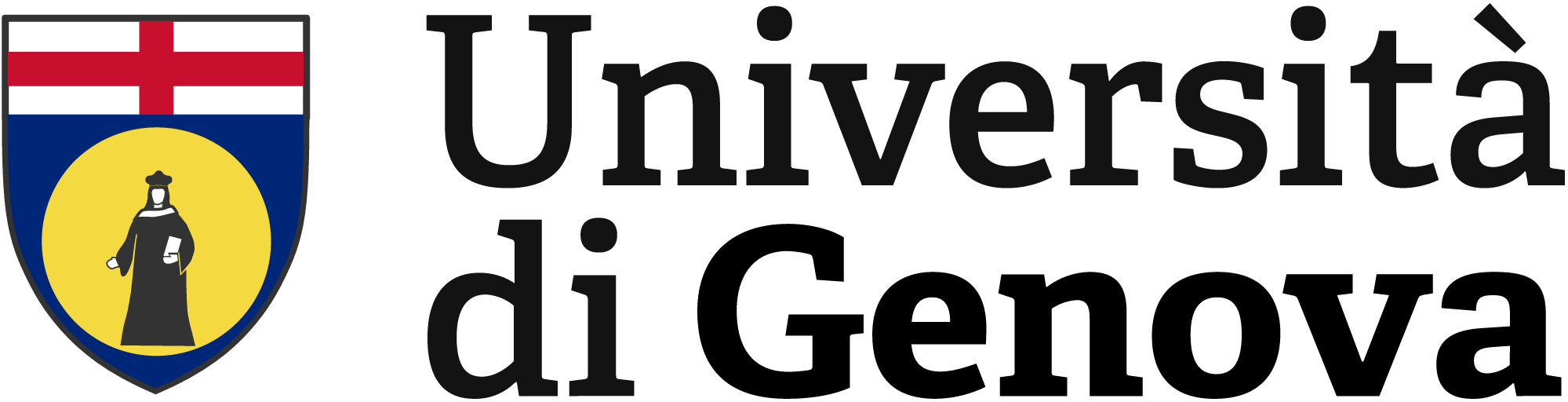}}\hfill
\end{figure}

\vspace*{1cm}

\begin{center}
\begin{LARGE}\textsc{Doctoral Thesis}\end{LARGE}
\end{center}
\noindent\hrulefill
\begin{center}
	{\fontsize{23}{30}\selectfont\textsc{On the Quantisation of Linear Gauge Theories on Lorentzian Manifolds}
	\par
	}\vspace*{0.5cm}
	\begin{LARGE} Maxwell's Theory via Complete Gauge Fixing
	\end{LARGE}
\end{center}
\hrulefill

\vspace*{1cm}

\begin{Large}
\noindent\begin{tabular}[t]{@{}l} 
\textit{Author:}\\ \textsc{Gabriel Schmid}
\end{tabular}
\hfill
\begin{tabular}[t]{l@{}}
\textit{Supervisors:}\\
\textsc{Dr.~Simone Murro}\\[-1mm]
{\small (University of Genoa)}\\
\textsc{Prof.~Dr.~Matteo Capoferri}
\\[-1mm]
{\small (University of Milan \& Heriot-Watt University)}
\end{tabular}
\end{Large}

\vspace*{2cm}

\begin{center}
\begin{Large}\textit{A thesis submitted for the degree of}\end{Large}\\[2mm]
\begin{Large}\textsc{Doctor of Philosophy (PhD)}\end{Large}\\[2mm]
\begin{Large}\textit{in}\end{Large}\\[2mm]
\begin{Large}\textsc{Mathematics and Applications}\end{Large}
\end{center}

\vfill

\begin{center}
	\begin{huge}\textsc{Department of Mathematics}\end{huge}\\[3mm]
	\noindent\begin{huge}\textsc{University of Genoa}\end{huge}
\end{center}

\vfill

\begin{center}
\begin{Large}Genoa, the 5$^{\mathrm{th}}$ of December 2025\end{Large}
\end{center}

\end{titlepage}
%
%
%
%
%
\newgeometry{top=3cm,bottom=1in,outer=1in,inner=1in}
\setlength{\headsep}{15pt}
\pagestyle{plain}
%
%
%
%
%
\newpage\thispagestyle{empty}

\begin{itemize}\item[]\end{itemize}
\vfill
\begin{Large}
\begin{tabular}[t]{@{}l} 
\\ 
\end{tabular}
\hfill
\begin{tabular}[t]{l@{}}
\underline{\textsc{Disputation:}}\\Genoa, the 24$^{\mathrm{th}}$ of February 2026\\[1em]
\underline{\textsc{External Reviewers:}}\\
Prof.~Dr.~Alexander Strohmaier\\[-1mm]
{\small (Leibniz University Hannover)}\\
Prof.~Dr.~Michał Wrochna\\[-1mm]
{\small (Utrecht University \& Vrije Universiteit Brussel)}
\\[1em]
\underline{\textsc{Defence Committee:}}
\\
Prof.~Dr.~Christian Gérard\\[-1mm]
{\small (Paris-Saclay University)}\\
Prof.~Dr.~Nicola Pinamonti\\[-1mm]
{\small (University of Genoa)}\\
Prof.~Dr.~Michał Wrochna\\[-1mm]
{\small (Utrecht University \& Vrije Universiteit Brussel)}
\end{tabular}
\end{Large}

\vspace*{1cm}

\begin{flushleft}
\begin{minipage}{1\textwidth} 
    \hrule
    \vspace{2mm}
    \footnotesize
    \textbf{Note on this Version:}\\
    This is an updated arXiv version of the doctoral thesis submitted on December 5, 2025, and defended on February 24, 2026. It incorporates minor corrections and typographical improvements over the original. The original version is available at IRIS UniGe under the permanent link \href{https://hdl.handle.net/11567/1286476}{hdl.handle.net/11567/1286476}.
    \\[2ex]
    \textbf{Current Revision:} 29$^{\mathrm{th}}$ of April, 2026
\end{minipage}
\end{flushleft}
\frontmatter
\newpage
\section*{Declaration of Authorship}
I hereby declare that the thesis submitted is my own unaided work and that all direct or indirect sources used are acknowledged as references.

\vspace*{0.5cm}

\noindent The main results of this doctoral thesis are based on the following two articles:
\begin{itemize}
	\item[$\bullet$]F.~Finster, S.~Murro and G.~Schmid.
\newblock {The Cauchy Problem for Symmetric Hyperbolic Systems with Nonlocal Potentials}. 2025. Preprint: \href{https://arxiv.org/abs/2507.05004}{	arXiv:2507.05004 [math.AP]}.
	\item[$\bullet$]S.~Murro and G.~Schmid. The Quantization of Maxwell Theory in the Cauchy Radiation Gauge: Hodge Decomposition and Hadamard States. {\em {Journal of the London Mathematical Society}}, 110(5), e70020, 2024. DOI:~\href{https://doi.org/10.1112/jlms.70020}{10.1112/jlms.70020}. Preprint:~\href{https://arxiv.org/abs/2401.08403}{arXiv:2401.08403 [math.AP]}.
\end{itemize}
To a much lesser extent, the thesis also draws on aspects of the following two publications:
\begin{itemize}
	\item[$\bullet$]L.~Andersson, B.~Moser, M.~A.~Oancea and C.~F.~Paganini and G.~Schmid: Pseudodifferential Weyl calculus on vector bundles. 2025. Preprint:~\href{https://arxiv.org/abs/2507.11965}{	arXiv:2507.11965 [math-ph]}.
	\item[$\bullet$]M.~Capoferri, S.~Murro and G.~Schmid. On boundary conditions for linearised Einstein's equations. \textit{Applied Mathematics Letters}, 158(109210), 2024. DOI:~\href{https://doi.org/10.1016/j.aml.2024.109210}{10.1016/j.aml.2024.109210}. Preprint:~\href{https://arxiv.org/abs/2407.07576}{arXiv:2407.07576 [math.AP]}.
\end{itemize}
 
\vspace*{1cm}

\begin{figure}[H]\includegraphics[width=0.3\textwidth,right]{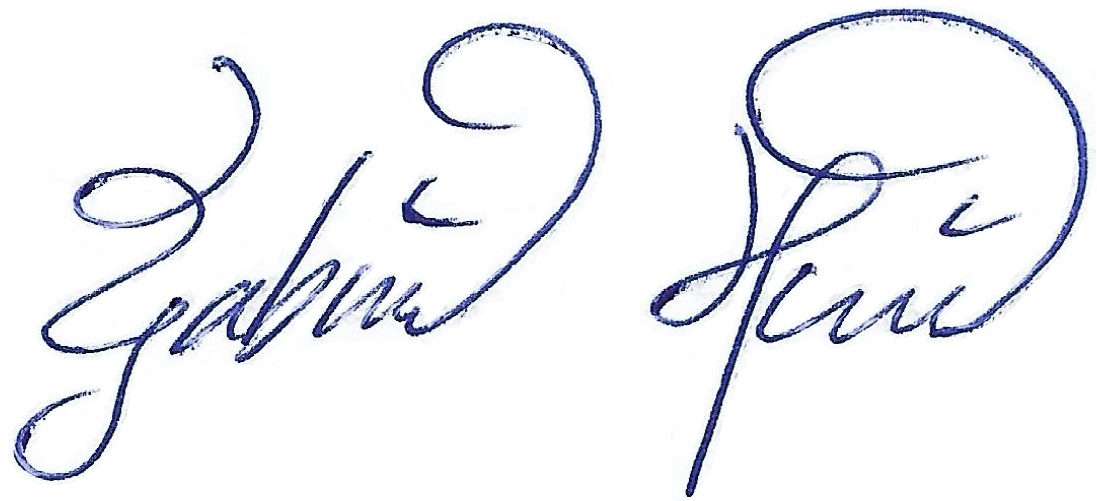} \end{figure}
Genoa, the 5$^{\mathrm{th}}$ of December 2025\hfill Gabriel Schmid\hspace*{10mm}
%
%
%
%
%
\newpage\null\newpage

\vspace*{3cm}
\begin{center}
\textit{To my family,}\\
\textit{my mother Renate with Markus,}\\
\textit{my father Bernhard,}\\
\textit{my brother Benjamin,}\\
\textit{and Flori,}\\
\textit{for all their overwhelming support throughout my life.}
\end{center}
\vfill
\setlength{\epigraphwidth}{2.7in}
\begin{minipage}[t]{0.5\textwidth}\raggedleft
\epigraph{\textit{``To those who do not know mathematics it is difficult to get across a real feeling as to the beauty, the deepest beauty, of nature. [${\dots}$]
If you want to learn about nature, to appreciate nature, it is necessary to understand the language that she speaks in.''}}{-- Richard P. Feynman \cite[p.~58]{FeynmanQuote}}
\end{minipage}
\begin{minipage}[t]{0.5\textwidth}\raggedleft
\epigraph{\textit{``In so far as theories of mathematics speak about reality, they are not certain, and in so far as they are certain, they do not speak about reality.''}}{-- Albert Einstein \cite[pp.~3-4]{EinsteinQuote}}
\end{minipage}
\begin{minipage}[t]{0.5\textwidth}\raggedleft
\epigraph{\textit{``The scientist does not study nature because it is useful to do so. He studies it because he takes pleasure in it, and he takes pleasure in it because it is beautiful.''}}{-- J. Henri Poincaré \cite[p.~22]{PoincareQuote}}
\end{minipage}
\begin{minipage}[t]{0.5\textwidth}\raggedleft
\epigraph{\textit{``The limit of man's knowledge in any subject possesses a high interest, which is perhaps increased by its close neighbourhood to the realms of imagination.''}}{-- Charles R.~Darwin \cite[p.~301]{DarwinQuote}}
\end{minipage}
\begin{minipage}[t]{0.5\textwidth}\raggedleft
\epigraph{\textit{``The more we learn about the world, and the deeper our learning, the more conscious, specific, and articulate will be our knowledge of \emph{what we do not know}, our knowledge of our ignorance.''}}{-- Karl R. Popper \cite[p.~28]{PopperQuote}}
\end{minipage}
\begin{minipage}[t]{0.5\textwidth}\raggedleft
\epigraph{\textit{``Science has fallen into many errors --  errors which have been fortunate and useful rather than otherwise, for they have been the steppingstones to truth.''}}{-- Prof.~Lidenbrock to Axel in Jules Verne’s\\ \textit{A Journey to the Center of the Earth} \cite{JulesVerne}}
\end{minipage}
\begin{minipage}[t]{0.5\textwidth}\raggedleft
\epigraph{\textit{``If you wish to make an apple pie from scratch, you must first invent the universe.''}}{-- Carl E.~Sagan \cite[p.~218]{SaganQuote}}
\end{minipage}
\begin{minipage}[t]{0.5\textwidth}\raggedleft
\epigraph{\textit{``This is the end of the beginning rather than the beginning of the end.''}}{-- Pierre Martinetti on writing a PhD thesis}
\end{minipage}
%
%
%
%
%
\newpage\null\newpage
\newpage
\textcolor{white}{ }
\vfill
\section*{Abstract}\vspace{-9pt}
This thesis is devoted to the study of hyperbolic differential operators on globally hyperbolic Lorentzian manifolds, linear field-theoretic models exhibiting a gauge symmetry, and their quantisation.

In the first part, we treat the Cauchy problem for symmetric hyperbolic systems and normally hyperbolic operators on globally hyperbolic manifolds from first principles, complemented by many examples and a discussion of Green hyperbolicity. Although hyperbolic equations are usually studied in the context of local interactions, there are strong motivations from several areas of mathematical physics to consider also nonlocal interactions. As an intermezzo, we therefore take a small deviation from the classical local theory and prove well-posedness of the Cauchy problem for symmetric hyperbolic systems coupled to a broad class of nonlocal potentials, with applications, for instance, to the Maxwell equations in linear dispersive media. 

The subsequent part presents a detailed exposition of linear gauge theories in globally hyperbolic spacetimes. Linear gauge theories are yet another deviation from the concept of hyperbolicity: the corresponding equations of motion are generically non-hyperbolic, however, can always be reduced to a constrained hyperbolic dynamics once an appropriate gauge fixing procedure has been applied. In particular, we give a thorough analysis of their Cauchy problems and construct the corresponding classical phase space. A central part of this section is the presentation of several examples of direct physical interest, some of which have not appeared yet in the literature in this context. Moreover, we explain how linear gauge theories are quantised following the algebraic approach to quantum field theory, which offers a mathematically rigorous quantisation scheme, ideally suited for field theories defined on Lorentzian backgrounds. In particular, we introduce the notion of Hadamard states, which are physically distinguished states whose two-point function has a specific singularity structure as a bidistribution, and provide a detailed bibliographic review of existence results for such states.

The final chapter of this thesis is devoted to the quantisation of Maxwell’s theory on globally hyperbolic spacetimes. After a detailed discussion of the Cauchy and gauge problems for Maxwell's theory on Lorentzian manifolds, the central goal is to prove the existence of Hadamard states for Maxwell's theory on any globally hyperbolic spacetime. The novelty of our approach lies in a new gauge-fixing procedure at the level of initial data, which allows us to suppress all the unphysical degrees of freedom. This gauge is achieved by means of a new Hodge decomposition for differential $k$-forms in Sobolev spaces on complete (possibly non-compact) Riemannian manifolds. Using tools from microlocal analysis, we explicitly construct Hadamard states on ultrastatic spacetimes for the completely gauge-fixed theory and define states obtained as the pull-back thereof along the gauge-fixing projector, while ensuring that this construction preserves the Hadamard property. For general spacetimes, we employ a deformation argument.

This thesis is complemented by an appendix that, alongside other supplementary materials, provides a self-contained introduction to microlocal analysis and pseudodifferential calculus.
%
%
%
%
%
\newpage
\textcolor{white}{ }
\vfill
\section*{Acknowledgements}\vspace{-9pt}
Much like any good proof in mathematics that is based on lemmata and proposition, this thesis was not built in isolation. It relies on many essential steps, both scientifically and personally, each contributed by the people I was, and still am, fortunate to learn from.

First and foremost, I would like to express my deepest gratitude to my supervisors Simone Murro and Matteo Capoferri for all their teachings, their patience and their unwavering support and guidance throughout this journey. Simone was my main supervisor in Genoa and I could not have asked for a better mentor and \emph{Sensei}. He invested an incredible amount of time, always finding a moment to discuss any doubts or questions, often leaving everything aside, no matter how busy he was. He supported me in every single step of my academic journey and gave me countless many opportunities. I am deeply grateful not only for his guidance and support, but also for his deep care that went far beyond what I could have expected. Whether through his broad interests and knowledge, or through his originality in approaching problems and many mathematical ideas, I learned an incredible amount from him, both in mathematics and about research in general. \textit{Grazie mille Simo!} 

I met Matteo during my three-month visit to the Heriot-Watt University in Edinburgh, and from the very first second he made me feel welcome and supported. With his gentle and genuinely kind soul, he always created an atmosphere in which I felt comfortable, encouraged, and inspired. He was constantly ready to help, whether with research questions, ideas, or simply with reassuring words when needed. His clarity, rigorous approach to science and the way in which he tackled every problem taught me a lot, both academically and personally. I am sincerely grateful for all his support, his patience, and last but certainly not least, all his academic advice and feedback on applications and related topics towards the end of my PhD. \textit{Grazie mille anche a te Matteo!}

I am not sure how these past three years in Genoa would have been without my best peers, fellow PhD students, and friends, Luca Fiorindo and Stefano Galanda. I was already virtually in touch with Stefano shortly before the start of our PhD. On the first day, the 2nd of November 2022, we decided to meet in front of the department. However, it was raining, so I let him know that I would rather wait inside. With his response, \textit{``Only a man with a degree in mathematical physics could have come up with such a clever decision''}, a wonderful friendship began. We met Luca on that very same day, and with his genuinely kind soul, unshakable positivity, and occasional singing in the office that echoed through the hallways, he would not only become a vital part of the department, but -- despite his algebras being \emph{commutative} -- also a wonderful friend for both of us. It was clear that nothing could separate the three of us, the very latest after our first hike on the 6th of November to the \emph{forti} around Genoa. Throughout the years, the \emph{(Superior-)Tirol, Südtirol, Veneto}-trio shared countless adventures, movie and gaming nights, holidays to each of our home towns, as well as constant and invaluable support both on an academic and personal level though up and downs during this journey, making even the toughest days memorable. Their friendship, humour, and unwavering encouragement were invaluable, both in my academic life and beyond. I will leave Genoa not only with  research experience gained, but also with many memories and with two friends for life. I could definitely not have asked for better companions for this journey and I am looking forward to the many years to come.

The community of PhD students in Genoa felt like a little family, bringing life to the department through many initiatives, such as the PhD seminars or informal welcoming meetings for new PhD students each year, but also through their friendship and positive energy. I was truly happy to be part of it. I want to especially mention Ludovica Buelli, Laura Carini, Edoardo D'Angelo, Emilien Ouvrard and Andrea Poggio, with whom I shared countless memorable moments filled with laughter, warmth, and kindness. I will always treasure them and they continue to brighten my life whenever I think back.

When I first arrived in Genoa in November 2022, I immediately felt at home and warmly welcomed by the wonderful members of the \emph{Mathematical Physics} research group, who, at various times over the past three years, included, Marco Benini, Edoardo D'Angelo, Nicolò Drago, Pietro Falzoni, Stefano Galanda, Roberta Iseppi, Pierre Martinetti, Simone Murro, Giorgio Musante, Gaston Nieuviarts, Emilien Ouvrard, Nicola Pinamonti, Andrea Pistolesi, Stefano Rosarin and Peem Ubonsri. I am deeply grateful to everyone for making these past three years such a memorable and joyful experience, through scientific discussions, our daily lunch and coffee breaks, and all the social dinners and other occasions we shared. Thank you for your kindness, support, and for creating such an inspiring environment. Moreover, I would like to extend my acknowledgements to the other department members I had the opportunity to interact with. In particular, I would like to mention Tommaso Bruno and Andrea Carbonaro for discussions and support.

During my time as a PhD student, I was very glad to have had the chance to travel extensively, both for conferences and research visits. In my second year, I spent three months at the Heriot-Watt University in Edinburgh, visiting Matteo Capoferri. I want to take this opportunity to thank the whole analysis group for their warm and kind hospitality. In this regards, I want especially express my gratitude to Lyonell Boulton for inspiring discussions at numerous occasions. In my final year, I had the opportunity to visit Felix Finster and his group in Regensburg. I felt immediately very welcome in his group, which at the time consisted of Marco van den Beld-Serrano, Patrick Fischer, Shane Farnsworth, and Claudio Paganini. I want to take this opportunity to thank Felix for his support and all the inspiring discussions, as well as his group for their warm hospitality and for the new friendships that made the visit truly unforgettable. Directly after Regensburg, I spent one month in Vienna working on a research-in-teams project together with Lars Andersson, Benjamin Moser, Marius Oancea, and Claudio Paganini. My time in Vienna was equally filled with countless enjoyable moments and stimulating discussions and I want to express my gratitude for the research team that made this possible. 

In addition, I undertook several shorter research visits, namely to Jan Sbierski in Edinburgh, as well as to the mathematical physics groups in Trento, in particular, to Nicolò Drago, Paolo Meda, Valter Moretti, Lorenzo Pettinari and Carmine De Rosa, and to Pavia, visiting Claudio Dappiaggi and his group consisting of Alberto Bonicelli, Beatrice Costeri, Paolo Rinaldi, Raman Deep Singh, and Luca Sinibaldi, among others. Thank you all for your hospitality, countless discussions and the personal kindness that made these visits truly memorable. Additionally, I had the chance to discuss with many other researchers, either at conferences or during their visits to Genoa, and I especially want to thank Christian Gérard and Nicolas Ginoux for inspiring discussions at several occasions.

Having attended many conferences, workshops, and summer schools over the past years, I have encountered many people I cannot mention individually. Nevertheless, I am deeply grateful to all of them for their support, inspiring conversations, and friendship throughout my PhD. I have made many friends along the way, and their encouragement and camaraderie have meant a great deal to me. I am always happy to meet them again at conferences, where seeing familiar faces over the years has made this community feel like a second family.

More broadly, I would also like to express my deep gratitude to the many people who supported my academic journey even before I began my PhD in Genoa. I would specifically like to mention my Bachelor thesis supervisor, Anita Reimer, as well as my Master thesis supervisors, Daniele Oriti and Christophe Goeller. Under their guidance, I gained my first research experience and all of them supported me wholeheartedly in my applications and academic career, ensuring that I could continue pursuing the path I wished to follow. At this stage, I would like to extend my gratitude to all the professors I have had the chance to learn from, and to the countless scientists who came before me, whose work I was able to study and build upon. We are truly standing on the shoulders of giants, and none of this work would have been possible as an individual.

I would like to sincerely thank the two referees of this thesis, Alexander Strohmaier and Michał Wrochna, for having kindly agreed to review this work, for their careful reading, and for their valuable suggestions and kind remarks. I would also like to express my gratitude, once again, to Christian Gérard, Nicola Pinamonti, and Michał Wrochna for having kindly agreed to be part of my defence committee.

Last, but certainly not least, I want to thank my family, to whom I dedicate this work, as well as all my friends in Tyrol and around the world for all their unwavering support during the last years of my academic career and in my life in general. In particular, I am deeply grateful to my mum, who has always been there for me at every stage of my life, offering unwavering support, encouragement, and love, and who continued to believe in me even when I did not.\bigskip

\hfill \textit{Grazie a tutti,} \textit{ein herzliches Dankeschön} \textit{\&} \textit{thank you all,}\\
\null\hfill Gabriel
%
%
%
%
{\pagestyle{plain}\hypersetup{linkcolor=black}
\tableofcontents}
\phantomsection
\chapter*{Conventions and Notations}\label{Conventions}
\addcontentsline{toc}{section}{\hspace{-16pt}\textbf{Conventions and Notations}}
\markboth{Conventions and Notations}{Conventions and Notations}

\fbox{\textbf{Basic Conventions:}}
\begin{itemize}
    \item[$\bullet$] The end of a proof is denoted by the \textit{tombstone} (or \emph{Halmos symbol}) \qedsymbol. The end of an example or a remark is denoted by a black triangle $\blacktriangle$. 
    \item[$\bullet$]The terms \textit{function} and \textit{map} will be used synonymously, as opposed to some authors, which use \textit{map} for specific types of functions (e.g. linear maps,${\dots}$).
\end{itemize}    

\vspace*{0.5cm}
\fbox{\textbf{Set Theory:}}
\begin{itemize}  
    \item[$\bullet$] Usually, we shall use the symbol $\bb{K}$ (from German \textit{Körper} for \textit{(algebraic) field}) when we specifically work with either $\bb{R}$ or $\bb{C}$, if not explicitly stated otherwise.
\item[$\bullet$] $\mathbb{N}:=\{1,2,\dots\}$ and $\mathbb{N}_{0}:=\mathbb{N}\cup\{0\}$.
\item[$\bullet$] If $\sf{X}$ is an arbitrary set, then we denote the identity function by $\mathrm{id}_{\sf{X}}\:\sf{X}\to\sf{X}, \, x\mapsto x$ or simply $\mathrm{id}$ if it is clear from the context which set is used.
\item[$\bullet$]$\sf{X}\subset \sf{Y} :\Leftrightarrow \forall x\in X:x\in\sf{Y}$, i.e.~$\sf{X}\subset\sf{Y}$ does also include the case $\sf{X}=\sf{Y}$. In other words, we do not distinguish between $\subseteq$ and $\subset$ unlike other authors. Instead, if we explicitly want to exclude the case $\sf{X}=\sf{Y}$, we use the notation $\sf{X} \subsetneq\sf{Y}$.
\item[$\bullet$]We will sometimes employ the \emph{multi-index notation}, i.e.~for $\alpha=(\alpha_{1},\dots,\alpha_{d})\in\bb{N}^{d}$ and $x=(x_{1},\dots,x_{d})\in\bb{R}^{d}$, $d\in\bb{N}$, we write
\begin{align*}
	\vert\alpha\vert:=\sum_{i=1}^{d}\alpha_{i}\,,\qquad \alpha!:=\prod_{i=1}^{d}\alpha_{i}!\,,\qquad x^{\alpha}:=x_{1}^{\alpha_{1}}\cdot\dots\cdot x_{d}^{\alpha_{d}}\,,\qquad \partial^{\alpha}:=\partial_{x_{1}}^{\alpha_{1}}\dots\partial_{x_{d}}^{\alpha_{d}}\, .
\end{align*}
\end{itemize}

\vspace*{0.5cm}
\fbox{\textbf{Differential and pseudo-Riemannian Geometry:}}
\begin{itemize}
    \item[$\bullet$]A \textit{manifold} is a second-countable locally Euclidean Hausdorff space and hence in particular regular, normal, paracompact and metrisable. Furthermore, all our manifolds are usually assumed to be smooth and without boundary if not explicitly stated otherwise.
    \item[$\bullet$]If $\sf{M}$ and $\sf{N}$ are two smooth manifolds and $f\in C^{\infty}(\sf{M},\sf{N})$ a smooth map, we denote its \emph{(total) differential} at a point $p\in\sf{M}$ by $\d_{p}f\:\sf{T}_{p}\sf{M}\to\sf{T}_{f(p)}\sf{N}$. 
    \item[$\bullet$]If $\sf{E}\xrightarrow{\pi}\sf{M}$ is a smooth vector bundle over $\sf{M}$, we denote the $C^{\infty}(\sf{M})$-module of smooth sections by $\Gamma^{\infty}(\sf{E})$. By $\Gamma^{k}(\sf{E})$ for $k\in\bb{N}_{0}$ we denote the module of $C^{k}$-sections. Additionally, we will use the following special notation for the modules of vector and covector fields:
    \begin{align*}
    		\mathfrak{X}(\sf{M}):=\Gamma^{\infty}(\sf{T}\sf{M}),\qquad \mathfrak{X}^{\ast}(\sf{M}):=\Gamma^{\infty}(\sf{T}^{\ast}\sf{M})\, .
    \end{align*}
    \item[$\bullet$]We denote by $\underline{\bb{K}}^{m}_{\sf{M}}:=\sf{M}\times\bb{K}^{m}$ the trivial $\bb{K}$-vector bundle of rank $m\in\bb{N}_{0}$ over $\sf{M}$.
    \item[$\bullet$]A \textit{Hermitian vector bundle} is a pair $(\sf{E},\langle\cdot,\cdot\rangle_{\sf{E}})$ consisting of a $\bb{C}$-vector bundle $\sf{E}\xrightarrow{\pi}\sf{M}$ and a bundle (pseudo-)metric $\langle\cdot,\cdot\rangle_{\sf{E}}\in\Gamma^{\infty}(\overline{\sf{E}}^{\ast}\otimes\sf{E}^{\ast})$, i.e.~a fibrewise non-degenerate Hermitian sesquilinear form $\langle\cdot,\cdot\rangle_{\sf{E}_{p}}\:\sf{E}_{p}\times\sf{E}_{p}\to\bb{C}$ (antilinear in the first argument) depending smoothly on the point $p\in\sf{M}$. Furthermore, if $\sf{M}$ is equipped with a pseudo-Riemannian metric $\sf{g}$, we use the notation
    \begin{align*}
    		(\psi,\varphi)_{\sf{E}}:=\int_{\sf{M}}\langle\psi_{p},\varphi_{p}\rangle_{\sf{E}_{p}}\,\d\mu_{\sf{g}}(p)\in\bb{C}
    \end{align*}
    for all $\psi,\varphi\in\Gamma^{\infty}(\sf{E})$ with $\mathrm{supp}(\psi)\cap\mathrm{supp}(\varphi)$ compact, where $\d\mu_{\sf{g}}$ denotes the pseudo-Riemannian volume measure of $(\sf{M},\sf{g})$. 
    \item[$\bullet$]For $r,s\in\bb{N}_{0}$, we denote the bundle of \textit{tensors of rank $(r,s)$} by $\sf{T}^{r,s}\sf{M}:=\sf{T}\sf{M}^{\otimes p}\otimes\sf{T}^{\ast}\sf{M}^{\otimes q}$.
    \item[$\bullet$]In this thesis, we usually employ \textit{Einstein's summation convention}, i.e.~repeated covariant and contravariant indices imply a summation over all their possible values.
    \item[$\bullet$]Let $(\sf{M},\sf{g})$ be a pseudo-Riemannian manifold. We denote the \emph{musical isomorphisms} as usual by $\sharp\:\sf{T}^{\ast}\sf{M}\to\sf{T}\sf{M}$ and $\flat\:\sf{T}\sf{M}\to\sf{T}^{\ast}\sf{M}$. These are vector bundle homomorphisms locally defined by $(\omega^{\sharp})^{\alpha}:=\sf{g}^{\alpha\beta}\omega_{\beta}$ for $\omega=\omega_{\alpha}\d x^{\alpha}\in\mathfrak{X}^{\ast}(\sf{M})$ and $(\sf{X}^{\flat})_{\alpha}:=\sf{g}_{\alpha\beta}\sf{X}^{\beta}$ for $\sf{X}=\sf{X}^{\alpha}\partial_{\alpha}\in\mathfrak{X}(\sf{M})$, where $\sf{g}^{\alpha\beta}$ denotes the (pointwise) inverse matrix of $\sf{g}_{\alpha\beta}:=\sf{g}(\partial_{\alpha},\partial_{\beta})$. Furthermore, we denote the induced bundle metric on $\sf{T}^{\ast}\sf{M}$ by
    	\begin{align*}
    		\sf{g}^{\sharp}\:\sf{T}^{\ast}_{p}\sf{M}\times\sf{T}^{\ast}_{p}\sf{M}\to\bb{R}\,,\qquad \sf{g}^{\sharp}(\omega,\eta):=\sf{g}(\omega^{\sharp},\eta^{\sharp})=\sf{g}^{\alpha\beta}\omega_{\alpha}\eta_{\beta}\, .
    	\end{align*}
    \item[$\bullet$]We shall use the following notation and sign conventions for the \textit{Riemann}, \textit{Ricci} and \textit{scalar curvature tensors} of some pseudo-Riemannian manifold ($\sf{M},\sf{g})$:
    \begin{align*}
    		[\nabla_{\alpha},\nabla_{\beta}]v^{\gamma}=\tensor{\mathrm{Riem}(\sf{g})}{_\alpha_\beta^\gamma_\delta}v^{\delta}\,,\quad \mathrm{Ric}(\sf{g})_{\alpha\beta}=\tensor{\mathrm{Riem}(\sf{g})}{^\lambda_\alpha_\lambda_\beta}\quad\text{and}\quad \mathrm{Scal}(\sf{g}):=\sf{g}^{\alpha\beta}\mathrm{Ric}(\sf{g})_{\alpha\beta}\, .
    \end{align*}
    These are the same conventions as in Choquet-Bruhat \cite{ChoquetBruhat}, Misner-Thorne-Wheeler \cite{MisnerThorneWheeler}, Ringström \cite{Ringstrom} and Wald \cite{Wald}. Note that the popular textbook of Lee~\cite{LeeRiemann} has a different convention for $\tensor{\mathrm{Riem}(g)}{_\alpha_\beta^\gamma_\delta}$, however, in the end the same convention for the Ricci tensor. In our conventions, the scalar curvature of the Riemannian sphere is positive-definite. In coordinates, our conventions imply the following expressions:
    \begin{align*}
    		\tensor{\mathrm{Riem}(\sf{g})}{_\alpha_\beta^\gamma_\delta}=\partial_{\alpha}\Gamma_{\beta\delta}^{\gamma}-\partial_{\beta}\Gamma_{\alpha\delta}^{\gamma}+\Gamma_{\alpha\lambda}^{\gamma}\Gamma_{\beta\delta}^{\lambda}-\Gamma_{\beta\lambda}^{\gamma}\Gamma_{\alpha\delta}^{\lambda}\\
    		\tensor{\mathrm{Ric}(\sf{g})}{_\alpha_\beta}=\partial_{\lambda}\Gamma_{\alpha\beta}^{\lambda}-\partial_{\alpha}\Gamma_{\beta\lambda}^{\lambda}+\Gamma_{\alpha\beta}^{\lambda}\Gamma_{\lambda\rho}^{\rho}-\Gamma_{\alpha\lambda}^{\rho}\Gamma_{\beta\rho}^{\lambda}
    \end{align*}
    \item[$\bullet$]Let $(\sf{M},\sf{g})$ be a pseudo-Riemannian manifold. For a $(0,k)$-tensor field $\sf{T}\in\Gamma^{\infty}(\sf{T}^{\ast}\sf{M}^{\otimes k})$, there is the well-known identity
    \begin{align}\label{eq:Identity}
    		[\nabla_{\alpha},\nabla_{\beta}]\sf{T}_{\mu_{1}\dots\mu_{k}}=\sum_{i}\tensor{\mathrm{Riem}(\sf{g})}{_\alpha_\beta_{\mu_{i}}^\lambda}\sf{T}_{\mu_{1}\dots\mu_{i-1}\lambda\mu_{i+1}\dots\mu_{k}}\, .
    \end{align}
    A fast way to prove this is to recall that there is a $C^{\infty}(\sf{M})$-module isomorphism
    \begin{align*}
    		\Gamma^{\infty}(\sf{T}^{\ast}\sf{M}^{\otimes k})\cong\mathfrak{X}^{\ast}(\sf{M})^{\otimes_{C^{\infty}(\sf{M})}k}
    \end{align*}
    allowing us to write any $\sf{T}\in\Gamma^{\infty}(\sf{T}^{\ast}\sf{M}^{\otimes k})$ as a finite linear combinations of tensors of the form $\sf{A}_{1}\otimes\dots\otimes\sf{A}_{k}$ for $\sf{A}_{i}\in\mathfrak{X}^{\ast}(\sf{M})$. The claim then follows from the a straightforward computation using the Leibniz rule and the defining relation for $\mathrm{Riem}(\sf{g})$.
    \item[$\bullet$]If $\sf{T}\in\Gamma^{\infty}(\sf{T}^{\ast}\sf{M}^{\otimes k})$ is a $(0,k)$-tensor field, we denote the symmetrisation and antisymmetrisation of indices by
    \begin{align*}
    		\sf{T}_{(\mu_{1}\dots\mu_{k})}:=\frac{1}{k!}\sum_{\sigma\in\mathfrak{S}_{k}}\sf{T}_{\sigma(\mu_{1})\dots\sigma(\mu_{k})}\,,\qquad \sf{T}_{[\mu_{1}\dots\mu_{k}]}:=\frac{1}{k!}\sum_{\sigma\in\mathfrak{S}_{k}}\mathrm{sign}(\sigma)\sf{T}_{\sigma(\mu_{1})\dots\sigma(\mu_{k})}\, ,
    \end{align*}
    where $\mathfrak{S}^{k}$ denotes the symmetric group of degree $k$. With our normalisation conventions, the (anti)symmetrisation of indices is hence an indempotent operation.
\end{itemize}

\vspace*{0.5cm}
\fbox{\textbf{Conventions and Notation for Differential Forms:}}\label{DiffForms}
\begin{itemize}
\item[$\bullet$]Let $\sf{M}$ be a smooth manifold of dimension $d\in\mathbb{N}$. Consider the \emph{$k$th exterior power} of the cotangent bundle, i.e.~the finite-rank $\bb{R}$-vector bundle
\begin{align*}
	\sf{A}_{k}:=\bigwedge^{k}\sf{T}^{\ast}\sf{M}
\end{align*}
whose space of sections is the $C^{\infty}(\sf{M})$-module of smooth differential $k$-forms on $\sf{M}$, i.e.\footnote{If $\mathscr{M}$ is a $\sf{R}$-module over a commutative ring $\sf{R}$, then we denote by $\mathrm{Alt}_{\sf{R}}^{k}(\mathscr{M},\sf{R})$ the $\sf{R}$-module of skew-symmetric multilinear maps of the form $f\:\underbrace{\mathscr{M}\times\dots\times\mathscr{M}}_{k \text{times}}\to\sf{R}$.}
\begin{align*}
	\Omega^{k}(\sf{M}):=\Gamma^{\infty}(\sf{A}_{k})\cong{\bigwedge^{k}}_{C^{\infty}(\sf{M})}\mathfrak{X}^{\ast}(\sf{M})\cong\mathrm{Alt}^{k}_{C^{\infty}(\sf{M})}(\mathfrak{X}(\sf{M}),C^{\infty}(\sf{M}))
\end{align*}
the $C^{\infty}(\sf{M})$-module of smooth $k$-forms. For the \textit{wedge product} $\wedge\:\Omega^{k}(\sf{M})\times\Omega^{l}(\sf{M})\to\Omega^{k+l}(\sf{M})$ we use the normalisation convention
\begin{align*}
	(\alpha\wedge\beta)(X_{1},\dots,X_{k+l}):=\frac{1}{k!l!}\sum_{\sigma\in\mathfrak{S}^{k+l}}\mathrm{sgn}(\sigma)\alpha(X_{\sigma(1)},\dots,X_{\sigma(k)})\beta(X_{\sigma(k+1)},\dots,X_{\sigma(k+l)})\, ,
\end{align*}
which is used in most modern textbooks such as \cite{LeeRiemann,LeeSmooth}. If $\omega\in\Omega^{k}(\sf{M})$ is a $k$-form, we write it in local coordinates $(\mathcal{U},\varphi=(x^{1},\dots,x^{d}))$ as
\begin{align*}
		\omega=\frac{1}{k!}\omega_{\mu_{1}\dots \mu_{k}}\mathrm{d}x^{\mu_{1}}\wedge\dots\wedge\mathrm{d}x^{\mu_{k}}=\sum_{0\leq \mu_{1}<\dots <\mu_{k}\leq d}\omega_{\mu_{1}\dots \mu_{k}}\mathrm{d}x^{\mu_{1}}\wedge\dots\wedge\mathrm{d}x^{\mu_{k}}\, ,
\end{align*}
where $\omega_{\mu_{1}\dots \mu_{k}}:=\omega(\partial_{\mu_{1}},\dots,\partial_{\mu_{k}})\in C^{\infty}(\mathcal{U})$ are the smooth coefficient functions. In particular, the wedge product for $\alpha\in\Omega^{k}(\sf{M})$ and $\beta\in\Omega^{l}(\sf{M})$ can locally be written as
\begin{align*}
	(\alpha\wedge\beta)_{\mu_{1}\dots\mu_{k+l}}=\frac{(k+l)!}{k!l!}\alpha_{[\mu_{1}\dots\mu_{k}}\beta_{\mu_{k+1}\dots\mu_{k+l}]}\, ,
\end{align*}
where we recall that $[\dots]$ is defined with a normalisation factor making it indempotent.
\item[$\bullet$]As usual, we define the \textit{exterior derivative} $\d\:\Omega^{k}(\sf{M})\to\Omega^{k+1}(\sf{M})$ by
\begin{align*}
	(\d\omega)(X_{1},\dots,X_{k+1})=&\sum_{i}(-1)^{i-1}X_{i}(\omega(X_{1},\dots,\hat{X_{i}},\dots,X_{k+1}))\\&+\sum_{i<j}(-1)^{i+j}\omega([X_{i},X_{j}],X_{1},\dots,\hat{X_{i}},\dots,\hat{X_{j}},\dots,X_{k+1})\\=&\sum_{i}^{k}(-1)^{i-1}(\nabla_{X_{i}}\omega)(X_{1},\dots,\hat{X}_{i},\dots,X_{k+1})
\end{align*}
for all $\omega\in\Omega^{k}(\sf{M})$ and $X_{1},\dots,X_{k+1}\in\mathfrak{X}(\sf{M})$, where the hat indicates that this entry is left out and where $\nabla$ is \emph{any} torsion-free connection on $\sf{T}^{\ast}\sf{M}$ ($\d$ is independent of that choice). In particular, in local coordinates we find
\begin{align*}
	(\d\omega)_{\mu_{1}\dots\mu_{k+1}}=(1+k)\nabla_{[\mu_{1}}\omega_{\mu_{2}\dots\mu_{k+1}]}=(1+k)\partial_{[\mu_{1}}\omega_{\mu_{2}\dots\mu_{k+1}]}\, .
\end{align*}
\item[$\bullet$]Now, let $(\sf{M},\sf{g})$ be an oriented pseudo-Riemannian manifold of dimension $d\in\mathbb{N}$ and arbitrary signature. We denote by $\mathrm{ind}(\sf{g})$ the dimension of the largest subspace of $\sf{T}\sf{M}$ on which $\sf{g}$ is \textit{negative} definite, i.e.~the number of minus' in the signature. Then, we set
\begin{align*}
	s:=(-1)^{\mathrm{ind}(\sf{g})}=\textit{``sign of the determinant of $\sf{g}$ in fixed but arbitrary coordinates''}\, .
\end{align*}
In particular, $s=1$ for Riemannian manifolds and $s=-1$ for Lorentzian manifolds with sign convention $(-,+,\dots,+)$. Now, we equip the vector bundle $\sf{A}_{k}$ with the bundle metric $\langle\cdot,\cdot\rangle_{\sf{A}_{k}}\in\Gamma^{\infty}(\sf{A}_{k}^{\ast}\otimes\sf{A}_{k}^{\ast})$ defined by
\begin{align*}
	\langle\alpha,\beta\rangle_{\sf{A}_{k}}:=\frac{1}{k!}(\sf{g}^{\sharp})^{\otimes k}(\alpha,\beta)=\frac{1}{k!}\alpha_{\mu_{1}\dots \mu_{k}}\beta^{\mu_{1}\dots \mu_{k}}=\sum_{0\leq\mu_{1}<\dots\mu_{k}\leq d}\alpha_{\mu_{1}\dots \mu_{k}}\beta^{\mu_{1}\dots \mu_{k}}\,,
\end{align*}
which gives rise to the $C^{\infty}(\sf{M})$-bilinear form $\langle\cdot,\cdot\rangle_{\sf{A}_{k}}\:\Omega^{k}(\sf{M})\times\Omega^{k}(\sf{M})\to C^{\infty}(\sf{M})$ on the level of sections. Furthermore, we define the \textit{Hodge $\ast$-operator} to be the linear operator $\ast:\Omega^{k}(\mathsf{M})\to\Omega^{d-k}(\mathsf{M})$ uniquely determined by the condition
\begin{align*}
	\alpha\wedge\ast\beta=\langle\alpha,\beta\rangle_{\sf{A}_{k}}\,\d\mu_{g}
\end{align*}
for all $\alpha,\beta\in\Omega^{k}(\sf{M})$, where $\d\mu_{\sf{g}}\in\Omega^{\mathrm{dim}(\sf{M})}(\sf{M})$ denotes the metric volume form $(\sf{M},\sf{g})$.\footnote{Note that the definition of $\ast$ depends on the choice of orientation of $(\sf{M},\sf{g})$ and the choice of the opposite orientation results affects the sign.} With this convention, the operator $\ast$ has the following properties:
\begin{align*}
\text{(i)}&\hspace*{1cm}\ast\ast=(-1)^{k(d-k)}s\hspace*{1cm}\text{acting on $\Omega^{k}(\sf{M})$}\\
\text{(ii)}&\hspace*{0.9cm}\ast^{-1}=(-1)^{k(d-k)}s\ast\hspace*{0.95cm}\text{acting on $\Omega^{k}(\sf{M})$}\\
\text{(iii)}&\hspace*{0.8cm}\ast \mathds{1}=\d\mu_{g}\\
\text{(iv)}&\hspace*{0.8cm}\ast\d\mu_{g}=s
\end{align*}
where $\mathds{1}\in C^{\infty}(\sf{M})$ denotes the constant function $\sf{M}\ni p\mapsto 1$. Next, we define the \textit{codifferential} to be the linear operator defined by
\begin{align*}
	\delta\:\Omega^{k}(\sf{M})\to\Omega^{k-1}(\sf{M}),\qquad \delta=(-1)^{d(k-1)+1}s\ast\mathrm{d}\ast=(-1)^{k}\ast^{-1}\mathrm{d}\ast\, .
\end{align*}
With the sign and normalisation conventions chosen, the codifferential is the formal adjoint of the exterior derivative with respect to the non-degenerate symmetric bilinear form
\begin{align*}
	(\alpha,\beta)_{\sf{A}_{k}}:=\int_{\sf{M}}\langle\alpha,\beta\rangle_{\sf{A}_{k}}\,\d\mu_{\sf{g}}=\int_{\sf{M}} \alpha\wedge\ast\beta
\end{align*} 
defined for all $\alpha,\beta\in\Omega^{k}(\sf{M})$ with compactly-overlapping supports, i.e.~it holds that $(\d\omega,\eta)_{\sf{A}_{k+1}}=(\omega,\delta\eta)_{\sf{A}_{k}}$ for all $\omega\in\Omega^{k}(\sf{M})$ and $\eta\in\Omega^{k+1}(\sf{M})$ with $\mathrm{supp}(\omega)\cap\mathrm{supp}(\eta)$ compact. In local coordinates, a straightforward computation\footnote{Instead of expressing the action of $\ast\d\ast$ on $\omega$ in local coordinates, it is in fact easier to compute $\delta$ directly using the explicit expression of $(\cdot,\cdot)_{\sf{A}_{k}}$ and $\d$ in coordinates. This way, the formula for $\delta$ becomes almost immediate.} shows that
\begin{align*}
(\delta\omega)_{\mu_{1}\dots\mu_{k-1}}=-g^{\alpha\beta}\nabla_{\alpha}\omega_{\beta\mu_{1}\dots\mu_{k-1}}=-\nabla^{\lambda}\omega_{\lambda\mu_{1}\dots\mu_{k-1}}.
\end{align*}
for all $\omega\in\Omega^{k}(\sf{M})$, where $\nabla$ denotes the Levi-Civita connection of $(\sf{M},\sf{g})$. Note that $\delta$, unlike the exterior derivative, depends on the choice of metric chosen.

Another important operation on differential forms is the \textit{interior product}, which is the $C^{\infty}(\sf{M})$-bilinear operation defined by
\begin{align*}
	\mathfrak{X}(\sf{M})\times\Omega^{k}(\sf{M})\to\Omega^{k-1}(\sf{M})\,,\qquad (X,\omega)\mapsto X\lrcorner\omega:=(i_{X}\omega):=\omega(X,\cdot,\dots,\cdot)\,,
\end{align*}
i.e.~the insertion of a vector field in the first slot of $\omega$, which is equivalently viewed as an element of $\mathrm{Alt}_{C^{\infty}(\sf{M})}^{k}(\mathfrak{X}(\sf{M}),C^{\infty}(\sf{M}))$. In local coordinates, it holds that 
\begin{align*}
	(X\lrcorner\omega)_{\mu_{1}\dots\mu_{k-1}}:=X^{\alpha}\omega_{\alpha\mu_{1}\dots\mu_{k-1}}\, .
\end{align*}

Last but not least, we mention that our conventions for the Hodge $\ast$-operator and codifferential agree with (John~M.)~Lee \cite{LeeSmooth,LeeRiemann} and Jost \cite{Jost} (which, however, only deal with the Riemannian case), as well as (Jeffrey~M.)~Lee \cite{LeeManifolds} and Nakahara \cite{Nakahara}. 

Within the mathematical gauge theory literature, our conventions coincide with Hamilton \cite{Hamilton}, but we stress that our convention for the Hodge $\ast$-operator differs by a factor of $(-1)^{s}$ from Rudolph-Schmidt~\cite{RudolphSchmidt2} and Baum~\cite{Baum}, since they define $\ast$ via $\alpha\wedge\ast\beta=(-1)^{s}\langle\alpha,\ast\beta\rangle_{\sf{A}_{k}}\d\mu_{\sf{g}}$. Of course, one obtains the same convention for $\delta$. 

\item[$\bullet$]Finally, let us review some definitions for bundle-valued differential forms. Let $\sf{E}\xrightarrow{\pi}\sf{M}$ be a finite-rank $\bb{R}$-vector bundle over $\sf{M}$ and consider the \emph{$\sf{E}$-twisted exterior bundle}
\begin{align*}
	\sf{E}_{k}:=\sf{E}\otimes\sf{A}_{k}=\sf{E}\otimes{\bigwedge^{k}}\sf{T}^{\ast}\sf{M}\,,
\end{align*}
whose space of sections is the $C^{\infty}(\sf{M})$-module of $\sf{E}$-valued smooth differential $k$-forms on $\sf{M}$, i.e.
\begin{align*}
	\Omega^{k}(\sf{M},\sf{E}):=\Gamma^{\infty}(\sf{E}_{k})\cong\Gamma^{\infty}(\sf{E})\otimes_{C^{\infty}(\sf{M})}\Omega^{k}(\sf{M})\cong\mathrm{Alt}^{k}_{C^{\infty}(\sf{M})}(\mathfrak{X}(\sf{M}),\Gamma^{\infty}(\sf{E}))\,,
\end{align*}
where we identify $\Omega^{0}(\sf{M},\sf{E})=\Gamma^{\infty}(\sf{E})$. If $\{e_{a}\}_{a=1,\dots,\mathrm{rank}_{\bb{R}}(\sf{E})}\subset\Gamma^{\infty}(\mathcal{U},\sf{E}\vert_{\mathcal{U}})$ is some local frame of $\sf{E}$ defined on some open set $\mathcal{U}\subset\sf{M}$, then every $\omega\in\Omega^{k}(\sf{M},\sf{E})$ can be locally written as $\omega=\omega^{a}\otimes e_{a}$ for coefficient forms $\omega^{a}\in\Omega^{k}(\mathcal{U})$. 

Now, every connection $\nabla^{\sf{E}}$ on $\sf{E}$ can naturally be viewed as a linear operator of the form $\nabla^{\sf{E}}\:\Omega^{0}(\sf{M},\sf{E})\to\Omega^{1}(\sf{M},\sf{E})$ and similarly to the situation of $\bb{R}$-valued forms, in which the total differential $\d\:\Omega^{0}(\sf{M})\to\Omega^{1}(\sf{M})$, $\d f(X):=X(f)$, gives rise to a family of operators $\d\:\Omega^{k}(\sf{M})\to\Omega^{k+1}(\sf{M})$, any connection $\nabla^{\sf{E}}$ defines a family of operators $\d^{\nabla^{\sf{E}}}\:\Omega^{k}(\sf{M},\sf{E})\to\Omega^{k+1}(\sf{M},\sf{E})$ defined by
\begin{align*}
	(\d^{\nabla^{\sf{E}}}\omega)(X_{1},\dots,X_{k+1})=&\sum_{i}(-1)^{i-1}\nabla_{X_{i}}^{\sf{E}}(\omega(X_{1},\dots,\hat{X_{i}},\dots,X_{k+1}))\\&+\sum_{i<j}(-1)^{i+j}\omega([X_{i},X_{j}],X_{1},\dots,\hat{X_{i}},\dots,\hat{X_{j}},\dots,X_{k+1})
\end{align*}
for all $\omega\in\Omega^{k}(\sf{M},\sf{E})$ and $X_{1},\dots,X_{k+1}\in\mathfrak{X}(\sf{M})$, called the \textit{exterior covariant derivative}. In fact, the exterior derivative for $\bb{R}$-valued forms is a special case of this construction, since the total derivative of functions, $\d f(X):=X(f)$, defines a connection on the trivial line bundle $\underline{\bb{R}}_{\sf{M}}:=\bb{R}\times\sf{M}$, called the \textit{trivial flat connection}. It is important to stress that the exterior covariant does \textit{not} satisfy the property $\d^{\nabla^{\sf{E}}}\circ\d^{\nabla^{\sf{E}}}=0$. In fact, it holds that
\begin{align*}
	\d^{\nabla^{\sf{E}}}\d^{\nabla^{\sf{E}}}\omega=\sf{F}^{\nabla^{\sf{E}}}\wedge_{\Phi}\omega
\end{align*}
for all $\omega\in\Omega^{k}(\sf{M},\sf{E})$, where $\sf{F}^{\nabla^{\sf{E}}}\in\Omega^{2}(\sf{M},\mathrm{End}(\sf{E}))$ denotes the \textit{curvature $2$-form} of the bundle $(\sf{E},\nabla^{\sf{E}})$ defined by 
\begin{align*}
	\sf{F}^{\nabla^{\sf{E}}}(X,Y,\psi):=\nabla_{X}^{\sf{E}}\nabla^{\sf{E}}_{Y}\psi-\nabla_{Y}^{\sf{E}}\nabla^{\sf{E}}_{X}\psi-\nabla^{\sf{E}}_{[X,Y]}\psi
\end{align*}
for all $X,Y\in\mathfrak{X}(\sf{M})$, $\psi\in\Gamma^{\infty}(\sf{E})$ and where $\wedge_{\Phi}\:\Omega^{k}(\sf{M},\mathrm{End}(\sf{E}))\times\Omega^{l}(\sf{M},\sf{E})\to\Omega^{k+l}(\sf{M},\sf{E})$ is the wedge-type product induced from the natural pairing $\Phi_{p}\:\mathrm{End}(\sf{E}_{p})\times\sf{E}_{p}\to\sf{E}_{p}$, $p\in\sf{M}$. Furthermore, let us mention the following useful fact: the connection $\nabla^{\sf{E}}$ induces a connection on the bundle $\mathrm{End}(\sf{E})$ defined by means of the Leibniz rule, i.e.
\begin{align*}
	\nabla_{X}^{\sf{E}}(\Phi(\varphi))=(\nabla_{X}^{\mathrm{End}(\sf{E})}\Phi)\varphi+\Phi(\nabla_{X}^{\sf{E}}\varphi)
\end{align*}
for all $X\in\mathfrak{X}(\sf{M})$, $\Phi\in\Gamma^{\infty}(\mathrm{End}(\sf{E}))\cong\mathrm{End}_{C^{\infty}(\sf{M})}(\Gamma^{\infty}(\sf{E}))$ and $\varphi\in\Gamma^{\infty}(\sf{E})$. Then, the curvature $\sf{F}^{\nabla^{\sf{E}}}\in\Omega^{2}(\sf{M},\mathrm{End}(\sf{E}))$ of $(\sf{E},\nabla^{\sf{E}})$ satisfies
\begin{align*}
	\d^{\nabla^{\mathrm{End}(\sf{E})}}\sf{F}^{\nabla^{\sf{E}}}=0\, .
\end{align*}
This equation is called the differential \textit{Bianchi identity} and reduces to the \textit{second Bianchi identity} of pseudo-Riemannian geometry in the special case of the tangent bundle.

Now, we define the Hodge $\ast$-operator $\ast\:\Omega^{k}(\sf{M},\sf{E})\to\Omega^{d-k}(\sf{M},\sf{E})$ for $\sf{E}$-valued form locally by $\ast\omega:=(\ast\omega^{a})\otimes e_{a}$. It is easy to check that this is independent of the chosen local frame. Furthermore, this operator can be characterised in a manner analogous to the $\mathbb{R}$-valued case: first of all, we equip the bundles $\sf{E}_{k}$ with the natural bundle metric
\begin{align*}
	\langle\omega,\eta\rangle_{\sf{E}_{k}}:=\langle\omega^{a},\eta^{b}\rangle_{\sf{A}_{k}}\langle e_{a},e_{b}\rangle_{\sf{E}}\,,
\end{align*}
where we wrote $\omega=\omega^{a}\otimes e_{a}$ and $\eta=\eta^{a}\otimes e_{b}$. Furthermore, let us choose a non-degenerate bundle metric $\langle\cdot,\cdot\rangle_{\sf{E}}$ on $\sf{E}$ and let us consider the natural wedge-type product $\langle\cdot\wedge\cdot\rangle_{\sf{E}}\:\Omega^{k}(\sf{M},\sf{E})\times\Omega^{l}(\sf{M},\sf{E})\to\Omega^{k+l}(\sf{M})$ induced from $\langle\cdot,\cdot\rangle_{\sf{E}}$. Then, it holds that
\begin{align*}
	\langle\omega\wedge\ast\eta\rangle_{\sf{E}}=\langle\alpha,\beta\rangle_{\sf{E}_{k}}\,\d\mu_{\sf{g}}\,,
\end{align*}
generalising the $\bb{R}$-valued case. With this notation, we define the corresponding non-degenerate symmetric bilinear form
\begin{align*}
	(\omega,\eta)_{\sf{E}_{k}}:=\int_{\sf{M}}\langle\omega,\eta\rangle_{\sf{E}_{k}}\,\d\mu_{\sf{g}}=\int_{\sf{M}}\langle\omega\wedge\ast\eta\rangle_{\sf{E}}
\end{align*}
for all $\omega,\eta\in\Omega^{k}(\sf{M},\sf{E})$ with compactly-overlapping supports. If the connection $\nabla^{\sf{E}}$ is compatible with the metric $\langle\cdot,\cdot\rangle_{\sf{E}}$, then the formal adjoint of $\d^{\nabla^{\sf{E}}}$ with respect to $(\cdot,\cdot)_{\sf{E}_{k}}$, the \textit{covariant codifferential}, is given by
\begin{align*}
	\delta^{\nabla^{\sf{E}}}\:\Omega^{k}(\sf{M},\sf{E})\to\Omega^{k-1}(\sf{M},\sf{E})\,,\qquad \delta=(-1)^{d(k-1)+1}s\ast\mathrm{d}^{\nabla^{\sf{E}}}\ast=(-1)^{k}\ast^{-1}\mathrm{d}^{\nabla^{\sf{E}}}\ast\, ,
\end{align*}
generalising the analogous formula in the case of $\bb{R}$-valued differential forms.
\end{itemize}
\mainmatter 
\pagestyle{fancy}
\fancyhead[RE]{\nouppercase{\leftmark}}
\fancyhead[LO]{\nouppercase{\rightmark}}
\fancyhead[RO,LE]{\thepage}
\fancyfoot[C]{}
\phantomsection
\chapter*{Introduction}
\addcontentsline{toc}{section}{\hspace{-16pt}\textbf{Introduction}}
\markboth{Introduction}{Introduction}

Hyperbolic partial differential equations play a fundamental role in mathematical physics, as they model many physical phenomena, particularly those involving \emph{wave propagation}, \emph{causality}, and \emph{finite propagation speed of information}.\footnote{For the sake of readability, this introductory chapter, kept at a general level, contains only minimal citations. A detailed bibliography with pointers to the literature is provided in the following chapters.} Their mathematical structure allows for the formulation of a well-posed \emph{initial value problem}, meaning that suitable initial data assigned at a fixed moment in time uniquely determine the evolution of the system in a stable manner. At the same time, hyperbolicity permits the construction of so-called \emph{advanced} and \emph{retarded Green operators}, which are fundamental solutions encoding the \emph{causal} propagation of disturbances. These features, in turn, are indispensable for the formulation of \emph{relativistic field theories}, where the causal structure of the underlying Lorentzian spacetime imposes strict constraints on how information may propagate. As a result, hyperbolic equations arise naturally in a wide range of \emph{classical} and \emph{quantum field theories}, such as the Dirac and Klein–Gordon equations.

Besides these basic examples, there is another fundamental class of field-theoretic models, which are mathematically different in nature, namely \emph{gauge theories}, characterised by a dynamics that is constrained by the invariance under a \emph{local} symmetry transformation. Being responsible for mediating the fundamental forces of nature via interactions with charged particles, those kind of theories play a major role in our modern understanding of the universe. Examples range from Maxwell's theory of electromagnetism, the celebrated Yang-Mills theory, with applications both to elementary particle physics and pure mathematics, from the Einstein equations of general relativity, to Chern-Simons theory with applications, for instance, in condensed matter physics. In particular, gauge theories are of indispensable importance in both of the two cornerstones of modern mathematical physics, namely \emph{Einstein's theory of general relativity}, the basic theory underlying the \emph{standard model of cosmology}, as well as the \emph{standard model of particle physics}, which lies within the realm of (relativistic) \emph{quantum field theory}. 

The presence of a \emph{gauge symmetry}, however, implies that the associated equations of motion are generically \emph{non}-hyperbolic in nature and hence fail to yield a well-posed Cauchy problem. In particular, without further modifications, one cannot construct advanced and retarded Green operators, objects that, as mentioned above, are indispensable for the analytic treatment of wave equations and for the construction of quantum field theories on curved Lorentzian spacetimes. This difficulty is typically resolved by imposing a suitable \emph{gauge-fixing condition}, such as the \emph{Lorenz gauge} in Maxwell's theory or the \emph{harmonic/de Donder gauge} in general relativity and linearised gravity. Upon fixing a suitable gauge, the equations of motion acquire a hyperbolic character, thereby restoring well-posedness and enabling the definition of the relevant Green operators. At the same time, the physical content of the classical theory is encoded in \emph{gauge-invariant observables}, ensuring that the resulting description remains consistent with the underlying symmetry principles and gauge redundancies. 

In the setting of \emph{linear} field-theoretic models, an abstract and axiomatic formalism for gauge theories on globally hyperbolic Lorentzian manifolds, containing essentially all the examples of interest for physical applications, has been initially developed by Hack-Schenkel \cite{HackSchenkel} and further refined by Gérard-Wrochna \cite{GerardWrochna}. Within this formalism, the dynamics of a linear gauge theory is described by two linear differential operators
\begin{align*}
	\sf{P}\:\Gamma^{\infty}(\sf{E}_{2})\to\Gamma^{\infty}(\sf{E}_{2})\,,\qquad \sf{K}\:\Gamma^{\infty}(\sf{E}_{1})\to\Gamma^{\infty}(\sf{E}_{2})\,,
\end{align*}
acting on smooth sections of the Hermitian vector bundles $(\sf{E}_{1},\langle\cdot,\cdot\rangle_{\sf{E}_{1}})$ and $(\sf{E}_{2},\langle\cdot,\cdot\rangle_{\sf{E}_{2}})$ defined over a globally hyperbolic Lorentzian manifold $(\sf{M},\sf{g})$, with the property that $\sf{P}\circ\sf{K}=0$. The operator $\sf{P}$, assumed to be formally self-adjoint, parametrises the dynamical equations of motion of the gauge theory in question, while the operator $\sf{K}$ encodes the gauge transformations via $\Gamma^{\infty}(\sf{E}_{2})\ni\psi\mapsto\psi+\sf{K}\omega$ for $\omega\in\Gamma^{\infty}(\sf{E}_{1})$. By definition, these transformations leave the equations of motion invariant, i.e.~$\sf{P}\psi=\sf{P}(\psi+\sf{K}\omega)$. One further assumes that the two operators
\begin{align*}
	\sf{D}_{1}:=\sf{K}^{\ast}\sf{K}\:\Gamma^{\infty}(\sf{E}_{1})\to\Gamma^{\infty}(\sf{E}_{1})\,,\qquad\sf{D}_{2}:=\sf{P}+\sf{K}\sf{K}^{\ast}\:\Gamma^{\infty}(\sf{E}_{2})\to\Gamma^{\infty}(\sf{E}_{2})
\end{align*}
are hyperbolic. The operators $\sf{D}_{2}$ is the \emph{gauge-fixed} operator with respect to the canonical \emph{gauge condition} $\sf{K}^{\ast}\psi=0$ for $\psi\in\Gamma^{\infty}(\sf{E}_{2})$, where $\sf{K}^{\ast}$ denotes the formal adjoint with respect to $(\cdot,\cdot)_{\sf{E}_{i}}=\int_{\sf{M}}\langle\cdot,\cdot\rangle_{\sf{E}_{i}}\,\d\mu_{\sf{g}}$, while the role of $\sf{D}_{1}$ is to parametrise the \emph{remaining gauge freedom} in the sense that a gauge transformation $\psi\mapsto\psi+\sf{K}\omega$ with $\omega\in\Gamma^{\infty}(\sf{E}_{1})$ preserves the condition $\sf{K}^{\ast}\psi=0$ if and only if $\omega\in\mathrm{ker}(\sf{D}_{1})$. Hyperbolicity of $\sf{D}_{1}$ ensures that the space of remaining gauge freedom is well-behaved and, moreover, guarantees that the canonical gauge condition can always be achieved. Indeed, for every $\psi\in\Gamma^{\infty}(\sf{E}_{2})$, finding a gauge transformation $\omega\in\Gamma^{\infty}(\sf{E}_{1})$ such that $\sf{K}^{\ast}(\psi+\sf{K}\omega)=0$ is equivalent to finding a solution to the hyperbolic equation $\sf{D}_{1}\omega=-\sf{K}^{\ast}\omega$.

Now, as already mentioned above, the operator $\sf{P}$ of a (non-trivial, i.e.~$\sf{K}\neq 0$) linear gauge theory is generically non-hyperbolic. Indeed, the presence of a gauge symmetry implies that solutions to the equation $\sf{P}\psi=\phi$ for a given source $\phi$ and initial data can never be unique, if they exist, as one can always perform a gauge transformation to obtain another, distinct, solution. Nevertheless, thanks to the gauge symmetry present, one can always reduce the discussion to the analysis of the hyperbolic operator $\sf{D}_{2}$. Indeed, from a physics perspective, we are interested in the quotient space in which we identify two solutions of $\sf{P}\psi=0$ whenever they differ by a gauge transformation and, continuing the above discussion, this space of \emph{gauge equivalent classes} can be identified with a suitable quotient space of solutions to the hyperbolic problem $\sf{D}_{2}\psi=0$ with the gauge constraints $\sf{K}^{\ast}\psi=0$. Schematically, we have
\begin{align*}
	\cfrac{\mathrm{ker}(\sf{P}\vert_{\Gamma^{\infty}_{\mathrm{sc}}})}{\mathrm{ran}(\sf{K}\vert_{\Gamma^{\infty}_{\mathrm{sc}}})}\cong\cfrac{\mathrm{ker}(\sf{D}_{2}\vert_{\Gamma^{\infty}_{\mathrm{sc}}})\cap\mathrm{ker}(\sf{K}^{\ast}\vert_{\Gamma^{\infty}_{\mathrm{sc}}})}{\sf{K}(\mathrm{ker}(\sf{D}_{1}\vert_{\Gamma^{\infty}_{\mathrm{sc}}}))}\, ,
\end{align*}
where the denominator of the space on the right-hand side describes the remaining gauge freedom present, as explained above. A corresponding Cauchy problem can then be formulated by restricting the class of initial data to those evolving to solutions $\psi$ satisfying the canonical gauge condition $\sf{K}^{\ast}\psi=0$ and by defining a suitable notion of gauge transformation on the level of initial data to account for the remaining gauge freedom.

From a physical perspective, we are interested in studying the \emph{classical phase space}, that is, the space of \emph{linear, on-shell, gauge-invariant observables}. In the context of the axiomatic formalism presented above, this is the space
\begin{align*}
	\mathcal{V}_{\mathrm{c}}:=\cfrac{\mathrm{ker}(\sf{K}^{\ast}\vert_{\Gamma^{\infty}_{\mathrm{c}}})}{\mathrm{ran}(\sf{P}\vert_{\Gamma^{\infty}_{\mathrm{c}}})}\,,\qquad\sigma\:\mathcal{V}_{\mathrm{c}}\times\mathcal{V}_{\mathrm{c}}\to\bb{C}\,,\qquad\sigma([\psi],[\phi]):=i\int_{\sf{M}}\langle\psi,\sf{G}_{2}\phi\rangle_{\sf{E}_{2}}\,\d\mu_{\sf{g}}\,,
\end{align*}
where $\sf{G}_{2}\:\Gamma^{\infty}_{\mathrm{c}}(\sf{E}_{2})\to \Gamma^{\infty}_{\mathrm{sc}}(\sf{E}_{2})$ denotes the \emph{causal propagator} of $\sf{D}_{2}$ (see Definition~\ref{Def.CausProp}), i.e.~the difference of the retarded and advanced Green operators. Indeed, every element $[\psi]\in\mathcal{V}_{\mathrm{c}}$ determines a corresponding linear observable by setting
\begin{align*}
	\mathcal{O}_{[\psi]}\:\mathrm{ker}(\sf{P}\vert_{\Gamma^{\infty}_{\mathrm{sc}}})\to\bb{C}\,,\qquad \mathcal{O}_{[\psi]}(\phi):=\int_{\sf{M}}\langle\psi,\phi\rangle_{\sf{E}_{2}}\,\d\mu_{\sf{g}}\,,
\end{align*}
which is gauge-invariant in the sense that $\mathcal{O}_{[\psi]}(\phi+\sf{K}\omega)=\mathcal{O}_{[\psi]}(\phi)$ for all $\omega\in\Gamma^{\infty}(\sf{E}_{1})$. The sesquilinear form $\sigma$ equips $\mathcal{V}_{\mathrm{c}}$ with a (in general degenerate) Hermitian structure.

Having understood the classical theory, the next step consists of \emph{quantisation}. Within this context, we employ the \emph{algebraic approach} to quantum field theory, which offers a mathematically rigorous quantisation scheme, ideally suited for describing field theories on curved Lorentzian backgrounds. In this framework, quantisation consists of a two-step procedure:
\begin{itemize}
	\item[(i)]First, one assigns a unital $\ast$-algebra $\mathrm{CCR}(\mathcal{V}_{\mathrm{c}},\sigma)$ of observables to the classical phase space $(\mathcal{V}_{\mathrm{c}},\sigma)$, which encodes structural
properties such as causality, dynamics, and the \emph{canonical commutation relations} $[\Phi(v),\Phi^{\ast}(w)]=\sigma(v,w)\mathds{1}$. Here the \emph{quantum fields} $\Phi(v),\Phi^{\ast}(v)$ with $v\in\mathcal{V}_{\mathrm{c}}$ are the generators of the algebra, and $\mathds{1}$ is the algebra unit.
	\item[(ii)]Next, one identifies a \emph{physical state} $\omega$, namely a positive, linear and normalised functional on the algebra $\mathrm{CCR}(\mathcal{V}_{\mathrm{c}},\sigma)$. Among the plethora of mathematically well-defined states, not all of them can be considered physical and hence, one usually considers only those satisfying the so-called \emph{Hadamard condition}. This condition ensures the correct short-distance behaviour of the $n$-point functions, mimicking the ultraviolet singularity structure of the \emph{Minkowski vacuum}, and plays a key role in the perturbative approach to quantum field theory. Indeed, it implies the finiteness of the quantum fluctuations of the expectation value of every observable, and it allows one to construct Wick polynomials and other nonlinear observables.
\end{itemize}

The Hadamard condition can be best understood by employing tools from \emph{microlocal analysis}. Indeed, already since the appearance of the seminal paper by Duistermaat-Hörmander in the 1970s on the study of partial differential equations via the wavefront set \cite{DuistermaatHormander}, it has become increasingly clear that \emph{microlocal analysis} is an indispensable tool for formulating quantum field theories in a mathematically rigorous way. On the one hand, microlocal analysis provides powerful techniques for solving the partial differential equations governing the dynamics of quantum field theory. On the other hand, it offers practical criteria for the existence of products of quantum observables, viewed as Hilbert space-valued distributions, thus enabling a rigorous formulation of Wick polynomials. In the 1990s, the seminal work of Radzikowski \cite{Radzikowski1,Radzikowski2} brought microlocal analysis to the \textit{center stage} of quantum field theory in curved spacetime by showing that the Hadamard condition can be reformulated using the wavefront set. In particular, in a modern formulation, the Hadamard condition is equivalent to the requirement
\begin{align*}
	\mathrm{WF}^{\prime}(\lambda^{\pm})\subset\mathcal{N}^{\pm}\times\mathcal{N}^{\pm}\,,
\end{align*}
where $\mathcal{N}^{\pm}$ are the \emph{positive/negative energy cones}, i.e.~the two connected components of the light cone in $\sf{T}^{\ast}\sf{M}$, and where $\lambda^{\pm}\:\Gamma^{\infty}_{\mathrm{c}}(\sf{E}_{2})\to\Gamma^{\infty}(\sf{E}_{2})$ are the distributional kernels of the $2$-point functions, i.e.~
\begin{align*}
	(\psi,\lambda^{+}\phi)_{\sf{E}_{2}}=\omega(\Phi([\phi])\Phi^{\ast}([\psi]))\,,\qquad (\psi,\lambda^{-}\phi)_{\sf{E}_{2}}=\omega(\Phi([\psi])\Phi^{\ast}([\phi]))\, .
\end{align*}
The \emph{wavefront set} $\mathrm{WF}(\sf{T})\in\sf{T}^{\ast}\sf{M}\backslash\{\textbf{0}\}$ of a distribution $\sf{T}$ is a refinement of the singular support of $\sf{T}$ that includes additional informations on the singular directions in Fourier space. In particular, the above definition makes precise the idea of \emph{``states with the same singular ultraviolet structure of the Minkowski vacuum''}.

Remarkably, following earlier works of Junker-Schrohe \cite{Junker1,Junker2,JunkerSchrohe}, Gérard-Wrochna demonstrated that Hadamard states can be constructed using pseudodifferential techniques on general curved spacetimes for Klein-Gordon fields \cite{GerardWrochnaPseudo}. The construction of Hadamard states for linear field theories that possess a gauge symmetry, such as Maxwell's theory or linearised gravity, however, presents much more subtle challenges, mainly due to the following reasons:
\begin{itemize}
	\item[(i)]First, as already indicated above, the presence of gauge symmetry implies that the differential operator governing the dynamics of a gauge theory is generically non-hyperbolic.
	\item[(ii)]Secondly, while the fibre metric on the space of solutions is typically non-degenerate, it is not, in general, positive-definite. Indeed, in essentially all the linear gauge theories coming from physical applications, such as Maxwell's theory, linearised Yang-Mills theory and linearised gravity, the fibre metric $\langle\cdot,\cdot\rangle_{\sf{E}_{2}}$ is usually constructed out of the Lorentzian metric $\sf{g}$ itself and hence Lorentzian in nature.
	\item[(iii)]Linear gauge theories often suffer from infrared problems. On ultrastatic spacetimes, the construction of Hadamard states is usually reduced to the construction of projections onto the subspace of positive frequency solutions. For \emph{massive} theories, this, in turn, is achieved as soon as the operator $\sqrt{\Delta+m^{2}}$ and its inverse are well-defined. For massless theories ($m=0$), however, there are a priori obstructions to the construction of these projector, since the Laplacian $\Delta$ is in general strictly non-positive. For ordinary field theories, such as massless Klein-Gordon theory, this problem is usually overcome by employing techniques from pseudodifferential calculus. More precisely, one aims to construct a suitable \emph{smoothing operator} $r_{-\infty}$ for which $\sqrt{\Delta+r_{-\infty}}$ consistently can be defined. Being a smoothing operator, this does not spoil the Hadamard property, however, in the case of gauge theories, it is in conflict with \emph{gauge invariance}, which requires the corresponding projections to commute with the operator $\sf{K}$ that parametrises the gauge transformations.
\end{itemize}
How to tackle the first issue is well-understood, namely by means of a suitable gauge-fixing procedure, as explained above. The second issue, namely the non-positivity of the fibre metric, however, is much more intricate and leads to a conflict between the Hadamard condition and the positivity of states, while infrared problem provide further complications to be taken into account. While it is well-known how to construct bidistributions satisfying the Hadamard condition, ensuring the positivity and gauge-invariance of states in linear gauge theories hence remains a highly non-trivial task at the forefront of current research. 

This thesis is devoted to the study of hyperbolic operators on Lorentzian manifolds as well as of linear gauge theories and their quantisation. We now provide a detailed overview of the main results as well as the overall content and structure of this doctoral thesis.

\section*{Content, Main Results and Organisation of this Thesis}
The thesis is organised into three chapters, devoted to the study of hyperbolic differential operators on Lorentzian manifolds and their Cauchy problem, a detailed analysis of linear gauge theories and their quantisation, as well as to the explicit quantisation of Maxwell's theory on general globally hyperbolic Lorentzian manifolds. Each chapter is divided into three sections and is supported by the necessary preliminary and expositional material, ensuring that the presentation is entirely self-contained. Moreover, the thesis is supplemented by an appendix containing, besides some additional technical results and auxiliary material, a self-contained exposition on microlocal analysis and pseudodifferential calculus.\bigskip

\underline{\textbf{{Chapter}~\ref{Chap:Hyp}.}} The first chapter of this thesis is devoted to the study of \emph{(linear) hyperbolic partial differential operators} on globally hyperbolic Lorentzian manifolds, thereby providing the essential analytic foundations for the investigation of  relativistic field theories. Section~\ref{Sec:Lorentzian} begins with a brief review of the basic notions of Lorentzian geometry. In particular, we discuss the notion of \emph{globally hyperbolic spacetimes}, complemented by several physically relevant examples, which constitute a particularly well-suited class of backgrounds for formulating and analysing the Cauchy problem for hyperbolic partial differential equations. In particular, we review the theorem of Bernal-Sánchez (Theorem~\ref{Thm:BernalSanchez}), which states that any such manifold takes the form
\begin{align*}
	\sf{M}=\bb{R}\times\Sigma\,,\qquad\sf{g}=-\beta^{2}\d t\otimes\d t+\sf{h}_{t}\,,
\end{align*}
where $\Sigma$ is a smooth and spacelike Cauchy hypersurface, $\beta\in C^{\infty}(\sf{M},(0,\infty))$ the lapse function and $(\sf{h}_{t})_{t\in\bb{R}}$ a one-parameter family of Riemannian metrics on $\bb{R}$.

In the subsequent section, Section~\ref{Sec:CauchyProb}, we recall the definition of linear differential operators on smooth manifolds, emphasising their characterisation through the concept of \emph{locality}. We then provide a detailed treatment of \emph{symmetric hyperbolic systems}, which are linear first-order differential operators $\sf{S}\:\Gamma^{\infty}(\sf{E})\to\Gamma^{\infty}(\sf{E})$ acting on the smooth sections of a Hermitian vector bundle $(\sf{E},\langle\cdot,\cdot\rangle_{\sf{E}})$ over a globally hyperbolic Lorentzian manifold $(\sf{M},\sf{g})$, whose principal symbol $\sigma_{\sf{S}}\:\sf{T}^{\ast}\sf{M}\to\mathrm{End}(\sf{E})$ is Hermitian with respect to the bundle metric $\langle\cdot,\cdot\rangle_{\sf{E}}$ and which is hyperbolic in the sense that $\langle\sigma_{\sf{S}}(\tau)\cdot,\cdot\rangle_{\sf{E}}$ is positive-definite for all future-directed timelike $\tau\in\sf{T}^{\ast}\sf{M}$. Symmetric hyperbolic systems are of fundamental importance in mathematical physics, as they include a wide range of hyperbolic operators that arise naturally in applications. In particular, the \emph{Dirac operator} of a globally hyperbolic \emph{spin} manifold falls into this class. Moreover, every \emph{normally hyperbolic operator}, that is, wave-type operator $\sf{N}=-\square^{\nabla^{\sf{E}}}+\sf{N}_{0}$, where $\square^{\nabla^{\sf{E}}}$ denotes the connection d’Alembertian and $\sf{N}_{0}$ a zero-order operator, can be rewritten as a symmetric hyperbolic system, as we will explain in some detail.

A central aim of this chapter is to establish the well-posedness of the Cauchy problem for symmetric hyperbolic systems from first principles, without assuming any prior familiarity with the subject. More precisely, the aim is to show that the Cauchy problem
\begin{align*}\begin{cases}
	\sf{S}\psi&=\phi\\
	\psi\vert_{t=0}&=\mathfrak{f}\end{cases}
\end{align*}
has a unique smooth solution $\psi\in\Gamma^{\infty}(\sf{E})$ for arbitrary Cauchy data $(\mathfrak{f},\phi)\in\Gamma^{\infty}(\sf{E}\vert_{\Sigma_{0}})\times\Gamma^{\infty}(\sf{E})$ that propagates at most with the speed of light, i.e.~$\mathrm{supp}(\psi)\subset\mathcal{J}(\mathrm{supp}(\phi)\cup\mathrm{supp}(\mathfrak{f}))$. This is the content of Theorem~\ref{Thm:Cauchy}. To this end, we provide a novel proof which has not previously appeared in the literature in the context of hyperbolic systems on manifolds. The strategy is as follows: we first restrict our attention to spatially compact manifolds, an admissible reduction in the curved setting, since one can always employ a compactification procedure to achieve spatial compactness. We then establish the existence of weak solutions on a finite time strip by suitably adapting the classical Fréchet–Riesz argument to the setting of manifolds. Finally, we prove regularity of solutions using a new approach based on formulating an extended system that acts simultaneously on the unknown field and its derivatives. As mentioned above, we obtain the well-posedness of the Cauchy problem for normally hyperbolic operators as a byproduct, summarised in Theorem~\ref{Thm:NHCauchy}. We conclude this section with a discussion of \emph{Green hyperbolicity}.

In the final part of this chapter, Section~\ref{Sec:SHSNonLoc}, we investigate a small but conceptually significant departure from the classical theory. As noted above, differential operators are intrinsically linked to the notion of locality, in the sense that their action is determined entirely by data in an arbitrarily small neighbourhood. Recent developments in mathematical physics and related areas, however, have highlighted the importance of incorporating also nonlocal potentials or interactions into hyperbolic equations, where effects at one point can depend on events occurring far away in space
and time. Examples range from Maxwell's equations in linear dispersive media, semiclassical models such as the semiclassical Einstein equations to relativistic field equations on noncommutative spacetimes. Following \cite{SchmidMurroFinster}, we investigate the initial value problem for symmetric hyperbolic systems on globally hyperbolic Lorentzian manifolds coupled to a large class of nonlocal potentials. More precisely, we consider a linear and continuous operator $\sf{B}\:\Gamma^{\infty}_{\mathrm{c}}(\sf{E})\to\Gamma^{\infty}(\sf{E})$. Under some additional assumptions, one can associate to it a \textit{time kernel}, namely a two-parameter family $\sf{B}_{t,\tau}\colon \Gamma^{\infty}_{\mathrm{c}}(\Sigma_{\tau},\sf{E}\vert_{\Sigma_{\tau}})\to \Gamma^{\infty}(\Sigma_{t},\sf{E}\vert_{\Sigma_{t}})$, where $(\Sigma_{t}:=\{t\}\times\Sigma)_{t\in\bb{R}}$ is a foliation of $\sf{M}$ by Cauchy surfaces, such that $(\sf{B}\psi)\vert_{\Sigma_{t}}$ is obtained as the integral of $\sf{B}_{t,\tau}\psi_{\tau}$ over $\tau\in\bb{R}$ with $\psi_{t}:=\psi\vert_{\Sigma_{t}}$. Now, consider the Cauchy problem
\begin{align*}\begin{cases}
	(\sf{S}-\sf{B})\psi&=\phi\\
	\psi\vert_{t=0}&=\mathfrak{f}\end{cases}
\end{align*}
on the time strip $\sf{M}_{\mathrm{T}}:=\bigcup_{t\in [0,\mathrm{T}]}\Sigma_{t}$ with Cauchy data $(\phi,\mathfrak{f}) \in \Gamma^\infty_{\mathrm{c}}(\sf{E})\times \Gamma^{\infty}_{\mathrm{c}}(\sf{E}\vert_{\Sigma_{0}})$. Then, denoting the future-directed timelike covector of the foliation by $\eta:=\beta\d t$, we prove the following:
\begin{itemize}
	\item[(i)]Suppose that $\sf{V}:=\beta\sigma_{\sf{S}}(\eta)^{-1}\sf{B}$ is uniformly bounded on $\sf{M}_{\mathrm{T}}$, i.e. 
	\begin{align*}
	\exists C_{\mathrm{T}}>0:\quad \Vert\mathcal{V}_{t,\tau}\psi_{\tau}\Vert_{t}\leq C_{\mathrm{T}}\Vert\psi_{\tau}\Vert_{\tau}\,.\end{align*}
	 Then, there exists a (strong) solution $\psi$ on $\sf{M}_{\mathrm{T}}$ to the Cauchy problem. Moreover, if all the derivatives of $\sf{B}$ satisfy the same properties (in a suitable sense), then $\psi$ is smooth, unique and propagates at most with the speed of light. This is the content of Theorem~\ref{Thm:Ret}.
	 \item[(ii)]Suppose that the zero-order operator $\sf{Z}_{\sf{S}}:=\beta\sigma(\eta)^{-1}(\sf{S}+\sf{S}^{\ast})$ is uniformly bounded, i.e.
	\begin{align*}
		\exists D>0:\quad\Vert\sf{Z}_{\sf{S}}\psi_{t}\Vert_{t}\leq D\Vert\psi_{t}\Vert_{t}
	\end{align*}
	and assume that $\sf{V}:=\beta\sigma_{\sf{S}}(\eta)^{-1}\sf{B}$ is uniformly bounded, small compared to $\delta$ and $D$, i.e.
	\begin{align*}
		\exists 0<C<(8e\delta^{2})^{-1}:\quad \Vert\sf{B}_{t,\tau}\psi_{\tau}\Vert_{t}\leq Ce^{-\frac{1}{2}D\vert\tau\vert}\Vert\psi_{t}\Vert_{t}\, .
	\end{align*}
	Then, there exists a (strong) solution $\psi$ on $\sf{M}_{\mathrm{T}}$ to the Cauchy problem. This is the content of Theorem~\ref{Thm:Small}. Moreover, we constructed counterexamples of nonlocal potentials, violating this
condition on the uniform bound, for which there are no (weak) solutions.
\end{itemize}

\underline{\textbf{{Chapter}~\ref{Chap:LinGaug}.}} The second chapter of this thesis provides a detailed exposition on linear gauge theories on globally hyperbolic Lorentzian manifolds and their quantisation. In the first part, Section~\ref{Sec:HackSchenkel}, we discuss the mathematical structure of \emph{classical} linear gauge theories employing the formalism initially described by Hack-Schenkel and further developed by Gérard-Wrochna, as outlined above. In particular, we provide a detailed discussion of the Cauchy problem for linear gauge theories as well as of the structure of the classical theory by studying several equivalent characterisations of the corresponding phase space.

Having established an abstract axiomatic framework for linear gauge theories and the well-posedness of the corresponding Cauchy problems, we come to the main part of the second chapter, namely Section~\ref{Sec:Exam}. Within this section, we will study several examples of linear gauge theories of direct physical interest in the context of the Hack-Schenkel formalism, some of which have not been discussed yet in the literature. To start with, we will discuss the following gauge theories on general globally hyperbolic Lorentzian manifolds:
\begin{itemize}
	\item[$\hookrightarrow$]Maxwell's theory of electromagnetism (see Proposition~\ref{Prop:MaxGauge}).
	\item[$\hookrightarrow$]Yang-Mills theory on arbitrary, possibly non-trivial, smooth principal fibre bundles and with possibly non-compact structure groups, linearised around general, possibly non-flat, background solutions to the non-linear Yang-Mills equations (see Proposition~\ref{Prop:YMGauge}).
	\item[$\hookrightarrow$]Einstein gravity, linearised around arbitrary (globally hyperbolic) solutions to the non-linear vacuum Einstein equations (see Proposition~\ref{eq:EinsteinHS}). Moreover, we will consider the equivalent theory by taking trace-reversed vacuum Einstein equations as background equations instead, which leads to a slightly different description of linearised gravity in the Hack-Schenkel formalism (see Proposition~\ref{Prop:EinsteinHS2}).
\end{itemize}
Through the prism of the quest of studying more realistic models from a physical perspective, we will further study the following coupled systems:
\begin{itemize}
	\item[$\hookrightarrow$]Linearised Yang-Mills theory coupled to a (real-valued) Klein-Gordon (or ``Higgs'') field taking values in an associated vector bundle and transforming under an arbitrary (finite-dimensional) Lie group representations of the structure group (see Proposition~\ref{Prop:YMKGGauge}).
	\item[$\hookrightarrow$]Linearised Einstein gravity coupled to a real-valued scalar field (see Proposition~\ref{Prop:GaugeEinKG}).
\end{itemize}
In particular, the main task is to show that all the aforementioned theories fit within the Hack-Schenkel formalism of linear gauge theories. As one can see by looking at the list of theories, we restrict ourselves to \emph{bosonic} theories. This, however, is purely done for convenience and it is strongly suggested that also other and more complicated coupled theories, such as a linearised \emph{Yang-Mills-Dirac-Higgs-Yukawa theory} eventually fits within this formalism.

In the final part of this chapter, Section~\ref{Sec.AQFT}, we will turn to the quantisation of linear gauge theories. In particular, we will give a detailed exposition of the Gérard-Wrochna formalism for quantisation of linear gauge theories. To start with, we provide the reader with a short review of the \emph{algebraic approach} to quantum field theory. As mentioned above, quantisation in this framework consists of two steps, namely the assignment of a $\ast$-algebra of observables and the construction of a suitable state satisfying the \emph{Hadamard condition}. We will explain both these ingredients in detail. We conclude this chapter by providing a detailed bibliography of existence results for Hadamard states for various different field theories.\bigskip

\underline{\textbf{{Chapter}~\ref{Chap:Maxwell}.}} The last chapter is devoted to the quantisation of Maxwell's theory on globally hyperbolic spacetimes. Let $(\sf{M},\sf{g})$ be a globally hyperbolic spacetime. Then, Maxwell's theory as a linear gauge theory in the formalism of Hack-Schenkel is described by the two operators
\begin{align*}
	\sf{P}\:=\delta\d\:\Omega^{1}(\sf{M})\to\Omega^{1}(\sf{M})\,,\qquad\sf{K}:=\d\:C^{\infty}(\sf{M})\to\Omega^{1}(\sf{M})\,,
\end{align*}
where $\delta$ is the codifferential and $\mathrm{d}$ the exterior derivative acting on differential forms $\Omega^{k}(\sf{M})$ on $\sf{M}$. The operator $\sf{P}$ is formally self-adjoint with respect to the Hodge inner product
\begin{align*}
	(\cdot,\cdot):= \int_{\sf{M}}\,\sf{g}^{\sharp}(\cdot,\cdot)\,\d\mu_{\sf{g}}\,,
	\end{align*}
which is \emph{not} positive definite, since $\sf{g}^{\sharp}$ is a Lorentzian metric. The canonical gauge condition $\sf{K}^{\ast}\sf{A}=\delta\sf{A}=0$ for $\sf{A}\in\Omega^{1}(\sf{M})$ corresponds to the \emph{Lorenz gauge} and the operators $\sf{D}_{i}$ are exactly the de Rham-Hodge d'Alembertians $\sf{D}_{i}=\d\delta+\delta\d$ on $\Omega^{i-1}(\sf{M})$ for $i=1,2$.

In the first section of this chapter, Section~\ref{Sec:MaxTheLorMa}, we provide a more detailed analysis of Maxwell's theory on globally hyperbolic manifolds. On the one hand, this section aims to prepare the ground for the construction of Hadamard states of Maxwell's theory to be performed in the subsequent sections. On the other hand, it provides a detailed and self-contained treatment of both the Cauchy problem for Maxwell's equations on general globally hyperbolic spacetimes as well as the corresponding gauge problem by describing several gauge conditions, their achievability and their relations among each other.

Afterwards, we will turn to the construction of Hadamard states for Maxwell's theory, following~\cite{MurroSchmid}. As mentioned above, the construction of Hadamard states for field-theoretic models admitting a gauge theory is mathematically more complicated compared to the ordinary field theories. To overcome the difficulties mentioned above, the main idea of our approach is to fix completely all the gauge degrees of freedom. This will be achieved by working in the so-called {\em Cauchy radiation gauge}, a novel gauge condition. More precisely, we consider those solutions of the Maxwell equation that satisfy the conditions
\begin{align*}
	\delta\sf{A}=0\,\qquad\sf{A}_{\mathrm{T}}\vert_{t=0}=\partial_{t}\sf{A}_{\mathrm{T}}\vert_{t=0}=0\,,
\end{align*}
where we decomposed $\sf{A}=\sf{A}_{\mathrm{T}}\eta+\sf{A}_{\Sigma}$ with $\eta:=\beta\d t$ and $\sf{A}_{\mathrm{T}}:=\eta\lrcorner\sf{A}$. On globally hyperbolic \emph{ultrastatic} spacetimes, such as Minkowski spacetime, this gauge is equivalent to the so-called \textit{radiation gauge}, i.e.~$\delta\sf{A}=\sf{A}_{\mathrm{T}}=0$ and, if $\Sigma$ is in addition non-compact, it is (on-shell the equations of motion) equivalent to the \textit{Coulomb gauge} $\delta_\Sigma \sf{A}_\Sigma =0$, under suitable decay assumptions on the involved fields. The gauge condition is chosen in such a way that it removes the negative contributions in the bundle metric at the initial-data stage, thereby enabling the construction of positive Cauchy covariances for the fully gauge-fixed theory.

In order to show that the Cauchy radiation gauge can always be achieved, we will need to solve the \emph{Poisson equation} $\Delta f=\delta_{\Sigma}\omega$ for a given source $\omega\in\Omega^{1}(\Sigma)$, where $\Delta=\delta_{\Sigma}\d_{\Sigma}$ denotes the de Rham-Hodge Laplacian acting on functions, on general \emph{complete} Riemannian manifolds. Now, for compact $\Sigma$, this can easily be seen to be equivalent to the celebrated \emph{Hodge decomposition theorem}. Indeed, if a solution $f$ exists to the above equation, then $\delta_{\Sigma}(\d_{\Sigma}f-\omega)=0$ and hence, $\omega=\d_{\Sigma}f+\eta$, where $\eta\in\mathrm{ker}(\delta_{\Sigma})$. Therefore, the idea is to generalise the Hodge decomposition to non-compact, complete Riemannian manifolds, for differential forms taking values in suitable \emph{Sobolev spaces}. This will be achieved in Section~\ref{Sec:HodgeDecomp}, in particular in Theorem~\ref{Thm:HodgeDecom} and in Theorem~\ref{Thm:HodgeDecomSmooth}.

In the last section of this chapter, Section~\ref{Sec:ConsHadMax}, we will prove the existence of Hadamard states for Maxwell theory on arbitrary globally hyperbolic manifolds. It is the first proof on general globally hyperbolic backgrounds, with earlier works imposing several technical analytic or topological restrictions. Building upon the results of the previous discussion, we will show that the Cauchy radiation gauge can always be achieved in suitable spacetimes and in suitable Sobolev spaces. Moreover, we will show that the Cauchy problem for differential forms is well-posed when working with the differential forms that are smooth in time and $\sf{H}^{\infty}$ in space, see Theorem~\ref{Thm:CauchySob}. The construction of Hadamard states in the Cauchy radiation gauge will be first performed explicitly in ultrastatic globally hyperbolic manifolds. As already emphasised above, the main benefit of working in the Cauchy radiation gauge is that the fibre metric becomes manifestly positive definite. States for Maxwell’s theory are will therefore be defined first on the completely gauge-fixed theory and then defined  on the original phase space via the pull-back along the gauge-fixing projector. It will be a central part to show that this procedure preserves the Hadamard property. The existence of Hadamard states on general globally hyperbolic manifolds is then deduced by a deformation argument, leading us the main theorem of this section, Theorem~\ref{Thm:Hadamard}. \bigskip

\underline{\textbf{{Appendix}~\ref{Chap:Appendix}.}} This thesis is supplemented with an appendix consisting of the following parts: the first two sections, Section~\ref{Appendix:Micro} and Section~\ref{App:PSIDO}, provide a detailed and self-contained introduction into \emph{microlocal analysis}, including a brief review of \emph{distribution theory}, both on $\bb{R}^{d}$ and on manifolds, the celebrated \emph{kernel theorem of Schwartz}, the notion of the \emph{wavefront set} of distributions, \emph{pseudodifferential operators} on $\bb{R}^{d}$ and on manifolds, with a particular focus on \emph{Riemannian manifolds of bounded geometry}, as well as into the celebrated \emph{propagation of singularities} theorem. The aim of this section is two-fold: on the one hand, it provides the basic technical tools needed in the formulation and construction of Hadamard states. On the other hand, it can be seen as an independent introduction to microlocal analysis on its own without assuming any background knowledge. 

Section~\ref{App:FuncAna} supplements the thesis with a short discussion of basic notions of \emph{functional analysis}. In particular, we provide a basic dictionary for unbounded operators on Hilbert spaces and review the spectral theorem, which we will need in our discussion of the Hodge decomposition in Section~\ref{Sec:HodgeDecomp}.

Last but not least, the final section of the appendix, Section~\ref{App:Quot}, includes a technical result on isomorphisms between quotients of vector spaces that will be useful when discussing various equivalent characterisations of the phase space for linear gauge theories in Section~\ref{Sec:HackSchenkel}.
\chapter{Hyperbolic Operators on Lorentzian Manifolds}\label{Chap:Hyp}
In this first chapter, we lay the essential geometric and analytic foundations for the study of linear field theories on Lorentzian manifolds. We begin by introducing the notion of globally hyperbolic spacetimes, which constitute a particularly well-suited class of Lorentzian backgrounds for the formulation and analysis of the Cauchy problem for hyperbolic partial differential equations. Subsequently, we give a concise overview of linear differential operators on manifolds acting on sections of vector bundles. Building up on this, we proceed to discuss various notions of hyperbolicity in the context of globally hyperbolic manifolds. In particular, we will prove the well-posedness of the Cauchy problem for symmetric hyperbolic systems and normally hyperbolic operators starting from first principles, along with the concept of Green operators, which will play a central role in the subsequent development of the theory. 

Last but not least, we study a small deviation from the classical \emph{local} theory, namely the Cauchy problem of symmetric hyperbolic systems coupled to potentials that are \emph{nonlocal} both in time and space. In particular, we show that the Cauchy problem is well-posed for a large class of nonlocal potentials, which is the first main result of this thesis.

\section{Globally Hyperbolic Lorentzian Manifolds}\label{Sec:Lorentzian}
In this preliminary section we shall discuss some basic concepts of Lorentzian geometry. In particular, we introduce the notion of \textit{globally hyperbolic spacetimes} that will provide a suitable background for analysing the Cauchy problem for hyperbolic equations.

\subsection{Preliminaries on Lorentzian Geometry}
Throughout this section, let $(\sf{M}, \sf{g})$ denote a smooth Lorentzian manifold of dimension $d=n+1$ with $n \in \mathbb{N}$ and signature convention $(-,+,\dots,+)$. It is well known that topological obstructions can prevent the existence of a Lorentzian metric: if $\sf{M}$ is non-compact, then a Lorentzian metric always exists, while in the case in which $\sf{M}$ is compact, a Lorentzian metric exists on $\sf{M}$ if and only if its Euler characteristic vanishes; see, e.g.,~\cite[Chap.~5, Prop.~37]{ONeill}.

As usual, the Lorentzian metric $\sf{g}$ allows us to classify tangent vectors into three different types: a tangent vector $v\in\sf{T}_{p}\sf{M}$ at some point $p\in\sf{M}$ is called
\begin{align*}
	\text{(i)}&\qquad\textit{spacelike} \quad\text{if}\quad \sf{g}_{p}(v,v)>0 \quad \text{or}\quad v=0\,;\\
	\text{(ii)}&\qquad\textit{lightlike} \quad\text{or}\quad \textit{null} \quad\text{if}\quad \sf{g}_{p}(v,v)=0 \quad\text{and}\quad v\neq 0\,;\\
	\text{(iii)}&\qquad\textit{timelike} \quad\text{if}\quad \sf{g}_{p}(v,v)<0\,.
\end{align*}
A vector that is either timelike or lightlike is also called \textit{causal}. The classification into spacelike, lightlike, timelike and causal objects extends to vector fields when applied pointwise and to curves when applied to their tangent vectors. Moreover, the classification extends to hypersurface when applied to their normal vector fields suitably, i.e.~a \emph{timelike} normal defines a \emph{spacelike} hypersurface and vice versa.

In order to introduce some notions of causality, we assume that $(\sf{M},\sf{g})$ is \textit{time-orientable}, that it, we assume that there exists a global continuous timelike vector field $\sf{X}\in\Gamma^{0}(\sf{T}\sf{M})$ on $\sf{M}$. The choice of such a vector field is called a \textit{time orientation} of $\sf{M}$ and allows to decompose the \textit{light cone} at $p\in\sf{M}$, i.e.~the linear subspace
\begin{align}\label{eq:LightCone}
	\sf{V}_{p}:=\{v\in\sf{T}_{p}\sf{M}\mid \sf{g}_{p}(v,v)<0\}\subset\sf{T}_{p}\sf{M}\, ,
\end{align}
into two connected components, namely the \textit{future} and \textit{past lightcones} at $p\in\sf{M}$ defined by
\begin{align}\label{eq:LightCone2}
	\sf{V}^{\pm}_{p}:=\{v\in\sf{T}_{p}\sf{M}\mid \sf{g}_{p}(v,v)<0\,\,\text{and}\,\,\sf{g}_{p}(\sf{X}_{p},v)\lessgtr 0\}\, .
\end{align}
Tangent vectors in $\sf{V}^{+}_{p}$ and $\sf{V}^{-}_{p}$ are also called \textit{future-directed} (or \textit{future-pointing}) and \textit{past-directed} (or \textit{past-pointing}), respectively. This classification extends to all causal vectors by considering the \textit{closed} lightcones. Furthermore, the classification extends to causal vector fields when applied pointwise and to causal curves when applied to their tangent vector fields.

\begin{remark}\label{Rem:ExTOLM}
	The existence of a time-orientable Lorentzian metric is again a purely topological question: the existence of a Lorentzian metric on a manifold $\sf{M}$ is equivalent to the existence of a nowhere vanishing vector field. Given such a vector field, it can always be chosen to be timelike with respect to some Lorentzian metric. Hence, the existence of a Lorentzian metric is equivalent to the existence of a \textit{time-orientable} Lorentzian metric, see~\cite[Chap.~5, Prop.~37]{ONeill}.

On the other hand, not every Lorentzian manifold is time-orientable. Nevertheless, every Lorentzian manifold admits a time-orientable Lorentzian double cover and if $\sf{M}$ is simply connected, then every Lorentzian metric on $\sf{M}$ is time-orientable, see~\cite[Chap.~7, Lemma~17]{ONeill}.

Lastly, we stress that time-orientability and  orientability are independent condition. For instance, the cylinder $S^{1}\times\bb{R}$ is an \textit{orientable} manifold, for which one can easily construct both a time-orientable and non-time-orientable Lorentzian metric. Similarly, one can find both a time-orientable and non-time-orientable Lorentzian metric on the \textit{non-orientable} Möbius strip.
\end{remark}

Now, if $(\sf{M},\sf{g})$ is a time-oriented Lorentzian manifold, we can define for every $p\in\sf{M}$ its \textit{chronological future} and \textit{past} by
\begin{align*}
	\mathcal{I}_{\sf{M}}^{\pm}(p):=\bigg\{q\in\sf{M}\,\,\bigg\vert\,\,\exists\begin{array}{l}\text{future-directed}\\\text{past-directed}\end{array} \text{ timelike curve }\gamma\:[0,1]\to\sf{M}\text{ with }\gamma(0)=p\text{ and }\gamma(1)=q\bigg\}\, .
\end{align*}
Replacing the word ``timelike'' with ``causal'', we obtain the \textit{causal future} and \textit{past} denoted by $\mathcal{J}^{\pm}_{\sf{M}}(p)$. For a subset $\Omega\subset\sf{M}$, we further define 
\begin{align*}
	\mathcal{I}_{\sf{M}}^{\pm}(\Omega):=\bigcup_{p\in\Omega}\mathcal{I}_{\sf{M}}^{\pm}(p)\qquad\text{and}\qquad \mathcal{J}_{\sf{M}}^{\pm}(\Omega):=\bigcup_{p\in\Omega}\mathcal{J}_{\sf{M}}^{\pm}(p)\, .
\end{align*}
Last but not least, we set $\mathcal{J}_{\sf{M}}(\Omega):=\mathcal{J}_{\sf{M}}^{+}(\Omega)\cup\mathcal{J}^{-}_{\sf{M}}(\Omega)$ and $\mathcal{I}_{\sf{M}}(\Omega):=\mathcal{I}_{\sf{M}}^{+}(\Omega)\cup\mathcal{J}^{-}_{\sf{M}}(\Omega)$.  If it is clear in the context which ambient Lorentzian manifold is used, we simply write $\mathcal{J}^{\pm}:=\mathcal{J}^{\pm}_{\sf{M}}$ and $\mathcal{I}^{\pm}:=\mathcal{I}^{\pm}_{\sf{M}}$ as well as $\mathcal{J}:=\mathcal{J}_{\sf{M}}$ and $\mathcal{I}:=\mathcal{I}_{\sf{M}}$. By definition, the chronological and causal future/past have the following important properties:
\begin{align*}
	\text{(i)}&\qquad\Omega\cup\mathcal{I}^{\pm}(\Omega)\subset\mathcal{J}^{\pm}(\Omega)\\
	\text{(ii)}&\qquad \mathcal{J}^{\pm}(\mathcal{J}^{\pm}(\Omega))=\mathcal{J}^{\pm}(\Omega)\\
	\text{(iii)}&\qquad \mathcal{J}^{\pm}(\Omega)\subset\overline{\mathcal{I}^{\pm}(\Omega)}\qquad\text{with ``='' iff and only if $\Omega$ is closed}
\end{align*}
for $\Omega\subset\sf{M}$ arbitrary. Furthermore, we note that $\mathcal{I}^{\pm}(\Omega)$ is an open set for arbitrary $\Omega$, while $\mathcal{J}^{\pm}(\Omega)$ can be both open or closed. For more details on causality in Lorentzian geometry, we refer to the detailed discussion in \cite[Chap.~14]{ONeill} and \cite[Chap.~3]{Beem}.

\subsection{Global Hyperbolicity}
After this preliminary discussion on Lorentzian geometry, we will now define and study a special class of time-orientable Lorentzian manifolds, namely those for which the Cauchy problem of hyperbolic operators is well-posed. To this end, we make the following definition.

\begin{definition}\label{Def:GlobHyp} (Spacetimes and Globally Hyperbolic Manifolds)\newline
	A \emph{spacetime} is a connected, oriented, time-oriented and smooth Lorentzian manifold. Furthermore, a spacetime $(\sf{M},\sf{g})$ is called \emph{globally hyperbolic} if the following holds true:
	\begin{itemize}
		\item[(i)]$(\sf{M},\sf{g})$ is \emph{causal}, i.e.~there are no closed causal curves;
		\item[(ii)]For every $p, q \in \sf{M}$, the set $\mathcal{J}^+ (p) \cap \mathcal{J}^{-}(q)$ is compact.
	\end{itemize}
\end{definition}

\begin{remarks} (Historical Remarks and Equivalent Definitions)
\begin{itemize}
\item[(i)]
	In earlier literature, one usually encounters Definition~\ref{Def:GlobHyp} with condition (i) replaced by the requirement that $(\sf{M},\sf{g})$ is \textit{strongly causal}, i.e.~there are no \textit{almost} causal curves, that is, for each point $p \in \sf{M}$ and every open neighbourhood $\mathcal{U} \subset \sf{M}$ of $p$, there exists an open neighbourhood $\mathcal{V} \subset \mathcal{U}$ of $p$ such that any causal curve starting and ending in $\mathcal{V}$ remains within $\mathcal{U}$. However, it has been shown by Bernal-Sánchez \cite[Thm.~3.2]{BernalSanchezCausal} that causality and strong causality are equivalent, provided the compactness condition (ii) holds.

In fact, it has recently been shown by Hounnonkpe-Minguzzi in \cite{Minguzzi} that for Lorentzian manifolds with $\mathrm{dim}(\sf{M}) \geq 3$, the causality assumption (i) can entirely be omitted, provided that $\sf{M}$ is additionally assumed to be non-compact.
\item[(ii)]Globally hyperbolic manifolds were originally introduced by Leray \cite{Leray} in the context of hyperbolic differential equations. However, Leray did not use the modern definition given above. Instead, he considered for each $p, q \in \sf{M}$ the set
\begin{align*}
\mathcal{C}(p,q) :=\{ C^0\text{-curves from } p \text{ to } q \text{ future-directed causal} \}/_{\sim}\,,
\end{align*}
where two curves are equivalent if they differ by a strictly monotonic reparametrisation,
endowed with the $C^{0}$-topology.\footnote{For a given $\gamma\in\mathcal{C}(p,q)$, the set of curves in $\mathcal{C}(p,q)$ contained in $\mathrm{ran}(\gamma)\subset\sf{M}$ is a neighbourhood basis.} Then, $(\sf{M},\sf{g})$ is \emph{globally hyperbolic} if it is causal and if $\mathcal{C}(p,q)$ is compact for all $p,q\in\sf{M}$. The equivalence of this definition with Definition~\ref{Def:GlobHyp} has been shown by Hawking-Ellis \cite[Prop.~6.6.2]{HawkingEllis}, based on some technical results of Geroch \cite{GerochCauchy} and Seifert \cite{Seifert}. A modern proof can also be found in \cite[Thm.~3.79]{SanchezMinguzzi}.
\end{itemize}
\end{remarks}

As a first observation, we note that global hyperbolicity imposes a topological restriction: any globally hyperbolic Lorentzian manifold must necessarily be non-compact.

\begin{lemma}\label{Lemma:NonCompactGH}
	A globally hyperbolic spacetime is non-compact.
\end{lemma}

\begin{proof}
	The collection $\{\mathcal{I}^{+}(p)\}_{p\in\sf{M}}$ defines an open cover of $\sf{M}$ and if $\sf{M}$ is compact, we can find a finite subcover $\{\mathcal{I}^{+}(p_{1}),\dots,\mathcal{I}^{+}(p_{n})\}$ for some $n\in\bb{N}$ and points $p_{1},\dots,p_{n}\in\sf{M}$. Without loss of generality, we assume that none of the sets $\mathcal{I}(p_{i})$ is entirely contained in another one. But now, note that $p_{i}\in\mathcal{I}^{+}(p_{i})$ for every $i$, which means that there is timelike curve starting and ending at $p_{i}$. Hence, $(\sf{M},\sf{g})$ violates the causality condition.
\end{proof}

It turns out that any globally hyperbolic manifold possesses  a particularly simple topological structure: it is diffeomorphic to a product $\bb{R}\times\Sigma$, where $\Sigma$ is some (compact or non-compact) spacelike hypersurface. To discuss this fact in more detail, we start with the following definition:

\begin{definition} (Cauchy Surfaces)\newline
	Let $(\sf{M},\sf{g})$ be a spacetime. A subset $\Sigma\subset\sf{M}$ is called a \emph{Cauchy (hyper)surface} if it is intersected exactly once by any inextensible timelike curve.
\end{definition}

It is not too hard to show that any Cauchy surface $\Sigma$ is in fact an embedded topological submanifold of codimension one that is intersected exactly once by every inextendible \textit{causal} curve, see e.g.~\cite[Chap.~14, Lemma~29]{ONeill}. 

\begin{theorem} \emph{(Geroch \cite{GerochCauchy})}\newline
	A spacetime is globally hyperbolic if and only if it admits a Cauchy surface.
\end{theorem}

As a by-product of Geroch's theorem, it follows that a globally hyperbolic spacetime $(\sf{M},\sf{g})$ admits a continuous foliation by Cauchy surfaces, i.e.~$\sf{M}$ is homeomorphic to $\bb{R}\times\Sigma$ for some Cauchy surface $\Sigma$. This can be established by finding a \textit{Cauchy time function}, i.e.~a continuous function $t\colon\sf{M}\rightarrow\mathbb{R}$ which is strictly increasing on any future-directed timelike curve and whose level sets $t^{-1}(\tau)$, $\tau\in \mathbb{R}$, are Cauchy surfaces homeomorphic to $\Sigma$. Geroch's splitting is topological in nature and the fact that it can be done in a smooth way has been part of the mathematical folklore (see e.g.~the proof of Prop.~6.6.8 in Hawking-Ellis \cite{HawkingEllis}), whilst remaining an open problem for many years (see also the introduction of \cite{bernal+sanchez0}). Only recently Bernal-Sánchez obtained a smooth version of Geroch's theorem by introducing the notion of a \emph{Cauchy temporal function}.

\begin{definition} (Cauchy Temporal Functions)\newline
	For a spacetime $(\sf{M},\sf{g})$, we call $t\in C^{\infty}(\sf{M},\bb{R})$ a \emph{Cauchy temporal function} if $t^{-1}(\tau)$ for all $\tau\in\bb{R}$ are smooth Cauchy hypersurfaces and $\mathrm{grad}_{\sf{g}}(t)=(\d t)^\sharp$ is past-directed timelike. 
\end{definition} 

With this definition, the central result of Bernal-Sánchez is as follows.

\begin{theorem}\label{Thm:BernalSanchez} \emph{(Bernal-Sánchez \cite[Thm.~1.1~and~1.2]{bernal+sanchez2}, \cite[Thm.~1.2]{bernal+sanchez})}\newline
	Let $(\sf{M},\sf{g})$ be a globally hyperbolic spacetime. Then, there exists a Cauchy temporal function $\tau\in C^{\infty}(\sf{M})$. Furthermore, there exists a smooth diffeomorphism $\psi\:\sf{M}^{\prime}\to\sf{M}$ such that
	\begin{align*}
		\sf{M}^{\prime}:=\bb{R}\times\Sigma\,,\qquad \psi^{\ast}\sf{g}=-\beta^{2}(t,\cdot) \d t\otimes \d t+\sf{h}_{t}\, ,
	\end{align*}
	where $t:=\psi^{\ast}\tau\:\sf{M}^{\prime}\to\bb{R}$ acts as the natural projection, $\beta\in C^{\infty}(\sf{M}^{\prime},(0,\infty))$, $\Sigma$ is a smooth, spacelike Cauchy surface and $\sf{h}\in \Gamma^{\infty}(\sf{M}^{\prime},\mathrm{pr}_{2}^{\ast}(\sf{T}\Sigma^{\otimes_{s}2}))$ with $\mathrm{pr}_{2}\:\sf{M}^{\prime}\to\Sigma$ being the natural projection, is such that $\sf{h}_{\tau}:=\sf{h}\vert_{\Sigma_{\tau}}$ is a Riemannian metric on $\Sigma_{\tau}:=\{\tau\}\times\Sigma$ for all $\tau\in\bb{R}$.

	Moreover, if $\Sigma \subset \sf{M}$ is any smooth and spacelike Cauchy hypersurface, there exists a Cauchy temporal function $t\in C^{\infty}(\sf{M},\bb{R})$ such that $\Sigma\cong t^{-1}(\tau)$ for some $\tau\in\bb{R}$.
\end{theorem}

The function $\beta$ is usually called the \textit{lapse function} and the tensor $\sf{h}$ can be thought of as a one-parameter family $(\sf{h}_{t})_{t\in\bb{R}}$ of Riemannian metrics on $\Sigma$, depending smoothly on $t$. If we choose local coordinates $(x^{i})_{i=1,\dots,k}$ of $\Sigma$ on some open subset $\mathcal{U}\subset\Sigma$, then $(x^{0}:=t,x^{1},\dots,x^{k})$ are local coordinates of $\sf{M}$ on $\bb{R}\times\mathcal{U}$, and
\begin{align*}
	(\psi^{\ast}\sf{g})_{00}(t,x^{i})=-\beta^{2}(t,x^{i})\,,\qquad (\psi^{\ast}\sf{g})_{i0}(t,x^{i})=0\qquad\text{and}\qquad (\psi^{\ast}\sf{g})_{ij}(t,x^{i})=(\sf{h}_{t})_{ij}(x^{i})\, .
\end{align*}
Note that there are no ``cross-terms'', i.e.~non-trivial metric components $\sf{g}_{0i}$, as for instance in ADM-type decompositions of the form
\begin{align*}
	\psi^{\ast}\sf{g}=(-\beta^{2}+\alpha^{i}\alpha_{i})\,\d t\otimes\d t+2\alpha_{i}\,\d t\otimes\d x^{i}+\sf{g}_{ij}\,\d x^{i}\otimes\d x^{j}
\end{align*}
for some \textit{shift vector field} $\alpha\in\Gamma^{\infty}(\sf{M}^{\prime},\mathrm{pr}_{2}^{\ast}\sf{T}\Sigma)$. In other words, the theorem of Bernal-Sánchez implies that one can always choose coordinates with a \textit{global} time coordinate such that $\alpha=0$, i.e.~the Bernal-Sánchez theorem provides a \textit{globally orthogonal} splitting.\footnote{It is worth to mention, however, that there are situations in which it is more useful to consider coordinates with a non-trivial shift vector, as in the case of \textit{stationary spacetimes}, for instance.} 

\begin{remark}\label{Rem:GHSBound}
	Global hyperbolicity can also be extended to include timelike boundaries. A \textit{globally hyperbolic spacetimes with timelike boundary} is a spacetime $(\sf{M},\sf{g})$ with $\partial\sf{M}\neq\emptyset$ timelike, that is causal and with compact diamonds, as in Definition~\ref{Def:GlobHyp}, where we use the notation in which $\partial\sf{M}\subset\sf{M}$. Both the Theorems of Geroch and Bernal-Sánchez generalise accordingly, where now the Cauchy surfaces are manifolds with boundary, see \cite{SanchezBoundary,Solis}.
\end{remark}

The class of globally hyperbolic spacetimes is non-empty and contains most of the spacetimes relevant for physical applications. We briefly mention some important examples.

\begin{examples}\label{Examples:GlobHyp} (Globally Hyperbolic Spacetimes)
	\begin{itemize}
		\item[(i)]Let $(\Sigma,\sf{h})$ be a Riemannian manifold. Then, a Lorentzian manifold of the form
		\begin{align*}
			\sf{M}:=\bb{R}\times\Sigma\,,\qquad\sf{g}:=-\d t\otimes\d t+\mathrm{pr}_{2}^{\ast}\sf{h}
		\end{align*}
		with $\mathrm{pr}_{2}\:\sf{M}\to\Sigma$ being the natural projection, is globally hyperbolic if and only if $(\Sigma,\sf{h})$ is complete (see e.g.~\cite[Thm.~3.1]{SanchezUltra} or \cite[Prop.~5.2]{KayUltra}). Spacetimes of this form are called \textit{ultrastatic}. An explicit example is, of course, Minkowski spacetime
		\begin{align*}
			\bb{M}^{1,n}:=\bb{R}\times\bb{R}^{n}\,,\qquad\eta:=-\d t\otimes\d t+\sum_{i=1}^{n}\d x^{i}\otimes\d x^{i}\, .
		\end{align*}
		Note, however, that for more general \textit{sliced spacetimes} $(\sf{M}=\bb{R}\times\Sigma,\sf{g}=-\d t\otimes\d t+\sf{h}_{t})$, completeness of the slices $(\Sigma_{t},\sf{h}_{t})$ is neither a sufficient nor necessary criterion for globally hyperbolicity, see \cite{SanchezSliced} for a detailed discussion including counterexamples.
		\item[(ii)]A general class of globally hyperbolic manifolds generalising the previous example can be constructed as follows: consider a smooth function $f\in C^{\infty}(\bb{R},(0,\infty))$ and a Riemannian manifold $(\Sigma,\sf{h})$. Then, the \textit{(Lorentzian) warped product} $\sf{M}:=\bb{R}\times_{f}\Sigma$ defined by
		\begin{align*}
			\sf{M}=\bb{R}\times\Sigma\,,\qquad\sf{g}=-\d t\otimes\d t+f(t)\mathrm{pr}^{\ast}_{2}\sf{h}
		\end{align*}
		is globally hyperbolic if and only if $(\Sigma,\sf{h})$ is complete \cite[Thm.~3.66]{Beem}. An example is \textit{Friedmann–Lemaître–Robertson–Walker spacetime} (after \cite{Friedmann1,Friedmann2,Lemaitre1,Lemaitre2,Robertson1,Robertson2,Robertson3,Walker1}; see e.g.~\cite[Chap.~6]{Kriele} for a modern discussion) given by $\sf{M}=\bb{R}\times\Sigma$ with metric
		\begin{align*}
			\sf{g}:=-\d t\otimes \d t+a(t)^{2}\bigg(\frac{1}{1-kr^2}\d r\otimes\d r+r^2\d\mathbb{S}^{2}(\theta,\varphi)\bigg)\,,\qquad 
			\begin{cases}
				\Sigma=\mathbb{S}^{3} & k=1\\
				\Sigma=\bb{R}^{3} & k=0\\
				\Sigma=\mathbb{H}^{3} & k=-1
			\end{cases}
		\end{align*}
		for some \textit{scale factor} $a\in C^{\infty}(\bb{R},(0,\infty))$, where $\d\mathbb{S}^{2}(\theta,\varphi):=\d\theta\otimes\d\theta+\sin(\theta)^{2}\d\varphi\otimes\d\varphi$ is the standard Riemannian metric on $\mathbb{S}^{2}$, and where the domain of $r$ is $(0,1)$ for $k=1$ and $(0,\infty)$ for $k=0,-1$, which plays a central role in cosmology. An important example (for $k=1$) is \textit{de Sitter spacetime} (after \cite{Sitter1,Sitter2}; see \cite[Sec.~5.2]{HawkingEllis}), the maximal symmetric solution to the Einstein's equations with a positive cosmological constant.
		
		The result on warped products can also be generalised: for a spacetime $(\sf{N},\sf{k})$, a Riemannian manifold $(\Sigma,\sf{h})$ and $f\in C^{\infty}(\sf{N},(0,\infty))$, the warped product $\sf{M}=\sf{N}\times_{f}\Sigma$, i.e.
		\begin{align*}
			\sf{M}=\sf{N}\times\Sigma\,,\qquad\sf{g}=\mathrm{pr}_{1}^{\ast}\sf{k}+(f\circ\mathrm{pr}_{1})\mathrm{pr}_{2}^{\ast}\sf{h}\, ,
		\end{align*}
		is globally hyperbolic iff $(\sf{N},\sf{k})$ is globally hyperbolic and $(\Sigma,\sf{h})$ complete \cite[Thm.~3.68]{Beem}.
		\item[(iii)]Another important example is provided by \textit{Schwarzschild spacetime} (after \cite{Schwarzschild}), which is a spherically symmetric solution to the vacuum Einstein equations describing a (non-rotating) star or black hole. Let $m\in\bb{R}_{>0}$ and consider $\sf{M}_{\mathrm{I}}:=\bb{R}\times (0,2m)\times \mathbb{S}^{2}$ as well as $\sf{M}_{\mathrm{II}}:=\bb{R}\times (2m,\infty)\times \mathbb{S}^{2}$ equipped with the \textit{Schwarzschild metric}
		\begin{align*}
			\sf{g}=-f(r)\d t\otimes\d t+f(r)^{-1}\d r+\d r+r^2\d\mathbb{S}^{2}(\theta,\varphi)\qquad\text{with}\qquad f(r):=\bigg(1-\frac{2m}{r}\bigg)^{\frac{1}{2}}\, .
		\end{align*}
		The spacetime $(\sf{M}_{\mathrm{I}},\sf{g})$ describes the interior region of a black hole and $(\sf{M}_{\mathrm{II}},\sf{g})$ the gravitational field of a star or black hole in the exterior. Both of them are globally hyperbolic. The metric singularity at $r=2m$ is a coordinate singularity and can be overcome by considering a different set of coordinates, such as \textit{Eddington-Finkelstein coordinates} (after \cite{Eddington,Finkelstein}; see \cite{Penrose}). In this way, one can show that also $\sf{M}_{\mathrm{I}\cup\mathrm{II}}:=\sf{M}_{\mathrm{I}}\cup\sf{M}_{\mathrm{II}}\cup\mathcal{H}_{\mathrm{ev}}$, where $\mathcal{H}_{\mathrm{ev}}$ denotes their common boundary, the \textit{event horizon}, is globally hyperbolic. A proof can be found in \cite{Roken}, where a Cauchy surface for $\sf{M}_{\mathrm{I}\cup\mathrm{II}}$ is constructed. In fact, also the \textit{Kruskal-Szekeres spacetime} (see e.g.~\cite[Sec.~6.4]{Wald}, \cite{Kruskal,Szekeres}), the maximal extension of Schwarzschild, is globally hyperbolic. Similar statements can be made for the \textit{Kerr spacetime} (after \cite{Kerr}; see \cite{ONeillKerr} for details on Kerr geometry), that describes \emph{rotating} black holes (or stars), outside the \textit{Cauchy horizon}, see \cite[Prop.~5.1, 5.3, 5.4]{GerardHafner}.
\end{itemize}
\end{examples}

Examples of spacetimes that are not globally hyperbolic are the \emph{anti de Sitter-spacetime}, see e.g.~\cite[Sec.~3.5]{BaerBook}, and the \emph{Gödel universe}, see e.g.~\cite{Godel} and \cite[Sec.~5.7]{HawkingEllis}. Note that both of there examples are solutions of the vacuum Einstein equations.

Last but not least, we introduce some further terminology of causality. Let $(\sf{M},\sf{g})$ be a globally hyperbolic spacetime and $\sf{A}\subset\sf{M}$ be closed. Then,
\begin{itemize}
	\item[$\bullet$]$\sf{A}$ is called \textit{past compact}, if $\sf{A}\cap\mathcal{J}^{-}(p)$ is compact for all $p\in\sf{M}$.
	\item[$\bullet$]$\sf{A}$ is called \textit{past compact}, if $\sf{A}\cap\mathcal{J}^{+}(p)$ is compact for all $p\in\sf{M}$.
	\item[$\bullet$]$\sf{A}$ is called \textit{temporally compact}, if it is both past and future compact.
	\item[$\bullet$]$\sf{A}$ is called \textit{spatially compact}, if there exists a compact set $\sf{K}\subset\sf{M}$ such that $\sf{A}\subset\mathcal{J}(\sf{M})$.
\end{itemize}
It is easy to see that $\sf{A}$ is past (resp.~future) compact if and only if there exists a Cauchy surface $\Sigma\subset\sf{M}$ such that $\sf{A}\in\mathcal{J}^{+}(\Sigma)$ (resp.~$\sf{A}\subset\mathcal{J}^{-}(\Sigma)$), see e.g.~\cite[Lemma~1.2]{BaerGreen} or \cite[Thm.~3.1]{SandersCompact}. Furthermore, as proven in \cite[Thm.~2.2]{SandersCompact}, $\sf{A}$ is spatially compact if and only if $\sf{A}\cap\Sigma$ is compact for every spacelike Cauchy surface $\Sigma\subset\sf{M}$. Last but not least, if $\sf{E}\xrightarrow{\pi}\sf{M}$ is some real or complex vector bundle over $\sf{M}$, we denote by \begin{align*}
	\Gamma_{\mathrm{c}}^{\infty}(\sf{E})\quad\subset\quad\Gamma_{\mathrm{pc}}^{\infty}(\sf{E}),\,\Gamma_{\mathrm{fc}}^{\infty}(\sf{E}),\, \Gamma_{\mathrm{tc}}^{\infty}(\sf{E}),\,\Gamma_{\mathrm{sc}}^{\infty}(\sf{E})\quad\subset\quad\Gamma^{\infty}(\sf{E})
\end{align*}
the linear subspaces of $\Gamma^{\infty}(\sf{E})$ consisting of sections whose support is a past compact, future compact, temporally compact and spatially compact set, respectively.

\section{Hyperbolic Operators and Cauchy Problem}\label{Sec:CauchyProb}
Having established a suitable geometric background for the study of hyperbolic equations on manifolds, we now turn our attention to differential operators. We begin with a brief review of linear partial differential operators on manifolds, following the approach of Peetre in which those kind of operators are purely characterised by their locality property. Afterwards, we discuss two important classes of hyperbolic operators, namely symmetric hyperbolic systems and normally hyperbolic operators. In particular, we demonstrate that the Cauchy problem for each of these classes is well-posed. Last but not least, we discuss retarded/advanced Green's operators and the notion of Green hyperbolicity.

\subsection{Preliminaries on Linear Differential Operators}\label{Sec:LinOp}
The goal of the first section is to briefly review the concept of linear differential operators on manifolds. To start with, we recall some basic definitions of the \textit{local theory}, i.e.~linear differential operators on $\bb{R}^{d}$. Let $d,m,n\in\bb{N}$ and $\mathcal{U}\subset\bb{R}^{d}$ be open. A \textit{linear partial differential operator on} $\mathcal{U}$ is a $\bb{K}$-linear map $\sf{D}\:C^{\infty}(\mathcal{U},\mathbb{K}^{m})\to C^{\infty}(\mathcal{U},\bb{K}^{n})$ of the form
\begin{align}\label{eq:LinDiffOp}
	(\sf{D}u)(x)=\sum_{\alpha\in\bb{N}_{0}^{d},\,\vert\alpha\vert\leq k}a_{\alpha}(x)\partial_{x}^{\alpha}u(x),\qquad \forall u \in C^{\infty}(\mathcal{U},\mathbb{K}^{m}),\,x\in \mathcal{U}\, ,
\end{align}
for some $k\in\bb{N}_{0}$ and some coefficients $a_{\alpha}\in C^{\infty}(\mathcal{U},\bb{K}^{n\times m})$ labelled by multi-indices $\alpha\in\mathbb{N}_{0}^{d}$. In principle one can of course also consider differential operators whose coefficients have lower regularity, however, we shall only consider the smooth case here. The highest number $k$ for which there is at least one $\alpha\in\mathbb{N}_{0}^{d}$ with $\vert\alpha\vert=k$ such that $\alpha\neq 0$ is called the \textit{order} of $\sf{D}$. Any linear partial differential operator $\sf{D}$ is a \textit{local operator}, which means that $\mathrm{supp}(\sf{D}u)\subset\mathrm{supp}(u)$ for all $u\in C^{\infty}(\mathcal{U},\mathbb{K}^{m})$, as one can easily see from the definition. 

More generally, let $\mathsf{M}$ be a smooth manifold of dimension $d$. Consider two smooth $\mathbb{K}$-vector bundles $\mathsf{E} \xrightarrow{\pi_{\mathsf{E}}} \mathsf{M}$ and $\mathsf{F} \xrightarrow{\pi_{\mathsf{F}}} \mathsf{M}$ over $\mathsf{M}$ of finite rank $\mathrm{rank}_{\bb{K}}(\sf{E})=m$ and $\mathrm{rank}_{\bb{K}}(\sf{F})=n$, respectively. A $\mathbb{K}$-linear operator $\sf{D}\:\Gamma^{\infty}(\sf{E})\to\Gamma^{\infty}(\sf{F})$ is called \textit{local}, if
\begin{align*}
	\mathrm{supp}(\sf{D}\psi)\subset\mathrm{supp}(\psi)\,\qquad\forall\psi\in\Gamma^{\infty}(\sf{E})\, .
\end{align*}
That is, for local operators, the value of $\mathsf{D}\psi$ at a point $p \in \mathsf{M}$ depends only on the behaviour of $\psi$ in an arbitrarily small neighbourhood of $p$. More precisely, if $\psi,\varphi \in \Gamma^{\infty}(\mathsf{E})$ are sections that agree on an open set $\mathcal{U} \subset \mathsf{M}$, then also $\mathsf{D}\psi = \mathsf{D}\varphi$ on $\mathcal{U}$. Indeed, linearity and locality imply
\begin{align*}
	\mathrm{supp}(\sf{D}\psi-\sf{D}\varphi)=\mathrm{supp}(\sf{D}(\psi-\varphi))\subset\mathrm{supp}(\psi-\varphi)\subset\sf{M}\backslash\mathcal{U}\, ,
\end{align*}
which shows that $\mathsf{D}\psi - \mathsf{D}\varphi=0$ on $\mathcal{U}$. In particular, locality allows us to define the \textit{restricted operator} $\sf{D}\vert_{\mathcal{U}}\:\Gamma^{\infty}(\mathcal{U},\sf{E})\to\Gamma^{\infty}(\mathcal{U},\sf{F})$ for every $\mathcal{U}\subset\sf{M}$ open in such a way that
\begin{align*}
	\sf{D}\vert_{\mathcal{U}}(\psi\vert_{\mathcal{U}})=(\sf{D}\psi)\vert_{\mathcal{U}}\,,\qquad\forall\psi\in\Gamma^{\infty}(\sf{E})\, .
\end{align*}

Now, let $(\mathcal{U}, \phi)$ be a local coordinate chart of $\mathsf{M}$, with $\phi \colon \mathcal{U} \xrightarrow{\cong} \mathcal{V} \subset \mathbb{R}^d$, and suppose that $\mathcal{U}$ has been chosen small enough to trivialise also both the vector bundles $\mathsf{E}$ and $\mathsf{F}$. Moreover, let us denote the corresponding local trivialisations by
\begin{align*}
	\Phi_{\sf{E}}\:\pi_{\sf{E}}^{-1}(\mathcal{U})\xrightarrow{\cong}\mathcal{U}\times\mathbb{K}^{m},\qquad \Phi_{\sf{F}}\:\pi_{\sf{F}}^{-1}(\mathcal{U})\xrightarrow{\cong}\mathcal{U}\times\mathbb{K}^{n}\, .
\end{align*}
As usual, these identifications induce bijections between the modules of smooth local sections and smooth functions on $\mathcal{U}$, which we denote by
\begin{align*}
	\Gamma_{\sf{E}}&\:\Gamma^{\infty}(\mathcal{U},\sf{E})\xrightarrow{\cong} C^{\infty}(\mathcal{U},\mathbb{K}^{m}),\qquad \psi\mapsto \mathrm{pr}_{2}\circ\Phi_{\sf{E}}\circ\psi\,,\\
	\Gamma_{\sf{F}}&\:\Gamma^{\infty}(\mathcal{U},\sf{F})\xrightarrow{\cong} C^{\infty}(\mathcal{U},\mathbb{K}^{n}),\qquad \psi\mapsto \mathrm{pr}_{2}\circ\Phi_{\sf{F}}\circ\psi\, .
\end{align*}
We then define the \textit{local representation} of $\mathsf{D}$ in coordinates to be the $\bb{K}$-linear operator 
\begin{align*}
	\sf{D}_{\mathcal{U}}\:C^{\infty}(\mathcal{V},\bb{K}^{m})\to C^{\infty}(\mathcal{V},\bb{K}^{n})\,,\qquad \sf{D}_{\mathcal{U}}u:=(\Gamma_{\sf{F}}\circ\sf{D}\vert_{\mathcal{U}}\circ\Gamma_{\sf{E}}^{-1})(u\circ\phi)\circ\phi^{-1}\, .
\end{align*}

Now, a natural way to define differential operators on manifolds is to consider linear and local operators whose local representations are differential operators as defined in~Eq.~\eqref{eq:LinDiffOp}. Surprisingly, the second condition is actually redundant: any local operator between spaces of smooth sections is automatically a differential operator in the local sense. This is exactly the content of Peetre's theorem, whose elegant proof we sketch here for completeness.

\begin{theorem}\label{Thm:Peetre} \emph{(Theorem of Peetre \cite{Peetre1,Peetre2})}\newline
	Let $\sf{D}\:\Gamma^{\infty}(\sf{E})\to\Gamma^{\infty}(\sf{F})$ be a $\bb{K}$-linear operator. Then, $\sf{D}$ is local if and only if for every $p\in\sf{M}$, there exists $m\in\mathbb{N}_{0}$ and a local chart $(\mathcal{U},\phi)$ around $p$, trivialising also $\sf{E},\sf{F}$, such that $\sf{D}_{\mathcal{U}}$ is a differential operator of order $m$, i.e.~there are $a_{\alpha}\in C^{\infty}(\phi(\mathcal{U}),\bb{K}^{n\times m})$ with
	\begin{align*}
		\sf{D}_{\mathcal{U}}=\sum_{\alpha\in\bb{N}_{0}^{d},\,\vert\alpha\vert\leq m}a_{\alpha}(\cdot)\partial_{x}^{\alpha}\, .
\end{align*}
\end{theorem}

\begin{proof}[Proof (sketch).]
	The direction ``$\Leftarrow$'' is obvious. For ``$\Rightarrow$'', we follow the proof of \cite[Thm.~6.1 and 6.2]{Kahn}. W.l.o.g., we consider the scalar case $\sf{D}\:C^{\infty}(\sf{M}) \to C^{\infty}(\sf{M})$, since the bundle-valued case can be reduced to this by working componentwise and shrinking open neighbourhoods sufficiently to trivialise the bundles involved.

	We begin with a local observation: let $\sf{Q}\:C^{\infty}(\mathcal{V}) \to C^{\infty}(\mathcal{V})$ be any $\bb{K}$-linear and local operator on an open set $\mathcal{V} \subset \bb{R}^{d}$. Then, for every $x \in \mathcal{V}$, there exists an open neighborhood $\mathcal{W} \subset \mathcal{V}$ of $x$, an integer $m \in \mathbb{N}$, and a constant $C > 0$ such that
	\begin{align*}
		\Vert\sf{Q}u\Vert_{\infty} \leq C \max_{\Vert\alpha\Vert \leq m} \Vert\partial^{\alpha}u\Vert_{\infty} =: C \Vert u \Vert_{C^{m}} \qquad \forall u \in C_{\mathrm{c}}^{\infty}(\mathcal{W} \setminus \{x\})\,.
	\end{align*}
	This estimate can be established via a contradiction argument (see, e.g.,~\cite[Lemma~6.3]{Kahn}). Next, suppose $\mathcal{W} \subset \mathcal{V}$ is open and satisfies the estimate
	\begin{align*}
		\Vert\sf{Q}u\Vert_{\infty} \leq C \Vert u \Vert_{C^{m}} \qquad \forall u \in C_{\mathrm{c}}^{\infty}(\mathcal{W})
	\end{align*}
	for some $C > 0$ and $m \in \mathbb{N}$. Define a distribution $\sf{Q}_{x}\in\mathcal{D}^{\prime}(\mathcal{W})$ by $\sf{Q}_{x}(u) := (\sf{Q}u)(x)$. Then,
	\begin{align*}
		\vert\langle \sf{Q}_{x}, u \rangle_{\mathcal{W}} \vert \leq C \Vert u \Vert_{C^{m}} \qquad \forall u \in C_{\mathrm{c}}^{\infty}(\mathcal{W})\,,
	\end{align*}
	where $\langle \cdot, \cdot \rangle_{\mathcal{W}}$ denotes the distributional pairing on $\mathcal{W}$. Hence, $\sf{Q}_{x}$ is a distribution of finite order $m$ supported at $\{x\}$. By the structure theorem for distributions \cite[Thm.~6.25]{Rudin}, it follows that $\sf{Q}_{x}$ is a finite linear combination of derivatives of the Dirac distribution $\delta_{x} \in \mathcal{D}^{\prime}(\mathcal{W})$. That is, there exist coefficients $a_{\alpha} \in C^{\infty}(\mathcal{W}, \bb{K})$ such that
	\begin{align*}
		(\sf{Q}u)(x) = \langle \sf{Q}_{x}, u \rangle_{\mathcal{W}} = \sum_{\vert\alpha\vert \leq m} a_{\alpha}(x) \partial^{\alpha}u(x) \qquad \forall u \in C_{\mathrm{c}}^{\infty}(\mathcal{W})\,.
	\end{align*}
	By locality, this formula extends to all $u \in C^{\infty}(\mathcal{W})$, e.g.~by using a suitable cut-off function.

	We now prove the theorem: let $p \in \sf{M}$ and let $(\varphi, \mathcal{U})$ be a relatively compact chart around $p$. Set $\mathcal{V} := \varphi(\mathcal{U})$. By relative compactness, we can cover $\mathcal{V}$ by finitely many open sets $(\mathcal{V}_{i})_{1 \leq i \leq k}$, each containing a point $x_{i} \in \mathcal{V}_{i}$ satisfying the above lemma with order $m_{i}$. Using a subordinate partition of unity, it follows that there exists $C > 0$ such that
	\begin{align*}
		\Vert\sf{Q}u\Vert_{\infty} \leq C \Vert u \Vert_{C^{m}} \qquad \forall u \in C_{\mathrm{c}}^{\infty}(\mathcal{V} \setminus \{x_{1}, \dots, x_{k}\})\,,
	\end{align*}
	where $m := \max_{i} m_{i}$. By locality and continuity, this inequality extends to all $u \in C^{\infty}(\mathcal{V})$.
\end{proof}

Hence, if $\sf{D}\:\Gamma^{\infty}(\sf{E})\to\Gamma^{\infty}(\sf{F})$ is a linear and local operator, it is locally a linear differential operator of finite order. From the Leibniz rule, it is clear that the order at some fixed point $p\in\sf{M}$ is independent of the chosen charts and trivialisations. We hence, define the \textit{order} of $\sf{D}$ to be the supremum of the order of the local representations at each point. We stress that this number might as well be infinite if $\sf{M}$ is non-compact. To sum up:

\begin{definition} (Linear Differential Operators)\newline
Let $\sf{E},\sf{F}$ be two finite-rank $\bb{K}$-vector bundles over a smooth manifold $\sf{M}$. A linear and local operator $\sf{D}\:\Gamma^{\infty}(\sf{E})\to\Gamma^{\infty}(\sf{F})$ of order $m$ is called a \emph{linear differential operator of order (at most) $m$}. We denote the space of all such operators by $\mathrm{DO}^{m}(\sf{E},\sf{F})$ and write
\begin{align*}
	\mathrm{DO}(\sf{E},\sf{F}):=\bigcup_{m\in\mathbb{N}_{0}}\mathrm{DO}^{m}(\sf{E},\sf{F})
\end{align*}
\end{definition}

In the case $\sf{E}=\sf{F}$, we will also use the simplified notations $\mathrm{DO}^{m}(\sf{E}):=\mathrm{DO}^{m}(\sf{E},\sf{E})$ for all $m\in\bb{N}_{0}$ and $\mathrm{DO}(\sf{E}):=\mathrm{DO}(\sf{E},\sf{E})$. We also observe that the space of zero-order operators $\mathrm{DO}^{0}(\sf{E},\sf{F})$ is a $C^{\infty}(\sf{M})$-module and
\begin{align*}
	\mathrm{DO}^{0}(\sf{E},\sf{F})=\mathrm{Hom}_{C^{\infty}(\sf{M})}(\Gamma^{\infty}(\sf{E}),\Gamma^{\infty}(\sf{F}))\cong\Gamma^{\infty}(\mathrm{Hom}(\sf{E},\sf{F}))\, .
\end{align*}

The highest-order coefficient in the local representation of a differential operator can easily be seen to behave behave tensorially. This gives rise to the \textit{principal symbol} defined as follows.

\begin{definition} (Principal Symbol)\newline
	For $\sf{D}\in\mathrm{DO}^{m}(\sf{E},\sf{F})$, the \emph{principal symbol} is the map $\sigma_{\sf{D}}\:\sf{T}^{\ast}\sf{M}\to\mathrm{Hom}(\sf{E},\sf{F})$ defined by 
\begin{align*}
	\sigma_{\sf{D}}(\xi)e:=\frac{1}{m!}\sf{D}(f^{m}\widetilde{e})\vert_{p}
\end{align*}
for any $e\in\sf{E}_{p}$, $p\in\sf{M}$ and $\xi\in\sf{T}_{p}\sf{M}$, where $f\in C^{\infty}(\sf{M})$ is such that $\xi=\d f(p)$ and $f(p)=0$, and where $\widetilde{e}\in\Gamma^{\infty}(\sf{E})$ is such that $\widetilde{e}(p)=e$.
\end{definition}

By the Leibnitz rule and locality, it is easy to check that the definition is independent of the choice of $f$ and extension $\widetilde{e}$. If $\mathcal{U}$ is a local coordinate chart around $p$ with local representative $\sf{D}_{\mathcal{U}}$ given by $\sf{D}_{\mathcal{U}}=\sum_{\alpha\in\bb{N}_{0}^{d},\,\vert\alpha\vert\leq k}a_{\alpha}(x)\partial_{x}^{\alpha}$, then it holds that
\begin{align*}
	\sigma_{\sf{D}}(\xi)=\sum_{\vert\alpha\vert=k}a_{\alpha}\xi^{\alpha}\, .
\end{align*}
By definition, the principal symbol can be viewed as a section of $\mathrm{Hom}(\pi^{\ast}\sf{E},\pi^{\ast}\sf{F})$, where $\pi\:\sf{T}^{\ast}\sf{M}\to\sf{M}$ denotes the bundle projection. Alternatively, the principal symbol can also be understood as a section of $\sf{T}\sf{M}^{\otimes_{s}k}\otimes\mathrm{Hom}(\sf{E},\sf{F})$, see e.g.~\cite[Rmk.~5.7.2]{RudolphSchmidt2}.

We need one more ingredient: let $(\sf{M},\sf{g})$ be a pseudo-Riemannian manifold and $(\sf{E},\langle\cdot,\cdot\rangle_{\sf{E}})$ be a Hermitian vector bundle. Following the notation section of this thesis, we denote by
\begin{align*}
	(\psi,\varphi)_{\sf{E}}:=\int_{\sf{M}}\langle\psi(p),\varphi(p)\rangle_{\sf{E}_{p}}\,\d\mu_{\sf{g}}(p)\,,
\end{align*}
with $\mathrm{supp}(\psi)\cap\mathrm{supp}(\varphi)$ compact, the induced non-degenerate sesquilinear form on $\Gamma^{\infty}(\sf{E})$.

\begin{definition} (Formal Adjoints and Self-Adjointness)\newline
	Let $(\sf{M},\sf{g})$ is a pseudo-Riemannian manifold and $(\sf{E},\langle\cdot,\cdot\rangle_{\sf{E}})$, $(\sf{F},\langle\cdot,\cdot\rangle_{\sf{F}})$ be two Hermitian vector bundles over $\sf{M}$. For $\sf{D}\in\mathrm{DO}^{m}(\sf{E},\sf{F})$, we call an operator $\sf{D}^{\ast}\in\mathrm{DO}^{m}(\sf{F},\sf{E})$ its \emph{formal adjoint} with respect to $(\cdot,\cdot)_{\sf{E}}$ and $(\cdot,\cdot)_{\sf{F}}$, if, for all $\psi\in\Gamma^{\infty}(\sf{E})$ and $\varphi\in\Gamma^{\infty}(\sf{F})$ with $\mathrm{supp}(\psi)\cap\mathrm{supp}(\varphi)$ compact,
\begin{align*}
	(\sf{D}\psi,\varphi)_{\sf{F}}=(\psi,\sf{D}^{\ast}\varphi)_{\sf{E}}\, .
\end{align*}
\end{definition}

If $\sf{F}=\sf{E}$ and $\sf{D}=\sf{D}^{\ast}$, we call $\sf{D}$ \emph{formally self-adjoint}, as usual. It is well known that the formal adjoint always exists and is unique, see e.g.~\cite[Sec.~1.1.3]{BaerLecture}.\newpage

In the remainder of this section, we introduce some important concepts that will later enable the application of functional-analytic methods to the study differential operators. Let $(\sf{M},\sf{g})$ be a \textit{Riemannian} manifold and $(\sf{E},\langle\cdot,\cdot\rangle_{\sf{E}})$, $(\sf{F},\langle\cdot,\cdot\rangle_{\sf{F}})$ be two Hermitian vector bundles over $\sf{M}$. Furthermore, let $\sf{D}\in\mathrm{DO}(\sf{E},\sf{F})$ be a linear differential operator. Now, we would like to view $\sf{D}$ operator as a \textit{closed} operator in some Hilbert space. The natural Hilbert space to consider is, of course, the space of $\sf{L}^{2}$-sections, which we denote by
\begin{align}\label{eq:L2}
	\sf{L}^{2}(\sf{M}):=\sf{L}^{2}(\sf{M},\sf{E}):=\overline{\Gamma_{\mathrm{c}}^{\infty}(\sf{E})}^{\Vert\cdot\Vert_{\sf{L}^{2}(\sf{E})}},\qquad \langle\cdot,\cdot\rangle_{\sf{L}^{2}(\sf{E})}:=(\cdot,\cdot)_{\sf{E}}=\int_{\sf{M}}\,\langle\cdot,\cdot\rangle_{\sf{E}}\,\d\mu_{g}\, .
\end{align}
Now, both $\sf{D}$ and its formal adjoint $\sf{D}^{\ast}$ can be viewed as densely-defined operators 
\begin{align*}
	\sf{D}\:\mathrm{dom}(\sf{D})\to\sf{L}^{2}(\sf{F})\qquad\text{and}\qquad \sf{D}^{\ast}\:\mathrm{dom}(\sf{D}^{\ast})\to\sf{L}^{2}(\sf{E})\, ,
\end{align*}
with\footnote{In principle, one could of course also take any domain $\mathrm{dom}(\sf{D})\subset\Gamma^{\infty}(\sf{E})\cap\sf{L}^{2}(\sf{E})$ with $\mathrm{ran}(\sf{D}\vert_{\mathrm{dom}(\sf{D})})\subset\sf{L}^{2}(\sf{F})$ that is dense in $\sf{L}^{2}(\sf{E})$, for instance, $\mathrm{dom}(\sf{D})=\{\omega\in\Gamma^{\infty}(\sf{E})\cap\sf{L}^{2}(\sf{E})\mid\sf{D}\omega\in\sf{L}^{2}(\sf{F})\}$.} $\mathrm{dom}(\sf{D}):=\Gamma_{\mathrm{c}}^{\infty}(\sf{E})$ and $\mathrm{dom}(\sf{D}^{\ast}):=\Gamma_{\mathrm{c}}^{\infty}(\sf{F})$. In order to distinguish the \textit{functional-analytic} $\sf{L}^{2}$-adjoint from the \textit{formal adjoint}, we denote the former by $\sf{D}^{\dagger}\:\mathrm{dom}(\sf{D}^{\dagger})\to\sf{L}^{2}(\sf{F})$, where
\begin{align}\label{eq:DomAdj}
	\mathrm{dom}(\sf{D}^{\dagger}):=\{\psi\in\sf{L}^{2}(\sf{F})\mid \exists\eta\in\sf{L}^{2}(\sf{E}):\forall\varphi\in\mathrm{dom}(\sf{D}):\langle \psi,\sf{D}\varphi\rangle_{\sf{L}^{2}(\sf{F})}=\langle \eta,\varphi\rangle_{\sf{L}^{2}(\sf{E})}\}\,,
\end{align}
as usual. Now, the operator $\sf{D}\:\mathrm{dom}(\sf{D})\to\sf{L}^{2}(\sf{F})$ is clearly not a \textit{closed} operator. However, it is clearly \textit{closable}, since an operator in a Hilbert space is closable if and only if its adjoint is densely defined, which is clearly the case for $\sf{D}$, since its $\sf{L}^{2}$-adjoint $\sf{D}^{\dagger}$ contains at least the domain of the formal adjoint $\sf{D}^{\ast}$, i.e.~$\Gamma_{\mathrm{c}}^{\infty}(\sf{F})\subset\mathrm{dom}(\sf{D}^{\dagger})$. Let us introduce the following terminology:

\begin{definition} (Minimal and Maximal Closed Extension)\newline
Let us view $\sf{D}\in\mathrm{DO}(\sf{E},\sf{F})$ as a densely-defined operator of the form $\sf{D}\:\mathrm{dom}(\sf{D})\to\sf{L}^{2}(\sf{F})$ with $\mathrm{dom}(\sf{D})=\Gamma_{c}^{\infty}(\sf{E})$. We define the following two closed extensions:
\begin{itemize}
	\item[(i)]$\sf{D}_{\mathrm{min}}:=\overline{\sf{D}}$ \hfill (\textit{minimal closed extension})
	\item[(ii)]$\sf{D}_{\mathrm{max}}:=(\sf{D}^{\ast})^{\dagger}$ \hfill (\textit{maximal closed extension})
\end{itemize}
\end{definition}

By definition, the \textit{minimal closed extension} $\sf{D}_{\mathrm{min}}$ is exactly the functional-analytic \textit{closure} of the operator $\sf{D}\:\mathrm{dom}(\sf{D})\to\sf{L}^{2}(\sf{E})$. In other words, it is the operator with domain
\begin{align*}
	\mathrm{dom}(\mathrm{D}_{\mathrm{min}}):=\{\omega\in\sf{L}^{2}(\sf{E})\mid \exists \{\omega_{n}\}_{n\in\bb{N}}\subset\mathrm{dom}(\sf{D}):\lim_{n\to\infty}\omega_{n}=\omega\text{ and }\{\sf{D}\omega_{n}\}\text{ convergent in }\sf{L}^{2}(\sf{F})\}
\end{align*}
defined by $\sf{D}\omega:=\lim_{n\to\infty}\sf{D}\omega_{n}$ for all $\omega\in \mathrm{dom}(\mathrm{D}_{\mathrm{min}})$ and $\{\omega_{n}\}_{n\in\mathbb{N}}$ as above. The fact that this is independent of the chosen sequence follows from the fact that $\sf{D}$ is closable.

The domain of the maximal closed extension is essentially the largest domain in which the operator $\sf{D}$ still acts as a differential operator in a suitable sense. Using distribution theory, its domain can be characterised as
\begin{align}\label{eq:MaxClExtDom}
	\mathrm{dom}(\mathrm{D}_{\mathrm{max}})=\{\omega\in\sf{L}^{2}(\sf{E})\mid \sf{D}\omega\in\sf{L}^{2}(\sf{F})\text{ in the distributional sense }\}\, .
\end{align}
To see this, we first observe that any differential operator $\sf{D}\in\mathrm{DO}(\sf{E},\sf{F})$ can be lifted to a linear operator $\sf{D}\:\mathcal{D}^{\prime}(\sf{F})\to\mathcal{D}^{\prime}(\sf{E})$ on the level of distributions (see Appendix~\ref{Appendix:Micro} for the relevant notation and terminology) by duality, i.e.~$\langle\sf{D}u,\varphi\rangle_{\sf{M}}=\langle u,\sf{D}^{\ast}\varphi\rangle_{\sf{M}}$ for all $u\in \sf{D}^{\prime}(\sf{E})$ and $ \varphi\in\Gamma^{\infty}_{\mathrm{c}}(\sf{F})$, where $\langle\cdot,\cdot\rangle_{\sf{M}}$ denotes the distributional pairing. Now, saying that $\sf{D}\omega\in\sf{L}^{2}(\sf{F})$ for $\omega\in\sf{L}^{2}(\sf{E})$ ``\textit{in the distributional sense}'' means, by definition, that there exists a section $\eta\in\sf{L}^{2}(\sf{F})$ such that $\sf{D}\sf{T}_{\omega}=\sf{T}_{\eta}$, where $\sf{T}_{\omega}\in\mathcal{D}^{\prime}(\sf{E})$ denotes the regular distribution $\sf{T}_{\omega}(\cdot):=\langle\omega,\cdot\rangle_{\sf{L}^{2}(\sf{E})}$, as usual. But this is equivalent to require
\begin{align*}
	\forall\varphi\in\Gamma_{\mathrm{c}}^{\infty}(\sf{F}):\qquad \langle\omega,\sf{D}^{\ast}\varphi\rangle_{\sf{L}^{2}(\sf{E})}=\langle\eta,\varphi\rangle_{\sf{L}^{2}(\sf{F})}\, .
\end{align*}
Hence, the set~\eqref{eq:MaxClExtDom} is exactly the domain $\mathrm{dom}((\sf{D}^{\ast})^{\dagger})$ defined in the usual way (see~Eq.~\eqref{eq:DomAdj}). \bigskip

As the name and definitions suggests, the closed operator $\sf{D}_{\mathrm{max}}$ is an extension of $\sf{D}_{\mathrm{min}}$. Indeed, since $\sf{D}^{\ast}\subset\sf{D}^{\dagger}$, it follows that 
\begin{align*}
	\sf{D}_{\mathrm{min}}=\sf{D}^{\dagger\dagger}\subset (\sf{D}^{\ast})^{\dagger}=\sf{D}_{\mathrm{max}}\, .
\end{align*}
In general, however, the two domain do not coincide and hence, $\sf{D}_{\mathrm{min}}$ and $\sf{D}_{\mathrm{max}}$ are just two of many possible closed extensions that one might define in between. So, one question one might ask is for which types of differential operators there exists a \textit{canonical} closed extension. Furthermore, having discussed suitable closed extensions of linear partial differential operators, the next question to answer is when a \textit{formally self-adjoint operator} $\sf{D}$ can be extended to an honest \textit{self-adjoint operator} in the Hilbert space $\sf{L}^{2}(\sf{E})$ in a unique way. Both of these questions are addressed in the following proposition:

\begin{proposition} \emph{(Essentially Self-Adjointness and Closed Extensions)}\newline
	Let $\sf{D}\in\mathrm{DO}(\sf{E})$ be a formally self-adjoint linear partial differential operator, which we view as a densely-defined operator $\sf{D}\:\mathrm{dom}(\sf{D})\to\sf{L}^{2}(\sf{E})$ with $\mathrm{dom}(\sf{D})=\Gamma_{c}^{\infty}(\sf{E})$. Then,
\begin{align*}
	\sf{D}\text{ is essentially self-adjoint }\quad\Leftrightarrow\quad\sf{D}_{\mathrm{min}}=\sf{D}_{\mathrm{max}}\quad\Leftrightarrow\quad\mathrm{dom}(\sf{D}_{\mathrm{max}}) \subset \mathrm{dom}(\sf{D}_{\mathrm{min}})\, . 
\end{align*}	
\end{proposition}

\begin{proof}
	By definition, $\sf{D}\:\mathrm{dom}(\sf{D})\to\sf{L}^{2}(\sf{E})$ is essentially self-adjoint if and only if $\overline{\sf{D}}$ is self-adjoint, i.e.~$\overline{\sf{D}}^{\dagger}=\overline{\sf{D}}$. Now, its a general fact that the adjoint of the closure coincides with the closure, since $\overline{\sf{D}}^{\dagger}=(\sf{D}^{\dagger\dagger})^{\dagger}=(\sf{D}^{\dagger})^{\dagger\dagger}=\overline{(\sf{D}^{\dagger})}=\sf{D}^{\dagger}$, where we used the fact that the adjoint of any operator is closed. Hence, essentially self-adjointness is equivalent to $\sf{D}^{\dagger}=\overline{\sf{D}}$. But this is equivalent to $\sf{D}_{\mathrm{min}}=\sf{D}_{\mathrm{max}}$, by definition. For the second equivalence, we recall that $\mathrm{dom}(\sf{D}_{\mathrm{max}}) \supset \mathrm{dom}(\sf{D}_{\mathrm{min}})$ since $\sf{D}_{\mathrm{min}}\subset\sf{D}_{\mathrm{max}}$, as mentioned above. Hence, $\sf{D}_{\mathrm{min}}=\sf{D}_{\mathrm{max}}$ if and only if also the other inclusion is true.
\end{proof}

On a practical level, showing $\mathrm{dom}(\sf{D}_{\mathrm{max}}) \subset \mathrm{dom}(\sf{D}_{\mathrm{min}})$ explicitly for a given operator be quite complicated, however, it turns out to be true for Laplace-type and Dirac-type operators. The concept of minimal and maximal closed extensions is standard in functional analysis and operator theory, see, for instance, \cite[Chap.~4]{Grubb} and \cite[pp.~73ff.]{EichhornBook} for general references.

\subsection{Symmetric Hyperbolic Systems}
\label{Sec:SHS}
Having established the general framework of linear differential operators on smooth manifolds, we now focus on the class of \textit{hyperbolic} operators defined on globally hyperbolic Lorentzian manifolds. Broadly speaking, the notion of \textit{hyperbolicity} is fundamentally tied to the well-posedness of the Cauchy problem for initial data prescribed on a non-characteristic hypersurface. We begin by examining two particularly important classes of hyperbolic equations. In Section~\ref{Sec:GreenCauchyHyp}, we extend our discussion to more general formulations. 

In this section, we consider a very general class of first-order hyperbolic operators, namely \textit{symmetric hyperbolic systems}. Throughout this section, we fix the following notation.
\begin{itemize}
	\item[$\bullet$]Let $(\sf{M},\sf{g})$ be a globally hyperbolic Lorentzian manifold of dimension $d=1+n$, $n\in\bb{N}$. 
	\item[$\bullet$]Let $(\sf{E}\xrightarrow{\pi}\sf{M},\langle\cdot,\cdot\rangle_{\sf{E}})$ be a Hermitian $\bb{K}$-vector bundle of rank $\mathrm{rank}_{\bb{K}}(\sf{E})=\mathrm{N}\in\bb{N}$, equipped with a metric-compatible connection $\nabla^{\sf{E}}$. 
\end{itemize}

\begin{definition}\label{Def:SHS} (Symmetric Hyperbolic System, \cite[Def.~5.1]{BaerGreen})\newline
A first-order operator $\sf{S}\in\mathrm{DO}^{1}(\sf{E})$ is called a \emph{symmetric hyperbolic system}, if its principal symbol $\sigma_{\sf{S}}\:\sf{T}^{\ast}\sf{M}\to\mathrm{End}(\sf{E})$ has the following properties.
\begin{itemize}
	\item[(i)]$\sigma_{\sf{S}}(\xi)$ is Hermitian with respect to $\langle\cdot,\cdot\rangle_{\sf{E}_{p}}$ for all $\xi\in\sf{T}^{\ast}_{p}\sf{M}$, $p\in\sf{M}$.
	\item[(ii)]$\langle\sigma_{\sf{S}}(\tau)\cdot,\cdot\rangle_{\sf{E}_{p}}$ is positive-definite for all future-directed\footnote{A \textit{co}vector $\xi\in\sf{T}^{\ast}\sf{M}$ is \textit{future/past-directed} if $\xi(v)>0$ for all future/past-directed $v\in\sf{T}\sf{M}$. In particular, the musical isomorphism map future-directed to past-directed objects and vice versa.} timelike $\tau\in\sf{T}_{p}^{\ast}\sf{M}$, $p\in\sf{M}$.
\end{itemize}
\end{definition}

The following example shows how this definition is related to the usual one found in classical PDE literature, see e.g.~the textbooks \cite{TaylorI,John} and the classical articles \cite{Friedrichs1,Friedrichs2}.

\begin{example}
	We consider $(1+n)$-dimensional Minkowski spacetime $(\bb{M}^{1,n}:=\bb{R}^{n+1},\eta)$ together with the trivial vector bundle $\sf{E}:=\underline{\bb{R}}_{\sf{M}}=\sf{M}\times\bb{R}^{\mathrm{N}}$ equipped with the standard complex inner product on each fibre. Any first-order operator on $\sf{M}$ can be written in the form
\begin{align*}
	\sf{S}:=\sf{A}^{0}\partial_{t}+\sf{A}^{j}\partial_{j}+\sf{S}_{0}\, ,
\end{align*}	
where $\sf{A}^{0},\sf{A}^{j},\sf{S}_{0}\in C^{\infty}(\bb{M}^{1,n},\bb{C}^{\mathrm{N}\times\mathrm{N}})$. Now, in this case, property (i) in Definition~\ref{Def:SHS} translates to the requirement that
\begin{align*}
	(\sf{A}^{0}(t,\vec{x}))^{\dagger}=\sf{A}^{0}(t,\vec{x})\,,\qquad (\sf{A}^{j}(t,\vec{x}))^{\dagger}=\sf{A}^{j}(t,\vec{x})\qquad\forall (t,\vec{x})\in\bb{M}^{1,n}\, .
\end{align*}
Furthermore, any future-directed timelike covector on $\bb{M}^{1,n}$ can (up to a positive function) be written as $\tau=\d t+\alpha_{i}\d x^{i}$ for some coefficient functions $\alpha_{i}$ satisfying $\sum_{j}\alpha_{i}^{2}<1$. Hence, condition (ii) is the requirement that
\begin{align*}
	(\sf{A}^{0}+\alpha_{j}\sf{A}^{j})(t,\vec{x})\qquad \text{ is positive definite }\qquad\forall (t,\vec{x})\in\bb{M}^{1,n}\, .
\end{align*}
This requirement seems a bit stronger than the one found in classical PDE literature, in which one usually just requires $\sf{A}^{0}$ to be positive-definite. However, since the principal symbol $\sigma_{\sf{S}}(\xi)=\sf{A}^{\mu}\xi_{\mu}$ depends smoothly on the coordinates, the positive definiteness of $\sigma_{\sf{S}}(\d t)=\sf{A}^{0}$ extends to small perturbations of $\d t$, i.e.~to $\sigma_{\sf{S}}(\d t+\alpha_{i}\d x^{i})=\sf{A}^{0}+\alpha_{i}\sf{A}^{i}$ for sufficiently small $\alpha_{i}$. 
\end{example} 

The preceding example shows that Definition~\ref{Def:SHS} generalises the standard notion of symmetric hyperbolic systems to Lorentzian manifolds in a covariant manner, i.e.~without the need of specifying a time function to define hyperbolicity. In the following, we will look into some more examples that are of direct interest for applications in theoretical and mathematical physics.

\begin{example}\label{Ex:Maxwell} (Maxwell's Equations on Ultrastatic Manifolds)\newline
	An example of a symmetric hyperbolic system is provided by Maxwell's equations in ultrastatic manifolds, which in this context has been discussed in \cite[Sec.~5]{SchmidMurroFinster}, \cite[Ex.~3.7.3]{BaerLecture} and \cite[App.~D]{CapoferriCurl}. To start with, consider a $(1+3)$-dimensional globally hyperbolic ultrastatic spacetime
	\begin{align*}
		\sf{M}:=\bb{R}\times\Sigma\,\qquad \sf{g}:=-\d t\otimes\d t+\sf{h}\, ,
	\end{align*}
	where $(\Sigma,\sf{h})$ is a complete three-dimensional Riemannian manifold. On $(\Sigma,\sf{h})$, we define the \textit{divergence} and \textit{curl operator} in local coordinates $(x^{i})_{i=1,2,3}$ of $\Sigma$ by
\begin{align*}
	\mathrm{div}_{\Sigma}(X):=\nabla_{i}X^{i},\qquad \mathrm{curl}_{\Sigma}(X)_{i}:=[(\ast\d X^{\flat})^{\sharp}]_{i}=\epsilon_{ijk}\nabla^{j}\vec{X}^{k}
\end{align*}
for all $X\in\mathfrak{X}(\Sigma)$, where $\nabla$ and $\epsilon$ denote the Levi-Civita connection and pseudotensor with sign convention $\epsilon_{123}=\sqrt{\mathrm{det}(h)}$ on $(\Sigma,\sf{h})$, respectively. Time-dependent vector fields on $\Sigma$ can be identified with sections of the pullback bundle $\sf{F}:=\pi_{2}^{\ast}(\sf{T}\Sigma)\to\sf{M}$, where $\pi_{2}\colon\sf{M}\to\Sigma$ denotes the natural projection. We equip $\sf{F}$ with the bundle metric induced by $\sf{h}$.

While Maxwell's equations on manifolds are often dealt with in the manifestly covariant formulation using differential forms, they can also be understood as a symmetric hyperbolic systems in terms of the electric and magnetic fields, once a time function has been fixed. We denote the \emph{electric} and \emph{magnetic fields} on $\sf{M}$ by $\mathscr{E},\mathscr{B}\in \Gamma^{\infty}(\sf{F})$. Now, given an external current $j\in \Gamma^{\infty}(\sf{F})$ and charge $\rho\in C^{\infty}(\sf{M})$ satisfying the \textit{continuity equations} $\partial_{t}\rho+\mathrm{div}_{\sf{h}}(j)=0$, \textit{Maxwell's equations in vacuum} (in Gaußian units and $c=1$) are given by
\begin{align}\label{eq:MaxSHS}
	\mathsf{S}\begin{pmatrix}
		\mathscr{E}\\\mathscr{B}
	\end{pmatrix}:=\bigg[\partial_{t}+\begin{pmatrix}
		0 & -\mathrm{curl}_{\Sigma}\\
		\mathrm{curl}_{\Sigma} &0
	\end{pmatrix}\bigg]
	\begin{pmatrix}
		\mathscr{E}\\\mathscr{B}
	\end{pmatrix}=-4\pi
	\begin{pmatrix}
	 j\\ 0
	\end{pmatrix}\,,\qquad\qquad \mathrm{div}_{\Sigma}\begin{pmatrix}
	\mathscr{E}\\ \mathscr{B}
	\end{pmatrix}=4\pi\begin{pmatrix}
	\rho\\ 0
\end{pmatrix}\,,
\end{align}
As proven for instance in \cite[Lemma~5.1]{SchmidMurroFinster}, $\sf{S}\in\mathrm{DO}^{1}(\sf{E})$ for $\sf{E}:=\sf{F}\oplus\sf{F}$ equipped with the obvious induced bundle metric is a symmetric hyperbolic system. To show this, we first note that the principal symbol $\sigma_{\mathrm{curl}}\colon \sf{T}^{\ast}\Sigma\to\mathrm{End}(\sf{F})$ of the operator $\mathrm{curl}_{\Sigma}$ is given by
	\begin{align*}
		(\sigma_{\mathrm{curl}}(\vec{\xi})\psi)_{i}=\epsilon_{ijk}\xi^{j}\psi^{k}\,,\qquad\forall\vec{\xi}=\xi_{i}\d x^{i}\in \sf{T}^{\ast}\Sigma\, ,\,\psi\in \Gamma^{\infty}(\sf{F})\, ,
	\end{align*}
	where $(x^{i})_{i=1,2,3}$ are local coordinates on $\Sigma$. Now, it is easy to see that $\sigma_{\mathrm{curl}}(\vec{\xi})$ is antisymmetric with respect to $\langle\cdot,\cdot\rangle_{\sf{F}}$, since
	\begin{align*}
		\langle	\sigma_{\mathrm{curl}}(\vec{\xi})\psi,\varphi\rangle_{\sf{F}}=\epsilon_{ijk}\xi^{j}\psi^{k}\varphi^{i}=-\epsilon_{kji}\xi^{j}\psi^{k}\varphi^{i}=-\langle\psi,\sigma_{\mathrm{curl}}(\vec{\xi})\varphi\rangle_{\sf{F}}
	\end{align*}		
	for all $\vec{\xi}\in \sf{T}^{\ast}\Sigma$ and $\psi,\varphi\in \Gamma^{\infty}(\sf{F})$. In particular, this shows that 
	\begin{align*}
		\sigma_{\sf{S}}(\xi)=\begin{pmatrix}
			\xi_{0}\mathrm{id}_{\sf{F}} & -\sigma_{\mathrm{curl}}(\vec{\xi})\\ \sigma_{\mathrm{curl}}(\vec{\xi}) & \xi_{0}\mathrm{id}_{\sf{F}}
		\end{pmatrix}\in\mathrm{End}(\sf{E})\,,\qquad\forall \xi=(\xi_{0},\vec{\xi})\in \sf{T}^{\ast}\sf{M}
	\end{align*}
	is symmetric with respect to $\langle\cdot,\cdot\rangle_{\sf{E}}$. For hyperbolicity, we take a general future-directed time-like vector of the form $\tau=\d t+\vec{\alpha}$, where $\vec{\alpha}=\alpha_{i}\d x^{i}\in \Gamma^{\infty}(\sf{F})$ is such that $\Vert\vec{\alpha}\Vert_{\sf{F}}^{2}=h^{ij}\alpha_{i}\alpha_{j}<1$. Then, for every $\omega=(\psi,\varphi)$ with $\psi,\varphi\in \Gamma^{\infty}(\sf{F})$, it holds that
	\begin{align*}
		\langle\omega, \sigma_{\sf{S}}(\tau)\omega\rangle_{\sf{E}}=\Vert\omega\Vert_{\sf{E}}^{2}-2\langle\psi,\sigma_{\mathrm{curl}}(\vec{\alpha})\varphi\rangle_{\sf{F}}&\geq \Vert\omega\Vert_{\sf{E}}^{2}-2\Vert\psi\Vert_{\sf{F}}\Vert\varphi\Vert_{\sf{F}}\Vert\sigma_{\mathrm{curl}}(\vec{\alpha})\Vert_{\mathrm{op}}\\&\geq\Vert\omega\Vert_{\sf{E}}^{2}(1-\Vert\sigma_{\mathrm{curl}}(\vec{\alpha})\Vert_{\mathrm{op}})\, .
	\end{align*}
	Hence, $\langle\cdot, \sigma_{\sf{S}}(\tau)\cdot\rangle_{\sf{E}}$ is positive-definite provided that the operator norm of $\sigma_{\mathrm{curl}}(\vec{\alpha})$ satisfies $\Vert\sigma_{\mathrm{curl}}(\vec{\alpha})\Vert_{\mathrm{op}}<1$, which is the case of our choice of $\alpha$: first, note that
	\begin{align*}
		\Vert \sigma_{\mathrm{curl}}(\vec{\alpha})\psi\Vert_{\sf{F}}^{2}=\epsilon_{ijk}\epsilon^{ilm}\alpha^{j}\alpha_{l}\psi^{k}\psi_{m}=\Vert\vec{\alpha}\Vert_{\sf{F}}^{2}\Vert\psi\Vert_{\sf{F}}^{2}-(\langle\vec{\alpha},\psi\rangle_{\sf{F}})^{2}\leq \Vert\vec{\alpha}\Vert_{\sf{F}}^{2}\Vert\psi\Vert_{\sf{F}}^{2}\, ,
	\end{align*}		
	where we used that $\epsilon_{ijk}\epsilon^{ilm}=\delta_{j}^{l}\delta_{k}^{m}-\delta_{j}^{m}\delta_{k}^{l}$. In particular, it follows that
	\begin{align*}
		\Vert\sigma_{\mathrm{curl}}(\vec{\alpha})\Vert_{\mathrm{op}}^{2}&=\sup_{\Vert\psi\Vert_{\sf{F}}=1}\Vert \sigma_{\mathrm{curl}}(\vec{\alpha})\psi\Vert_{\sf{F}}^{2}\leq\Vert\vec{\alpha}\Vert_{\sf{F}}^{2}<1\, .
	\end{align*}
	
The full Maxwell equations in~\eqref{eq:MaxSHS} include also the divergence conditions. However, those equations are not dynamical and simply provide constraints on the initial data. More precisely, consider the Cauchy problem
\begin{align*}
		\begin{cases}
			\sf{S}\psi &=\phi\\
			\psi\vert_{\Sigma_{0}}&=\mathfrak{f}
		\end{cases}\, ,\qquad\text{where}\qquad \psi:=\begin{pmatrix}
			\mathscr{E}\\ \mathscr{B}
		\end{pmatrix}\,,\quad\phi:=\begin{pmatrix}
			-4\pi j\\ 0
		\end{pmatrix}\,,\qquad\mathfrak{f}:=\begin{pmatrix}
		\mathfrak{f}_{1}\\\mathfrak{f}_{2}
		\end{pmatrix}
	\end{align*}	
	for given initial data $\mathfrak{f}_{1},\mathfrak{f}_{2}\in \Gamma^{\infty}(\Sigma_{0},\sf{F}\vert_{\Sigma_{0}})\cong\mathfrak{X}(\Sigma_{0})$ on $\Sigma_{0}:=\{0\}\times\Sigma\cong\Sigma$. Now, taking the divergence of these equations yields $\partial_{t}\mathrm{div}_{\Sigma}(\mathscr{E})=4\pi\partial_{t}\rho$ and $\partial_{t}\mathrm{div}_{\Sigma}(\mathscr{B})=0$. Hence, we see that the constraint equations $\mathrm{div}_{\Sigma}(\mathscr{E})=4\pi\rho$ and $\mathrm{div}_{\Sigma}(\mathscr{B})=0$ are satisfied if and only if we chooses the initial data $\mathfrak{f}=(\mathfrak{f}_{1},\mathfrak{f}_{2})$ such that $\mathrm{div}_{\Sigma}(\mathfrak{f}_{1})=4\pi\rho\vert_{\Sigma_{0}}$ and $\mathrm{div}_{\Sigma}(\mathfrak{f}_{2})=0$.
\end{example}

A generalisation of the previous example can be found in \cite{MurroMaxwell}, in which it is shown that the Maxwell's equations for $k$-forms in terms of the electric and magnetic part of the field strength tensor define a symmetric hyperbolic system on arbitrary globally hyperbolic manifolds.

\begin{example}\label{Ex:Dirac} (Spin Dirac Operator)\newline
Let $(\sf{M},\sf{g})$ be a $(1+n)$-dimensional globally hyperbolic \textit{spin} manifold equipped with a \textit{spin structure}, i.e.~a pair $(\sf{P},\Lambda)$ consisting of a $\mathrm{Spin}^{+}(1,n)$-principal bundle $\pi_{\sf{P}}\:\sf{P}\to\sf{M}$ together with a twofold covering map $\Lambda$ from $\sf{P}$ to the frame bundle $\mathrm{Fr}_{\mathrm{SO}^{+}}(\sf{M})$ such that
\begin{equation*}
	\begin{tikzcd}
		\sf{P}\times \mathrm{Spin}^{+}(1,n) \arrow[r,"\cdot"]\arrow[d,swap,"\Lambda\times\lambda"] & \sf{P} \arrow[rd,"\pi_{\sf{P}}"]\arrow[d,"\Lambda"] & \\
		\mathrm{Fr}_{\mathrm{SO}^{+}}(\sf{M}) \times \mathrm{SO}^{+}(1,n)\arrow[r,"\cdot"]  & \mathrm{Fr}_{\mathrm{SO}^{+}}(\sf{M}) \arrow[r,swap,"\pi_{\mathrm{Fr}}"] & \sf{M}
	\end{tikzcd}
\end{equation*}
commutes, where $\lambda$ denotes the double covering at the group level, $\pi_{\sf{P}}$ and $\pi_{\mathrm{Fr}}$ refer to the respective bundle projections, and the horizontal arrows represent the right group actions on the corresponding principal bundles. The question whether a given (pseudo-)Riemannian manifold admits a spin stricture is a purely topological one. It can be shown that a spin structure exists if and only if the \textit{second Stiefel-Whitney class} $w_{2}(\sf{M})\in\sf{H}^{2}(\sf{M},\bb{Z}_{2})$ vanishes. Note that the first Stiefel-Whitney class $w_{1}(\sf{M})\in\sf{H}^{1}(\sf{M},\bb{Z}_{2})$ is an obstruction for orientability of a manifold. Hence, the existence of a spin structure can be understood as the existence
of a ``higher orientation'' for $\sf{M}$. Furthermore, there is a bijection between the set of isomorphism classes of spin structures\footnote{Two spin structures $(\sf{P},\Lambda)$ and $(\sf{P}^{\prime},\Lambda^{\prime})$ on a spin manifold are isomorphic if there exists a $\mathrm{Spin}^{+}(1,n)$-principal bundle isomorphism $f\:\sf{P}\to\sf{P}^{\prime}$ such that $\Lambda^{\prime}\circ f=\Lambda$.} and $\sf{H}^{1}(\sf{M},\mathbb{Z}_{2})$. A sufficient (but not necessary) condition for the existence of a spin structure is parallelisability of $\sf{M}$. In particular, since any $3$-dimensional orientable manifold is parallelisable, Theorem~\ref{Thm:BernalSanchez} implies that any $(1+3)$-dimensional globally hyperbolic manifold $\sf{M}=\bb{R}\times\Sigma$ admits a spin structure. Furthermore, since $\sf{H}^{1}(\sf{M},\mathbb{Z}_{2})\cong\sf{H}^{1}(\Sigma,\mathbb{Z}_{2})\cong \mathrm{Hom}(\pi_{1}(\Sigma),\mathbb{Z}_{2})$, the spin structure is unique if $\Sigma$ is simply connected. We refer to \cite{RudolphSchmidt2,Lawson} for details on spin geometry in the Riemannian setting and to \cite{Hamilton,Baum2} for the pseudo-Riemannian and Lorentzian case.

Using the \textit{spin representation} $\rho\colon \mathrm{Spin}^{+}(1,n)\to \mathrm{Aut}(\bb{C}^{\mathrm{N}})$ with $\mathrm{N}:= 2^{\lfloor \frac{n+1}{2}\rfloor}$, where $\lfloor x\rfloor:=\mathrm{max}\{k\in\bb{Z}\mid k\leq x\}$ for $x\in\bb{R}$ denotes the \emph{floor function}, we construct the \textit{spinor bundle}, that is, the associated $\bb{C}$-vector bundle
\begin{align*}
	\sf{S}\sf{M}:=\sf{P}\times_\rho \mathbb{C}^\mathrm{N}\,.
\end{align*}
On this bundle, we have the following additional structures:
\begin{itemize}
\item[$\bullet$]A natural $\mathrm{Spin}^{+}(1,n)$-invariant non-degenerate fibre metric $\langle\cdot,\cdot\rangle_{\sf{S}\sf{M}}$ (defined up to a sign).
\item[$\bullet$]A \textit{Clifford multiplication}, i.e.~a vector bundle homomorphism $\gamma\colon \sf{T}\sf{M}\to \mathrm{End}(\sf{S}\sf{M})$ such that $\langle\gamma(X)\psi,\varphi\rangle_{\sf{S}\sf{M}}=\langle\psi,\gamma(X)\varphi\rangle_{\sf{S}\sf{M}}$ for all $\varphi,\psi\in\Gamma^{\infty}(\sf{S}\sf{M})$ and $X\in\mathfrak{X}(\sf{M})$ and such that 
\begin{align*}
	\gamma(X)\gamma(Y) + \gamma(Y)\gamma(X) = -2\sf{g}(X, Y)\mathds{1}_{\sf{S}\sf{M}}\,,\qquad\forall X,Y\in\mathfrak{X}(\sf{M})\, .
\end{align*}
\item[$\bullet$]A connection $\nabla^{\sf{S}\sf{M}}$ on $\sf{S}\sf{M}$, the \textit{spin connection}, defined as a lift of the Levi-Civita connection on $(\sf{M},\sf{g})$ that is compatible with the metric $\langle\cdot,\cdot\rangle_{\sf{S}\sf{M}}$. 
\end{itemize}
After this preliminary discussion, we define the \emph{(spin) Dirac operator} $\sf{D}\in\mathrm{DO}^{1}(\sf{S}\sf{M})$ by
\begin{align*}
	\sf{D}\:\Gamma^{\infty}(\sf{S}\sf{M})\xrightarrow{\nabla^{\sf{S}\sf{M}}}\Gamma^{\infty}(\sf{S}\sf{M}\otimes\sf{T}^{\ast}\sf{M})\xrightarrow{\mathrm{id}\otimes\sf{g}^{\sharp}}\Gamma^{\infty}(\sf{S}\sf{M}\otimes\sf{T}\sf{M})\xrightarrow{\gamma}\Gamma^{\infty}(\sf{S}\sf{M})\, .
\end{align*}
We claim that $\sf{D}$ is a symmetric hyperbolic system. Its principal symbol is $\sigma_{\sf{D}}(\xi)=\gamma(\xi^{\sharp})$, which is Hermitian with respect to $\langle\cdot,\cdot\rangle_{\sf{S}\sf{M}}$. For hyperbolicity, we refer to \cite[Prop.~1.1]{DimockDirac}, in which it is shown that $\sf{D}$ is hyperbolic provided $\langle\cdot,\cdot\rangle_{\sf{S}\sf{M}}$ has been chosen with the ``correct'' sign.
\end{example}

\begin{remark}\label{Rem:Dirac} (Clifford Dirac Operators)\newline
	The previous example is a special case of a more general construction: let $(\sf{M},\sf{g})$ be a globally hyperbolic Lorentzian manifold and $(\sf{E},\gamma,\langle\cdot,\cdot\rangle_{\sf{E}},\nabla^{\sf{E}})$ be a \emph{Dirac bundle} (see e.g.~\cite[Sec.~5.5]{RudolphSchmidt2}, \cite[Chap.~5]{Lawson}), that is,
	\begin{itemize}
		\item[(i)]a \textit{Clifford module bundle} $(\sf{E},\gamma)$, i.e.~a real or complex vector bundle together with a bundle morphism $\gamma\:\sf{T}\sf{M}\to\mathrm{End}(\sf{E})$ satisfying
		\begin{align*}
			\gamma(X)^{2}=-\sf{g}(X,X)\mathrm{id}_{\sf{E}}\,,\qquad\forall X\in\mathfrak{X}(\sf{M})\, .
		\end{align*}
		The universal property of the Clifford bundle $\mathrm{Cl}(\sf{M},\sf{g})$ and $\sf{T}\sf{M}\subset \mathrm{Cl}(\sf{M},\sf{g})$ imply that the morphism $\gamma$ admits a unique extension on the Clifford bundle $\mathrm{Cl}(\sf{M},\sf{g})$. One can hence think of the bundle $\sf{E}$ as the vector bundle whose fibre at a point $p\in\sf{M}$ is isomorphic to a given left module of the Clifford algebra $\mathrm{Cl}(\sf{T}_{p}\sf{M},\sf{g}_{p})$, see~\cite{AtiyahClifford}.	
		\item[(ii)]a bundle metric $\langle\cdot,\cdot\rangle_{\sf{E}}$ such that $\langle\gamma(X)\psi,\varphi\rangle_{\sf{E}}=\langle\psi,\gamma(X)\varphi\rangle_{\sf{E}}$ for all $X\in\mathfrak{X}(\sf{M})$.
	\item[(iii)]a \textit{Clifford connection} $\nabla^{\sf{E}}$, i.e.~a connection on $\sf{E}$ satisfying
	\begin{align*}
		\nabla^{\sf{E}}(\gamma(X)\psi)=\gamma(\nabla X)\psi+\gamma(X)(\nabla^{\sf{E}}\psi)\, ,
	\end{align*}
	or equivalently $\nabla_{X}^{\mathrm{End}(\sf{E})}(\gamma(Y))=\gamma(\nabla_{X}Y)$ for all $X,Y\in\mathfrak{X}(\sf{M})$, where $\nabla^{\mathrm{End}(\sf{E})}$ is the induced connection on $\mathrm{End}(\sf{E})$. Another equivalent characterisation is to require that $\gamma$, viewed as a section of $\mathrm{End}(\sf{E})\otimes\sf{T}^{\ast}\sf{M}$, is \emph{parallel}, i.e.~$\nabla^{\mathrm{End}(\sf{E})\otimes\sf{T}^{\ast}\sf{M}}\gamma=0$, where $\nabla^{\mathrm{End}(\sf{E})\otimes\sf{T}^{\ast}\sf{M}}$ denotes connection on $\mathrm{End}(\sf{E})\otimes\sf{T}^{\ast}\sf{M}$ induced by $\nabla^{\sf{E}}$ and the Levi-Civita connection.
	\end{itemize}
	In this context, one can define the \emph{(Clifford) Dirac operator} $\sf{D}\in\mathrm{DO}^{1}(\sf{E})$ similarly as in Example~\ref{Ex:Dirac}. In fact, Example~\ref{Ex:Dirac} is the special case in which, for a spin manifold, the Clifford module is chosen to be the irreducible spin representation. Furthermore, we remark that on an even-dimensional spin manifold, every Clifford module bundle is obtained by twisting the spinor bundle, see \cite[Prop.~3.35]{GetzlerBook}.
	
	Note, however, that not every Clifford Dirac operator is a symmetric hyperbolic system. For instance, consider the Clifford module bundle $\sf{E}:=\bigwedge\sf{T}^{\ast}\sf{M}_{\bb{C}}\cong\bigoplus_{r=0}^{\infty}\bigwedge^{r}\sf{T}^{\ast}\sf{M}_{\bb{C}}$, the complex Graßmann bundle, whose sections are (complex) differential forms, i.e.~$\Gamma^{\infty}(\sf{E}):=\Omega(\sf{M},\bb{C})\cong\bigoplus_{r=0}^{\infty}\Omega^{r}(\sf{M},\bb{C})$. We equip this bundle with the natural non-degenerate sesquilinear form $\langle\cdot,\cdot\rangle_{\sf{E}}$ induced by the Hodge forms $\langle\cdot,\cdot\rangle_{r}$ on $\Omega^{r}(\sf{M},\bb{C})$ for each $r\in\bb{N}_{0}$, i.e.
	\begin{align*}
		\langle\alpha,\beta\rangle_{r}:=\frac{1}{r!}(\sf{g}^{\sharp})^{\otimes r}(\overline{\alpha},\beta),\qquad (\alpha,\beta)_{r}:=\int_{\sf{M}}\langle\alpha,\beta\rangle_{r}\,\d\mu_{\sf{g}}=\int_{\sf{M}}\overline{\alpha}\wedge\ast\beta\,,\qquad\alpha,\beta\in\Omega^{r}(\sf{M},\bb{C})\, .
	\end{align*}
	A Clifford connection is provided by the Levi-Civita connection and the Clifford multiplication $\gamma\:\sf{T}\sf{M}\to\mathrm{End}(\sf{E})$ is given by
	\begin{align*}
		\gamma(X)\omega=X^{\flat}\wedge\omega+X\lrcorner\omega
	\end{align*}
	for all $X\in\sf{T}\sf{M}$ and $\omega\in\bigwedge\sf{T}^{\ast}\sf{M}$. Then, the Clifford Dirac operator for $(\sf{E},\gamma,\langle\cdot,\cdot\rangle_{\sf{E}},\nabla)$ is exactly the \textit{de Rham-Hodge Dirac operator} $\sf{D}=\d+\delta$, where $\d$ denotes the exterior derivative and $\delta$ its formal adjoint, the codifferential. Now, the principal symbol $\sigma_{\sf{D}}(\xi)=\gamma(\xi^{\sharp})$ can easily be seen to be Hermitian with respect to $\langle\cdot,\cdot\rangle_{\sf{E}}$. However, it is clearly not hyperbolic: take a future-directed timelike covector $\tau\in\Omega^{1}(\sf{M})$. Then, for every $\omega=(\omega_{i})_{i=0,\dots,k+1}\in\Omega(\sf{M},\bb{C})$,
	\begin{align*}
		\langle\sigma_{\sf{S}}(\tau)\omega,\omega\rangle_{\sf{E}}=2\sum_{i=0}^{d}\langle\tau^{\sharp}\lrcorner\omega_{i+1},\omega_{i}\rangle_{i}\, .
	\end{align*}
	Now, choose $\omega=(\omega_{i})_{i=0,\dots,k+1}\in\Omega(\sf{M},\bb{C})$ such that $\omega_{0}=f$ for some $f\in C^{\infty}(\sf{M},(0,\infty))$ arbitrary, $\omega_{1}:=\tau$ and $\omega_{i}=0$ otherwise. Then, $\langle\sigma_{\sf{S}}(\tau)\omega,\omega\rangle_{\sf{E}}=2\sf{g}^{\sharp}(\tau,\tau)f<0$, since $\tau$ is assumed to be timelike, contradicting hyperbolicity (Definition~\ref{Def:SHS}(ii)).
	
	Another special case of a Clifford Dirac operator that is not a symmetric hyperbolic system is the \textit{Buchdahl operator}, see e.g.~\cite[Sec.~2.5]{GinouxBaer}. However, we note that both the Buchdahl and de Rham-Hodge Dirac operators are \textit{Green hyperbolic}, see Section~\ref{Sec:GreenCauchyHyp}.
\end{remark}

Now, the goal of the following discussion is the prove that the Cauchy problem for symmetric hyperbolic systems is well-posed. More precisely, the aim is to prove the following theorem.

\begin{theorem}\label{Thm:Cauchy} \emph{(The Cauchy Problem for Symmetric Hyperbolic Systems)}\newline
	Let $(\sf{M},\sf{g})$ be a globally hyperbolic manifold and $t\:\sf{M}\to\bb{R}$ be a Cauchy temporal function with corresponding foliation $(\Sigma_{t})_{t\in\bb{R}}$. Furthermore, let $\sf{S}\in\mathrm{DO}^{1}(\sf{E})$ be a symmetric hyperbolic system on a Hermitian vector bundle $(\sf{E},\langle\cdot,\cdot\rangle_{\sf{E}})$. Then, the following holds true:
\begin{itemize}
	\item[\emph{(i)}]For every pair $(\phi,\mathfrak{f})\in \Gamma^{\infty}(\sf{M},\sf{E})\times \Gamma^{\infty}(\Sigma_{0},\sf{E}\vert_{\Sigma_{0}})$, there exists a unique solution $\psi\in \Gamma^{\infty}(\sf{M},\sf{E})$ to the Cauchy problem
	\begin{align*}
		\begin{cases}
			\sf{S}\psi &=\phi\\
			\psi\vert_{\Sigma_{0}}&=\mathfrak{f}
		\end{cases}\, .
	\end{align*}	
\item[\emph{(ii)}]The unique solution $\psi$ in (i) propagates at most with the speed of light, i.e.
	\begin{align*}
		\mathrm{supp}(\psi)\cap\mathcal{J}^{\pm}(\Sigma_{0})\subset \mathcal{J}^{\pm}\big((\mathrm{supp}(\phi)\cap\mathcal{J}^{\pm}(\Sigma_{0}))\cup \mathrm{supp}(\mathfrak{f})\big)
	\end{align*}
	and hence in particular $\mathrm{supp}(\psi)\subset \mathcal{J}(\mathrm{supp}(\phi)\cup \mathrm{supp}(\mathfrak{f}))$.
	\item[\emph{(iii)}]The Cauchy problem is stable, i.e.~the solution maps
	\begin{align*}
		&\Gamma^{\infty}(\sf{M},\sf{E})\times \Gamma^{\infty}(\Sigma_{0},\sf{E}\vert_{\Sigma_{0}})\ni (\phi,\mathfrak{f})\mapsto\psi \in \Gamma^{\infty}(\sf{M},\sf{E})\,,\\
		&\Gamma^{\infty}_{\mathrm{c}}(\sf{M},\sf{E})\times \Gamma_{\mathrm{c}}^{\infty}(\Sigma_{0},\sf{E}\vert_{\Sigma_{0}})\ni (\phi,\mathfrak{f})\mapsto\psi \in \Gamma^{\infty}_{\mathrm{sc}}(\sf{M},\sf{E})
	\end{align*}
	are continuous in the corresponding Fréchet- and LF-topologies, respectively.
	\end{itemize}
\end{theorem}

\begin{remark} Extensions to Theorem~\ref{Thm:Cauchy} to globally hyperbolic spacetimes with timelike boundary (see Remark~\ref{Rem:GHSBound}) can, for instance, be found in \cite{GinouxMurro}.
\end{remark}

There are several ways to prove the well-posedness of the Cauchy problem for symmetric hyperbolic systems on globally hyperbolic manifolds. First, in \cite{BaerGreen}, with an earlier sketch in \cite{GerochQuasi}, the existence of solutions is obtained by gluing together local solutions obtained by standard PDE theory. An alternative proof, based on the construction of weak solutions via a suitable mollification procedure, is presented in the lecture notes \cite{BaerLecture}. In this thesis, we adopt a different strategy, in which weak solutions are first constructed via the representation theorem of Fréchet-Riesz, i.e.~translating the standard approach of classical PDE theory to a global setting, followed by proof of regularity based on the construction of an extended system acting on sections and their first (covariant) derivatives.

To start with, we fix a globally hyperbolic spacetime $(\sf{M},\sf{g})$ with Levi-Civita connection $\nabla$ and a Cauchy temporal function $t\:\sf{M}\to\bb{R}$, which allows us to identify
\begin{align*}
	\sf{M}:=\bb{R}\times\Sigma\,,\qquad\sf{g}=-\beta^{2}\d t\otimes \d t+\sf{h}_{t}\,,
\end{align*}
as in Theorem~\ref{Thm:BernalSanchez}. We denote by $(\Sigma_{t}:=\{t\}\times\Sigma)_{t\in\mathbb{R}}$ the corresponding foliation of $\sf{M}$ by Cauchy hypersurfaces. Furthermore, let $\sf{S}\in\mathrm{DO}^{1}(\sf{E})$ be a symmetric hyperbolic system on a Hermitian vector bundle $(\sf{E},\langle\cdot,\cdot\rangle_{\sf{E}})$. Now, let $\nabla^{\sf{E}}\:\Gamma^{\infty}(\sf{E})\to\Gamma^{\infty}(\sf{E}\otimes\sf{T}^{\ast}\sf{M})$ be a connection on $\sf{E}$, assumed to be compatible with the bundle metric $\langle\cdot,\cdot\rangle_{\sf{E}}$. Then, in local coordinates $(x^{\mu})_{\mu=0,\dots,n}$ on some open subset $\mathcal{U}\subset\sf{M}$, $\sf{S}$ can be written in the form 
\begin{align*}
	\sf{S}=\sigma_{\sf{S}}(\d x^{\mu})\nabla^{\sf{E}}_{\partial_{\mu}}+\sf{S}_{0}
\end{align*} 
for some zero-order operator $\sf{S}_{0}\in\mathrm{DO}^{0}(\sf{E})$. We stress that $\sf{S}_{0}$ depends on the choice of connection and choosing a different (metric-compatible) connection $\widetilde{\nabla}^{\sf{E}}$ on $\sf{E}$ results into a change of zero-order terms $\sf{S}_{0}\mapsto\widetilde{\sf{S}}_{0}=\sf{S}_{0}+\sigma_{\sf{S}}(\d x^{\mu})(\nabla^{\sf{E}}_{\partial_{\mu}}-\widetilde{\nabla}^{\sf{E}}_{\partial_{\mu}})$.

Moreover, since $\sf{S}$ is a \textit{first-order} operator, we observe that $\sigma_{\sf{S}}(\xi)$ is also $\bb{R}$-linear in $\xi\in\sf{T}^{\ast}_{p}\sf{M}$ for all $p\in\sf{M}$, i.e.~$\sigma_{\sf{S}}$ can be viewed as a section of $\mathrm{End}(\sf{E})\otimes \sf{T}\sf{M}$ locally given by $\sigma_{\sf{S}}=\sigma_{\sf{S}}(\d x^{\mu})\otimes\partial_{\mu}$. 

\begin{lemma} \emph{(Formal Adjoint, \cite[Prop.~2.8]{SchmidMurroFinster})}\label{Lemma:AdjointSHS} \newline
	Let $(\sf{M},\sf{g})$ and $(\sf{E},\langle\cdot,\cdot\rangle_{\sf{E}})$ be as above and consider a symmetric hyperbolic system 
	\begin{align*}
		\sf{S}=\sigma_{\sf{S}}(\d x^{\mu})\nabla^{\sf{E}}_{\partial_{\mu}}+\sf{S}_{0}\, .
	\end{align*}
	Furthermore, denote by $\mathscr{D}$ the divergence-like operator defined as the composition\footnote{$\nabla^{\mathrm{End}(\sf{E})\otimes\sf{T}	\sf{M}}:=\nabla^{\mathrm{End}(\sf{E})}\otimes\mathrm{id}+\mathrm{id}\otimes\nabla$ denotes the \textit{tensor product connection} on $\mathrm{End}(\sf{E})\otimes\sf{T}\sf{M}$.} 
	\begin{align*}
		\mathscr{D}\colon \Gamma^{\infty}(\sf{M},\mathrm{End}(\sf{E})\otimes \sf{T}\sf{M})\xrightarrow{\nabla^{\mathrm{End}(\sf{E})\otimes\sf{T}\sf{M}}} \Gamma^{\infty}(\sf{M},\mathrm{End}(\sf{E})\otimes \sf{T}\sf{M}\otimes \sf{T}^{\ast}\sf{M})\xrightarrow{c}\Gamma^{\infty}(\sf{M},\mathrm{End}(\sf{E}))\, ,
	\end{align*}
where $\nabla^{\mathrm{End}(\sf{E})}$ denotes the connection on $\mathrm{End}(\sf{E})$ induced by $\nabla^{\sf{E}}$ and $c\colon \sf{T}\sf{M}\otimes \sf{T}^{\ast}\sf{M}\ni v\otimes \xi\mapsto \xi(v)$ the duality pairing. Then, the formal adjoint $\sf{S}^{\ast}$ of $\sf{S}$ with respect to $(\cdot,\cdot)_{\sf{E}}$ is given by
	\begin{align*}
		\sf{S}^{\ast}=-\sigma_{\sf{S}}(\d x^{\mu})\nabla^{\sf{E}}_{\partial_{\mu}}+\sf{S}_{0}^{\ast}-\mathscr{D}\sigma_{\sf{S}}\, .
	\end{align*}
\end{lemma}

As a by-product of the previous lemma, we note that $-\sf{S}^{\ast}\in\mathrm{DO}^{1}(\sf{E})$ is itself a symmetric hyperbolic system and that the sum $\sf{S}+\sf{S}^{\ast}$ is a differential operator of order zero.

\begin{proof}[Proof of Lemma~\ref{Lemma:AdjointSHS}.]
	Fix $\psi,\varphi\in \Gamma^{\infty}_{\mathrm{c}}(\sf{E})$ and consider the $n$-form $\omega\in\Omega^{n}(\sf{M})$ locally defined by
	\begin{align*}
		\omega:=\langle\sigma_{\sf{S}}(\d x^{\mu})\psi,\varphi\rangle_{\sf{E}} \partial_{\mu}\lrcorner \d\mu_{\sf{g}}\,.
	\end{align*}
	Defining the vector field $\sf{V}:=\langle\sigma_{\sf{S}}(\d x^{\mu})\psi,\varphi\rangle_{\sf{E}} \partial_{\mu}\in \mathfrak{X}(\sf{M})$, then $\sf{V}\lrcorner d\mu_{\sf{g}}$ denotes the insertion of $\sf{V}$ into the first slot of the volume form $\d\mu_{\sf{g}}$. \emph{Cartan's magic formula} implies 
	\begin{align*}
		\d\omega=\d(\sf{V}\lrcorner\d\mu_{\sf{g}})=\mathcal{L}_{\sf{V}}\d\mu_{\sf{g}}=\mathrm{div}_{\sf{g}}(\sf{V})\d\mu_{\sf{g}}\, ,
	\end{align*}	
	where $\mathcal{L}_{\sf{V}}$ denotes the Lie derivative along $\sf{V}$ and $\mathrm{div}_{\sf{g}}(\sf{V})$ the divergence of $\sf{V}$. Furthermore, using the fact that the connection $\nabla^{\sf{E}}$ is compatible with $\langle\cdot,\cdot\rangle_{\sf{E}}$, we find
	\begin{align*}
		\mathrm{div}_{\sf{g}}(\sf{V})=&\partial_{\mu}\langle \sigma_{\sf{S}}(dx^{\mu})\psi,\varphi\rangle_{\sf{E}}+\Gamma_{\mu\nu}^{\mu}\langle \sigma_{\sf{S}}(dx^{\nu})\psi,\varphi\rangle_{\sf{E}}\\=&\langle \nabla^{\sf{E}}_{\partial_{\mu}}(\sigma_{\sf{S}}(dx^{\mu})\psi),\varphi\rangle_{\sf{E}}+\langle \sigma_{\sf{S}}(dx^{\mu})\psi,\nabla^{\sf{E}}_{\partial_{\mu}}\varphi\rangle_{\sf{E}}+\Gamma_{\mu\nu}^{\mu}\langle \sigma_{\sf{S}}(dx^{\nu})\psi,\varphi\rangle_{\sf{E}}\\=&\langle \underbrace{(\nabla^{\mathrm{End}(\sf{E})}_{\partial_{\mu}}\sigma_{\sf{S}}(dx^{\mu})+\Gamma_{\mu\nu}^{\mu}\sigma_{\sf{S}}(dx^{\nu}))\psi}_{=\mathscr{D}\sigma_{\sf{S}}}+\sigma_{\sf{S}}(dx^{\mu})\nabla_{\partial_{\mu}}^{\sf{E}}\psi,\varphi\rangle_{\sf{E}}+\langle \psi,\sigma_{\sf{S}}(dx^{\mu})\nabla^{\sf{E}}_{\partial_{\mu}}\varphi\rangle_{\sf{E}}\\=&\langle\psi,\sf{S}\varphi\rangle_{\sf{E}}+\langle(\sigma_{\sf{S}}(d x^{\mu})\nabla^{\sf{E}}_{\partial_{\mu}}-\sf{S}_{0}^{\ast}+\mathscr{D}\sigma_{\sf{S}})\psi,\varphi\rangle_{\sf{E}} \, ,
	\end{align*}
	where we added and subtracted $\langle\psi,\sf{S}_{0}\varphi\rangle_{\sf{E}}$ in the last step and used the fact that the adjoint relation for $\sf{S}_{0}$ actually holds pointwise, i.e.~$\langle\psi,\sf{S}_{0}\varphi\rangle_{\sf{E}}=\langle\sf{S}_{0}^{\ast}\psi,\varphi\rangle_{\sf{E}}$, since it is an operator of order zero. Combining everything, we have shown that
	\begin{align}\label{eq:AdjointProof}
		\d\omega=\langle\psi,\sf{S}\varphi\rangle_{\sf{E}}-\langle\sf{S}^{\ast}\psi,\varphi\rangle_{\sf{E}}\, .
	\end{align}
	where $\sf{S}^{\ast}:=-\sigma_{\sf{S}}(\d  x^{\mu})\nabla^{\sf{E}}_{\mu}+\sf{S}_{0}^{\ast}-\mathscr{D}\sigma_{\sf{S}}$. By integrating equation~\eqref{eq:AdjointProof} over $\sf{M}$ and using Stokes' theorem, we obtain $(\psi,\sf{S}\varphi)_{\sf{E}}=(\sf{S}^{\ast}\psi,\varphi)_{\sf{E}}$. Hence, $\sf{S}^{\ast}$ is indeed the formal adjoint of $\sf{S}$. 
\end{proof}

\begin{example}\label{Ex:DiracASA}
	Consider the Dirac operator from Example~\ref{Ex:Dirac}. Its principal symbol is $\sigma_{\sf{D}}(\xi)=\gamma(\xi^{\sharp})$ and since the spin connection is a \textit{Clifford connection}, $\gamma$ viewed as a section of $\mathrm{End}(\sf{S}\sf{M})\otimes\sf{T}^{\ast}\sf{M}$ satisfies $\nabla^{\mathrm{End}(\sf{S}\sf{M})\otimes\sf{T}^{\ast}\sf{M}}\gamma=0$. Hence, $\mathscr{D}\sigma_{\sf{D}}=0$ and $\sf{D}^{\ast}=-\sf{D}$.
\end{example}

Next, let us define suitable function spaces for the subsequent analysis. We write $\eta:=\beta\d t$. By definition, $\eta_{t}:=(\beta\d t)\vert_{\Sigma_{t}}$ is the future-directed timelike unit covector field normal to $\Sigma_{t}$ in $\sf{M}$. With this notation, we consider the \textit{positive-definite} bundle metric 
\begin{align*}
	\langle\cdot,\cdot\rangle_{\sf{E},\beta}:=\langle\sigma_{\sf{S}}(\eta)\cdot,\cdot\rangle_{\sf{E}}
\end{align*}
with $\sigma_{\sf{S}}(\eta)$ viewed as an element of $\mathrm{End}_{C^{\infty}(\sf{M})}(\Gamma^{\infty}(\sf{E}))\cong\Gamma^{\infty}(\mathrm{End}(\sf{E}))$ and define the natural $\sf{L}^{2}$-Hilbert space on the Riemannian manifold $(\Sigma_{t},\sf{h}_{t})$ taking values in $(\sf{E},\langle\cdot,\cdot\rangle_{\sf{E},\beta})$ by
\begin{align*}
	\sf{L}^{2}(\Sigma_{t},\sf{E}\vert_{\Sigma_{t}}):=\overline{\Gamma^{\infty}_{\mathrm{c}}(\Sigma_{t},\sf{E}\vert_{\Sigma_{t}})}^{\Vert\cdot\Vert_{t}}\,,\qquad \langle\mathfrak{f},\mathfrak{g}\rangle_{\sf{L}^{2}(\Sigma_{t})}:=\int_{\Sigma_{t}}\langle \mathfrak{f},\mathfrak{g}\rangle_{\sf{E},\beta}\,\d\mu_{\sf{h}_{t}}\, ,
\end{align*}
for all $t\in\bb{R}$, cf.~Eq.~\eqref{eq:L2}. Furthermore, for a given \textit{time-strip} $\sf{M}_{\mathrm{T}}:=t^{-1}([0,\mathrm{T}])=[0,\mathrm{T}]\times \Sigma$, we define the natural Hilbert space
\begin{align*}
	\sf{L}^{2}(\sf{M}_{\mathrm{T}},\sf{E}):=\int^{\oplus}_{[0,\mathrm{T}]}\sf{L}^{2}(\Sigma_{t},\sf{E}\vert_{\Sigma_{t}})\,\d t\,,\qquad\langle\psi,\varphi\rangle_{\sf{L}^{2}(\sf{M}_{\mathrm{T}})}:=\int_{0}^{\mathrm{T}}\,\langle\psi_{t},\varphi_{t}\rangle_{\sf{L}^{2}(\Sigma_{t})}\,\d t\,,
\end{align*}
which can be identified with the $\sf{L}^{2}$-Hilbert space on $\sf{M}_{\mathrm{T}}$ equipped with the \textit{auxiliary Riemannian metric} $\sf{g}_{0}:=\d t\otimes\d t+\sf{h}_{t}$ taking values in the Hermitian bundle $(\sf{E},\langle\cdot,\cdot\rangle_{\sf{E},\beta})$.

Last but not least, let $p\in\sf{M}$ be an arbitrary point. For the following proposition, we also introduce the notation
\begin{align*}
	\langle\mathfrak{f},\mathfrak{g}\rangle_{\sf{L}^{2}(\Sigma_{t}\cap\mathcal{J}^{-}(p))}:=\int_{\Sigma_{t}\cap\mathcal{J}^{-}(p)}\langle\mathfrak{f},\mathfrak{g}\rangle_{\sf{E},\beta}\,\d\mu_{\sf{h}_{t}}\,,
\end{align*}
the $\sf{L}^{2}$-inner product on the \textit{compact} Riemannian manifold $(\Sigma_{t}\cap\mathcal{J}^{-}(p),\sf{h}_{t})$ valued in $(\sf{E},\langle\cdot,\cdot\rangle_{\sf{E},\beta})$.

\begin{proposition}\label{Prop:En1} \emph{(Energy Estimates I, \cite[Thm.~5.3]{BaerGreen})}\newline
	Let $\sf{S}\in\mathrm{DO}^{1}(\sf{E})$ be a symmetric hyperbolic system on $(\sf{M},\sf{g})$ and $t\in C^{\infty}(\sf{M})$ be a Cauchy temporal function. Then, for every $p\in\sf{M}$ and $t_{0}\in\bb{R}$, there exists a constant $C=C(t_{0},p)>0$ such that
	\begin{align*}
		\Vert\psi_{t}\Vert_{\sf{L}^{2}(\Sigma_{t}\cap\mathcal{J}^{-}(p))}^{2}\leq e^{C(t-t_{0})}\bigg(C\int_{t_{0}}^{t}\Vert(\sf{S}\psi)_{\tau}\Vert_{\sf{L}^{2}(\Sigma_{\tau}\cap\mathcal{J}^{-}(p))}^{2}\,\d\tau+\Vert\psi_{t_{0}}\Vert_{\sf{L}^{2}(\Sigma_{t_{0}}\cap\mathcal{J}^{-}(p))}^{2}\bigg)
	\end{align*}
	for all $\psi\in\Gamma^{\infty}(\sf{E})$ and $t\geq t_{0}$, where we wrote $\psi_{t}:=\psi\vert_{\Sigma_{t}}\in\Gamma^{\infty}(\Sigma_{t},\sf{E}\vert_{\Sigma_{t}})$ for all $t\in\bb{R}$.
\end{proposition}

\begin{proof}
	Let $\psi\in \Gamma^{\infty}(\sf{E})$ be arbitrary. As in the proof of Lemma~\ref{Lemma:AdjointSHS}, we consider the real-valued $n$-form $\omega\in\Omega^{n}(\sf{M})$ defined by $\omega:=\langle\sigma_{\sf{S}}(\d x^{\mu})\psi,\psi\rangle_{\sf{E}}\,\partial_{\mu}\lrcorner \d\mu_{\sf{g}}$. As explained in more details in the proof of Lemma~\ref{Lemma:AdjointSHS}, its exterior derivative can be written as
	\begin{align}\label{eq:Leibniz}
		\d\omega=\langle\psi,\sf{S}\psi\rangle_{\sf{E}}-\langle\sf{S}^{\ast}\psi,\psi\rangle_{\sf{E}}=2\mathrm{Re}\langle\sf{S}\psi,\psi\rangle_{\sf{E}}-\langle (\sf{S}+\sf{S}^{\ast})\psi,\psi\rangle_{\sf{E}}\, .
	\end{align}
	Now, consider a closed time strip $\sf{M}_{t_{0},t}:=t^{-1}([t_{0},t])$ for $t\geq t_{0}$ and set $\sf{K}:=\sf{M}_{t,t_{0}}\cap\mathcal{J}^{-}(p)$. Stokes' theorem (note that $\partial\mathcal{J}^{-}(p)$ is a Lipschitz surface, see e.g.~\cite[pp.~413ff.]{ONeill}) implies
	\begin{align}\label{eq:Stokes}
		\int_{\sf{K}}\d\omega=\int_{\Sigma_{t}}i_{t}^{\ast}\omega-\int_{\Sigma_{t_{0}}}i_{t_0}^{\ast}\omega+\int_{\sf{M}_{t_{0},t}\cap\partial\mathcal{J}^{-}(p)}\omega\, ,
	\end{align}
	where $i_{\tau}\colon\Sigma_{\tau}\hookrightarrow\sf{M}$ denote the natural embeddings. Using Fubini's theorem and Equation~\eqref{eq:Leibniz}, we rewrite the left-hand side of~\eqref{eq:Stokes} as
	\begin{align}\label{eq:LHSProof}
		\int_{\sf{K}}\d\omega&=\int_{t_{0}}^{t}\int_{\Sigma_{\tau}\cap\mathcal{J}^{-}(p)}\bigg\{2\mathrm{Re} \langle\sf{S}\psi,\psi\rangle_{\sf{E}}-\langle (\sf{S}+\sf{S}^{\ast})\psi,\psi\rangle_{\sf{E}} \bigg\}\beta_{\tau}\,\d\mu_{\sf{h}_{\tau}} \d\tau\nonumber\\ &=\int_{t_{0}}^{t}\int_{\Sigma_{\tau}\cap\mathcal{J}^{-}(p)}\bigg\{2\mathrm{Re} \langle\sigma_{\sf{S}}(\eta)^{-1}\sf{S}\psi,\psi\rangle_{\sf{E},\beta}-\langle \sigma_{\sf{S}}(\eta)^{-1}(\sf{S}+\sf{S}^{\ast})\psi,\psi\rangle_{\sf{E},\beta} \bigg\}\beta_{\tau}\,\d\mu_{\sf{h}_{\tau}} \d\tau\nonumber\\&\leq \widetilde{C}\int_{t_{0}}^{t}\bigg\{2\Vert(\sf{S}\psi)_{\tau}\Vert_{\sf{L}^{2}(\Sigma_{\tau}\cap\mathcal{J}^{-}(p))}\Vert\psi_{\tau}\Vert_{\sf{L}^{2}(\Sigma_{\tau}\cap\mathcal{J}^{-}(p))}+\Vert\psi_{\tau}\Vert_{\sf{L}^{2}(\Sigma_{\tau}\cap\mathcal{J}^{-}(p))}^{2}\bigg\}d\tau\nonumber\\&\leq C\int_{t_{0}}^{t}(\Vert(\sf{S}\psi)_{\tau}\Vert_{\sf{L}^{2}(\Sigma_{\tau}\cap\mathcal{J}^{-}(p))}^{2}+\Vert\psi_{\tau}\Vert_{\sf{L}^{2}(\Sigma_{\tau}\cap\mathcal{J}^{-}(p))}^{2})\,\d\tau
	\end{align}
	for suitable constants $\widetilde{C},C>0$ only depending on $\sf{K}$, i.e.~on $t_{0}$, $t$ and $p$. In the equality from the first to the second line, we used that $\sigma(\eta)\in\Gamma^{\infty}(\mathrm{End}(\sf{E}))$ is pointwise invertible. In the first estimate, we used the Cauchy-Schwartz inequality and the facts that $\beta\sigma_{\sf{S}}(\eta)^{-1}$ and $\beta\sigma_{\sf{S}}^{-1}(\eta)(\sf{S}+\sf{S}^{\ast})$ are operators of order zero and hence in particular bounded on a compact domain. Last but not least, we used the algebraic inequality $2ab\leq a^{2}+b^{2}$ for all $a,b\in\bb{R}$.
	
	Note that we can choose $C$ to be independent of $t$. Indeed, $t\mapsto C(t)$ is continuous by construction since it is defined in terms of $\beta$ and the coefficients of $\sf{S}$. Hence, we can take the maximal constant on $[t_{0},t_{\mathrm{max}}]$ where $t_{\mathrm{max}}\geq t_{0}$ is sufficiently large so that $p\in\mathcal{J}^{-}(\Sigma_{t_{\mathrm{max}}})$.
	
	For the right-hand side of~Eq.~\eqref{eq:Stokes}, we first observe that any vector field $\sf{X}\in\mathfrak{X}(\sf{M})$ can be decomposed into a \textit{normal} $\sf{X}^{\perp}:=-\sf{g}(\sf{X},\nu)\nu$ and \textit{tangential component} $\sf{X}^{\top}:=\sf{X}-\sf{X}^{\perp}$, where $\nu$ denotes the future-directed timelike normal vector of $\Sigma_{\tau}$ in $\sf{M}$. It follows that
	\begin{align*}
		i_{\tau}^{\ast}(\sf{X}\lrcorner \d\mu_{\sf{g}})=i_{\tau}^{\ast}(X^{\perp}\lrcorner \d\mu_{\sf{g}})=-\sf{g}(\sf{X},\nu)i_{\tau}^{\ast}(\nu\lrcorner \d\mu_{\sf{g}})=-\sf{g}(\sf{X},\nu)\d\mu_{\sf{h}_{\tau}}\, .
	\end{align*}
	In particular, when choosing coordinates adapted to the splitting obtained from our Cauchy temporal function $t\colon\sf{M}\to\bb{R}$, i.e.~$(x^{\mu})_{\mu=0,\dots,k}$ with $x^{0}=t$, where $(x^{i})_{i=1,\dots,k}$ are local coordinates on some open subset of $\Sigma$, we find $\nu=\beta^{-1}\partial_{t}$ and hence, we conclude that $-\sf{g}(\sf{X},\nu)=\langle\sigma_{\sf{S}}(\beta dt)\psi,\psi\rangle_{\sf{E}}$ for $\sf{X}:=\langle\sigma_{\sf{S}}(dx^{\mu})\psi,\psi\rangle_{\sf{E}}\, \partial_{\mu}$. To sum up, we have shown that 
	\begin{align}\label{eq:RHSProof}
		\int_{\Sigma_{\tau}\cap\mathcal{J}^{-}(p)}i_{\tau}^{\ast}\omega=\int_{\Sigma_{\tau}\cap\mathcal{J}^{-}(p)}\langle\sigma_{\sf{S}}(\eta)\psi,\psi\rangle_{\sf{E}}\d\mu_{\sf{h}_{\tau}}=\Vert\psi_{\tau}\Vert_{\sf{L}^{2}(\Sigma_{\tau}\cap\mathcal{J}^{-}(p))}^{2}
	\end{align}
	for all $\tau\in\bb{R}$. Now, as argued in~\cite[Thm.~5.3]{BaerGreen}, the last integral in Eq.~\eqref{eq:Stokes} over $\sf{M}_{t_{0},t}\cap\partial\mathcal{J}^{-}(p)$ is non-negative and hence, using~\eqref{eq:LHSProof} and~\eqref{eq:RHSProof}, we obtain from~\eqref{eq:Stokes} the estimate
	\begin{align*}
		\Vert\psi_{t}\Vert_{\sf{L}^{2}(\Sigma_{t}\cap\mathcal{J}^{-}(p))}^{2}-\Vert\psi_{t_{0}}\Vert_{\sf{L}^{2}(\Sigma_{t_{0}}\cap\mathcal{J}^{-}(p))}^{2}\leq C\int_{t_{0}}^{t}\bigg(\Vert(\sf{S}\psi)_{\tau}\Vert_{\sf{L}^{2}(\Sigma_{\tau}\cap\mathcal{J}^{-}(p))}^{2}+\Vert\psi_{\tau}\Vert_{\sf{L}^{2}(\Sigma_{\tau}\cap\mathcal{J}^{-}(p))}^{2}\bigg)\,\d\tau
	\end{align*}
	Using \textit{Gronwall's lemma} (see~\cite{Gronwall} or \cite[Lem.~1.5.1]{BaerLecture}), we obtain the claimed result.
\end{proof}

\vspace*{-0.3cm}
\begin{corollary}\label{Cor:SHSUni} \emph{(Uniqueness and Finite Propagation Speed, \cite[Cor.~5.4]{BaerGreen}))}\newline
Consider the set-up of Theorem~\ref{Thm:Cauchy}. There it at most one smooth solution $\psi\in\Gamma^{\infty}(\sf{E})$ to
	\begin{align*}
		\begin{cases}
			\sf{S}\psi &=\phi\\
			\psi\vert_{\Sigma_{0}}&=\mathfrak{f}
		\end{cases}\, .
	\end{align*}	
	for given $(\phi,\mathfrak{f})\in \Gamma^{\infty}(\sf{M},\sf{E})\times \Gamma^{\infty}(\Sigma_{0},\sf{E}\vert_{\Sigma_{0}})$. Furthermore, the solution, if it exists, satisfies
	\begin{align*}
		\mathrm{supp}(\psi)\cap\mathcal{J}^{\pm}(\Sigma_{0})\subset \mathcal{J}^{\pm}\big((\mathrm{supp}(\phi)\cap\mathcal{J}^{\pm}(\Sigma_{0}))\cup \mathrm{supp}(\mathfrak{f})\big)\, .
	\end{align*}
\end{corollary}

\begin{proof}
	We first show that second claim. Let $\psi\in\Gamma^{\infty}(\sf{E})$ be such that $\sf{S}\psi=\phi$ and $\psi\vert_{\Sigma_{0}}=\mathfrak{f}$. Now, let $p\in\mathcal{J}^{+}(\Sigma)$ be such that $p\in\sf{M}\backslash\mathcal{J}^{+}\big((\mathrm{supp}(\phi)\cap\mathcal{J}^{+}(\Sigma_{0}))\cup \mathrm{supp}(\mathfrak{f})\big)$. It follows that $\mathcal{J}^{-}(p)\cap\mathcal{J}^{+}(\Sigma_{0})$ does not intersect $\mathrm{supp}(\phi)\cap\mathrm{supp}(\mathfrak{f})$, as depicted in Figure~\ref{Fig:Uni}. Therefore, the energy estimates from Proposition~\ref{Prop:En1} imply $\psi=0$ on $\mathcal{J}^{-}(p)\cap\mathcal{J}^{+}(\Sigma_{0})$. It follows that $\psi(p)=0$, which proves the claimed inclusion for $\mathcal{J}^{+}$. 
\begin{figure}[H]
\centering
\includegraphics[scale=2.4]{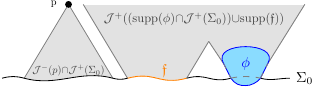}
\caption{Finite Speed of Propagation and Uniqueness.}\label{Fig:Uni}
\end{figure}	
	The same inclusion with $\mathcal{J}^{-}$ follows from a time reversal argument. Uniqueness is a direct consequence of this: let $\psi^{\prime}\in\Gamma^{\infty}(\sf{E})$ be another solution to the same Cauchy problem. Then, $\psi-\psi^{\prime}$ satisfies the \textit{homogeneous problem}, i.e.~$\sf{S}(\psi-\psi^{\prime})=0$ and $(\psi-\psi^{\prime})\vert_{\Sigma_{0}}=0$. But then, finite speed of propagation implies $\mathrm{supp}(\psi-\psi^{\prime})=\emptyset$ and hence $\psi=\psi^{\prime}$.
\end{proof}

Let us now turn to existence of solutions. We will start with a more simple situations, namely we will show the existence of \textit{weak solution} on a time strip for a \textit{spatially compact} globally hyperbolic spacetime.  We then establish regularity of these solutions and conclude with the existence of globally defined smooth solutions. The compactness assumption on the Cauchy surfaces will be removed lateron by means of a suitable compactification procedure.

\begin{lemma}\label{Lem:EE} \emph{(Energy Estimates II)}\newline
Let $\sf{S}\in\mathrm{DO}^{1}(\sf{E})$ be a symmetric hyperbolic system on a spatially compact globally hyperbolic manifold $(\sf{M},\sf{g})$ and $t\in C^{\infty}(\sf{M})$ be a Cauchy temporal function. Furthermore, consider a time strip $\sf{M}_{\mathrm{T}}=t^{-1}([0,\mathrm{T}])$. Then, there is a constant $C=C(\mathrm{T})>0$ such that for all $\psi\in\Gamma^{\infty}(\sf{E})$ , it holds
\begin{align*}
	\Vert\psi\Vert_{\sf{L}^{2}(\sf{M}_{\mathrm{T}})}\leq C(\Vert\sf{S}\psi\Vert_{\sf{L}^{2}(\sf{M}_{\mathrm{T}})}+\Vert\psi\vert_{\Sigma_{0}}\Vert_{\sf{L}^{2}(\Sigma_{0})})\, .
\end{align*}
\end{lemma}

\begin{proof}
	The first step of the proof follows similar arguments of the proof of Proposition~\ref{Prop:En1}. We consider the real-valued $n$-form $\omega\in\Omega^{n}(\sf{M})$ defined by $\omega:=\langle\sigma_{\sf{S}}(\d x^{\mu})\psi,\psi\rangle_{\sf{E}}\,\partial_{\mu}\lrcorner \d\mu_{\sf{g}}$ and integrate over $\sf{M}_{t}=t^{-1}([0,t])$ with $t\in [0,\mathrm{T}]$ to obtain
	\begin{align*}
		\int_{\sf{M}_{t}}\d\omega=\int_{\Sigma_{t}}i_{t}^{\ast}\omega-\int_{\Sigma_{t_{0}}}i_{0}^{\ast}\omega\, ,
	\end{align*}
	where $i_{t}\:\Sigma_{t}\hookrightarrow\sf{M}$ are the canonical embeddings. Following the same steps as in the proof of Proposition~\ref{Prop:En1}, relying on the fact that $\Sigma$ is assumed to be compact, we obtain the estimate
	\begin{align*}
		\Vert\psi_{t}\Vert_{\sf{L}^{2}(\Sigma_{t})}^{2}\leq e^{\widetilde{C}t}\bigg(\widetilde{C}\int_{0}^{t}\Vert(\sf{S}\psi)_{\tau}\Vert_{\sf{L}^{2}(\Sigma_{\tau})}^{2}\,\d\tau+\Vert\psi_{0}\Vert_{\sf{L}^{2}(\Sigma_{0})}^{2}\bigg)\, .
	\end{align*}
	for some constant $\widetilde{C}=\widetilde{C}(\mathrm{T})>0$. In particular, we obtain the estimate
	\begin{align*}
	\Vert\psi_{t}\Vert_{\sf{L}^{2}(\Sigma_{t})}^{2}\leq C\bigg(\Vert\sf{S}\psi\Vert^{2}_{\sf{L}^{2}(\sf{M}_{\mathrm{T}})}+\Vert\psi_{0}\Vert_{\sf{L}^{2}(\Sigma_{0})}^{2}\bigg)
	\end{align*}
	for some suitable constant $C=C(\mathrm{T})>0$. Integrating over the time variable $t$ from $0$ to $\mathrm{T}$ yields the claimed result. Last but not least, we take the square root of the this inequality and use the fact that $\sqrt{a^{2}+b^{2}}\leq \vert a\vert+\vert b\vert$ for all $a,b\in\bb{R}$.
\end{proof}

Next, let us discuss the notion of \emph{weak solutions} on a time strip $\sf{M}_{\mathrm{T}}:=t^{-1}([0,\mathrm{T}])$ with $\mathrm{T}>0$. To start with, let us define the operator $\sf{S}^{t}\in\mathrm{DO}^{1}(\sf{E})$ by 
\begin{align*}
	\sf{S}^{t}:=\sigma_{\sf{S}}(\d t)^{-1}\circ\sf{S}^{\ast}\circ\sigma_{\sf{S}}(\d t)\,,
\end{align*}
where we recall that $\sigma(\d t)\in\Gamma^{\infty}(\mathrm{End}(\sf{E}))\cong\mathrm{End}_{C^{\infty}(\sf{M})}(\Gamma^{\infty}(\sf{E}))$ is (pointwise) invertible and where $\sf{S}^{\ast}$ denotes the formal adjoint with respect to $(\cdot,\cdot)_{\sf{E}}=\int_{\sf{M}}\langle\cdot,\cdot\rangle_{\sf{E}}\,\d\mu_{\sf{g}}$, as usual. By definition, $\sf{S}^{t}$ is the formal adjoint with respect to $\int_{\sf{M}}\langle\cdot,\cdot\rangle_{\sf{E},\beta}\,\d\mu_{\sf{g}_{0}}$ with $\sf{g}_{0}=\d t\otimes\d t+\sf{h}_{t}$. Now, consider the Cauchy problem
\begin{align}\label{eq:Cauchy2}
	\begin{cases}
		\sf{S}\psi &=\phi\\
		\psi\vert_{\Sigma_{0}}&=\mathfrak{f}
	\end{cases}
\end{align}
for arbitrary Cauchy data $(\mathfrak{f},\phi)\in\Gamma^{\infty}(\Sigma_{0},\sf{E}\vert_{\Sigma_{0}})\times\Gamma^{\infty}(\sf{M},\sf{E})$. At this stage, it is useful to note that we can always assume $\mathfrak{f}=0$ without loss of generality. Indeed, if $\psi$ is a solution of the Cauchy problem~\eqref{eq:Cauchy2}, then $\widetilde{\psi}:=\psi-\pi_{2}^{\ast}\mathfrak{f}$ is a solution to Cauchy problem
\begin{align*}
	\begin{cases}
		\sf{S}\widetilde{\psi} &=\phi_{\mathfrak{f}}\\
		\psi\vert_{\Sigma_{0}}&=0
	\end{cases}
\end{align*}
with source $\phi_{\mathfrak{f}}:=\phi-\sf{S}(\pi_{2}^{\ast}\mathfrak{f})$, where $\pi_{2}\:\sf{M}\to\Sigma$ denotes the obvious projection. Now, we call $\psi\in\sf{L}^{2}(\sf{M}_{\mathrm{T}},\sf{E})$ a \textit{weak solution} to the Cauchy problem~\eqref{eq:Cauchy2} with $\mathfrak{f}=0$, if 
\begin{align*}
	\langle\psi,\sf{S}^{t}\varphi\rangle_{\sf{L}^{2}(\sf{M}_{\mathrm{T}})}=\langle\phi,\varphi\rangle_{\sf{L}^{2}(\sf{M}_{\mathrm{T}})}
\end{align*}
for all text sections $\varphi\in\mathcal{D}:=\{\varphi\in\Gamma^{\infty}(\sf{M}_{\mathrm{T}},\sf{E})\mid \varphi_{\mathrm{T}}=0\}\subset\Gamma^{\infty}(\sf{M}_{\mathrm{T}},\sf{E})$. Integration by parts and standard density arguments show that any \emph{smooth} weak solution $\psi$, is also a classical solution, i.e.~$\sf{S}\psi=\phi$ and $\psi\vert_{t=0}=0$. 

Using the energy estimates derived in Lemma~\ref{Lem:EE}, we obtain the following result.

\begin{proposition}\label{Prop:ExWeakSHS} \emph{(Existence of Weak Solutions in the Spatially Compact Case)}\newline
Let $\sf{S}\in\mathrm{DO}^{1}(\sf{E})$ be a symmetric hyperbolic system on $(\sf{M},\sf{g})$ and $t\in C^{\infty}(\sf{M})$ be a Cauchy temporal function. Furthermore, assume that $\Sigma$ is compact and consider a time strip $\sf{M}_{\mathrm{T}}=t^{-1}([0,\mathrm{T}])$. Then, there exists a \emph{weak solution} $\psi\in\sf{L}^{2}(\sf{M}_{\mathrm{T}},\sf{E})$ to the Cauchy problem~\eqref{eq:Cauchy2} with initial datum $\mathfrak{f}=0$ and source $\phi\in\Gamma^{\infty}(\sf{E})$.
\end{proposition}

\begin{proof}
	Since the formal adjoint $-\sf{S}^{\ast}$ is a symmetric hyperbolic system on its own right, by Lemma~\ref{Lemma:AdjointSHS}, and since $\Sigma$ is assumed to be compact, which implies that $\sigma(\d t)\in\Gamma^{\infty}(\mathrm{End}(\sf{E}))$ is bounded on $\sf{M}_{\mathrm{T}}$, we obtain from Lemma~\ref{Lem:EE}
	\begin{align}\label{eq:EstimateWeak}
		\Vert\varphi\Vert_{\sf{L}^{2}(\sf{M}_{\mathrm{T}})}\leq C(\Vert \sf{S}^{t}\varphi\Vert_{\sf{L}^{2}(\sf{M}_{\mathrm{T}})}+\Vert\varphi_{\mathrm{T}}\Vert_{\sf{L}^{2}(\Sigma_{\mathrm{T}})})=C\Vert \sf{S}^{t}\varphi\Vert_{\sf{L}^{2}(\sf{M}_{\mathrm{T}})}
	\end{align}
	for all $\varphi\in\mathcal{D}$, where we used a time reversal argument to assign the initial data on $\Sigma_{\mathrm{T}}$ rather than $\Sigma_{0}$. Now, consider the linear functional $l_{\phi}\:\mathrm{ran}(\sf{S}^{t}\vert_{\mathcal{D}})\to\bb{C}$ defined by 
	\begin{align*}
		l_{\phi}(\omega):=\langle\phi,\varphi\rangle_{\sf{L}^{2}(\sf{M}_{\mathrm{T}})}\,,
	\end{align*}
	where $\varphi\in\mathcal{D}$ is such that $\sf{S}^{t}\varphi=\omega$. Note that this definition is independent of the choice of $\varphi$: for any other $\widetilde{\varphi}\in\mathcal{D}$ with $\sf{S}^{t}\widetilde{\varphi}=\omega$, the section $\theta=\varphi-\widetilde{\varphi}$ satisfies $\sf{S}^{t}\theta=0$ and hence $\theta=0$ by the energy estimate Eq.~\eqref{eq:EstimateWeak}. Moreover, the estimate~\eqref{eq:EstimateWeak} implies that $l_{\phi}$ is bounded, since
	\begin{align*}
		\vert l_{\phi}(\omega)\vert =\vert(\langle\phi,\varphi\rangle_{\sf{L}^{2}(\sf{M}_{\mathrm{T}})}\vert\leq  \Vert\phi\Vert_{\sf{L}^{2}(\sf{M}_{\mathrm{T}})}\cdot\Vert\varphi\Vert_{\sf{L}^{2}(\sf{M}_{\mathrm{T}})}\leq C_{\phi}\Vert\omega\Vert_{\sf{L}^{2}(\sf{M}_{\mathrm{T}})}
	\end{align*}
	for all $\omega\in\mathrm{ran}(\sf{S}^{t}\vert_{\mathcal{D}})$ with constant $C_{\phi}:=C\Vert\phi\Vert_{\sf{L}^{2}(\sf{M}_{\mathrm{T}})}$. Now, consider the Hilbert space 
	\begin{align*}
		\mathcal{H}:=\overline{\mathrm{ran}(\sf{S}^{t}\vert_{\mathcal{D}})}^{\Vert\cdot\Vert_{\sf{L}^{2}(\sf{M}_{\mathrm{T}})}}\,,
	\end{align*}
	which naturally can be viewed as a closed subspace of $\sf{L}^{2}(\sf{M}_{\mathrm{T}},\sf{E})$. Then, since $l_{\phi}\:\mathrm{ran}(\sf{S}^{t}\vert_{\mathcal{D}})\to\bb{C}$ is bounded, it extends to a linear and continuous functional $l_{\phi}\:\mathcal{H}\to\bb{C}$. Furthermore, the \emph{representation theorem of Fréchet-Riesz} (see e.g.~\cite[Thm.~V.3.6]{Werner}) implies that there exists a unique $\psi\in\mathcal{H}$ such that $l_{\phi}(\omega)=\langle\psi,\omega\rangle_{\sf{L}^{2}(\sf{M}_{\mathrm{T}})}$ for all $\omega\in\mathcal{H}$. In particular, we obtain
	\begin{align*}
		\langle\phi,\varphi\rangle_{\sf{L}^{2}(\sf{M}_{\mathrm{T}})}=l_{\phi}(\omega)=\langle\psi,\omega\rangle_{\sf{L}^{2}(\sf{M}_{\mathrm{T}})}=\langle\psi,\sf{S}^{t}\varphi\rangle_{\sf{L}^{2}(\sf{M}_{\mathrm{T}})}
	\end{align*}
	for all test sections $\varphi\in\mathcal{D}$. In other words, $\psi$ is a weak solution to the Cauchy problem~\eqref{eq:Cauchy2} with initial datum $\mathfrak{f}=0$.
\end{proof}

Note that we only obtain \emph{uniqueness} of weak solutions on $\mathcal{H}$, which in general is only a closed subspace of $\sf{L}^{2}(\sf{M}_{\mathrm{T}},\sf{E})$, since it is not clear a priori that $\sf{S}^{t}$ defined on $\mathcal{D}$ has a dense image. Nevertheless, we will show below that the solution $\psi$ constructed abstractly in the previous proposition is actually \emph{smooth} and hence, uniqueness will follow from Corollary~\ref{Cor:SHSUni}.

As a next step, as already indicated above, we prove that the weak solutions constructed above are in fact smooth. To this end, we follow the approach of \cite[Sec.~4.2]{SchmidMurroFinster}, inspired by the discussion in~\cite[Sec.~5.3]{John} for symmetric hyperbolic systems on $\mathbb{R}^{n}$ (see also~\cite[Sec.~13.3]{FinsterBook}) to the global framework of manifolds considered in this section.

Let us introduce the following notation. We consider the bundle $\mathcal{E}:=\sf{E}\oplus(\sf{E}\otimes\sf{T}^{\ast}\sf{M})$ equipped with the bundle metric $\langle\cdot,\cdot\rangle_{\mathcal{E}}$ induced from $\langle\cdot,\cdot\rangle_{\sf{E}}$ and $\sf{g}_{0}=\d t\otimes\d t+\sf{h}_{t}$ and define
\begin{align*}
	\Gamma^{\infty}(\sf{E})\ni\quad\psi\mapsto\Theta_{\psi}:=(\psi,\nabla^{\sf{E}}\psi)\quad\in\Gamma^{\infty}(\mathcal{E})\, .
\end{align*}
Now, the goal of the following is to show that there exists symmetric hyperbolic system $\mathfrak{S}$ on $(\mathcal{E},\langle\cdot,\cdot\rangle_{\mathcal{E}})$, such that for every $\psi,\varphi\in \Gamma^{\infty}(\sf{E})$, the equation $\sf{S}\psi=\phi$ is equivalent to $\mathfrak{S}\Theta_{\psi}=\Theta_{\phi}$.

\begin{lemma}\label{Lemma:BiggerSystem} \emph{(Following \cite[Prop.~4.9]{SchmidMurroFinster})}\newline
Let $\sf{S}$ be a symmetric hyperbolic system over $(\sf{E},\langle\cdot,\cdot\rangle_{\sf{E}})$. There exists a unique symmetric hyperbolic system $\mathfrak{S}$ on $(\mathcal{E},\langle\cdot,\cdot\rangle_{\mathcal{E}})$ with (diagonal) principal symbol
\begin{align*}
	\sigma_{\mathfrak{S}}(\xi):=\begin{pmatrix}
		\sigma_{\sf{S}}(\xi) &0\\ 0 & \sigma_{\sf{S}}(\xi)\otimes\mathrm{id}_{\sf{T}^{\ast}\sf{M}}
	\end{pmatrix}\in \mathrm{End}(\mathcal{E}),\qquad\forall \xi\in \sf{T}^{\ast}\sf{M}\, ,
\end{align*}
such that $\sf{S}\psi=\phi$ if and only if $\mathfrak{S}\Theta_{\psi}=\Theta_{\phi}$ for all $\psi,\phi\in \Gamma^{\infty}(\sf{E})$.
\end{lemma}

\begin{proof}
	
	Let $\nabla^{\sf{E}}\:\Gamma^{\infty}(\sf{E})\to\Gamma^{\infty}(\sf{E}\otimes\sf{T}^{\ast}\sf{M})$ be a (fixed) connection on $\sf{E}$ and consider the \emph{commutator} of $\sf{S}$ and $\nabla^{\sf{E}}$ defined by
	\begin{align*}
		\sf{C}:=\nabla^{\sf{E}}\circ\sf{S}-(\sf{S}\otimes\mathrm{id})\circ\nabla^{\sf{E}}\:\Gamma^{\infty}(\sf{E})\to\Gamma^{\infty}(\sf{E}\otimes\sf{T}^{\ast}\sf{M})\, ,
	\end{align*}
	where $(\sf{S}\otimes\mathrm{id})(\psi\otimes\xi):=\sf{S}\psi\otimes\xi$ for all $\psi\in\Gamma^{\infty}(\sf{E})$ and $\xi\in\mathfrak{X}^{\ast}(\sf{M})$. This defines a linear differential operator of order one, i.e.~$\sf{C}\in\mathrm{DO}^{1}(\sf{E},\sf{E}\otimes\sf{T}^{\ast}\sf{M})$, as one can easily see by noting that
	\begin{align*}
		(\sigma_{\nabla^{\sf{E}}}(\xi)\circ\sigma_{\sf{S}}(\xi)-\sigma_{\sf{S}\otimes\mathrm{id}}(\xi)\circ\sigma_{\nabla^{\sf{E}}}(\xi))\psi=\sigma_{\sf{S}}(\xi)\psi\otimes\xi-\sigma_{\sf{S}}(\xi)\psi\otimes\omega=0
	\end{align*}
	for all $\xi\in\sf{T}^{\ast}\sf{M}$ and $\psi\in\sf{E}$. As a consequence, we can find two linear zero-order operators $\sf{R}_{1}\:\Gamma^{\infty}(\sf{E}\otimes\sf{T}^{\ast}\sf{M})\to\Gamma^{\infty}(\sf{E}\otimes\sf{T}^{\ast}\sf{M})$ and $\sf{R}_{2}\:\Gamma^{\infty}(\sf{E})\to\Gamma^{\infty}(\sf{E}\otimes\sf{T}^{\ast}\sf{M})$, uniquely determined by $\sf{S}$ and $\nabla^{\sf{E}}$, such that
	\begin{align*}
		\sf{C}=\sf{R}_{1}\circ\nabla^{\sf{E}}+\sf{R}_{2}\, .
	\end{align*}
	The precise expressions of $\sf{R}_{1}$ and $\sf{R}_{2}$ are not important for the sake of this proof. In general, they are the linear combinations of several terms constructed out of the principal symbol $\sigma_{\sf{S}}$, the zero-order terms $\sf{S}_{0}:=\sf{S}-\sigma_{\sf{S}}(\d x^{\mu})\nabla_{\partial_{\mu}}^{\sf{E}}$, the (covariant) derivatives of $\sigma_{\sf{S}}$ and $\sf{S}_{0}$, as well as bundle curvature terms of $(\sf{E},\nabla^{\sf{E}})$ coming from the commutator of $\nabla^{\sf{E}}$. Explicit expressions can be found in Remark~\ref{Rem:BigSys} below. Now, let us define the linear differential operator $\mathfrak{S}\in\mathrm{DO}^{1}(\mathcal{E})$ by
	\begin{align}\label{eq:ExtSHS}
		\mathfrak{S}:=
		\begin{pmatrix}
			\mathsf{S} & 0\\
			\sf{R}_{2} & \mathsf{S}\otimes\mathrm{id}+\sf{R}_{1}
		\end{pmatrix}\in\mathrm{DO}^{1}(\mathcal{E})\, .
	\end{align}
	By construction, it is clear that $\mathfrak{S}$ has the desired principal symbol. Moreover, $\mathfrak{S}$ has the required action on sections in the subspace $\mathrm{ran}(\Theta_{\bullet}\vert_{\Gamma^{\infty}})\subset\Gamma^{\infty}(\mathcal{E})$. Indeed, for all $\psi\in\Gamma^{\infty}(\sf{E})$, it holds that
	\begin{align*}
		\mathfrak{S}\Theta_{\psi}=\begin{pmatrix}
			\mathsf{S} & 0\\
			\sf{R}_{2} & \mathsf{S}\otimes\mathrm{id}+\sf{R}_{1}
		\end{pmatrix}\begin{pmatrix}\psi\\\nabla^{\sf{E}}\psi\end{pmatrix}=\begin{pmatrix}\sf{S}\psi\\\sf{R}_{1}\nabla^{\sf{E}}\psi+\sf{R}_{2}\psi+(\sf{S}\otimes\mathrm{id})\nabla^{\sf{E}}\psi\end{pmatrix}=\begin{pmatrix}\sf{S}\psi\\\nabla^{\sf{E}}\sf{S}\psi\end{pmatrix}=\Theta_{\sf{S}\psi}\,,
	\end{align*}
	where we used that $\sf{R}_{1}\nabla^{\sf{E}}\psi+\sf{R}_{2}\psi=\nabla^{\sf{E}}\sf{S}\psi-(\sf{S}\otimes\mathrm{id})\nabla^{\sf{E}}\psi$. In particular, $\sf{S}\psi=\phi$ if and only if $\mathfrak{S}\Theta_{\psi}=\Theta_{\phi}$ for all $\psi,\phi\in \Gamma^{\infty}(\sf{E})$.
	
	Now, by construction, it is clear that $\mathfrak{S}$ defines a symmetric hyperbolic system over $(\mathcal{E},\langle\cdot,\cdot\rangle_{\mathcal{E}})$, i.e.~$\sigma_{\mathfrak{S}}$ satisfies both conditions in Definition~\ref{Def:SHS}. For uniqueness, we note that the leading order terms are fixed by the prescribed principal symbol, while the zero-order terms are fixed by the requirement that $\sf{S}\psi=\phi$ if and only if $\mathfrak{S}\Theta_{\psi}=\Theta_{\phi}$ for all $\psi,\phi\in \Gamma^{\infty}(\sf{E})$. Indeed, if $\mathfrak{S}^{\prime}$ is any other symmetric hyperbolic system with the same principal symbol and the same property, then its difference $\mathfrak{S}-\mathfrak{S}^{\prime}$ is an operator of order zero that is identically zero on $\mathrm{ran}(\Theta_{\bullet})\subset\Gamma^{\infty}(\mathcal{E})$. Now, for every $p\in\sf{M}$ and $(\omega,\xi)\in\sf{E}_{p}\oplus (\sf{E}_{p}\otimes\sf{T}^{\ast}_{p}\sf{M})$, there is a section $\psi\in\Gamma^{\infty}(\sf{E})$ such that $\psi(p)=\omega$ and $(\nabla^{\sf{E}}\psi)(p)=\xi$, as one can easily check in local coordinates. Hence, since the difference $\mathfrak{S}-\mathfrak{S}^{\prime}=0$ is an endomorphism, we find $\mathfrak{S}-\mathfrak{S}^{\prime}=0$ on all of $\Gamma^{\infty}(\mathcal{E})$.
\end{proof}

As discussed above, $\mathfrak{S}$ is uniquely determined by its principal symbol and the requirement that $\sf{S}\psi=\phi$ if and only if $\mathfrak{S}\Theta_{\psi}=\Theta_{\phi}$ for all $\psi,\phi\in\Gamma^{\infty}(\sf{E})$. Let us stress, however, that $\Theta_{\bullet}$ and hence also $\mathfrak{S}$ depends explicitly on the choice of connection $\nabla^{\sf{E}}$. Choosing a different connection on $\sf{E}$ results into different zero-order components $\sf{R}_{1}$ and $\sf{R}_{2}$.

\begin{remark}\label{Rem:BigSys} (Explicit Expressions of the Extended System)\newline
Let $\sf{S}$ be a symmetric hyperbolic system. For completeness, let us derive a more explicit expression of the extended system $\mathfrak{S}$. We will provide two such expressions. First, we derive explicit expressions of the zero-order operators $\sf{R}_{1}$ and $\sf{R}_{2}$ appearing in Eq.~\eqref{eq:ExtSHS}. Secondly, we write $\mathfrak{S}$ in the form $\mathfrak{S}=\sigma_{\mathfrak{S}}(\d x^{\mu})\nabla^{\mathcal{E}}_{\partial_{\mu}}+\mathfrak{S}_{0}$ for a suitable connection $\nabla^{\mathcal{E}}$ on $\mathcal{E}$.
\begin{itemize}
	\item[(i)]First of all, we derive explicit expressions of the zero-order operators $\sf{R}_{1}$ and $\sf{R}_{2}$ defined by $\sf{R}_{1}\circ\nabla^{\sf{E}}+\sf{R}_{2}=\nabla^{\sf{E}}\circ\sf{S}-(\sf{S}\otimes\mathrm{id})\circ\nabla^{\sf{E}}$. Let us choose local coordinates $(x^{\mu})_{\mu=0,\dots,n}$ on some open subset $\mathcal{U}$ of $\sf{M}$ and write $\sf{S}=\sigma_{\sf{S}}(\d x^{\mu})\nabla^{\sf{E}}_{\partial_\mu}+\sf{S}_{0}$ for some zero-order operator $\sf{S}_{0}$. Now, the commutator $[\nabla_{\partial_{\mu}}^{\sf{E}},\sf{S}]$ for some fixed basis vector $\partial_{\mu}$ is an operator of order one acting on $\Gamma^{\infty}(\mathcal{U},\sf{E})$ and a straightforward computation shows that it is explicitly given by
	\begin{align}\label{eq:SHSCommu}
		[\nabla_{\partial_{\mu}}^{\sf{E}},\sf{S}]\psi=(\nabla^{\mathrm{End}(\sf{E})}_{\partial_{\mu}}\sigma_{\sf{S}}(\d x^{\alpha}))\nabla_{\partial_{\alpha}}^{\sf{E}}\psi+\sigma_{\sf{S}}(\d x^{\alpha})\sf{F}^{\nabla^{\sf{E}}}_{\mu\alpha}(\psi)-(\nabla^{\mathrm{End}(\sf{E})}_{\partial_{\mu}}\sf{S}_{0})(\psi)
	\end{align}
	 for all $\psi\in \Gamma^{\infty}(\mathcal{U},\sf{E})$, where $\nabla^{\mathrm{End}(\sf{E})}$ denotes the induced connection on $\mathrm{End}(\sf{E})$, as usual, and where $\sf{F}^{\nabla^{\sf{E}}}\in\Omega^{2}(\sf{M},\mathrm{End}(\sf{E}))$ is the curvature $2$-form of $(\sf{E},\nabla^{\sf{E}})$ (see pp.~\pageref{Conventions}ff.). Then, for all $\psi\in\Gamma^{\infty}(\mathcal{U},\sf{E})$, we write $\nabla^{\sf{E}}\psi=\nabla^{\sf{E}}_{\partial_{\mu}}\psi\otimes\d x^{\mu}$ and obtain from Eq.~\eqref{eq:SHSCommu}
	\begin{align*}
		\nabla^{\sf{E}}\sf{S}\psi-&(\sf{S}\otimes\mathrm{id})\nabla^{\sf{E}}\psi=\nabla^{\sf{E}}_{\partial_{\mu}}(\sf{S}\psi)\otimes\d x^{\mu}-\sf{S}\nabla^{\sf{E}}_{\partial_{\mu}}\psi\otimes\d x^{\mu}=[\nabla^{\sf{E}}_{\partial_{\mu}},\sf{S}]\psi\otimes\d x^{\mu}=\\&=\bigg\{(\nabla^{\mathrm{End}(\sf{E})}_{\partial_{\mu}}\sigma_{\sf{S}}(\d x^{\alpha}))\nabla_{\partial_{\alpha}}^{\sf{E}}\psi+\sigma_{\sf{S}}(\d x^{\alpha})\sf{F}^{\nabla^{\sf{E}}}_{\mu\alpha}(\psi)-(\nabla^{\mathrm{End}(\sf{E})}_{\partial_{\mu}}\sf{S}_{0})(\psi)\bigg\}\otimes\d x^{\mu}\, .
	\end{align*}
	From this computation, we can immediately read of the operators $\sf{R}_{1}$ and $\sf{R}_{2}$ and obtain
	\begin{align*}
		&\sf{R}_{1}\:\Gamma^{\infty}(\sf{E}\otimes\sf{T}^{\ast}\sf{M})\to\Gamma^{\infty}(\sf{E}\otimes\sf{T}^{\ast}\sf{M})\,,\quad  &&\psi_{\alpha}\otimes\d x^{\alpha}\mapsto \big\{(\nabla^{\mathrm{End}(\sf{E})}_{\partial_{\mu}}\sigma_{\sf{S}}(\d x^{\alpha}))\psi_{\alpha}\big\}\otimes\d x^{\mu}\\
		&\sf{R}_{2}\:\Gamma^{\infty}(\sf{E})\to\Gamma^{\infty}(\sf{E}\otimes\sf{T}^{\ast}\sf{M})\,,\quad &&\psi \mapsto \big\{\sigma_{\sf{S}}(\d x^{\alpha})\sf{F}^{\nabla^{\sf{E}}}_{\mu\alpha}(\psi)-(\nabla^{\mathrm{End}(\sf{E})}_{\partial_{\mu}}\sf{S}_{0})(\psi)\big\}\otimes\d x^{\mu}\,,
	\end{align*}
	where we used the fact that every section of $\Gamma^{\infty}(\sf{E}\otimes\sf{T}^{\ast}\sf{M})$ can locally be written as $\psi_{\alpha}\otimes\d x^{\alpha}$ with local coefficient sections $\{\psi_{\alpha}\}_{\alpha=0,\dots,n}\subset\Gamma^{\infty}(\mathcal{U},\sf{E})$. 
	\item[(ii)]Using the connection $\nabla^{\sf{E}}$ on $\sf{E}$ and the Levi-Civita connection $\nabla^{0}$ on $(\sf{M},\sf{g}_{0})$, let us consider the induced connection $\nabla^{\mathcal{E}}$ on $\mathcal{E}=\sf{E}\oplus(\sf{E}\otimes\sf{T}^{\ast}\sf{M})$. Then, writing $\sf{S}=\sigma_{\sf{S}}(\d x^{\mu})\nabla^{\sf{E}}_{\partial_\mu}+\sf{S}_{0}$, we find
\begin{align*}
	\sigma_{\mathfrak{S}}(\d x^{\mu})\nabla_{\partial_\mu}^{\mathcal{E}}\Theta_{\psi}&=\begin{pmatrix}
	\sigma_{\sf{S}}(\d x^{\mu})\nabla^{\sf{E}}_{\partial_\mu}\psi\\
	(\sigma_{\sf{S}}(\d x^{\mu})\otimes\mathrm{id})\nabla^{\mathcal{E}}_{\partial_\mu}(\nabla^{\sf{E}}\psi)
	\end{pmatrix}\\&=\begin{pmatrix}
	(\sf{S}-\sf{S}_{0})\psi\\
	\sigma_{\sf{S}}(\d x^{\mu})\nabla^{\sf{E}}_{\partial_\mu}\nabla^{\sf{E}}_{\partial_\alpha}\psi\otimes \d x^{\alpha}+ \sigma_{\sf{S}}(\d x^{\mu})\nabla^{\sf{E}}_{\partial_\alpha}\psi\otimes \nabla^{0}_{\partial_\mu}\d x^{\alpha}
	\end{pmatrix}\\&=\begin{pmatrix}
	(\sf{S}-\sf{S}_{0})\psi\\
	(\sf{S}-\sf{S}_{0})(\nabla^{\sf{E}}_{\partial_\alpha}\psi)\otimes \d x^{\alpha}-(\Gamma^{0})_{\mu\beta}^{\alpha}\sigma_{\sf{S}}(\d x^{\mu})\nabla_{\partial_\alpha}^{\sf{E}}\psi\otimes \d x^{\beta}
	\end{pmatrix}\\&=
	\Theta_{\sf{S}\psi}-\begin{pmatrix}
	\sf{S}_{0}\psi\\
	([\nabla^{\sf{E}}_{\partial_\mu},\sf{S}]\psi+\sf{S}_{0}\nabla^{\sf{E}}_{\partial_\mu}\psi+(\Gamma^{0})_{\mu\beta}^{\alpha}\sigma_{\sf{S}}(\d x^{\mu})\nabla_{\partial_\alpha}^{\sf{E}}\psi)\otimes \d x^{\mu}
	\end{pmatrix}\, ,
\end{align*}
for all $\psi\in\Gamma^{\infty}(\sf{E})$, where $\Gamma^{0}$ are the Christoffel symbols corresponding to $\nabla^{0}$. In particular, 
\begin{align*}
	\mathfrak{S}=\sigma_{\mathfrak{S}}(\d x^{\mu})\nabla^{\mathcal{E}}_{\partial_{\mu}}+\mathfrak{S}_{0}\,,\qquad\mathfrak{S}_{0}:=
	\begin{pmatrix}
		\sf{S}_{0} & 0 \\
		\sf{R}_{2} & \sf{R}_{1}+(\sf{S}_{0}\otimes\mathrm{id})+\sf{G}
	\end{pmatrix}\,,
\end{align*}
where $\sf{G}\:\Gamma^{\infty}(\sf{E}\otimes\sf{T}^{\ast}\sf{M})\to\Gamma^{\infty}(\sf{E}\otimes\sf{T}^{\ast}\sf{M}),\,\psi_{\alpha}\otimes\d x^{\alpha}\mapsto \{(\Gamma^{0})_{\mu\beta}^{\alpha}\sigma_{\sf{S}}(\d x^{\mu})\psi_{\alpha}\}\otimes \d x^{\mu}$ is an operator of order zero. Let us stress that $\mathfrak{S}_{0}$ depends, of course, on the choice of connection on $\mathscr{E}$, which here is chosen to be the one induced from $\nabla^{\sf{E}}$ and $\nabla^{0}$.
\end{itemize}
\end{remark}

Now, consider the (open) time strip $\sf{M}_{\mathrm{T}}^{\circ}=(0,\mathrm{T})\times\Sigma$ equipped with the positive-definite auxiliary metric $\sf{g}_{0}=\d t\otimes\d t+\sf{h}_{t}$, as above. We denote the corresponding $\sf{L}^{2}$-\emph{Sobolev space} of order $k\in\bb{N}_{0}$ by
\begin{align*}
	\sf{H}^{k}(\sf{M}_{\mathrm{T}},\sf{E}):=\overline{\{\varphi\in\Gamma^{\infty}(\sf{M}_{\mathrm{T}}^{\circ},\sf{E})\mid \Vert\varphi\Vert_{\sf{H}^{k}(\sf{M}_{\mathrm{T}})}<\infty\}}^{\Vert\cdot\Vert_{\sf{H}^{k}(\sf{M}_{\mathrm{T}})}}\,,\quad \Vert\varphi\Vert_{\sf{H}^{k}(\sf{M}_{\mathrm{T}})}:=\sum_{i=0}^{k}\Vert\nabla^{\sf{E},i}\varphi\Vert_{\sf{L}^{2}(\sf{M}_{\mathrm{T}})}\,,
\end{align*}
where we denote by $\nabla^{\sf{E},i}\:\Gamma^{\infty}(\sf{E})\to\Gamma^{\infty}(\sf{E}\otimes\sf{T}^{\ast}\sf{M}^{\otimes k})$ the $k$th covariant derivative induced by $\nabla^{\sf{E}}$, $\nabla^{0}$ and by $\Vert\cdot\Vert_{\sf{L}^{2}(\sf{M}_{\mathrm{T}})}$ the norm of $\sf{L}^{2}(\sf{M}_{\mathrm{T}},\sf{E}\otimes\sf{T}^{\ast}\sf{M}^{\otimes k})$ with bundle metric on $\sf{E}\otimes\sf{T}^{\ast}\sf{M}^{\otimes k}$ induced by $\langle\cdot,\cdot\rangle_{\sf{E},\beta}$ and $\sf{g}_{0}$. With this notation, it is clear that 
\begin{align*}
	\Vert\Theta_{\psi}\Vert_{\sf{L}^{2}(\sf{M}_{\mathrm{T}})}=\Vert\psi\Vert_{\sf{H}^{1}(\sf{M}_{\mathrm{T}})}
\end{align*}
for all $\psi\in\Gamma^{\infty}(\sf{M},\sf{E})$. With this notation, we are finally able to prove the following proposition:

\begin{proposition}\label{Prop:RegSol} \emph{(Regularity of Solution in the Spatially Compact Case)}\newline
Let $\sf{S}\in\mathrm{DO}^{1}(\sf{E})$ be a symmetric hyperbolic system on $(\sf{M},\sf{g})$ and $t\in C^{\infty}(\sf{M})$ be a Cauchy temporal function. Furthermore, assume that $\Sigma$ is compact. Then, there exists a \emph{smooth solution} $\psi\in\Gamma^{\infty}(\sf{M},\sf{E})$ to the Cauchy problem~\eqref{eq:Cauchy2} for any source $\phi\in\Gamma^{\infty}_{\mathrm{c}}(\sf{M},\sf{E})$.
\end{proposition}

\begin{proof}
	Consider a time strip $\sf{M}_{\mathrm{T}}:=t^{-1}([0,\mathrm{T}])$. We first show that the weak solution constructed in Proposition~\ref{Prop:ExWeakSHS} is contained in $\sf{H}^{k}(\sf{M}_{\mathrm{T}},\sf{E})$ for all $k\in\bb{N}_{0}$. Consider the extended system $\mathfrak{S}$ on $(\mathcal{E},\langle\cdot,\cdot\rangle_{\mathcal{E}})$ with $\mathcal{E}=\sf{E}\oplus (\sf{E}\otimes\sf{T}^{\ast}\sf{M})$, as defined in Lemma~\ref{Lemma:BiggerSystem}. Now, suppose that there exist a smooth solution $\psi\in\Gamma^{\infty}(\sf{E})$ to the Cauchy problem in Eq.~\eqref{eq:Cauchy2} for Cauchy data $(\mathfrak{f},\phi)\in\Gamma^{\infty}(\Sigma_{0},\sf{E}\vert_{\Sigma_{0}})\times\Gamma^{\infty}(\sf{M},\sf{E})$. Then, $\Theta_{\psi}:=(\psi,\nabla^{\sf{E}}\psi)$ is the unique solution to $\mathfrak{S}\Theta_{\psi}=\Theta_{\phi}$ with initial data $\mathfrak{F}:=(\mathfrak{f},\mathfrak{g})$, where $\mathfrak{g}:=\mathfrak{g}(\phi,\mathfrak{f}):=\nabla^{\sf{E}}\psi\vert_{t=0}\in\Gamma^{\infty}(\Sigma_{0},(\sf{E}\otimes\sf{T}^{\ast}\sf{M})\vert_{\Sigma_{0}})$ is completely fixed by $\sf{S}$, $\phi\vert_{t=0}$ and $\mathfrak{f}$ through the equation $(\sf{S}\psi)\vert_{t=0}=\phi\vert_{t=0}$ restricted to $t=0$.\footnote{Choosing local coordinates $(x^{i})_{i=1,\dots,n}$ on $\Sigma$ and local coordinates $(x^{0}:=t,x^{i})$ on $\sf{M}$, we write $\sf{S}=\sigma_{\sf{S}}(\d x^{\mu})\nabla^{\sf{E}}_{\partial_{\mu}}+\sf{S}_{0}$ for some $\sf{S}_{0}\in\mathrm{DO}^{0}(\sf{E})$. Then, $\mathfrak{g}=\{\sigma_{\sf{S}}(\d t)^{-1}(\phi-\sigma_{\sf{S}}(\d x^{i})\mathfrak{f}-\sf{S}_{0}\mathfrak{f})\otimes\d t+\nabla^{\sf{E}}_{\partial_{i}}\mathfrak{f}\otimes\d x^{i}\}\vert_{t=0}$.} Now, in the following, we assume $\mathfrak{f}=0$ and that the source $\phi$ is supported far enough in the future of $\Sigma_{0}$, so that also $\mathfrak{g}=0$. This is, of course, no loss of generality, since we assumed the source $\phi\in\Gamma^{\infty}(\sf{E})$ to be compactly supported and hence, we only need to adjust the time strip accordingly.
	
	Since $\mathfrak{S}$ is a symmetric hyperbolic systems on its own right, Proposition~\ref{Prop:ExWeakSHS} implies that there exists a weak solution $\Psi=(\psi,\hat{\psi})\in\sf{L}^{2}(\sf{M}_{\mathrm{T}},\mathcal{E})$ to the Cauchy problem $\mathfrak{S}\Psi=\Theta_{\phi}$ with source $\Theta_{\phi}=(\phi,\nabla^{\sf{E}}\phi)$ and initial datum $\Psi\vert_{t=0}=0$, i.e.
	\begin{align}\label{eq:WeakSolProof}
		\langle\Psi,\mathfrak{S}^{t}\Phi\rangle_{\sf{L}^{2}(\sf{M}_{\mathrm{T}})}=\langle\Theta_{\phi},\Phi\rangle_{\sf{L}^{2}(\sf{M}_{\mathrm{T}})}
	\end{align}
	for all test sections $\Phi=(\varphi,\hat{\varphi})\in\Gamma^{\infty}(\sf{M}_{\mathrm{T}},\mathcal{E})$ with $\Phi_{t=\mathrm{T}}=0$. Now, we claim that $\psi$ is the weak solution of $\sf{S}\psi=\phi$ with $\psi\vert_{t=0}=0$, as constructed in Proposition~\ref{Prop:ExWeakSHS}, and that $\psi$ is weakly differentiable with weak derivative $\nabla^{\sf{E}}\psi=\hat{\psi}$. Indeed, both of this facts are encoded in Eq.~\eqref{eq:WeakSolProof} and the definition of the extended system $\mathfrak{S}$: choosing test sections with $\hat{\varphi}=0$, we recover the defining equation 
	\begin{align}\label{eq:Proof:Reg1}
		\langle\psi,\sf{S}^{t}\varphi\rangle_{\sf{L}^{2}(\sf{M}_{\mathrm{T}})}=\langle\phi,\varphi\rangle_{\sf{L}^{2}(\sf{M}_{\mathrm{T}})}
	\end{align}
	and hence, by uniqueness of weak solutions on the Hilbert space $\mathcal{H}\subset\sf{L}^{2}(\sf{M}_{\mathrm{T}},\sf{E})$, we conclude that $\psi$ coincides with the weak solution constructed in Proposition~\ref{Prop:ExWeakSHS}. On the other hand, choosing test sections with $\varphi=0$, we obtain
	\begin{align}\label{eq:Proof:Reg2}
		\langle\hat{\psi},(\sf{S}\otimes\mathrm{id})^{t}\hat{\varphi}\rangle_{\sf{L}^{2}(\sf{M}_{\mathrm{T}})}+\langle\hat{\psi},\sf{R}_{1}^{t}\hat{\varphi}\rangle_{\sf{L}^{2}(\sf{M}_{\mathrm{T}})}+&\langle\psi,\sf{R}_{2}^{t}\hat{\varphi}\rangle_{\sf{L}^{2}(\sf{M}_{\mathrm{T}})}=\langle\nabla^{\sf{E}}\phi,\hat{\varphi}\rangle_{\sf{L}^{2}(\sf{M}_{\mathrm{T}})}\, ,
	\end{align}
	where we denote by $\sf{A}^{t}$ the formal adjoint of some operator $\sf{A}$ with respect to $\int_{\sf{M}}\langle\cdot,\cdot\rangle_{\sf{E},\beta}\,\d\mu_{\sf{g}_{0}}$, as usual, and where $\sf{R}_{1}$ and $\sf{R}_{2}$ are the zero-order operators uniquely determined by $\sf{R}_{1}\circ\nabla^{\sf{E}}+\sf{R}_{2}=\nabla^{\sf{E}}\circ\sf{S}-(\sf{S}\otimes\mathrm{id})\circ\nabla^{\sf{E}}$. Now, consider the distributional derivative $\nabla^{\sf{E}}\psi$ of $\psi$. Then, using Eq.~\eqref{eq:Proof:Reg1}, we obtain
	\begin{align*}
		\langle\nabla^{\sf{E}}\psi,(\sf{S}+\mathrm{id})^{t}\hat{\varphi}\rangle_{\sf{L}^{2}(\sf{M}_{\mathrm{T}})}&=\langle\psi,(\nabla^{\sf{E}})^{t}(\sf{S}+\mathrm{id})^{t}\hat{\varphi}\rangle_{\sf{L}^{2}(\sf{M}_{\mathrm{T}})}=\\&=\langle\psi,(\sf{S}^{t}\circ (\nabla^{\sf{E}})^{t}-(\nabla^{\sf{E}})^{t}\circ\sf{R}_{1}^{t}-\sf{R}_{2}^{t})\hat{\varphi}\rangle_{\sf{L}^{2}(\sf{M}_{\mathrm{T}})}=\\&=\langle\nabla^{\sf{E}}\phi,\hat{\varphi}\rangle_{\sf{L}^{2}(\sf{M}_{\mathrm{T}})}-\langle\nabla^{\sf{E}}\psi,\sf{R}_{1}^{t}\hat{\varphi}\rangle_{\sf{L}^{2}(\sf{M}_{\mathrm{T}})}-\langle\psi,\sf{R}_{2}^{t}\hat{\varphi}\rangle_{\sf{L}^{2}(\sf{M}_{\mathrm{T}})}
	\end{align*}
	in the distributional sense. In particular, we see that $\nabla^{\sf{E}}\psi$ satisfies the same equation as $\hat{\psi}$ in Eq.~\eqref{eq:Proof:Reg2} and hence, by uniqueness, we conclude that $\hat{\psi}=\nabla^{\sf{E}}\psi$. It follows that $\psi\in\sf{H}^{1}(\sf{M}_{\mathrm{T}},\sf{E})$.
	
	Applying the same reasoning inductively, we eventually obtain $\psi\in\sf{H}^{k}(\sf{M}_{\mathrm{T}},\sf{E})$ for all $k\in\bb{N}_{0}$. Standard arguments employing the (local) Sobolev embedding theorems then imply $\psi\in C^{\infty}(\sf{M}_{\mathrm{T}},\sf{E})$. Taking the limit $\mathrm{T}\to\infty$ and using a time reversal argument, we hence conclude that there exists a smooth solution $\psi\in\Gamma^{\infty}(\sf{E})$ to the Cauchy problem in Eq.~\eqref{eq:Cauchy2} with $\mathfrak{f}=0$. Existence of solutions to the case $\mathfrak{f}\neq 0$ then follows as explained above, i.e., by considering the modified source $\phi_{\mathfrak{f}}:=\phi-\sf{S}(\pi_{2}^{\ast}\mathfrak{f})$.
\end{proof}

Collecting all the previous results, we are finally in the position to prove Theorem~\ref{Thm:Cauchy}, i.e.~the well-posedness of the Cauchy problem for symmetric hyperbolic systems.

\begin{proof}[Proof of Theorem~\ref{Thm:Cauchy}] 
	Claim (ii), uniqueness and finite speed of propagation, has been shown in Corollary~\ref{Cor:SHSUni}. 
	
	Let us now turn to (i), the existence of (smooth) solutions. Let $(\sf{M},\sf{g})$ be an arbitrary globally hyperbolic spacetime with $\sf{M}=\bb{R}\times\Sigma$ and $(\mathfrak{f},\phi)\in\Gamma^{\infty}_{\mathrm{c}}(\Sigma_{0},\sf{E}\vert_{\Sigma_{0}})\times\Gamma^{\infty}_{\mathrm{c}}(\sf{M},\sf{E})$ be compactly-supported Cauchy data. In the case in which $\Sigma$ is compact, we have shown in Proposition~\ref{Prop:RegSol} that there exists a smooth solution $\psi\in\Gamma^{\infty}(\sf{E})$ to the Cauchy problem $\sf{S}\psi=\phi$, $\psi\vert_{t=0}=\mathfrak{f}$, which is unique by Corollary~\ref{Cor:SHSUni}. Let us now extend this result to the case in which $\Sigma$ is non-compact. To this end, we follow the strategy outlined in the lecture notes \cite[Proof of Thm.~3.7.7]{BaerLecture}, in which the problem is reduced to the spatially compact case: set $\sf{K}:=\mathrm{supp}(\mathfrak{f})\cup\mathrm{supp}(\phi)$, which, by definition, is a compact subset of $\sf{M}$. Furthermore, we choose $\mathrm{T}>0$ big enough so that $\sf{K}\subset\sf{M}_{\mathrm{T}}^{\circ}:=(-\mathrm{T},\mathrm{T})\times\Sigma$. Clearly, $\sf{M}_{\mathrm{T}}^{\circ}$ is itself a globally hyperbolic spacetime. Now, consider the compact set $\mathcal{J}(\sf{K})\cap\sf{M}_{\mathrm{T}}$ with $\sf{M}_{\mathrm{T}}=[-\mathrm{T},\mathrm{T}]\times\Sigma$ and define $\hat{\Sigma}:=\mathrm{\pi}_{1}(\mathcal{J}(\sf{K})\cap\sf{M}_{\mathrm{T}})\subset\Sigma_{0}$, where $\pi_{1}\:\sf{M}\to\Sigma_{0}$ denotes the projection $(t,\vec{x})\mapsto (0,\vec{x})$, see Figure~\ref{Fig:Proof}(a). Next, we choose a relatively compact open set $\mathcal{U}\subset\Sigma_{0}$ with smooth boundary $\partial\mathcal{U}$ such that $\hat{\Sigma}\subset\mathcal{U}$. Now, consider a disjoint copy of $\hat{\Sigma}$ and glue it to $\hat{\Sigma}$ smoothly along the boundary $\partial\mathcal{U}$ to obtain a manifold $\Sigma^{\prime}=\hat{\Sigma}\coprod_{\partial\mathcal{U}}\hat{\Sigma}$, see Figure~\ref{Fig:Proof}(b). Without loss of generality, we may assume that the metric $\sf{h}_{t}$ near $\partial\mathcal{U}$ is of product type, so that we obtain a smooth Riemannian metric on $\Sigma^{\prime}$. In this way, we have constructed a \textit{spatially-compact} manifold $\sf{M}_{\mathrm{T}}^{\prime}=(-\mathrm{T},\mathrm{T})\times\Sigma^{\prime}$ and since the supports of $\phi$ and $\mathfrak{f}$ are contained in $(-\mathrm{T},\mathrm{T})\times\hat{\Sigma}$, we may regard them as sections defined on $\sf{M}_{\mathrm{T}}^{\prime}$. Now, applying Proposition~\ref{Prop:RegSol}, we obtain a smooth solution in $\sf{M}_{\mathrm{T}}^{\prime}$, which, by finite speed of propagation, is contained in the subspace $\mathcal{J}(\sf{K})\cap\sf{M}_{\mathrm{T}}\subset(-\mathrm{T},\mathrm{T})\times\hat{\Sigma}$ and hence defines a smooth section on the original spacetime $\sf{M}$. Taking the limit $\mathrm{T}\to\infty$ and using uniqueness of solutions, we obtain existence of a global smooth solution $\Gamma^{\infty}(\sf{E})$.
	
	\begin{figure}[H] 
	\centering
    \subfloat[Definition of $\hat{\Sigma}$.]{\includegraphics[width=0.4\textwidth]{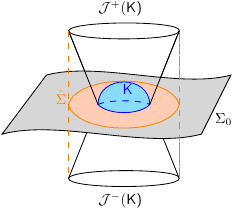}}\hspace*{1cm}
    \subfloat[Definition of $\Sigma^{\prime}$.]{\includegraphics[width=0.5\textwidth]{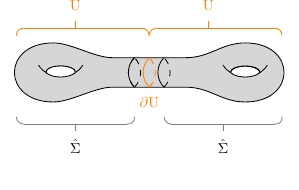}}
\caption{Sketch for the definition of $\hat{\Sigma}$ and $\Sigma^{\prime}$.\label{Fig:Proof}} 
\end{figure} 
	
	It remains to get rid of the support assumptions on Cauchy data: let $(\mathfrak{f},\phi)\in\Gamma^{\infty}(\Sigma_{0},\sf{E}\vert_{\Sigma_{0}})\times\Gamma^{\infty}(\sf{M},\sf{E})$ be arbitrary and consider an exhaustion of $\sf{M}$ by compact sets, i.e.~a sequence of subsets $(\sf{K}_{n})_{n\in\bb{N}}$ with $\sf{K}_{n}\subset\sf{M}$ compact, such that $\sf{K}_{n}\subset\sf{K}_{n+1}^{\circ}$ for all $n\in\bb{N}$. Moreover, choose cut-off functions $\chi_{n}\in C^{\infty}_{\mathrm{c}}(\sf{M})$ with $\chi_{n}=1$ on $\sf{K}_{n}$ for all $n\in\bb{N}$. Now, consider the unique solutions $\psi_{n}\in\Gamma^{\infty}(\sf{E})$ to the Cauchy problems
	\begin{align*}
		\begin{cases}
			\sf{S}\psi_{n}&=\chi_{n}\phi\\
			\psi_{n}\vert_{t=0}&=\chi_{n}\vert_{t=0}\mathfrak{f}
		\end{cases}\, ,
	\end{align*}
	which exist by the arguments above. By finite speed of propagation, i.e.~Corollary~\ref{Cor:SHSUni}, the solution $\psi_{n}$ on any compact region $\sf{K}\subset\sf{M}$ depends only on the initial data and source within $\mathcal{D}(\sf{K}):=\mathcal{J}(\sf{K})\cap\Sigma_{0}$. Now, for every compact $\sf{K}$, we choose $\mathrm{N}\in\bb{N}$ large enough so that $\mathcal{D}(\sf{K})\subset\sf{K}_{\mathrm{N}}$. Then, for any $n\geq\mathrm{N}$, we have $\psi_{n}\vert_{\sf{K}}=\psi_{\mathrm{N}}\vert_{\sf{K}}$, by uniqueness. With this observation, we define $\psi\in\Gamma^{\infty}(\sf{E})$ globally by setting $\psi\vert_{\sf{K}}:=\psi_{\mathrm{N}}\vert_{\sf{K}}$ for every compact set $\sf{K}\subset\sf{M}$ and $\mathrm{N}\in\bb{N}$ large enough. By construction, clearly $\sf{S}\psi=\phi$ and $\psi\vert_{t=0}=\mathfrak{f}$, which concludes the proof.
	
	Last but not least, we prove (iii), the stability of the Cauchy problem. To start with, consider the  map $\mathcal{R}_{\sf{S}}\:\Gamma^{\infty}(\sf{E})\to\Gamma^{\infty}(\sf{E})\times\Gamma^{\infty}(\sf{E}\vert_{\Sigma_{0}})$ defined by $\mathcal{R}_{\sf{S}}(\psi):=(\sf{S}\psi,\mathfrak{f})$. Clearly, $\mathcal{R}_{\sf{S}}$ is linear and continuous with respect to the Fréchet topologies. Moreover, by the discussion above, also bijective. Now, let $\sf{A}\subset\sf{M}$ be any closed subset and consider the closed subset $\Gamma^{\infty}_{\sf{A}}(\sf{M}):=\{\psi\in\Gamma^{\infty}(\sf{M})\mid \mathrm{supp}(\psi)\subset\sf{A}\}$. By continuity, also the preimage 
	\begin{align*}
		\sf{V}_{\sf{A}}:=\mathcal{R}_{\sf{S}}^{-1}(\Gamma^{\infty}_{\sf{A}}(\sf{E})\times\Gamma^{\infty}_{\sf{A}\cap\Sigma_{0}}(\sf{E}\vert_{\Sigma_{0}}))
	\end{align*}
	is closed and hence a Fréchet space on its own. Now, by finite speed of  propagation, $\sf{R}_{\sf{S}}\vert_{\sf{V}_{\sf{A}}}$ is well-defined and bijective as a map from $\sf{V}_{\sf{A}}$ to $\Gamma^{\infty}_{\sf{A}}(\sf{E})\times\Gamma^{\infty}_{\sf{A}\cap\Sigma_{0}}(\sf{E}\vert_{\Sigma_{0}})$ and by the \emph{open mapping theorem} (see e.g.~\cite[Thm.~2.11 and Cor.~2.12(b)]{Rudin}), $(\sf{R}_{\sf{S}}\vert_{\sf{V}_{\sf{A}}})^{-1}\:\Gamma^{\infty}_{\sf{A}}(\sf{E})\times\Gamma^{\infty}_{\sf{A}\cap\Sigma_{0}}(\sf{E}\vert_{\Sigma_{0}})\to\sf{V}_{\sf{A}}$ is continuous. Hence, using the continuous embeddings $\Gamma_{\sf{A}}^{\infty}(\sf{A})\hookrightarrow\Gamma^{\infty}_{\mathrm{c}}(\sf{E})$ and $\sf{V}_{\sf{A}}\hookrightarrow\Gamma^{\infty}_{\mathrm{sc}}(\sf{E})$ for all $\sf{A}$, we obtain continuity of the restricted solution map $\Gamma_{\mathrm{c}}^{\infty}(\sf{E})\times\Gamma_{\mathrm{c}}^{\infty}(\sf{E}\vert_{\Sigma_{0}})\to\Gamma_{\mathrm{sc}}^{\infty}(\sf{E})$. Continuity of the solution map $\Gamma^{\infty}(\sf{E})\to\Gamma^{\infty}(\sf{E})\times\Gamma^{\infty}(\sf{E}\vert_{\Sigma_{0}})$ then follows by density.
\end{proof}

\subsection{Normally Hyperbolic Operators}
In this section we shall study yet another important class of linear hyperbolic differential operators, namely those generalising the wave operator (d'Alembertian) on Minkowski spacetime. We start with the central definition of this section.

\begin{itemize}
	\item[$\bullet$]Throughout this section, let $(\sf{M},\sf{g})$ be a $(1+n)$-dimensional globally hyperbolic spacetime with Levi-Civita connection $\nabla$ and $\sf{E}\xrightarrow{\pi}\sf{M}$ be a smooth and finite-rank $\bb{K}$-bundle over $\sf{M}$. 
\end{itemize}

\begin{definition}[Normally Hyperbolic Operators]
	A linear differential operator $\sf{N}\in\mathrm{DO}^{2}(\sf{E})$ called \emph{normally hyperbolic} if its principal symbol is given by
\begin{align*}
	\sigma_{\sf{N}}(\xi)=-g^{\sharp}(\xi,\xi)\mathds{1}_{\sf{E}}\qquad\forall\xi\in\sf{T}\sf{M}\,.
\end{align*}
\end{definition} 

\begin{example}\label{Ex:ConDA} (Connection and Bochner d'Alembertians)\newline
	Let $\nabla^{\sf{E}}\:\Gamma^{\infty}(\sf{E})\to\Gamma^{\infty}(\sf{E}\otimes\sf{T}^{\ast}\sf{M})$ be a connection on $\sf{E}$ and define the differential operator
	\begin{align*}
		\square^{\nabla^{\sf{E}}}\:\Gamma^{\infty}(\sf{E})\xrightarrow{\nabla^{\sf{E}}}\Gamma^{\infty}(\sf{E}\otimes\sf{T}^{\ast}\sf{M})\xrightarrow{\nabla^{\sf{E}\otimes\sf{T}^{\ast}\sf{M}}}\Gamma^{\infty}(\sf{E}\otimes\sf{T}^{\ast}\sf{M}^{\otimes 2})\xrightarrow{\mathds{1}_{\sf{E}}\otimes\mathrm{tr}_{\sf{g}}}\Gamma^{\infty}(\sf{E})\, ,
	\end{align*}
	where $\nabla^{\sf{E}\otimes\sf{T}^{\ast}\sf{M}}$ is the connection on $\sf{E}\otimes\sf{T}^{\ast}\sf{M}$ induced by $\nabla^{\sf{E}}$ and $\nabla$. Then, $\sf{N}:=-\square^{\nabla^{\sf{E}}}$ is clearly normally hyperbolic. It is usually called the \textit{connection} or \textit{rough d'Alembertian}. If $(x^{\mu})_{\mu=0,\dots,n}$ are local coordinates of $\sf{M}$ and denote by $\Gamma^{\gamma}_{\alpha\beta}$ the Christoffel symbols of $(\sf{M},\sf{g})$, then $\sf{N}$ can locally be written in the form
\begin{align*}
	\sf{N}=-\square^{\nabla^{\sf{E}}}=-\sf{g}^{\alpha\beta}(\nabla^{\sf{E}}_{\partial_{\alpha}}\nabla^{\sf{E}}_{\partial_{\beta}}-\nabla^{\sf{E}}_{\nabla_{\partial_{\alpha}}\partial_{\beta}})=-\sf{g}^{\alpha\beta}(\nabla^{\sf{E}}_{\partial_{\alpha}}\nabla^{\sf{E}}_{\partial_{\beta}}-\Gamma_{\alpha\beta}^{\gamma}\nabla^{\sf{E}}_{\partial_{\gamma}})\, .
\end{align*}
	 Now, choose any bundle metric $\langle\cdot,\cdot\rangle_{\sf{E}}$ on $\sf{E}$ such that $\nabla^{\sf{E}}$ is metric-compatible (if it exists, see Remark~\ref{Rem:ExBM} below). Then, the formal adjoint $(\nabla^{\sf{E}})^{\ast}\:\Gamma^{\infty}(\sf{E}\otimes\sf{T}^{\ast}\sf{M})\to\Gamma^{\infty}(\sf{E})$ of $\nabla^{\sf{E}}$, where $\sf{E}\otimes\sf{T}^{\ast}\sf{M}$ is equipped with the obvious bundle metric induced by $\langle\cdot,\cdot\rangle_{\sf{E}}$ and $\sf{g}$, is given by
	\begin{align*}
		(\nabla^{\sf{E}})^{\ast}(\psi\otimes X^{\flat})=-\nabla^{\sf{E}}_{X}\psi-\mathrm{div}_{\sf{g}}(X)\psi
	\end{align*}
	for all $X\in\mathfrak{X}(\sf{M})$ and $\psi\in\Gamma^{\infty}(\sf{E})$, where $\mathrm{div}_{\sf{g}}(X):=-(\delta X^{\flat})=\nabla_{\mu}X^{\mu}=\partial_{\mu}X^{\mu}+\Gamma^{\mu}_{\mu\nu}X^{\nu}$ denotes the divergence and where we identified $\Gamma^{\infty}(\sf{E}\otimes\sf{T}^{\ast}\sf{M})\cong \Gamma^{\infty}(\sf{E})\otimes_{C^{\infty}(\sf{M})}\mathfrak{X}^{\ast}(\sf{M})$. In particular, we find $\square^{\nabla^{\sf{E}}}=-(\nabla^{\sf{E}})^{\ast}\nabla^{\sf{E}}$. The operator $(\nabla^{\sf{E}})^{\ast}\nabla^{\sf{E}}$ is usually called the \textit{Bochner d'Alembertian} and hence agrees with the connection d'Alembertian up to a sign.
	
	 An important special case is provided by the tensor bundle $\sf{T}^{p,q}\sf{M}:=\sf{T}\sf{M}^{\otimes p}\otimes\sf{T}^{\ast}\sf{M}^{q}$ for $(p,q)\in\bb{N}_{0}^{2}$ defined via the Levi-Civita connection $\nabla$. In this case, we usually simply write $\square:=\square^{\nabla}$. In a local coordinate chart $(\mathcal{U},(x^{i})_{0=1,\dots,n})$ of $\sf{M}$ it is given by
	\begin{align*}
		(\square\sf{T})^{\alpha_{1}\dots\alpha_{p}}_{\beta_{1}\dots\beta_{q}}=g^{\mu\nu}(\nabla_{\partial_{\mu}}\nabla_{\partial_{\nu}}-\nabla_{\nabla_{\partial_{\mu}}\partial_{\nu}}\sf{T})^{\alpha_{1}\dots\alpha_{p}}_{\beta_{1}\dots\beta_{q}}=g^{\mu\nu}\nabla_{\mu}\nabla_{\nu}\sf{T}^{\alpha_{1}\dots\alpha_{p}}_{\beta_{1}\dots\beta_{q}}\, ,
	\end{align*}
	where the last expression is written in \textit{abstract index notation}\footnote{In this notation, indices do \emph{not} correspond to coordinates, but abstractly represent the \emph{type} of the tensor field. Thus, $\nabla_{\partial_{\mu}}\sf{T}\in\Gamma^{\infty}(\sf{T}^{p,q}\sf{M})$ for $\sf{T}\in\Gamma^{\infty}(\sf{T}^{p,q}\sf{M})$ is the covariant derivative along $\partial_{\mu}$ with fixed $\mu$, while $(\nabla_{\mu}\sf{T})^{\alpha_{1}\dots\alpha_{p}}_{\beta_{1}\dots\beta_{q}}$ are the components of $\nabla\sf{T}\in\Gamma^{\infty}(\sf{T}^{p,q+1}\sf{M})$. When dealing with \emph{higher-order covariant derivatives}, one has to carefully distinguish expressions such as $\nabla_{\partial_\mu}\nabla_{\partial\nu}$, in which $\mu,\nu$ are fixed and which maps $\Gamma^{\infty}(\sf{T}^{p,q}\sf{M})$ onto itself, and $\nabla_{\mu}\nabla_{\nu}$, where $\mu,\nu$ are free, which are the components of $\nabla\nabla\:\Gamma^{\infty}(\sf{T}^{p,q}\sf{M})\to\Gamma^{\infty}(\sf{T}^{p,q+2}\sf{M})$. They are related via $(\nabla_{\partial_{\mu}}\nabla_{\partial_{\nu}}-\nabla_{\nabla_{\partial_{\mu}}\partial_{\nu}}\sf{T})^{\alpha_{1}\dots\alpha_{p}}_{\beta_{1}\dots\beta_{q}}=\nabla_{\mu}\nabla_{\nu}\sf{T}^{\alpha_{1}\dots\alpha_{p}}_{\beta_{1}\dots\beta_{q}}$, as one can easily check.} (cf.~\cite[Sec.~2.4]{Wald}, and \cite[Chap.~2]{PenroseAI}). In the case $p=q=0$, this operator is also known as the \textit{Beltrami d'Alembertian}.
\end{example}	

The previous example is actually the generic case, as the following well-known lemma shows (see e.g.~\cite[Prop.~3.1]{BaumKath} or \cite[Lem.~1.5.5]{BaerBook}):

\begin{lemma}\label{Lem.NH}
	For every normally hyperbolic operator $\sf{N}\in\mathrm{DO}^{2}(\sf{E})$, there exists a unique connection $\nabla^{\sf{E}}\:\Gamma^{\infty}(\sf{E})\to\Gamma^{\infty}(\sf{E}\otimes\sf{T}^{\ast}\sf{M})$ and zero-order operator $\sf{N}_{0}\in\mathrm{DO}^{0}(\sf{E})$ such that
	\begin{align*}
		\sf{N}=-\square^{\nabla^{\sf{E}}}+\sf{N}_{0}\, .
	\end{align*}
	The connection $\nabla^{\sf{E}}$ is called the $\sf{E}$-\emph{compatible connection} of $\sf{D}$ and characterised by
	\begin{align*}
		\nabla^{\sf{E}}_{\d f}\psi=\frac{1}{2}\bigg(f\sf{N}\psi-\sf{N}(f\psi)-(\square f)\psi\bigg)\qquad\forall f\in C^{\infty}(\sf{M}),\,\psi\in\Gamma^{\infty}(\sf{E})\, .
	\end{align*}
\end{lemma}

\begin{proof}
	Assume that there exists a connection $\nabla^{\sf{E}}$ and a zero-order operator $\sf{N}_{0}$ such that $\sf{N}=-\square^{\nabla^{\sf{E}}}+\sf{N}_{0}$. For every $f\in C^{\infty}(\sf{M})$ and $\psi\in\Gamma^{\infty}(\sf{E})$, it holds that $\sf{N}_{0}(f\psi)=f\sf{N}_{0}\psi$ and hence
	\begin{align*}
		f(\sf{N}\psi+\square^{\nabla^{\sf{E}}}\psi)=\sf{N}(f\psi)+\square^{\nabla^{\sf{E}}}(f\psi)=\sf{N}(f\psi)+f(\square^{\nabla^{\sf{E}}}\psi)+2\nabla^{\sf{E}}_{\d f}\psi+(\square f)\psi\, ,
	\end{align*}
	where we used the Leibniz rule. In particular, we obtain the claimed relation
	\begin{align*}
		\nabla^{\sf{E}}_{\d f}\psi=\frac{1}{2}\bigg(f\sf{N}\psi-\sf{N}(f\psi)-(\square f)\psi\bigg)\, .
	\end{align*}
	Now, at a given point $p\in\sf{M}$, every $X\in\sf{T}\sf{M}$ can be written as $X_{p}=(\d f)_{p}$ for some suitable function $f$. Hence, the connection $\nabla^{\sf{E}}$, if it exists, is uniquely specified by the formula above. For existence, we use the above formula to explicitly define a connection locally.
\end{proof}

\begin{remark}\label{Rem:ExBM}
	For a normally hyperbolic operator $\mathsf{M}$ on $\mathsf{E}$ with  $\mathsf{E}$-compatible connection $\nabla^{\mathsf{E}}$, one may ask whether there exists a bundle metric 
$\langle \cdot, \cdot \rangle_{\mathsf{E}}$ 
with respect to which $\nabla^{\mathsf{E}}$ is metric-compatible. In general, this need not be possible and the existence of such a metric is determined by the holonomy group of the connection 
(see e.g.~\cite{Schmidt}). For normally hyperbolic operators, this question is in fact equivalent to formal self-adjointness: we can find a bundle metric with respect to which $\mathsf{N}$ is formally self-adjoint if and only if $\nabla^{\mathsf{E}}$ preserves this fibre metric.
\end{remark}

Besides the connection d'Alembertian, we mention two other important classes of examples of normally hyperbolic operators. By the previous lemma, each of these can be expressed as a connection d'Alembertian plus a zeroth-order operator.

\begin{example}\label{Example:HodgeLapl} (de Rham-Hodge d'Alembertian)\newline
	Recall the notation and conventions used for differential forms, as explained in the notation section of this thesis (see page~\pageref{DiffForms}ff.). We consider the exterior bundle $\sf{A}_{k}:=\bigwedge^{k}\sf{T}^{\ast}\sf{M}$, whose sections are smooth differential forms $\Omega^{k}(\sf{M})=\Gamma^{\infty}(\sf{A}_{k})$, equipped with the bundle metric 
	\begin{align*}
		\langle\alpha,\beta\rangle_{\sf{A}_{k}}:=\frac{1}{k!}(\sf{g}^{\sharp})^{\otimes k}(\alpha,\beta)=\frac{1}{k!}\alpha^{\mu_{1}\dots\mu_{k}}\beta_{\mu_{1}\dots\mu_{k}}\, .
	\end{align*}
	Furthermore, we denote the exterior derivative by $\d\:\Omega^{k}(\sf{M})\to\Omega^{k+1}(\sf{M})$ and its formal adjoint, the codifferential, by $\delta\:\Omega^{k+1}(\sf{M})\to\Omega^{k}(\sf{M})$. Then, the \textit{de Rham-Hodge d'Alembertian}
	\begin{align*}
		\square_{\mathrm{dRH}}:=\d\delta+\delta\d\:\Omega^{k}(\sf{M})\to\Omega^{k}(\sf{M})
	\end{align*}
	is normally hyperbolic. Indeed, a computation in coordinates shows that $\square_{\mathrm{dRH}}$ and the connection d'Alembertian $\square:=\sf{g}^{\alpha\beta}\nabla_{\alpha}\nabla_{\beta}$ are related by means of the \textit{Weitzenböck identity}
	\begin{align*}
		\square_{\mathrm{dRH}}=-\square+\mathcal{R}_{k}\qquad\text{on}\quad\Omega^{k}(\sf{M})\,,
	\end{align*}
	where $\mathcal{R}_{k}$ a $0$th-order operator constructed out of the curvature of $(\sf{M},\sf{g})$. For $f\in\Omega^{0}(\sf{M})$, we have $\square_{\mathrm{dRH}}f=-\square f$, while for $\omega\in\Omega^{1}(\sf{M})$, the formula reads $\square_{\mathrm{dRH}}\omega-\square \omega+\mathrm{Ric}_{\sf{g}}(\omega)$ with $\mathrm{Ric}_{\sf{g}}(\omega)_{\alpha}:=\mathrm{Ric}(\sf{g})_{\alpha}^{\beta}\omega_{\beta}$. For an explicit expression of $\mathcal{R}_{k}$ in coordinates, we refer to \cite[\S 26]{deRhamBook}.

	More generally, consider the bundle $\sf{E}_{k}:=\sf{E}\otimes\bigwedge^{k}\sf{T}^{\ast}\sf{M}$ whose space of sections are $\sf{E}$-valued differential forms $\Omega^{k}(\sf{M},\sf{E}):=\Gamma^{\infty}(\sf{E}_{k})$ and equip $\sf{E}$ with a bundle metric $\langle\cdot,\cdot\rangle_{\sf{E}}$ and a metric-compatible connection $\nabla^{\sf{E}}\:\Gamma^{\infty}(\sf{E})\to\Gamma^{\infty}(\sf{E}\otimes\sf{T}^{\ast}\sf{M})$. We denote the \textit{exterior covariant derivative} induced by $\nabla^{\sf{E}}$ by $\d^{\nabla^{\sf{E}}}\:\Omega^{k}(\sf{M},\sf{E})\to\Omega^{k+1}(\sf{M},\sf{E})$. Now, choosing a local frame $\{e_{a}\}_{a}\subset\Gamma^{\infty}(\mathcal{U},\sf{E})$ on some open set $\mathcal{U}\subset\sf{M}$, any $k$-form $\omega\in\Omega^{k}(\sf{M},\sf{E})$ can locally be written as $\omega=\omega^{a}\otimes e_{a}$ for coefficient forms $\omega^{a}\in\Omega^{k}(\mathcal{U},\bb{K})$. We equip the bundles $\sf{E}_{k}$ with the natural bundle metric
	\begin{align*}
		\langle\alpha,\beta\rangle_{\sf{E}_{k}}:=\langle\alpha^a,\beta^b\rangle_{\sf{A}_{k}}\langle e_{a},e_{b}\rangle_{\sf{E}}
	\end{align*}		
	and denote the formal adjoint of $\d^{\nabla^{\sf{E}}}$ with respect to $(\cdot,\cdot)_{\sf{E}_{k}}$, the \textit{covariant codifferential}, by $\delta^{\nabla^{\sf{E}}}$. With this notation, we define the $\sf{E}$-valued \textit{de Rham-Hodge connection d'Alembertian} by
	\begin{align*}
		\square^{\nabla^{\sf{E}}}_{\mathrm{dRH}}:=\d^{\nabla^{\sf{E}}}\delta^{\nabla^{\sf{E}}}+\delta^{\nabla^{\sf{E}}}\d^{\nabla^{\sf{E}}}\qquad\text{on}\quad\Omega^{k}(\sf{M},\sf{E})\, .
	\end{align*}
	As before, a straightforward computation in coordinates shows that $\square^{\nabla^{\sf{E}}}_{\mathrm{dRH}}$ satisfies the \textit{(generalised) Weitzenböck identity} 
	\begin{align*}
		\square^{\nabla^{\sf{E}}}_{\mathrm{dRH}}=-\square^{\nabla^{\sf{E}}}+\mathcal{R}_{k}^{\nabla^{\sf{E}}}\, ,
	\end{align*}
	where $\mathcal{R}_{k}^{\nabla^{\sf{E}}}$ is a zeroth-order operator constructed out of both the pseudo-Riemannian curvature of $(\sf{M},\sf{g})$ and bundle curvature of $(\sf{E},\nabla^{\sf{E}})$. We refer to \cite[Cor.~2.7.21]{RudolphSchmidt2} for further details.
\end{example}

\begin{example} (Spin d'Alembertian)\newline
	Let $(\sf{E},\gamma,\langle\cdot,\cdot\rangle_{\sf{E}},\nabla^{\sf{E}})$ be a Dirac bundle as defined in Remark.~\ref{Rem:Dirac}. As a special case, one can think of the spin Dirac operator with $\sf{E}=\sf{S}\sf{M}$ as defined in Example~\ref{Ex:Dirac}. Then, the operator $\sf{N}:=\sf{D}^{2}$ is normally hyperbolic. Indeed, $\sigma_{\sf{D}}(\xi)=\gamma(\xi^{\sharp})$ and hence
	\begin{align*}
		\sigma_{\sf{N}}(\xi)=\gamma(\xi^{\sharp})^{2}=-\sf{g}(\xi^{\sharp},\xi^{\sharp})\mathrm{id}_{\sf{E}}=-\sf{g}^{\sharp}(\xi,\xi)\mathrm{id}_{\sf{E}}\, ,
	\end{align*}
	where we used the fact that $\gamma$ is a Clifford multiplication. As a special case, we obtain the de Rham-Hodge d'Alembertian from the de Rham-Hodge Dirac operator $\sf{D}=\d+\delta$. 
\end{example}

We now discuss the Cauchy problem for normally hyperbolic operators.

\begin{theorem}\label{Thm:NHCauchy} \emph{(The Cauchy Problem for Normally Hyperbolic Operators)}\newline
	Let $(\sf{M},\sf{g})$ be a globally hyperbolic manifold and fix a Cauchy temporal function $t\:\sf{M}\to\bb{R}$ with corresponding foliation $(\Sigma_{t})_{t\in\bb{R}}$. Furthermore, let $\sf{N}\in\mathrm{DO}^{2}(\sf{E})$ be a normally hyperbolic operator with associated connection $\nabla^{\sf{E}}$. Then, the following holds true:
\begin{itemize}
	\item[\emph{(i)}]For every Cauchy data $(\phi,\mathfrak{f}_{1},\mathfrak{f}_{2})\in \Gamma^{\infty}(\sf{M},\sf{E})\times \Gamma^{\infty}(\Sigma_{0},\sf{E}\vert_{\Sigma_{0}})\times \Gamma^{\infty}(\Sigma_{0},\sf{E}\vert_{\Sigma_{0}})$, there exists a unique solution $\psi\in \Gamma^{\infty}(\sf{M},\sf{E})$ to the Cauchy problem
	\begin{align*}
		\begin{cases}
			\sf{N}\psi &=\phi\\
			\psi\vert_{\Sigma_{0}}&=\mathfrak{f}_{1}\\
			(\nabla_{\nu}^{\sf{E}}\psi)\vert_{\Sigma_{0}}&=\mathfrak{f}_{2}
		\end{cases}\, ,
	\end{align*}	
	where $\nu$ denotes the future-directed timelike normal of $\Sigma_{0}$ in $\sf{M}$.
\item[\emph{(ii)}]The unique solution $\psi$ in (i) satisfies
	\begin{align*}
		\mathrm{supp}(\psi)\cap\mathcal{J}^{\pm}(\Sigma_{0})\subset \mathcal{J}^{\pm}\big((\mathrm{supp}(\phi)\cap\mathcal{J}^{\pm}(\Sigma_{0}))\cup \mathrm{supp}(\mathfrak{f}_{1})\cup \mathrm{supp}(\mathfrak{f}_{2})\big)
	\end{align*}
	and hence in particular $\mathrm{supp}(\psi)\subset \mathcal{J}(\mathrm{supp}(\phi)\cup \mathrm{supp}(\mathfrak{f}_{1})\cup \mathrm{supp}(\mathfrak{f}_{2}))$, i.e.~$\psi$ propagates at most with the speed of light.
	\item[\emph{(iii)}]The Cauchy problem is stable, i.e.~the solution maps
	\begin{align*}
		&\Gamma^{\infty}(\sf{M},\sf{E})\times \Gamma^{\infty}(\Sigma_{0},\sf{E}\vert_{\Sigma_{0}})\times \Gamma^{\infty}(\Sigma_{0},\sf{E}\vert_{\Sigma_{0}})\in (\phi,\mathfrak{f}_{1},\mathfrak{f}_{2})\mapsto\psi \in \Gamma^{\infty}(\sf{M},\sf{E})\,,\\
		&\Gamma^{\infty}_{\mathrm{c}}(\sf{M},\sf{E})\times \Gamma^{\infty}_{\mathrm{c}}(\Sigma_{0},\sf{E}\vert_{\Sigma_{0}})\times \Gamma^{\infty}_{\mathrm{c}}(\Sigma_{0},\sf{E}\vert_{\Sigma_{0}})\in (\phi,\mathfrak{f}_{1},\mathfrak{f}_{2})\mapsto\psi \in \Gamma_{\mathrm{sc}}^{\infty}(\sf{M},\sf{E})
	\end{align*}
	are continuous in the corresponding Fréchet- and LF-topologies, respectively.
	\end{itemize}
\end{theorem}

Well-posedness of the Cauchy problem for normally hyperbolic operators can be shown in various different ways. A modern proof based on an explicit construction of fundamental solutions via \emph{Riesz distributions} can be found in \cite{BaerBook} and \cite{GinouxCauchy}. Earlier proofs in the global setting can be found in Leray's unpublished lecture notes \cite{Leray} and in \cite{ChoquetBruhatCauchy}. Here we follow a different strategy, which has been discussed, for example, in \cite[Rem.~3.7.11.]{BaerLecture} and \cite[Sec.~6.3.3]{GinouxMurro}, namely by rewriting a normally hyperbolic operator as an equivalent symmetric hyperbolic system and proving well-posedness as a corollary of Theorem~\ref{Thm:Cauchy}. 

\begin{proof}[Proof of Theorem~\ref{Thm:NHCauchy}] As explained above, the idea is to find a symmetric hyperbolic system with equivalent Cauchy problem. We follow the procedure outlined in~\cite[Rem.~3.7.11.]{BaerLecture} and \cite[Sec.~6.3.3]{GinouxMurro}, which generalises the analogues well-known procedure from $\bb{R}^{n}$ to the global setting. However, the approach here is slightly modified and made more precise by filling in some details and by defining suitable bundles and connections.

First of all, let $\nabla^{\sf{E}}\:\Gamma^{\infty}(\sf{E})\to\Gamma^{\infty}(\sf{E}\otimes\sf{T}^{\ast}\sf{M})$ be the $\sf{E}$-compatible connection of $\sf{N}$ and write $\sf{N}=-\square^{\nabla^{\sf{E}}}+\sf{N}_{0}$ for some $\sf{N}_{0}\in\mathrm{DO}^{0}(\sf{E})$, as in Lemma~\ref{Lem.NH}. Furthermore, we choose an arbitrary \emph{positive-definite} bundle metric $\langle\cdot,\cdot\rangle_{\sf{E}}$ on $\sf{E}$, for which $\nabla^{\sf{E}}$ is not necessarily compatible, see Remark~\ref{Rem:ExBM}, and identify
	\begin{align*}
		\sf{M}=\bb{R}\times\Sigma\,,\qquad\sf{g}=-\beta^{2}\d t\otimes\d t+\sf{h}_{t}\, ,
	\end{align*}
	as usual. Now, consider the natural projection $\pi_{2}\:\sf{M}\to\Sigma$, which allows us to decompose the tangent bundle as $\sf{T}^{\ast}\sf{M}=\underline{\bb{R}}_{\sf{M}}\oplus\pi_{2}^{\ast}(\sf{T}\Sigma)$, where $\underline{\bb{R}}_{\sf{M}}$ denotes the trivial $\bb{R}$-line bundle over $\sf{M}$, and consider the corresponding projection $\sf{P}_{2}\:\sf{T}^{\ast}\sf{M}\to\pi_{2}^{\ast}(\sf{T}^{\ast}\Sigma)$. We note that sections of $\pi_{2}^{\ast}(\sf{T}^{\ast}\Sigma)$ can be understood as \emph{time-dependent $1$-forms on $\Sigma$}, i.e.~if we choose local coordinates $(x^{i})_{i=1,\dots,k}$ on an open subset $\mathcal{U}\subset\Sigma$ and supplement them by $x^{0}:=t$ to obtain local coordinates $(x^{\mu})_{\mu=0,\dots,k}$ of $\sf{M}$ on $\bb{R}\times\mathcal{U}$, then any section of $X\in\Gamma^{\infty}(\pi_{2}^{\ast}(\sf{T}^{\ast}\Sigma))$ can locally be written as $X(t,\vec{x})=X^{i}(t,\vec{x})\partial_{i}$ for $(t,\vec{x})\in\bb{R}\times\Sigma$. Now, we introduce the linear operator
	\begin{align*}
		\nabla^{\sf{E},\Sigma}\:\Gamma^{\infty}(\sf{E})\xrightarrow{\nabla^{\sf{E}}}\Gamma^{\infty}(\sf{E}\otimes\sf{T}^{\ast}\sf{M})\xrightarrow{\mathrm{id}_{\sf{E}}\otimes\sf{P}_{2}}\Gamma^{\infty}(\sf{E}\otimes\pi^{\ast}_{2}(\sf{T}^{\ast}\Sigma))\,.
	\end{align*}		
	At this stage, it is important to stress that this does \emph{not} define a connection on $\sf{E}$, as one can already see from the mapping properties. Indeed, $\nabla^{\sf{E},\Sigma}$ satisfies the Leibniz rule only at fixed time. In local coordinates as above, it holds that 
	\begin{align*}
		\nabla^{\sf{E}}\psi=\nabla_{\partial_{\mu}}^{\sf{E}}\psi\otimes\d x^{\mu}=\nabla^{\sf{E}}_{\partial_{t}}\psi\otimes\d t+\nabla^{\sf{E},\Sigma}\psi\,,\qquad\qquad \nabla^{\sf{E},\Sigma}\psi=\nabla_{\partial_{i}}^{\sf{E}}\psi\otimes\d x^{i}\, .
	\end{align*} 
	In other words, $\nabla^{\sf{E},\Sigma}$ describes the \emph{spatial} covariant derivatives. Now, the idea of the following is to define a symmetric hyperbolic system $\sf{S}\in\mathrm{DO}^{1}(\mathcal{E})$ on the bundle $\mathcal{E}:=\sf{E}\oplus\sf{E}\oplus (\sf{E}\otimes\pi_{2}^{\ast}(\sf{T}^{\ast}\Sigma))$, equipped with the bundle metric $\langle\cdot,\cdot\rangle_{\mathcal{E}}$ induced from $\langle\cdot,\cdot\rangle_{\sf{E}}$ and $\sf{h}(\cdot,\cdot)$, such that
	\begin{align}\label{eq:AnsatzNHSHS}
		\sf{S}
		\begin{pmatrix}
			\psi\\\vspace{2pt}
			\nabla_{\partial_{t}}^{\sf{E}}\psi\\\vspace{2pt}
			\nabla^{\sf{E},\Sigma}\psi
		\end{pmatrix}=
		\begin{pmatrix}
			0\\\vspace{2pt}\phi\\\vspace{2pt} 0
		\end{pmatrix}\qquad\Leftrightarrow\qquad \sf{N}\psi=\phi
	\end{align}
	for all $\psi,\phi\in\Gamma^{\infty}(\sf{E})$. Now, to start with, let us equip the bundle $\mathcal{E}$ with the connection $\nabla^{\mathcal{E}}\:\Gamma^{\infty}(\mathcal{E})\to\Gamma^{\infty}(\mathcal{E}\otimes\sf{T}^{\ast}\sf{M})$ induced by $\nabla^{\sf{E}}$ and the connection $\nabla^{\Sigma}$ on $\pi_{2}^{\ast}(\sf{T}^{\ast}\Sigma)$ defined by\footnote{To get a better feeling of how this connections acts, let us denote by $\Gamma^{\gamma}_{\alpha\beta}$ the Christoffel symbols of $(\sf{M},\sf{g})$ so that $(\nabla_{\partial_{\alpha}}\omega)_{\beta}=\partial_{\alpha}\omega_{\beta}-\Gamma^{\gamma}_{\alpha\beta}\omega_{\gamma}$ for $\omega\in\Gamma^{\infty}(\sf{T}^{\ast}\sf{M})$. Then, for $\eta\in\Gamma^{\infty}(\pi_{2}^{\ast}(\sf{T}^{\ast}\Sigma))$, we obtain $(\nabla^{\Sigma}_{\partial_{\alpha}}\eta)_{i}=\partial_{\alpha}\omega_{i}-\Gamma^{k}_{\alpha i}\omega_{k}$, where the sum is just over spatial variables $k=1,\dots,n$.}
	\begin{align*}
		\nabla^{\Sigma}:=(\sf{P}_{2}\otimes\mathrm{id}_{\sf{T}^{\ast}\sf{M}})\circ\nabla\vert_{\Gamma^{\infty}(\pi^{\ast}(\sf{T}^{\ast}\Sigma))}\:\Gamma^{\infty}(\pi^{\ast}(\sf{T}^{\ast}\Sigma))\to\Gamma^{\infty}(\pi^{\ast}(\sf{T}^{\ast}\Sigma)\otimes\sf{T}^{\ast}\sf{M})\, .
	\end{align*}
	Moreover, let $\nu_{t}:=\beta^{-1}\partial_{t}\vert_{\Sigma_{t}}$ denote the future-directed timelike normal vector field of $\Sigma_{t}$ in $\sf{M}$ and consider the second fundamental form $\sf{k}_{t}\in\Gamma^{\infty}(\sf{T}^{\ast}\Sigma_{t}^{\otimes_{s}2})$ defined by $\sf{k}_{t}(X,Y):=\sf{g}(\nu_{t},\nabla_{X}Y)$ for all $X,Y\in\mathfrak{X}(\Sigma_{t})$. In local coordinates as above, it is straightforward to verify that $(\sf{k}_{t})_{ij}=-\nabla_{j}(\nu_{t})_{i}=-\frac{1}{2\beta}\partial_{t}(\sf{h}_{t})_{ij}$. Collecting these tensors for all $t\in\bb{R}$, we obtain the time-dependent symmetric $(0,2)$-tensor field 
	\begin{align*}
		\sf{k}:=-\frac{1}{2\beta}\partial_{t}\sf{h}\in\Gamma^{\infty}(\pi^{\ast}(\sf{T}^{\ast}\Sigma^{\otimes_{s}2}))\, .
	\end{align*}
	Having fixed the relevant notation, let us decompose $\sf{N}$ into its temporal and spatial parts. A straightforward computation in local coordinates shows that
	\begin{align}\label{eq:dfdsafadsa}
		\sf{N}-\sf{N}_{0}&=-\square^{\nabla^{\sf{E}}}=-\sf{g}^{\alpha\beta}(\nabla^{\sf{E}}_{\partial_{\alpha}}\nabla^{\sf{E}}_{\partial_{\beta}}-\nabla^{\sf{E}}_{\nabla_{\partial_{\alpha}}\partial_{\beta}})=\nonumber\\&=\frac{1}{\beta^{2}}\nabla^{\sf{E}}_{\partial_{t}}\nabla^{\sf{E}}_{\partial_{t}}-\sf{h}^{ij}(\nabla^{\sf{E}}_{\partial_{i}}\nabla^{\sf{E}}_{\partial_{j}}-\Gamma^{k}_{ij}\nabla^{\sf{E}}_{\partial_{k}})+\bigg(\sf{h}^{ij}\Gamma^{0}_{ij}-\frac{1}{\beta^{2}}\Gamma^{0}_{00}\bigg)\nabla^{\sf{E}}_{\partial_{t}}-\frac{1}{\beta^{2}}\Gamma^{k}_{00}\nabla^{\sf{E}}_{\partial_{k}}=\nonumber\\&=\frac{1}{\beta^{2}}\nabla_{\partial_{t}}^{\sf{E}}\nabla_{\partial_{t}}^{\sf{E}}-\Delta^{\sf{E},\Sigma}+f\nabla^{\sf{E}}_{\partial_{t}}+\nabla^{\sf{E}}_{\sf{X}}
	\end{align}
	where $f\in C^{\infty}(\sf{M})$ and the time-dependent vector field $\sf{X}\in\Gamma^{\infty}(\pi_{2}^{\ast}(\sf{T}\Sigma))$ are given by
	\begin{align*}
		&f:=-\frac{1}{\beta}\bigg(\mathrm{tr}_{\sf{h}}(\sf{k})+\frac{\partial_{t}\beta}{\beta^{2}}\bigg)\,,\qquad \sf{X}:=-\frac{1}{\beta}(\d_{\Sigma}\beta)^{\sharp}=-\frac{1}{\beta}(\sf{h}^{ij}\partial_{i}\beta)\partial_{j}\,,
	\end{align*}
	with $\d_{\Sigma}$ being the (time-dependent) exterior derivative on $(\Sigma,\sf{h}_{t})$, and where the Laplace-type operator $\Delta^{\sf{E},\Sigma}\:\Gamma^{\infty}(\sf{E})\to\Gamma^{\infty}(\sf{E})$ is defined as the composition\footnote{In local coordinates, $\Delta^{\sf{E},\Sigma}\psi=\sf{h}^{ij}(\nabla_{\partial_{i}}^{\sf{E}}\nabla_{\partial_{j}}^{\sf{E}}-\Gamma_{ij}^{k}\nabla^{\sf{E}}_{\partial_{k}})\psi$, where $\Gamma^{\gamma}_{\alpha\beta}$ are the Christoffel symbols of $(\sf{M},\sf{g})$ and where Latin indices $i,j,k$ denote spatial indices, as usual.}
	\begin{align*}
		\Delta^{\sf{E},\Sigma}\:\Gamma^{\infty}(\sf{E})&\xrightarrow{\nabla^{\sf{E},\Sigma}}\Gamma^{\infty}(\sf{E}\otimes\pi^{\ast}_{2}(\sf{T}^{\ast}\Sigma))\xrightarrow{\nabla^{\sf{E}\otimes\Sigma}}\Gamma^{\infty}(\sf{E}\otimes\pi_{2}^{\ast}(\sf{T}^{\ast}\Sigma)\otimes\sf{T}^{\ast}\sf{M})\\&\xrightarrow{\mathrm{id}\otimes\mathrm{id}\otimes\sf{P}_{2}}\Gamma^{\infty}(\sf{E}\otimes\pi_{2}^{\ast}(\sf{T}^{\ast}\Sigma)\otimes\pi_{2}^{\ast}(\sf{T}^{\ast}\Sigma))\xrightarrow{\mathrm{id}\otimes\mathrm{tr}_{\sf{h}}}\Gamma^{\infty}(\sf{E})\,,
	\end{align*}
	with $\nabla^{\sf{E}\otimes\Sigma}$ being the natural connection on $\sf{E}\otimes\pi_{2}^{\ast}(\sf{T}^{\ast}\Sigma)$ induced by $\nabla^{\sf{E}}$ and $\nabla^{\Sigma}$. With this notation, we define $\sigma_{\sf{S}}\:\sf{T}^{\ast}\sf{M}\to\mathrm{End}(\mathcal{E})$ for all $\xi=\xi_{0}\d t+\xi_{\Sigma}\in\sf{T}^{\ast}\sf{M}$ with $\xi_{\Sigma}\in\pi^{\ast}_{2}(\sf{T}^{\ast}\Sigma)$ by 
	\begin{align*}
	\sigma_{\sf{S}}(\xi)=
	\begin{pmatrix}
		\xi_{0} & 0 & 0 \\ 0 & \beta^{-2}\xi_{0} & -\xi_{\Sigma}^{\sharp}\lrcorner\bullet \\ 0 & -\bullet\otimes\xi_{\Sigma} & \xi_{0}
	\end{pmatrix}\, ,
\end{align*}
where $\lrcorner\:\Gamma^{\infty}(\pi^{\ast}_{2}(\sf{T}\Sigma))\times\Gamma^{\infty}(\sf{E}\otimes\pi^{\ast}_{2}(\sf{T}^{\ast}\Sigma))\to\Gamma^{\infty}(\sf{E})$ is defined as the usual interior product on the level of (time-dependent) forms on $\Sigma$, i.e.~$(\sf{X}\lrcorner (\omega\otimes\xi)):=\xi(\sf{X})\omega=(\xi_{i}\sf{X}^{i})\omega$ for $\sf{X}\in\Gamma^{\infty}(\pi^{\ast}_{2}(\sf{T}\Sigma))$, $\xi\in\Gamma^{\infty}(\pi^{\ast}_{2}(\sf{T}^{\ast}\Sigma))$ and $\omega\in\Gamma^{\infty}(\sf{E})$. 

It is easy to see that $\sigma_{\sf{S}}(\xi)$ is Hermitian with respect to $\langle\cdot,\cdot\rangle_{\mathcal{E}}$. Now, consider the first-order operator $\sigma_{\sf{S}}(\d x^{\mu})\nabla^{\mathcal{E}}_{\partial_{\mu}}$. By definition, $\sigma_{\sf{S}}$ is defined in such a way that it is Hermitian and such that the operator $\sigma_{\sf{S}}(\d x^{\mu})\nabla^{\mathcal{E}}_{\partial_{\mu}}$ acting on $(\psi,\nabla^{\sf{E}}_{\partial_{t}},\nabla^{\sf{E},\Sigma}\psi)^{\mathrm{T}}$ for $\psi\in\Gamma^{\infty}(\sf{E})$ gives, $(0,\sf{N}\psi,0)$ up to zeroth-order terms. Indeed, a straightforward computation using Eq.~\eqref{eq:dfdsafadsa} shows
\begin{align*}
	\sigma_{\sf{S}}(\d x^{\mu})\nabla^{\mathcal{E}}_{\partial_{\mu}}\begin{pmatrix}
			\psi\\\vspace{2pt}
			\nabla_{\partial_{t}}^{\sf{E}}\psi\\\vspace{2pt}
			\nabla^{\sf{E},\Sigma}\psi
		\end{pmatrix}-\begin{pmatrix}
			0\\\vspace{2pt}\sf{N}\psi\\\vspace{2pt} 0
		\end{pmatrix}&=
		\begin{pmatrix}
			\nabla^{\sf{E}}_{\partial_{t}}\psi\\\vspace{2pt}
			\beta^{-2}\nabla_{\partial_{t}}^{\sf{E}}\nabla_{\partial_{t}}^{\sf{E}}\psi-\Delta^{\sf{E},\Sigma}\psi\\\vspace{2pt} \nabla^{\sf{E}\otimes\Sigma}_{\partial_{t}}\nabla^{\sf{E},\Sigma}\psi-\nabla^{\sf{E},\Sigma}\nabla_{\partial_{t}}^{\sf{E}}\psi
		\end{pmatrix}-\begin{pmatrix}
			0\\\vspace{2pt}\sf{N}\psi\\\vspace{2pt} 0
		\end{pmatrix}=\\&=
		-\begin{pmatrix}
			0 & -\mathrm{id} & 0 \\\vspace*{2pt}
			\sf{N}_{0} & f\cdot\mathrm{id} & \sf{X}\lrcorner\bullet\\\vspace*{2pt}
			\sf{R}_{2} & 0 & \sf{R}_{1}
		\end{pmatrix}\begin{pmatrix}
			\psi\\\vspace{2pt}
			\nabla_{\partial_{t}}^{\sf{E}}\psi\\\vspace{2pt}
			\nabla^{\sf{E},\Sigma}\psi
		\end{pmatrix}=:-\sf{S}_{0}\begin{pmatrix}
			\psi\\\vspace{2pt}
			\nabla_{\partial_{t}}^{\sf{E}}\psi\\\vspace{2pt}
			\nabla^{\sf{E},\Sigma}\psi
		\end{pmatrix}\,,
\end{align*}
where $\sf{R}_{1}\:\Gamma^{\infty}(\sf{E}\otimes\pi_{2}^{\ast}(\sf{T}^{\ast}\Sigma))\to\Gamma^{\infty}(\sf{E}\otimes\pi_{2}^{\ast}(\sf{T}^{\ast}\Sigma))$ and $\sf{R}_{2}\:\Gamma^{\infty}(\sf{E})\to\Gamma^{\infty}(\sf{E}\otimes\pi_{2}^{\ast}(\sf{T}^{\ast}\Sigma))$ are differential operators of order zero uniquely determined by the commutator 
\begin{align}\label{eq:DefComBl}
	(\nabla^{\sf{E}\otimes\Sigma}_{\partial_{t}}\nabla^{\sf{E},\Sigma}-\nabla^{\sf{E},\Sigma}\nabla_{\partial_{t}}^{\sf{E}})\psi=-\sf{R}_{1}\nabla^{\sf{E},\Sigma}\psi-\sf{R}_{2}\psi\, .
\end{align}
A straightforward computation in coordinates shows that $\sf{R}_{1}$ and $\sf{R}_{2}$ are explicitly given by
\begin{align*}
		&\sf{R}_{1}\:\Gamma^{\infty}(\sf{E}\otimes\pi_{2}^{\ast}(\sf{T}^{\ast}\Sigma))\to\Gamma^{\infty}(\sf{E}\otimes\pi_{2}^{\ast}(\sf{T}^{\ast}\Sigma))\,,\quad  &&\omega_{i}\otimes\d x^{i}\mapsto -\beta\sf{k}^{j}_{i}\omega_{j}\otimes\d x^{i}\,,\\
		&\sf{R}_{2}\:\Gamma^{\infty}(\sf{E})\to\Gamma^{\infty}(\sf{E}\otimes\pi_{2}^{\ast}(\sf{T}^{\ast}\Sigma))\,,\quad &&\psi \mapsto -\sf{F}^{\nabla^{\sf{E}}}_{0i}(\psi)\otimes\d x^{\mu}\,,
	\end{align*}
	where $\sf{F}^{\nabla^{\sf{E}}}\in\Omega^{2}(\sf{M},\mathrm{End}(\sf{E}))$ denotes the curvature locally defined by $\sf{F}^{\nabla^{\sf{E}}}_{\alpha\beta}(\psi):=[\nabla_{\partial_{\alpha}}^{\sf{E}},\nabla_{\partial_{\beta}}^{\sf{E}}]\psi$. By definition, $\sf{S}_{0}$ is an operator of order zero and we consider the first order operator defined by $\sf{S}:=\sigma_{\sf{S}}(\d x^{\mu})\nabla^{\sf{E}}_{\partial_{\mu}}+\sf{S}_{0}$.  By construction, $\sf{S}$ has the required property as described in Eq.~\eqref{eq:AnsatzNHSHS}. It remains to show hyperbolicity. Choose $\tau:=\d t+\xi_{\Sigma}$ such that $\sf{h}^{\sharp}(\xi_{\Sigma},\xi_{\Sigma})<\beta^{-2}$. Then, for all $\Psi=(\psi,\psi^{t},\psi^{\Sigma})\in\Gamma^{\infty}(\mathcal{E})$, we have
\begin{align*}
	\langle\sigma_{\mathcal{S}}(\xi)\Psi,\Psi\rangle_{\mathcal{E}}&=\Vert\psi\Vert_{\sf{E}}^{2}+\beta^{-2}\Vert\psi^{t}\Vert_{\sf{E}}^{2}+\Vert\psi^{\Sigma}\Vert_{\sf{E}\otimes\pi_{2}(\sf{T}^{\ast}\Sigma)}^{2}-2\langle\xi_{\Sigma}^{\sharp}\lrcorner\psi^{\Sigma},\psi^{t}\rangle_{\sf{E}}\\&\geq\Vert\psi\Vert_{\sf{E}}^{2}+\beta^{-2}\Vert\psi^{t}\Vert_{\sf{E}}^{2}+\Vert\psi^{\Sigma}\Vert_{\sf{E}\otimes\pi_{2}(\sf{T}^{\ast}\Sigma)}^{2}-2\sf{h}^{\sharp}(\xi_{\Sigma},\xi_{\Sigma})^{\frac{1}{2}}\Vert\psi^{t}\Vert_{\sf{E}}\Vert\psi^{\Sigma}\Vert_{\sf{E}\otimes\pi_{2}(\sf{T}^{\ast}\Sigma)}\\&\geq\Vert\psi\Vert_{\sf{E}}^{2}+\beta^{-2}\Vert\psi^{t}\Vert_{\sf{E}}^{2}+\Vert\psi^{\Sigma}\Vert_{\sf{E}\otimes\pi_{2}(\sf{T}^{\ast}\Sigma)}^{2}-2\beta^{-1}\Vert\psi^{t}\Vert_{\sf{E}}\Vert\psi^{\Sigma}\Vert_{\sf{E}\otimes\pi_{2}(\sf{T}^{\ast}\Sigma)}\geq 0\,, 
\end{align*}
where in the last step we used that $-2ab\geq -a^{2}-b^{2}$ applied to $a=\beta^{-1}\Vert\psi_{1}\Vert_{\sf{E}}$ and $b=\Vert\psi_{2}\Vert_{\sf{E}\otimes\pi_{2}(\sf{T}^{\ast}\Sigma)}$. We conclude that $\sf{S}$ is a symmetric hyperbolic system on $(\mathcal{E},\langle\cdot,\cdot\rangle_{\mathcal{E}})$. 

Last but not least, let us compare the Cauchy problems for $\sf{N}$ and $\sf{S}$. To this end, consider the Cauchy problems 
\begin{align*}
\begin{cases}
		\sf{N}\psi &=\phi\\
		\psi\vert_{\Sigma_{0}}&=\mathfrak{f}\\
		\nabla^{\sf{E}}_{\partial_{t}}\psi\vert_{\Sigma_{0}}&=\mathfrak{g}
	\end{cases}\,,\qquad
	\begin{cases}
		\sf{S}\Psi &=\Phi_{\phi}\\
		\Psi\vert_{\Sigma_{0}}&=\mathfrak{F}_{\mathfrak{f},\mathfrak{g}}
	\end{cases}\,,\qquad\text{where}\qquad\Phi_{\phi}:=\begin{pmatrix}
			0\\\vspace{2pt}\phi\\\vspace{2pt} 0
		\end{pmatrix}\,,\,\mathfrak{F}_{\mathfrak{f},\mathfrak{g}}:=\begin{pmatrix}
			\mathfrak{f}\\\vspace{2pt}\mathfrak{g}\\\vspace{2pt} \nabla^{\sf{E},\Sigma}\mathfrak{f}
		\end{pmatrix}
\end{align*}
for $\phi\in\Gamma^{\infty}(\sf{E})$ and $\mathfrak{f},\mathfrak{g}\in\Gamma^{\infty}(\sf{E}\vert_{\Sigma_{0}})$. Clearly, if $\psi\in\Gamma^{\infty}(\sf{E})$ is a solution to the former, then $\Psi_{\psi}:=(\psi,\nabla_{\partial_{t}}^{\sf{E}}\psi,\nabla^{\sf{E},\Sigma}\psi)$ is the unique solution to the latter. 

For the other direction, let $\Psi=(\psi,\psi^{t},\psi^{\Sigma})\in\Gamma^{\infty}(\mathcal{E})$ be the unique solution to the latter problem. By definition, the first row of $\sf{S}\Psi=\Phi_{\phi}$ encodes the relation $\psi^{t}=\nabla^{\sf{E}}_{\partial_{t}}\psi$. Moreover, we already know that $\psi^{\Sigma}\vert_{\Sigma_{0}}=\nabla^{\sf{E},\Sigma}\psi\vert_{\Sigma_{0}}=\nabla^{\sf{E},\Sigma}\mathfrak{f}$, by assumption, and we claim that in fact $\psi^{\Sigma}=\nabla^{\sf{E},\Sigma}\psi$ globally. Indeed, the third row in $\sf{S}\Psi=\Phi_{\phi}$ gives $\nabla^{\sf{E}\otimes\Sigma}_{\partial_{t}}\psi^{\Sigma}-\nabla^{\sf{E},\Sigma}\psi^{t}+\sf{R}_{1}\psi^{\Sigma}+\sf{R}_{2}\psi=0$ and, using that $\psi^{t}=\nabla^{\sf{E}}_{\partial_{t}}\psi$ as well as the definitions of $\sf{R}_{1}$ and $\sf{R}_{2}$ in Eq.~\eqref{eq:DefComBl}, we obtain
\begin{align*}
	0=&\nabla^{\sf{E}\otimes\Sigma}_{\partial_{t}}\psi^{\Sigma}-\nabla^{\sf{E},\Sigma}\nabla^{\sf{E}}_{\partial_{t}}\psi+\sf{R}_{1}\psi^{\Sigma}+\sf{R}_{2}\psi=(\nabla^{\sf{E}\otimes\Sigma}_{\partial_{t}}+\sf{R}_{1})(\psi^{\Sigma}-\nabla^{\sf{E},\Sigma}\psi)\, .
\end{align*}
Hence, $\psi^{\Sigma}-\nabla^{\sf{E},\Sigma}\psi$ solves a first order equation in time with zero initial data, which shows that $\psi^{\Sigma}=\nabla^{\sf{E},\Sigma}\psi$, as claimed. We conclude that $\psi$ is a solution to $\sf{N}\psi=\phi$ with initial data $(\mathfrak{f},\mathfrak{g})$.

To sum up, the Cauchy problems for $\sf{N}$ and $\sf{S}$ are equivalent provided one restricts the class of initial data of the latter to the subspace $\{(\mathfrak{f},\mathfrak{g},\mathfrak{h})^{\mathrm{T}}\in\Gamma^{\infty}(\mathcal{E}\vert_{\Sigma_{0}})\mid\mathfrak{h}=\nabla^{\sf{E},\Sigma}\mathfrak{f}\}$ of $\Gamma^{\infty}(\mathcal{E}\vert_{\Sigma_{0}})$. All the claims then follow from Theorem~\ref{Thm:Cauchy}.
\end{proof}

\subsection{Green Hyperbolic Operators}
\label{Sec:GreenCauchyHyp}
Having established well-posedness of the Cauchy problem for symmetric hyperbolic systems and normally hyperbolic operators, we now turn to a slightly more general notion of hyperbolicity. The concept of \textit{Green hyperbolicity}, introduced by Bär-Ginoux in \cite{GinouxBaer} and further developed by Bär in \cite{BaerGreen}, focuses on another important property of hyperbolic operators, namely the existence of \textit{retarded and advanced Green operators}, rather than the Cauchy problem itself.

\begin{definition}\label{Def:GreenHyp} (Green Hyperbolicity)\newline
    Let $(\sf{E},\langle\cdot,\cdot\rangle_{\sf{E}})$ be a Hermitian vector bundle over a globally hyperbolic spacetime $(\sf{M},\sf{g})$. An operator $\sf{D}\in\mathrm{DO}(\sf{E})$ is called \emph{Green hyperbolic}, if there exist linear operators $\sf{G}^{\pm}\:\Gamma^{\infty}_{\mathrm{c}}(\sf{E})\to\Gamma^{\infty}(\sf{E})$, called \emph{retarded and advanced Green operators}, satisfying the following conditions:
    \begin{align*}
        &\text{(i)}\hspace*{1cm}\sf{G}^{+}\circ \sf{D}\vert_{\Gamma^{\infty}_{\mathrm{c}}}=\sf{D}\circ \sf{G}^{+}=\mathrm{id}_{\Gamma^{\infty}_{\mathrm{c}}}\\
        &\text{(ii)}\hspace*{0.9cm}\sf{G}^{-}\circ \sf{D}\vert_{\Gamma^{\infty}_{\mathrm{c}}}= \sf{D}\circ \sf{G}^{-}=\mathrm{id}_{\Gamma^{\infty}_{\mathrm{c}}}\\
        &\text{(iii)}\hspace*{0.8cm}\mathrm{supp}(\sf{G}^{\pm}\psi)\subset \mathcal{J}^{\pm}(\mathrm{supp}(s))\hspace*{1cm}\forall \psi\in\Gamma^{\infty}_{\mathrm{c}}(\sf{E})\, .
\end{align*}
\end{definition}

\begin{remark}
	In the original articles \cite{GinouxBaer,BaerGreen}, it is also required that the \textit{dual operator} $\sf{D}^{t}\in\mathrm{DO}(\sf{E}^{\ast})$ admits retarded/advanced Green operators. The main advantage of this is that $\sf{G}^{\pm}$ then continuously extend to distributions with the extensions satisfying similar properties as (i)-(iii) above. In other works, such as~\cite{HackSchenkel}, it is required that the formal adjoint $\sf{D}^{\ast}$ with respect to some fixed bundle metric on $\sf{E}$ admits Green operators instead.
\end{remark}

\begin{example}\label{Example:GreenHyp} (Symmetric Hyperbolic Systems and Normally Hyperbolic Operators)\newline
	Let $\sf{S}\in\mathrm{DO}^{1}(\sf{E})$ be a symmetric hyperbolic system and $\varphi\in\Gamma_{\mathrm{c}}^{\infty}(\sf{E})$ be arbitrary. Now, choose a Cauchy hypersurface such that $\Sigma\cap\mathcal{J}^{+}(\mathrm{supp}(\varphi))=\emptyset$ and define $\psi\in\Gamma_{\mathrm{sc}}^{\infty}(\sf{E})$ to be the unique solution to the Cauchy problem
	\begin{align*}	
		\begin{cases}
			\sf{S}\psi &=\varphi\\
			\psi\vert_{\Sigma}&=0\, ,
		\end{cases}
	\end{align*}
	which exists by Theorem~\ref{Thm:Cauchy}. We set $\sf{G}^{+}\varphi:=\psi$. It is easy to see that this defines a linear operator $\sf{G}^{+}\:\Gamma^{\infty}_{\mathrm{c}}(\sf{E})\to\Gamma^{\infty}(\sf{E})$ such that $\sf{G}^{+}\circ\sf{S}\vert_{\Gamma_{\mathrm{c}}^{\infty}}=\sf{S}\circ\sf{G}^{+}=\mathrm{id}_{\Gamma^{\infty}_{\mathrm{c}}}$. Furthermore, by finite speed of propagation, it holds that 
	\begin{align*}
		\mathrm{supp}(\sf{G}^{+}\varphi)=\mathrm{supp}(\psi)\subset \mathcal{J}^{+}(\mathrm{supp}(\varphi))\, .
	\end{align*}
	Choosing $\Sigma$ such that $\Sigma\cap\mathcal{J}^{-}(\mathrm{supp}(\varphi))=\emptyset$, we construct an operator $\sf{G}^{-}$ satisfying $\mathrm{supp}(\sf{G}^{-}\varphi)\subset\mathcal{J}^{-}(\mathrm{supp}(\varphi))$. Hence, every symmetric hyperbolic system is Green hyperbolic. Similarly, one shows that normally hyperbolic operators are Green hyperbolic.
\end{example}

\begin{example}\label{Ex:PreNor} (Pre-Normally Hyperbolic Operators)\newline
	A linear differential operator $\sf{D}\in\mathrm{DO}(\sf{E})$ is called \textit{pre-normally hyperbolic} if there exists another differential operator $\sf{Q}\in\mathrm{DO}(\sf{E})$ such that $\sf{N}:=\sf{D}\circ\sf{Q}$ is normally hyperbolic. In this case, it is easy to see that $\sf{D}$ is Green hyperbolic with Green operators given by $\sf{G}^{\pm}_{\sf{D}}:=\sf{Q}\circ\sf{G}_{\sf{N}}^{\pm}$, see e.g.~\cite[Thm.~2.4.6]{WrochnaPhD} or \cite{Muhlhoff}. An explicit example is provided by the \textit{Proca operator} $\sf{P}=\delta\d+m^{2}\:\Omega^{1}(\sf{M})\to\Omega^{1}(\sf{M})$ for some mass parameter $m>0$. In this case, we set $\sf{Q}:=\mathrm{id}+m^{-2}\d\delta\:\Omega^{2}(\sf{M})\to\Omega^{2}(\sf{M})$. Then, the composition $\sf{N}:=\sf{P}\sf{Q}=\square_{\mathrm{dRH}}+m^{2}$ is normally hyperbolic by Example~\ref{Example:HodgeLapl}.
\end{example}

For later applications, we record the following properties of $\sf{G}^{\pm}$.

\begin{proposition}\label{Prop:GreenProp}	
	Let $\sf{D}\in\mathrm{DO}(\sf{E})$ be a Green hyperbolic operator. Then,
	\begin{itemize}
		\item[\emph{(i)}]The Green operators $\sf{G}^{+}\:\Gamma^{\infty}_{\mathrm{c}}(\sf{E})\to\Gamma^{\infty}(\sf{E})$ are continuous and unique.
		\item[\emph{(ii)}]$\sf{G}^{\pm}$ have continuous extensions $\sf{G}^{+}\:\Gamma^{\infty}_{\mathrm{pc}}(\sf{E})\to\Gamma^{\infty}_{\mathrm{pc}}(\sf{E})$ and $\sf{G}^{-}\:\Gamma^{\infty}_{\mathrm{fc}}(\sf{E})\to\Gamma^{\infty}_{\mathrm{fc}}(\sf{E})$ satisfying the same properties (i)-(iii) in Definition~\ref{Def:GreenHyp} on their respective domains.
		\item[\emph{(iii)}]Suppose that also the formal adjoint $\sf{D}^{\ast}$ of $\sf{D}$ with respect to some bundle metric $\langle\cdot,\cdot\rangle_{\sf{E}}$ is Green hyperbolic and denote its Green operators by $\sf{G}^{\pm}_{\ast}$. Then, $(\sf{G}^{\pm})^{\ast}=\sf{G}_{\ast}^{\mp}$.
		\end{itemize}
\end{proposition}

\begin{proof}
	Continuity is a direct application of the \textit{open mapping theorem} for Fréchet spaces, see \cite[Cor.~3.11]{BaerGreen}. Claim (ii) follows easily from the mapping properties of $\sf{G}^{\pm}$, see \cite[Thm.~3.8]{BaerGreen}. By definition, $\sf{G}^{+}$ defined on $\Gamma^{\infty}_{\mathrm{pc}}(\sf{E})$ is the inverse of $\sf{D}\vert_{\Gamma^{\infty}_{\mathrm{pc}}}$, while $\sf{G}^{-}$ defined on $\Gamma^{\infty}_{\mathrm{fc}}(\sf{E})$ is the inverse of $\sf{D}\vert_{\Gamma^{\infty}_{\mathrm{fc}}}$, which shows uniqueness of $\sf{G}^{\pm}$, cf.~\cite[Cor.~3.12]{BaerGreen}. For (iii), we compute
	\begin{align*}
		(\sf{G}^{\pm}\psi,\varphi)_{\sf{E}}=(\sf{G}^{\pm}\psi,\sf{D}^{\ast}\sf{G}^{\mp}_{\ast}\varphi)_{\sf{E}}=(\sf{D}\sf{G}^{\pm}\psi,\sf{G}^{\mp}_{\ast}\varphi)_{\sf{E}}=(\psi,\sf{G}^{\mp}_{\ast}\varphi)_{\sf{E}}\,,
	\end{align*}
	for all $\psi,\varphi\in\Gamma^{\infty}_{\mathrm{c}}(\sf{E})$, where we used the defining properties of $\sf{G}^{\pm}$ and $\sf{G}^{\pm}_{\ast}$.
\end{proof}

A central object in the solution theory of Green hyperbolic operators and also for applications in quantum field theory is the so-called \textit{causal propagator} (or also \textit{Pauli-Jordan distribution/function} \cite{PauliJordan} in the physics literature), which is defined as follows.

\begin{definition}\label{Def.CausProp} (Causal Propagator)\newline
	Let $\sf{D}\in\mathrm{DO}(\sf{E})$ be a Green hyperbolic operator. The linear operator
	\begin{align*}
		\sf{G}:=\sf{G}^{+}-\sf{G}^{-}\:\Gamma^{\infty}_{\mathrm{c}}(\sf{E})\to\Gamma^{\infty}_{\mathrm{sc}}(\sf{E})
	\end{align*}
	is called the \textit{causal propagator} of $\sf{D}$.
\end{definition}

For completeness and later applications, let us record the following properties.

\begin{proposition}\label{Prop:ExactSeq}
Let $\sf{D}\in\mathrm{DO}(\sf{E})$ be a Green hyperbolic operator.
\begin{itemize}
\item[\emph{(i)}]The causal propagator $\sf{G}\:\Gamma^{\infty}_{\mathrm{c}}(\sf{E})\to\Gamma^{\infty}_{\mathrm{sc}}(\sf{E})$ is continuous and has a linear and continuous extension $\sf{G}\:\Gamma^{\infty}_{\mathrm{tc}}(\sf{E})\to\Gamma^{\infty}(\sf{E})$.
\item[\emph{(ii)}]$\sf{G}\circ\sf{D}\vert_{\Gamma^{\infty}_{\mathrm{tc}}}=\sf{D}\circ\sf{G}=0$.
\item[\emph{(iii)}]Suppose that also the formal adjoint $\sf{D}^{\ast}$ of $\sf{D}$ with respect to some bundle metric $\langle\cdot,\cdot\rangle_{\sf{E}}$ is Green hyperbolic and denote its causal propagator by $\sf{G}_{\ast}$. Then, $\sf{G}^{\ast}=-\sf{G}_{\ast}$.
\item[\emph{(iv)}]The following sequences are exact complexes:
\begin{align*}
	& 0\xrightarrow{}\Gamma_{\mathrm{c}}^{\infty}(\sf{E})\xrightarrow{\sf{D}}\Gamma_{\mathrm{c}}^{\infty}(\sf{E})\xrightarrow{\sf{G}}\Gamma_{\mathrm{sc}}^{\infty}(\sf{E})\xrightarrow{\sf{D}}\Gamma_{\mathrm{sc}}^{\infty}(\sf{E})\\
	& 0\xrightarrow{}\Gamma_{\mathrm{tc}}^{\infty}(\sf{E})\xrightarrow{\sf{D}}\Gamma_{\mathrm{tc}}^{\infty}(\sf{E})\xrightarrow{\sf{G}}\Gamma^{\infty}(\sf{E})\xrightarrow{\sf{D}}\Gamma^{\infty}(\sf{E})
\end{align*}
\end{itemize}
\end{proposition}

\begin{proof}
	Claims (i)-(iii) are a direct consequence of Proposition~\ref{Prop:GreenProp}. For (iv), we show exactness of the first sequence, following \cite[Thm.~3.5]{GinouxBaer}. The second one can be shown analogously. First of all, by (ii), the sequence is clearly a complex. Now, let $\psi\in\mathrm{ker}(\sf{D}\vert_{\Gamma^{\infty}_{\mathrm{c}}})$. Then, $\psi=\sf{G}^{+}\sf{D}\psi=0$, which shows exactness at the first $\Gamma^{\infty}_{\mathrm{c}}(\sf{E})$. For exactness at the second $\Gamma^{\infty}_{\mathrm{c}}(\sf{E})$, we note that the inclusion $\mathrm{ran}(\sf{D}\vert_{\Gamma^{\infty}_{\mathrm{c}}})\subset\mathrm{ker}(\sf{G}\vert_{\Gamma^{\infty}_{\mathrm{c}}})$ is trivial. For the other direction, let $\psi\in\mathrm{ker}(\sf{G}\vert_{\Gamma^{\infty}_{\mathrm{c}}})$ and set $\varphi:=\sf{G}^{+}\psi=\sf{G}^{-}\psi$. Then, 
	\begin{align*}
        \mathrm{supp}(\varphi)=\mathrm{supp}(\sf{G}^{+}\psi)\cap\mathrm{supp}(\sf{G}^{-}\psi)\subset \mathcal{J}^{+}(\mathrm{supp}(\psi))\cap\mathcal{J}^{-}(\mathrm{supp}(\psi))\,,
    \end{align*}
    which is compact by global hyperbolicity.\footnote{By~\cite[Thm.~8.3.11]{Wald}, the sets $\mathcal{J}^{\pm}(\sf{K})$ are closed for any $\sf{K}\subset\sf{M}$ compact, whenever $(\sf{M},\sf{g})$ is globally hyperbolic. It follows that $\mathcal{J}^{+}(\sf{K})\cap\mathcal{J}^{-}(\sf{K})$ is compact.} It follows that $\varphi\in\Gamma^{\infty}_{c}(\sf{E})$ and $\sf{D}\varphi=\sf{D}\sf{G}^{+}\psi=\psi$. Therefore, $\psi\in\mathrm{ran}(\sf{D}\vert_{\Gamma^{\infty}_{\mathrm{c}}})$. For exactness at the first $\Gamma^{\infty}_{\mathrm{sc}}(\sf{E})$, we note that the inclusion $\mathrm{ran}(\sf{G}\vert_{\Gamma^{\infty}_{\mathrm{c}}})\subset\mathrm{ker}(\sf{D}\vert_{\Gamma^{\infty}_{\mathrm{sc}}})$ is trivial. For the other direction, let $\psi\in \mathrm{ker}(\sf{D}\vert_{\Gamma^{\infty}_{\mathrm{sc}}})$. Now, decompose $\psi=\psi^{+}-\psi^{-}$ for $\psi^{\pm}$ such that $\mathrm{supp}(\psi^{\pm})\subset\mathcal{J}^{\pm}(\sf{K})$ for some compact set $\sf{K}\subset\sf{M}$. Set $\varphi:=\sf{D}\psi^{+}=\sf{D}\psi^{-}$. By construction, $\mathrm{supp}(\varphi)\subset\mathcal{J}^{+}(\sf{K})\cap\mathcal{J}^{-}(\sf{K})$ and hence $\varphi\in\Gamma^{\infty}_{\mathrm{c}}(\sf{E})$. It follows that $\varphi^{\pm}=\sf{G}^{\pm}\sf{D}\varphi^{\pm}$ and hence $\psi=\sf{G}\psi$, which concludes the proof.
\end{proof}

Following the same line of reasoning as in Example~\ref{Example:GreenHyp}, one can see that every linear differential operator that admits a well-posed Cauchy problem is necessarily Green hyperbolic. As it turns out, the converse statement also holds true, at least in a suitable sense: every differential operator $\mathsf{D}\in\mathrm{DO}^{m}(\mathsf{E})$ of order $m\in\mathbb{N}$ is \textit{Cauchy hyperbolic}, which means that for any smooth spacelike Cauchy hypersurface $\Sigma\subset \mathsf{M}$, there exists a smooth finite-rank vector bundle $\mathsf{E}_{\rho}\to\Sigma$ over $\Sigma$ together with a linear map $\rho:\Gamma^{\infty}_{\mathrm{sc}}(\mathsf{E}) \to \Gamma^{\infty}_{\mathrm{c}}(\mathsf{E}_{\rho})$, called the \textit{initial data map}, defined as the composition of a differential operator of order $\leq m-1$ with the pullback along the embedding $i\:\Sigma\hookrightarrow \mathsf{M}$, that is invertible when restricted to $\ker(\mathsf{D}\vert_{\Gamma^{\infty}_{\mathrm{sc}}})$. In other words, for every initial datum $\mathfrak{f}\in\Gamma^{\infty}_{\mathrm{c}}(\mathsf{E}_{\rho})$, there exists a unique solution $\psi\in\Gamma^{\infty}_{\mathrm{sc}}(\mathsf{E})$ to the Cauchy problem
\begin{align*}
   \begin{cases}
      \mathsf{D}\psi &= 0, \\
      \rho(\psi) &= \mathfrak{f}.
   \end{cases}
\end{align*}
We refer to \cite[Sec.~4.4]{KhavkineCone} for details. The inverse of $\rho\vert_{\ker(\mathsf{D}\vert_{\Gamma^{\infty}_{\mathrm{sc}}})}$ is the \textit{Cauchy evolution operator} $\mathcal{U}\:\Gamma^{\infty}_{\mathrm{c}}(\sf{E}_{\rho})\to\ker(\mathsf{D}\vert_{\Gamma^{\infty}_{\mathrm{sc}}})$. Of course, the pair $(\rho,\sf{E}_{\rho})$ is not unique. If $\sf{S}\in\mathrm{DO}^{1}(\sf{E})$ is a symmetric hyperbolic system, then the obvious choice for $(\sf{E}_{\rho},\rho)$ is given by
\begin{align*}
	\sf{E}_{\rho}:=\sf{E}\vert_{\Sigma}\,,\qquad \Gamma^{\infty}(\sf{E})\ni\psi\mapsto\rho(\psi):=\psi\vert_{\Sigma}
\end{align*}
For a normally hyperbolic operator $\sf{N}=-\square^{\nabla^{\sf{E}}}+\sf{N}_{0}$, the natural choice consists of 
\begin{align}\label{eq:IDM.NH}
			\sf{E}_{\rho}:=\sf{E}\vert_{\Sigma}\oplus\sf{E}\vert_{\Sigma}\,,\qquad \Gamma^{\infty}(\sf{E})\ni\psi\mapsto\rho(\psi):=(\psi\vert_{\Sigma},\nabla^{\sf{E}}_{\nu}\psi\vert_{\Sigma})\,,
\end{align}
where $\nu$ denotes the future-directed timelike unit vector of $\Sigma$ in $\sf{M}$. The fact that these two maps for $\sf{S}$ and $\sf{N}$ are indeed invertible when restricted to the kernel of the respective operators, then follows from Theorem~\ref{Thm:Cauchy} and Theorem~\ref{Thm:NHCauchy}.

\begin{remark}
Of course, $(\rho,\sf{E}_{\rho})$ has to be chosen suitably. At first sight, one might naively think that for a given Green hyperbolic operator $\sf{D}\in\mathrm{DO}^{m}(\sf{E})$ of order $m$, one can simply choose $\sf{E}_{\rho}:=\sf{E}\vert_{\Sigma}^{\oplus m}$ together with the map $\rho$ that assigns to a given section $\psi\in\Gamma_{\mathrm{sc}}^{\infty}(\sf{E})$ the restriction of $\psi$ and all its normal derivatives up to order $m-1$ to $\Sigma$. However, this is clearly not the case: consider two Hermitian vector bundles $(\sf{E}_{i},\langle\cdot,\cdot\rangle_{\sf{E}_{i}})$ for $i=1,2$ together with a connection $\nabla^{\sf{E}_{1}}$ and a symmetric hyperbolic system $\sf{S}\in\mathrm{DO}^{1}(\sf{E}_{2})$ on $(\sf{E}_{2},\langle\cdot,\cdot\rangle_{\sf{E}_{2}})$. Then,
\begin{align*}
	\begin{pmatrix}-\square^{\nabla^{\sf{E}}}& 0 \\ 0 &\sf{S}\end{pmatrix}\in\mathrm{DO}^{2}(\mathcal{E})\,,\qquad\mathcal{E}:=\sf{E}_{1}\oplus\sf{E}_{2}
\end{align*}
is clearly Green hyperbolic, see e.g.~\cite[Lem.~2.29]{GinouxBaer} and \cite[Lem.~3.17.]{BaerGreen}. In this case, we need \emph{two} initial data for $-\square^{\nabla^{\sf{E}}}$, but only \emph{one} for $\sf{S}$. Hence, the natural pair $(\rho,\mathcal{E}_{\rho})$ to consider is
\begin{align*}
	\mathcal{E}_{\rho}:=\mathcal{E}\vert_{\Sigma}\oplus\sf{E}_{2}\vert_{\Sigma}\,,\qquad \Gamma^{\infty}(\mathcal{E})\ni (\psi,\varphi)\mapsto \rho((\psi,\varphi)):=(\psi\vert_{\Sigma},\nabla^{\sf{E}}_{\nu}\psi\vert_{\Sigma},\varphi\vert_{\Sigma})\,,
\end{align*}
where we identified $\Gamma^{\infty}(\mathcal{E})\cong\Gamma^{\infty}(\sf{E}_{1})\oplus\Gamma^{\infty}(\sf{E}_{2})$, as usual.
\end{remark}

\section{Symmetric Hyperbolic Systems with Nonlocal Potentials}\label{Sec:SHSNonLoc}
As we have discussed in detail in the previous sections, hyperbolic partial differential equations model many physical phenomena, particularly those involving wave propagation, causality, and finite speed of propagation. They arise, for instance, naturally in systems governed by classical and quantum field theories. While they are traditionally studied in the context of local interactions, it has become increasingly clear in many areas of mathematics that it would be highly desirable to extend this type of equations by incorporating also nonlocal interactions, thereby providing a more flexible framework for modelling complex phenomena, where long-range correlations or memory effects play a significant role. Examples range from \emph{linear response theory}, in which a system's response does not only depend on the instantaneous perturbation but also on its entire history, hence leading to nonlocal temporal kernels, \emph{viscoelastic materials}, in which memory-dependent stress–strain relations are naturally described by kernels of the same type (see e.g.~\cite{viscoelastic1,viscoelastic2}), classical and quantum fields on \emph{noncommutative spacetimes} \cite{VerchNonLoc,LechnerVerch}, as well as causal fermion systems, a non-perturbative framework unifying general relativity and quantum field theory (see e.g.~\cite{FinsterBook}), where nonlocalities arise in the causal action principle \cite{FinsterMinkowski} and the quantum field theoretic limit~\cite{fockdynamics}, with applications, e.g., to collapse models discussed in \cite{FinsterCollapse1,FinsterCollapse2}. Beyond these physical applications, nonlocal PDEs have also become a vibrant area of research in other applied areas of mathematical analysis, for instance, in biology and life science (see e.g.~the recent review \cite{PalMelnik}).

As seen above, there is a plethora of motivations arising from diverse areas of mathematical physics and various areas of applied analysis. Before turning to the main content of this section, let us highlight two specific examples and sources of motivation within the area of classical and quantum field theory in more detail.

\paragraph{Maxwell Equations in Linear Dispersive Media.} The first example falls into the general category of \emph{linear response theory} briefly mentioned above and provides an explicit example of the general type of nonlocal equations that we will consider in the forthcoming discussion.

Following Example~\ref{Ex:Maxwell}, we consider an ultrastatic globally hyperbolic spacetime of the form $(\sf{M}=\bb{R}\times\Sigma,\,\sf{g}=-\d t\otimes\d t+\sf{h})$, where $(\Sigma,\sf{h})$ denotes a complete Riemannian manifold. Now, in the presence of matter, the electric and magnetic fields $\mathscr{E},\mathscr{B}\in\Gamma^{\infty}(\pi^{\ast}(\sf{T}\Sigma))$, where $\pi\:\sf{M}\to\Sigma$ denotes the projection, are not entirely independent of the charge and current distributions; rather, they interact with bound charges and currents within the medium. These interactions give rise to \emph{induced fields}, making the total sources inseparable from the fields themselves.  To handle this coupling in a tractable way, the standard approach is to decompose the total charge and current densities into external and induced parts, i.e.~$\rho = \rho^{(\mathrm{ext})} + \rho^{(\mathrm{ind})}$ and $j = j^{(\mathrm{ext})} + j^{(\mathrm{ind})}$, and to introduce the auxiliary fields $\mathscr{D}$ (\emph{electric displacement}) and  $\mathscr{H}$ (\emph{magnetising field/magnetic field strength}), in Gaußian units and with speed of light $c=1$, by
\begin{align}\label{eq:Aux}
\mathscr{D} := \mathscr{E} + 4\pi \mathscr{P}\,, \qquad
\mathscr{H} := \mathscr{B} - 4\pi \mathscr{M}\, ,
\end{align}
where $\mathscr{P}, \mathscr{M}\in\Gamma^{\infty}(\pi^{\ast}(\sf{T}\Sigma))$ are the \emph{polarisation} and \emph{magnetisation} vector fields, defined in terms of the induced charge and current by the system\footnote{Note that $\mathscr{P}$ and $\mathscr{M}$ are only specified \emph{up to gauge}, that is, up to $(\mathscr{P},\mathscr{M})\mapsto (\mathscr{P}+\mathrm{curl}_{\Sigma}(\mathscr{G}),M-\partial_{t}\mathscr{G})$ for any vector field $\mathscr{G}\in\Gamma^{\infty}(\pi^{\ast}(\sf{T}\Sigma))$, as one can easily see from the defining equations.}
\begin{align*}
\begin{cases}
\rho^{(\mathrm{ind})} &= -\mathrm{div}_{\Sigma}(\mathscr{P}) \\
j^{(\mathrm{ind})} &= \partial_t \mathscr{P} + \mathrm{curl}_{\Sigma}(\mathscr{M})
\end{cases}\, .
\end{align*}
In other words, $\mathscr{P}$ and $\mathscr{M}$ describe how the materials reacts to the electric and magnetic fields present and how the induced charges and currents contributes to the overall electromagnetic fields. With these definitions, the \emph{macroscopic Maxwell's equations} can be written in the form
\begin{align}\label{eq:Max}
	\begin{cases}
			\partial_{t}\mathscr{D}-\mathrm{curl}_{\Sigma}(\mathscr{H})&=-4\pi j^{(\mathrm{ext})}\\
			\partial_{t}\mathscr{B}+\mathrm{curl}_{\Sigma}(\mathscr{E})&=0
		\end{cases},\qquad \begin{cases}
			\mathrm{div}_{\Sigma}(\mathscr{D})=4\pi\rho^{(\mathrm{ext})}\\
			\mathrm{div}_{\Sigma}(\mathscr{B})=0
		\end{cases}\, .
	\end{align}
As written, Maxwell's equations~\eqref{eq:Max}, viewed as a system for the fields $\mathscr{E}, \mathscr{B}, \mathscr{D}, \mathscr{H}$ in terms of the external sources $\rho^{(\mathrm{ext})}$ and $j^{(\mathrm{ext})}$, are of course underdetermined. To resolve this issue, one must supplement the system with a suitable \emph{matter model} in which $\mathscr{P}$ and $\mathscr{M}$ are given functions of $\mathscr{E}$ and $\mathscr{B}$, hence determining the auxiliary fields $\mathscr{D}$ and $\mathscr{H}$ in terms of $\mathscr{E}$ and $\mathscr{B}$ via Eq.~\eqref{eq:Aux}. The precise form of these \emph{constitutive equations} is determined by the physical properties of the medium and when chosen appropriately, ensure a well-posed formulation of Maxwell's equations. In the most simplest case, one finds $\mathscr{P}\propto\mathscr{E}$ and $\mathscr{M}\propto\mathscr{B}$ (``\emph{linear and isotropic materials}''), in which case the Maxwell equations take a similar form as in vacuum. 

In many cases, however, the dependency of $(\mathscr{P},\mathscr{M})$ on $(\mathscr{E},\mathscr{B})$ is more complicated. A very important class of materials is provided by \emph{linear dispersive media}, see e.g.~\cite{LandLifMax,Cessenat}. In this case, there is a \emph{linear response function} $\chi\in C^{\infty}(\bb{R})$ with the property that $\chi(0)=0$, such that
\begin{align}\label{eq:Response}
	\mathscr{P}(t,\vec{x})=\frac{1}{4\pi}\int_{-t_{0}}^{t}\chi(t-\tau)\mathscr{E}(\tau,\vec{x})\,d\tau\,,\quad \mathscr{M}(t,\vec{x})=0\,,
\end{align}
where $t_{0}\geq 0$ is fixed. An explicit example of such a model is the well-known \emph{Lorentz-Drude model}, which has been used to describe the response of electrons in a material to external electromagnetic fields. In this case, the response function $\chi$ is given by $\chi(t)\propto e^{-c_{1}t}\mathrm{sin}(c_{2}t)$ for some constants $c_{1},c_{2}>0$, see e.g.~\cite[Sec.~5]{SchmidMurroFinster}. The dynamical part of the Maxwell equations~\eqref{eq:Max} together with~the constitutive equations~\eqref{eq:Response} can be written as
\begin{align}\label{eq:MaxwellNonLoc}
	\bigg[\partial_{t}+\begin{pmatrix}
		0 & -\mathrm{curl}\\
		\mathrm{curl} &0
	\end{pmatrix}\bigg]
	\begin{pmatrix}
		\mathscr{E}(t,\cdot)\\\mathscr{B}(t,\cdot)
	\end{pmatrix}+\int_{-t_{0}}^{t}
	\begin{pmatrix}
		\dot{\chi}(t-\tau)\mathscr{E}(\tau,\cdot)\\  0
	\end{pmatrix}
	\,\d\tau=
	\begin{pmatrix}
	-4\pi j^{(\mathrm{ext})}(t,\cdot)\\ 0
	\end{pmatrix}\, ,
\end{align}
where we used that $\chi(0)=0$, which implies that there is no boundary term when taking the $t$-derivative of the nonlocal part in the definition of $\mathscr{P}$. Following the discussion in Example~\ref{Ex:Maxwell}, we conclude that the Maxwell equations on ultrastatic spacetimes in linear dispersive media are a symmetric hyperbolic system with a nonlocal potential.

\paragraph{Semiclassical Models.} Another important motivational example for the general theme of hyperbolic equations with nonlocal interactions are \emph{semiclassical models}, in which nonlocalities naturally emerge due to the interaction of classical and quantum matter. An important example in this direction is provided by semiclassical gravity. General relativity has proven remarkably successful within its domain, yet there are situations in which quantum effects cannot be neglected, such as in the vicinity of black holes. The dynamics of particles in the microscopic domain, on the other hand, is described by quantum field theory, in which in particular the \emph{algebraic approach} (see e.g.~Sec.~\ref{Sec.AQFT}) offers a mathematically rigorous framework, well-suited to describe fields on curved Lorentzian backgrounds. While, despite numerous efforts, a consistent theory of quantum gravity remains elusive, a major step in this direction is the study of backreactions, i.e., the effects of quantum matter on the classical geometry, captured by the \emph{semiclassical Einstein equations} (see for instance~\cite[Sec.~2.6]{HackBook} for a review),
\begin{align}\label{eq:SEE}
	\mathrm{Ric}(\sf{g})-\frac{1}{2}\mathrm{Scal}(\sf{g})\sf{g}+\Lambda\sf{g}=\omega_{\sf{g}}(\mathopen{:}\sf{T}(\sf{g},\Phi)\mathclose{:})
\end{align}
where $\omega_{\sf{g}}(\mathopen{:}\sf{T}(\sf{g},\Phi)\mathclose{:})$ is the normal-ordered energy-momentum tensor of a quantised field $\Phi$, $\Lambda$ a cosmological constant and $\omega_{\sf{g}}$ a suitable quantum state, we refer to Sec.~\ref{Sec.AQFT} for details. The unknowns of Eq.~\eqref{eq:SEE} are the state $\omega_{\sf{g}}$ and the Lorentzian metric $\sf{g}$. To define the right-hand side of Eq.~\eqref{eq:SEE} rigorously, a renormalisation procedure is required, essentially due to the singular nature of quantum fields as operator-valued distributions, which makes nonlinear expressions, such as $\Phi^{2}$ contained in $\sf{T}(\sf{g},\Phi)$ ill-defined. It is well known, however, that this obstacle can be overcome by considering a \emph{Hadamard state} that mimics the singular structure of the two-point function of the Minkowski vacuum, we refer to Sec.~\ref{Sec.AQFT} for details. As a by-product of this procedure, the right-hand side of Eq.~\eqref{eq:SEE} involves terms that are nonlocal both in time and space, reflecting the inherent nonlocality of quantum correlation functions. These features cannot be fully absorbed into the renormalisation freedom present \cite{HollandsWald1,HollandsWald2}, hence resulting in a nonlinear and nonlocal equation. While the classical Einstein equations can be understood as a quasilinear wave equation, ensuring a well-posed Cauchy problem, the combination of nonlinearities, higher-order derivative terms, and terms that are
nonlocal both in time and space, makes the mathematical analysis of the semiclassical equations  both mathematically intriguing and significantly
more challenging to analyse than their classical counterpart. Besides some recent progress in highly symmetric spacetimes by various different authors, e.g. \cite{Pinamonti1,Pinamonti2,Pinamonti3,Gottschalk1,Gottschalk2,Gottschalk3,Sanders,Benito1,Benito2}, establishing a well-posed Cauchy problem for the semiclassical Einstein equations on general globally hyperbolic Lorentzian manifolds remains an important open problem.

To obtain a better understanding of semiclassical models, we remark that it has recently been shown in \cite{Meda} that the \emph{linearised} semiclassical Einstein equations, coupled to a Klein-Gordon field of mass $m$, in a cosmological spacetime can be written in the form
\begin{align*}
	\sf{N}_{1}\sf{K}\psi+\sf{N}_{2}\psi=\phi
\end{align*}
acting on the unknown scalar degree of freedom $\psi$, where $\sf{N}_{1}$ and $\sf{N}_{2}$ are normally hyperbolic operators and $\phi$ a source term. The operator $\sf{K}$ is a continuous operator with Schwartz kernel
\begin{align}\label{eq:NonLocSemi}
	\sf{K}(x,y)=\int_{4m^{2}}^{\infty}\frac{\rho(\sf{M}^{2})}{\sf{M}^{2}+a}(\square+a)\sf{G}^{+}_{\sf{M}^{2}}(x,y)\,,
\end{align}
where $a$ is a (renormalisation) constant, $\rho$ a suitable smooth function and $\sf{G}^{+}_{\sf{M}^{2}}(x,y)$ the kernel of the retarded Green's operator of the Klein-Gordon equation with mass $\sf{M}$. Note that the nonlocality enters in fact at the leading order.

Another example in the context of semiclassical models are the \emph{semiclassical Maxwell equations}, i.e.~the electromagnetic analogue of the semiclassical Einstein equations, see e.g.~\cite{Galanda}.\bigskip

Despite the diversity of hyperbolic equations with nonlocal potentials/interactions, a unifying mathematical treatment based on symmetric hyperbolic systems with nonlocal potentials that provides a coherent and powerful language to investigate their dynamics is still missing. Notable exceptions include the recent work \cite{FewsterNonLoc}, which considers modifications of Green hyperbolic operators through the addition of a possibly nonlocal operator acting within a (fixed) compact subset in spacetime, as well as \cite{Bachelot} on semilinear hyperbolic systems with nonlocal potentials in Minkowski spacetime. In the present section, following \cite{SchmidMurroFinster}, we aim to provide a systematic study of the Cauchy problem for a symmetric hyperbolic system coupled to a large class of nonlocal potentials.

\subsection{Nonlocal Potentials} 
As a first part in our analysis, we need to establish a general framework for nonlocal potentials on globally hyperbolic Lorentzian manifolds. Furthermore, we need to define a suitable class of nonlocal potentials for which the Cauchy problem when coupled to a symmetric hyperbolic system can be studied. Throughout this section, let $(\sf{M},\sf{g})$ be a $(1+n)$-dimensional globally hyperbolic spacetime and $t\in C^{\infty}(\sf{M},\bb{R})$ a fixed Cauchy temporal function such that
\begin{align*}
	\sf{M}=\bb{R}\times\Sigma\,,\qquad \sf{g}=-\beta^{2}\d t\otimes\d t+\sf{h}_{t}\,,
\end{align*}
where $\Sigma$ is a smooth spacelike Cauchy hypersurface, $\beta\in C^{\infty}(\sf{M},(0,\infty))$ a lapse function and $(\sf{h}_{t})_{t\in\bb{R}}$ a one-parameter family of Riemannian metrics on $\Sigma$ depending smoothly on $t$. The corresponding foliation of $\sf{M}$ by spacelike Cauchy surfaces is denoted by $(\Sigma_{t}:=\{t\}\times\Sigma)_{t\in\bb{R}}$.

\paragraph{A Brief Recap on Distributions and Kernels.} Let $(\sf{E},\langle\cdot,\cdot\rangle_{\sf{E}})$ be a Hermitian vector bundle. In Appendix~\ref{Appendix:Micro}, we introduced the relevant notation and terminology for \emph{distributions} and \emph{kernels} and we refer the reader to that discussion for further details on the following material. We equip $\Gamma^{\infty}_{\mathrm{c}}(\sf{E})$ and $\Gamma^{\infty}(\sf{E})$ with their natural LF and Fréchet topologies, respectively, and define the space of \emph{distributions} by $\mathcal{D}^{\prime}(\sf{E}):=(\Gamma^{\infty}_{\mathrm{c}}(\sf{E}^{\ast}))^{\prime}$, equipped with its strong dual topology. Note that many authors define the space $\mathcal{D}^{\prime}(\sf{E})$ instead as the dual space of \emph{densitised} test sections, which has the advantage that one can define a natural embedding $\Gamma^{\infty}(\sf{E})\hookrightarrow\mathcal{D}^{\prime}(\sf{E})$ on general manifolds without the need of specifying a metric structure beforehand (see e.g.~\cite{Tarkhanov}). In our case, however, we have a fixed Lorentzian metric from the beginning and hence, it is natural to embed $\Gamma^{\infty}(\sf{E})$ into $\mathcal{D}^{\prime}(\sf{E})$ using the Lorentzian volume measure, i.e.
\begin{align*}
	\Gamma^{\infty}(\sf{E})\hookrightarrow\mathcal{D}^{\prime}(\sf{E}),\qquad \psi\mapsto\bigg(\Gamma^{\infty}_{\mathrm{c}}(\sf{E}^{\ast})\ni\varphi\mapsto \int_{\sf{M}}\varphi(\psi)\,d\mu_{\sf{g}}\bigg)\, .
\end{align*}
As usual, we shall call distributions contained in the image of this map \textit{regular}.

Now, if $\sf{B}\:\Gamma^{\infty}_{\mathrm{c}}(\sf{E})\to\mathcal{D}^{\prime}(\sf{E})$ is any linear operator, the celebrated \emph{kernel theorem of Schwartz} (see Theorem~\ref{Thm:Kernel}) states that $\sf{B}$ is continuous if and only if there exists a bidistributional \emph{kernel} $k_{\sf{B}}\in \mathcal{D}^{\prime}(\sf{M}\times\sf{M},\sf{E}\boxtimes\sf{E}^{\ast})$, i.e.
\begin{align}\label{Eq:Schwartz} 
		\langle \sf{B}\varphi,\psi\rangle_{\sf{M}}=\langle k_{\sf{B}},\psi\boxtimes\varphi\rangle_{\sf{M}\times\sf{M}}\,,
	\end{align}
or all $\varphi\in \Gamma_{\mathrm{c}}^{\infty}(\sf{E})$ and $\psi\in \Gamma^{\infty}_{\mathrm{c}}(\sf{E}^{\ast})$, where $\langle\cdot,\cdot\rangle_{\sf{M}}\:\mathcal{D}^{\prime}(\sf{E})\times\Gamma^{\infty}(\sf{E})\to\bb{C}$ denotes the bilinear \emph{distributional pairings} on $\sf{M}$ and where $\sf{E}\boxtimes\sf{E}^{\ast}$ denotes the \emph{outer tensor product}, i.e.~the bundle over $\sf{M}\times\sf{M}$ whose fibre at a point $(p,q)\in\sf{M}\times\sf{M}$ is given by $\sf{E}_{p}\otimes\sf{E}^{\ast}_{q}$.

Usually, we shall restrict ourselves to \emph{semiregular operators}, that are, operator whose Schwartz kernel $k_{\sf{B}}$ lies in the subspace $\Gamma^{\infty}(\sf{E})\hat{\otimes}\mathcal{D}^{\prime}(\sf{E})$, where $\hat{\otimes}$ denotes the projective/injective tensor product of nuclear spaces. On a more practical level, this means that $\sf{B}$ is continuous as a linear operator of the form $\sf{B}\:\Gamma^{\infty}_{\mathrm{c}}(\sf{E})\to\Gamma^{\infty}(\sf{E})$.

If the operator $\sf{B}$ happens to be a \textit{smoothing operator}, which means that its kernel $k_{\sf{B}}$ is a regular distribution, i.e.~$k_{\sf{B}}\in \Gamma^{\infty}(\sf{E}\boxtimes\sf{E}^{\ast})$, we can identify $k_{\sf{B}}$ with a family of $\bb{C}$-linear maps $k_{\sf{B}}(x,y)\colon\sf{E}_{y}\to\sf{E}_{x}$, depending smoothly on $x,y\in\sf{M}$. Furthermore, using the (anti-linear) bijection $\varphi\mapsto\langle\varphi,\cdot\rangle_{\sf{E}}$ to identify sections of $\sf{E}$ with sections of $\sf{E}^{\ast}$ and vice versa, the defining relation~\eqref{Eq:Schwartz} takes the form
\begin{align*}
	(\psi,\sf{B}\varphi)_{\sf{M}}=\int_{\sf{M}}\langle\psi(x),\sf{B}\varphi(x)\rangle_{\sf{E}_{x}}\,\d\mu_{\sf{g}}(x)=\int_{\sf{M}\times\sf{M}}\langle\psi(x),k_{\sf{B}}(x,y)\varphi(y)\rangle_{\sf{E}_{x}}\,(\d\mu_{\sf{g}}\times\d\mu_{\sf{g}})(x,y)
\end{align*}
for all $\varphi,\psi\in \Gamma^{\infty}_{\mathrm{c}}(\sf{E})$. For this reason, it is usually customary to use the informal notation
\begin{align*}
	\sf{B}\varphi(x)=\int_{\sf{M}}k_{\sf{B}}(x,y)\varphi(y)\,\d\mu_{\sf{g}}(y)
\end{align*} 
also in the case in which $k_{\sf{B}}$ is a (non-regular) distribution. Such expressions must, of course, be interpreted in the distributional sense: in general $k_{\sf{B}}$ cannot be identified with an ordinary function, and even in the regular case the integral above is not well defined, as the integrand takes values in a vector bundle and must first be contracted with the corresponding bundle metric. While this notation is useful in applications for understanding the underlying structures and is widely employed in the literature, we will avoid it largely in the following discussion.

Now, in full generality, we will consider semiregular operators $\sf{B}\:\Gamma^{\infty}_{\mathrm{c}}(\sf{E})\to\Gamma^{\infty}(\sf{E})$ as nonlocal potentials. Of course, at this level, the framework is far too general to make sense of the Cauchy problem, and some restrictions are required. Since we will typically work with energy methods, it is natural to begin by assuming that $\sf{B}$ admits a \emph{time kernel}. 

\paragraph{Time Kernels.}
We consider a globally hyperbolic spacetime $(\sf{M},\sf{g})$ with a fixed Cauchy temporal function $t\in C^{\infty}(\sf{M})$ and corresponding foliation by Cauchy surfaces $(\Sigma_{t})_{t\in\bb{R}}$, as above. 
The goal of the following discussion is to introduce the notion of a \textit{time kernel}, which loosely speaking means a two-parameter family of operator $\sf{B}_{t,\tau}\:\Gamma_{\mathrm{c}}^{\infty}(\Sigma_{\tau},\sf{E}\vert_{\Sigma_{\tau}})\to\Gamma^{\infty}(\Sigma_{t},\sf{E}\vert_{\Sigma_{t}})$ such that a semiregular operator $\sf{B}\:\Gamma^{\infty}_{\mathrm{c}}(\sf{E})\to\Gamma^{\infty}(\sf{E})$ can (informally) be written in the form
\begin{align*}
	\sf{B}\varphi(t,\cdot)=\int_{-\infty}^{\infty}\sf{B}_{t,\tau}\varphi_{\tau}(\cdot)\,\d\tau
\end{align*}
for all $t\in\bb{R}$ and $\varphi\in\Gamma^{\infty}_{\mathrm{c}}(\sf{E})$, where $\varphi_{t}:=\varphi\vert_{\Sigma_{t}}\in\Gamma^{\infty}(\Sigma_{t},\sf{E}\vert_{\Sigma_{t}})$. It is the goal of this section to make the notion of time kernels more precise and to establish criteria for their existence.

To start with, let us consider the following simple case as a motivational example:
\begin{itemize}
	\item[$\bullet$]The trivial line bundle $\sf{E}:=\underline{\bb{C}}_{\sf{M}}=\sf{M}\times\bb{C}$ with space of sections $C^{\infty}(\sf{M})\cong\Gamma^{\infty}(\underline{\bb{C}}_{\sf{M}})$, equipped with standard complex inner product $\langle\varphi,\psi\rangle_{\sf{E}_{p}}:=\varphi\cdot\psi$ on its fibres.
	\item[$\bullet$]A smooth kernel $k_{\sf{B}}\in C^{\infty}(\sf{M}\times\sf{M})$ with corresponding integral operator 
	\begin{align*}
		\sf{B}\psi(x):=\int_{\sf{M}}k_{\sf{B}}(x,y)\psi(y)\,\d\mu_{\sf{g}}(y)\,,\qquad\forall\psi\in C_{\mathrm{c}}^{\infty}(\sf{M})\, .
	\end{align*}
\end{itemize}
Now, if instead of integrating $k_{\sf{B}}(x,y)$ over \textit{spacetime} we just integrate over the \textit{spatial} variables, we obtain a family of linear and continuous operators $\sf{B}_{t,\tau}:C^{\infty}_{\mathrm{c}}(\Sigma_\tau)\to C^{\infty}(\Sigma_{t})$ defined by
\begin{align*}
	(\sf{B}_{t,\tau}\varphi)(\vec{x}):=\int_{\Sigma_{\tau}}k_{\sf{B}}(t,\vec{x};\tau,\vec{y})\varphi(\vec{y})\,\beta(\tau,\vec{y})\d\mu_{\sf{h}_{\tau}}(\vec{y})\,,\quad\forall\varphi\in C^{\infty}_{\mathrm{c}}(\Sigma_{\tau})\, .
\end{align*}
In other words, the operator $\sf{B}_{t,\tau}$ for $t,\tau\in\bb{R}$ is the integral operator with kernel 
\begin{align*}
	k_{t,\tau}:=(\mathrm{pr}_{2}^{\ast}\beta_{\tau})(i_{t}\times i_{\tau})^{\ast}k_{\sf{B}}\in C^{\infty}(\Sigma_{t}\times\Sigma_{\tau})\,,\end{align*} 
where $\beta_{\tau}(\cdot):=\beta(\tau,\cdot)$ and where $\mathrm{pr}_{2}\colon \Sigma_{t}\times\Sigma_{\tau}\to\Sigma_{\tau}$ denotes the projection onto the second factor. With this definition, it follows from Fubini's theorem as well as from the fact that $\d\mu_{\sf{g}}(t,\vec{x})=\beta(t,\vec{x})\d\mu_{\sf{h}_{t}}(\vec{x})\d t$ that the nonlocal potential $\sf{B}$ is related to $\sf{B}_{t,\tau}$ via
\begin{align*}
	(\sf{B}\psi)_{t}(\vec{x})=\int_{\bb{R}} (\sf{B}_{t,\tau}\psi_{\tau})(\vec{x})\,\d\tau\, ,
 \end{align*}		
for all $\psi\in C^{\infty}_{\mathrm{c}}(\sf{M})$, where we used the usual notation $\psi_{\tau}(\cdot):=\psi(\tau,\cdot)$. The two-parameter family of operators $(\sf{B}_{t,\tau})_{t,\tau\in\bb{R}}$ will be referred to as the \textit{time kernel} of the nonlocal potential $\sf{B}$.

Now, the aim of the following discussion is to generalise the notion of a \emph{time kernel} in a broader distributional context. To this end, we make the following definition.

\begin{definition}\label{Def:NonLocPot} (Nonlocal Potentials and Time Kernels)\newline
	Let $(\sf{M},\sf{g})$ be a globally hyperbolic manifold with Cauchy temporal function $t\in C^{\infty}(\sf{M})$ and corresponding foliation $(\Sigma_{t})_{t\in\bb{R}}$. A \emph{nonlocal potential} is a linear, continuous and semi-regular operator $\sf{B}\:\Gamma^{\infty}_{\mathrm{c}}(\sf{E})\to\Gamma^{\infty}(\sf{E})$ with Schwarz kernel $k_{\sf{B}}\in\mathcal{D}^{\prime}(\sf{E}\boxtimes\sf{E}^{\ast})$ such that there exists a two-parameter family of distributional kernels $k_{t,\tau}\in\mathcal{D}^{\prime}(\Sigma_{t}\times\Sigma_{\tau},\sf{E}\vert_{\Sigma_{t}}\boxtimes\sf{E}^{\ast}\vert_{\Sigma_{\tau}})$ with corresponding linear and continuous operators $\sf{B}_{t,\tau}\:\Gamma_{\mathrm{c}}^{\infty}(\Sigma_{\tau},\sf{E}\vert_{\Sigma_{\tau}})\to\Gamma^{\infty}(\Sigma_{t},\sf{E}\vert_{\Sigma_{t}})$, called the \emph{time kernel}, such that the following holds for all $\psi\in \Gamma^{\infty}_{\mathrm{c}}(\sf{E}^{\ast})$ and $\varphi\in \Gamma^{\infty}_{\mathrm{c}}(\sf{E})$:
	\begin{itemize}
		\item[(i)]$\bb{R}\times\bb{R}\ni (t,\tau)\mapsto \langle k_{t,\tau},\psi_{t}\boxtimes\varphi_{\tau}\rangle_{\Sigma_{t}\times\Sigma_{\tau}}=\langle \sf{B}_{t,\tau}\varphi_{\tau},\psi_{t}\rangle_{\Sigma_{t}}\in\bb{C}$ is locally integrable.
		\item[(ii)]Denoting by $\langle\cdot,\cdot\rangle_{\Sigma_{t}}$ the distributional pairing on $\Sigma_{t}$, then 
		\begin{align*}
			\langle(\sf{B}\varphi)_{t},\psi_{t}\rangle_{\Sigma_{t}}=\int_{\bb{R}}\langle \sf{B}_{t,\tau}\varphi_{\tau},\psi_{t}\rangle_{\Sigma_{t}}\,\d\tau\, .
\end{align*}
	\end{itemize}
\end{definition}

\begin{example}
	Generalising the motivational example at the beginning of this discussion, consider $k_{\sf{B}}\in \Gamma^{\infty}(\sf{M}\times\sf{M},\sf{E}\boxtimes\sf{E}^{\ast})$. Then, the time kernel can directly be defined by
		\begin{align*}
			\langle\sf{B}_{t,\tau}\psi_{\tau},\varphi_{t}\rangle_{\Sigma_{t}}:=\int_{\Sigma_{t}}\int_{\Sigma_{\tau}}\langle\varphi_{t}(\vec{x}),k_{\sf{B}}(t,\vec{x};\tau,\vec{y})\psi_{\tau}(\vec{y})\rangle_{\sf{E}_{(t,\vec{x})}}\beta(\tau,\vec{y})\d\mu_{\sf{h}_{\tau}}(\vec{y})\,\d\mu_{\sf{h}_{t}}(\vec{x})
		\end{align*}
		for all $\varphi,\psi\in \Gamma_{\mathrm{c}}^{\infty}(\sf{E})$, where we identified $\sf{E}$ with $\sf{E}^{\ast}$ using the bundle metric $\langle\cdot,\cdot\rangle_{\sf{E}}$, as usual. More generally, a similar definition can be made for every kernel that can be pointwise defined. For instance, for continuous kernels $k_{\sf{B}}\in \Gamma^{0}(\sf{M}\times\sf{M},\sf{E}\boxtimes\sf{E}^{\ast})$ (provided it is in addition semiregular, i.e.~smooth in the first variable, to fit in our framework).
\end{example}

The notion of a time kernel can, of course, be defined for more general distributional kernels. To discuss this point in greater detail, let us first consider a simpler situation as a \emph{toy model}. Let $u\in\mathcal{D}^{\prime}(\sf{M},\sf{E})$ be a distribution on $\sf{M}$. The guiding question in the following discussion is for which class of distribution the following can be achieved:

\begin{center}
	\textit{\underline{Question}: Under which conditions on $u\in\mathcal{D}^{\prime}(\sf{M},\sf{E})$ does there exist a one-parameter family $u_{t}\in\mathcal{D}^{\prime}(\Sigma_{t},\sf{E}\vert_{\Sigma_{t}})$, labelled by $t\in\bb{R}$, such that}
	\begin{align*}
		\langle u,\varphi\rangle_{\sf{M}}=\int_{-\infty}^{\infty}\langle u_{t},\varphi_{t}\rangle_{\Sigma_{t}}\,\d t
	\end{align*}
	\textit{for all test sections $\varphi\in \Gamma^{\infty}_{\mathrm{c}}(\sf{E}^{\ast})$, where we wrote $\varphi_{t}:=\varphi\vert_{\Sigma_{t}}$, as usual?}
\end{center}

Similarly as in our discussions of time kernels above, the answer to the previous question is clearly positive whenever $u$ is sufficiently regular. For instance, if we consider a regular distribution $u\in\Gamma^{\infty}(\sf{E})$, or more generally $u\in\Gamma^{0}(\sf{E})$, we can set
\begin{align}\label{eq:Pullback}
	u_{t}:=\beta_{t}(i_{t}^{\ast}u)\in\Gamma^{0}(\Sigma_ {t},\sf{E}\vert_{\Sigma_{t}})\, ,
\end{align}	
where $i_{t}\:\Sigma_{t}\to\sf{M}$ denotes the natural embedding and $i_{t}^{\ast}u:=u\circ i_{t}$ the pullback of a continuous function. Then, Fubini's theorem implies that
\begin{align}\label{eq:Fubini}
	\langle u,\varphi\rangle_{\sf{M}}=\int_{\bb{R}}\langle u_{t},\varphi_{t}\rangle_{\Sigma_{t}}\,\d t\, ,
\end{align}
which is the desired property. More generally, for $u\in\mathcal{D}^{\prime}(\sf{E})$, we can always find a sequence $\{u_{n}\}_{n\in\bb{N}}\subset \Gamma^{\infty}_{\mathrm{c}}(\sf{E})$ converging to $u$ in $\mathcal{D}^{\prime}(\sf{E})$, by density. Furthermore, since $\Gamma^{\infty}_{\mathrm{c}}(\sf{E})$ is a \textit{Montel space}, convergence in the strong topology is equivalent to convergence in the weak topology, which implies that
\begin{align*}
	\langle u_{n},\varphi\rangle_{\sf{M}}\xrightarrow{n\to\infty} \langle u,\varphi\rangle_{\sf{M}}\qquad\forall\varphi\in \Gamma^{\infty}_{\mathrm{c}}(\sf{E}^{\ast})\, .
\end{align*}
Now, suppose that we can arrange $(u_{n})_{n\in\mathbb{N}}$ in such way that also the sequence $(i^{\ast}_{t}u_{n})_{n\in\bb{N}}$ converges in $\mathcal{D}^{\prime}(\Sigma_{t},\sf{E}\vert_{\Sigma_{t}})$. In this case, we denote the corresponding limit by
\begin{align*}
	i_{t}^{\ast}u:=\lim_{n\to\infty}i_{t}^{\ast}u_{n}\qquad\text{in}\quad\mathcal{D}^{\prime}(\Sigma_{t},\sf{E}\vert_{\Sigma_{t}})\, .
\end{align*}
Now, it is clear that Equality~\eqref{eq:Fubini} will still hold in this more general situation whenever the right-hand side is well-defined. For instance, if for some arbitrary $u\in\mathcal{D}^{\prime}(\sf{E})$, the \emph{pullback} $i_{t}^{\ast}u$ can be defined in a suitable way, the map $t\mapsto \langle u_{t},\psi_{t}\rangle_{\Sigma_{t}}$ with $u_{t}:=\beta_{t}i_{t}^{\ast}u$ is locally integrable and it holds that
\begin{align*}
	\int_{\bb{R}}\langle \beta_{t}i^{\ast}_{t}u_{n},\psi_{t}\rangle_{\Sigma_{t}}\,\d t\xrightarrow{n\to\infty} \int_{\bb{R}}\langle u_{t},\psi_{t}\rangle_{\Sigma_{t}}\,\d t\, ,
\end{align*}
then, since $u_{n}$ is smooth for all $n\in\bb{N}$, the left-hand side is nothing else then 
\begin{align*}
\int_{\bb{R}}\langle \beta_{t} i^{\ast}_{t}u_{n},\psi_{t}\rangle_{\Sigma_{t}}\,\d t =\langle u_{n},\psi\rangle_{\sf{M}}\xrightarrow{n\to\infty} \langle u,\psi\rangle_{\sf{M}}\, ,
\end{align*}
which establishes the validity of Equation~\eqref{eq:Fubini} by uniqueness of limits. Hence, the answer to question imposed above is positive whenever the pull-back of $u$ via $i_{t}$ can be defined by means of an approximate sequence such that in addition $t\mapsto \langle u_{t},\psi_{t}\rangle_{\Sigma_{t}}$ is at least locally integrable. This also matches and motivates our definition in the case of bidistributional kernels, i.e.~Definition~\ref{Def:NonLocPot}.

Hence, we are essentially left to ask under which conditions on $u$, the pullbacks $i_{t}^{\ast}u$ are well-defined. An example of such a condition is provided by a certain condition on the \textit{wavefront set}. The wavefront set of a distribution $u\in\mathcal{D}^{\prime}(\sf{E})$ is a set $\mathrm{WF}(u)\subset\sf{T}^{\ast}\sf{M}\backslash\{\textbf{0}\}$, where $\textbf{0}$ denotes the zero section, that can be seen as a refinement of the singular support of a distribution in the sense that it also contains informations on the singular directions in Fourier space. We refer to Appendix~\ref{Appendix:Micro} for a discussion. Now, let $u \in \mathcal{D}^{\prime}(\sf{E})$ be a distribution on $\sf{M}$ satisfying the condition
\begin{align}\label{eq:WF}
   \bigg(\bigcup_{t\in\bb{R}}\sf{N}^{\ast}\Sigma_t \bigg)\cap \mathrm{WF}(u) = \emptyset \, ,
\end{align}
where $\sf{N}^{\ast}\Sigma_t \to \Sigma_t$ denotes the \textit{co-normal bundle} over $\Sigma_{t}$ in $\sf{M}$. If this condition is satisfied, then there is a unique way to define the pull-back $i_{t}^{\ast}u\in \mathcal{D}^{\prime}(\Sigma_t, \sf{E}\vert_{\Sigma_t})$ in such a way that 
\begin{align*}
	\mathrm{WF}(i^{\ast}_{t}u)\subset i_{t}^{\ast}\mathrm{WF}(u)\,,
\end{align*}	
which is a well-known theorem due to Hörmander, see e.g.~\cite[Theorem~8.2.4, Corollary 8.2.7]{HormanderI}. The way $i^{\ast}_{t}u$ is defined in the proof of this theorem is as follows: if $\Gamma\subset \sf{T}^{\ast}\sf{M}\backslash\{\textbf{0}\}$ is an arbitrary closed cone such that $\Gamma\cap \sf{N}^{\ast}\Sigma_{t}=\emptyset$, we consider the subspace 
\begin{align*}
	\mathcal{D}_{\Gamma}^{\prime}(\sf{E})=\{u\in\mathcal{D}^{\prime}(\sf{E})\mid \mathrm{WF}(u)\subset\Gamma\}
\end{align*}
equipped with the \textit{Hörmander topology}, see e.g.~\cite[Sec.~8.2]{HormanderI}.\footnote{The Hörmander topology is finer than the weak topology on $\mathcal{D}^{\prime}(\sf{E})$. In particular, if $u_{n}\xrightarrow{n\to\infty} u$ in the Hörmander topology, then also $\langle u_{n},\varphi\rangle_{\sf{M}}\to\langle u,\varphi\rangle_{\sf{M}}$ for all $\varphi\in \Gamma^{\infty}_{\mathrm{c}}(\sf{E}^{\ast})$.}  Now, if $(u_{n})_{n\in\bb{N}}$ is a sequence in $\Gamma^{\infty}_{\mathrm{c}}(\sf{E})$ converging to $u$ in $\mathcal{D}^{\prime}_{\Gamma}(\sf{E})$, then one can shown that $i_{t}^{\ast}u_{n}$ has a limit in the space $\mathcal{D}^{\prime}_{i_{t}^{\ast}\Gamma}(\sf{E})$, which provides a definition of the pull-back $i_{t}^{\ast}u$. Note that this is closely related to the general considerations above.

Now, if $u$ satisfies the condition~\eqref{eq:WF}, we can define the one-parameter family of distributions as in~\eqref{eq:Pullback} as
\begin{align}\label{eq:Family}
	u_{t}:=\beta_{t}(i_t^{\ast} u) \in \mathcal{D}^{\prime}(\Sigma_t, \sf{E}\vert_{\Sigma_t})\, .
\end{align}
In this case, Equation~\eqref{eq:Fubini} can be obtained as follows:

\begin{proposition}\label{Prop:Distr} 
	Let $u\in\mathcal{D}^{\prime}(\sf{E})$ be such that condition~\eqref{eq:WF} is satisfied and define $(u_{t})_{t\in\bb{R}}$ as above in~\eqref{eq:Family}. Then, $\bb{R}\ni t\mapsto \langle u_{t},\psi_{t}\rangle_{\Sigma_{t}}\in\bb{C}$ is continuous for all $\psi\in \Gamma^{\infty}_{\mathrm{c}}(\sf{E}^{\ast})$ and it holds that
	\begin{align*}
		\langle u,\psi\rangle_{\sf{M}}=\int_{\bb{R}}\langle u_{t},\psi_{t}\rangle_{\Sigma_{t}}\,\d t\, .
	\end{align*}
\end{proposition}

\begin{proof}
	As a first step, let us denote by $\delta_{t}\in\mathcal{D}(\sf{M})$ the $\delta$-distribution on $\Sigma_{t}\subset\sf{M}$, that is, the distribution defined by
	\begin{align*}
		\langle\delta_{t},\varphi\rangle_{\sf{M}}:=\int_{\Sigma_{t}}\varphi(t,\vec{x})\,\d\mu_{\sf{h}_{t}}(\vec{x})
	\end{align*}
	for all $\varphi\in C^{\infty}_{\mathrm{c}}(\sf{M})$. It follows from standard arguments that the wavefront set of $\delta_{t}$ is given by $\mathrm{WF}(\delta_{t})=\sf{N}^{\ast}\Sigma_{t}\backslash\{\textbf{0}\}=:\Gamma$. Now, let $\Gamma^{\prime}\subset \sf{T}^{\ast}\sf{M}\backslash\{\textbf{0}\}$ be a closed cone such that $\Gamma^{\prime}\cap \sf{N}^{\ast}\Sigma_{t}=\emptyset$ for all $t\in\bb{R}$. The \emph{multiplication theorem of Hörmander} (see~\cite[Theorem~8.2.10]{HormanderI}) implies that the point-wise multiplication of smooth function extends to a continuous bilinear map $\mathcal{D}^{\prime}_{\Gamma^{\prime}}(\sf{M})\times\mathcal{D}^{\prime}_{\Gamma}(\sf{M},\sf{E})\to\mathcal{D}^{\prime}(\sf{M},\sf{E})$. In particular, the product $\delta_{t} u$ is well-defined and it holds that
	\begin{align*}
		\langle u_{t},\psi_{t}\rangle_{\Sigma_{t}}=\langle \beta_{t}(\delta_{t} u),\psi\rangle_{\sf{M}}\, .
	\end{align*}
	for all $\psi\in \Gamma^{\infty}_{\mathrm{c}}(\sf{E}^{\ast})$, by uniqueness of the pullback. The map $t\mapsto \delta_{t}\in\mathcal{D}^{\prime}_{\Gamma}(\sf{M})$ can easily be seen to be continuous, which shows continuity of $t\mapsto \langle u_{t},\psi_{t}\rangle_{\Sigma_{t}}$. The fact that the integral of $\langle u_{t},\psi_{t}\rangle_{\Sigma_{t}}$ coincides with $\langle u,\psi\rangle_{\sf{M}}$ then follows from the same arguments as above.
\end{proof}

Hence, condition~\eqref{eq:WF} provides another class of distribution for which the answer to the question raised above is positive. However, we stress that \eqref{eq:WF} is a rather strong and it is clear that the pull-back can be defined for much more general distributions. For instance, whenever $u\in \Gamma^{0}(\sf{E})$, there is an obvious way to define $i_{t}^{\ast}u$, as discussed above, even though Condition~\eqref{eq:WF} is not necessarily satisfied. Furthermore, the pullback theorem of Hörmander can be generalised by replacing the wavefront set with the \emph{Sobolev wavefront set}.

\begin{remark} (A Measure-Theoretic Point of View)\newline
	The previous discussion is closely related to the \emph{disintegration of measures}. Let $u\in\mathcal{D}^{\prime}(\sf{M})$ be a distribution of order zero, i.e.~$u$ continuously extends to a linear form on $C^{0}_{\mathrm{c}}(\sf{M})$. By the \emph{theorem of Riesz-Markov}, there exists a complex Radon measure $\nu$ on the Borel $\sigma$-algebra of $\sf{M}$ such that 
	\begin{align*}
		\langle u,\varphi\rangle_{\sf{M}}=\int_{\sf{M}}\,\varphi\,\d\nu
	\end{align*}
	for all $\varphi\in C^{0}_{\mathrm{c}}(\sf{M})$, see e.g.~\cite[Thm.~6.19]{Rudin2}. As an example, the $\delta$-distribution centred at some point $p\in\sf{M}$, i.e.~the distribution defined by $\langle\delta_{p},\varphi\rangle_{\sf{M}}:=\varphi(p)$, is of order zero and induces the \emph{Dirac measure} defined by $\delta_{p}(\mathrm{S}):=\chi_{\mathrm{S}}(p)$ for all Borel sets $\mathrm{S}\in\mathcal{B}(\bb{R})$, where $\chi_{\mathrm{S}}$ denotes the characteristic function.
	
	 Now, by the \emph{disintegration theorem} (see e.g.~\cite[Chap.~45]{FremlinMeasures} for a detailed review on that subject), one can find a (non-unique) family of measures $\nu_{t}$ on $\Sigma_{t}$ such that
	\begin{align*}
		\langle u,\varphi\rangle_{\sf{M}}=\int_{\sf{M}}\,\varphi\,\d\nu=\int_{\bb{R}}\int_{\Sigma_{t}}\varphi(t,\vec{x})\,\d\nu_{t}(\vec{x})\,\d\pi(t)\, ,
	\end{align*}
	where $\pi:=(\mathrm{pr}_{\bb{R}})_{\ast}\nu$ denotes the complex measure on $\bb{R}$ obtained by the pushforward via the projection $\mathrm{pr}_{\bb{R}}\colon\sf{M}\to\mathbb{R},\,(t,\vec{x})\mapsto t$. Using the \emph{Lebesgue decomposition theorem}, we can uniquely write $\pi=\pi_{0}+\pi_{\mathrm{sing}}$, where $\pi_{0}$ is absolutely continuous and $\pi_{\mathrm{sing}}$ singular with respect to $\d t$, i.e.~$\pi_{0}\ll \d t$ and $\pi_{\mathrm{sing}}\perp \d t$. In particular, by the \emph{Radon-Nikodým theorem}, there is a unique $h\in \sf{L}^{1}(\bb{R},\d t)$, the \emph{Radon-Nikodým derivative} of $\pi$ with respect to $\d t$, such that
	\begin{align*}
		\pi(\mathrm{S})=\int_{\mathrm{S}} h\,\d t+\pi_{\mathrm{sing}}(\mathrm{S})
	\end{align*}
	for all Borel sets $\mathrm{S}\in\mathcal{B}(\bb{R})$, see e.g.~\cite[Thm.~6.10]{Rudin2}. Hence, in this case, a positive answer to the question raised above for some given (zeroth-order) distribution $u\in\mathcal{D}^{\prime}(\sf{M})$ is related to the requirement that the singular part $\pi_{\mathrm{sing}}$ vanishes.	
\end{remark}

After this general discussion, let us return to nonlocal potentials. Adapting the previous discussion to the setting of \emph{bi}distributions, we obtain the following two examples:

\begin{example}
	Let $\sf{B}\:\Gamma^{\infty}_{\mathrm{c}}(\sf{E})\to \Gamma^{\infty}(\sf{E})$ be linear, continuous and semiregular. If its Schwartz kernel $k_{\sf{B}}\in\mathcal{D}^{\prime}(\sf{M}\times\sf{M},\sf{E}\boxtimes\sf{E}^{\ast})$ satisfies the condition 
	\begin{align*}
   		\bigg( \bigcup_{t,\tau\in\bb{R}}\sf{N}^{\ast}(\Sigma_t\times \Sigma_{\tau})\bigg) \cap \mathrm{WF}(k_{\sf{B}}) = \emptyset \, ,
	\end{align*}
	then $\sf{B}$ is a nonlocal potential with time kernel defined by $k_{t,\tau}:=(\mathrm{pr}_{2}^{\ast}\beta_{\tau})(i_{t}\times i_{\tau})^{\ast}k_{\sf{B}}$ for $t,\tau\in\bb{R}$, where $\mathrm{pr}_{2}\colon \Sigma_{t}\times\Sigma_{\tau}\to\Sigma_{\tau}$ denotes the projection onto the second factor.
\end{example}

\begin{example} Following detailed section on pseudodifferential operators in Appendix~\ref{App:PSIDO}, we recall that a \emph{pseudodifferential operator} on $\sf{M}$ with values in $\sf{E}$ is a linear operator $\mathrm{Op}(b)\colon \Gamma^{\infty}_{\mathrm{c}}(\sf{E})\to\Gamma^{\infty}(\sf{E})$ whose Schwartz kernel is locally of the form
	\begin{align*}
		k_{b}(x,y)=\int_{\mathbb{R}^{n+1}}e^{i(x-y)k}b(x,y,\xi)\,\d^{k+1}\xi
	\end{align*}
	for some \emph{symbol} $b\in\mathcal{S}^{m}(\mathcal{U}\times\mathcal{U},\bb{C}^{\mathrm{N}\times\mathrm{N}})$ with \emph{order} $m\in\bb{R}$ and $\mathcal{U}\subset\sf{M}$ open, where the integral makes sense as an oscillatory integral, i.e.~with an appropriate regularisation. Any such operator is \emph{pseudolocal}, which means that $\mathrm{sing}\,\mathrm{supp}(k_{b})$ is contained in the diagonal. Furthermore, 
	\begin{align*}
		\mathrm{WF}(k_{b})=\mathrm{WF}(\mathrm{Op}(b))\cap \sf{N}^{\ast}\Delta\backslash\{\textbf{0}\}\, ,
	\end{align*}
	where $\Delta:=\{(x,x)\mid x\in\sf{M}\}$ denotes the diagonal and $\mathrm{WF}(\mathrm{Op}(b))$ the \emph{operator wavefront set}, which is locally defined as the subset of covectors $(x,\xi)\in\sf{T}^{\ast}\sf{M}\backslash\{\textbf{0}\}$ around which the symbol $b$ fails to be a smoothing symbol. Now, note that
	\begin{align*}
		\bigg(\bigcup_{t,\tau\in\bb{R}} \sf{N}^{\ast}(\Sigma_{t}\times\Sigma_{\tau})\bigg) \cap\sf{N}^{\ast}\Delta=\{((t,\vec{x};\xi_{t},0),(t,\vec{x};-\xi_{t},0))\mid (t,\vec{x})\in \sf{M}, \xi_{t}\in \sf{T}^{\ast}_{t}\bb{R}\}\, .
	\end{align*}
	Hence, if the operator wavefront set $\mathrm{WF}(\mathrm{Op}(b))$ does not intersect this set, or in other words, if the operator $\mathrm{Op}(b)$ is a smoothing operator in time, then $\mathrm{Op}(b)$ defines a nonlocal potential with time kernel. This is in particular includes the case for time-dependent pseudodifferential operators on space $\Sigma$ as a special case. 
\end{example}

\paragraph{Two Classes of nonlocal Potentials.} Let $\sf{B}\:\Gamma^{\infty}_{\mathrm{c}}(\sf{E})\to\Gamma^{\infty}(\sf{E})$ be a nonlocal potential with time kernel $(\sf{B}_{t,\tau})_{t,\tau\in\bb{R}}$ as in Definition~\ref{Def:NonLocPot}. In full generality, there is no hope of establishing well-posedness of the Cauchy problem for a symmetric hyperbolic system on $\sf{M}$ over $(\sf{E},\langle\cdot,\cdot\rangle_{\sf{E}})$ coupled to the nonlocal potential $\sf{B}$. The main obstruction, as one might expect, is the nonlocality in time. Indeed, the Cauchy problem for operators of the form $\partial_{t}-\sf{A}(t)$ on $\bb{R}^{1+n}$, where $t\mapsto \sf{A}(t)\in\Psi^{1}(\bb{R}^{n})$ is a time-dependent pseudodifferential operator of order one (see Appendix~\ref{App:PSIDO} for details on pseudodifferential calculus) is classical and can be treated using techniques closely analogous to those in the fully local setting; see, for example, \cite[Sec.~7.1]{Hintz} and \cite[Sec.~7.7]{TaylorII}. In that situation, where the nonlocality is purely spatial, the standard energy-estimate arguments still yield well-posedness, under suitable regularity assumptions on $\sf{A}(t)$. When a potential is also nonlocal in time, however, this fundamentally interferes with the energy methods and prevents a direct extension of these arguments.

Therefore, we need to impose some restrictions on the nonlocality in time. To this end, we consider two classes of nonlocal potentials, both of which are motivated by purely mathematical reasons, but also by application to mathematical physics.

\begin{definition}\label{Def:Potentials} 
	Let $\sf{B}\colon\Gamma^{\infty}_{\mathrm{c}}(\sf{E})\to \Gamma^{\infty}(\sf{E})$ be a nonlocal potential with time kernel.
	\begin{itemize}
		\item[(R)]$\sf{B}$ is called \emph{retarded}, if $\mathrm{supp}(\sf{B}\varphi)\subset \mathcal{J}^{+}(\mathrm{supp}(\varphi))$ for all $\varphi\in\Gamma^{\infty}_{\mathrm{c}}(\sf{E})$ and hence also $\sf{B}_{t,\tau}=0$ for $\tau>t$. If, in addition, there exists a $t_{0}\in\bb{R}$ such that $\sf{B}_{t,\tau}=0$ for $\tau<t_{0}$, we say that $\sf{B}$ is \emph{past compact with switch-on time $t_{0}$}.
		\item[(S)]$\sf{B}$ has \emph{short time range}, if there exists a $\delta>0$ such that $\sf{B}_{t,\tau}=0$ for $\vert t-\tau\vert>\delta$. 
	\end{itemize}
\end{definition}

Retarded nonlocal potentials are particularly well-suited for an initial mathematical investigation of the Cauchy problem due to their inherent causal structure and compatibility with energy methods. Moreover, many of the motivating examples presented at the beginning of this section fall into this class, for instance, the Maxwell equations in linear dispersive media and semiclassical models. Considering nonlocal potentials with a small time range is a natural next step in the mathematical analysis: in this regime one can examine the effects from allowing a controlled, localised influence from future points, restricted to a small temporal neighbourhood. From a physical point of view, nonlocal potentials of this type appear, for instance, in the aforementioned collapse models from causal fermion systems, in which one studies the Dirac equation coupled to a nonlocal potential with small time range.

\begin{figure}[H] 
	\centering
    \subfloat[Retarded Nonlocal Potentials.]{\includegraphics[width=0.4\textwidth]{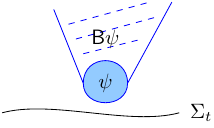}}\hspace*{2cm}
    \subfloat[Potentials with Short Time Range.]{\includegraphics[width=0.4\textwidth]{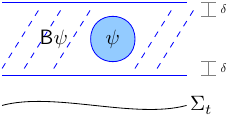}}
\caption{A schematic sketch of the support properties of the two classes of nonlocal potentials.\label{Fig:NonLocPot}} 
\end{figure}

\begin{remark} \label{Rem:MappingProp}
	The two types of nonlocal potentials introduced in Definition~\ref{Def:Potentials} have the following mapping properties, cf.~\cite[Lemma~2.13]{BaerGreen}:
	\begin{itemize}
		\item[(i)]If $\sf{B}$ is a retarded nonlocal potential, then it is easy to see that
		\begin{align*}
			\mathrm{supp}(k_{\sf{B}})\subset\{(x,y)\in\sf{M}\times\sf{M}\mid y\in \mathcal{J}^{-}(x)\}\, .
		\end{align*}
		In particular, $\sf{B}$ can be extended uniquely to an operator $\sf{B}\colon \Gamma^{\infty}_{\mathrm{pc}}(\sf{E})\to \Gamma^{\infty}_{\mathrm{pc}}(\sf{E})$.	
		\item[(ii)]If $\sf{B}$ has short time range $\delta>0$, then 
		\begin{align*}
			\mathrm{supp}(k_{\sf{B}})\subset\{(t,\vec{x};\tau,\vec{y})\in\sf{M}\times\sf{M}\mid \vert t-\tau\vert\leq \delta\}\, .
		\end{align*}	
		In particular, $\sf{B}$ can be extended uniquely to an operator $\sf{B}\colon \Gamma^{\infty}_{\mathrm{sc}}(\sf{E})\to \Gamma^{\infty}(\sf{E})$. Furthermore, $\sf{B}$ is well-defined as a linear operator of the form $\sf{B}\colon \Gamma^{\infty}((0,\mathrm{T})\times\Sigma,\sf{E})\to \Gamma^{\infty}((-\delta,\mathrm{T}+\delta)\times\Sigma,\sf{E})$ for all $\mathrm{T}>0$.
	\end{itemize}
	Figure~\ref{Fig:NonLocPot} represents the support properties of retarded nonlocal potentials and nonlocal potentials with small time range schematically.
\end{remark}

\begin{example}\label{Exam:RetUltraStat}
	Consider an ultrastatic globally hyperbolic manifold
	\begin{align*}
		\sf{M}=\bb{R}\times\Sigma\,,\qquad\sf{g}=-\d t\otimes\d t+\sf{h}\,.\end{align*} Furthermore, let $k\in\mathcal{D}(\Sigma\times\Sigma,\sf{E}\vert_{\Sigma}\times\sf{E}\vert_{\Sigma}^{\ast})$ be an arbitrary kernel on $(\Sigma,\sf{h})$ and consider a map $\chi\in C^{0}(\bb{R})$ with the property that $\chi(t)=0$ for $t<0$. Then,
	\begin{align*}
		k_{\sf{B}}(t,\vec{x};\tau,\vec{y}):=\chi(t-\tau)k(\vec{x},\vec{y})\in\mathcal{D}^{\prime}(\sf{M}\times\sf{M},\sf{E}\boxtimes\sf{E}^{\ast})
	\end{align*}
	is an example of a retarded nonlocal potential. Kernels of this form appear, for instance, in \emph{linear response theory}, see e.g.~the discussion of the Maxwell equations in linear dispersive media at the beginning of this section, cf.~Eq.~\eqref{eq:MaxwellNonLoc}.
\end{example}

\begin{example} Consider $(1+n)$-dimensional Minkowski spacetime $\sf{M}:=\bb{R}^{1+n}$ and a linear evolution equation of the form 
	\begin{align*}
		\sf{S}:=\partial_{t}-\sf{P}(t)\colon C^{\infty}(\sf{M})\to C^{\infty}(\sf{M})\, ,
	\end{align*}
	where $\sf{P}(t)$ is one-parameter family of linear and elliptic first-order differential operators on $\Sigma_{t}\cong \bb{R}^{n}$. We denote by 
	\begin{align*}
		\mathcal{U}(t,\tau)\colon C^{\infty}_{\mathrm{c}}(\Sigma_{\tau})\to C^{\infty}_{\mathrm{c}}(\Sigma_{t})\, ,\qquad \mathfrak{f}\mapsto \psi\vert_{\Sigma_{t}}
	\end{align*}
	the corresponding \emph{unitary evolution operator}, where $\psi\in C^{\infty}_{\mathrm{sc}}(\Sigma)$ is the unique solution to $\sf{S}\psi=0$ with initial datum $\psi\vert_{\Sigma_{\tau}}=\mathfrak{f}$. Then, its retarded Green operator $\sf{G}^{+}\colon C^{\infty}_{\mathrm{c}}(\sf{M})\to C^{\infty}_{\mathrm{sc}}(\sf{M})$ is a retarded nonlocal potential in the sense of Definition~\ref{Def:Potentials} with time kernel
	\begin{align*}
		(\sf{G}^{+})_{t,\tau}:=\theta(t-\tau)\mathcal{U}(t,\tau)\, ,
	\end{align*}
	where $\theta$ denotes the Heaviside function. Indeed, noting that the Cauchy evolution operator satisfies the operator equations	
	\begin{align*}
		\begin{cases}
		\partial_{t}\mathcal{U}(t,\tau)&=\sf{P}(t)\mathcal{U}(t,\tau)\\
		\partial_{\tau}\mathcal{U}(t,\tau)&=-\mathcal{U}(t,\tau)\sf{P}(\tau)
		\end{cases}\,,
	\end{align*}
	one immediately verifies the defining properties of $\sf{G}^{+}$ (see~Definition~\ref{Def:GreenHyp}). For instance, for all $\psi\in C^{\infty}_{\mathrm{c}}(\sf{M})$, it holds that 
	\begin{align*}
		\sf{S}(\sf{G}^{+}\psi)(t,\vec{x})&=(\partial_{t}-\sf{P}(t))\int_{-\infty}^{t}(\mathcal{U}(t,\tau)\psi_{\tau})(\vec{x})\,\d\tau=\\&=(\underbrace{\mathcal{U}(t,t)}_{=\mathrm{id}}\psi_{t})(\vec{x})+\int_{-\infty}^{t}\underbrace{(\partial_{t}-\sf{P}(t))(\mathcal{U}(t,\tau)\psi_{\tau})(\vec{x})}_{=0}\,\d\tau=\psi(t,\vec{x})\, ,
	\end{align*}
	where we used the \emph{Leibniz integral rule}. With a similar computation, one shows that $\sf{G}^{+}(\sf{S}\psi)=\psi$. Last but not least, by definition, it is clear that $\mathrm{supp}(\sf{G}^{+}\psi)\subset\mathcal{J}^{+}(\mathrm{supp}(\psi))$, which proves the claim by uniqueness of the retarded and advances Green operators, cf.~Proposition~\ref{Prop:GreenProp}(i).
	
	We remark that examples of hyperbolic equations with nonlocal operators constructed out of the retarded Green's operator are relevant for semiclassical equations, see Eq.~\eqref{eq:NonLocSemi}.
\end{example}

\subsection{The Cauchy Problem of Nonlocal Symmetric Hyperbolic Systems}
Having discussed a suitable notion of nonlocal potentials, let us now turn to the analysis of the Cauchy problem. Throughout this section, we consider the following set-up:

\begin{setup}\label{SetUp:Nonloc} We consider the following set-up:
\begin{itemize}
	\item[(i)]A globally hyperbolic manifold $(\sf{M},\sf{g})$ with fixed Cauchy temporal function, i.e.
	  \begin{align*}
	\sf{M}=\bb{R}\times\Sigma\,,\qquad \sf{g}=-\beta^{2}\d t\otimes\d t+\sf{h}_{t}\,,
\end{align*}
with corresponding foliation by spacelike Cauchy surfaces $(\Sigma_{t}:=\{t\}\times\Sigma)_{t\in\bb{R}}$.
\item[(ii)]A Hermitian vector bundle $(\sf{E},\langle\cdot,\cdot\rangle_{\sf{E}})$ over $\sf{M}$ with compatible connection $\nabla^{\sf{E}}$.
\item[(iii)]A symmetric hyperbolic system $\sf{S}=\sigma_{\sf{S}}(\d x^{\mu})\nabla^{\sf{E}}_{\partial_{\mu}}+\sf{S}_{0}\in\mathrm{DO}^{1}(\sf{E})$ on $\sf{M}$.
\item[(iv)]A nonlocal potential $\sf{B}\:\Gamma_{\mathrm{c}}^{\infty}(\sf{E})\to\Gamma^{\infty}(\sf{E})$ with time kernel $(\sf{B}_{t,\tau})_{t,\tau\in\bb{R}}$.
\end{itemize}
\end{setup}

Now, if $(\sf{M},\sf{g})$ and $\sf{S},\sf{B}$ are as in Set-up~\ref{SetUp:Nonloc}, the goal is to study the Cauchy problem
\begin{align}\label{eq:CauchyNonloc}
	\begin{cases}(\sf{S}-\sf{B})\psi&=\phi\\
	\psi\vert_{t=0}&=\mathfrak{f}
	\end{cases}
\end{align}
for a given source $\phi\in\Gamma_{\mathrm{c}}^{\infty}(\sf{M}_{\mathrm{T}},\sf{E})$ and initial datum $\mathfrak{f}\in\Gamma_{\mathrm{c}}^{\infty}(\Sigma_{0},\sf{E}\vert_{\Sigma_{0}})$. In particular, we shall prove that for retarded nonlocal potentials and nonlocal potentials with short time range the Cauchy problem~\eqref{eq:CauchyNonloc} admits strong solutions (as in \cite{Friedrichs1}) on a time strip $\sf{M}_{\mathrm{T}}=:t^{-1}([0,\mathrm{T}])$ for $\mathrm{T}>0$, clearly under suitable assumptions. 

As in Section~\ref{Sec:SHS}, we set $\eta:=\beta\d t$ and consider the Hilbert spaces $\sf{L}^{2}(\Sigma_{t},\sf{E}\vert_{\Sigma_{t}})$ and $\sf{L}^{2}(\sf{M}_{\mathrm{T}},\sf{E})$. To keep the notation light, we simply write 
\begin{align*}
	\langle\cdot,\cdot\rangle_{t}:=\langle\cdot,\cdot\rangle_{\sf{L}^{2}(\Sigma_{t})}=\int_{\Sigma_{t}}\langle\sigma_{\sf{S}}(\eta)\cdot,\cdot\rangle_{\sf{E},\beta}\,\d\mu_{\sf{h}_{t}}\, ,\qquad \langle\cdot,\cdot\rangle_{\sf{M}_{\mathrm{T}}}:=\langle\cdot,\cdot\rangle_{\sf{L}^{2}(\sf{M}_{\mathrm{T}})}=\int_{\bb{R}}\langle\sigma_{\sf{S}}(\eta)\cdot,\cdot\rangle_{t}\,\d t\, .
\end{align*}
In the context of nonlocal symmetric hyperbolic systems, strong solutions are defined as follows.
			
\begin{definition} (Strong Solutions)
	\begin{itemize}
		\item[(i)]Let $\sf{B}$ be a retarded and past-compact nonlocal potential with switch-on time $-t_{0}\geq 0$. In addition, assume that $\sf{B}$ is uniformly bounded in time on $[-t_{0},\mathrm{T}]$, i.e.~there is a constant $C_{\mathrm{T}}>0$ such that $\Vert\sf{B}_{t,\tau}\varphi\Vert_{t}\leq C_{\mathrm{T}}\Vert\varphi\Vert_{\tau}$ for all $\varphi\in\Gamma^{\infty}_{\mathrm{c}}(\sf{E}\vert_{\Sigma_{\tau}})$ and $t,\tau\in [-t_{0},\mathrm{T}]$. Then, a \emph{strong solution on $\sf{M}_{\mathrm{T}}$} is an element $\psi\in\sf{L}^{2}([-t_{0},\mathrm{T}]\times\Sigma,\sf{E})$ for which there exists a sequence $\psi_{k}\in \Gamma^{\infty}([-t_{0},\mathrm{T}]\times\Sigma,\sf{E})\cap \sf{L}^{2}([-t_{0},\mathrm{T}]\times\Sigma,\sf{E})$ such that
		\begin{align*}
			\psi_{k}\vert_{\Sigma_{0}}\xrightarrow{k\to\infty}\mathfrak{f}&\quad\text{in}\quad\sf{L}^{2}(\Sigma_{0},\sf{E}\vert_{\Sigma_{0}})\,,\qquad\qquad\psi_{k}\xrightarrow{k\to\infty}\psi\quad\text{in}\quad \sf{L}^{2}([-t_{0},\mathrm{T}]\times\Sigma,\sf{E})\,,\\ &\quad(\sf{S}-\sf{B})\psi_{k}\xrightarrow{k\to\infty} \phi\quad\text{in}\quad \sf{L}^{2}(\sf{M}_{\mathrm{T}},\sf{E})\, .
		\end{align*}
		\item[(ii)]Let $\sf{B}$ be a nonlocal potential with small time range $\delta>0$. In addition, assume that $\sf{B}$ is uniformly bounded in time, i.e.~there is a constant $C_{\mathrm{T}}>0$ such that $\Vert\sf{B}_{t,\tau}\varphi\Vert_{t}\leq C_{\mathrm{T}}\Vert\varphi\Vert_{\tau}$ for all $\varphi\in\Gamma^{\infty}_{\mathrm{c}}(\sf{E}\vert_{\Sigma_{\tau}})$ and $t,\tau\in [-\delta,\mathrm{T}+\delta]$. Then, a \emph{strong solution on $\sf{M}_{\mathrm{T}}$} is an element $\psi\in\sf{L}^{2}([-\delta,\mathrm{T}+\delta]\times\Sigma,\sf{E})$ for which there exists a sequence $\psi_{k}\in \Gamma^{\infty}([-\delta,\mathrm{T}+\delta]\times\Sigma,\sf{E})\cap \sf{L}^{2}([-\delta,\mathrm{T}+\delta]\times\Sigma,\sf{E})$ such that
		\begin{align*}
			\psi_{k}\vert_{\Sigma_{0}}\xrightarrow{k\to\infty}\mathfrak{f}&\quad\text{in}\quad\sf{L}^{2}(\Sigma_{0},\sf{E}\vert_{\Sigma_{0}})\,,\qquad\qquad\psi_{k}\xrightarrow{k\to\infty}\psi\quad\text{in}\quad \sf{L}^{2}([-\delta,\mathrm{T}+\delta]\times\Sigma,\sf{E})\,,\\ &\quad(\sf{S}-\sf{B})\psi_{k}\xrightarrow{k\to\infty} \phi\quad\text{in}\quad \sf{L}^{2}(\sf{M}_{\mathrm{T}},\sf{E})\, .
		\end{align*}
	\end{itemize}
\end{definition}

Note that, it is important in the above definition that the approximating sequence $\psi_{k}$ is defined not just on the time strip $\sf{M}_{\mathrm{T}}=[0,\mathrm{T}]\times\Sigma$, but on a bigger domain. For example, in the retarded case, if we want to know $\sf{B}\psi_{k}$ on $\sf{M}_{\mathrm{T}}$, we need to specify $\psi_{k}$ on $[-t_{0},\mathrm{T}]\times\Sigma$ due to the nonlocal character of the potential. Similar considerations hold true in the case of nonlocal potentials with short time range. Moreover, we note that the boundedness assumption in time ensures that $\sf{B}$ is well-defined on smooth sections that are $\sf{L}^{2}$ on each time slice.  Let us also remark that in principle, one could also define notion of \emph{weak solutions}. Also in this case one needs to use different time strips in the definition to account for the nonlocality. Last but not least, we define \emph{classical} solutions as follows.

\begin{definition} (Classical Solutions)
	\begin{itemize}
		\item[(i)]Let $\sf{B}$ be a retarded and past-compact nonlocal potential with switch-on time $-t_{0}\geq 0$. A \emph{classical solution on $\sf{M}_{\mathrm{T}}$} is a section $\psi \in \Gamma^{\infty}([-t_{0},\mathrm{T}] \times \Sigma,\sf{E})$ such that
		\begin{align*}
			(\sf{S}-\sf{B})\psi = \phi \quad \text{pointwise for all } t \in [0,\mathrm{T}]\,, \quad \text{and} \quad \psi\vert_{\Sigma_{0}} = \mathfrak{f}\,.
		\end{align*}
		\item[(ii)]Let $\sf{B}$ be a nonlocal potential with small time range $\delta>0$. A \emph{classical solution on $\sf{M}_{\mathrm{T}}$} is a section $\psi \in \Gamma^{\infty}([-\delta,\mathrm{T}+\delta] \times \Sigma,\sf{E})$ such that
		\begin{align*}
			(\sf{S}-\sf{B})\psi = \phi \quad \text{pointwise for all } t \in [0,\mathrm{T}]\,, \quad \text{and} \quad \psi\vert_{\Sigma_{0}} = \mathfrak{f}\,.
		\end{align*}
	\end{itemize}
\end{definition}

Clearly, if a strong solution is in addition smooth, then it is also a classical solution. Indeed, any strong solution is a weak solution and if it is also smooth then it is classical.

Now, before we are able to prove the main theorems of this section, we need one last ingredient. In Lemma~\ref{Lemma:BiggerSystem}, we have shown that for every symmetric hyperbolic system $\sf{S}$ on $(\sf{E},\langle\cdot,\cdot\rangle_{\sf{E}})$, there exists an extended system $\mathfrak{S}$ on $(\mathcal{E},\langle\cdot,\cdot\rangle_{\mathcal{E}})$, where $\mathcal{E}:=\sf{E}\oplus (\sf{E}\otimes\sf{T}^{\ast}\sf{M})$ and $where \langle\cdot,\cdot\rangle_{\mathcal{E}}$ denotes bundle metric induced by $\langle\cdot,\cdot\rangle_{\sf{E}}$ and $\sf{g}_{0}=\d t\otimes\d t+\sf{h}_{t}$, such that
\begin{align*}
	\sf{S}\psi=\phi\qquad\Leftrightarrow\qquad\mathfrak{S}\Theta_{\psi}=\Theta_{\phi}
\end{align*}
for all $\psi,\phi\in\Gamma^{\infty}(\sf{E})$ with $\Theta_{\psi}:=(\psi,\nabla^{\sf{E}}\psi)$. Now, in the presence of a nonlocal potential, we obtain the following extension of this result:

\begin{lemma}\label{Lemma:BiggerSystem2} \emph{(Following \cite[Prop.~4.11]{SchmidMurroFinster})}\newline
Let $\sf{S}$ be a symmetric hyperbolic system over $(\sf{E},\langle\cdot,\cdot\rangle_{\sf{E}})$ and $\mathfrak{S}$ be the extended system over $(\mathcal{E},\langle\cdot,\cdot\rangle_{\mathcal{E}})$ as per Lemma~\ref{Lemma:BiggerSystem}. Define
\begin{align*}
	\mathfrak{B}\:\Gamma^{\infty}_{\mathrm{c}}(\mathcal{E})\to\Gamma^{\infty}(\mathcal{E})\,,\qquad\mathfrak{B}:=
	\begin{pmatrix}
		\sf{B} & 0 \\
		\nabla^{\sf{E}}\circ\sf{B}-(\sf{B}\otimes\mathrm{id})\circ\nabla^{\sf{E}} & \sf{B}\otimes\mathrm{id}
	\end{pmatrix}
\end{align*}
Then, $(\sf{S}-\sf{B})\psi=\phi$ if and only if $(\mathfrak{S}-\mathfrak{B})\Theta_{\psi}=\Theta_{\phi}$ for all $\psi,\phi\in\Gamma^{\infty}_{\mathrm{c}}(\sf{E})$.
\end{lemma}

\begin{proof}
	This follows from Lemma~\ref{Lemma:BiggerSystem} and the definition of $\mathfrak{B}$.
\end{proof}

Now, if $\sf{B}$ is a nonlocal potential on $\sf{E}$, we set $\mathfrak{B}_{0}:=\sf{B}$ and $\mathfrak{B}_{1}:=\mathfrak{B}$. Moreover, we define inductively nonlocal potentials $\mathfrak{B}_{k}$ for $k\in\bb{N}$ on $\mathcal{E}_{k}:=\mathcal{E}_{k-1}\oplus (\mathcal{E}_{k-1}\otimes\sf{T}^{\ast}\sf{M})$ with $\mathcal{E}_{0}:=\sf{E}$ by means of Lemma~\ref{Lemma:BiggerSystem2}.

\begin{remark}
	Note that for arbitrary nonlocal potentials $\sf{B}$ with time kernel in the sense of Definition~\ref{Def:NonLocPot}, it is in general not the case that $\mathfrak{B}$ is again a nonlocal potential with time kernel in the sense of Definition~\ref{Def:NonLocPot}, since its kernel might not be locally integrable in time and hence not posses a time kernel according to our definition. In the end, requiring that also $\mathfrak{B}$ admits a time kernel can be understood as an assumption on the regularity of the kernel of $\sf{B}$.
\end{remark}

Having introduced the relevant definitions and terminology, we are now in the position to state the central results of this section. To start with, for a given symmetric hyperbolic system $\sf{S}$, recall that $\sf{S}+\sf{S}^{\ast}$, where $\sf{S}^{\ast}$ denotes the formal adjoint with respect to $(\cdot,\cdot)_{\sf{E}}=\int_{\sf{M}}\langle\cdot,\cdot\rangle_{\sf{E}}\,\d\mu_{\sf{g}}$, is an operator of order zero that is build out of the zeroth-order terms of $\sf{S}$ and the (covariant) derivatives of the principal symbol $\sigma_{\sf{S}}$. With this in mind, we introduce the zeroth-order operator 
\begin{align*}
	\sf{Z}_{\sf{S}}:=\beta\sigma_{\sf{S}}(\eta)^{-1}(\sf{S}+\sf{S}^{\ast})\,,
\end{align*}
where $\eta:=\beta\d t$ denotes the future-directed timelike normal covector of the foliation $(\Sigma_{t})_{t\in\bb{R}}$. Moreover, for a given nonlocal potential $\sf{B}$ with time kernel, we write
\begin{align*}
	\sf{V}:=\beta\sigma_{\sf{S}}(\eta)^{-1}\sf{B}\,,
\end{align*}
which is again a nonlocal potential with time kernel given by $\sf{V}_{t,\tau}=\beta_{t}\sigma_{\sf{S}}(\eta_{t})^{-1}\sf{B}_{t,\tau}$.

The first theorem concerns the well-posedness of the Cauchy problem~\ref{eq:CauchyNonloc} for \emph{retarded} nonlocal potentials.

\begin{theorem}\label{Thm:Ret} \emph{(Cauchy Problem for Retarded Nonlocalities)}\newline
	Let $(\sf{M},\sf{g})$, $(\sf{E},\langle\cdot,\cdot\rangle_{\sf{E}},\nabla^{\sf{E}})$, $\sf{S}$ and $\sf{B}$ be as in~Set-up~\ref{SetUp:Nonloc}. Furthermore, consider a time strip $\sf{M}_{\mathrm{T}}:=t^{-1}([0,\mathrm{T}])$ with $\mathrm{T}>0$, Cauchy data $(\phi,\mathfrak{f})\in\Gamma^{\infty}(\sf{M}_{\mathrm{T}},\sf{E})\times\Gamma^{\infty}(\Sigma_{0},\sf{E}\vert_{\Sigma_{0}})$ and assume the following:
	\begin{itemize}
		\item[\emph{(i)}]$\sf{B}$ is a \emph{retarded} nonlocal potential, cf.~Definition~\ref{Def:Potentials}(R).
		\item[\emph{(ii)}]$\sf{B}$ is \emph{past compact} with switch-on time $t_{0}=0$.
		\item[\emph{(iii)}]$\mathsf{V}:=\beta\sigma_{\sf{S}}(\eta)^{-1}\sf{B}$ is uniformly bounded in the time strip, i.e.
\begin{align*}
	\exists C_{\mathrm{T}}>0:\qquad \Vert\sf{V}_{t,\tau}\varphi\Vert_{t}\leq C_{\mathrm{T}}\Vert\varphi\Vert_{\tau}\qquad\qquad \forall \varphi\in \Gamma^{\infty}_{\mathrm{c}}(\Sigma_{\tau},\sf{E}\vert_{\Sigma_{\tau}}),\,t,\tau\in [0,\mathrm{T}]
\end{align*}		
		\end{itemize}
		Then, there exists a strong solution of the Cauchy problem~\ref{eq:CauchyNonloc} on $\sf{M}_{\mathrm{T}}$. If additionally
		\begin{itemize}
			\item[\emph{(iv)}]the extended nonlocal potentials $\mathfrak{B}_{k}$ admit time kernels and satisfy assumptions (i)-(iii) above for $k\in\bb{N}$,
		\end{itemize}
		then the solution is smooth, unique and propagates at most with the speed of light, i.e.~
		\begin{align*}
			\mathrm{supp}(\psi)\cap\mathcal{J}^{+}(\Sigma_{0})\subset\mathcal{J}^{+}(\mathrm{supp}(\phi)\cup\mathrm{supp}(\mathfrak{f}))\,.
		\end{align*}
\end{theorem}

A proof of this theorem will follow below. The second theorem concerns the well-posedness of the Cauchy problem~\ref{eq:CauchyNonloc} for nonlocal potentials with a \emph{short time range}.

\begin{theorem}\label{Thm:Small} \emph{(Cauchy Problem for Nonlocal Potentials with Short Time Range)}\newline
	Let $(\sf{M},\sf{g})$, $(\sf{E},\langle\cdot,\cdot\rangle_{\sf{E}},\nabla^{\sf{E}})$, $\sf{S}$ and $\sf{B}$ be as in~Set-up~\ref{SetUp:Nonloc}. Furthermore, consider a time strip $\sf{M}_{\mathrm{T}}:=t^{-1}([0,\mathrm{T}])$ with $\mathrm{T}>0$, Cauchy data $(\phi,\mathfrak{f})\in\Gamma^{\infty}(\sf{M}_{\mathrm{T}},\sf{E})\times\Gamma^{\infty}(\Sigma_{0},\sf{E}\vert_{\Sigma_{0}})$ and assume the following:
	\begin{itemize}
		\item[\emph{(i)}]$\sf{B}$ has \emph{short time range} $\delta>0$, cf.~Definition~\ref{Def:Potentials}(S).
		\item[\emph{(ii)}]The zeroth-order operator $\sf{Z}_{\sf{S}}:=\beta\sigma_{\sf{S}}(\eta)^{-1}(\sf{S}+\sf{S}^{\ast})$ is uniformly bounded in time, i.e.
		\begin{align*}
	\exists D>0:\qquad \Vert\sf{Z}_{\sf{S}}(t,\cdot)\varphi\Vert_{t}\leq D\Vert\varphi\Vert_{\tau}\qquad\qquad \forall \varphi\in \Gamma^{\infty}_{\mathrm{c}}(\Sigma_{t},\sf{E}\vert_{\Sigma_{t}}),\,t\in\bb{R}
\end{align*}	
		\item[\emph{(iii)}]$\mathsf{V}:=\beta\sigma_{\sf{S}}(\eta)^{-1}\sf{B}$ is uniformly bounded in time with bound small compared to $\delta$ and $\sf{D}$ in the sense that
		\begin{align*}
	\exists 0<C<(8e\delta^{2})^{-1}:\qquad \Vert\sf{V}_{t,\tau}\varphi\Vert_{t}\leq Ce^{-\frac{1}{2}D\vert\tau\vert}\Vert\varphi\Vert_{\tau}\qquad \forall \varphi\in \Gamma^{\infty}_{\mathrm{c}}(\Sigma_{\tau},\sf{E}\vert_{\Sigma_{\tau}}),\,t,\tau\in\bb{R}
\end{align*}	
		\end{itemize}
		Then, there exists a strong solution of the Cauchy problem~\ref{eq:CauchyNonloc} on $\sf{M}_{\mathrm{T}}$.
\end{theorem}

Let us highlight the main differences in the assumptions of Theorem~\ref{Thm:Ret} and Theorem~\ref{Thm:Small}. First of all, note that in the retarded case we only assume uniform boundedness on the time interval $[0,T]$, whereas in the short-time-range case we require uniform boundedness on the entire time domain. The reason for this distinction is as follows: the idea of the proof is to construct local sections $\psi^{(n)}\in\Gamma^{\infty}(\sf{E})$, labelled by $n\in\bb{N}_{0}$, which are inductively defined as the solutions of the local problems
\begin{align*}
	\begin{cases}
		\sf{S}\psi^{(0)}&=\phi\\
		\psi^{(0)}\vert_{t=0}&=\mathfrak{f}
	\end{cases}\,,\qquad \begin{cases}
		\sf{S}\psi^{(n+1)}&=\sf{B}\psi^{(n)}\\
		\psi^{(n+1)}\vert_{t=0}&=0
	\end{cases}\qquad\forall n\in\bb{N}\, .
\end{align*}
The goal is then to prove that the series $\psi:=\sum_{n=0}^{\infty}\psi^{(n)}$ converges (absolutely) in a suitable Hilbert space and defines a (strong) solution to the Cauchy problem~\ref{eq:CauchyNonloc}. In the case of Theorem~\ref{Thm:Ret}, the assumptions on the nonlocal potential ensure that the source term for each of these equations remains within the future of $\Sigma_{0}$. In other words, due to the retarded property, if we want to solve the Cauchy problem on the time strip $\sf{M}_{\mathrm{T}}$, it is sufficient to study each of the local Cauchy problems on the same time strip $\sf{M}_{\mathrm{T}}$. In contrast, in the short-time-range case, i.e.~Theorem~\ref{Thm:Small}, the support of the source increases at each inductive step by $\pm\delta$ in time. Consequently, we must have full control over the solutions $\psi^{(n)}$ at all times. For this reason, also an additional uniform boundedness assumption on the zeroth-order terms is required that in addition has to be compatible with the bounds on the nonlocal potential. The bound $C<(8e\delta^{2})^{-1}$, which will become clear in the proof, is needed in the short-time-range case to guarantee convergence of the series.

Before we enter into the details of the proof, let us observe that at least the scaling $C\lesssim\delta^{-2}$ is actually optimal, i.e.~there exist examples of symmetric hyperbolic systems and uniformly bounded nonlocal potentials with short time range violating this bound, for which the Cauchy problem does not for every source admit a solution.

\begin{example} (A simple Counterexample)\newline
	We consider $(1+1)$-dimensional Minkowski spacetime $\sf{M}:=\bb{R}_{t}\times\bb{R}_{x}$ and the trivial line bundle $\underline{\bb{C}}_{\sf{M}}$ equipped with the standard complex inner product on its fibres. As a symmetric hyperbolic system, we consider the simple operator
	\begin{align*}
		\sf{S}\colon C^{\infty}(\sf{M})\to C^{\infty}(\sf{M})\,,\qquad \sf{S}:=\partial_{t}\, .
	\end{align*}
	Moreover, let us choose a real-valued function $f\in C^{\infty}_{\mathrm{c}}(\sf{M})$ such that $\Vert f\Vert_{\sf{L}^{2}(\sf{M})}=1$. As a nonlocal potential, we choose the integral operator $\sf{B}:C^{\infty}_{\mathrm{c}}(\sf{M})\to C^{\infty}(\sf{M})$ defined by
	\begin{align*}
		(\sf{B}\psi)(t,x):=-f(t,x)\int_{\bb{R}}\,\d\tau\,\int_{\bb{R}}\,\d y\,\dot{f}(\tau,y)\psi(\tau,y)\,,\qquad\forall \psi\in C^{\infty}_{\mathrm{c}}(\sf{M})\, ,
	\end{align*}
	or in \emph{bra/ket notation}, $\sf{B}=|f\rangle\langle\sf{S}^\dagger f|=-|f\rangle\langle \dot{f}|$, where $\langle\cdot|\cdot\rangle:=\langle\cdot,\cdot\rangle_{\sf{L}^{2}(\sf{M})}$ denotes the inner product induced by the bundle metric, as usual, which in this case coincides with the $\sf{L}^{2}$-inner product on $\sf{M}$. The Schwartz kernel $k_{\sf{B}}\in C^{\infty}(\sf{M}\times\sf{M})$ of $\sf{B}$ is regular and given by $k_{\sf{B}}(t,x;\tau,y)=-f(t,x)\dot{f}(\tau,y)\in C^{\infty}(\sf{M}\times\sf{M})$ and the corresponding time kernel $\sf{B}_{t,\tau}\:C^{\infty}_{\mathrm{c}}(\bb{R}_{x})\to C^{\infty}(\bb{R}_{x})$ can be written as
	\begin{align*}
		\sf{B}_{t,\tau}\psi_{\tau}(x)=-f(t,x)\int_{\bb{R}}\,\d y\,\dot{f}(\tau,y)\psi(\tau,y)\,,\qquad\forall \psi\in C^{\infty}_{\mathrm{c}}(\sf{M})\, .
	\end{align*}
	Now, in order for $\sf{B}$ to satisfy the smallness condition, we choose $f$ such that $\mathrm{supp}(f)\subset (0,\delta)\times\sf{K}$, where $\sf{K}\subset\bb{R}_{x}$ is some compact subset. In this way, it trivially holds that $\sf{B}_{t,\tau}=0$ for $\vert t-\tau\vert >\delta$, since the kernel $k_{\sf{B}}$ is supported in $((0,\delta)\times \sf{K})\times ((0,\delta)\times \sf{K})$. 
	
	Next, let us show that $\sf{B}_{t,\tau}$ is uniformly bounded in time, i.e.~that there exists a constant $C>0$ such that $\Vert\sf{B}_{t,\tau}\psi_{\tau}\Vert_{t}\leq C\Vert\psi_{\tau}\Vert_{\tau}$ for all $\psi\in C^{\infty}_{\mathrm{c}}(\sf{M})$ and all $t,\tau\in\bb{R}$. For this, let $\psi\in C^{\infty}_{\mathrm{c}}(\sf{M})$ be arbitrary. Then
	\begin{align*}
		\Vert\sf{B}_{t,\tau}\psi_{\tau}\Vert_{\tau}^{2}&=\int_{\bb{R}}\,\d x\,\vert \sf{B}_{t,\tau}\psi_{\tau}(x)\vert^{2}=\int_{\bb{R}}\,\d x\,\bigg\vert f(t,x)\int_{\bb{R}}\,\d y\,\dot{f}(\tau,y)\psi(\tau,y)\bigg\vert^{2}\stackrel{\text{Hölder ineq.}}{\leq}\\&\leq \bigg(\int_{\bb{R}}\,\d x\,\vert f(t,x)\vert^{2}\bigg)\bigg(\int_{\bb{R}}\,\d y\,\vert \dot{f}(\tau,y)\vert^{2}\bigg)\Vert \psi_{\tau}\Vert_{\tau}^{2}\leq \underbrace{\bigg(\max_{t,t^{\prime}\in (0,\delta/2)}\Vert f_{t}\Vert_{t}\Vert \dot{f}_{\tau}\Vert_{\tau}\bigg)}_{=:C^{2}}\Vert \psi_{\tau}\Vert_{\tau}^{2}
	\end{align*} 
	where the constant $C$ depends on $\delta$ and $f$, but not on time. Next, we will show that there exists a $\psi\in C^{\infty}_{\mathrm{c}}(\sf{M})$ and a choice of $t,\tau\in\bb{R}$ such that $\Vert\sf{B}_{t,\tau}\psi_{\tau}\Vert_{t}\gtrsim\delta^{-2}\Vert\psi_{\tau}\Vert_{\tau}$, which shows that the constant $C$ cannot be chosen smaller than $\delta^{-2}$ and hence that assumption (ii) in Theorem~\ref{Thm:Small} is not fulfilled for the nonlocal potential $\sf{B}$. For this, we choose $\psi:=\dot{f}$ and compute
	\begin{align*}
		\Vert\sf{B}_{t,t}\dot{f}_{t}\Vert_{t}^{2}&=\int_{\bb{R}}\,\d x\,\bigg\vert f(t,x)\int_{\bb{R}}\,\d y\,\vert\dot{f}(t,y)\vert^{2}\bigg\vert^{2}=\\&=\bigg(\int_{\bb{R}}\,\d x\,\vert f(t,x)\vert^{2}\bigg)\bigg(\int_{\bb{R}}\,\d y\,\vert \dot{f}(t,y)\vert^{2}\bigg)\Vert\dot{f}_{t}\Vert_{t}^{2}\stackrel{\text{Hölder ineq.}}{\geq}\\&\geq \bigg(\int_{\bb{R}}\,\d x\,\vert f(t,x)\dot{f}(t,x)\vert\bigg)^{2}\Vert\dot{f}_{t}\Vert_{t}^{2}\geq\bigg(\frac{1}{2}\partial_{t}\int_{\bb{R}}\,\d x\,\vert f(t,x)\vert^{2}\bigg)^{2}\Vert\dot{f}_{t}\Vert_{t}^{2}\, ,
	\end{align*}
	which holds true for any value of $t\in\bb{R}$. Now, let us set $g(t):=\int_{\bb{R}}\d x\,\vert f(t,x)\vert^{2}$. By definition, we note that $g(t)\geq 0$ for all $t$ and $\Vert g\Vert_{\sf{L}^{1}(0,\delta)}=1$. Let $t_{\mathrm{max}}\in (0,\delta)$ be such that $g(t)\leq g(t_{\mathrm{max}})$ for all $t\in (0,\delta)$. The normalisation $\int_{\bb{R}}\d t\,g(t)=\int_{0}^{\delta}\,\d t\,g(t)=1$ implies that $g(t_{\mathrm{max}})\geq \frac{1}{\delta}$. By the mean value theorem we conclude that there exists a $t_{0}\in (0,t_{\mathrm{max}})$ such that
	\begin{align*}
		\partial_{t}g(t_{0})=\frac{g(t_{\mathrm{max}})-g(0)}{t_{\mathrm{max}}-0}\geq \frac{1}{\delta^{2}}\,, 
	\end{align*} 
	since $t_{\mathrm{max}}\leq \delta$. Hence, we have shown that there exists a $t_{0}\in (0,\delta)$ such that $\Vert\sf{B}_{t_{0},t_{0}}\dot{f}_{t_{0}}\Vert_{t_{0}}\geq (2\delta^{2})^{-1}\Vert \dot{f}_{t_{0}}\Vert_{t_{0}}$. In particular, our nonlocal potential violates the condition $C<(8e\delta^{2})^{-1}$ of Theorem~\ref{Thm:Small}. We will now show that the nonlocal equation $(\sf{S}-\sf{B})\psi=\phi$ does not always admit a solution. Suppose there exists a weak solution $\psi\in\sf{L}^{2}(\sf{M})$ to the nonlocal equation $(\sf{S}-\sf{B})\psi=f$ with source $f$. Then, for any test function $\varphi\in C^{\infty}_{\mathrm{c}}(\sf{M})$ we must have $\langle\psi,(\sf{S}-\sf{B})^{t}\varphi\rangle_{\sf{L}^{2}(\sf{M})}=\langle f,\varphi\rangle_{\sf{L}^{2}(\sf{M})}$. Choosing as a test function $\varphi=f$, we obtain
	\begin{align*}
		 \Vert f\Vert_{\sf{L}^{2}(\sf{M})}^{2}=\langle\psi,(\sf{S}-\sf{B})^{t}\varphi\rangle_{\sf{L}^{2}(\sf{M})}=\langle\psi,\sf{S}^{t}f\rangle_{\sf{L}^{2}(\sf{M})}(1-\Vert f\Vert_{\sf{L}^{2}(\sf{M})}^{2})=0\,,
	\end{align*}
which is a contradiction. Hence, such a weak solution $\psi$ cannot exist.
\end{example}

Let us now turn to the proofs of Theorem~\ref{Thm:Ret} and Theorem~\ref{Thm:Small}. To start with, we need suitable energy estimates. 

\begin{proposition}\label{Prop:EnEstNonLoc} \emph{(Energy Estimates)}\newline
	Consider Set-up~\ref{SetUp:Nonloc} and let $\psi\in \Gamma^{\infty}_{\mathrm{sc}}(\sf{E})$ and $\phi\in \Gamma^{\infty}_{\mathrm{c}}(\sf{E})$ be such that $\sf{S}\psi=\sf{B}\phi$. Then,
	\begin{align*}
		\bigg\vert\frac{\d}{\d t}\Vert\psi_{t}\Vert_{t}\bigg\vert\leq \int_{\bb{R}}\Vert \sf{V}_{t,\tau}\phi_{\tau}\Vert_{t}\,\d\tau+\frac{1}{2}\Vert(\sf{Z}_{\sf{S}}\psi)_{t}\Vert_{t}\, ,
	\end{align*}
	for all $t\in\bb{R}$, where $\mathsf{Z}_{\sf{S}}:=\beta\sigma_{\sf{S}}(\eta)^{-1}(\sf{S}+\sf{S}^{\ast})$ is an operator of order zero.
\end{proposition}

\begin{proof}
	To start with, let $\sf{S}$ be a symmetric hyperbolic system and $\psi\in\Gamma^{\infty}_{\mathrm{sc}}(\sf{E})$ be arbitrary. Then, we claim that the energy $\Vert\psi_{t}\Vert_{t}^{2}$, where $\psi_{t}:=\psi\vert_{\Sigma_{t}}$ as usual, satisfies the evolution equation
	\begin{align}\label{eq:EnEv}
		\frac{\d}{\d t}\Vert\psi_{t}\Vert_{t}^{2}=2\mathrm{Re}\int_{\Sigma_{t}}\langle\sf{S}\psi,\psi\rangle_{\sf{E}}\beta\,\d\mu_{\sf{h}_{t}}-\int_{\Sigma_{t}}\langle (\sf{S}+\sf{S}^{\ast})\psi,\psi\rangle_{\sf{E}}\beta\,\d\mu_{\sf{h}_{t}}\, .
	\end{align}
	In the case in which $\sf{S}=-\sf{S}^{\ast}$, as in the case of the Dirac operator (see~Example~\ref{Ex:DiracASA}), we obtain the well-known fact that $\Vert\psi_{t}\Vert_{t}$ is a preserved quantity in time, i.e.~that the norm $\Vert\cdot\Vert_{t}$ is independent of time whenever $\psi$ is a solution to the equations $\sf{S}\psi=0$. More generally, there is a loss term that is related to the zeroth-order term of the system.
	
	Now, the proof of Eq.~\eqref{eq:EnEv} follows exactly the same steps as the first part in the proof of Proposition~\ref{Prop:En1}: following the proof Lemma~\ref{Lemma:AdjointSHS}, we consider the real-valued $n$-form $\omega\in\Omega^{n}(\sf{M})$ defined by $\omega:=\langle\sigma_{\sf{S}}(\d x^{\mu})\psi,\psi\rangle_{\sf{E}}\,\partial_{\mu}\lrcorner \d\mu_{\sf{g}}$. As explained in more details in the proof of Lemma~\ref{Lemma:AdjointSHS}, its exterior derivative can be written as
	\begin{align*}
		\d\omega=\langle\psi,\sf{S}\psi\rangle_{\sf{E}}-\langle\sf{S}^{\ast}\psi,\psi\rangle_{\sf{E}}=2\mathrm{Re}\langle\sf{S}\psi,\psi\rangle_{\sf{E}}-\langle (\sf{S}+\sf{S}^{\ast})\psi,\psi\rangle_{\sf{E}}\, .
	\end{align*}
	Now, consider a closed time strip $\sf{M}_{t}:=t^{-1}([0,t])$ for $t\geq 0$. Then, Stokes' theorem implies
	\begin{align*}
		\int_{0}^{t}\int_{\Sigma_{t}}\bigg( 2\mathrm{Re}\langle\sf{S}\psi,\psi\rangle_{\sf{E}}-&\langle (\sf{S}+\sf{S}^{\ast})\psi,\psi\rangle_{\sf{E}}\bigg)\,\beta\d\mu_{\sf{h}_{\tau}}\,\d\tau=\\&=\int_{\sf{M}_{t}}\d\omega=\int_{\Sigma_{t}}i_{t}^{\ast}\omega-\int_{\Sigma_{0}}i_{0}^{\ast}\omega=\Vert\psi_{t}\Vert_{t}^{2}-\Vert\psi_{0}\Vert_{0}^{2}\, ,
	\end{align*}
	where $i_{\tau}\colon\Sigma_{\tau}\hookrightarrow\sf{M}$ denote the natural embeddings and where we used the fact that $\int_{\Sigma_{\tau}}i_{\tau}^{\ast}\omega=\Vert\psi_{\tau}\Vert^{2}_{\tau}$ for all $\tau\in\bb{R}$, as explained in the proof of Proposition~\ref{Prop:En1} (see Eq.~\eqref{eq:RHSProof} therein). Taking the time-derivative of this equation proves the validity of Eq.~\eqref{eq:EnEv} for $t\geq 0$. For $t\leq 0$, we choose a time strip $[t,0]\times\Sigma$ for $t\leq 0$ to obtain
	\begin{align*}
		\int_{t}^{0}\int_{\Sigma_{t}}\bigg( 2\mathrm{Re}\langle\sf{S}\psi,\psi\rangle_{\sf{E}}-\langle (\sf{S}+\sf{S}^{\ast})\psi,\psi\rangle_{\sf{E}}\bigg)\,\beta\d\mu_{\sf{h}_{\tau}}\,\d\tau=\Vert\psi_{0}\Vert_{0}^{2}-\Vert\psi_{t}\Vert_{t}^{2}
	\end{align*}
	with similar steps, which proves Eq.~\eqref{eq:EnEv} for $t\leq 0$ by taking the time derivative.
	
	Now, the claim of the proposition is a consequence of Eq.~\eqref{eq:EnEv} applied to $\sf{S}\psi=\sf{B}\phi$: using the defining relation of the time kernel in Definition~\ref{Def:NonLocPot} as well as the definition of the modified time kernel $\sf{V}$, we can write		
	\begin{align*}
		\int_{\Sigma_{t}}\langle\sf{B}\phi,\psi\rangle_{\sf{E}}\beta\,\d\mu_{\sf{h}_{t}}&=\int_{\bb{R}}\int_{\Sigma_{t}}\langle\beta_{t}\sf{B}_{t,\tau}\phi_{\tau},\psi_{\tau}\rangle_{\sf{E}}\,\d\mu_{\sf{h}_{t}}\,\d\tau=\\&=\int_{\bb{R}}\int_{\Sigma_{t}}\langle\sigma_{\sf{S}}(\eta_{t})\sf{V}_{t,\tau}\phi_{\tau},\psi_{t}\rangle_{\sf{E}}\,\d\mu_{\sf{h}_{t}}\,\d\tau=\int_{\bb{R}}\, (\sf{V}_{t,\tau}\phi_{\tau},\psi_{t})_{t}\,\d\tau\, .
	\end{align*}
	Using this equation as well as the notation $\sf{Z}_{\sf{S}}:=\beta\sigma_{\sf{S}}(\eta)^{-1}(\sf{S}+\sf{S}^{\ast})$, we take the absolute value in Eq.~\eqref{eq:EnEv} applied to $\psi$ to obtain the estimate
	\begin{align*}
		\bigg\vert\frac{\d}{\d t}\Vert\psi_{t}\Vert^{2}_{t}\bigg\vert\leq \Vert\psi_{t}\Vert_{t}\bigg(2\int_{\bb{R}}\Vert \sf{V}_{t,\tau}\phi_{\tau}\Vert_{t}\,\d\tau+\Vert(\sf{Z}_{\sf{S}}\psi)_{t}\Vert_{t}\bigg)\, ,
	\end{align*}
	where we used the triangle and Cauchy-Schwartz inequalities. Rewriting the left-hand side using the Leibniz rule yields the claimed result.
\end{proof}

Let us now turn to the proof of the retarded case.

\begin{proof}[Proof of Theorem~\ref{Thm:Ret}.]
We will split the proof of the theorem in four parts.\bigskip

\textit{Step 1.} Assume Set-up.~\ref{SetUp:Nonloc} together with assumptions (i)-(iii) in Theorem~\ref{Thm:Ret} and let $\psi^{(n)}\in \Gamma^{\infty}(\sf{E})$ be inductively defined as the (unique) solutions to the local Cauchy problems
	\begin{align*}
		\begin{cases}
			\sf{S}\psi^{(0)} &=\phi\\
			\psi^{(0)}\vert_{\Sigma_{0}}&=\mathfrak{f}
		\end{cases}\qquad\text{and}\qquad 
		\begin{cases}
			\sf{S}\psi^{(n+1)} &=\sf{B}\psi^{(n)}\\
			\psi^{(n+1)}\vert_{\Sigma_{0}}&=0
		\end{cases},\quad n\geq 0\, .
	\end{align*}
	Then, we claim that $\psi_{t}:=\sum_{n=0}^{\infty}\psi^{(n)}_{t}$ is absolutely convergent in the Hilbert space $\sf{L}^{2}(\Sigma_{t},\sf{E}\vert_{\Sigma_{t}})$ for every $t\in [0,\mathrm{T}]$. Furthermore, we claim that the map $\psi:t\mapsto\psi_{t}$ is a well-defined element in the space $\sf{L}^{2}(\sf{M}_{\mathrm{T}},\sf{E})$.
	
	To start with, let $\psi^{(n)}$ be the solutions to the local Cauchy problems as defined above. By finite speed of propagation, see~Theorem~\ref{Thm:Cauchy}(ii), together with assumptions (i) and (ii) on the nonlocal potential $\sf{B}$, it holds that 
	\begin{align*}
		\mathrm{supp}(\psi^{(0)})\cap \mathcal{J}^{+}(\Sigma_{0}) \subset \mathcal{J}^{+}(\mathrm{supp}(\phi)\cup\mathrm{supp}(\mathfrak{f}))\quad\text{and}\quad \mathrm{supp}(\psi^{(n)}) \subset \mathcal{J}^{+}(\mathrm{supp}(\phi)\cup\mathrm{supp}(\mathfrak{f}))
	\end{align*}
	 for all $n\geq 1$. In particular, note that the supports of all the sections $\psi^{(n)}$ are contained within the same set in the future of $\Sigma_{0}$. Furthermore, note that $\sf{B}\psi^{(n)}$ is well-defined, since $\sf{B}$ extends to an operator acting on smooth sections, which are in $\sf{L}^{2}(\Sigma_{t},\sf{E}\vert_{\Sigma_{t}})$ at any fixed $t\in \bb{R}$, as a consequence of assumptions (i), (ii) and the standard energy estimates. Now, the goal of the following discussion is to show that the perturbative Ansatz
	\begin{align}\label{eq:SeriesAnsatz}
		\psi=\sum_{n=0}^{\infty}\psi^{(n)}
	\end{align}
	is absolutely convergent in $\sf{L}^{2}(\Sigma_{t},\sf{E}\vert_{\Sigma_{t}})$ for every $t\in [0,\mathrm{T}]$. Using the energy estimates from Proposition~\ref{Prop:EnEstNonLoc}, we have
	\begin{align*}
		\frac{\d}{\d t}\Vert\psi^{(n+1)}_{t}\Vert_{t}\leq \int_{\bb{R}}\Vert \sf{V}_{t,\tau}\psi^{(n)}_{\tau}\Vert_{t}\,\d\tau+\frac{1}{2}\Vert(\sf{Z}_{\sf{S}}\psi^{(n+1)})_{t}\Vert_{t}
	\end{align*}
	for all $n\in\bb{N}_{0}$ and $t\in [0,\mathrm{T}]$, where $\sf{Z}_{\sf{S}}:=\beta\sigma_{\sf{S}}(\eta)^{-1}(\sf{S}+\sf{S}^{\ast})$ is an operator of order zero. Since all the sections $\psi^{(n)}$ are contained within the same compact set, there exists a constant $D_{\mathrm{T}}>0$ such that
	\begin{align*}
		\Vert (\sf{Z}_{\sf{S}}\psi^{(n)})_{t}\Vert_{t}\leq D_{\mathrm{T}}\Vert\psi^{(n)}_{t}\Vert_{t}
	\end{align*}
	for all $t\in [0,\mathrm{T}]$ and $n\in\bb{N}_{0}$. Using the assumptions (i)-(iii) as well as the fact that $\psi^{(n)}=0$ for $t<0$ and $n\geq 1$, we obtain the estimates
	\begin{align*}
		\frac{\d}{\d t}\Vert\psi^{(n+1)}_{t}\Vert_{t}&\leq C_{\mathrm{T}}\int_{0}^{t}\Vert \psi^{(n)}_{\tau}\Vert_{\tau}\, \d\tau+\frac{1}{2}D_{\mathrm{T}}\Vert\psi_{t}^{(n+1)}\Vert_{t}\, ,
	\end{align*}
	for all $n\in\mathbb{N}_{0}$ and $t\in [0,\mathrm{T}]$. Multiplying by $e^{-\frac{1}{2}D_{\mathrm{T}}t}$, this can be written in the compact form
	\begin{align*}
		\frac{\d}{\d t}(e^{-\frac{1}{2}D_{\mathrm{T}}t}\Vert\psi^{(n+1)}_{t}\Vert_{t})&\leq C_{\mathrm{T}}e^{-\frac{1}{2}D_{\mathrm{T}}t}\int_{0}^{t}\Vert \psi^{(n)}_{\tau}\Vert_{\tau}\, \d\tau\leq C_{\mathrm{T}}\int_{0}^{t}\Vert \psi^{(n)}_{\tau}\Vert_{\tau}\, \d\tau
	\end{align*}
	and integrating over time from $0$ to $t\in [0,\mathrm{T}]$, we obtain the inequality
	\begin{align*}
		\Vert\psi^{(n+1)}_{t}\Vert_{t} \leq C_{\mathrm{T}}e^{\frac{1}{2}D_{\mathrm{T}}t}\int_{0}^{t}\int_{0}^{\tau}\Vert \psi^{(n)}_{s}\Vert_{s}\, \d s\,\d\tau\,,
	\end{align*}
	where we used again the fact that $\psi^{(n)}=0$ for $t\leq 0$. Last but not least, we bound $e^{\frac{1}{2}D_{\mathrm{T}}t}\leq e^{\frac{1}{2}D_{\mathrm{T}}\mathrm{T}}$ for all $t\in [0,\mathrm{T}]$ to obtain the estimates
	\begin{align}\label{eq:Estimate}
		\Vert\psi^{(n+1)}_{t}\Vert_{t} \leq K_{\mathrm{T}}\int_{0}^{t}(t-\tau)\Vert \psi^{(n)}_{\tau}\Vert_{\tau}\,\d\tau
	\end{align}
	for all $n\in\mathbb{N}_{0}$ and $t\in [0,\mathrm{T}]$, where we set $K_{\mathrm{T}}:=C_{\mathrm{T}}e^{\frac{1}{2}D_{\mathrm{T}}\mathrm{T}}>0$.

	Now, consider the induction hypothesis $\Vert\psi_{t}^{(n)}\Vert_{t}\leq \kappa(n)t^{2n}M_{\mathrm{T}}$ for all $t\in [0,\mathrm{T}]$, where $M_{\mathrm{T}}:=\max_{t\in [0,\mathrm{T}]}\Vert\psi^{(0)}_{t}\Vert_{t}$. By finite speed of propagation, $M_{\mathrm{T}}$ is clearly finite, and the Ansatz trivially holds true for $n=0$ with $\kappa(0)=1$. By induction, we assume that it is true for $n$. Using the estimate~\eqref{eq:Estimate}, we obtain
	\begin{align*}
		\Vert\psi^{(n+1)}_{t}\Vert_{t} &\leq K_{\mathrm{T}}\int_{0}^{t}(t-\tau)\Vert \psi^{(n)}_{t}\Vert_{\tau}\, d\tau\leq K_{\mathrm{T}}M_{\mathrm{T}}\kappa(n)\frac{t^{2n+2}}{(2n+1)(2n+2)}\, .
	\end{align*}
	This completes the inductive step and allows us to determine the coefficients $\kappa(n)$ in $\Vert\psi_{t}^{(n)}\Vert_{t}\leq \kappa(n)t^{2n}M_{\mathrm{T}}$ as $\kappa(n)=\frac{K_{\mathrm{T}}^n}{(2n)!}$. To sum up, we have shown that 
	\begin{align*}
		\forall t\in [0,\mathrm{T}]\,,n\in\bb{N}_{0}:\quad \Vert\psi_{t}^{(n)}\Vert_{t}\leq \frac{K_{\mathrm{T}}^n}{(2n)!}t^{2n}\, .
	\end{align*}
	In particular, this implies that the perturbative series~\eqref{eq:SeriesAnsatz} is absolutely convergent in the $\sf{L}^{2}(\Sigma_{t},\sf{E}\vert_{\Sigma_{t}})$-topology for every $t\in [0,\mathrm{T}]$, since
	\begin{align*}
	\sum_{n=0}^{\infty}\Vert\psi^{(n)}_{t}\Vert_{t}\leq \sum_{n=0}^{\infty}\frac{M_{\mathrm{T}}K_{\mathrm{T}}^{n}t^{2n}}{(2n)!}=M_{\mathrm{T}}\mathrm{cosh}(\sqrt{K_{\mathrm{T}}}t)<\infty\, .
	\end{align*}
	
	Now, we have constructed a map $\psi:t\mapsto\psi_{t}$, where $\psi_{t}$ is the limit of the series~\ref{eq:SeriesAnsatz} in the $\sf{L}^{2}(\Sigma_{t},\sf{E}\vert_{\Sigma_{t}})$-topology. Now, it is clear that $t\mapsto \psi_{t}$ is measurable, since it is the pointwise limit of the smooth (and in particular strongly measurable) functions $t\mapsto \sum_{k=0}^{n}\psi^{(k)}_{t}$ for $n\to\infty$, see e.g.~\cite[Thm.~1.14]{AmannEscherIII}. To see that $\psi$ is an element of $\sf{L}^{2}([0,\mathrm{T}],\sf{L}^{2}(\Sigma_{\bullet},\sf{E}\vert_{\Sigma_{\bullet}}))$, we note that the above estimates imply
\begin{align*}
		\Vert\psi\Vert_{\sf{M}_{\mathrm{T}}}^{2}=\int_{0}^{\mathrm{T}}\Vert\psi_{t}\Vert_{t}^{2}\,\d t\leq M_{\mathrm{T}}^{2}\int_{0}^{T} \mathrm{cosh}(\sqrt{K_{\mathrm{T}}}t)^{2}\,\d t=M_{\mathrm{T}}^{2}\bigg(\frac{\mathrm{T}}{2}+\frac{\mathrm{sinh}(2\sqrt{K_{\mathrm{T}}}T)}{4\sqrt{K_{\mathrm{T}}}}\bigg)<\infty\,.
	\end{align*}
	It is easy to see that the series~\eqref{eq:SeriesAnsatz} converges also in the $\sf{L}^{2}(\sf{M}_{\mathrm{T}},\sf{E})$-topology and by the dominant convergence theorem for series (see e.g.~\cite[Thm.~3.16]{AmannEscherIII}), we conclude that the corresponding limit coincides with $\psi$.\bigskip
	
\textit{Step 2:} Next, we show that the element $\psi$ constructed in the previous proposition defines a strong solution to the Cauchy problem~\eqref{eq:CauchyNonloc}.

Consider the sequence $(\psi^{(n)})_{n\in\bb{N}_{0}}$ as constructed in Step 1. As proven above, the corresponding series converges absolutely in the $\sf{L}^{2}(\sf{M}_{\mathrm{T}},\sf{E})$-topology to some element $\psi$. We will now show that $\psi$ is a {strong} solution to the Cauchy problem~\eqref{eq:CauchyNonloc}. For this, we write
	\begin{align*}
		\psi_{k}:=\sum_{n=0}^{k}\psi^{(n)}\in \Gamma^{\infty}(\sf{M},\sf{E})
	\end{align*}
	for $k\in\bb{N}_{0}$. As shown above, it holds that $\Vert\psi_{k}-\psi\Vert_{\sf{M}_{\mathrm{T}}}\to 0$ as $k\to\infty$. Furthermore, it is clear that also $\Vert\psi_{k}\vert_{\Sigma_{0}}-\mathfrak{f}\Vert_{0}\xrightarrow{k\to\infty} 0$, since $\psi_{k}\vert_{\Sigma_{0}}=\mathfrak{f}$ for all $k\in\bb{N}_{0}$, by construction. Note also that $\Vert\psi^{(n)}\Vert_{\sf{M}_{\mathrm{T}}}\xrightarrow{n\to\infty}0$, since otherwise the series~\eqref{eq:SeriesAnsatz} would not converge absolutely. 

	On the other hand, it holds that $(\sf{S}-\sf{B})\psi_{k}=\phi-\sf{B}\psi^{(k)}$, by construction, and hence, we get
	\begin{align*}
		\Vert(\sf{S}-\sf{B})\psi_{k}-\phi\Vert^2_{\sf{M}_{\mathrm{T}}}&=\langle\sf{B}\psi^{(k)},\sf{B}\psi^{(k)}\rangle_{\sf{M}_{\mathrm{T}}}=\int_{0}^{\mathrm{T}}\,\d t\,\int_{0}^{t}\,\d\tau\,\int_{0}^{t}\,\d s\,(\sf{B}_{t,\tau}\psi_{\tau}^{(k)},\sf{B}_{t,s}\psi_{s}^{(k)})_{t}\\
		&\leq (C^{\prime}_{\mathrm{T}})^{2}\int_{0}^{\mathrm{T}}\,\d t\,\int_{0}^{t}\,\d\tau\,\int_{0}^{t}\,\d s\,\Vert\psi_{\tau}^{(k)}\Vert_{\tau}\Vert\psi_{s}^{(k)}\Vert_{s} \\
		 &\leq \mathrm{T}^{2}(C^{\prime}_{\mathrm{T}})^{2} \Vert \psi^{(k)}\Vert_{\sf{M}_{\mathrm{T}}}^2\xrightarrow{k\to\infty}0\, ,
	\end{align*}
	where we used Hölder's inequality. Furthermore, we used the following: by assumption, it holds that $\Vert\sf{V}_{t,\tau}\psi^{(k)}_{\tau}\Vert_{t}\leq C_{\mathrm{T}}\Vert\psi^{(k)}_{\tau}\Vert_{\tau}$ for all $k\in\mathbb{N}$ and $t,\tau\in [0,\mathrm{T}]$, where $\sf{V}=\beta\sigma_{\sf{S}}(\eta)^{-1}\sf{B}$. Now, since $\psi^{(k)}$ for arbitrary $k\in\mathbb{N}$ are supported within the same set as explained in the Step 1 above, the norm $\Vert\sf{B}_{t,\tau}\psi^{(k)}_{\tau}\Vert_{t}$ can be bounded by $\Vert\sf{V}_{t,\tau}\psi^{(k)}_{\tau}\Vert_{t}$ on $[0,\mathrm{T}]$ and vice versa. Hence, we obtain a similar estimate for $\sf{B}_{t,\tau}$ for some constant $C^{\prime}_{\mathrm{T}}>0$. \bigskip

\textit{Step 3:} Next, assume that assumption (iv) holds true. Then, we claim that the strong solution constructed above is smooth. To start with, let $\psi$ be the strong solution constructed above, i.e.~$\psi=\sum_{k=0}^{\infty}\psi^{(k)}$, where $\psi^{(k)}$ are the unique solutions to the local Cauchy problems
	\begin{align*}
		\begin{cases}
			\sf{S}\psi^{(0)} &=\phi\\
			\psi^{(0)}\vert_{\Sigma_{0}}&=\mathfrak{f}
		\end{cases}\qquad\text{and}\qquad 
		\begin{cases}
			\sf{S}\psi^{(n+1)} &=\sf{B}\psi^{(n)}\\
			\psi^{(n+1)}\vert_{\Sigma_{0}}&=0
		\end{cases},\quad n\geq 0\, .
	\end{align*}
	Now, consider the extended system $\mathfrak{S}$ defined on $\mathcal{E}:=\sf{E}\oplus (\sf{E}\otimes\sf{T}^{\ast}\sf{M})$, as defined in Lemma~\ref{Lemma:BiggerSystem}, and the extended nonlocal potential $\mathfrak{B}$ defined on $\mathcal{E}$ as defined in Lemma~\ref{Lemma:BiggerSystem2}.  If the nonlocal potential $\mathfrak{B}$ satisfies the same conditions as $\sf{B}$, we can construct a strong solution $\Psi$ of the nonlocal Cauchy problem
	\begin{align*}
		\begin{cases}
			(\mathfrak{S}-\mathfrak{B})\Psi=\Phi\\
			\Psi\vert_{\Sigma_{0}}=\mathfrak{F}
		\end{cases}
	\end{align*}
	with $\Phi:=(\phi,\nabla^{\sf{E}}\phi)$ and where the initial data $\mathfrak{F}$ is uniquely determined by $\sf{S}\psi=\phi$ restricted to $t=0$ and hence depends on $\mathfrak{f}$ and $\phi\vert_{t=0}$, where we note that $(\sf{B}\psi)\vert_{t=0}=0$, due to the past-compactness assumption on $\sf{B}$. As above, the strong solution $\Psi$ is constructed by considering the perturbative Ansatz $\Psi=\sum_{k=0}^{\infty}\Psi^{(k)}$ with $\Psi^{(k)}$ being the unique solutions to the local problems
	\begin{align*}
		\begin{cases}
			\mathfrak{S}\Psi^{(0)} &=\Phi\\
			\Psi^{(0)}\vert_{\Sigma_{0}}&=\mathfrak{F}
		\end{cases}
		\qquad\text{and}\qquad  
		\begin{cases}
			\mathfrak{S}\Psi^{(n+1)} &=\mathfrak{B}\Psi^{(n)}\\
			\Psi^{(n+1)}\vert_{\Sigma_{0}}&=0
		\end{cases},\quad n\geq 0\, .
	\end{align*} 
	Now, due to the past-compactness assumption on $\sf{B}$, it holds that $(\sf{B}\psi^{(k)})\vert_{\Sigma_{0}}=0$ for all $k\in\bb{N}_{0}$. In particular, this implies that the local solutions $\psi^{(n)}$ satisfy $\nabla^{\sf{E}}\psi^{(n)}\vert_{\Sigma_{0}}=0$ and, by uniqueness of solutions to the local equations, we conclude that in fact $\Psi^{(n)}=(\psi^{(n)},\nabla^{\sf{E}}\psi^{(n)})$ for all $n\in\bb{N}$.
	
	Now, let us write $\eta:=\mathrm{pr}_{2}\Psi$. By construction, $\nabla^{\sf{E}}\psi_{n}\to\eta$ where $\psi_{n}:=\sum_{k=0}^{n}\psi^{(k)}$ as $k\to\infty$ in $\sf{L}^{2}(\sf{M}_{\mathrm{T}},\sf{E})$. In particular, this shows that the sequence $(\psi_{n})_{n}$ converges also in $\sf{H}^{1}(\sf{M}_{\mathrm{T}},\sf{E})$, since 
	\begin{align*}
		\Vert \psi_{n}-\psi_{m}\Vert_{\sf{H}^{1}(\sf{M}_{\mathrm{T}})}^{2}=\Vert \psi_{n}-\psi_{m}\Vert_{\sf{L}^{2}(\sf{M}_{\mathrm{T}})}^{2}+\Vert \nabla^{\sf{E}}\psi_{n}-\nabla^{\sf{E}}\psi_{m}\Vert_{\sf{L}^{2}(\sf{M}_{\mathrm{T}})}^{2}\to 0\, .
	\end{align*}
	Now, it is clear that the limit of the sequence $(\psi_{n})_{n}$ in $\sf{H}^{1}(\sf{M}_{\mathrm{T}},\sf{E})$ coincides with $\psi$, which is the limit of $(\psi_{n})_{n}$ in $\sf{L}^{2}(\sf{M}_{\mathrm{T}},\sf{E})$, since $\Vert\cdot\Vert_{\sf{L}^{2}(\sf{M}_{\mathrm{T}})}\leq \Vert\cdot\Vert_{\sf{H}^{1}(\sf{M}_{\mathrm{T}},\sf{E})}$. Hence, we conclude that $\psi\in \sf{H}^{1}(\sf{M}_{\mathrm{T}},\sf{E}_{\beta})$. Proceeding by induction, we obtain $\psi\in \sf{H}^{k}(\sf{M}_{\mathrm{T}},\sf{E})$ and, by using the (local) Sobolev embedding theorems, we obtain the claimed result.\bigskip
	
	\textit{Step 4:} Last but not least, let us proof uniqueness and finite speed of propagation. Suppose that $\psi\in \Gamma^{\infty}(\sf{M}_{\mathrm{T}},\sf{E})$ is a solution of the homogeneous nonlocal equation $(\sf{S}-\sf{B})\psi=0$ with initial condition $\psi\vert_{\Sigma_{0}}=0$. Then, following exactly the same steps as Step 1, we obtain
	\begin{align*}
		\Vert\psi_{t}\Vert_{t} \leq K_{\mathrm{T}}\int_{0}^{t}(t-\tau)\Vert \psi_{\tau}\Vert_{\tau}\,\d\tau \leq K_{\mathrm{T}}T\int_{0}^{t}\Vert \psi_{\tau}\Vert_{\tau}\,\d\tau=:K_{\mathrm{T}}\mathrm{T}F(t)
	\end{align*}
	for all $t\in [0,\mathrm{T}]$, where $K_{\mathrm{T}}>0$ is some constant only depending on $\mathrm{T}$. Now, observe that $\dot{F}(t)=\Vert\psi\Vert_{t}$, where the dot denotes the time-derivative, which allows us to rewrite the above inequality as $\dot{F}(t)\leq K_{\mathrm{T}}\mathrm{T}F(t)$. In other words, we find that
	\begin{align*}
		\frac{\d}{\d t}(e^{-K_{\mathrm{T}}\mathrm{T}t}F(t))\leq 0
	\end{align*}
	for all $t\in [0,\mathrm{T}]$. Setting $G(t):=e^{-K_{\mathrm{T}}\mathrm{T}t}F(t)$, we observe that $G(t)\geq 0$ and $\dot{G}(t)\leq 0$ as well as $G(0)=0$. The only such $C^{1}$-function is given by $G(t)=0$ and hence $\psi=0$ on $\sf{M}_{\mathrm{T}}$.
	
	Now, if $\psi$ is a smooth solution to the Cauchy problem $(\sf{S}-\sf{B})\psi=\phi$ with $\psi\vert_{\Sigma_{0}}=\mathfrak{f}$, uniqueness implies that $\psi$ is constructed via the perturbative series $\psi=\sum_{n=0}^{\infty}\psi^{(n)}$, which is absolutely convergent in $\sf{L}^{2}(\sf{M}_{\mathrm{T}},\sf{E})$, where $\psi^{(n)}$ are the solutions to the local problems $\sf{S}\psi^{(0)}=\phi$ and $\sf{S}\psi^{(n+1)}=\sf{B}\psi^{(n)}$ with $\psi^{(n)}\vert_{\Sigma_{0}}=0$. As explained in Step 1, by the assumptions on $\sf{B}$, we know that 
\begin{align*}
	\mathrm{supp}(\psi^{(n)})\cap \mathcal{J}^{+}(\Sigma_{0})\subset \mathcal{J}^{+}(\mathrm{supp}(\phi)\cup\mathrm{supp}(\mathfrak{f}))
\end{align*}
for all $n\in\bb{N}_{0}$. Hence, also $\psi$, assumed to be smooth, is supported within the same set. 
\end{proof}

Having proven the well-posedness of the Cauchy problem for symmetric hyperbolic systems with \emph{retarded} nonlocal potentials, it remains to prove well-posedness in the short-time-range case.

\begin{proof}[Proof of Theorem~\ref{Thm:Small}.]
	Assume Set-up.~\ref{SetUp:Nonloc} together with assumptions (i)-(iii) in Theorem~\ref{Thm:Small} and let $\psi^{(n)}\in \Gamma^{\infty}(\sf{E})$ be inductively defined as the (unique) solutions to the local Cauchy problems
	\begin{align*}
		\begin{cases}
			\sf{S}\psi^{(0)} &=\phi\\
			\psi^{(0)}\vert_{\Sigma_{0}}&=\mathfrak{f}
		\end{cases}\qquad\text{and}\qquad 
		\begin{cases}
			\sf{S}\psi^{(n+1)} &=\sf{B}\psi^{(n)}\\
			\psi^{(n+1)}\vert_{\Sigma_{0}}&=0
		\end{cases},\quad n\geq 0\, .
	\end{align*}
	Then, as in the proof of the retarded case above, we claim that $\psi_{t}:=\sum_{n=0}^{\infty}\psi^{(n)}_{t}$ is absolutely convergent in the Hilbert space $\sf{L}^{2}(\Sigma_{t},\sf{E}\vert_{\Sigma_{t}})$ for every $t\in [0,\mathrm{T}]$.
	
To start with, finite speed of propagation for (local) symmetric hyperbolic systems, see~Theorem~\ref{Thm:Cauchy}(ii), implies
	\begin{align*}
		\mathrm{supp}(\psi^{(0)}) \subset\mathcal{J}(\mathrm{supp}(\phi)\cup\mathrm{supp}(\mathfrak{f}))\quad\text{and}\quad \mathrm{supp}(\psi^{(n+1)}) \subset \mathcal{J}(\mathrm{supp}(\sf{B}\psi^{(n)}))
	\end{align*}
	 for all $n\in\bb{N}_{0}$. Note that $\sf{B}\psi^{(n)}$ is well-defined, since $\sf{B}$ extends to an operator acting on smooth sections, which are in $\sf{L}^{2}(\Sigma_{t},\sf{E}\vert_{\Sigma_{t}})$ at any fixed $t\in\bb{R}$, as a consequence of assumption (iii) and the standard energy estimates. Moreover, unlike in the retarded case, observe that $\psi^{(n)}$ will in general also include a component propagating into the past, both due to the non-zero initial datum $\mathfrak{f}$ for $\psi^{(0)}$ and the nonlocality, which increases the support of the source at each step $n$ by $\pm\delta$ in time, and hence, at some order, the source will thus intersect the initial slice $\Sigma_{0}$.
	
	Now, the goal of the following discussion is to show that the perturbative ansatz
\begin{align}\label{ansatzpertur2}
		\psi=\sum_{n=0}^{\infty}\psi^{(n)}
	\end{align}	
	is absolutely convergent in $\sf{L}^{2}(\Sigma_{t},\sf{E}\vert_{\Sigma_{t}})$ for all $t\in\bb{R}$. Using Proposition~\ref{Prop:EnEstNonLoc}, we have
\begin{align}\label{eq:EnergyEstSmall}
		\bigg\vert\frac{\d}{\d t}\Vert\psi^{(n+1)}_{t}\Vert_{t}\bigg\vert\leq C\int_{t-\delta}^{t+\delta}e^{-\frac{1}{2}D\vert\tau\vert}\Vert \psi^{(n)}_{\tau}\Vert_{\tau}\, \d\tau+\frac{1}{2}D\Vert\psi^{(n+1)}_{t}\Vert_{t}
	\end{align}
	for all $n\in\bb{N}_{0}$ and $t\in\bb{R}$, where we used assumptions (i)-(iii). As a first step, we absorb the last term in~\eqref{eq:EnergyEstSmall} using an appropriate exponential factor, by noting that~\eqref{eq:EnergyEstSmall} imply
	\begin{subequations}
	\begin{align}
		\frac{\d}{\d t}\bigg(e^{-\frac{1}{2}D t}\Vert\psi^{(n+1)}_{t}\Vert_{t}\bigg)\leq Ce^{-\frac{1}{2}D t}\int_{t-\delta}^{t+\delta}e^{-\frac{1}{2}D\vert\tau\vert}\Vert \psi^{(n)}_{\tau}\Vert_{\tau}\, \d\tau\,,\quad &\text{for }t\geq 0\label{eq:geqt}\\
		-\frac{\d}{\d t}\bigg(e^{\frac{1}{2}D t}\Vert\psi^{(n+1)}_{t}\Vert_{t}\bigg)\leq Ce^{\frac{1}{2}D t}\int_{t-\delta}^{t+\delta}e^{-\frac{1}{2}D\vert\tau\vert}\Vert \psi^{(n)}_{\tau}\Vert_{\tau}\, \d\tau\,,\quad &\text{for }t\leq 0\label{eq:leqt}
	\end{align}
	\end{subequations}
	for all $n\in\bb{N}_{0}$. Now, we distinguish between to cases. If $t\geq 0$, we integrate~\eqref{eq:geqt} from $0$ to $t$ and, using the fact that $\psi^{(n+1)}_{t=0}=0$, we obtain the estimate\footnote{In the first estimate, we changed the order of integration by using that \begin{align*}\begin{cases}0 \leq \tau \leq t\\ \tau-\delta \leq s\leq\tau+\delta\end{cases}\qquad\Leftrightarrow\qquad\begin{cases}-\delta\leq s \leq t+\delta\\ \mathrm{max}\{s-\delta,0\} \leq \tau\leq\mathrm{min}\{s+\delta,t\}\end{cases}\, \end{align*}
	for all $\delta>0$, $t\geq 0$. Note also that $\mathrm{max}\{s-\delta,0\}\leq \mathrm{min}\{s+\delta,t\}$ on the range $s\in [-\delta,t+\delta]$ for $\delta>0$, $t\geq 0$.}
		\begin{align}
		\Vert\psi_{t}^{(n+1)}\Vert_{t}&=Ce^{\frac{1}{2}Dt}\int_{0}^{t}\underbrace{e^{-\frac{1}{2}D\tau}}_{\leq 1}\int_{\tau-\delta}^{\tau+\delta}e^{-\frac{1}{2}D\vert s\vert}\Vert\psi_{s}\Vert_{s}\,\d s\,\d\tau\nonumber\\&\leq Ce^{\frac{1}{2}Dt}\int_{-\delta}^{t+\delta}\bigg(\underbrace{\mathrm{min}\{s+\delta,t\}-\mathrm{max}\{s-\delta,0\}}_{\leq 2\delta}\bigg)e^{-\frac{1}{2}D\vert s\vert}\Vert\psi^{(n)}_{s}\Vert_{s}\,\d s\nonumber\\&\leq (2\delta C)e^{\frac{1}{2}Dt}\int_{-\delta}^{t+\delta}e^{-\frac{1}{2}D\vert s\vert}\Vert\psi^{(n)}_{s}\Vert_{s}\,\d s\label{eq:EstPost}
	\end{align}
	for all $t\geq 0$ and $n\in\mathbb{N}_{0}$. Similarly, in the case $t\leq 0$, we integrate over the estimate~\eqref{eq:leqt} from $t$ to $0$. Using again that $\psi^{(n+1)}_{t=0}=0$, we obtain the similar estimate
	\begin{align}
		\Vert\psi_{t}^{(n+1)}\Vert_{t}&=Ce^{-\frac{1}{2}Dt}\int_{t}^{0}\underbrace{e^{\frac{1}{2}D\tau}}_{\leq 1}\int_{\tau-\delta}^{\tau+\delta}e^{-\frac{1}{2}D\vert s\vert}\Vert\psi_{s}\Vert_{s}\,\d s\,\d\tau\nonumber\\&\leq Ce^{-\frac{1}{2}Dt}\int_{t-\delta}^{\delta}\bigg(\underbrace{\mathrm{min}\{s+\delta,0\}-\mathrm{max}\{s-\delta,t\}}_{\leq 2\delta}\bigg)e^{-\frac{1}{2}D\vert s\vert}\Vert\psi^{(n)}_{s}\Vert_{s}\,\d s\nonumber\\&\leq (2\delta C)e^{-\frac{1}{2}Dt}\int_{t-\delta}^{\delta}e^{-\frac{1}{2}D\vert s\vert}\Vert\psi^{(n)}_{s}\Vert_{s}\,\d s\label{eq:EstNegt}
	\end{align}
	for all $t\leq 0$ and $n\in\mathbb{N}_{0}$. Now, note that the previous two estimates~\eqref{eq:EstPost} and~\eqref{eq:EstNegt} for $t\geq 0$ and $t\leq 0$ can be combined and written in the compact form
	\begin{align}
		\forall t\in\bb{R},\,n\in\bb{N}_{0}:\quad\Vert\psi^{(n+1)}_{t}\Vert_{t}\leq (2\delta C)e^{\frac{1}{2}D\vert t\vert}\int_{-\delta}^{\vert t\vert+\delta}e^{-\frac{1}{2}D\vert s\vert}\Vert\psi^{(n)}_{\mathrm{sgn}(t)s}\Vert_{\mathrm{sgn}(t)s}\, \d s\, . \label{ites}
	\end{align}
	
	After deriving the relevant estimates for the $\sf{L}^{2}(\Sigma_{t},\sf{E}\vert_{\Sigma_{t}})$-norm of $\psi^{(n+1)}$ in terms of the $\sf{L}^{2}(\Sigma_{t},\sf{E}\vert_{\Sigma_{t}})$-norm of $\psi^{(n)}$, we derive suitable estimates for $\Vert\psi^{(n)}_{t}\Vert_{t}$. As a first step, by standard energy estimates, we can find a constant $M>0$ (depending on $\mathfrak{f}$ and $\phi$) such that $\Vert\psi^{(0)}_{t}\Vert_{t}\leq e^{\frac{1}{2}D\vert t\vert}M$ for all $t\in\bb{R}$. Now, for given $n\in\bb{N}_{0}$, we consider the induction hypothesis 
	\begin{align*}
		\Vert\psi^{(n)}_{t}\Vert_{t}\leq M\kappa(n)e^{\frac{1}{2}D\vert t\vert}(\vert t\vert+2n\delta)^{n},\qquad\forall t\in\bb{R}\, .
	\end{align*}	
	 Clearly, this inequality holds true for $n=0$ with $\kappa(0)=1$. By induction, we assume that it is true for $n$. Using the estimate~\eqref{ites} and the induction hypothesis, we obtain
	\begin{align*}
		\Vert\psi^{(n+1)}_{t}\Vert_{t}&\leq (2\delta C)e^{\frac{1}{2}D\vert t\vert}\int_{-\delta}^{\vert t\vert+\delta}e^{-\frac{1}{2}D\vert s\vert}\Vert\psi^{(n)}_{\mathrm{sgn}(t)s}\Vert_{\mathrm{sgn}(t)s}\, \d s\\&\leq (2\delta C)M\kappa(n)e^{\frac{1}{2}D\vert t\vert}\int_{-\delta}^{\vert t\vert+\delta}(\vert s\vert+2n\delta)^{n}\, \d s\\&= (2\delta C)M\kappa(n)e^{\frac{1}{2}D\vert t\vert}\frac{1}{n+1}\bigg\{(\vert t\vert+2\delta n+\delta)^{n+1}-2(2n\delta)^{n+1}+(2\delta n+\delta)^{n+1}\bigg\}\\&\leq (4\delta C)M\kappa(n)e^{\frac{1}{2}D\vert t\vert}\frac{(\vert t\vert+2(n+1)\delta)^{n+1}}{n+1}\, .
	\end{align*}
	This allows us to determine the coefficients $\kappa(n)$ in $\Vert\psi^{(n)}_{t}\Vert_{t}\leq\kappa(n)e^{\frac{1}{2}D\vert t\vert}(\vert t\vert+n\delta)^{n}M$ and we obtain $\kappa(n)=\frac{1}{n!}(4C\delta)^{n}$. To sum up, we have shown that
	\begin{align}\label{eq:EstimatesSmallTR}
		\forall t\in\bb{R}\,,n\in\bb{N}_{0}:\quad \Vert\psi^{(n)}_{t}\Vert_{t} \leq \frac{M}{n!}(4C\delta)^{n}e^{\frac{1}{2}D\vert t\vert}(\vert t\vert+2n\delta)^{n}\, .
	\end{align}

	To show that the perturbative Ansatz~\eqref{ansatzpertur2} is absolutely convergent in the $\sf{L}^{2}(\Sigma_{t},\sf{E}\vert_{\Sigma_{t}})$-topology, we have to study the bounds on the $\sf{L}^{2}(\Sigma_{t},\sf{E}\vert_{\Sigma_{t}})$-norms of $\psi^{(n)}$ for for large $n$. In this case, using Stirling's formula for the factorial, one obtains
	\begin{align*}
		\Vert\psi^{(n)}_{t}\Vert_{t} &\leq \frac{M}{n!}(4C\delta)^{n}e^{\frac{1}{2}D\vert t\vert}(\vert t\vert +2n\delta)^{n}= \frac{M}{n!}(8C\delta^2)^{n}n^{n}e^{\frac{1}{2}D\vert t\vert}\left(1+\frac{\vert t\vert}{2n\delta}\right)^n \\&\sim\frac{Me^{\frac{1}{2\delta}\vert t\vert(D\delta+1)}}{\sqrt{2\pi n}}(8e\delta^{2}C)^{n}\left(1 + \mathcal{O}\left(\frac{1}{n}\right)\right) \quad \text{as} \quad n \to \infty.
	\end{align*}
	In particular, using the asymptotic expansion as written in the last line, it follows that 
	\begin{align*}
		\lim_{n\to\infty}\frac{\Vert\psi^{(n+1)}_{t}\Vert_{t}}{\Vert\psi^{(n)}_{t}\Vert_{t}}=8e\delta^{2}C
	\end{align*}
	for all $t\in\bb{R}$, which shows that the series $\sum_{n=0}^{\infty}\Vert\psi^{(n)}_{t}\Vert_{t}$ is convergent whenever $8e\delta^{2}C<0$, by the ratio test of convergence of series. 
	
	Now, we have constructed a map $\psi:t\mapsto\psi_{t}$, where $\psi_{t}$ is the limit of the series~\ref{ansatzpertur2} in the $\sf{L}^{2}(\Sigma_{t},\sf{E}\vert_{\Sigma_{t}})$-topology. Now, it is clear that $t\mapsto \psi_{t}$ is measurable, since it is the pointwise limit of the smooth (and in particular strongly measurable) functions $t\mapsto \sum_{k=0}^{n}\psi^{(k)}_{t}$ for $n\to\infty$, see e.g.~\cite[Thm.~1.14]{AmannEscherIII}. Now, consider the time interval $[0,\mathrm{T}]$ for $\mathrm{T}>0$. Using~\eqref{eq:EstimatesSmallTR}, we obtain
	\begin{align*}
		\Vert\psi_{t}^{(n)}\Vert_{\sf{M}_{\mathrm{T}}}^2=\int_{0}^{\mathrm{T}}\Vert\psi^{(n)}_{t}\Vert^2_{t}\,\d t \leq \frac{M^2}{(n!)^2}(4C\delta)^{2n}\int_{0}^{\mathrm{T}}e^{D\vert t\vert}(\vert t\vert+2n\delta)^{2n}\,\d t\leq \frac{\mathrm{T}M^2e^{D\mathrm{T}}}{(n!)^2}(4C\delta)^{2n}(\mathrm{T}+2n\delta)^{2n}\, .
	\end{align*}
	Using similar arguments as before, we conclude that the series of $\Vert\psi_{t}^{(n)}\Vert_{\sf{L}^{2}(\sf{M}_{\mathrm{T}})}^2$ is convergent. Hence, by the dominant convergence theorem applied to series (see e.g.~\cite[Thm.~3.16]{AmannEscherIII}), we conclude that the series~\eqref{ansatzpertur2} is also absolutely convergent in the $\sf{L}^{2}(\sf{M}_{\mathrm{T}},\sf{E})$-topology and 
	\begin{align*}
		\Vert\psi\Vert_{\sf{M}_{\mathrm{T}}}^2=\int_{0}^{\mathrm{T}}\Vert\psi_{t}\Vert_{t}^2\,\d t=\int_{0}^{\mathrm{T}}\sum_{n=0}^{\infty}\Vert\psi_{t}^{(n)}\Vert_{t}^2\,\d t=\sum_{n=0}^{\infty}\int_{0}^{\mathrm{T}}\Vert\psi_{t}^{(n)}\Vert_{t}^2\,\d t=\sum_{n=0}^{\infty}\Vert\psi^{(n)}\Vert_{\sf{M}_{\mathrm{T}}}^2<\infty\, .
	\end{align*}
	
	The claim that $\psi$ is indeed a strong solution to the Cauchy problem~\eqref{eq:CauchyNonloc} follows from similar arguments as in the proof of Theorem~\ref{Thm:Ret} above (see Step 2 therein).
\end{proof}	

To come back to the motivational examples considered at the beginning of our discussion on symmetric nonlocal hyperbolic systems, we show how Theorem~\ref{Thm:Ret} can be applied to the Maxwell equations in linear dispersive media.

\begin{example} (Maxwell's Equations in Linear Dispersive Media)\newline
	Let $(\sf{M}=\bb{R}\times\Sigma,\sf{g}=-\d t\otimes\d t+\sf{h})$ be a globally hyperbolic ultrastatic manifold. Following the notation used in Example~\ref{Ex:Maxwell}, we write $(\sf{F}:=\pi_{2}^{\ast}(\sf{T}\Sigma),\langle\cdot,\cdot\rangle_{\sf{F}}=:\sf{h}(\cdot,\cdot))$ with $\pi_{2}\:\sf{M}\to\Sigma$ being the natural projection. Moreover, we set $\sf{E}:=\sf{F}\oplus\sf{F}$ equipped with the natural induced bundle metric $\langle\cdot,\cdot\rangle_{\sf{E}}$. Then, following Example~\ref{Ex:Maxwell}, we consider the symmetric hyperbolic system 
	\begin{align*}
		\sf{S}:=\partial_{t}+\begin{pmatrix}
		0 & -\mathrm{curl}_{\Sigma}\\
		\mathrm{curl}_{\Sigma} &0
	\end{pmatrix}
	\end{align*}
	acting on the electric and magnetic field $(\mathscr{E},\mathscr{B})\in\Gamma^{\infty}(\sf{E})$. Moreover, following the general discussion of linear dispersive media at the beginning of Section~\ref{Sec:SHSNonLoc}, let $\chi\in C^{\infty}(\bb{R})$ be such that $\chi(0)=0$. Then, the piecewise continuous kernel $k_{\chi}(x,y)\colon \sf{E}_{y}\to\sf{E}_{x}$ defined by
\begin{align*}
	k_{\chi}((t,\vec{x}),(\tau,\vec{y}))=\theta(\tau)\theta(t-\tau)\begin{pmatrix}\dot{\chi}(t-\tau)\delta(\vec{x}-\vec{y}) & 0 \\ 0& 0\end{pmatrix}
\end{align*}
is the kernel of a retarded and past compact nonlocal potential $\sf{B}_{\chi}\colon \Gamma^{\infty}_{\mathrm{c}}(\sf{E})\to \Gamma^{\infty}(\sf{E})$.

Now, consider $\mathfrak{f}_{1},\mathfrak{f}_{2}\in \Gamma^{\infty}_{\mathrm{c}}(\Sigma_{0},\sf{F}\vert_{\Sigma_{0}})$ and external sources supported on a time strip $\sf{M}_{\mathrm{T}}$, i.e. $j^{(\mathrm{ext})}\in \Gamma^{\infty}_{\mathrm{c}}(\sf{M}_{\mathrm{T}},\sf{F})$ and $\rho^{(\mathrm{ext})}\in C^{\infty}_{\mathrm{c}}(\sf{M}_{\mathrm{T}})$ such that $\partial_{t}\rho^{(\mathrm{ext})}+\mathrm{div}_{\Sigma}(j^{(\mathrm{ext})})=0$. Then, the Cauchy problem for Maxwell's equations in linear dispersive media with response function $\chi$, see Eq.~\eqref{eq:MaxwellNonLoc}, can be written in the form
	\begin{align}\label{eq:dfsdfgr}
		\begin{cases}
			(\sf{S}-\sf{B}_{\chi})\psi&=\phi\\
			\psi\vert_{\Sigma_{0}}&=\mathfrak{f}
		\end{cases}\, ,\qquad\text{where}\qquad \psi:=\begin{pmatrix}
			\mathscr{E}\\ \mathscr{B}
		\end{pmatrix}\,,\quad\phi:=\begin{pmatrix}
			-4\pi j^{(\mathrm{ext})}\\ 0
		\end{pmatrix}\,,\qquad\mathfrak{f}:=\begin{pmatrix}
		\mathfrak{f}_{1}\\\mathfrak{f}_{2}
		\end{pmatrix}
	\end{align}
	Now, we claim that all the assumptions of Theorem~\ref{Thm:Ret} are fulfilled. First of all, note that the condition $\Vert\sf{V}_{t,\tau}\psi_{\tau}\Vert\leq C_{\mathrm{T}}\Vert\psi_{\tau}\Vert_{\tau}$ on $[0,\mathrm{T}]$ is automatically satisfied given that $\dot{\chi}$ is continuous. Moreover, if we look at the extended operator $\mathfrak{B}$ acting on $(\psi,\nabla^{\sf{E}}\psi)$ (see Lemma~\ref{Lemma:BiggerSystem2}), the only relevant terms to consider is the commutator $[\partial_{t},\sf{B}]$. Now, for any $\varphi\in \Gamma^{\infty}_{\mathrm{pc}}(\sf{F})$, it holds that
	\begin{align*}
		&\partial_{t}\int_{0}^{t}\dot{\chi}(t-\tau)\varphi_{\tau}(\cdot)\,\d\tau-\int_{0}^{t}\dot{\chi}(t-\tau)\partial_{\tau}\varphi_{\tau}(\cdot)=\\&\dot{\chi}(0)\varphi_{t}(\cdot)+\int_{0}^{t}\ddot{\chi}(t-\tau)\varphi_{\tau}(\cdot)\,\d\tau-\bigg(\dot{\chi}(0)\varphi_{t}(\cdot)-\dot{\chi}(t)\varphi_{0}(\cdot)-\int_{0}^{t}(\partial_{\tau}\dot{\chi}(t-\tau))\varphi_{\tau}(\cdot)\,\d\tau\bigg)=\dot{\chi}(t)\varphi_{0}(\cdot)
	\end{align*}
	Hence, up to a local operator that can be added to the zero-order terms of $\mathfrak{S}$, the nonlocal potential $\mathfrak{B}$ just acts component-wise as $\sf{B}$ on $\psi$. Hence, if $\chi$ is regular enough, the extended systems can be solved in the same way and we obtain in fact a smooth solution. In other words, by Theorem~\ref{Thm:Ret}, there exists a unique smooth solution $\psi\in\Gamma^{\infty}(\sf{E})$ to the Cauchy problem~\eqref{eq:dfsdfgr} that propagates at most with the speed of light.
	
	Now, the Maxwell equations also include constraints, namely the conditions $\mathrm{div}_\Sigma(\mathscr{D})=4\pi\rho^{(\mathrm{ext})}$ and $\mathrm{div}_{\Sigma}(\mathscr{B})=0$. We claim that this only give rise to constraints on the level of initial data, as in the vacuum case in Example~\ref{Ex:Maxwell}. Taking the divergence of $(\sf{S}-\sf{B}_{\chi})\psi=\phi$ implies
	\begin{align*}
		\partial_{t}\bigg(\mathrm{div}_{\Sigma}(\mathscr{E}(t,\cdot))+\int_{0}^{t}\chi(t-\tau)\,\mathrm{div}_{\Sigma}(\mathscr{E}(\tau,\cdot))\,\d\tau-4\pi\partial_{t}\rho^{(\mathrm{ext})}(t,\cdot)\bigg)=0\,,\qquad \partial_{t}\mathrm{div}_{\Sigma}(\mathscr{B})=0\, ,
	\end{align*} 
	where we used the continuity equation $\partial_{t}\rho^{(\mathrm{ext})}+\mathrm{div}_{\Sigma}(j^{(\mathrm{ext})})=0$. Hence, due to the choice of switch-on time $t_{0}=0$, we see that the constraint equations are fulfilled whenever we choose initial data satisfying $\mathrm{div}\,\mathfrak{f}_{1}=\rho^{(\mathrm{ext})}\vert_{t=0}$ and $\mathrm{div}\,\mathfrak{f}_{2}=0$.	
\end{example}
\chapter{Linear Gauge Theories on Manifolds}\label{Chap:LinGaug}
We now turn to \textit{linear gauge theories}, which are linear field-theoretic models admitting local gauge symmetries, and their quantisation. First of all, we discuss the mathematical structure of \textit{classical} linear gauge theories by focusing on an axiomatic framework introduced by Hack-Schenkel \cite{HackSchenkel}, which generalises and abstracts the so-called \textit{subsidiary condition formalism}. In particular, we discuss the Cauchy problem for linear gauge theories and provide various equivalent descriptions of the their classical phase space. Afterwards, we provide the reader with many examples of direct physical interest, namely Maxwell's theory, linearised Yang-Mills theory, linearised Einstein gravity, as well as the coupled linearised Yang-Mills-Klein-Gordon and Einstein-Klein-Gordon systems. Having studied the classical theory in some detail, we turn to their quantisation in the last part of this chapter. For this, we will adopt the so-called \textit{algebraic approach} to quantum field theory, which offers a mathematically rigorous quantisation scheme, ideally suited for field theories defined on globally hyperbolic spacetimes. Within this formalism, quantisation is essentially a two-step procedure: first, one assigns a non-commutative \textit{$\ast$-algebra of observables} to the classical phase space, which encodes structural
properties such as the \textit{canonical commutation relations}. The second step consists in choosing a \textit{state} on that algebra, that is, a positive, linear and normalised functional describing the physical preparation of the system. Among the plethora of admissible mathematical states, not every one can be considered \emph{physical} and hence, we will introduce the notion of \textit{Hadamard states}, which are physically distinguished states whose two-point function has a specific singularity structure as a bidistribution. This class of states plays a crucial role in ensuring the physical viability of a given quantum theory by allowing for a well-defined notion of renormalisation and the formulation of physically relevant observables.

\section{The Formalism of Hack-Schenkel and Cauchy Problem}\label{Sec:HackSchenkel}
In this section, we discuss linear field-theoretic models admitting local gauge symmetries on general globally hyperbolic spacetimes. As mentioned above, we follow the framework developed by Hack-Schenkel in \cite{HackSchenkel} (see also \cite[Sec.~2.2.2]{HackBook} for a textbook treatment), which axiomatises the \textit{subsidiary condition formalism} and generalises earlier treatments of Maxwell's theory \cite{DimockMaxwell} and linearised gravity \cite{FewsterHunt} on curved backgrounds to more general linear gauge theories.  Related ideas have also been explored by Khavkine \cite{KhavkinePhaseSpace} in the context of the \textit{covariant phase space formalism}. Within this formalism, the classical theory underlying quantisation is both well-understood and rigorously controlled, making it perfectly suited for the quantisation procedure to be described in the subsequent sections. We stress that our focus is restricted to \textit{linear} gauge theories and a \textit{simple} class of observables. In particular, there is no need to introduce \textit{auxiliary fields} (e.g.~\textit{(anti)ghosts}, \textit{Nakanishi–Lautrup fields}, \textit{antifields}, etc.)~as in the widely applied and highly successful Faddeev-Popov \cite{FaddeevPopov} (see \cite[Chap.~9]{RudolphSchmidt2} for a mathematically oriented treatment), BRST \cite{BRST1,BRST2,BRST3,BRST4} or BV formalisms \cite{Batalin1,Batalin2,Batalin3} (see also the reviews in \cite{Henneaux} and \cite[Chap.~7]{RejznerBook}) for gauge theories. In fact, the equivalence of the formalism of Hack-Schenkel and the BRST approach in the linear setting can straightforwardly be established and has been studied in some detail by Wrochna-Zahn in \cite{WrochnaZahn}.

\subsection{The Axiomatic Framework}
To begin with, we introduce the general definition of linear gauge theories within the Hack-Schenkel formalism \cite{HackSchenkel}. However, we shall adopt a slightly more refined version thereof introduced and discussed in detail by Gérard-Wrochna in \cite{GerardWrochna} and subsequently discussed and applied in  \cite{GerardMurroWrochna,BeniniMurro,WrochnaZahn,GerardWrochna2,Gerard,MurroSchmid} in various different contexts.

\begin{definition}\label{Def:LinGaugeTh} (Linear Gauge Theory)\newline
	Let $(\sf{M},\sf{g})$ be a globally hyperbolic spacetime and consider a quadruple $(\sf{E}_{1},\sf{E}_{2},\sf{P},\sf{K})$ consisting of the following data:
	\begin{itemize}
		\item[$\bullet$]Two Hermitian vector bundles $(\sf{E}_{1}\xrightarrow{\pi_{1}}\sf{M},\langle\cdot,\cdot\rangle_{\sf{E}_{1}})$ and $(\sf{E}_{2}\xrightarrow{\pi_{2}}\sf{M},\langle\cdot,\cdot\rangle_{\sf{E}_{2}})$.
		\item[$\bullet$]An operators $\sf{K}\in\mathrm{DO}(\sf{E}_{1},\sf{E}_{2})$ and a formally self-adjoint operator $\sf{P}\in\mathrm{DO}(\sf{E}_{2})$.
	\end{itemize}
	We call $(\sf{E}_{1},\sf{E}_{2},\sf{P},\sf{K})$ a \emph{linear gauge theory}, if it satisfies the following two properties:
	\begin{itemize}
		\item[(i)]$\sf{P}\circ\sf{K}=0$
		\item[(ii)]$\sf{D}_{1}:=\sf{K}^{\ast}\circ\sf{K}\in\mathrm{DO}(\sf{E}_{1})$ and $\sf{D}_{2}:=\sf{P}+\sf{K}\circ\sf{K}^{\ast}\in\mathrm{DO}(\sf{E}_{2})$ are Green-hyperbolic.
	\end{itemize}
\end{definition}

Let $(\sf{E}_{1},\sf{E}_{2},\sf{P},\sf{K})$ be a linear gauge theory on a globally hyperbolic spacetime $(\sf{M},\sf{g})$. Then, the various quantities involved in this setup admit the following interpretation:
\begin{itemize}
	\item[$\bullet$]The bundle $\sf{E}_{2}$ is the bundle whose sections correspond to the \textit{gauge fields} on $\sf{M}$, while sections of $\sf{E}_{1}$ encode the \textit{gauge transformation parameters}.
	\item[$\bullet$]The operator $\sf{P}$ determines the \textit{equations of motion} and hence the dynamics of the gauge fields, while $\sf{K}$ describes the \textit{linear gauge transformations} via
	\begin{align*}
		\Gamma^{\infty}(\sf{E}_{2})\ni \psi\mapsto \psi+\sf{K}\omega\qquad\omega\in\Gamma^{\infty}(\sf{E}_{1})\, .
	\end{align*}
	Condition (i) in Definition~\ref{Def:LinGaugeTh} then encodes \textit{gauge invariance} of $\sf{P}$ in the sense that
\begin{align*}
    \sf{P}\psi=\varphi \quad \Leftrightarrow \quad \sf{P}(\psi+\sf{K}\omega)=\varphi \qquad \forall\, \psi,\varphi\in\Gamma^{\infty}(\sf{E}_{2})\,,\,\omega \in \Gamma^{\infty}(\sf{E}_{1})\, .
\end{align*}
\item[$\bullet$]The operator $\sf{D}_{2}$ is the \textit{gauge-fixed operator} with respect to the \textit{canonical gauge condition} (or \textit{subsidiary condition}) $\sf{K}^{\ast}u=0$. Green hyperbolicity of $\sf{D}_{2}$ ensures that the Cauchy problem of the gauge-fixed operator is well-posed.
\item[$\bullet$]Green hyperbolicity of $\sf{D}_{1}$ ensures that the canonical gauge condition can always be achieved: let $\psi\in\Gamma^{\infty}(\sf{E}_{2})$. Finding a gauge transformation $\omega\in\Gamma^{\infty}(\sf{E}_{1})$ such that $\psi^{\prime}:=\psi+\sf{K}\omega$ satisfies $\sf{K}^{\ast}\psi^{\prime}=0$ is equivalent to solving the differential equations $\sf{D}_{1}\omega=-\sf{K}^{\ast}\psi$, which can always be done if $\sf{D}_{1}$ is hyperbolic.
\end{itemize}

\begin{remark}\label{Rem:HackSchenkel}
	In the original formulation of Hack-Schenkel, one considers three operators, $\sf{P}\in\mathrm{DO}(\sf{E}_{2})$ and $\sf{K},\sf{T}\in\mathrm{DO}(\sf{E}_{1},\sf{E}_{2})$. Condition (ii) in Definition~\ref{Def:LinGaugeTh} is then replaced by the requirements that $\widetilde{\sf{D}}_{2}:=\sf{P}+\sf{T}\sf{K}^{\ast}$, $\sf{D}_{1}:=\sf{K}^{\ast}\sf{K}$ and $\widetilde{\sf{D}}_{1}:=\sf{K}^{\ast}\sf{T}$ are Green hyperbolic. However, in virtually all examples of interest from a physical perspective, it turns out that the operators $\sf{T}$ and $\sf{K}$ are equivalent up to a constant and by choosing the bundle metrics on $\sf{E}_{1}$ and $\sf{E}_{2}$ appropriately, one can arrange $\sf{T}=\sf{K}$. To the best of my knowledge, the only example of some physical interest in which $\sf{T}$ and $\sf{K}$ are distinct and not proportional to each other is provided by the fermionic \textit{Rarita-Schwinger gauge theory}, in which case $\sf{P}$ and $\sf{K}$ are operators of order one, while $\sf{T}$ an operator of order zero, see~\cite[Ex.~3.10]{HackSchenkel}.
\end{remark}

Before turning to the Cauchy problem of linear gauge theories, we collect some useful properties of linear gauge theories following~\cite[Thm.~3.12]{HackSchenkel} and \cite[Prop.~2.5]{GerardWrochna}.

\begin{proposition}\label{Prop:PropertiesLinGaugTh} Let $(\sf{E}_{1},\sf{E}_{2},\sf{P},\sf{K})$ be a linear gauge theory with Green hyperbolic operators $\sf{D}_{1}:=\sf{K}^{\ast}\sf{K}$ and $\sf{D}_{2}:=\sf{P}+\sf{K}\sf{K}^{\ast}$. Denote the corresponding Green operators by $\sf{G}_{1,2}^{\pm}$. Then
	\begin{align*}	
		\text{(i)}\quad &\sf{K}\circ\sf{D}_{1}=\sf{D}_{2}\circ\sf{K}\qquad\text{and}\qquad \sf{D}_{1}\circ\sf{K}^{\ast}=\sf{K}^{\ast}\circ\sf{D}_{2}\,;\\
		\text{(ii)}\quad & \sf{K}\circ\sf{G}^{\pm}_{1}=\sf{G}_{2}^{\pm}\circ\sf{K}\qquad\text{and}\qquad\sf{G}^{\pm}_{1}\circ\sf{K}^{\ast}=\sf{K}^{\ast}\circ\sf{G}_{2}^{\pm}\quad\text{on}\quad\Gamma_{\mathrm{c}}^{\infty}(\sf{E}_{i})\, .
	\end{align*}
\end{proposition}

\begin{proof}
	Claim (i) follows directly from the definitions and the fact that $\sf{P}\circ\sf{K}=0$. For the first claim in (ii), let $\psi\in\Gamma^{\infty}_{\mathrm{c}}(\sf{E}_{2})$ and $\varphi\in\Gamma^{\infty}_{\mathrm{c}}(\sf{E}_{1})$. Then,
	\begin{align*}
		(\psi,\sf{K}\sf{G}_{1}^{\pm}\varphi)_{\sf{E}_{2}}&=(\sf{D}_{2}\sf{G}_{2}^{\mp}\psi,\sf{K}\sf{G}_{1}^{\pm}\varphi)_{\sf{E}_{2}}=(\sf{G}_{2}^{\mp}\psi,\sf{D}_{2}\sf{K}\sf{G}_{1}^{\pm}\varphi)_{\sf{E}_{2}}\stackrel{(i)}{=}(\sf{G}_{2}^{\mp}\psi,\sf{K}\varphi)_{\sf{E}_{2}}=(\psi,\sf{G}_{2}^{\pm}\sf{K}\varphi)_{\sf{E}_{2}}\,,
	\end{align*}
	where we used (i) and Proposition~\ref{Prop:GreenProp}(iii). By non-degeneracy of $(\cdot,\cdot)_{\sf{E}_{2}}$, we conclude that $\sf{K}\sf{G}_{1}^{\pm}=\sf{G}_{2}^{\pm}\sf{K}$. The second claim now follows by taking the formal adjoint of the latter identity.
\end{proof}

\subsection{The Cauchy Problem for Linear Gauge Theories}\label{Sec:CauchyGauge}
Let $(\sf{M},\sf{g})$ be a globally hyperbolic spacetime and $(\sf{E}_{1},\sf{E}_{2},\sf{P},\sf{K})$ be a linear gauge theory as per Definition~\ref{Def:LinGaugeTh}. The case $\sf{K}=0$ corresponds to the special case in which there is no gauge symmetry present. In this case, the equations of motion described by the operator $\sf{P}$ are themselves hyperbolic. Examples are provided, for instance, by Klein-Gordon, Dirac or Proca fields. When $\sf{K}\neq 0$, however, it is not too hard to see that $\sf{P}$ is generically \textit{non}-hyperbolic. Intuitively, this can be easily seen as follows. Suppose $\psi$ is a solution to the equation $\sf{P}\psi=0$ for some initial datum $\psi\vert_{\Sigma}=\mathfrak{f}$ on some spacelike Cauchy hypersurface $\Sigma$. Now, choose any $\omega\in\Gamma^{\infty}(\sf{E}_{1})$ whose support lies sufficiently far in the future of $\Sigma$. Then, $\psi^{\prime}:=\psi+\sf{K}\omega$ is also a solution of $\sf{P}\psi^{\prime}=0$ with the same initial datum. Hence, the Cauchy problem is not well-posed. As it turns out, the operator $\sf{P}$ is in fact not even Green hyperbolic.

\begin{lemma} Let $(\sf{E}_{1},\sf{E}_{2},\sf{P},\sf{K})$ be a linear gauge theory. Then
\begin{align*}
	\sf{K}=0\qquad\Leftrightarrow\qquad \sf{P}\text{ is Green hyperbolic}\, .
\end{align*}
\end{lemma}

\begin{proof}
	Implication ``$\Rightarrow$'' is obvious. For ``$\Leftarrow$'', assume that $\sf{K}\neq 0$. Now, if $\sf{P}$ were Green hyperbolic, then we would have $\mathrm{ker}(\sf{P}\vert_{\Gamma^{\infty}_{\mathrm{c}}})=\{0\}$ by Proposition~\ref{Prop:ExactSeq}(iv). However, this is clearly not the case, since any $\psi:=\sf{K}\omega$ with $\omega\in\Gamma^{\infty}_{\mathrm{c}}(\sf{E}_{1})\backslash\{0\}$ is an element of $\mathrm{ker}(\sf{P}\vert_{\Gamma^{\infty}_{\mathrm{c}}})$.
\end{proof}

Hence, in order to discuss the solution space of a linear gauge theory, we must properly account for the gauge redundancy: rather than considering all solutions to the equations of motion, we are interested in \textit{gauge equivalent classes} of solutions. More precisely the space we want to work with is the quotient space 
\begin{align*}
	\mathrm{Sol}_{\mathrm{sc}}:=\cfrac{\mathrm{ker}(\sf{P}\vert_{\Gamma^{\infty}_{\mathrm{sc}}})}{\mathrm{ran}(\sf{K}\vert_{\Gamma^{\infty}_{\mathrm{sc}}})}\, .
\end{align*}
In other words, two solutions of the equation $\sf{P}\psi=0$, or \textit{on-shell configurations} in the physics terminology, are identified whenever they differ by a linear gauge transformation. In the following exposition, we follow the discussion from \cite{GerardWrochna,GerardMurroWrochna}.

Let $\psi\in\mathrm{ker}(\sf{P}\vert_{\Gamma^{\infty}_{\mathrm{sc}}})$. Then, we can always find a gauge-transformation $\omega\in\Gamma^{\infty}_{\mathrm{sc}}(\sf{E}_{1})$ such that $\psi^{\prime}:=\psi-\sf{K}\omega$ satisfies the canonical gauge condition $\sf{K}^{\ast}\psi^{\prime}=0$ and hence also  $\sf{D}_{2}\psi^{\prime}=0$. Indeed, the requirement $\sf{K}^{\ast}\psi^{\prime}=0$ is equivalent to the differential equation $\sf{D}_{1}\omega=\sf{K}^{\ast}\psi$, which always admits a solution in $\Gamma^{\infty}_{\mathrm{sc}}(\sf{E}_{1})$, by hyperbolicity of $\sf{D}_{1}$. Hence, any equivalence class in~$\mathrm{Sol}_{\mathrm{sc}}$ has a representative in the subspace $\mathrm{ker}(\sf{D}_{2}\vert_{\Gamma^{\infty}_{\mathrm{sc}}})\cap\mathrm{ker}(\sf{K}^{\ast}\vert_{\Gamma^{\infty}_{\mathrm{sc}}})$. In other words, the canonical gauge condition is always \textit{achievable} due to the requirement that $\sf{D}_{1}$ is hyperbolic. However, there is still some residual gauge freedom: if we perform another gauge transformation $\psi^{\prime\prime}:=\psi^{\prime}-\sf{K}\omega^{\prime}$, then we still have that $\psi^{\prime\prime}\in \mathrm{ker}(\sf{D}_{2}\vert_{\Gamma^{\infty}_{\mathrm{sc}}})\cap\mathrm{ker}(\sf{K}^{\ast}\vert_{\Gamma^{\infty}_{\mathrm{sc}}})$ provided that the gauge transformation parameter $\omega^{\prime}$ is chosen in such a way that $\omega^{\prime}\in \sf{K}(\mathrm{ker}(\sf{D}_{1}\vert_{\Gamma^{\infty}_{\mathrm{sc}}}))$. 

The previous arguments allow us to identify the solution space $\mathrm{Sol}_{\mathrm{sc}}$ with a quotient space defined in terms of the constrained solution space of the hyperbolic operator $\sf{D}_{2}$.

\begin{proposition}\label{Proposition2}
    Let $(\sf{E}_{1},\sf{E}_{2},\sf{P},\sf{K})$ be a linear gauge theory with Green hyperbolic operators $\sf{D}_{1}:=\sf{K}^{\ast}\sf{K}$ and $\sf{D}_{2}:=\sf{P}+\sf{K}\sf{K}^{\ast}$. Then, we have the following linear isomorphism:
    \begin{equation*}
        \begin{tikzcd}
        		\cfrac{\mathrm{ker}(\sf{D}_{2}\vert_{\Gamma^{\infty}_{\mathrm{sc}}})\cap\mathrm{ker}(\sf{K}^{\ast}\vert_{\Gamma^{\infty}_{\mathrm{sc}}})}{\sf{K}(\mathrm{ker}(\sf{D}_{1}\vert_{\Gamma^{\infty}_{\mathrm{sc}}}))}\arrow[r,hookrightarrow,swap,"{[\mathrm{id}]}"]\arrow[r,hookrightarrow,"\cong"] &\mathrm{Sol}_{\mathrm{sc}}=\cfrac{\mathrm{ker}(\sf{P}\vert_{\Gamma^{\infty}_{\mathrm{sc}}})}{\mathrm{ran}(\sf{K}\vert_{\Gamma^{\infty}_{\mathrm{sc}}})}
        \end{tikzcd}
    \end{equation*}
\end{proposition}

\begin{proof}
	First of all, we note that $[\mathrm{id}]$ is clearly well-defined, as it maps both the numerator and the denominator into the corresponding numerator and denominator of the respective quotient spaces. By Proposition~\ref{App:Quot}, injectivity amounts to show that
		\begin{align*}
			\textit{injectivity:}\quad \mathrm{ker}(\sf{D}_{2}\vert_{\Gamma^{\infty}_{\mathrm{sc}}})\cap\mathrm{ker}(\sf{K}^{\ast}\vert_{\Gamma^{\infty}_{\mathrm{sc}}})\cap \mathrm{ran}(\sf{K}\vert_{\Gamma^{\infty}_{\mathrm{sc}}})\subset\sf{K}(\mathrm{ker}(\sf{D}_{1}\vert_{\Gamma^{\infty}_{\mathrm{sc}}}))\, .
		\end{align*}
		Take $\psi=\sf{K}\varphi$ for some $\varphi\in\Gamma^{\infty}_{\mathrm{sc}}(\sf{E}_{1})$ such that $\sf{D}_{2}\psi=\sf{K}^{\ast}\psi=0$. Then, $\sf{D}_{1}\varphi=\sf{K}^{\ast}\sf{K}\varphi=\sf{K}^{\ast}\psi=0$, which shows that $\varphi\in\mathrm{ker}(\sf{D}_{1}\vert_{\Gamma^{\infty}_{\mathrm{sc}}})$. For surjectivity, we need to show that 
		\begin{align*}
			\textit{surjectivity:}\quad \mathrm{ker}(\sf{P}\vert_{\Gamma^{\infty}_{\mathrm{sc}}})=(\mathrm{ker}(\sf{D}_{2}\vert_{\Gamma^{\infty}_{\mathrm{sc}}})\cap\mathrm{ker}(\sf{K}^{\ast}\vert_{\Gamma^{\infty}_{\mathrm{sc}}}))+\mathrm{ran}(\sf{K}\vert_{\Gamma^{\infty}_{\mathrm{sc}}})\, .
		\end{align*}
		This is exactly the statement that the canonical gauge-condition $\sf{K}^{\ast}\psi=0$ can always be achieved on-shell. Indeed, take $\psi\in\mathrm{ker}(\sf{P}\vert_{\Gamma^{\infty}_{\mathrm{sc}}})$. Then, we need to find a gauge transformation $\omega\in\Gamma^{\infty}_{\mathrm{sc}}(\sf{E}_{1})$ such that $\psi^{\prime}=\psi+\sf{K}\omega$ satisfies $\sf{K}^{\ast}\psi^{\prime}=0$ (and hence also $\sf{D}_{2}\psi^{\prime}=0$ on account of $\sf{P}\psi=0$ and $\sf{P}\circ\sf{K}=0$). This is equivalent to finding a solution of the equation $\sf{D}_{1}\omega=-\sf{K}^{\ast}\psi$, which is always possible due to the hyperbolicity of $\sf{D}_{1}$.
\end{proof}

\begin{remark} The inverse of the isomorphism $[\mathrm{id}]$ is given by the quotient map induced from \textit{gauge-fixing}, i.e.
    \begin{align*}
       \frac{\mathrm{ker}(\sf{P}\vert_{\Gamma^{\infty}_{\mathrm{sc}}})}{\mathrm{ran}(\sf{K}\vert_{\Gamma^{\infty}_{\mathrm{sc}}})}\ni [\psi]\mapsto [\psi-\sf{K}\omega]\in \cfrac{\mathrm{ker}(\sf{D}_{2}\vert_{\Gamma^{\infty}_{\mathrm{sc}}})\cap\mathrm{ker}(\sf{K}^{\ast}\vert_{\Gamma^{\infty}_{\mathrm{sc}}})}{\sf{K}(\mathrm{ker}(\sf{D}_{1}\vert_{\Gamma^{\infty}_{\mathrm{sc}}}))}\,.
    \end{align*}
    where $\omega\in\Gamma^{\infty}_{\mathrm{sc}}(\sf{E}_{1})$ is such that $\sf{D}_{1}\omega=\sf{K}^{\ast}\psi$. By hyperbolicity of $\sf{D}_{1}$ such a $\omega$ always exists and the equivalence class on the right-hand side is clearly independent of this choice.
\end{remark}

By the previous discussion, we have reduced the analysis of gauge-equivalence classes of solutions of a linear gauge theory to the study of hyperbolic operators with constraints. In order to study the corresponding Cauchy problem, we have to introduce notation for initial data as well as for gauge transformations acting thereon.

First of all, since the operators $\sf{D}_{1}:=\sf{K}^{\ast}\sf{K}$ and $\sf{D}_{2}:=\sf{P}+\sf{K}\sf{K}^{\ast}$ of are Green hyperbolic, we can find pairs $(\rho_{1},\sf{E}_{\rho_{1}})$ and $(\rho_{2},\sf{E}_{\rho_{2}})$ where $\sf{E}_{\rho_{i}}\xrightarrow{}\Sigma$ are vector bundles over a (smooth and spacelike) Cauchy hypersurface $\Sigma$ of $\sf{M}$ and $\rho_{i}$ the corresponding initial data maps
\begin{align*}
	\rho_{i}\:\Gamma^{\infty}_{\mathrm{sc}}(\sf{E}_{i})\to\Gamma^{\infty}_{\mathrm{c}}(\sf{E}_{\rho_{i}})
\end{align*}
for $i=1,2$, namely the composition of the restriction map $r\:\sf{M}\to\Sigma$ and a differential operator of lower order than $\sf{D}_{i}$. By definition, the restriction of $\rho_{i}$ to $\mathrm{ker}(\sf{D}_{i}\vert_{\Gamma^{\infty}_{\mathrm{sc}}})$ is invertible and the corresponding inverse are the Cauchy evolution operators
\begin{align*}
	\mathcal{U}_{i}:=(\rho\vert_{\mathrm{ker}(\sf{D}_{i}\vert_{\Gamma^{\infty}_{\mathrm{sc}}})})^{-1}\:\Gamma_{\mathrm{c}}^{\infty}(\sf{E}_{\rho_{i}})\to\mathrm{ker}(\sf{D}_{i}\vert_{\Gamma^{\infty}_{\mathrm{sc}}})\subset\Gamma^{\infty}_{\mathrm{sc}}(\sf{E}_{i})\, .
\end{align*}
The operators $\rho_{i}$ and $\mathcal{U}_{i}$ allow us to describe gauge transformations and the canonical gauge condition on the level of initial data by introducing the following operators.

\begin{definition} \label{Def:KSigma}
	Let $(\sf{E}_{1},\sf{E}_{2},\sf{P},\sf{K})$ be a linear gauge theory, Then, we define
	\begin{align*}
		\sf{K}_{\Sigma}:=\rho_{2}\circ\sf{K}\circ\mathcal{U}_{1}\qquad\text{and}\qquad \sf{K}_{\Sigma}^{\dagger}:=\rho_{1}\circ\sf{K}^{\ast}\circ\mathcal{U}_{2}\, .
	\end{align*}
\end{definition}

The origin of the notation $\sf{K}_{\Sigma}^{\dagger}$ will become clear in the next section, where we will show that $\sf{K}^{\dagger}_{\Sigma}$ is the formal adjoint with respect to a natural Hermitian form on $\sf{E}_{\rho_{i}}$ (see Corollary~\ref{Cor:KSigmaAdj} below). For later convenience, we note the following properties.

\begin{proposition}\label{Lem:PropKSigma} \emph{(Properties of $\sf{K}_{\Sigma}$ and $\sf{K}_{\Sigma}^{\dagger}$)}\newline
	The operators $\sf{K}_{\Sigma}$ and $\sf{K}_{\Sigma}^{\dagger}$ introduced in Definition~\ref{Def:KSigma} have the following properties:
	\begin{align*}
		\text{\emph{(i)}}\quad &\sf{K}_{\Sigma}\in\mathrm{DO}(\sf{E}_{\rho_{1}},\sf{E}_{\rho_{2}})\qquad\text{and}\qquad\sf{K}_{\Sigma}^{\dagger}\in\mathrm{DO}(\sf{E}_{\rho_{2}},\sf{E}_{\rho_{1}})\,;\\
		\text{\emph{(ii)}}\quad &\sf{K}\circ\mathcal{U}_{1}=\mathcal{U}_{2}\circ\sf{K}_{\Sigma}\qquad\text{and}\qquad \sf{K}^{\ast}\circ\mathcal{U}_{2}=\mathcal{U}_{1}\circ\sf{K}_{\Sigma}^{\dagger}\,;\\
		\text{\emph{(iii)}}\quad & \rho_{2}\circ\sf{K}=\sf{K}_{\Sigma}\circ\rho_{1}\qquad\text{and}\qquad\rho_{1}\circ\sf{K}^{\ast}=\sf{K}_{\Sigma}^{\dagger}\circ\rho_{2}\quad\text{on}\quad\mathrm{ker}(\sf{D}_{i}\vert_{\Gamma^{\infty}_{\mathrm{sc}}})\,;\\
		\text{\emph{(iv)}}\quad &\sf{K}_{\Sigma}^{\dagger}\circ\sf{K}_{\Sigma}=0\, .
	\end{align*}
\end{proposition}

\begin{proof}
	For (i), it is enough to show that $\sf{K}_{\Sigma}$ and $\sf{K}_{\Sigma}^{\dagger}$ are \textit{local} operators, see~Theorem~\ref{Thm:Peetre}, which follows from the fact that $\rho_{i}$ is the composition of the restriction map and a differential operator. Both (ii) and (iii) follow immediately from the definition and the fact that $\rho_{i}$ is invertible as an operator from $\mathrm{ker}(\sf{D}_{i}\vert_{\Gamma^{\infty}_{\mathrm{sc}}})$ to $\Gamma^{\infty}_{\mathrm{c}}(\sf{E}_{\rho_{i}})$. For (iv), we observe that
	\begin{align*}
		\sf{K}_{\Sigma}^{\dagger}\sf{K}_{\Sigma}=\rho_{1}\sf{K}^{\ast}\mathcal{U}_{2}\rho_{2}\sf{K}\mathcal{U}_{1}=\rho_{1}\sf{K}^{\ast}\sf{K}\mathcal{U}_{1}=\rho_{1}\sf{D}_{1}\mathcal{U}_{1}=0\, ,
	\end{align*}
	where in the last step we used the fact that $\sf{D}_{1}\circ\mathcal{U}_{i}=0$.
\end{proof}

Now, the operator $\sf{K}_{\Sigma}^{\dagger}$ parametrises the canonical gauge condition $\sf{K}^{\ast}\psi=0$ on the level of initial data. More precisely, if we examine the constrained Cauchy problem
\begin{align*}
	\begin{cases}
		\sf{D}_{2}\psi&=\varphi\\
		\sf{K}^{\ast}\psi&=0\\
		\rho_{2}(\psi)&=\mathfrak{f}
	\end{cases}
\end{align*}
with Cauchy data $(\varphi,\mathfrak{f})\in\Gamma^{\infty}(\sf{E}_{2})\times\Gamma^{\infty}(\sf{E}_{\rho_{2}})$, we see that the condition $\sf{K}^{\ast}\psi=0$ is satisfied if and only if the initial data satisfy $\sf{K}_{\Sigma}^{\dagger}\mathfrak{f}=0$. Taking into account also the gauge transformations, which on the level of initial data are described by the operator $\sf{K}_{\Sigma}$, one obtains the following description of the Cauchy problem for linear gauge theories.

\begin{proposition}\label{Prop:CauchyGauge} \emph{(The Cauchy Problem of Linear Gauge Theories)}\newline
	 Let $(\sf{E}_{1},\sf{E}_{2},\sf{P},\sf{K})$ be a linear gauge theory on a globally hyperbolic spacetime $(\sf{M},\sf{g})$ with Green hyperbolic operators $\sf{D}_{1}:=\sf{K}^{\ast}\sf{K}$ and $\sf{D}_{2}:=\sf{P}+\sf{K}\sf{K}^{\ast}$. Furthermore, let $\Sigma$ be any smooth, spacelike Cauchy hypersurface of $\sf{M}$ and $\rho_{i}\:\Gamma^{\infty}_{\mathrm{sc}}(\sf{E}_{i})\to\Gamma^{\infty}_{\mathrm{c}}(\sf{E}_{i})$ be the initial data maps of $\sf{D}_{i}$ for $i=1,2$. Then, we have the following isomorphisms:
    \begin{equation*}
        \begin{tikzcd}
        		\mathrm{Sol}_{\mathrm{sc}}=\cfrac{\mathrm{ker}(\sf{P}\vert_{\Gamma^{\infty}_{\mathrm{sc}}})}{\mathrm{ran}(\sf{K}\vert_{\Gamma^{\infty}_{\mathrm{sc}}})}\cong\cfrac{\mathrm{ker}(\sf{D}_{2}\vert_{\Gamma^{\infty}_{\mathrm{sc}}})\cap\mathrm{ker}(\sf{K}^{\ast}\vert_{\Gamma^{\infty}_{\mathrm{sc}}})}{\sf{K}(\mathrm{ker}(\sf{D}_{1}\vert_{\Gamma^{\infty}_{\mathrm{sc}}}))}\arrow[r,swap,"{[\rho_{2}]}"]\arrow[r,"\cong"] &\cfrac{\mathrm{ker}(\sf{K}_{\Sigma}^{\dagger}\vert_{\Gamma_{\mathrm{c}}^{\infty}})}{\mathrm{ran}(\sf{K}_{\Sigma}\vert_{\Gamma_{\mathrm{c}}^{\infty}})}
        \end{tikzcd}
    \end{equation*}
\end{proposition}

\begin{proof}
	First of all, if $\psi\in\Gamma_{\mathrm{sc}}^{\infty}(\sf{E}_{2})$ is such that $\sf{D}_{2}\psi=\sf{K}^{\ast}\psi=0$, then it holds that $\sf{K}_{\Sigma}^{\dagger}\rho_{2}\psi=\rho_{1}\sf{K}^{\ast}\psi=0$, by Lemma~\ref{Lem:PropKSigma}(iii). On the other hand, if we consider $\psi:=\sf{K}\omega$ for some $\omega\in\Gamma_{\mathrm{sc}}^{\infty}(\sf{E}_{1})$ satisfying $\sf{D}_{1}\omega=0$, Lemma~\ref{Lem:PropKSigma} implies that $\rho_{2}\psi=\rho_{2}\sf{K}\omega=\sf{K}_{\Sigma}\rho_{1}\omega$. Hence, $[\rho_{2}]$ is well-defined. By Proposition~\ref{App:Quot}, to show injectivity, we need to show that
	\begin{align*}
			\textit{injectivity:}\quad \rho_{2}^{-1}(\mathrm{ran}(\sf{K}_{\Sigma}\vert_{\Gamma_{\mathrm{c}}^{\infty}}))\subset\sf{K}(\mathrm{ker}(\sf{D}_{1}\vert_{\Gamma^{\infty}_{\mathrm{sc}}}))\, .
		\end{align*}
		Let $\psi\in\Gamma^{\infty}_{\mathrm{sc}}(\sf{E}_{2})$ be such that $\rho_{2}\psi=\sf{K}_{\Sigma}\mathfrak{f}$ for some $\mathfrak{f}\in\Gamma^{\infty}_{\mathrm{c}}(\sf{E}_{\rho_{1}})$. Then, $\psi=\mathcal{U}_{2}\rho_{2}\psi=\mathcal{U}_{2}\sf{K}_{\Sigma}\mathfrak{f}=\sf{K}\mathcal{U}_{1}\mathfrak{f}$, which proves the inclusion, since $\mathcal{U}_{1}\mathfrak{f}\in \mathrm{ker}(\sf{D}_{1}\vert_{\Gamma^{\infty}_{\mathrm{sc}}})$. For surjectivity, we need to show that
		\begin{align*}
			\textit{surjectivity:}\quad \mathrm{ker}(\sf{K}_{\Sigma}^{\dagger}\vert_{\Gamma_{\mathrm{c}}^{\infty}})=\rho_{2}(\mathrm{ker}(\sf{D}_{2}\vert_{\Gamma^{\infty}_{\mathrm{sc}}})\cap\mathrm{ker}(\sf{K}^{\ast}\vert_{\Gamma^{\infty}_{\mathrm{sc}}}))+\mathrm{ran}(\sf{K}_{\Sigma}\vert_{\Gamma_{\mathrm{c}}^{\infty}})\, .
		\end{align*}
		However, this is essentially trivial: the inclusion ``$\supset$'' follows from $\sf{K}_{\Sigma}^{\dagger}\sf{K}_{\Sigma}=0$ and the fact that $\rho_{1}\circ\sf{K}^{\ast}=\sf{K}^{\dagger}_{\Sigma}\circ\rho_{2}$ on $\mathrm{ker}(\sf{D}_{2}\vert_{\Gamma_{\mathrm{sc}}^{\infty}})$ by Lemma~\ref{Lem:PropKSigma}(iii), whereas the inclusion ``$\subset$'' follows from the identity $\mathfrak{f}=\rho_{2}\mathcal{U}_{2}\mathfrak{f}$ and the fact that $\psi:=\mathcal{U}_{2}\mathfrak{f}\in\mathrm{ker}(\sf{D}_{2}\vert_{\Gamma^{\infty}_{\mathrm{sc}}})\cap\mathrm{ker}(\sf{K}^{\ast}\vert_{\Gamma^{\infty}_{\mathrm{sc}}})$ if $\mathfrak{f}\in \mathrm{ker}(\sf{K}_{\Sigma}^{\dagger}\vert_{\Gamma_{\mathrm{c}}^{\infty}})$.
\end{proof}

\subsection{Classical Phase Space of a Linear Gauge Theory}\label{Sec:PhaseGauge}
Following the mathematical treatment of the Cauchy problem for linear gauge theories, we now turn to the description of their \textit{classical phase space}, namely the space of observables. For pedagogical reasons, we first discuss the case of \textit{ordinary field theories}, i.e.~the special case $\sf{K}=0$ in Definition~\ref{Def:LinGaugeTh}. For more details in the context of Klein-Gordon, Proca and Dirac fields, we refer to~\cite{BeniniDappiaggi}. Let $(\sf{M},\sf{g})$ be a globally hyperbolic spacetime and $\sf{D}\in\mathrm{DO}(\sf{E})$ be a Green hyperbolic operator on a Hermitian vector bundle $(\sf{E}\xrightarrow{\pi}\sf{M},\langle\cdot,\cdot\rangle_{\sf{E}})$. We denote by $\sf{G}^{\pm}$ the advanced/retarded Green operators of $\sf{D}$ and the causal propagator by $\sf{G}:=\sf{G}^{+}-\sf{G}^{-}$. 

The exact sequence described in Proposition~\ref{Prop:ExactSeq}(iv) implies $\mathrm{ran}(\sf{G}\vert_{\Gamma^{\infty}_{\mathrm{c}}})=\mathrm{ker}(\sf{D}\vert_{\Gamma^{\infty}_{\mathrm{sc}}})$ and $\mathrm{ker}(\sf{G}\vert_{\Gamma^{\infty}_{\mathrm{c}}})=\mathrm{ran}(\sf{D}\vert_{\Gamma^{\infty}_{\mathrm{c}}})$. Therefore, the causal propagator $\sf{G}$ of $\sf{D}$ induces a linear isomorphism
\begin{align}\label{eq:IsoPhaseOFT}
	\mathcal{V}_{\mathrm{c}}:=\cfrac{\Gamma^{\infty}_{\mathrm{c}}(\sf{E})}{\mathrm{ran}(\sf{D}\vert_{\Gamma^{\infty}_{\mathrm{c}}})}\xrightarrow[\cong]{\quad[\sf{G}]\quad}\mathrm{ker}(\sf{D}\vert_{\Gamma_{\mathrm{sc}}^{\infty}})=:\mathrm{Sol}_{\mathrm{sc}}\, .
\end{align}
The vector space $\mathcal{V}_{\mathrm{c}}$ is called the \textit{classical phase space} of $\sf{D}$. Now, for a given $f\in\Gamma^{\infty}_{\mathrm{c}}(\sf{E})$, a \textit{linear observable} is a $\bb{C}$-linear functional of the form
\begin{align}\label{eq:Obs}
	\mathcal{O}_{f}\:\Gamma^{\infty}(\sf{E})\to\bb{C}\,,\qquad \psi\mapsto (f,\psi)_{\sf{E}}=\int_{\sf{M}}\langle f,\psi\rangle_{\sf{E}}\,\d\mu_{\sf{g}}\, .
\end{align}
In other words, the observable labelled by $f$ assigns to every \textit{field configuration} $\psi\in\Gamma^{\infty}(\sf{E})$ a complex number. Now, since $\langle\cdot,\cdot\rangle_{\sf{E}}$ is assumed to be non-degenerate, the space of labels $\Gamma^{\infty}_{\mathrm{c}}(\sf{E})$ is in one-to-one correspondence with the space of observables of the form~\eqref{eq:Obs}. In order to account for the dynamics of the field theory in question, we would like to restrict our observables to the space of \textit{on-shell field configurations} $\mathrm{Sol}_{\mathrm{sc}}=\mathrm{ker}(\sf{D}\vert_{\Gamma_{\mathrm{sc}}^{\infty}})$. However, observables on $\mathrm{Sol}_{\mathrm{sc}}$ are no longer uniquely labelled by sections in $\Gamma^{\infty}_{\mathrm{c}}(\sf{E})$. For instance, every $f=\sf{D}g$ for some $g\in\Gamma^{\infty}_{\mathrm{c}}(\sf{E})$ induces the \textit{trivial observable} on $\mathrm{Sol}_{\mathrm{sc}}$, since
\begin{align}\label{eq:TrivObs}
	\mathcal{O}_{f}(\psi)=\mathcal{O}_{\sf{D}g}(\psi)=(\sf{D}g,\psi)_{\sf{E}}=(g,\sf{D}\psi)=0\qquad\forall\psi\in\mathrm{Sol}_{\mathrm{sc}}\, .
\end{align}
Hence, we have to consider a suitable quotient space of $\Gamma^{\infty}_{\mathrm{c}}(\sf{E})$ that identifies those labels that encode the same linear observables on $\mathrm{Sol}_{\mathrm{sc}}$. It turns out that this space is exactly given by $\mathcal{V}_{\mathrm{c}}$.

\begin{proposition}
	Let $\sf{D}$ be a Green hyperbolic operator. Then, we have the isomorphism
	\begin{align*}
		\mathcal{V}_{\mathrm{c}}:=\cfrac{\Gamma^{\infty}_{\mathrm{c}}(\sf{E})}{\mathrm{ran}(\sf{D}\vert_{\Gamma^{\infty}_{\mathrm{c}}})}\xrightarrow[\cong]{[f]\,\,\mapsto\,\, \mathcal{O}_{f}\vert_{\mathrm{Sol}_{\mathrm{sc}}}}\{\mathcal{O}_{f}\vert_{\mathrm{Sol}_{\mathrm{sc}}}\mid f\in\Gamma_{\mathrm{c}}^{\infty}(\sf{E})\}\,,
	\end{align*}
	where $\mathcal{O}_{f}(\bullet):=(f,\bullet)_{\sf{E}}\:\Gamma^{\infty}(\sf{E})\to\bb{C}$ is defined in accordance with Eq.~\eqref{eq:Obs}.
\end{proposition}

\begin{proof}
	To show that $[f]\mapsto\mathcal{O}_{f}\vert_{\mathrm{Sol}_{\mathrm{sc}}}$ is well-defined and bijective it is enough to show that
	\begin{align*}
		\mathrm{ran}(\sf{D}\vert_{\Gamma^{\infty}_{\mathrm{c}}})=\{f\in\Gamma^{\infty}_{\mathrm{c}}(\sf{E})\mid \mathcal{O}_{f}=0\text{ on }\mathrm{Sol}_{\mathrm{sc}}\}\, .
	\end{align*}
	The inclusion ``$\subset$'' has been shown in Eq.~\eqref{eq:TrivObs} above. For ``$\supset$'', let $f\in\Gamma^{\infty}_{\mathrm{c}}(\sf{E})$ be such that $\mathcal{O}_{f}(\psi)=0$ for all $\psi\in\mathrm{Sol}_{\mathrm{sc}}$. In other words, $(f,\psi)_{\sf{E}}=0$ for all $\psi\in\mathrm{Sol}_{\mathrm{sc}}$. Now, by Proposition~\ref{Prop:ExactSeq}(iv), every $\psi\in\mathrm{Sol}_{\mathrm{sc}}$ can be written as $\psi=\sf{G}\varphi$ for some $\varphi\in\Gamma^{\infty}_{\mathrm{c}}(\sf{E})$. Hence, this is equivalent to the fact that $(f,\sf{G}\varphi)_{\sf{E}}=0$ for all $\varphi\in\Gamma^{\infty}_{\mathrm{c}}(\sf{E})$. But now, recall that $(f,\sf{G}\varphi)_{\sf{E}}=-(\sf{G}f,\varphi)_{\sf{E}}$, by Proposition~\ref{Prop:ExactSeq}(iii). Hence, non-degeneracy of $(\cdot,\cdot)_{\sf{E}}$ implies $\sf{G}f=0$. Using again Proposition~\ref{Prop:ExactSeq}(iv), we conclude that $f=\sf{D}g$ for some $g\in\Gamma^{\infty}_{\mathrm{c}}(\sf{E})$.
\end{proof}

\begin{remark}
	While the class of observables we consider has a relatively simple structure at first sight, it is rich enough to distinguish between different configurations: for two \textit{off-shell} configurations, i.e.~sections $\psi,\varphi\in\Gamma^{\infty}(\sf{E})$, we can always find an observable $\mathcal{O}_{f}$ labelled by some $f\in\Gamma_{\mathrm{c}}^{\infty}(\sf{E})$ such that $\mathcal{O}_{f}(\psi)\neq\mathcal{O}_{f}(\varphi)$. A similar statement is true for linear \textit{on-shell} observables, since taking the quotient does not alter this property. From a physical perspective, this means that there is no need to consider a larger class of observables than $\{\mathcal{O}_{f}\}_{f\in\Gamma^{\infty}_{\mathrm{c}}(\sf{E})}$.
\end{remark}

In summary, the classical phase space $\mathcal{V}_{\mathrm{c}}$ is in one-to-one correspondence with the space of \textit{linear on-shell observables} of the type~\eqref{eq:Obs}. Before moving on to gauge theories, we conclude this preliminary discussion with the observation that the causal propagator equips the phase space $\mathcal{V}_{\mathrm{c}}$ naturally with the structure of a Hermitian vector space.

\begin{proposition}\label{Prop:SympOFT} Let $\sf{D}$ be a Green hyperbolic operator and $\mathcal{V}_{\mathrm{c}}$ as in Eq.~\eqref{eq:IsoPhaseOFT}. Then, the map $\sigma\:\mathcal{V}_{\mathrm{c}}\times\mathcal{V}_{\mathrm{c}}\to\bb{C}$ defined by
\begin{align*}
	\sigma([\psi],[\varphi]):=(\psi,i\sf{G}\varphi)_{\sf{E}}
\end{align*}
for all $\psi,\varphi\in\Gamma^{\infty}_{\mathrm{c}}(\sf{E})$ is a well-defined weakly non-degenerate\footnote{A Hermitian sesquilinear form $\sigma$ on a $\bb{C}$-vector space $\sf{V}$ is \textit{weakly non-degenerate}, if $\sf{V}\to\sf{V}^{\ast},v\mapsto\sigma(v,\bullet)$ is injective. If it is bijective, $\sigma$ is \textit{non-degenerate}. For $\mathrm{dim}_{\bb{C}}(\sf{V})<\infty$, these notions coincide.} Hermitian sesquilinear form.
\end{proposition}

\begin{proof}
	Let $\psi=\sf{D}\varphi$ for some $\varphi\in\Gamma^{\infty}_{\mathrm{c}}(\sf{E})$ and let $\omega\in\Gamma^{\infty}_{\mathrm{c}}(\sf{E})$ be arbitrary. Then,
    \begin{align*}
        (\psi,\sf{G}\omega)_{\sf{E}}&=(\sf{D}\varphi,\sf{G}\omega)_{\sf{E}}=(\varphi,\sf{D}\sf{G}\omega)_{\sf{E}}=0\, .
    \end{align*}
    Hence, the definition of $\sigma$ is independent of the chosen representatives. It is clear that $\sigma$ is sesquilinear and Hermitian on account of $\sf{G}^{\ast}=-\sf{G}$. For non-degeneracy, let $[\psi]\in\mathcal{V}_{\mathrm{c}}$ be such that $\sigma([\psi],[\varphi])=0$ for all $\varphi\in\Gamma^{\infty}_{\mathrm{c}}(\sf{E})$. By non-degeneracy of $(\cdot,\cdot)_{\sf{E}}$, it follows that $\sf{G}\psi=0$. Now, since $\mathrm{ker}(\sf{G}\vert_{\Gamma^{\infty}_{\mathrm{c}}})=\mathrm{ran}(\sf{D}\vert_{\Gamma^{\infty}_{\mathrm{c}}})$ by Proposition~\ref{Prop:ExactSeq}(iv), we conclude that $[\psi]=0$.
\end{proof}

After this preliminary discussion, let us turn to the more general setting of linear gauge theories. For the remainder of this section, let $(\sf{M},\sf{g})$ be a globally hyperbolic spacetime and $(\sf{E}_{1},\sf{E}_{2},\sf{P},\sf{K})$ be a linear gauge theory on $(\sf{M},\sf{g})$ with Green hyperbolic operators $\sf{D}_{1}:=\sf{K}^{\ast}\sf{K}$ and $\sf{D}_{2}:=\sf{P}+\sf{K}\sf{K}^{\ast}$. Furthermore, we denote the Green operators of $\sf{D}_{i}$ by $\sf{G}_{i}^{\pm}$ and the corresponding causal propagators by $\sf{G}_{i}:=\sf{G}_{i}^{+}-\sf{G}_{i}^{-}$.

As a generalisation of Proposition~\ref{eq:IsoPhaseOFT}, we obtain the following isomorphism (see~\cite[Thm.~3.12]{HackSchenkel} and \cite[Prop.~2.7]{GerardWrochna}) for linear gauge theories.

\begin{proposition}\label{Prop:IsoGFT}
    Let $(\sf{E}_{1},\sf{E}_{2},\sf{P},\sf{K})$ be a linear gauge theory with Green hyperbolic operators $\sf{D}_{1}:=\sf{K}^{\ast}\sf{K}$ and $\sf{D}_{2}:=\sf{P}+\sf{K}\sf{K}^{\ast}$ and corresponding causal propagators $\sf{G}_{1,2}$. Then,
    \begin{align*}
            \mathcal{V}_{\mathrm{c}}:=\cfrac{\mathrm{ker}(\sf{K}^{\ast}\vert_{\Gamma^{\infty}_{c}})}{\mathrm{ran}(\sf{P}\vert_{\Gamma^{\infty}_{c}})}\xrightarrow[\cong]{\quad[\sf{G}_{2}]\quad}\cfrac{\mathrm{ker}(\sf{P}\vert_{\Gamma^{\infty}_{\mathrm{sc}}})}{\mathrm{ran}(\sf{K}\vert_{\Gamma^{\infty}_{\mathrm{sc}}})}= \mathrm{Sol}_{\mathrm{sc}}\, .
    \end{align*}
\end{proposition}

\begin{proof} If $\psi\in\mathrm{ker}(\sf{K}^{\ast}\vert_{\Gamma^{\infty}_{\mathrm{sc}}})$, then $\sf{P}\sf{G}_{2}\psi=\sf{D}_{2}\sf{G}_{2}\psi-\sf{K}\sf{K}^{\ast}\psi=0$ and hence $\sf{G}_{2}\psi\in\mathrm{ker}(\sf{P}\vert_{\Gamma^{\infty}_{\mathrm{sc}}})$. Furthermore, if $\psi=\sf{P}\varphi$ for some $\varphi\in\Gamma^{\infty}_{\mathrm{c}}(\sf{E}_{2})$, then Proposition~\ref{Prop:PropertiesLinGaugTh}(ii) implies that
		\begin{align*}
			\sf{G}_{2}\psi=\sf{G}_{2}\sf{D}_{2}\varphi-\sf{G}_{2}\sf{K}\sf{K}^{\ast}\varphi=-\sf{K}(\sf{G}_{1}\sf{K}^{\ast}\varphi)\, ,
		\end{align*}
		which shows that $\sf{G}_{2}\psi\in\mathrm{ran}(\sf{K}\vert_{\Gamma^{\infty}_{\mathrm{sc}}})$. We conclude that the quotient map $[\sf{G}_{2}]$ is well-defined. To show that $[\sf{G}_{2}]\:\mathcal{V}\to\mathrm{Sol}_{\mathrm{sc}}$ is injective, we need to show that
		\begin{align*}
			\textit{injectivity:}\quad (\sf{G}_{2}\vert_{\mathrm{ker}(\sf{K}^{\ast}\vert_{\Gamma_{\mathrm{c}}^{\infty}})})^{-1}(\mathrm{ran}(\sf{K}\vert_{\Gamma^{\infty}_{\mathrm{sc}}}))\subset\mathrm{ran}(\sf{P}\vert_{\Gamma_{\mathrm{c}}^{\infty}})\, ,
		\end{align*}
		see~Proposition~\ref{App:Quot}. Take $\psi\in\mathrm{ker}(\sf{K}^{\ast}\vert_{\Gamma_{\mathrm{c}}^{\infty}})$ such that there exists a $\varphi\in\Gamma_{\mathrm{sc}}^{\infty}(\sf{E}_{1})$ with $\sf{G}_{2}\psi=\sf{K}\varphi$. Then,
		\begin{align*}
			\sf{D}_{1}\varphi=\sf{K}^{\ast}\sf{K}\varphi=\sf{K}^{\ast}\sf{G}_{2}\psi=\sf{G}_{1}\sf{K}^{\ast}\psi=0\, ,
		\end{align*}
		which implies that there exists a $\omega\in\Gamma_{\mathrm{c}}^{\infty}(\sf{E}_{1})$ such that $\varphi=\sf{G}_{1}\omega$. But then, it follows that
		\begin{align*}
			\sf{G}_{2}(\psi-\sf{K}\omega)=\sf{K}\varphi-\sf{K}\sf{G}_{1}\omega=\sf{K}\varphi-\sf{K}\varphi=0
		\end{align*}
		and hence, we can find a section $\chi\in\Gamma_{\mathrm{c}}^{\infty}(\sf{E}_{2})$ such that $\psi-\sf{K}\omega=\sf{D}_{2}\chi$. Combining everything, we conclude that 
		\begin{align*}
			\sf{D}_{1}(\sf{K}^{\ast}\chi+\omega)=\sf{K}^{\ast}\sf{D}_{2}\chi+\sf{D}_{1}\omega=\sf{K}^{\ast}(\psi-\sf{K}\omega)+\sf{D}_{1}\omega=-\sf{D}_{1}\omega+\sf{D}_{1}\omega=0
		\end{align*}
		and since $\sf{K}^{\ast}\chi+\omega$ is compactly supported, we conclude from $\sf{D}_{1}(\sf{K}^{\ast}\chi+\omega)=0$ that $\sf{K}^{\ast}\chi+\omega=0$. To sum up, we have shown that
		\begin{align*}
			\psi=\sf{D}_{2}\chi+\sf{K}\omega=(\sf{D}_{2}-\sf{K}\sf{K}^{\ast})\chi=\sf{P}\chi
		\end{align*}
		and hence $\psi\in \mathrm{ran}(\sf{P}\vert_{\Gamma_{\mathrm{c}}^{\infty}})$. This proves injectivity. For surjectivity, we need to show that
		\begin{align*}
			\textit{surjectivity:}\quad\mathrm{ker}(\sf{P}\vert_{\Gamma^{\infty}_{\mathrm{sc}}})\stackrel{!}{=}\sf{G}_{2}(\mathrm{ker}(\sf{K}^{\ast}\vert_{\Gamma^{\infty}_{\mathrm{c}}}))+\mathrm{ran}(\sf{K}\vert_{\Gamma^{\infty}_{\mathrm{sc}}})\, ,
		\end{align*}
		which essentially means that any $\psi\in \mathrm{ker}(\sf{P}\vert_{\Gamma^{\infty}_{\mathrm{sc}}})$ is gauge-equivalent to an element of the form $\sf{G}_{2}\varphi$ for some $\varphi\in\mathrm{ker}(\sf{K}^{\ast}\vert_{\Gamma^{\infty}_{\mathrm{c}}})$. Let $\psi\in \mathrm{ker}(\sf{P}\vert_{\Gamma^{\infty}_{\mathrm{sc}}})$ be arbitrary. First, as explained in Section~\ref{Sec:CauchyGauge}, we can always find a gauge transformation $\omega\in\Gamma^{\infty}_{\mathrm{sc}}(\sf{E}_{1})$ such that $\psi^{\prime}:=\psi+\sf{K}\omega$ satisfies $\sf{K}^{\ast}\psi^{\prime}=0$. It follows that $\sf{D}_{2}\psi^{\prime}=0$ and hence, there is a $\varphi\in\Gamma_{\mathrm{c}}^{\infty}(\sf{E}_{2})$ such that $\psi^{\prime}=\sf{G}_{2}\varphi$. Furthermore, since $\sf{K}^{\ast}\psi^{\prime}=\sf{G}_{1}\sf{K}^{\ast}\varphi=0$, we can find a $\chi\in\Gamma_{\mathrm{c}}^{\infty}(\sf{E}_{1})$ such that $\sf{K}^{\ast}\varphi=\sf{D}_{1}\chi$. Combining everything, we have that
		\begin{align*}
			\psi^{\prime}-\sf{K}\sf{G}_{1}\chi=\psi^{\prime}-\sf{G}_{2}\sf{K}\chi=\sf{G}_{2}(\varphi-\sf{K}\chi)\, .
		\end{align*}
		Hence, noting that $\sf{K}^{\ast}(\varphi-\sf{K}\chi)=\sf{K}^{\ast}\varphi-\sf{D}_{1}\chi=0$, we find $\psi=\sf{G}_{2}(\varphi-\sf{K}\chi)+\sf{K}(\sf{G}_{1}\chi-\omega)$, which is the required decomposition.
\end{proof}

Motivated from the discussion of ordinary field theories, we equip the complex vector space $\mathcal{V}_{\mathrm{c}}$ with an appropriate Hermitian form (cf.~\cite[Prop.~4.9]{HackSchenkel}). 

\begin{proposition}\label{Prop:PSSym}
	Let $(\sf{E}_{1},\sf{E}_{2},\sf{P},\sf{K})$ be a linear gauge theory, $\sf{G}_{2}$ be the causal propagator of $\sf{D}_{2}:=\sf{P}+\sf{K}\sf{K}^{\ast}$ and $\mathcal{V}_{\mathrm{c}}$ as in Proposition~\ref{Prop:IsoGFT}. Then, $\sigma\:\mathcal{V}_{\mathrm{c}}\times\mathcal{V}_{\mathrm{c}}\to\bb{C}$ defined by
\begin{align*}
	\sigma([\psi],[\varphi]):=(\psi,i\sf{G}_{2}\varphi)_{\sf{E}_{2}}
\end{align*}
for all $\psi,\varphi\in\Gamma^{\infty}(\sf{E}_{2})$ is a well-defined Hermitian sesquilinear form on $\mathcal{V}_{\mathrm{c}}$.
\end{proposition}

\begin{proof}
	Let $\psi=\sf{P}\varphi$ for some $\varphi\in\Gamma^{\infty}_{\mathrm{c}}(\sf{E}_{2})$ and $\omega\in\mathrm{ker}(\sf{K}^{\ast}\vert_{\Gamma^{\infty}_{c}})$. Then,
	\begin{align*}
		(\psi,\sf{G}_{2}\omega)_{\sf{E}_{2}}&=(\sf{P}\varphi,\sf{G}_{2}\omega)_{\sf{E}_{2}}=(\varphi,\sf{P}\sf{G}_{2}\omega)_{\sf{E}_{2}}=(\varphi,(\sf{D}_{2}-\sf{K}\sf{K}^{\ast})\sf{G}_{2}\omega)_{\sf{E}_{2}}=\\&=-(\varphi,\sf{K}\sf{G}_{1}\sf{K}^{\ast}\omega)_{\sf{E}_{2}}=-(\sf{K}^{\ast}\varphi,\sf{G}_{1}\sf{K}^{\ast}\omega)_{\sf{E}_{2}}=0\,,
	\end{align*}
	where we used Proposition~\ref{Prop:PropertiesLinGaugTh}. The fact that $\sigma$ is Hermitian and sesquilinear is clear from the definition.
\end{proof}

To sum up, we define the classical phase space of a linear gauge theory as follows: 

\begin{definition}\label{def.phasspace} (Classical Phase Space of a Linear Gauge Theory)\newline
	Let $(\sf{E}_{1},\sf{E}_{2},\sf{P},\sf{K})$ be a linear gauge theory. The pair $(\mathcal{V}_{\mathrm{c}},\sigma)$ defined by
	\begin{align*}
		 \mathcal{V}_{\mathrm{c}}:=\cfrac{\mathrm{ker}(\sf{K}^{\ast}\vert_{\Gamma^{\infty}_{c}})}{\mathrm{ran}(\sf{P}\vert_{\Gamma^{\infty}_{c}})}\,,\qquad \sigma\:\mathcal{V}_{\mathrm{c}}\times\mathcal{V}_{\mathrm{c}}\to\bb{C}\,,\quad([\psi],[\varphi])\mapsto\sigma([\psi],[\varphi]):=(\psi,i\sf{G}\varphi)_{\sf{E}_{2}}
	\end{align*}
	is called the \textit{classical phase space} of $(\sf{E}_{1},\sf{E}_{2},\sf{P},\sf{K})$.
\end{definition}

\begin{remark}\label{Rem:Deg} (Degeneracy of $\sigma$)\newline
	We stress that $\sigma$ is in general degenerate, as opposed to ordinary field theories. This point has been discussed in some detailed by Hack-Schenkel in \cite[Sec.~5]{HackSchenkel}. In fact, it has been shown therein that $\sigma$ is degenerate whenever $\mathrm{ran}(\sf{K})\cap\Gamma^{\infty}_{\mathrm{sc}}(\sf{E}_{2})\subset\mathrm{ran}(\sf{K}\vert_{\Gamma_{\mathrm{sc}}})$ is a \textit{proper} subspace. However, let us emphasise that this is not an if-and-only-if statement. For Maxwell's theory, it has been shown in \cite[Prop.~3.4]{DappiaggiSiemssen} (generalising \cite[Prop.~2.1]{Dappiaggi}) that $\sigma$ is weakly non-degenerate if either $\sf{H}^{1}(\sf{M})=0$ or $\sf{H}^{2}(\sf{M})=0$. More generally, a similar statement for linearised Yang-Mills theory with trivial principal bundles has been established in \cite[Ex.~5.7]{HackSchenkel}. For linearised gravity, $\sigma$ is weakly non-degenerate on spatially compact globally hyperbolic manifolds, as shown in \cite[Thm.~4.3]{FewsterHunt}. The case of spatially non-compact spacetimes remains open.
	
	Let us also remark, however, that the degeneracy of $\sigma$ is of physical significance. In fact, it has been argued that the degeneracy of $\sigma$ for Maxwell's theory in general is related to the celebrated Aharonov-Bohm effect, see \cite{DappiaggiAharonov}.
\end{remark}

\begin{remark} (Gauge-Invariant Observables)\newline
	The interpretation of the space $\mathcal{V}_{\mathrm{c}}$ follows the same lines of arguments as in the case of ordinary field theories without a gauge symmetry. First of all, note that a linear observable $\mathcal{O}_{f}\:\Gamma^{\infty}(\sf{E}_{2})\to\bb{C}$ is \textit{gauge-invariant} in the sense that
	\begin{align*}
		\mathcal{O}_{f}(\psi+\sf{K}\omega)=\mathcal{O}_{f}(\psi)\qquad\forall\psi\in\Gamma^{\infty}(\sf{E}_{2}),\,\omega\in\Gamma^{\infty}(\sf{E}_{1})
	\end{align*}
	if and only if the label $f$ is contained in the subspace $\mathrm{ker}(\sf{K}^{\ast}\vert_{\Gamma^{\infty}_{\mathrm{c}}})\subset\Gamma^{\infty}_{\mathrm{c}}(\sf{E}_{2})$, by the fact that $\mathcal{O}_{f}(\sf{K}\omega)=(f,\sf{K}\omega)_{\sf{E}_{2}}=(\sf{K}^{\ast}f,\omega)_{\sf{E}_{1}}$ and non-degeneracy of $(\cdot,\cdot)_{\sf{E}_{1}}$. Hence, for linear gauge theories, it makes sense to consider the linear subspace $\mathrm{ker}(\sf{K}^{\ast}\vert_{\Gamma^{\infty}_{\mathrm{c}}})/\mathrm{ran}(\sf{P}\vert_{\Gamma^{\infty}_{c}})$ of $\Gamma^{\infty}(\sf{E}_{2})/\mathrm{ran}(\sf{P}\vert_{\Gamma^{\infty}_{c}})$ as a phase space. As a by-product, we observe that even though $\sf{K}^{\ast}$ has also the interpretation of a gauge-fixing operator, we do \textit{not} perform any explicit gauge fixing, but rather work completely in terms of gauge invariant quantities when
discussing algebras of observables.
	
	 However, the space $\mathcal{V}_{\mathrm{c}}$ is in general \textit{not} isomorphic to the space of linear, on-shell, gauge-invariant observables, since the map
	\begin{align}\label{eq:IsoGaugeObs}
		\mathcal{V}_{\mathrm{c}}:=\cfrac{\mathrm{ker}(\sf{K}^{\ast}\vert_{\Gamma^{\infty}_{\mathrm{c}}})}{\mathrm{ran}(\sf{P}\vert_{\Gamma^{\infty}_{\mathrm{c}}})}\xrightarrow{[f]\,\,\mapsto\,\, \mathcal{O}_{f}\vert_{\mathrm{Sol}_{\mathrm{sc}}}}\{\mathcal{O}_{f}\vert_{\mathrm{Sol}_{\mathrm{sc}}}\mid f\in\mathrm{ker}(\sf{K}^{\ast}\vert_{\Gamma^{\infty}_{\mathrm{c}}})\}
	\end{align}
	is surjective but in general not injective. In other words, there may be some \textit{redundant observables} in $\mathcal{V}_{\mathrm{c}}$. On the other hand, it is not clear whether there are enough observables in $\mathcal{V}_{\mathrm{c}}$ to distinguish between different configurations, since it is unclear whether for two configurations $\psi,\varphi\in\Gamma^{\infty}(\sf{E}_{2})$ there exists a $f\in\mathrm{ker}(\sf{K}^{\ast}\vert_{\Gamma^{\infty}_{\mathrm{c}}})$ such that $\mathcal{O}_{f}(\psi)\neq\mathcal{O}_{f}(\varphi)$.
	
	If $\sigma$ is weakly non-degenerate, then both the previous issues are resolved. Indeed, in this case, $\mathcal{V}_{\mathrm{c}}$ is in one-to-one correspondence to the space of linear, gauge-invariant, on-shell observables, i.e.~the map in Eq.~\eqref{eq:IsoGaugeObs} is an isomorphism, and there are also enough observables in $\mathcal{V}$ to distinguish between different configurations. Furthermore, even though $\sigma$ is degenerate for Maxwell's theory on non-spatially-compact manifolds in general, the previous statement turns out to be still true in this case, as shown in \cite[Thm.~7.6]{Benini} by using more refined techniques from cohomology theory. For other gauge theories, this remains an important open question.
\end{remark}

In the last part of this section, we establish some equivalent phase spaces, which will be useful for the subsequent analysis. Let us denote by $\rho_{i}\:\Gamma^{\infty}_{\mathrm{sc}}(\sf{E}_{i})\to\Gamma^{\infty}_{\mathrm{c}}(\sf{E}_{i})$ the initial data maps of $\sf{D}_{i}$ for $i=1,2$, where $\Sigma$ is a fixed smooth and spacelike Cauchy hypersurface of $\sf{M}$. Furthermore, let us equip the bundles $\sf{E}_{\rho_{i}}$ with non-degenerate bundle metrics $\langle\cdot,\cdot\rangle_{\sf{E}_{\rho_{i}}}$. We stress that the choice of these metrics is completely arbitrary and serves only auxiliary purposes. For practical reasons, it is useful to choose these metrics to be positive-definite, however, non-degeneracy is enough for most of the subsequent constructions. Now, following the discussion in \cite[Sec.~2.4]{GerardWrochna} (see also \cite[Sec.~2.2]{GerardMurroWrochna}), we define the linear operator $\sf{G}_{\Sigma,i}\:\Gamma^{\infty}(\sf{E}_{\rho_{i}})\to\Gamma^{\infty}(\sf{E}_{\rho_{i}})$ by 
\begin{align*}
	\sf{G}_{i}=(\rho_{i}\sf{G}_{i})^{\ast}\sf{G}_{\Sigma,i}(\rho_{i}\sf{G}_{i})\, ,
\end{align*}
where the adjoint $\ast$ is taken with respect to $(\cdot,\cdot)_{\sf{E}_{i}}$ and $(\cdot,\cdot)_{\sf{E}_{\rho_{i}}}$. These operators have the following properties (cf.~\cite[Sec.~2.4]{GerardWrochna}, see also \cite[Sec.~5.5]{GerardBook}).

\begin{proposition}\label{Prop:PropGSigma}
	Let $(\sf{E}_{1},\sf{E}_{2},\sf{P},\sf{K})$ be a linear gauge theory and $\sf{G}_{\Sigma,i}$ as above. Then
	\begin{align*}
		\text{\emph{(i)}}\quad &\sf{G}_{\Sigma,i}\in\mathrm{DO}(\sf{E}_{\rho_{i}})\\
		\text{\emph{(ii)}}\quad & (\rho_{i}\sf{G}_{i})^{\ast}\sf{G}_{\Sigma,i}=\mathcal{U}_{i}\\
		\text{\emph{(iii)}}\quad &\sf{G}_{\Sigma,1}\sf{K}_{\Sigma}^{\dagger}=\sf{K}_{\Sigma}^{\ast}\sf{G}_{\Sigma,2}
	\end{align*}
	where $\sf{K}_{\Sigma}^{\ast}$ denotes the formal adjoint of $\sf{K}_{\Sigma}$ with respect to $(\cdot,\cdot)_{\sf{E}_{\rho_{i}}}$.
\end{proposition}

\begin{proof}	
	Claim (i) follows from \textit{Green's formula}, see \cite[Lem.~2.2]{GerardMurroWrochna}. For (ii), set $\mathcal{U}_{i}^{\prime}:=(\rho_{i}\sf{G}_{i})^{\ast}\sf{G}_{\Sigma,i}$. By Proposition~\ref{Prop:ExactSeq}(iv), any $\psi\in\mathrm{ker}(\sf{D}_{i}\vert_{\Gamma^{\infty}_{\mathrm{sc}}})$ is of the form $\psi=\sf{G}_{i}\varphi$ for some $\varphi\in\Gamma^{\infty}_{\mathrm{c}}(\sf{E}_{i})$. Hence,
	\begin{align*}
		\mathcal{U}_{i}^{\prime}\rho_{i}\psi=(\rho_{i}\sf{G}_{i})^{\ast}\sf{G}_{\Sigma,i}\rho_{i}\sf{G}_{i}\varphi=\sf{G}_{i}\varphi=\psi\, .
	\end{align*}
	Similarly, one shows that $\rho_{i}\mathcal{U}_{i}^{\prime}=\mathrm{id}$ on $\Gamma^{\infty}_{\mathrm{c}}(\sf{E}_{\rho_{i}})$. Hence, $\mathcal{U}_{i}^{\prime}$ is the inverse of $\rho_{i}\:\Gamma^{\infty}_{\mathrm{c}}(\sf{E}_{\rho_{i}})\to\mathrm{ker}(\sf{D}_{i}\vert_{\Gamma^{\infty}_{\mathrm{sc}}})$ and we conclude that $\mathcal{U}_{i}=\mathcal{U}_{i}^{\prime}$. Claim (iii) follows from (ii), since
	\begin{align*}
		\sf{G}_{\Sigma,1}\sf{K}_{\Sigma}^{\dagger}&=\sf{G}_{\Sigma,1}\rho_{1}\sf{K}^{\ast}\mathcal{U}_{2}=\sf{G}_{\Sigma,1}\rho_{1}\sf{K}^{\ast}(\rho_{2}\sf{G}_{2})^{\ast}\sf{G}_{\Sigma,2}=\sf{G}_{\Sigma,1}\rho_{1}(\rho_{2}\sf{G}_{2}\sf{K})^{\ast}\sf{G}_{\Sigma,2}=\sf{G}_{\Sigma,1}\rho_{1}(\rho_{2}\sf{K}\sf{G}_{1})^{\ast}\sf{G}_{\Sigma,2}=\\&=\sf{G}_{\Sigma,1}\rho_{1}(\sf{K}_{\Sigma}\rho_{1}\sf{G}_{1})^{\ast}\sf{G}_{\Sigma,2}=\sf{G}_{\Sigma,1}\rho_{1}(\rho_{1}\sf{G}_{1})^{\ast}\sf{K}_{\Sigma}^{\ast}\sf{G}_{\Sigma,2}=(\rho_{1}(\rho_{1}\sf{G}_{1})^{\ast}\sf{G}_{\Sigma,1})^{\ast}\sf{K}_{\Sigma}^{\ast}\sf{G}_{\Sigma,2}=\sf{K}_{\Sigma}^{\ast}\sf{G}_{\Sigma,2}\, ,
	\end{align*}
	where we used Propositions~\ref{Prop:PropertiesLinGaugTh} and Proposition~\ref{Prop:CauchyGauge}.
\end{proof}


\begin{corollary}\label{Cor:KSigmaAdj} $\sf{K}_{\Sigma}^{\dagger}$ is the formal adjoint of $\sf{K}_{\Sigma}$ for the Hermitian sesquilinear forms 
\begin{align*}
	\sigma_{\Sigma,i}\:\Gamma^{\infty}_{\mathrm{c}}(\sf{E}_{\rho_{i}})\times\Gamma^{\infty}_{\mathrm{c}}(\sf{E}_{\rho_{i}})\to\bb{C}\,,\qquad \sigma_{\Sigma,i}(\mathfrak{f},\mathfrak{g}):=(\mathfrak{f},i\sf{G}_{\Sigma,i}\mathfrak{g})_{\sf{E}_{\rho_{i}}}\, .
\end{align*}
\end{corollary}

\begin{remark}\label{Remark:PhysicalCharge}
Note that the sesquilinear forms $\sigma_{\Sigma,i}$ do not depend on the choice of auxiliary bundle metrics $\langle\cdot,\cdot\rangle_{\sf{E}_{\rho_{i}}}$, as opposed to the operators $\sf{G}_{\Sigma,i}$. Let $\mathfrak{f}\in\Gamma^{\infty}_{\mathrm{c}}(\sf{E}_{\rho_{i}})$ and write $\mathfrak{f}=\rho_{i}\mathcal{U}_{i}\mathfrak{f}$. Since $\mathcal{U}_{i}\mathfrak{f}\in\mathrm{ker}(\sf{D}_{i}\vert_{\Gamma^{\infty}_{\mathrm{sc}}})$, there is a $\psi\in\Gamma^{\infty}_{\mathrm{c}}(\sf{E}_{i})$ such that $\mathcal{U}_{i}\mathfrak{f}=\sf{G}_{i}\psi$. Similarly, let us write $\mathfrak{g}\in\Gamma^{\infty}_{\mathrm{c}}(\sf{E}_{\rho_{i}})$ as $\mathfrak{g}=\rho_{i}\sf{G}_{i}\varphi$ for some $\varphi\in\Gamma^{\infty}_{\mathrm{c}}(\sf{E}_{i})$. Then, 
\begin{align*}
	\sigma_{\Sigma}(\mathfrak{f},\mathfrak{g})=(\mathfrak{f},i\sf{G}_{\Sigma,i}\mathfrak{g})_{\sf{E}_{\rho_{i}}}=(\rho_{i}\sf{G}_{i}\psi,i\sf{G}_{\Sigma,i}\rho_{i}\sf{G}_{i}\mathfrak{g})_{\sf{E}_{\rho_{i}}}=(\psi,i(\rho_{i}\sf{G}_{i})^{\ast}\sf{G}_{\Sigma,i}\rho_{i}\sf{G}_{i}\mathfrak{g})_{\sf{E}_{i}}=(\psi,i\sf{G}_{i}\varphi)_{\sf{E}_{i}}\,,
\end{align*}
where the right-hand side is independent of the auxiliary bundle metrics $\langle\cdot,\cdot\rangle_{\sf{E}_{\rho_{i}}}$.
\end{remark}

Now, before providing several equivalent descriptions of the classical phase space $(\mathcal{V}_{\mathrm{c}},\sigma)$, we introduce the following notation: for any $\psi\in\Gamma^{\infty}_{\mathrm{sc}}(\sf{E}_{2})$ we can find a compact set $\sf{K}\subset\sf{M}$ such that $\mathrm{supp}(\psi)\subset\mathcal{J}(\sf{K})$. Choosing appropriate cut-off functions, we can split $\psi$ as 
\begin{align}\label{eq:Splitting}
	\psi=\psi^{+}+\psi^{-}
\end{align}
for some $\psi^{\pm}\in\Gamma^{\infty}_{\mathrm{sc}}(\sf{E}_{2})$ with $\mathrm{supp}(\psi^{\pm})\subset\mathcal{J}^{\pm}(\sf{K})$. This choice is, of course, not unique. However, if $\varphi^{\pm}$ is another splitting such that $\psi=\varphi^{+}+\varphi^{-}$ and $\mathrm{supp}(\varphi^{\pm})\subset\mathcal{J}^{\pm}(\sf{K})$, then $\psi^{+}-\varphi^{+}=\psi^{-}-\varphi^{-}\in\Gamma^{\infty}_{\mathrm{c}}(\sf{E}_{2})$. Hence, the splitting~\eqref{eq:Splitting} is unique up to a compactly supported section.

With this notation, we obtain the following diagram (see~\cite[Sec.2.4]{GerardWrochna} and \cite[Prop.~5.1]{HackSchenkel}).

\begin{proposition}\label{Prop:EqPS} \emph{(Equivalent Phase Spaces for Linear Gauge Theories)}\newline
	Let $(\sf{E}_{1},\sf{E}_{2},\sf{P},\sf{K})$ be a linear gauge theory. Then, the following diagram commutes and every map is a unitary isomorphism:
	\begin{equation*}
        \begin{tikzcd}[column sep=1.2ex]
			\bigg(\mathcal{V}_{\mathrm{c}}=\cfrac{\mathrm{ker}(\sf{K}^{\ast}\vert_{\Gamma^{\infty}_{\mathrm{c}}})}{\mathrm{ran}(\sf{P}\vert_{\Gamma^{\infty}_{\mathrm{c}}})} ,\sigma\bigg)\arrow[r,"{[\sf{G}_{2}]}"]\arrow[r,swap,"\cong"]\arrow[rd,swap,"{[\sf{G}_{2}]}"]\arrow[rd,"\cong"] & \bigg(\mathrm{Sol}_{\mathrm{sc}}=\cfrac{\mathrm{ker}(\sf{P}\vert_{\Gamma^{\infty}_{\mathrm{sc}}})}{\mathrm{ran}(\sf{K}\vert_{\Gamma^{\infty}_{\mathrm{sc}}})},\tau\bigg)\arrow[r,"{[\rho_{2}]}"]\arrow[r,swap,"\cong"]&\bigg(\mathcal{V}_{\Sigma}:=\cfrac{\mathrm{ker}(\sf{K}_{\Sigma}^{\dagger}\vert_{\Gamma_{\mathrm{c}}^{\infty}})}{\mathrm{ran}(\sf{K}_{\Sigma}\vert_{\Gamma_{\mathrm{c}}^{\infty}})},\sigma_{\Sigma,2}\bigg)\\
			& \bigg(\cfrac{\mathrm{ker}(\sf{D}_{2}\vert_{\Gamma^{\infty}_{\mathrm{sc}}})\cap\mathrm{ker}(\sf{K}^{\ast}\vert_{\Gamma^{\infty}_{\mathrm{sc}}})}{\sf{K}(\mathrm{ker}(\sf{D}_{1}\vert_{\Gamma^{\infty}_{\mathrm{sc}}}))},\tau\bigg)\arrow[u,hookrightarrow,swap,"{[\mathrm{id}]}"]\arrow[u,"\cong"]\arrow[ur,swap,"{[\rho_{2}]}"]\arrow[ur,"\cong"]&
    		\end{tikzcd}
    \end{equation*}
    where the Hermitian sesquilinear forms $\tau$ and $\sigma_{\Sigma,2}$ are given by
    \begin{align*}
    		&\tau\:\mathrm{Sol}_{\mathrm{sc}}\times\mathrm{Sol}_{\mathrm{sc}}\to\bb{C}\,,\qquad \tau([\psi],[\varphi]):=i(\sf{P}(\psi^{+}),\varphi)_{\sf{E}_{2}}\\
    		&\sigma_{\Sigma,2}\:\mathcal{V}_{\Sigma}\times\mathcal{V}_{\Sigma}\to\bb{C}\,,\qquad \sigma_{\Sigma,2}([\mathfrak{f}],[\mathfrak{g}]):=(\mathfrak{f},i\sf{G}_{\Sigma,2}\mathfrak{g})_{\sf{E}_{\rho_{2}}}\, .
    \end{align*}
\end{proposition}

\begin{proof}
	The isomorphisms in the first line have been shown in Proposition~\ref{Prop:CauchyGauge} and the vertical isomorphism in Proposition~\ref{Proposition2}. The diagonal maps are such that the diagram commutes.
	
	Let us now turn to the various sesquilinear forms. First of all, to show that $\sigma_{\Sigma,2}$ is well-defined, choose $\mathfrak{f}:=\sf{K}_{\Sigma}\mathfrak{a}$ for some $\mathfrak{a}\in\Gamma^{\infty}_{\mathrm{sc}}(\sf{E}_{\rho_{1}})$ and $\mathfrak{g}\in\mathrm{ker}(\sf{K}^{\dagger}_{\Sigma}\vert_{\Gamma^{\infty}_{\mathrm{c}}})$. Then,
	\begin{align*}
		(\mathfrak{f},\sf{G}_{\Sigma,2}\mathfrak{g})_{\sf{E}_{\rho_{2}}}=(\sf{K}_{\Sigma}\mathfrak{a},\sf{G}_{\Sigma,2}\mathfrak{g})_{\sf{E}_{\rho_{2}}}=(\mathfrak{a},\sf{K}_{\Sigma}^{\ast}\sf{G}_{\Sigma,2}\mathfrak{g})_{\sf{E}_{\rho_{1}}}=(\mathfrak{a},\sf{G}_{\Sigma,1}\sf{K}_{\Sigma}^{\dagger}\mathfrak{g})_{\sf{E}_{\rho_{1}}}=0\,,
	\end{align*}
	where we used Proposition~\ref{Prop:PropGSigma}(iii). From the definition, it is clear that $\sf{G}_{\Sigma,2}^{\dagger}=-\sf{G}_{\Sigma,2}$, which shows that $\sigma_{\Sigma,2}$ is a well-defined Hermitian sesquilinear form on $\mathcal{V}_{\Sigma}$. Furthermore, it holds that
	\begin{align*}
		\sigma_{\Sigma,2}([\rho_{2}\sf{G}_{2}\psi],[\rho_{2}\sf{G}_{2}\varphi])=(\rho_{2}\sf{G}_{2}\psi,i\sf{G}_{\Sigma,2}\rho_{2}\sf{G}_{2}\varphi)_{\sf{E}_{\rho_{2}}}=(\psi,i\sf{G}_{2}\varphi)_{\sf{E}_{2}}=\sigma([\psi],[\varphi])\,,
	\end{align*}
	which proves that $[\rho_{2}\sf{G}_{2}]\:\mathcal{V}_{\mathrm{c}}\to\mathcal{V}_{\Sigma}$ is unitary. For $\tau$ defined on $\mathrm{Sol}_{\mathrm{sc}}$, we first show that it is Hermitian. Let $\psi,\varphi\in\mathrm{ker}(\sf{P}\vert_{\Gamma^{\infty}_{\mathrm{sc}}})$ and let us split $\psi=\psi^{+}+\psi^{-}$, $\varphi=\varphi^{+}+\varphi^{-}$. Then,
	\begin{align*}
		(\sf{P}\psi^{+},\varphi)_{\sf{E}_{2}}&=(\sf{P}\psi^{+},\varphi^{+})_{\sf{E}_{2}}+(\sf{P}\psi^{+},\varphi^{-})_{\sf{E}_{2}}=-(\sf{P}\psi^{-},\varphi^{+})_{\sf{E}_{2}}+(\sf{P}\psi^{+},\varphi^{-})_{\sf{E}_{2}}=\\&=-(\psi^{-},\sf{P}\varphi^{+})_{\sf{E}_{2}}-(\sf{P}\psi^{+},\varphi^{+})_{\sf{E}_{2}}=-(\psi,\sf{P}\varphi^{+})_{\sf{E}_{2}}\,,
	\end{align*}
	which proves the claim. Now, if $\varphi=\sf{K}\omega$ for some $\omega\in\Gamma^{\infty}_{\mathrm{sc}}(\sf{E}_{1})$, then clearly $(\sf{P}\psi^{+},\varphi)_{2}=(\sf{K}^{\ast}\sf{P}\psi^{+},\omega)_{\sf{E}_{1}}=0$. Hence, $\tau$ is a well-defined Hermitian sesquilinear form on $\mathrm{Sol}_{\mathrm{sc}}$. To show unitarity of the map $[\sf{G}_{2}]\:\mathcal{V}_{\mathrm{c}}\to\mathrm{Sol}_{\mathrm{sc}}$, we first observe that $\sf{G}_{2}\psi=\sf{G}_{2}^{+}\varphi-\sf{G}_{2}^{-}\varphi$ provides an allowed decomposition of $\sf{G}_{2}\psi$ with $\varphi\in\Gamma^{\infty}_{\mathrm{c}}(\sf{E}_{2})$ . Then, for $\psi,\varphi\in\mathrm{ker}(\sf{K}^{\ast}\vert_{\Gamma^{\infty}_{\mathrm{c}}})$, we find
	\begin{align*}
		\tau([\sf{G}_{2}\psi],[\sf{G}_{2}\varphi])&=i(\sf{P}\sf{G}_{2}^{+}\psi,\sf{G}_{2}\varphi)_{\sf{E}_{2}}=i((\sf{D}_{2}-\sf{K}\sf{K}^{\ast})\sf{G}_{2}^{+}\psi,\sf{G}_{2}\varphi)_{\sf{E}_{2}}=\\&=i(\psi,\sf{G}_{2}\varphi)_{\sf{E}_{2}}-i(\sf{K}\sf{G}_{2}^{+}\sf{K}^{\ast}\psi,\sf{G}_{2}\varphi)_{\sf{E}_{2}}=i(\psi,\sf{G}_{2}\varphi)_{\sf{E}_{2}}=\sigma([\psi],[\varphi])\, .
	\end{align*}
	Unitarity of the remaining maps can be shown analogously.
\end{proof}

\section{Examples of Linear Gauge Theories}\label{Sec:Exam}
Having established an abstract axiomatic framework for linear gauge theories, we now turn to a detailed discussion of some concrete examples of physical relevance. Specifically, we will study Maxwell’s theory of electromagnetism, linearised Yang-Mills theory on arbitrary principal bundles with possibly non-compact Lie groups as well as Einstein gravity in vacuum linearised around general Einstein manifolds. Additionally, we will consider the linearised Yang-Mills-Klein-Gordon and linearised Einstein-Klein-Gordon theories as examples of coupled systems. In principle, one could of course also study fermionic theories, such as the linearised Yang-Mills-Dirac system relevant to quantum electro- and chromodynamics, or at the very end, a linearised version of the classical standard model of particle physics. Furthermore, one could also consider pure fermionic gauge theories such as Rarita-Schwinger theory (see~\cite[Exp.~3.10]{HackSchenkel}). However, we will restrict our analysis to bosonic fields in the subsequent analysis.

\subsection{Maxwell's Theory}\label{Sec:Maxwell}
One of the most basic examples of a linear gauge theory and in fact one of the main motivational examples for the Hack-Schenkel formalism following Dimock's work \cite{DimockMaxwell} is Maxwell's theory of electrodynamics, which provides a unified geometric description of classical electromagnetism. Let $(\sf{M},\sf{g})$ be a $d$-dimensional globally hyperbolic spacetime. In the notation section (see~pp.~\pageref{Conventions}\,ff.), we have explained the conventions used for differential forms and operators acting thereon. We denote the \textit{exterior bundles} equipped with their natural bundle metrics by
\begin{align}\label{eq:ExtBundles}
	\sf{A}_{k}:=\bigwedge^{k}\sf{T}^{\ast}\sf{M},\qquad\qquad \langle\alpha,\beta\rangle_{\sf{A}_{k}}:=\frac{1}{k!}(\sf{g}^{\sharp})^{\otimes k}(\alpha,\beta)=\frac{1}{k!}\alpha^{\mu_{1}\dots\mu_{k}}\beta_{\mu_{1}\dots\mu_{k}}\, .
\end{align}
By definition, $\Omega^{k}(\sf{M}):=\Gamma^{\infty}(\sf{A}_{k})$ is the $C^{\infty}(\sf{M})$-module of smooth differential forms of degree $k$. The non-degenerate symmetric bilinear form on the level of sections induced by the bundle metric $\langle\cdot,\cdot\rangle_{\sf{A}_{k}}$ coincides, by definition, with the usual \textit{Hodge bilinear form}, i.e.
\begin{align}\label{eq:ExtBundles2}
	(\alpha,\beta)_{\sf{A}_{k}}:=\int_{\sf{M}}\langle\alpha,\beta\rangle_{\sf{A}_{k}}\,\mathrm{vol}_{g}=\int_{\sf{M}}\alpha\wedge\ast\beta\, ,
\end{align}
for all $\alpha,\beta\in\Omega^{k}_{\mathrm{c}}(\sf{M})$, where $\ast\:\Omega^{k}(\sf{M})\to\Omega^{d-k}(\sf{M})$ denotes the Hodge $\ast$-operator. We denote by $\d\:\Omega^{k}(\sf{M})\to\Omega^{k+1}(\sf{M})$ the \textit{exterior derivative} and by $\delta:=(-1)^{k}\ast^{-1}\d\ast\:\Omega^{k}(\sf{M})\to\Omega^{k-1}(\sf{M})$ its formal adjoint with respect to~\eqref{eq:ExtBundles2}, the \textit{codifferential}. In our conventions,
\begin{align*}
	(\d\omega)_{\mu_{1}\dots\mu_{k+1}}&=(k+1)\nabla_{[\alpha_{1}}\omega_{\mu_{2}\dots\mu_{k+1}]}\\\
	(\delta\omega)_{\mu_{1}\dots\mu_{k-1}}&=-\nabla^{\lambda}\omega_{\lambda\mu_{1}\dots\mu_{k-1}}
\end{align*}
for $\omega\in\Omega^{k}(\sf{M})$ in local coordinates,
where $\nabla$ denotes the Levi-Civita connection on $(\sf{M},\sf{g})$. We stress that $\d$ is independent of $\nabla$ and we could replace $\nabla$ in the above formula for the exterior derivative by any torsion-free connection on $\sf{T}^{\ast}\sf{M}$. On the other hand, the $\delta$ strictly depends on the choice of metric. With this notation, we denote the \textit{de Rham-Hodge d'Alembertian} by
\begin{align*}
	\square_{\mathrm{dRH}}:=\d\delta+\delta\d\colon\Omega^{k}(\sf{M})\to\Omega^{k}(\sf{M})\, ,
\end{align*}
which up to a sign is related to the connection (or rough) d'Alembertian $\square:=\sf{g}^{\alpha\beta}\nabla_{\alpha}\nabla_{\beta}$ restricted to $\Omega^{k}(\sf{M})\subset\Gamma^{\infty}(\sf{T}^{\ast}\sf{M}^{\otimes 2})$ by means of the Weitzenböck identity, see~Example~\ref{Example:HodgeLapl}.

With this notation, the dynamics of \textit{Maxwell's theory} in its manifestly covariant formalism is determined by the \textit{Maxwell operator}, which is the second-order linear differential operator
\begin{align*}
	\sf{P}:=\delta\d\in\mathrm{DO}^{2}(\sf{A}_{1})\, .
\end{align*}
This operator is clearly not hyperbolic and it is invariant under the linear gauge transformations
\begin{align*}
	\Omega^{1}(\sf{M})\ni \sf{A}\mapsto \sf{A}+\d f,\qquad\forall f\in C^{\infty}(\sf{M})\, .
\end{align*}
Using the aforementioned Weitzenböck identity to relate the de Rham-Hodge d'Alembertian acting on $1$-forms to the connection d'Alembertian $\square=\sf{g}^{\alpha\beta}\nabla_{\alpha}\nabla_{\beta}$, we can write
\begin{align*}
	\sf{P}=\delta\d=\square_{\mathrm{dRH}}-\d\delta=-\square-\d\delta+\mathrm{Ric}_{\sf{g}}\, ,
\end{align*}
where $\mathrm{Ric}_{\sf{g}}\:\Gamma^{\infty}(\sf{T}^{\ast}\sf{M})\to\Gamma^{\infty}(\sf{T}^{\ast}\sf{M})$ denotes the zeroth-order operator $\mathrm{Ric}_{\sf{g}}(\sf{A})_{\lambda}:=\mathrm{Ric}(\sf{g})_{\alpha}^{\lambda}\sf{A}_{\lambda}$. In particular, we see that $\sf{P}$ acts as an normally hyperbolic operator on co-closed forms. This choice of gauge condition, i.e.~$\delta \sf{A}=0$, is usually called the \textit{Lorenz gauge} (named after \cite{Lorenz}). 


\begin{proposition}\label{Prop:MaxGauge} \emph{(Maxwell's Theory as a Linear Gauge Theory)}\newline
	Let $(\sf{M},\sf{g})$ be a globally hyperbolic spacetime. Then, $(\sf{A}_{0},\sf{A}_{1},\sf{P},\sf{K})$ with 
	\begin{align*}
		\sf{P}:=\delta\d\in\mathrm{DO}^{2}(\sf{A}_{1})\,, \qquad \sf{K}:=\d\in\mathrm{DO}^{1}(\sf{A}_{0},\sf{A}_{1})
	\end{align*}
	is a linear gauge theory in the sense of Definition~\ref{Def:LinGaugeTh}. The operators $\sf{D}_{1}:=\sf{K}^{\ast}\sf{K}\in\mathrm{DO}^{2}(\sf{A}_{0})$ and $\sf{D}_{2}:=\sf{P}+\sf{K}\sf{K}^{\ast}\in\mathrm{DO}^{2}(\sf{A}_{1})$ are the de Rham Hodge d'Alembertians
	\begin{align*}
		\sf{D}_{1}=\square_{\mathrm{dRH}}=\delta\d\,,\qquad\sf{D}_{2}=\square_{\mathrm{dRH}}=\d\delta+\delta\d\, .
	\end{align*}
\end{proposition}

\begin{proof}
	The operator $\sf{P}$ is clearly formally self-adjoint and clearly $\sf{P}\sf{K}=0$ on account of $\d\d=0$. Moreover, since $\sf{K}^{\ast}=\delta$, it immediately follows that $\sf{D}_{i}=\square_{\mathrm{dRH}}\:\Omega^{i-1}(\sf{M})\to\Omega^{i-1}(\sf{M})$.
\end{proof}

Note that $\sf{P}$ is more generally invariant under $\sf{A}\mapsto \sf{A}+\omega$, where $\omega\in\mathrm{ker}(\d\vert_{\Omega^{1}})$. On general manifolds, not every \emph{closed} $1$-form is (globally) \emph{exact}, with the obstructions classified by the \emph{de Rham cohomology group} $\sf{H}^{1}(\sf{M})=\mathrm{ker}(\d\vert_{\Omega^{1}})/\mathrm{ran}(\d\vert_{\Omega^{0}})$. Nevertheless, in the context of gauge theories, one usually restricts to transformations of the form $\omega=\d f$, in which case Maxwell's theory emerges as a special case of \emph{Yang–Mills theory} with Abelian gauge group $\sf{U}(1)$.

\begin{remark} Maxwell's equations were first introduced by Maxwell, scattered throughout a series of papers published\footnote{One of Maxwell's main contributions was the addition of the term $\partial_{t}\mathscr{E}$ to Ampére's law $\mathrm{rot}_{\Sigma}(\mathscr{B})=0$, which is crucial for the derivation of the electromagnetic wave equation and the unification of electricity and magnetism.} in 1861--1862 \cite{MaxwellIa,MaxwellIb,MaxwellIc,MaxwellId}, building upon the works of Ampère, Coulomb, Faraday, Lenz, Neumann, Ørsted and others. Notably, these papers already contained a derivation of the speed of light. The unified form of Maxwell's equations appeared in 1865 \cite{MaxwellII} and were later summarised in a book from 1873 \cite{MaxwellIII}. Maxwell originally wrote the equations in component form. Only later they were recast in the language of vector analysis (as in Example~\ref{Ex:Maxwell}) by Heaviside \cite{Heaviside} with independent work by Gibbs and Hertz. The formulation in terms of differential forms emerged at the beginning of the 20th century, through the work of Cartan (e.g.~\cite{CartanMax}), Weyl (e.g.~\cite{WeylMax}) and others.
\end{remark}

\subsection{Linearised Yang-Mills Theory}\label{Sec:LinYM}
One of the most foundational examples of a gauge theory in physics is \emph{Yang-Mills theory}. It was initially formulated by Yang and Mills in 1954 \cite{YangMills} in order to generalise gauge theories with abelian gauge groups, which were well understood at the time, to the non-abelian case. In particular, they were interested in the case of $\mathrm{SU}(2)$ in an attempt to explain isospin conservation in strong interaction processes.\footnote{An equivalent theory was discovered by R.~Shaw, a PhD student of A.~Salam in Cambridge, in January 1954. However, it remained unpublished and only appeared as a chapter in his 1955 PhD thesis \cite{ShawPhD}. In a footnote, Shaw acknowledges Yang-Mills, stating: ``\textit{But although their} [Yang and Mills] \textit{publication date was in 1954, Yang and Mills must have priority since it seems that their research was completed in 1953}.'' \cite[p.~117]{Fraser}} The modern mathematical formulation using connections and principal fibre bundles goes back to Wu-Yang \cite{WuYang1} (see~\cite{WuYang2} for a historical account of the same authors) and Atiyah-Hitchin-Singer \cite{AtiyahYM1,AtiyahYM2}, which also marks the beginning of a time of intense mathematical interest and work on gauge theories. From the physical point of view, Yang-Mills theory is a pure gauge theory and serves as the foundation of the celebrated \textit{standard model of particle physics} (see e.g.~\cite{Hamilton,RudolphSchmidt2,Baum} for a mathematically oriented introduction to the \textit{classical} theory), which is essentially a Yang-Mills theory with gauge group $\mathrm{SU}(3)\times\mathrm{SU}(2)\times\mathrm{U}(1)$, complemented by a suitable matter sector containing the quarks, leptons and the Higgs boson. Besides its importance in theoretical and mathematical physics, it also has profound applications in pure mathematics, as already indicated above. For instance, the study of instantons and the moduli space of solutions to the Yang-Mills equations has led to deep new insights into the topology of smooth $4$-manifolds (e.g.~\cite{Donaldson1,Donaldson2}, see also the monographs of Donaldson-Kronheimer \cite{Donaldson}, Freed-Uhlenbeck \cite{FreedUhlenbeck} and Scorpan \cite{Scorpan} on that subject and related topics).

While a fully non-perturbative quantum Yang-Mills theory remains out of reach (see, e.g., the \textit{Millennium Prize} problem statement of Jaffe-Witten on the \textit{mass gap problem} \cite{JaffeWitten}), one can study the \textit{linearised} theory as a simpler toy model to capture low-energy effects of the expected quantum theory. Linearised Yang-Mills theory as a linear gauge theory in the sense of Definition~\ref{Def:LinGaugeTh} has been considered in the original article of Hack-Schenkel \cite[Ex.~3.7]{HackSchenkel} as well as in the work of Gérard-Wrochna on Hadamard states for linearised Yang-Mills theory \cite{GerardWrochna}, both in the case of \textit{trivial} principal bundles and with \textit{semisimple compact} Lie groups. The non-trivial case was briefly sketched in Wrochna-Zahn \cite[Sec.~4.2]{WrochnaZahn}.

In this section, we explore linearised Yang-Mills theory as a linear gauge theory within the Hack-Schenkel formalism, working with \emph{non-trivial} principal bundles and (possibly \emph{non-compact} and \emph{not necessarily semisimple}) Lie groups whose Lie algebra admits an invariant bilinear form.

\begin{setup}\label{SetUp:YangMills}
We consider the following general setting:
\begin{itemize}
	\item[$\bullet$]A $d$-dimensional globally hyperbolic spacetime $(\sf{M},\sf{g})$.
    \item[$\bullet$]A (finite-dimensional, real) Lie group $\sf{G}$ with Lie algebra $(\mathfrak{g},[\cdot,\cdot]_{\mathfrak{g}})$ and a choice of $\mathrm{Ad}$-invariant\footnote{i.e.~$\langle\mathrm{Ad}_{g}X,\mathrm{Ad}_{g}Y\rangle_{\mathfrak{g}}=\langle X,Y\rangle_{\mathfrak{g}}$ for all $X,Y\in\mathfrak{g}$ and $g\in\sf{G}$. Note that this also implies that $\langle\cdot,\cdot\rangle_{\mathfrak{g}}$ is \emph{skew-symmetric} with respect to $\mathrm{ad}\:\mathfrak{g}\to\mathrm{End}(\mathfrak{g})$, i.e.~$\langle\mathrm{ad}_{Z}(X),Y\rangle_{\mathfrak{g}}=-\langle X,\mathrm{ad}_{Z}(Y)\rangle_{\mathfrak{g}}$ for all $X,Y,Z\in\mathfrak{g}$.} non-degenerate symmetric bilinear form $\langle\cdot,\cdot\rangle_{\mathfrak{g}}$ on $\mathfrak{g}$.
    \item[$\bullet$]A principal $\sf{G}$-bundle $\sf{P}\xrightarrow{\pi}\sf{M}$ with corresponding right action $r_{g}\:\sf{P}\to\sf{G},\, p\cdot g:=r_{g}(p)$.
\end{itemize}
\end{setup}

For the sake of keeping this thesis self-contained, let us remind the reader about the following notation and terminology used in the set-up above: if $\sf{G}$ is a Lie group with Lie algebra $(\mathfrak{g},[\cdot,\cdot]_{\mathfrak{g}})$, we denote its \emph{adjoint representation} by
    \begin{align*}
        \mathrm{Ad}\:\sf{G}\to\mathrm{Aut}(\mathfrak{g})\,,\quad g\mapsto\mathrm{Ad}_{g}:=\mathrm{d}_{e}(c_{g})\in\mathrm{Aut}(\mathfrak{g})\,,
    \end{align*}
    where $e\in\sf{G}$ denotes the neutral element of $\sf{G}$, where we identified $\sf{T}_{e}\sf{G}\cong\mathfrak{g}$, as usual, and where $c_{g}\:\sf{G}\to\sf{G}$ denotes the conjugation map, i.e.~$c_{g}(h)=ghg^{-1}$ for $g,h\in\sf{G}$. In the special case in which $\sf{G}$ is a \emph{matrix Lie group}, i.e.~a closed subgroup of $\mathrm{Gl}(n,\mathbb{C})$, we have $\mathrm{Ad}_{g}(X)=gXg^{-1}$ for all $X \in \mathfrak{g}\subset\mathfrak{gl}(n,\bb{C})$. Moreover, we denote the induced Lie algebra representation by 
    \begin{align*}
        \mathrm{ad}\:\mathfrak{g}\to\mathrm{End}(\mathfrak{g})\,,\qquad X\mapsto \mathrm{ad}_{X}:=(\mathrm{d}_{e}\mathrm{Ad})(X)\in\mathrm{End}(\mathfrak{g})\,.
    \end{align*}
    For a general Lie group $\sf{G}$, it is easy to verify that $\mathrm{ad}_{X}(Y) = [X,Y]_{\mathfrak{g}}$ for all $X,Y \in \mathfrak{g}$.

\begin{remark}\label{Rem:YMAd}
	Requiring the existence of an $\mathrm{Ad}$-invariant bilinear form $\langle \cdot, \cdot \rangle_{\mathfrak{g}}$ imposes a restriction on the choice of Lie group $\sf{G}$. Nevertheless, the class of admissible Lie groups is fairly general and $\sf{G}$ does not need to be a \textit{matrix} Lie group. First, every \textit{abelian} Lie group trivially satisfies this assumption. Next, if $\sf{G}$ is \textit{semisimple}, then such a bilinear form always exists, since the Killing form
$\kappa\: \mathfrak{g} \times \mathfrak{g} \to \mathbb{R}$, defined by $\kappa(X,Y) = \mathrm{tr}(\mathrm{ad}_X \circ \mathrm{ad}_Y)$, provides an example in this case. Moreover, also \textit{compact} Lie groups are contained in this class, since in this case, one can use the Haar measure on $\sf{G}$ to average any inner product on $\mathfrak{g}$ and obtain an $\mathrm{Ad}$-invariant one. 

The examples above also cover the most important class of Lie groups for applications in physics: if $\sf{G}$ is compact and simple (such as $\mathrm{SU}(n)$, $n\in\bb{N}$), and if $\langle \cdot, \cdot \rangle_{\mathfrak{g}}$ is positive-definite, then $\langle\cdot,\cdot\rangle_{\mathfrak{g}}$ is in fact necessarily a (negative) multiple of the Killing form of $\mathfrak{g}$. In particular, if $\sf{G}$ is a simple, compact \textit{matrix} Lie group, the bilinear form takes the form $\langle X, Y \rangle_{\mathfrak{g}} \propto \mathrm{tr}(XY)$. The constant of proportionality corresponds to the \textit{coupling constant} in physics terminology.
\end{remark}

Before stating the Yang-Mills equations, we need to briefly review the relevant definitions of \textit{mathematical gauge theory}, that is, the differential geometry of principal fibre bundles. We refer to the textbooks \cite{Hamilton,Baum,RudolphSchmidt2} for details and proofs for the following well-known constructions. 

\paragraph{Connections and Curvature on Principal Bundles.} First of all, we recall that a \textit{connection 1-form} is a $\mathfrak{g}$-valued $1$-form $\sf{A}\in\Omega^{1}(\sf{P},\mathfrak{g})$ satisfying the following two conditions:
\begin{itemize}
\item[(i)]$\sf{A}$ is \textit{of type $\mathrm{Ad}$}, i.e. $(r_{g}^{\ast}\sf{A})_{p}=\mathrm{Ad}_{g^{-1}}\circ \sf{A}_{p}$ for all $g\in\sf{G}$ and $p\in\sf{P}$.
\item[(ii)]$\sf{A}(X^{\#})=X$ for all $X\in\mathfrak{g}$, where $X^{\#}\in\mathfrak{X}(\sf{P})$ is the \textit{fundamental vector field} of $X$, i.e. \begin{align*}
	X^{\#}_{p}:=\frac{\mathrm{d}}{\mathrm{d}t}\bigg\vert_{t=0}r_{\mathrm{exp}_{\sf{G}}(tX)}(p)=\frac{\mathrm{d}}{\mathrm{d}t}\bigg\vert_{t=0}(p\cdot\mathrm{exp}_{\sf{G}}(tX))\,,
\end{align*}
with $\mathrm{exp}_{\sf{G}}\:\mathfrak{g}\to\sf{G}$ denoting the \textit{exponential map} of $\sf{G}$.
\end{itemize} 
We denote the space of all connection $1$-forms by $\mathcal{C}(\sf{P})$. Now, it is also useful to introduce the following: let $\sf{V}_{p}:=\mathrm{ker}(\d\pi)$ denote the \textit{vertical tangent space at $p$}. An \textit{Ehresmann connection} (after \cite{Ehresmann}) is a vector subbundle $\sf{H}\subset \sf{T}\sf{P}$ such that 
\begin{align*}
	\sf{T}_{p}\sf{P}=\sf{H}_{p}\oplus \sf{V}_{p}
\end{align*}	
for all $p\in\sf{M}$. Once an Ehresmann connection is fixed, a tangent vector in $\sf{H}_{p}$ is called \textit{horizontal}. If, in addition, it holds that $(r_{g})_{\ast}(\sf{H}_{p})=\sf{H}_{p\cdot g}$ for all $p\in\sf{P}$ and $g\in\sf{G}$, $\sf{H}$ is called a \textit{principal Ehresmann connection}. Now, any connection $1$-form $\sf{A}\in\mathcal{C}(\sf{P})$ defines a principal Ehresmann connection by $\sf{H}_{p}:=\mathrm{ker}(\sf{A}_{p})$. On the other hand, if $\sf{H}$ is a principal Ehresmann connection, then setting $\sf{A}_{p}(X^{\sharp}_{p}+Y_{p}):=\sf{X}$ for all $X\in\mathfrak{g}$ and $Y_{p}\in\sf{H}_{p}$ defines a connection $1$-form.  We remark that the concept of Ehresmann connections is very general and allows to define the notion of \textit{connections} and \textit{curvature} even on arbitrary \textit{fibre} bundles, containing both vector and principal bundle connections as special cases, see e.g.~\cite{Michor1,Michor2} for more details.

Now, for any connection $1$-form $\sf{A}\in\mathcal{C}(\sf{P})$, we define the corresponding \textit{curvature 2-form} by means of \textit{Cartan's structure equation}, i.e.
\begin{align}\label{eq:CartanSE}
	\sf{F}^{\sf{A}}:=\d\sf{A}+\frac{1}{2}[\sf{A}\wedge\sf{A}]_{\mathfrak{g}}\in\Omega^{2}(\sf{P},\mathfrak{g})\, ,
\end{align}
where $[\cdot\wedge\cdot]_{\mathfrak{g}}\:\Omega^{k}(\sf{P},\mathfrak{g})\times\Omega^{l}(\sf{P},\mathfrak{g})\to\Omega^{k+l}(\sf{P},\mathfrak{g})$ denotes the natural wedge product induced from the Lie algebra bracket $[\cdot,\cdot]_{\mathfrak{g}}$ on $\mathfrak{g}$, as defined in Remark~\ref{Rem:Wedge} below.

\begin{remark}\label{Rem:Wedge} (Wedge Products)\newline
In texts on mathematical gauge theory, one frequently encounters various types of wedge products for vector- and bundle-valued forms, which at first sight can be quite confusing. It is often not sufficiently appreciated that all of these are, in fact, special cases of the following: if $\sf{V},\sf{W},\sf{Z}$ are finite-dimensional $\bb{R}$-vector spaces and $\mu\:\sf{V}\times\sf{W}\to\sf{Z}$ a bilinear map, then we define
\begin{align*}
	\wedge_{\mu}\:\Omega^{k}(\sf{M},\sf{V})\times\Omega^{l}(\sf{M},\sf{W})\to\Omega^{k+l}(\sf{M},\sf{Z}),\qquad \alpha\wedge_{\mu}\beta:=(\alpha^{a}\wedge\beta^{b})\otimes\mu(e_{a},f_{b})\, ,
\end{align*}
where $(e_{a})_{a}$ and $(f_{b})_{b}$ are bases of $\sf{V}$ and $\sf{W}$, respectively, and where we wrote $\alpha=\alpha^{a}\otimes e_{a}$ and $\beta=\beta^{b}\otimes f_{b}$ for coefficient forms $\alpha^{a}\in\Omega^{k}(\sf{M})$ and $\beta^{a}\in\Omega^{l}(\sf{M})$. As an example, for a Lie algebra $(\mathfrak{g},[\cdot,\cdot]_{\mathfrak{g}})$, we obtain the wedge product $[\cdot\wedge\cdot]_{\mathfrak{g}}:=\wedge_{[\cdot,\cdot]_{\mathfrak{g}}}$ mentioned above.

More generally, if $\sf{E},\sf{F},\sf{L}$ are $\bb{R}$-bundles over $\sf{M}$ and we are given a section $\mu\in\Gamma^{\infty}(\sf{L}\otimes (\sf{E}\otimes\sf{F})^{\ast})$, i.e.~a collection of bilinear maps $\mu_{p}\:\sf{E}_{p}\times\sf{F}_{p}\to\sf{L}_{p}$ depending smoothly on $p\in\sf{M}$, we set
\begin{align*}
	\wedge_{\mu}\:\Omega^{k}(\sf{M},\sf{E})\times\Omega^{l}(\sf{M},\sf{F})\to\Omega^{k+l}(\sf{M},\sf{L}),\qquad \alpha\wedge_{\mu}\beta:=(\alpha^{a}\wedge\beta^{b})\otimes\mu(e_{a},f_{b})\, ,
\end{align*}
where $\alpha=\alpha^{a}\otimes e_{a}$ and $\beta=\beta^{b}\otimes f_{b}$ for coefficients $\alpha^{a}\in\Omega^{k}(\mathcal{U})$ and $\beta^{a}\in\Omega^{l}(\mathcal{U})$ and local frames $(e_{a})_{a}$ and $(f_{b})_{b}$ of $\sf{E}$ and $\sf{F}$ defined on some open subset $\mathcal{U}\subset\sf{M}$, respectively, and where $\mu(e_{a},f_{b})\in\Gamma^{\infty}(\mathcal{U},\sf{L})$ is defined as $\mu(e_{a},f_{b})(p):=\mu_{p}(e_{a}(p),f_{b}(p))$ for all $p\in\sf{M}$, as usual.
\end{remark}

\paragraph{Associated Vector Bundles and Gauge Covariant Derivative.}
Now, suppose that we are given a Lie group representation $\rho\:\sf{G}\to\mathrm{Aut}(\sf{W})$ on some $\bb{R}$-vector space $\sf{W}$. Furthermore, assume that $\sf{W}$ is equipped with a $\rho$-invariant non-degenerate symmetric bilinear form $\langle\cdot,\cdot\rangle_{\sf{W}}$.

Now, consider the \textit{associated vector bundle} $\sf{E}:=\sf{P}\times_{\rho}\sf{W}$. In other words, for a given point $x\in\sf{M}$, the fibre of $\sf{E}$ at $x$ is the quotient space $\sf{E}_{x}:=(\sf{P}_{x}\times\sf{W})/\sf{G}$ with respect to the group action $(\sf{P}_{x}\times\sf{E})\times\sf{G}\to (\sf{P}_{x}\times\sf{E})$ defined by
\begin{align*}
	(p,v)\cdot g:=(p\cdot g,\rho(g^{-1})v)
\end{align*}
for all $p\in\sf{P}_{x}$, $v\in\sf{W}$ and $g\in\sf{G}$. We write $[p,v]\in \sf{E}_{x}$ for the corresponding equivalence class. An important special case is provided by the \textit{adjoint bundle} $\mathrm{Ad}(\sf{P}):=\sf{P}\times_{\mathrm{Ad}}\mathfrak{g}$, which will play a central role in Yang-Mills theory. At this point, we remark that the bilinear form $\langle\cdot,\cdot\rangle_{\sf{W}}$ induces a natural bundle metric on $\sf{E}$ via
\begin{align}\label{eq:IndBundMet}
	\langle [p,v],[p,w]\rangle_{\sf{E}_{x}}:=\langle v,w\rangle_{\sf{W}}
\end{align}
for all $x\in\sf{M}$, $p\in\sf{P}_{x}$ and $v,w\in\sf{W}$. Indeed, $\rho$-invariance of $\langle\cdot,\cdot\rangle_{\sf{W}}$ shows that this definition is independent of the representatives chosen.

Now, it is a well-known fact that a suitable linear subspace of $\Omega^k(\sf{P},\sf{W})$ can naturally be identified with $\sf{E}$-valued forms on the spacetime manifold $\sf{M}$. This identification is crucial for formulating Yang-Mills theory, since it allows us to formulate Yang-Mills fields and their curvature, which are originally defined on the principal bundle $\sf{P}$, \textit{globally}\footnote{Locally, this can, of course, always be done by simply choosing a \emph{local gauge}, i.e.~a local section $s\in\Gamma^{\infty}(\mathcal{U},\sf{P})$ defined on some open subset $\mathcal{U}\subset\sf{M}$, namely by considering the local gauge representative $\sf{A}_{s}:=s^{\ast}\sf{A}\in\Omega^{1}(\sf{P},\mathfrak{g})$. } in terms of geometric data living on the base manifold $\sf{M}$. More precisely, there is natural identification\footnote{Explicitly, $(\mathrm{I}\omega)_{x}(v_{1},\dots,v_{k}):=[p,\omega_{p}(X_{1},\dots,X_{k})]$ for all $x\in\sf{M}$, $p\in\sf{P}_{x}$ and $v_{1},\dots,v_{k}\in\sf{T}_{x}\sf{M}$, where $X_{i}\in\sf{T}_{p}\sf{P}$ are such that $\d\pi_{p}(X_{i})=v_{i}$. It is easy to see that this definition is independent of the choice of $X_{i}$.}
\begin{align}\label{eq:IsoPrinc}
	\mathrm{I}\:\Omega^{k}_{\mathrm{hor},\rho}(\sf{P},\sf{W})\xrightarrow{\cong}\Omega^{k}(\sf{M},\sf{E})\,,\qquad\omega\mapsto\omega_{\sf{M}}:=\mathrm{I}\omega\, ,
\end{align}
where $\Omega^{k}_{\mathrm{hor},\rho}(\sf{P},\sf{W})$ is the linear subspace consisting of all $\omega\in\Omega^{k}(\sf{P},\sf{W})$ that are 
\begin{itemize}
	\item[(i)]\textit{of type $\rho$}, i.e.~$(r_{g}^{\ast}\omega)_{p}=\rho(g^{-1})\circ \omega_{p}$ for all $p\in\sf{P}$ and $g\in\sf{G}$, and
	\item[(ii)]\textit{horizontal} in the sense that $\omega_{p}(v_{1},\dots,v_{k})=0$ if at least one vector $v_{i}\in\sf{T}_{p}\sf{P}$ is vertical.
\end{itemize}
The following two statements are well known can readily be shown:
\begin{itemize}
	\item[$\bullet$]$\sf{F}^{\sf{A}}\in\Omega^{2}_{\mathrm{hor},\mathrm{Ad}}(\sf{P},\mathfrak{g})$. Hence, it can be identified with a $2$-form $\sf{F}^{\sf{A}}_{\sf{M}}\in\Omega^{2}(\sf{M},\mathrm{Ad}(\sf{P}))$.
	\item[$\bullet$]If $\sf{A}_{1},\sf{A}_{2}\in\mathcal{C}(\sf{P})$, then $\sf{A}_{1}-\sf{A}_{2}\in\Omega^{1}_{\mathrm{hor},\mathrm{Ad}}(\sf{P},\mathfrak{g})$. Furthermore, $\sf{A}_{1}+\alpha\in\mathcal{C}(\sf{P})$ for every $\alpha\in\Omega^{1}_{\mathrm{hor},\mathrm{Ad}}(\sf{P},\mathfrak{g})$. Hence, $\mathcal{C}(\sf{P})$ is an affine space modelled over $\Omega^{1}_{\mathrm{hor},\mathrm{Ad}}(\sf{P},\mathfrak{g})\cong\Omega^{1}(\sf{M},\mathrm{Ad}(\sf{P}))$.
\end{itemize}

Now, any $\sf{A}\in\mathcal{C}(\sf{P})$ defines a natural connection $\nabla^{\sf{A}}$ on the vector bundle $\sf{E}$, called the \textit{gauge covariant derivative}, as follows: first, we define $\sf{D}_{\sf{A}}\:\Omega^{k}(\sf{P},\mathfrak{g})\to\Omega^{k+1}(\sf{P},\mathfrak{g})$ by
\begin{align*}
	(\sf{D}_{\sf{A}}\alpha)_{p}(v_{1},\dots,v_{k+1}):=(\d\alpha)_{p}(\mathrm{pr}_{\sf{H}}(v_{1}),\dots,\mathrm{pr}_{\sf{H}}(v_{k+1}))
\end{align*}
for all $p\in\sf{P}$ and $v_{1},\dots,v_{k+1}\in\sf{T}_{p}\sf{P}$, where $\mathrm{pr}_{\sf{H}}\:\sf{T}_{p}\sf{P}\to\sf{H}_{p}$ denotes the horizontal projection. In other words, $\sf{D}_{\sf{A}}$ is an exterior derivative on $\sf{P}$ preserving horizontality. Now, it is easy to see that $\sf{D}_{\sf{A}}$ acts on the linear subspace $\Omega^{k}_{\mathrm{hor},\rho}(\sf{P},\sf{W})$ as
\begin{align}\label{eq:CovDerHor}
	\sf{D}_{\sf{A}}\omega=\d\omega+\sf{A}\wedge_{\rho}\omega\,,
\end{align}
where $\wedge_{\rho}\:\Omega^{k}(\sf{P},\mathfrak{g})\times\Omega^{l}(\sf{P},\sf{W})\to\Omega^{k+l}(\sf{P},\sf{W})$ is the obvious wedge product induced from the bilinear map $\mathfrak{g}\times\sf{W}\to\mathsf{V},(X,v)\mapsto\rho_{\ast}(X)v$ with $\rho_{\ast}\:\mathfrak{g}\to\mathrm{End}(\mathfrak{g})$ denoting the induced Lie algebra representation\footnote{Every Lie group homomorphism $\phi\:\sf{G}\to\sf{H}$ induces a Lie algebra homomorphism $\phi_{\ast}:=\d_{e}\phi\:\mathfrak{g}\to\mathfrak{h}$, where $e$ denotes the neutral element of $\sf{G}$. By definition, it holds that $\phi(\mathrm{exp}_{\sf{G}}(X))=\mathrm{exp}_{\sf{H}}(\phi_{\ast}(X))$ for all $X\in\mathfrak{g}$.} (see Remark~\ref{Rem:Wedge}). In the special case in which $(\sf{W},\rho)=(\mathfrak{g},\mathrm{Ad})$, this coincides with $[\cdot\wedge\cdot]_{\mathfrak{g}}$. Furthermore, it is easy to show that $\sf{F}^{\sf{A}}=\sf{D}_{\sf{A}}\sf{A}$ for all $\sf{A}\in\mathcal{C}(\sf{P})$.

Now, looking at Eq.~\eqref{eq:CovDerHor}, we observe that $\sf{D}_{\sf{A}}$  maps the subspace $\Omega^{k}_{\mathrm{hor},\rho}(\sf{P},\sf{W})$ onto $\Omega^{k+1}_{\mathrm{hor},\rho}(\sf{P},\sf{W})$ and hence gives rise to a family of operators 
\begin{align*}
	\d_{\sf{A}}\:\Omega^{k}(\sf{M},\sf{E})\to\Omega^{k+1}(\sf{M},\sf{E})\,,\qquad \d_{\sf{A}}\omega_{\sf{M}}:=(\sf{D}_{\sf{A}}\omega)_{\sf{M}}
\end{align*}
in view of the isomorphism~\eqref{eq:IsoPrinc}. The restriction to $\Omega^{0}(\sf{M},\sf{E})=\Gamma^{\infty}(\sf{E})$ defines a connection $\nabla^{\sf{A}}\:\Gamma^{\infty}(\sf{E})\to\Gamma^{\infty}(\sf{E}\otimes\sf{T}^{\ast}\sf{M})$ on $\sf{E}$ and it turns out that the full family $\d_{\sf{A}}$ is exactly the \textit{exterior covariant derivative} induced by $\nabla^{\sf{A}}$, i.e.~$\d_{\sf{A}}=\d^{\nabla^{\sf{A}}}$. Note also that $\nabla^{\sf{A}}$ is metric with respect to the induced bundle metric $\langle\cdot,\cdot\rangle_{\sf{E}}$ defined in Eq.~\eqref{eq:IndBundMet}.

\paragraph{Bundle Metrics, Hodge Star Operator and Codifferential.} At this stage, we have almost introduced all the required notation and we now turn to discuss the relevant bundle metrics and operators for non-linear and linearised Yang-Mills theory. 

As before, we consider an arbitrary Lie group representation $\rho\:\sf{G}\to\mathrm{Aut}(\sf{W})$ on some $\bb{R}$-vector space $\sf{W}$ equipped with an $\rho$-invariant non-degenerate symmetric bilinear form $\langle\cdot,\cdot\rangle_{\sf{W}}$. An important special case, which we will mainly use in the following discussion, is $(\rho,\sf{W})=(\mathrm{Ad},\mathfrak{g})$ with $\langle\cdot,\cdot\rangle_{\mathfrak{g}}$. Different representation will be considered in the next section in which we describe the coupled Yang-Mills-Klein-Gordon system.

Recall the notation $(\sf{A}_{k}:=\bigwedge^{k}\sf{T}^{\ast}\sf{M},\langle\cdot,\cdot\rangle_{\sf{A}_{k}})$ from the previous section. In the following, we choose a local frame $(e_{a})_{a}$ of $\sf{E}$ on some open subset $\mathcal{U}\subset\sf{M}$ and write every $\alpha\in\Omega^{k}(\sf{M},\sf{E})$ locally as $\alpha=\alpha^{a}\otimes e_{a}$ for coefficient forms $\alpha^{a}\in\Omega^{k}(\mathcal{U})$. Now, as a generalisation of Eq.~\eqref{eq:ExtBundles}, we consider the Hermitian bundles 
\begin{align}\label{eq:ExtBundlesTwist}
	\sf{E}_{k}:=\sf{E}\otimes\sf{A}_{k}\,,\qquad \langle\alpha,\beta\rangle_{\sf{E}_{k}}:=\langle\alpha^{a},\beta^{b}\rangle_{\sf{A}_{k}}\langle e_{a},e_{b}\rangle_{\sf{E}}\, ,
\end{align}
where $\langle\cdot,\cdot\rangle_{\sf{E}}$ denotes the induced bundle metric as defined in Eq.~\eqref{eq:IndBundMet}. By definition, we have that $\Omega^{k}(\sf{M},\sf{E})=\Gamma^{\infty}(\sf{E}_{k})$. The \textit{Hodge $\ast$-operator} for $\sf{E}$-valued forms is defined, as usual, by
\begin{align*}
	\ast\:\Omega^{k}(\sf{M},\sf{E})\to\Omega^{d-k}(\sf{M},\sf{E}),\qquad \ast\omega:=(\ast\omega^{a})\otimes e_{a}\, .
\end{align*}
Last but not least, we introduce two wedge-type products for $\sf{E}$-valued forms (cf.~Remark~\ref{Rem:Wedge}):
\begin{itemize}
	\item[$\bullet$]First, the bundle metric $\langle\cdot,\cdot\rangle_{\sf{E}}\in\Gamma^{\infty}(\sf{E}^{\ast}\otimes\sf{E}^{\ast})$ induces a wedge product denoted by
	\begin{align*}
		\langle\cdot\wedge\cdot\rangle_{\sf{E}}\:\Omega^{k}(\sf{M},\sf{E})\times\Omega^{l}(\sf{M},\sf{E})\to\Omega^{k+l}(\sf{M})\, .
	\end{align*}
	We note that in the literature, this is usually denoted by $\mathrm{tr}(\cdot \wedge \cdot)$ in the special case $\mathrm{Ad}(\sf{P}) = \sf{E}$, reflecting the fact that the pairing $\langle \cdot, \cdot \rangle_{\mathfrak{g}}$ is typically chosen to be the Killing form whenever $\sf{G}$ is a simple matrix Lie group, cf.~Remark~\ref{Rem:YMAd}.
	\item[$\bullet$]The wedge product $\wedge_{\rho}\:\Omega^{k}(\sf{P},\mathfrak{g})\times\Omega^{l}(\sf{P},\sf{W})\to\Omega^{k+l}(\sf{P},\sf{W})$, induced by the pairing $\mathfrak{g}\times\sf{W}\ni (X,v)\mapsto \rho_{\ast}(X)v\in\sf{W}$, as above, is well-defined as a map
	\begin{align*}
		\wedge_{\rho}\:\Omega^{k}_{\mathrm{hor},\mathrm{Ad}}(\sf{P},\mathfrak{g})\times\Omega_{\mathrm{hor},\rho}^{l}(\sf{P},\sf{W})\to\Omega_{\mathrm{hor},\rho}^{k+l}(\sf{P},\sf{W})\,.
	\end{align*}
	Hence, using the isomorphism in Eq.~\eqref{eq:IsoPrinc}, we obtain a well-defined wedge product
	\begin{align}\label{eq:StrangeWedge}
		\wedge_{\sf{E}}\:\Omega^{k}(\sf{M},\mathrm{Ad}(\sf{P}))\times\Omega^{l}(\sf{M},\sf{E})\to\Omega^{k+l}(\sf{M},\sf{E})\, .
	\end{align}
	In other words, $\wedge_{\sf{E}}$ is the wedge-product induced by the map fibre-wise bilinear pairing $\mathrm{Ad}(\sf{P})	_{x}\times\sf{E}_{x}\to\sf{E}_{x}$ defined by $([p,X],[p,v])\mapsto [p,\rho_{\ast}(X)v]$ for all $x\in\sf{M}$, $p\in\sf{P}_{x}$, $X\in\mathfrak{g}$ and $v\in\sf{W}$. In the special case in which $\sf{E}=\mathrm{Ad}(\sf{P})$, we use the according notation $[\cdot\wedge\cdot]_{\mathrm{Ad}(\sf{P})}$.
\end{itemize}
With these definitions, we also obtain a corresponding generalisation of Eq.~\eqref{eq:ExtBundles2}, i.e.
\begin{align*}
	(\alpha,\beta)_{\sf{E}_{k}}:=\int_{\sf{M}}\langle\alpha,\beta\rangle_{\sf{E}_{k}}\,\d\mu_{\sf{g}}=\int_{\sf{M}}\langle\alpha\wedge\ast\beta\rangle_{\sf{E}}
\end{align*}
for all $\alpha,\beta\in\Omega^{k}(\sf{M},\sf{E})$ with compactly overlapping supports. Last but not least, we denote the formal adjoint of $\d_{\sf{A}}$ with respect to $(\cdot,\cdot)_{\sf{E}_{k}}$, the \textit{covariant codifferential}, by \begin{align*}
	\delta_{\sf{A}}\:\Omega^{k+1}(\sf{M},\sf{E})\to\Omega^{k}(\sf{M},\sf{E})\,,\qquad \delta_{\sf{A}}=(-1)^{k+1}\ast^{-1}\d_{\sf{A}}\ast\, .
\end{align*}

\paragraph{Linearised Yang-Mills Theory.}  The \textit{Yang-Mills equation} is the non-linear equation
\begin{align}\label{eq:YangMills}
	\delta_{\sf{A}}\sf{F}^{\sf{A}}_{\sf{M}}=0
\end{align}
for a given $\sf{A}\in\mathcal{C}(\sf{P})$. Together with the \textit{Bianchi identity} $\d_{\sf{A}}\sf{F}^{\sf{A}}_{\sf{M}}=0$, the Yang-Mills equation can be seen as a generalisation of Maxwell's theory, which is obtained as the special case with abelian gauge group $\mathrm{U}(1)$. (Global) Existence and uniqueness of solutions to~Eq.~\eqref{eq:YangMills} has, for instance, been studied by Segal \cite{SegalYangMills}, Choquet-Bruhat-Christodoulou \cite{Christodoulou}, Eardley-Moncrief \cite{EardleyM1,EardleyM2} and Klainerman-Machedon \cite{Klainerman} in Minkowski spacetime, and by Choquet-Bruhat \cite{ChoquetBruhatYangMills} (in 1+2 dimensions) and Chrusciel-Shatah \cite{Chrusciel} (1+3 dimensions) in globally hyperbolic spacetimes, with earlier works on the Einstein cylinder $\bb{R}\times S^{3}$ in \cite{ChristodoulouEC,ChoquetBruhatEC}. Global existence of the Yang-Mills equations has also been established for anti-de Sitter space in \cite{ChoquetBruhatADS}.\footnote{It should be noted that compactness of the Lie group seems to play a role. For instance, in \cite{Yang}, it is shown that for $\sf{G}=\mathrm{SU}(2,\mathbb{C})$ there exist finite-time blowing-up solutions over Minkowski spacetime.}

Let us now turn to the linearised theory. Since the space of connections $\mathcal{C}(\sf{P})$ is an affine space modelled over $\Omega^{1}_{\mathrm{hor},\mathrm{Ad}}(\sf{P},\mathfrak{g})$, we define a \textit{linearisation} or \textit{perturbation} of $\sf{A}\in\mathcal{C}(\sf{P})$ to be a $1$-form in $\Omega^{1}_{\mathrm{hor},\mathrm{Ad}}(\sf{P},\mathfrak{g})\cong\Omega^{1}(\sf{M},\mathrm{Ad}(\sf{P}))$. In fact, this can also be made more precise from a geometric perspective: being an affine space, $\mathcal{C}(\sf{P})$ can be equipped with the structure of an $\infty$-dimensional Fréchet manifold and its tangent space at $\sf{A}$ is exactly given by $\sf{T}_{\sf{A}}\mathcal{C}(\sf{P})\cong\Omega^{1}_{\mathrm{hor},\mathrm{Ad}}(\sf{P},\mathfrak{g})\cong\Omega^{1}(\sf{M},\mathrm{Ad}(\sf{P}))$, see for instance \cite[Sec.~B.2]{Naber}, \cite[Sec.~6.1]{RudolphSchmidt2} or \cite{AtiyahBott}.

\begin{proposition}\label{prop:LinYangMi} \emph{(Linearised Yang-Mills Equations)}\newline
	Let $\sf{A}\in\mathcal{C}^{1}(\sf{P})$ be a connection $1$-form and consider a linear perturbation $\alpha\in\Omega^{1}_{\mathrm{hor},\mathrm{Ad}}(\sf{P},\mathfrak{g})$. Writing $\alpha_{\sf{M}}:=\mathrm{I}\alpha\in\Omega^{1}(\sf{M},\mathrm{Ad}(\sf{P}))$, as usual, it holds that
	\begin{align*}
		\delta_{\sf{A}+\lambda\alpha}\sf{F}_{\sf{M}}^{\sf{A}+\lambda\alpha}=\delta_{\sf{A}}\sf{F}_{\sf{M}}^{\sf{A}}+\lambda\bigg\{\delta_{\sf{A}}\d_{\sf{A}}\alpha_{\sf{M}}-\ast[\ast\sf{F}_{\sf{M}}^{\sf{A}}\wedge\alpha_{\sf{M}}]_{\mathrm{Ad}(\sf{P})}\bigg\}+\mathcal{O}(\lambda^{2})\, .
	\end{align*}
\end{proposition} 

\begin{proof}
	It will be easier to work first with forms defined on $\sf{P}$ rather than on $\sf{M}$, since in this case we have explicit formulas we can use. As a first step, the structure equation~\eqref{eq:CartanSE} implies 
	\begin{align*}
		\sf{F}^{\sf{A}+\lambda\alpha}&=\d(\sf{A}+\lambda\alpha)+\frac{1}{2}[(\sf{A}+\lambda\alpha)\wedge (\sf{A}+\lambda\alpha)]_{\mathfrak{g}}=\\&=\sf{F}^{\sf{A}}+\lambda (\d\alpha+[\sf{A}\wedge\alpha]_{\mathfrak{g}})+\mathcal{O}(\lambda^{2})=\sf{F}^{\sf{A}}+\lambda \sf{D}_{\sf{A}}\alpha+\mathcal{O}(\lambda^{2})\, ,
	\end{align*}
	where we used Equation~\eqref{eq:CovDerHor} in the last step. Hence, using the isomorphism~\eqref{eq:IsoPrinc}, we obtain the corresponding expansion on $\sf{M}$, i.e.
	\begin{align}\label{eq:ExpCurv}
		\sf{F}_{\sf{M}}^{\sf{A}+\lambda\alpha}=\sf{F}_{\sf{M}}^{\sf{A}}+\lambda(\d_{\sf{A}}\alpha_{\sf{M}})+\mathcal{O}(\lambda^{2})\, .
	\end{align}
	Now, let $\omega\in\Omega^{k}_{\mathrm{hor},\mathrm{Ad}}(\sf{P},\mathfrak{g})$ be arbitrary and let us denote the corresponding form on $\sf{M}$ by $\omega_{\sf{M}}\in\Omega^{k}(\sf{M},\mathrm{Ad}(\sf{P}))$. Then, using again Eq.~\eqref{eq:CovDerHor}, we obtain $\sf{D}_{\sf{A}+\lambda\alpha}\omega=\sf{D}_{\sf{A}}\omega+\lambda [\alpha\wedge\omega]_{\mathfrak{g}}$ and hence, via the isomorphism~\eqref{eq:IsoPrinc},
	\begin{align}\label{eq:ExpExt}
		\d_{\sf{A}+\lambda\alpha}\omega_{\sf{M}}=\d_{\sf{A}}\omega_{\sf{M}}+\lambda [\alpha_{\sf{M}}\wedge\omega_{\sf{M}}]_{\mathrm{Ad}(\sf{P})}\, .
	\end{align}
	Combining the expansions in Eq.~\eqref{eq:ExpCurv} and Eq.~\eqref{eq:ExpExt}, we obtain the following expansion:
	\begin{align*}
		\delta_{\sf{A}+\lambda\alpha}\sf{F}_{\sf{M}}^{\sf{A}+\lambda\alpha}&=\ast^{-1}\sf{d}_{\sf{A}+\lambda\alpha}\ast (\sf{F}_{\sf{M}}^{\sf{A}}+\lambda\d_{\sf{A}}\alpha)+\mathcal{O}(\lambda^{2})=\\&=\delta_{\sf{A}}\sf{F}_{\sf{M}}^{\sf{A}}+\lambda\bigg\{\ast^{-1}[\alpha_{\sf{M}}\wedge\ast\sf{F}^{\sf{A}}_{\sf{M}}]_{\mathrm{Ad}(\sf{P})}+\delta_{\sf{A}}\d_{\sf{A}}\alpha_{\sf{M}}\bigg\}+\mathcal{O}(\lambda^{2})\, .
	\end{align*}
	Last but not least, let us arrange the zeroth-order term. In Lorentzian signature, it holds that $\ast^{-1}=-(-1)^{k(d-k)}\ast$ when acting on $k$-forms. On the other hand, the wedge-product $[\cdot\wedge\cdot]_{\mathrm{Ad}(\sf{P})}$ satisfies $[\alpha\wedge\beta]_{\mathrm{Ad}(\sf{P})}=-(-1)^{kl}[\beta\wedge\alpha]_{\mathrm{Ad}(\sf{P})}$ for all $\alpha\in\Omega^{k}(\sf{M},\mathrm{Ad}(\sf{P}))$ and $\beta\in\Omega^{l}(\sf{M},\mathrm{Ad}(\sf{P}))$, where the additional minus sign comes from the antisymmetry of $[\cdot,\cdot]_{\mathfrak{g}}$. Hence, we find
	\begin{align*}
		\ast^{-1}[\alpha_{\sf{M}}\wedge\ast\sf{F}^{\sf{A}}_{\sf{M}}]_{\mathrm{Ad}(\sf{P})}=-(-1)^{d-1}\ast[\alpha_{\sf{M}}\wedge\ast\sf{F}^{\sf{A}}_{\sf{M}}]_{\mathrm{Ad}(\sf{P})}=-\ast[\ast\sf{F}^{\sf{A}}_{\sf{M}}\wedge \alpha_{\sf{M}}]_{\mathrm{Ad}(\sf{P})}\,,
	\end{align*}
	which concludes the proof.
\end{proof}

The Yang-Mills equations are gauge-invariant in the following sense: a \textit{(global) gauge transformation} is a \textit{principal bundle automorphism}, i.e.~a diffeomorphism $f\in\mathrm{Diff}(\sf{P},\sf{P})$ that preserves the fibres, i.e.~$\pi\circ f=\pi$, and that is $\sf{G}$-equivariant, i.e.~$f(p\cdot g)=f(p)\cdot g$ for all $p\in\sf{P}$ and $g\in\sf{G}$. The group of (global) gauge transformations, denoted by $\mathcal{G}(\sf{P})$, is naturally isomorphic to the group of $\sf{G}$-equivariant smooth maps from $\sf{P}$ to $\sf{G}$, i.e.
\begin{align*}
	\mathcal{G}(\sf{G})\cong C^{\infty}(\sf{P},\sf{G})^{\sf{G}}:=\{\sigma\in C^{\infty}(\sf{P},\sf{G})\mid \forall p\in\sf{P},\,g\in\sf{G}:\sigma(p\cdot g)=g^{-1}\sigma(p)g\},\quad f\mapsto \sigma_{f}\, ,
\end{align*}
where $\sigma_{f}(p)\in\sf{G}$ is uniquely defined by $f(p)=p\cdot\sigma_{f}(p)$ for all $p\in\sf{P}$. A gauge transformation $f$ acts on a connection $1$-form $\sf{A}\in\mathcal{C}(\sf{P})$ via pull-back, i.e.
\begin{align}\label{eq:GaugTrafoCon}
	\sf{A}\mapsto f^{\ast}\sf{A}=\mathrm{Ad}_{\sigma_{f}^{-1}}\circ\sf{A}+\sigma_{f}^{\ast}\mu_{\sf{G}}\, ,
\end{align}
where $\mu_{\sf{G}}\in\Omega^{1}(\sf{G},\mathfrak{g})$ denotes the Maurer-Cartan form\footnote{If $\sf{G}$ is a Lie group with Lie algebra $(\mathfrak{g},[\cdot,\cdot]_{\mathfrak{g}})$, the \emph{Maurer-Cartan form} is defined by $\mu_{g}(v):=\d_{g}(l_{g^{-1}})(v)$ for all $g\in\sf{G}$ and $v\in\sf{T}_{g}\sf{G}$, where $l_{g}\:\sf{G}\to\sf{G}$ with $l_{g}(h):=gh$ denotes the left group multiplication. By definition, $\mu$ satisfies the \emph{Maurer-Cartan equation} $\d\mu+\frac{1}{2}[\mu\wedge\mu]_{\mathfrak{g}}=0$.} of $\sf{G}$. In the associate bundle $\sf{E}=\sf{P}\times_{\rho}\sf{W}$, gauge transformations act as
\begin{align*}
	\mathcal{G}(\sf{P})\times\sf{E}_{x}\to\sf{E}_{x}\,,\qquad (f,[p,v])\mapsto f\cdot [p,v]:=[f(p),v]=[p\cdot\sigma_{f}(p),v]=[p,\rho(\sigma_{f}(p)^{-1})v]\,,
\end{align*}
which is clearly well-defined by $\sf{G}$-equivariancy of $\sigma_{f}$. On sections of $\sf{E}$, or more generally forms in $\Omega^{k}(\sf{M},\sf{E})$, gauge transformations are defined pointwise.\footnote{We note that this is also compatible with the isomorphism~\eqref{eq:IsoPrinc}, i.e.~$(f^{\ast}\omega)_{\sf{M}}=f^{-1}\cdot\omega_{\sf{M}}$ for all $\omega\in\Omega^{k}_{\mathrm{hor},\rho}(\sf{P},\sf{W})$.} Now, it is easy to see that
\begin{align*}
	\sf{d}_{f^{\ast}\sf{A}}(f^{-1}\cdot\omega)=f^{-1}\cdot(\sf{d}_{\sf{A}}\omega)\quad\text{and}\quad \delta_{f^{\ast}\sf{A}}(f^{-1}\cdot\omega)=f^{-1}\cdot(\delta_{\sf{A}}\omega)\,,
\end{align*}
for all $\omega\in\Omega^{k}(\sf{M},\sf{E})$ and $\sf{A}\in\mathcal{C}(\sf{P})$, see e.g.~\cite[Lemma~7.5.8]{Hamilton}. Furthermore, $\sf{F}^{f^{\ast}\sf{A}}=f^{\ast}\sf{F}^{\sf{A}}$ and hence $\sf{F}_{\sf{M}}^{f^{\ast}\sf{A}}=f^{-1}\cdot\sf{F}_{\sf{M}}^{\sf{A}}$, which implies that the non-linear Yang-Mills equations~\eqref{eq:YangMills} are gauge invariant in the sense that
\begin{align}\label{eq:YMGaugeInv}
	\delta_{f^{\ast}\sf{A}}\sf{F}^{f^{\ast}\sf{A}}_{\sf{M}}=f^{-1}\cdot\delta_{\sf{A}}\sf{F}^{\sf{A}}_{\sf{M}}
\end{align}
for all $f\in\mathcal{G}(\sf{P})$ and $\sf{A}\in\mathcal{C}(\sf{P})$. It can be shown that $\mathcal{G}(\sf{P})$ has the structure of an infinite-dimensional Fréchet Lie group with Lie algebra $\Gamma^{\infty}(\mathrm{Ad}(\sf{P}))=\Omega^{0}(\sf{M},\mathrm{Ad}(\sf{P}))$, see e.g.~\cite{Schmid}.

\begin{proposition}\label{Prop:GaugInvYM} \emph{(Gauge Invariance of Linearised Yang-Mills Theory)}\newline
	 Let $\sf{A}\in\mathcal{C}^{1}(\sf{P})$ be solution to the non-linear Yang-Mills equations~\eqref{eq:YangMills}. Then, the linearised operator $\sf{P}:=\delta_{\sf{A}}\d_{\sf{A}}-\ast[\ast\sf{F}_{\sf{M}}^{\sf{A}}\wedge\cdot]_{\mathrm{Ad}(\sf{P})}$ is invariant under the transformations 
	\begin{align*}
			\Omega^{1}(\sf{M},\mathrm{Ad}(\sf{P}))\ni\alpha\mapsto\alpha+\d_{\sf{A}}\varepsilon\,,\qquad\forall\varepsilon\in\Gamma^{\infty}(\mathrm{Ad}(\sf{P}))\, .
	\end{align*}
\end{proposition}

\begin{proof}
	Instead of computing the action of the $\sf{P}$ on $\d_{\sf{A}}\varepsilon$ explicitly, we will argue in a more abstract way: denote by $\mathrm{exp}_{\mathcal{G}(\sf{P})}\:\Gamma^{\infty}(\mathrm{Ad}(\sf{P}))\to \mathcal{G}(\sf{P})$ the exponential map of the infinite-dimensional Lie group $\mathcal{G}(\sf{P})$. Any $\varepsilon\in\Gamma^{\infty}(\mathrm{Ad}(\sf{P}))$ can be identified with a function $\widetilde{\varepsilon}\in C^{\infty}(\sf{P},\mathfrak{g})$ satisfying $\widetilde{\varepsilon}(p\cdot g)=\mathrm{Ad}_{g^{-1}}(\varepsilon^{\prime}(p))$ and $\varepsilon(x)=[p,\widetilde{\varepsilon}(p)]$ for all $x\in\sf{M},\,p\in\sf{P}_{x}$, as a special case of the isomorphism~\ref{eq:IsoPrinc}. With this notation, the exponential map is given by $(\mathrm{exp}_{\mathcal{G}(\sf{P})}(\varepsilon))(p)=p\cdot\mathrm{exp}_{\sf{G}}(\widetilde{\varepsilon}(p))$, see~e.g.~the discussion in \cite[Sec.~3.5]{Schmid}.
	
	Now, we define a curve $\Phi_{t}$ in $\mathcal{G}(\sf{P})$ and set $\Phi_{t}:=\mathrm{exp}_{\mathcal{G}(\sf{P})}(t\varepsilon)$. By definition, it holds that $\Phi_{0}=\mathrm{id}_{\sf{P}}$ and $\widetilde{\varepsilon}=\frac{\d}{\d t}\big\vert_{t=0}\Phi_{t}$. Using Eq.~\eqref{eq:GaugTrafoCon} and $\sigma_{\Phi_{t}}(p)=\mathrm{exp}_{\sf{G}}(t\widetilde{\varepsilon}(p))$, we find
	\begin{align*}
		&\frac{\d}{\d t}\bigg\vert_{t=0}\Phi_{t}^{\ast}\sf{A}=\frac{\d}{\d t}\bigg\vert_{t=0}(\mathrm{Ad}_{\sigma_{\Phi_{t}}^{-1}}\circ\sf{A}+\sigma_{\Phi_{t}}^{\ast}\mu_{\sf{G}})=\frac{\d}{\d t}\bigg\vert_{t=0}(\mathrm{Ad}_{\mathrm{exp}_{\sf{G}}(-t\widetilde{\varepsilon})}\circ\sf{A}+\sigma_{\Phi_{t}}^{\ast}\mu_{\sf{G}})=\\&=-[\widetilde{\varepsilon}\wedge\sf{A}]_{\mathfrak{g}}+\frac{\d}{\d t}\bigg\vert_{t=0}(L_{\sigma_{\Phi_{t}}^{-1}})_{\ast}\d\sigma_{\Phi_{t}}=-[\widetilde{\varepsilon}\wedge\sf{A}]_{\mathfrak{g}}+\d\bigg(\frac{\d}{\d t}\bigg\vert_{t=0}\mathrm{exp}_{\sf{G}}(t\widetilde{\varepsilon})\bigg)=[\sf{A}\wedge\widetilde{\varepsilon}]_{\mathfrak{g}}+\d\widetilde{\varepsilon}=\sf{D}_{\sf{A}}\widetilde{\varepsilon}\, ,
	\end{align*}
	where $L_{g}\:\sf{G}\to\sf{G},h\mapsto hg$. Using the definition of the linearised operator $\sf{P}$, it follows that
	\begin{align*}
		\sf{P}\sf{d}_{\sf{A}}\varepsilon=\frac{\d}{\d t}\bigg\vert_{t=0}\delta_{\Phi_{t}^{\ast}\sf{A}}\sf{F}^{\Phi_{t}^{\ast}\sf{A}}_{\sf{M}}=\frac{\d}{\d t}\bigg\vert_{t=0}\Phi_{t}^{-1}\cdot\delta_{\sf{A}}\sf{F}^{\sf{A}}_{\sf{M}}=0\, ,
	\end{align*}
	where we used gauge-invariance (see Eq.~\eqref{eq:YMGaugeInv}) and $\delta_{\sf{A}}\sf{F}^{\sf{A}}_{\sf{M}}=0$ in the last step.
\end{proof}

\begin{remark}
	Gauge-invariance can of course also be shown directly, using that $\sf{D}_{\sf{A}}\sf{D}_{\sf{A}}\omega=[\sf{F}^{\sf{A}}\wedge\omega]_{\mathfrak{g}}$ for all $\omega\in\Omega^{k}_{\mathrm{hor},\mathrm{Ad}}(\sf{P},\mathfrak{g})$ and hence $\d_{\sf{A}}\d_{\sf{A}}\varepsilon=[\sf{F}_{\sf{M}}^{\sf{A}}\wedge\varepsilon]_{\mathrm{Ad}(\sf{P})}$ for all $\varepsilon\in\Gamma^{\infty}(\mathrm{Ad}(\sf{P}))$.
\end{remark}

We summarise all the previous findings in the following proposition.

\begin{proposition}\label{Prop:YMGauge} \emph{(Linearised Yang-Mills Theory as a Linear Gauge Theory)}\newline
Consider Set-up~\ref{SetUp:YangMills} and let $\sf{A}\in\mathcal{C}(\sf{P})$ be a solution to the non-linear Yang-Mills equations, i.e.~$\delta_{\sf{A}}\sf{F}^{\sf{A}}_{\sf{M}}=0$. Then, $(\mathrm{Ad}(\sf{P}),\mathrm{Ad}(\sf{P})_{1}=\mathrm{Ad}(\sf{P})\otimes\sf{T}^{\ast}\sf{M},\sf{P},\sf{K})$ with 
	\begin{align*}
		\sf{P}:=\delta_{\sf{A}}\d_{\sf{A}}-\ast[\ast\sf{F}_{\sf{M}}^{\sf{A}}\wedge\cdot]_{\mathrm{Ad}(\sf{P})}\in\mathrm{DO}^{2}(\mathrm{Ad}(\sf{P})_{1}), \qquad \sf{K}:=\d_{\sf{A}}\in\mathrm{DO}^{1}(\mathrm{Ad}(\sf{P}),\mathrm{Ad}(\sf{P})_{1})
	\end{align*}
	is a linear gauge theory in the sense of Definition~\ref{Def:LinGaugeTh}. The corresponding operators $\sf{D}_{1}:=\sf{K}^{\ast}\sf{K}$ and $\sf{D}_{2}:=\sf{P}+\sf{K}\sf{K}^{\ast}$ are the normally hyperbolic operators
	\begin{align*}
		\sf{D}_{1}=\square_{\mathrm{dRH}}^{\nabla^{\sf{A}}}=\delta_{\sf{A}}\d_{\sf{A}}\,,\qquad\sf{D}_{2}=\square_{\mathrm{dRH}}^{\nabla^{\sf{A}}}-\ast[\ast\sf{F}_{\sf{M}}^{\sf{A}}\wedge\cdot]_{\mathrm{Ad}(\sf{P})}=\delta_{\sf{A}}\d_{\sf{A}}+\d_{\sf{A}}\delta_{\sf{A}}-\ast[\ast\sf{F}_{\sf{M}}^{\sf{A}}\wedge\cdot]_{\mathrm{Ad}(\sf{P})}\, .
	\end{align*}
\end{proposition}

\begin{proof}
	In Proposition~\ref{Prop:GaugInvYM} we have shown that $\sf{P}\circ\sf{K}=0$. Now, clearly $\sf{K}^{\ast}=\delta_{\sf{A}}$, which implies that the operators $\sf{D}_{1}:=\sf{K}^{\ast}\sf{K}$ and $\sf{D}_{2}:=\sf{P}+\sf{K}\sf{K}^{\ast}$ are given by
	\begin{align*}
		\sf{D}_{1}=\delta_{\sf{A}}\d_{\sf{A}},\qquad\sf{D}_{2}=\delta_{\sf{A}}\d_{\sf{A}}+\d_{\sf{A}}\delta_{\sf{A}}-\ast[\ast\sf{F}_{\sf{M}}^{\sf{A}}\wedge\cdot]_{\mathrm{Ad}(\sf{P})}\, .
	\end{align*}
	Up to zeroth-order terms, these are the de Rham-Hodge d'Alembertians induced by $\nabla^{\sf{A}}$ and hence normally hyperbolic, see~Example~\ref{Example:HodgeLapl}. It remains to show that $\sf{P}$ is formally self-adjoint. By assumption, $\langle\cdot,\cdot\rangle_{\mathfrak{g}}$ is $\mathrm{Ad}$-invariant, and hence $\langle [X,Y]_{\mathfrak{g}},Z\rangle_{\mathfrak{g}}+\langle Y,[X,Z]_{\mathfrak{g}}\rangle_{\mathfrak{g}}=0$ for all $X,Y,Z\in\mathfrak{g}$. This, in turn translates on the level of wedge products to the fact that
	\begin{align*}
		\langle\alpha\wedge [\beta\wedge\gamma]_{\mathrm{Ad}(\sf{P})}\rangle_{\mathrm{Ad}(\sf{P})}=\langle[\alpha\wedge \beta]_{\mathrm{Ad}(\sf{P})}\wedge\gamma\rangle_{\mathrm{Ad}(\sf{P})}
	\end{align*}
	for all $\alpha,\beta,\gamma\in\Omega^{\bullet}(\sf{M},\mathrm{Ad}(\sf{P}))$. Now, writing $\mathcal{F}_{\sf{A}}:=\ast[\ast\sf{F}_{\sf{M}}^{\sf{A}}\wedge\cdot]_{\mathrm{Ad}(\sf{P})}\in\mathrm{Diff}^{0}(\mathrm{Ad}(\sf{P})_{1})$, we find
	\begin{align*}
		\langle\alpha\wedge\ast\mathcal{F}_{\sf{A}}(\beta)\rangle_{\mathrm{Ad}(\sf{P})}&=-(-1)^{d-1}\langle\alpha\wedge[\ast\sf{F}_{\sf{M}}^{\sf{A}}\wedge\beta]_{\mathrm{Ad}(\sf{P})}\rangle_{\mathrm{Ad}(\sf{P})}=(-1)^{d}\langle[\alpha\wedge \ast\sf{F}_{\sf{M}}^{\sf{A}}]_{\mathrm{Ad}(\sf{P})}\wedge\beta\rangle_{\mathrm{Ad}(\sf{P})}=\\&=-\langle[\ast\sf{F}_{\sf{M}}^{\sf{A}}\wedge\alpha]_{\mathrm{Ad}(\sf{P})}\wedge\beta\rangle_{\mathrm{Ad}(\sf{P})}=-(-1)^{d}\langle\ast\mathcal{F}_{\sf{A}}(\alpha)\wedge\beta\rangle_{\mathrm{Ad}(\sf{P})}=\\&=\langle\beta\wedge \ast\mathcal{F}_{\sf{A}}(\alpha)\rangle_{\mathrm{Ad}(\sf{P})}=\langle\mathcal{F}_{\sf{A}}(\alpha)\wedge\ast\beta\rangle_{\mathrm{Ad}(\sf{P})}
	\end{align*}
	for all $\alpha,\beta\in\Omega^{1}(\sf{M},\mathrm{Ad}(\sf{P}))$. Hence, $\mathcal{F}_{\sf{A}}$ is formally self-adjoint with respect to $(\cdot,\cdot)_{\mathrm{Ad}(\sf{P})_{1}}$.
\end{proof}

\subsection{Linearised Yang-Mills-Klein-Gordon System}
So far, we have only considered \textit{pure} gauge theories. As a first step towards more realistic models, we now study the coupled \textit{linearised Yang–Mills–Klein–Gordon system}. At the nonlinear level, this system is of direct physical interest, for example in the celebrated \textit{Higgs mechanism} within the standard model of particle physics, see e.g.~\cite[Chap.~8]{Hamilton}, \cite[Sec.~7.3]{RudolphSchmidt2} and \cite[Sec.10.3]{Bleecker} for a mathematical discussion. 

As in the previous section on Yang-Mills theory, we will consider the following general set-up.

\begin{setup}\label{SetUp:YangMillsKG}
We consider the following general setting:
\begin{itemize}
	\item[$\bullet$]A $d$-dimensional globally hyperbolic spacetime $(\sf{M},\sf{g})$.
    \item[$\bullet$]A (real, finite-dimensional) Lie group $\sf{G}$ with Lie algebra $(\mathfrak{g},[\cdot,\cdot]_{\mathfrak{g}})$ and a choice of $\mathrm{Ad}$-invariant and non-degenerate  symmetric bilinear form $\langle\cdot,\cdot\rangle_{\mathfrak{g}}$ on $\mathfrak{g}$.
    \item[$\bullet$]A principal $\sf{G}$-bundle $\sf{P}\xrightarrow{\pi}\sf{M}$.
    \item[$\bullet$]A Lie group representation $\rho\:\sf{G}\to\mathrm{Aut}(\sf{W})$ of $\sf{G}$ on some finite-dimensional $\bb{R}$-vector space $\sf{W}$ equipped with an $\rho$-invariant non-degenerate symmetric bilinear form $\langle\cdot,\cdot\rangle_{\sf{W}}$ and the corresponding associated vector bundle $\sf{E}:=\sf{P}\times_{\rho}\sf{W}$. 
    \item[$\bullet$]A \textit{potential} $\sf{V}\in C^{\infty}(\bb{R})$.
\end{itemize}
\end{setup}

The natural bundle metrics on $\mathrm{Ad}(\sf{P})$ and $\sf{E}$ induced by $\langle\cdot,\cdot\rangle_{\mathfrak{g}}$ and $\langle\cdot,\cdot\rangle_{\sf{W}}$, cf.~Eq.~\eqref{eq:IndBundMet}, are denoted by $\langle\cdot,\cdot\rangle_{\mathrm{Ad}(\sf{P})}$ and $\langle\cdot,\cdot\rangle_{\sf{E}}$, as before. Now, for a given \textit{potential} $\sf{V}$, we will use the suggestive notation $\sf{V}(\phi):=\sf{V}\circ\Vert\phi\Vert_{\sf{E}}^{2}\in C^{\infty}(\sf{M})$ for $\phi\in\Gamma^{\infty}(\sf{E})$. An important example relevant for the Higgs mechanism is the \textit{Mexican hat potential} 
\begin{align*}
	\sf{V}(\phi)=-\mu^{2}\Vert\phi\Vert_{\sf{E}}^{2}+\lambda\Vert\phi\Vert_{\sf{E}}^{4}
\end{align*}
for parameters $\mu^{2},\lambda\geq 0$, however, for our purpose it is enough to assume that $\sf{V}$ is smooth. With this notation, the \textit{Yang-Mills-Klein-Gordon equations} are the coupled system
\begin{align}\label{eq:YMKG}
	0=\mathrm{YMKG}_{\sf{V}}(\sf{A},\phi):=\begin{cases}
		\delta_{\sf{A}}\sf{F}^{\sf{A}}_{\sf{M}}-\sf{J}(\sf{A},\phi)\\
		\square_{\mathrm{dRH}}^{\nabla^{\sf{A}}}\phi+2\sf{V}^{\prime}(\phi)\phi
	\end{cases}\, ,
\end{align}
for $(\sf{A},\phi)\in\mathcal{C}(\sf{P})\times\Gamma^{\infty}(\sf{E})$, where $\sf{J}(\sf{A},\phi)\in\Omega^{1}(\sf{M},\mathrm{Ad}(\sf{P}))$ denotes the \textit{current} describing the interaction of $\phi$ with the gauge field $\sf{A}$. The operator $\square_{\mathrm{dRH}}^{\nabla^{\sf{A}}}:=\delta_{\sf{A}}\d_{\sf{A}}\:\Gamma^{\infty}(\sf{E})\to\Gamma^{\infty}(\sf{E})$ is the de Rham-Hodge d'Alembertian of the connection $\nabla^{\sf{A}}$ and hence in particular normally hyperbolic. Existence and uniqueness of solutions to~Eq.~\eqref{eq:YMKG} has for example been studied by Ginibre-Velo \cite{GinibreVelo1,GinibreVelo2}, Eardley-Moncrief \cite{EardleyM1,EardleyM2}, Choquet-Bruhat-Christodoulou \cite{Christodoulou} and Goganov-Kapitanskii \cite{Goganov} in Minkowski spacetime.

Now, before linearising the theory, we need an explicit expression of the current $\sf{J}(\sf{A},\phi)$. The derivation in a set-up that is similarly general than the one considered in this section is briefly sketched in \cite[Exercise 7.9.10]{Hamilton} and \cite[Sec.~7.2]{RudolphSchmidt2}. However, an explicit expression of $\sf{J}(\sf{A},\phi)$ in full generality is rarely found in the literature. Let us introduce the following notation: fix $v,w\in\sf{W}$. Then, by non-degeneracy of $\langle\cdot,\cdot\rangle_{\mathfrak{g}}$, there exists a unique $\mathfrak{J}(v,w)\in\mathfrak{g}$ such that
\begin{align}\label{Def:J}
	\langle X,\mathfrak{J}(v,w)\rangle_{\mathfrak{g}}=\langle v,\rho_{\ast}(X)w\rangle_{\sf{W}}
\end{align}
for all $X\in\mathfrak{g}$. This defines a map $\mathfrak{J}\:\sf{W}\times\sf{W}\to\mathfrak{g}$ with the following properties.

\begin{lemma}\label{Lemma:J}
The map $\mathfrak{J}\:\sf{W}\times\sf{W}\to\mathfrak{g}$ defined by~\eqref{Def:J} has the following properties.
\begin{itemize} 
	\item[\emph{(i)}]$\mathfrak{J}$ is antisymmetric and bilinear.
	\item[\emph{(ii)}]$\mathfrak{J}(\rho(g)v,\rho(g)w)=\mathrm{Ad}_{g}(\mathfrak{J}(v,w))$ for all $v,w\in\sf{W}$ and $g\in\sf{G}$, i.e.~$\mathfrak{J}$ is $\sf{G}$-equivariant.
	\item[\emph{(iii)}]$\mathfrak{J}(\rho_{\ast}(X)v,w)+\mathfrak{J}(v,\rho_{\ast}(X)w)=[X,\mathfrak{J}(v,w)]_{\mathfrak{g}}$ for all $v,w\in\sf{W}$ and $X\in\mathfrak{g}$.
	\item[\emph{(iv)}]$\mathfrak{J}$ is explicitly given by
\begin{align*}
		\mathfrak{J}\:\sf{W}\times\sf{W}\xrightarrow{(v,w)\mapsto (X\mapsto \langle v,\rho_{\ast}(X)w\rangle_{\sf{W}})}\mathfrak{g}^{\ast}\xrightarrow{\cong}\mathfrak{g}\, ,
	\end{align*}
	where the second map is the canonical isomorphism $\mathfrak{g}\cong\mathfrak{g}^{\ast}$ induced by $\langle\cdot,\cdot\rangle_{\mathfrak{g}}$.
\end{itemize}
\end{lemma}

\begin{proof}
	The map $\mathfrak{J}$ is clearly bilinear. Antisymmetry follows directly from $\rho$-invariance and symmetry of $\langle\cdot,\cdot\rangle_{\sf{W}}$, which implies $\langle v,\rho_{\ast}(X)w\rangle_{\sf{W}}=-\langle \rho_{\ast}(X)v,w\rangle_{\sf{W}}=-\langle w,\rho_{\ast}(X)v\rangle_{\sf{W}}$. For (ii), we first note that $\rho(g^{-1})\circ\rho_{\ast}(X)\circ\rho(g)=\rho_{\ast}(\mathrm{Ad}_{g^{-1}}(X))$ for all $g\in\sf{G}$ and $X\in\mathfrak{g}$, which follows directly from the definition of $\mathrm{Ad}_{g}$ and $\rho_{\ast}$, see e.g.~\cite[Prop.~2.1.46]{Hamilton}. It follows that
	\begin{align*}
		\langle X,\mathfrak{J}(\rho(g)v,\rho(g)w)\rangle_{\mathfrak{g}}&=\langle\rho(g)v,\rho_{\ast}(X)\rho(g)w\rangle_{\sf{W}}=\langle v,\rho(g^{-1})\rho_{\ast}(X)\rho(g)w\rangle_{\sf{W}}=\\&=\langle v,\rho_{\ast}(\mathrm{Ad}_{g^{-1}}(X))w\rangle_{\sf{W}}=\langle \mathrm{Ad}_{g^{-1}}(X),\mathfrak{J}(v,w)\rangle_{\mathfrak{g}}=\langle X,\mathrm{Ad}_{g}(\mathfrak{J}(v,w))\rangle_{\mathfrak{g}}\,,
	\end{align*}
	which shows that $\mathfrak{J}(\rho(g)v,\rho(g)w)=\mathrm{Ad}_{g}(\mathfrak{J}(v,w))$, by non-degeneracy of $\langle\cdot,\cdot\rangle_{\mathfrak{g}}$. Claim (iii) follows from (ii) and (iv) follows directly from the definition.
\end{proof}

Now, Lemma~\ref{Lemma:J}(ii) implies that $\mathfrak{J}$ gives rise to a well-defied bilinear fibre-wise map on the level of associated bundles, i.e.~$\sf{E}_{x}\times\sf{E}_{x}\to\mathrm{Ad}(\sf{P})_{x}\,, ([p,v],[p,w])\mapsto [p,\mathfrak{J}(v,w)]$. Let us denote the corresponding wedge-type product induced by this pairing by 
\begin{align*}
	\wedge_{\mathfrak{J}}\:\Omega^{k}(\sf{M},\sf{E})\times\Omega^{l}(\sf{M},\sf{E})\to\Omega^{k+l}(\sf{M},\mathrm{Ad}(\sf{P}))\,,
\end{align*}
see Remark~\ref{Rem:Wedge} for details. With this notation, the current $\sf{J}(\sf{A},\phi)$ can be written as follows.

\begin{lemma} 
	The current $\sf{J}(\sf{A},\phi)\in\Omega^{1}(\sf{M},\mathrm{Ad}(\sf{P}))$ is explicitly given by
	\begin{align*}
		\sf{J}(\sf{A},\phi)=\phi\wedge_{\mathfrak{J}}\d_{\sf{A}}\phi\in\Omega^{1}(\sf{M},\mathrm{Ad}(\sf{P}))\,.
	\end{align*}
\end{lemma}

\begin{proof}
Let us define the \textit{Lagrangian density} of the Yang-Mills-Klein-Gordon system by
\begin{align*}
	\mathcal{L}\:\mathcal{C}(\sf{P})\times\Gamma^{\infty}(\sf{E})\to\bb{R}\,,\qquad \mathcal{L}(\sf{A},\phi):=-\frac{1}{2}\langle\sf{F}_{\sf{M}}^{\sf{A}},\sf{F}^{\sf{A}}_{\sf{M}}\rangle_{\mathrm{Ad}(\sf{P})_{2}}-\frac{1}{2}\langle\d_{\sf{A}}\phi,\d_{\sf{A}}\phi\rangle_{\sf{E}_{1}}-\sf{V}(\phi)\, ,
\end{align*}
where the bundle metrics $\langle\cdot,\cdot\rangle_{\mathrm{Ad}(\sf{P})_{2}}$ and $\langle\cdot,\cdot\rangle_{\sf{E}_{1}}$ are as in Eq.~\eqref{eq:ExtBundlesTwist}. By definition, the solutions of the Yang-Mills-Klein-Gordon equations~\eqref{eq:YMKG} are the critical points of the corresponding \textit{action} $\sf{S}[\sf{A},\phi]$. Now, since $\sf{M}$ is non-compact, the following derivation is just a formal one, which can be made precise by working in suitable Sobolev spaces, see e.g.~\cite[Chap.~6,7]{RudolphSchmidt2}, or by inserting a suitable cut-off function as in the formalism discussed in \cite[Chap.~4]{RejznerBook}. By definition, $\delta_{\sf{A}}\sf{F}_{\sf{M}}^{\sf{A}}=\sf{J}(\sf{A},\phi)$ for $(\sf{A},\phi)\in\mathcal{C}(\sf{P})\times\Gamma^{\infty}(\sf{E})$ if and only if 
\begin{align}\label{eq:Variation}
	\frac{\d}{\d t}\bigg\vert_{t=0}\sf{S}[\sf{A}+t\alpha,\phi]:=&\frac{\d}{\d t}\bigg\vert_{t=0}\int_{\sf{M}}\mathcal{L}[\sf{A}+t\alpha,\phi]\,\d\mu_{\sf{g}}=0
\end{align}
for all $\alpha\in\Omega^{1}_{\mathrm{hor},\mathrm{Ad}}(\sf{P},\mathfrak{g})$. Now, as in the proof of Proposition~\ref{prop:LinYangMi}, $\sf{F}_{\sf{M}}^{\sf{A}+t\alpha}=\sf{F}_{\sf{M}}^{\sf{A}}+t(\d_{\sf{A}}\alpha_{\sf{M}})+\mathcal{O}(t^{2})$ and $\d_{\sf{A}+t\alpha}\phi=\d_{\sf{A}}\phi+t(\alpha_{\sf{M}}\wedge_{\sf{E}}\phi)$, where $\wedge_{\sf{E}}$ has been defined in~Eq.~\eqref{eq:StrangeWedge}. Hence,
\begin{align*}
	\frac{\d}{\d t}\bigg\vert_{t=0}\sf{S}[\sf{A}+t\alpha,\phi]=\int_{\sf{M}}\bigg(-\langle\alpha_{\sf{M}},\delta_{\sf{A}}\sf{F}_{\sf{M}}^{\sf{A}}\rangle_{\mathrm{Ad}(\sf{P})_{1}}-\langle\d_{\sf{A}}\phi,\alpha_{\sf{M}}\wedge_{\sf{E}}\phi\rangle_{\sf{E}_{1}}\bigg)\,\d\mu_{\sf{g}}\, .
\end{align*}
Now, let us define $\sf{J}(\sf{A},\phi)\in\Omega^{1}(\sf{M},\mathrm{Ad}(\sf{P}))$ to be the unique $\mathrm{Ad}(\sf{P})$-valued $1$-form such that
\begin{align}\label{eq:Darwin}
	\langle\alpha_{\sf{M}},\sf{J}(\sf{A},\phi)\rangle_{\mathrm{Ad}(\sf{P})_{1}}=\langle-\d_{\sf{A}}\phi,\alpha_{\sf{M}}\wedge_{\sf{E}}\phi\rangle_{\sf{E}_{1}}
\end{align} 
for all $\alpha_{\sf{M}}\in\Omega^{1}(\sf{M},\mathrm{Ad}(\sf{P}))$. Then, \eqref{eq:Variation} yields the required equations $\delta_{\sf{A}}\sf{F}^{\sf{A}}_{\sf{M}}=\sf{J}(\sf{A},\phi)$ by non-degeneracy of $(\cdot,\cdot)_{\mathrm{Ad}(\sf{P})_{1}}$. Let us choose a basis of $(T_{a})_{a}$ of $\mathfrak{g}$ and a basis $(v_{i})_{i}$ of $\sf{W}$. Then, if we choose a \textit{local gauge}, i.e.~a local section $s\in\Gamma^{\infty}(\mathcal{U},\sf{P})$ for some open set $\mathcal{U}\subset\sf{M}$, we obtain a local frame $t_{a}(\cdot):=[s(\cdot),T_{a}]$ of $\mathrm{Ad}(\sf{P})$ and $e_{i}(\cdot):=[s(\cdot),v_{i}]$ of $\sf{E}$. As usual, we write a differential form $\omega\in\Omega^{k}(\sf{M},\sf{E})$ locally as $\omega=\omega^{a}\otimes e_{a}$ for coefficient forms $\omega^{a}\in\Omega^{k}(\mathcal{U})$ and similarly for $\mathrm{Ad}(\sf{P})$-valued forms. Now, using the definitions of all the relevant bundle metrics and wedge products, the right-hand side of~\eqref{eq:Darwin} is locally given by
\begin{align*}
	\langle\d_{\sf{A}}\phi,\alpha_{\sf{M}}\wedge_{\sf{E}}\phi\rangle_{\sf{E}_{1}}&=\langle(\d_{\sf{A}}\phi)^{i},\alpha_{\sf{M}}^{a}\wedge\phi^{j}\rangle_{\sf{A}_{1}}\langle v_{i},\rho_{\ast}(T_{a})v_{j}\rangle_{\sf{W}}=\langle(\d_{\sf{A}}\phi)^{i},\alpha_{\sf{M}}^{a}\wedge\phi^{j}\rangle_{\sf{A}_{1}}\langle T_{a},\mathfrak{J}(v_{i},v_{j})\rangle_{\mathfrak{g}}=\\&=\langle\alpha_{\sf{M}}^{a},(\d_{\sf{A}}\phi)^{i}\wedge\phi^{j}\rangle_{\sf{A}_{1}}\langle T_{a},\mathfrak{J}(v_{i},v_{j})\rangle_{\mathfrak{g}}=\langle\alpha_{\sf{M}},\d_{\sf{A}}\phi\wedge_{\mathfrak{J}}\phi\rangle_{\mathrm{Ad}(\sf{P})_{1}}\, .
\end{align*}
By non-degeneracy of $(\cdot,\cdot)_{\mathrm{Ad}(\sf{P})}$, we conclude that $\sf{J}(\sf{A},\phi)=-(\d_{\sf{A}}\phi\wedge_{\mathfrak{J}}\phi)=\phi\wedge_{\mathfrak{J}}\d_{\sf{A}}\phi$.
\end{proof}

\begin{remark}
	The equation $\delta_{\sf{A}}\d_{\sf{A}}\phi=2\sf{V}^{\prime}(\phi)\phi$ is obtained by variation of the action $\sf{S}[\sf{A},\phi]$ in $\phi$, i.e.~$\delta_{\sf{A}}\d_{\sf{A}}\phi=2\sf{V}^{\prime}(\phi)\phi$ if and only if $\frac{\d}{\d t}\big\vert_{t=0}\sf{S}[\sf{A},\phi+t\varphi]=0$ for all $\varphi\in\Gamma^{\infty}(\sf{E})$.
\end{remark}

We now turn to the linearisation of the system. Following the notation employed in the previous section, we introduce the Hermitian vector bundle
\begin{align*}
	\mathcal{E}_{1}:=\mathrm{Ad}(\sf{P})_{1}\oplus\sf{E}\,,\qquad\langle\cdot,\cdot\rangle_{\mathcal{E}_{1}}=\langle\cdot,\cdot\rangle_{\mathrm{Ad}(\sf{P})_{1}}\oplus\langle\cdot,\cdot\rangle_{\sf{E}}\,,
\end{align*}
whose space of sections is given by $\Gamma^{\infty}(\mathcal{E}_{1})\cong\Omega^{1}(\sf{M},\mathrm{Ad}(\sf{P}))\oplus\Gamma^{\infty}(\sf{E})$.

\begin{proposition} \emph{(Linearised Yang-Mills-Klein-Gordon Operator)}\newline
	Consider background fields $(\sf{A},\phi)\in\mathcal{C}(\sf{P})\times\Gamma^{\infty}(\sf{E})$. Then, the \emph{linearised Yang-Mills-Klein-Gordon operator} $\mathscr{P}\in\mathrm{DO}(\mathcal{E}_{1})$ defined by
\begin{align*}
	\mathrm{YMKG}_{\sf{V}}(\sf{A}+\lambda\alpha,\phi+\lambda\varphi)=\mathrm{YMKG}_{\sf{V}}(\sf{A},\phi)+\lambda\mathscr{P}(\alpha_{\sf{M}},\varphi)+\mathcal{O}(\lambda^{2})
\end{align*}
for all perturbations $(\alpha,\varphi)\in\Omega^{1}_{\mathrm{hor},\mathrm{Ad}}(\sf{P},\mathfrak{g})\times\Gamma^{\infty}(\sf{E})$ and a parameter $\lambda\in\bb{R}$, is given by
	\begin{align*}
		\mathscr{P}=
		\begin{pmatrix}
			\mathscr{P}_{11} &\mathscr{P}_{12}\\
			\mathscr{P}_{21} &\mathscr{P}_{22}
		\end{pmatrix},\qquad
		\begin{cases}
			\mathscr{P}_{11}:=\delta_{\sf{A}}\d_{\sf{A}}-\ast[\ast\sf{F}_{\sf{M}}^{\sf{A}}\wedge\cdot]_{\mathrm{Ad}(\sf{P})}-(\cdot\wedge_{\sf{E}}\phi)\wedge_{\mathfrak{J}}\phi \\
			\mathscr{P}_{12}:=(\d_{\sf{A}}\phi\wedge_{\mathfrak{J}}\cdot)-(\phi\wedge_{\mathfrak{J}}\d_{\sf{A}}\cdot) \\
			\mathscr{P}_{21}:=\delta_{\sf{A}}(\cdot\wedge_{\sf{E}}\phi)+\ast(\cdot\wedge_{\sf{E}}\ast\d_{\sf{A}}\phi) \\
			\mathscr{P}_{22}:=\delta_{\sf{A}}\d_{\sf{A}}+2\sf{V}^{\prime}(\phi)+4\sf{V}^{\prime\prime}(\phi)\langle\phi,\cdot\rangle_{\sf{E}}\phi
		\end{cases}
	\end{align*}
\end{proposition}

\begin{proof}
	Let us start with the equation $\delta_{\sf{A}}\d_{\sf{A}}\phi+2\sf{V}^{\prime}(\phi)\phi=0$. Similarly as in the proof of Proposition~\ref{prop:LinYangMi}, we obtain the expansion
	\begin{align*}
		\d_{\sf{A}+\lambda\alpha}\omega=\d_{\sf{A}}\omega+\lambda(\alpha_{\sf{M}}\wedge_{\sf{E}}\omega)
	\end{align*}
	In particular, using that $\delta_{\sf{A}}=\ast\d_{\sf{A}}\ast$ when acting on $1$-forms, we obtain the expansion:
	\begin{align*}
		\delta_{\sf{A}+\lambda\alpha}\d_{\sf{A}+\lambda\alpha}(\phi+\lambda\varphi)&=\ast\d_{\sf{A}+\lambda\alpha}\ast\d_{\sf{A}+\lambda\alpha}(\phi+\lambda\varphi)=\\&\delta_{\sf{A}}\d_{\sf{A}}\phi+\lambda\bigg\{\delta_{\sf{A}}\d_{\sf{A}}\varphi+\delta_{\sf{A}}(\alpha_{\sf{M}}\wedge_{\sf{E}}\phi)+\ast(\alpha_{\sf{M}}\wedge_{\sf{E}}\ast\d_{\sf{A}}\phi)\bigg\}+\mathcal{O}(\lambda^{2})\, .
	\end{align*}
	Furthermore, the linearisation of $2\sf{V}^{\prime}(\phi)\phi$ is given by $4\sf{V}^{\prime\prime}(\phi)\langle\phi,\varphi\rangle_{\sf{E}}\phi+2\sf{V}^{\prime}(\phi)\varphi$. This gives the components $\mathscr{P}_{21}$ and $\mathscr{P}_{22}$ of the linearised operators $\mathscr{P}$. For the other components, we compute
	\begin{align*}
		\delta_{\sf{A}+\lambda\alpha}&\sf{F}^{\sf{A}+\lambda\alpha}_{\sf{M}}-\sf{J}(\sf{A}+\lambda\alpha,\phi+\lambda\varphi)=\delta_{\sf{A}}\sf{F}^{\sf{A}}_{\sf{M}}-\sf{J}(\sf{A},\phi)+\\&+\lambda\bigg\{\delta_{\sf{A}}\d_{\sf{A}}\alpha_{\sf{M}}-\ast[\ast\sf{F}_{\sf{M}}^{\sf{A}}\wedge\alpha_{\sf{M}}]_{\mathrm{Ad}(\sf{P})}-\varphi\wedge_{\mathfrak{J}}\d_{\sf{A}}\phi-\phi\wedge_{\mathfrak{J}}\d_{\sf{A}}\varphi-(\alpha_{\sf{M}}\wedge_{\sf{E}}\phi)\wedge_{\mathfrak{J}}\phi\bigg\}+\mathcal{O}(\lambda^{2})
	\end{align*}
	where we used Proposition~\ref{Prop:GaugInvYM}. Using that $\varphi\wedge_{\mathfrak{J}}\d_{\sf{A}}\phi=-\d_{\sf{A}}\phi\wedge_{\mathfrak{J}}\varphi$, we obtain $\mathscr{P}_{11}$ and $\mathscr{P}_{12}$. 
\end{proof}

Let us now turn to gauge invariance. Using the notation and terminology introduced in the previous section, the nonlinear Yang–Mills–Klein–Gordon equations~\eqref{eq:YMKG} are invariant under gauge transformations in the following sense:
\begin{align*}
    \forall f \in \mathcal{G}(\mathsf{P}): \quad 
    \mathrm{YMKG}_{\mathsf{V}}(f^{\ast}A, f^{-1} \cdot \phi) 
    = f^{-1} \cdot \mathrm{YMKG}_{\mathsf{V}}(A, \phi)\,,
\end{align*}
where the action on the right-hand side is understood componentwise on 
$\Gamma^\infty(\mathrm{Ad}(\mathsf{P})) \times \Gamma^\infty(\mathsf{E})$.

\begin{proposition}\label{Prop:GaugInvYMKG} \emph{(Gauge Invariance of Linearised Yang-Mills-Klein-Gordon Theory)}\newline
	 Let $(\sf{A},\phi)\in\mathcal{C}(\sf{P})\times\Gamma^{\infty}(\sf{E})$ be a solution to the Yang-Mills-Klein-Gordon equations~\eqref{eq:YMKG}. Then, the linearised operator $\mathscr{P}\in\mathrm{DO}^{2}(\mathcal{E}_{2})$ is invariant under the transformations 
	\begin{align*}
			\Omega^{1}(\sf{M},\mathrm{Ad}(\sf{P}))\times\Gamma^{\infty}(\sf{E})\ni \begin{pmatrix}\alpha\\\varphi\end{pmatrix}\mapsto  \begin{pmatrix}\alpha\\\varphi\end{pmatrix}+ \begin{pmatrix}\d_{\sf{A}}\varepsilon\\-\varepsilon\wedge_{\sf{E}}\phi\end{pmatrix}\qquad\forall\varepsilon\in\Omega^{0}(\sf{M},\mathrm{Ad}(\sf{P}))\, .
	\end{align*}
\end{proposition}

\begin{proof}
	Consider the exponential map $\mathrm{exp}_{\mathcal{G}(\sf{P})}\:\Gamma^{\infty}(\mathrm{Ad}(\sf{P}))\to \mathcal{G}(\sf{P})$ and define $\Phi_{t}:=\mathrm{exp}_{\mathcal{G}(\sf{P})}(t\varepsilon)$ for some $\varepsilon\in\Gamma^{\infty}(\mathrm{Ad}(\sf{P}))$. Now, as a special case of the isomorphism~\eqref{eq:IsoGaugeObs}, we can identify $\varepsilon$ with a function $\widetilde{\varepsilon}\in C^{\infty}(\sf{P},\mathfrak{g})$ such that $\widetilde{\varepsilon}(p\cdot g)=\mathrm{Ad}_{g^{-1}}(\widetilde{\varepsilon}(p))$ defined by $\varepsilon(x)=[p,\widetilde{\varepsilon}(p)]$ for all $x\in\sf{M}$ and $p\in\sf{P}_{x}$. Similarly, every $\phi\in\Gamma^{\infty}(\sf{E})$ can be identified with a function $\widetilde{\phi}\in C^{\infty}(\sf{P},\sf{W})$ satisfying $\widetilde{\phi}(p\cdot g)=\rho(g^{-1})\widetilde{\phi}(p)$. Furthermore, as explained in the proof of Proposition~\ref{Prop:GaugInvYM}, $\Phi_{t}(p)=p\cdot\sigma_{\Phi_{t}}(p)$ with $\sigma_{\Phi_{t}}(p)=\mathrm{exp}_{\sf{G}}(t\widetilde{\varepsilon}(p))$. Then,
	\begin{align*}
		(\Phi_{t}^{-1}\cdot\phi)(x)&=[p,(\Phi_{t}^{\ast}\widetilde{\phi})(p)]=[p,\widetilde{\phi}(\Phi_{t}(p))]=[p,\widetilde{\phi}(p\cdot\sigma_{\Phi_{t}}(p))]=\\&=[p,\rho(\sigma_{\Phi_{t}}(p)^{-1})\widetilde{\phi}(p)]=[p,\rho(\mathrm{exp}_{\sf{G}}(-t\widetilde{\varepsilon}(p)))\widetilde{\phi}(p)]
	\end{align*}		
	for all $x\in\sf{M}$ and $p\in\sf{P}_{x}$. Using that $\rho_{\ast}(X)=\frac{\d}{\d t}\big\vert_{t=0}\rho(\mathrm{exp}_{\sf{G}}(tX))$ for $X\in\mathfrak{g}$, we conclude that 	
	\begin{align*}
	\frac{\d}{\d t}\bigg\vert_{t=0}\Phi_{t}^{-1}\cdot\phi=-\varepsilon\wedge_{\sf{E}}\phi\,,
\end{align*}
which shows how the gauge transformation act on the scalar field in the linearised setting. Now, by definition of the linearised equations, we have that
\begin{align*}
	\mathscr{P}(\d_{\sf{A}}\varepsilon,-\varepsilon\wedge_{\sf{E}}\phi)=\frac{\d}{\d t}\bigg\vert_{t=0}\mathrm{YMKG}_{\sf{V}}(\Phi_{t}^{\ast}\sf{A},\Phi_{t}^{-1}\cdot\phi)=\frac{\d}{\d t}\bigg\vert_{t=0}\Phi_{t}^{-1}\cdot\mathrm{YMKG}_{\sf{V}}(\sf{A},\phi)=0\,,
\end{align*}
where we used gauge-invariance and the assumption $\mathrm{YMKG}_{\sf{V}}(\sf{A},\phi)=0$.
\end{proof}

\begin{proposition}\label{Prop:YMKGGauge} \emph{(Linearised Yang-Mills-Klein-Gordon Theory)}\newline
Consider Set-up~\ref{SetUp:YangMillsKG} and let $(\sf{A},\phi)\in\mathcal{C}(\sf{P})\times\Gamma^{\infty}(\sf{E})$ be a solution to the non-linear Yang-Mills-Klein-Gordon equations~\eqref{eq:YMKG}. Then, $(\mathcal{E}_{0}:=\mathrm{Ad}(\sf{P}),\mathcal{E}_{1}:=\mathrm{Ad}(\sf{P})_{1}\oplus\sf{E},\mathscr{P},\mathscr{K})$ with
	\begin{align*}
		\mathscr{P}&:=\begin{pmatrix}
		\delta_{\sf{A}}\d_{\sf{A}}-\ast[\ast\sf{F}_{\sf{M}}^{\sf{A}}\wedge\cdot]_{\mathrm{Ad}(\sf{P})}-(\cdot\wedge_{\sf{E}}\phi)\wedge_{\mathfrak{J}}\phi & (\d_{\sf{A}}\phi\wedge_{\mathfrak{J}}\cdot)-(\phi\wedge_{\mathfrak{J}}\d_{\sf{A}}\cdot)\\ \delta_{\sf{A}}(\cdot\wedge_{\sf{E}}\phi)+\ast(\cdot\wedge_{\sf{E}}\ast\d_{\sf{A}}\phi)  &\delta_{\sf{A}}\d_{\sf{A}}+\mathcal{V}_{\phi}(\cdot)
	\end{pmatrix}\in\mathrm{DO}^{2}(\mathcal{E}_{1})\\
	\mathscr{K}&:=\begin{pmatrix} \d_{\sf{A}}\cdot\\-(\cdot\wedge_{\sf{E}}\phi)\end{pmatrix}\in\mathrm{DO}^{1}(\mathcal{E}_{0},\mathcal{E}_{1})\,,
	\end{align*}
	where $\mathcal{V}_{\phi}(\varphi):=2\sf{V}^{\prime}(\phi)\varphi+4\sf{V}^{\prime\prime}(\phi)\langle\phi,\varphi\rangle_{\sf{E}}\phi$, is a linear gauge theory in the sense of Definition~\ref{Def:LinGaugeTh} with normally hyperbolic operators $\mathscr{D}_{1}:=\mathscr{K}^{\dagger}\mathscr{K}$ and $\mathscr{D}_{2}:=\mathscr{P}+\mathscr{K}\mathscr{K}^{\dagger}$,
	\begin{align*}
		\mathscr{D}_{1}&=\square^{\nabla^{\sf{A}}}_{\mathrm{dRH}}+(\cdot\wedge_{\sf{E}}\phi)\wedge_{\mathfrak{J}}\phi\\
		\mathscr{D}_{2}&=\begin{pmatrix}
		\square^{\nabla^{\sf{A}}}_{\mathrm{dRH}}-\ast[\ast\sf{F}_{\sf{M}}^{\sf{A}}\wedge\cdot]_{\mathrm{Ad}(\sf{P})}-(\cdot\wedge_{\sf{E}}\phi)\wedge_{\mathfrak{J}}\phi & 2(\d_{\sf{A}}\phi\wedge_{\mathfrak{J}}\cdot)\\ 2\ast(\cdot\wedge_{\sf{E}}\ast\d_{\sf{A}}\phi)  &\square_{\mathrm{dRH}}^{\nabla^{\sf{A}}}+\mathcal{V}_{\phi}(\cdot)-(\phi\wedge_{\mathfrak{J}}\cdot)\wedge_{\sf{E}}\phi
	\end{pmatrix}
	\end{align*}
\end{proposition}

\begin{proof}
First of all, we show that $\mathscr{P}$ is formally self-adjoint with respect to the natural bundle metrics $\langle\cdot,\cdot\rangle_{\mathcal{E}_{0}}:=\langle\cdot,\cdot\rangle_{\mathrm{Ad}(\sf{P})}$ and $\langle\cdot,\cdot\rangle_{\mathcal{E}_{1}}=\langle\cdot,\cdot\rangle_{\mathrm{Ad}(\sf{P})_{1}}\oplus\langle\cdot,\cdot\rangle_{\sf{E}}$. In the proof of Proposition~\ref{Prop:GaugInvYM}, we have already shown that $\delta_{\sf{A}}\d_{\sf{A}}-\ast[\ast\sf{F}_{\sf{M}}^{\sf{A}}\wedge\cdot]_{\mathrm{Ad}(\sf{P})}$ is formally self-adjoint. The only remaining term in $\mathscr{P}_{11}$ is the zeroth-order operator $\mathscr{Z}:=(\cdot\wedge_{\sf{E}}\phi)\wedge_{\mathfrak{J}}\phi$. Choosing bases $(v_{i})_{i}$ of $\sf{W}$ and $(T_{a})_{a}$ of $\mathfrak{g}$, as well as a local gauge $s\in\Gamma^{\infty}(\mathcal{U},\sf{P})$ on $\mathcal{U}\subset\sf{M}$ open, we obtain local frames $t_{a}(\cdot):=[s(\cdot),T_{a}]$ of $\mathrm{Ad}(\sf{P})$ and $e_{i}(\cdot):=[s(\cdot),v_{i}]$ of $\sf{E}$. Then, for $\alpha,\beta\in\Omega^{1}(\sf{M},\mathrm{Ad}(\sf{P}))$, locally written as $\alpha=\alpha^{a}\otimes t_{a}$ and $\beta=\beta^{b}\otimes t_{b}$ for $\alpha^{a},\beta^{b}\in\Omega^{1}(\mathcal{U})$, and $\phi=\phi^{i}\otimes e_{i}$ with $\phi^{i}\in C^{\infty}(\mathcal{U})$, we obtain
	\begin{align*}
		\langle\mathscr{Z}\alpha,\beta\rangle_{\mathrm{Ad}(\sf{P})_{1}}&=\langle (\alpha\wedge_{\sf{E}}\phi)\wedge_{\mathfrak{J}}\phi,\beta\rangle_{\mathrm{Ad}(\sf{P})_{1}}=\langle \alpha^{a}\phi^{i}\phi^{j},\beta^{b}\rangle_{\sf{A}_{1}}\langle \mathfrak{J}(\rho_{\ast}(T_{a})v_{i},v_{j}),T_{b}\rangle_{\mathfrak{g}}=\\&=\langle \alpha^{a},\beta^{b}\phi^{i}\phi^{j}\rangle_{\sf{A}_{1}}\langle \rho_{\ast}(T_{a})v_{i},\rho_{\ast}(T_{b})v_{j}\rangle_{\sf{W}}=\langle \alpha^{a},\beta^{b}\phi^{i}\phi^{j}\rangle_{\sf{A}_{1}}\langle T_{a},\mathfrak{J}(\rho_{\ast}(T_{b})v_{j},v_{i})\rangle_{\sf{W}}=\\&=\langle \alpha,(\beta\wedge_{\sf{E}}\phi)\wedge_{\mathfrak{J}}\phi\rangle_{\mathrm{Ad}(\sf{P})_{1}}=\langle\alpha,\mathscr{Z}\beta\rangle_{\mathrm{Ad}(\sf{P})_{1}}\,.
	\end{align*}
	We conclude that $\mathscr{P}_{11}^{\ast}=\mathscr{P}_{11}$ with respect to $\langle\cdot,\cdot\rangle_{\mathrm{Ad}(\sf{P})_{1}}$. Furthermore, by definition, it is clear that $\mathscr{P}_{22}^{\ast}=\mathscr{P}_{22}$ with respect to $\langle\cdot,\cdot\rangle_{\sf{E}}$. It remains to show $\mathscr{P}_{12}^{\ast}=\mathscr{P}_{21}$. Let $\varphi=\varphi^{i}\otimes e_{i}\in\Gamma^{\infty}(\sf{E})$ and $\alpha=\alpha^{a}\otimes t_{a}\in\Omega^{1}(\sf{M},\mathrm{Ad}(\sf{P}))$. Then,
	\begin{align*}
		\langle \d_{\sf{A}}\varphi\wedge_{\mathfrak{J}}\phi,\alpha\rangle_{\mathrm{Ad}(\sf{P})_{1}}&=\langle (\d_{\sf{A}}\varphi)^{i}\phi^{j},\alpha^{a}\rangle_{\sf{A}_{1}}\langle \mathfrak{J}(v_{i},v_{j}),T_{a}\rangle_{\mathfrak{g}}=\\&=\langle (\d_{\sf{A}}\varphi)^{i},\alpha^{a}\phi^{j}\rangle_{\sf{A}_{1}}\langle v_{i},\rho_{\ast}(T_{a})v_{j}\rangle_{\sf{W}}=\langle\d_{\sf{A}}\varphi,\alpha\wedge_{\sf{E}}\phi\rangle_{\sf{E}_{1}}
	\end{align*}
	and hence $-(\phi\wedge_{\mathfrak{J}}\d_{\sf{A}}\varphi,\alpha)_{\mathrm{Ad}(\sf{P})_{1}}=(\d_{\sf{A}}\varphi\wedge_{\mathfrak{J}}\phi,\alpha)_{\mathrm{Ad}(\sf{P})_{1}}=(\varphi,\delta_{\sf{A}}(\alpha\wedge_{\sf{E}}\phi))_{\sf{E}}$, which shows that $(-\phi\wedge_{\mathfrak{J}}\d_{\sf{A}}\cdot)^{\ast}=\delta_{\sf{A}}(\cdot\wedge_{\sf{E}}\phi)$. Furthermore, 
	\begin{align*}
		\langle \d_{\sf{A}}\phi\wedge_{\mathfrak{J}}\varphi,&\alpha\rangle_{\mathrm{Ad}(\sf{P})_{1}}\,\d\mu_{\sf{g}}=\langle(\d_{\sf{A}}\phi)^{i}\varphi^{j},\alpha^{a}\rangle_{\sf{A}_{1}}\langle\mathfrak{J}(v_{i},v_{j}),T_{a}\rangle_{\mathfrak{g}}\,\d\mu_{\sf{g}}=\\&=\langle(\d_{\sf{A}}\phi)^{i}\varphi^{j},\alpha^{a}\rangle_{\sf{A}_{1}}\langle v_{i},\rho_{\ast}(T_{a})v_{j}\rangle_{\sf{W}}\,\d\mu_{\sf{g}}=-\langle(\d_{\sf{A}}\phi)^{i}\varphi^{j},\alpha^{a}\rangle_{\sf{A}_{1}}\langle v_{j},\rho_{\ast}(T_{a})v_{i}\rangle_{\sf{W}}\,\d\mu_{\sf{g}}=\\&=-\varphi^{j}\wedge(\alpha^{a}\wedge\ast(\d_{\sf{A}}\phi)^{i})\langle v_{j},\rho_{\ast}(T_{a})v_{i}\rangle_{\sf{W}}=\langle\varphi^{j},\ast (\alpha^{a}\wedge \ast(\d_{\sf{A}}\phi)^{i}\rangle_{\sf{E}}\langle v_{j},\rho_{\ast}(T_{a})v_{i}\rangle_{\sf{W}}\,\d\mu_{\sf{g}}=\\&=\langle\varphi,\ast(\alpha\wedge_{\sf{E}}\ast\d_{\sf{A}}\phi)\rangle_{\sf{E}}\,\d\mu_{\sf{g}}\,,
	\end{align*}
	which shows $(\d_{\sf{A}}\phi\wedge_{\mathfrak{J}}\cdot)^{\ast}=\ast(\cdot\wedge_{\sf{E}}\ast\d_{\sf{A}}\phi)$ and hence $\mathscr{P}_{12}^{\ast}=\mathscr{P}_{21}$ with respect to $\langle\cdot,\cdot\rangle_{\mathrm{Ad}(\sf{P})_{1}}$, $\langle\cdot,\cdot\rangle_{\sf{E}}$.

By Proposition~\ref{Prop:GaugInvYMKG}, it holds that $\mathscr{P}\circ\mathscr{K}=0$, provided the background field $(\sf{A},\phi)$ satisfy the non-linear Yang-Mills-Klein-Gordon equations. It remains to show Green hyperbolicity of $\mathscr{D}_{1}$ and $\mathscr{D}_{2}$.  First of all, a straightforward computation shows
	\begin{align*}
		\mathscr{K}^{\ast}\begin{pmatrix}\alpha\\\varphi\end{pmatrix}=
		\delta_{\sf{A}}\alpha +\phi\wedge_{\mathfrak{J}}\varphi\, .
	\end{align*}
	for all $(\alpha,\varphi)\in\Gamma^{\infty}(\mathcal{E}_{1})$. In particular, for a given $\varepsilon\in\Gamma^{\infty}(\mathrm{Ad}(\sf{P}))$, it follows that
	\begin{align*}
		\mathscr{D}_{1}:=\mathscr{K}^{\ast}\mathscr{K}\varepsilon=\delta_{\sf{A}}\d_{\sf{A}}\varepsilon-\phi\wedge_{\mathfrak{J}} (\varepsilon\wedge_{\sf{E}}\phi)=\delta_{\sf{A}}\d_{\sf{A}}\varepsilon+ (\varepsilon\wedge_{\sf{E}}\phi)\wedge_{\mathfrak{J}}\phi\, ,
	\end{align*}
	which is normally hyperbolic and hence in particular Green hyperbolic. Similarly, we obtain
	\begin{align*}
		\mathscr{K}\mathscr{K}^{\ast}\begin{pmatrix}
		\alpha\\\varphi
		\end{pmatrix}=\mathscr{K}(\delta_{\sf{A}}\alpha +\phi\wedge_{\mathfrak{J}}\varphi)=\begin{pmatrix}
				\d_{\sf{A}}\delta_{\sf{A}} & \d_{\sf{A}}(\phi\wedge_{\mathfrak{J}}\cdot)\\ -(\delta_{\sf{A}}\cdot)\wedge_{\sf{E}}\phi & -(\phi\wedge_{\mathfrak{J}}\cdot)\wedge_{\sf{E}}\phi
			\end{pmatrix}\begin{pmatrix}
		\alpha\\\varphi
		\end{pmatrix}\, .
	\end{align*}
	Using the Leibniz rule $\d_{\sf{A}}(\omega\wedge_{\mathfrak{J}}\eta)=(\d_{\sf{A}}\omega)\wedge_{\mathfrak{J}}\eta+(-1)^{k}\omega\wedge_{\mathfrak{J}}\d_{\sf{A}}\eta$ for all $\omega\in\Omega^{k}(\sf{M},\sf{E})$ and $\eta\in\Omega^{l}(\sf{M},\sf{E})$, which follows from a straightforward computation in local coordinates, we obtain
	\begin{align*}
		\mathscr{D}_{2}=\begin{pmatrix}
		\square^{\nabla^{\sf{A}}}_{\mathrm{dRH}}-\ast[\ast\sf{F}_{\sf{M}}^{\sf{A}}\wedge\cdot]_{\mathrm{Ad}(\sf{P})}-(\cdot\wedge_{\sf{E}}\phi)\wedge_{\mathfrak{J}}\phi & 2(\d_{\sf{A}}\phi\wedge_{\mathfrak{J}}\cdot)\\ 2\ast(\cdot\wedge_{\sf{E}}\ast\d_{\sf{A}}\phi)  &\square_{\mathrm{dRH}}^{\nabla^{\sf{A}}}+\mathcal{V}_{\phi}(\cdot)-(\phi\wedge_{\mathfrak{J}}\cdot)\wedge_{\sf{E}}\phi
	\end{pmatrix}\,,
	\end{align*}
	which is normally hyperbolic as well.
\end{proof}

\subsection{Linearised Einstein Gravity}\label{Sec:LinGrav}
General relativity is one of the cornerstones of modern physics. Since it was firstly published by Einstein around 1915 \cite{Einstein1,Einstein2,Einstein3,Einstein4,Einstein5}, many experiments have shown the great success of this theory in many of its aspects, two of the most recent ones being the detection of gravitational waves in September 2015 \cite{Ligo} as well as the first \textit{in situ} observations of black holes, with the central black hole of the elliptic galaxy M87 announced in 2019 \cite{M871,M872,M873,M874,M875,M876} and the Milky Way's central black hole Sagittarius A$^{\ast}$ following in 2023 \cite{SigA1,SigA2,SigA3,SigA4,SigA5,SigA6}.

In Einstein's theory of general relativity, spacetime and gravity is modelled by a Lorentzian manifold $(\sf{M},\sf{g})$ and the dynamics is governed by \textit{Einstein's field equations}
\begin{align}\label{eq:EinEq}
	\mathrm{Ein}_{\Lambda}(\sf{g})=\kappa\sf{T}\qquad\text{with}\qquad\mathrm{Ein}_{\Lambda}(\sf{g}):=\mathrm{Ric}(\sf{g})-\frac{1}{2}\mathrm{Scal}(\sf{g})\sf{g}+\Lambda\sf{g}
\end{align}
for a \textit{cosmological constant} $\Lambda\in\bb{R}$, where $\sf{T}\in\Gamma^{\infty}(\sf{T}^{\ast}\sf{M}^{\otimes_{s}2})$ is a divergence-free tensor, called the \textit{energy-momentum tensor}, modelling the matter content of the theory and where $\kappa:=8\pi G/c^{4}$. Exact solutions to~\eqref{eq:EinEq} include Minkowski spacetime, black hole spacetimes such as Schwarzschild and Kerr solutions for $\sf{T}=0$, as well as Reissner-Nordström and Kerr-Newman spacetimes for $\sf{T}\neq 0$. Another family of solutions are the FLRW-metrics relevant for cosmology, see Example~\ref{Examples:GlobHyp}(ii). From a mathematical perspective, Equation~\eqref{eq:EinEq} can be formulated as a hyperbolic problem subject to constraint equations whenever $(\sf{M},\sf{g})$ is globally hyperbolic, ensuring a well-posed Cauchy problem. This result is originally due to Choquet-Bruhat and Geroch \cite{FouresBruhat,Geroch}, building up on the observation of Lanczos that the Ricci tensor can be written as a nonlinear wave operator acting on $\sf{g}$ in suitable coordinates \cite{Lanczos}. For a more in-depth treatment on this subject, see the monographs of Choquet-Bruhat \cite{ChoquetBruhat} and Ringström \cite{Ringstrom}, the latter being based on a coordinate-free approach developed by DeTurck \cite{DeTurck1, DeTurck2}; for a comparison, see \cite{GrahamLee}. An alternative prove, in which the Einstein equations are equivalently viewed as a first-order quasilinear hyperbolic system, is due to Fischer-Marsden \cite{FischerMarsdenCauchy}.

Besides its clear mathematical structure, it is very challenging in practise to find \textit{exact} solutions to \eqref{eq:EinEq}. Consequently, numerical methods are often employed, or some form of approximation is required. In particular, when the gravitational field is \textit{weak} in a certain sense, linearising Einstein’s field equations provides an effective approach for modelling gravitational effects. Historically, this approach has been used by Einstein himself to predict the existence of gravitational waves in vacuum \cite{Einstein6,Einstein7} and has since then found many applications. In particular, in the absence of a fully satisfactory theory of quantum gravity, the linearised theory serves as a useful framework for describing low-energy effects of \textit{quantum} gravity, where linearised gravity is treated as a quantum field theory in a fixed background spacetime. In this section, we will see that linearised gravity provides in fact an example of a linear gauge theory in the sense of Definition~\ref{Def:LinGaugeTh} and hence can be quantised in a similar fashion as Maxwell's theory.

The structure of the classical theory of linearised gravity needed for quantisation is nowadays relatively well understood, see e.g.~\cite{FewsterHunt,RejznerGravity,KhavkineGrav}. It what follows, we will provide the reader with a systematic treatment of the theory starting from Einstein's equation~\eqref{eq:EinEq}. In order to derive the linearised Einstein equations, we will consider a one-parameter family of metrics $\hat{\sf{g}}(\lambda)$ depending \textit{smoothly} on a parameter $\lambda\in\bb{R}$, which then allows us to formally expand
\begin{align*}
	\hat{\sf{g}}(\lambda)=\sf{g}+\lambda\sf{h}+\mathcal{O}(\lambda^{2})\, ,
\end{align*}
where $\sf{g}:=\hat{\sf{g}}(\lambda=0)$ is the \textit{background metric} and $\sf{h}:=\frac{\d}{\d\lambda}\big\vert_{\lambda=0}\hat{\sf{g}}(\lambda)$ a symmetric $(0,2)$-tensor field, the \textit{linearisation}. For the sole purpose of deriving the linearised expressions, it will be sufficient to work on a formal level. We remark, however, that there are various ways to make this precise. For instance, Stewart-Walker \cite{StewartWalker} considered a family of Lorentzian manifolds $(\sf{M}_{\lambda},\sf{g}_{\lambda})$ embedded in a higher-dimensional ambient manifold $\widetilde{\sf{M}}$ as a set-up for rigorous perturbation theory. Alternatively, one might give the space of Lorentzian metrics $\mathcal{L}(\sf{M})$ on a given manifold $\sf{M}$ the structure of an $\infty$-dimensional Fréchet manifold with tangent space $\sf{T}_{\sf{g}}\mathcal{L}(\sf{M})\cong\Gamma^{\infty}(\sf{T}^{\ast}\sf{M}^{\otimes_{s}2})$ (see e.g.~the classical article \cite{Ebin}), in order to consider $\hat{\sf{g}}(\lambda)$ as a \textit{curve} in this infinite-dimensional space. This approach is for example taken in Fischer-Marsden's work on the ADM formalism of gravity and linearised stability \cite{FischerMarsden0}; see also the reviews \cite{FischerMarsden1,FischerMarsden2}.

To obtain the linearised Einstein equations, we first have to linearise the Ricci tensor and the scalar curvature. While this computation is straightforward and the obtained results are widely applied and well-documented in the literature (see, e.g.,~\cite[Sec.~I.11]{ChoquetBruhat}), explicit derivations are surprisingly rare. Notable exceptions include \cite[Sec.~1.K]{Besse} and the classical article of Lichnerowicz~\cite{Lichnerowicz}. This warrants the detailed proof of the following Lemma:

\begin{lemma}\label{Lem:LinRic} \emph{(Perturbation of Ricci and Scalar Curvature)}\newline
Let $\sf{M}$ be a smooth $d$-manifold and $\hat{\sf{g}}(\lambda)$ be a family of pseudo-Riemannian metrics on $\sf{M}$ such that $\hat{\sf{g}}(\lambda)=\sf{g}+\lambda \sf{h}+\mathcal{O}(\lambda^{2})$ with $\sf{g}:=\sf{g}(0)$ and $\sf{h}:=\frac{\d}{\d\lambda}\big\vert_{\lambda=0}\sf{g}(\lambda)\in\Gamma^{\infty}(\sf{T}^{\ast}\sf{M}^{\otimes_{s}2})$. Then,
	\begin{align*}\begin{aligned}
		\text{\emph{(i)}}\quad &\mathrm{Ric}(\hat{\sf{g}}(\lambda))_{\alpha\beta}=\mathrm{Ric}(\sf{g})_{\alpha\beta}+\frac{1}{2}\bigg\{-\nabla^{\rho}\nabla_{\rho}\sf{h}_{\alpha\beta}+2\nabla_{\rho}\nabla_{(\alpha}\sf{h}_{\beta)}^{\rho}-\nabla_{\alpha}\nabla_{\beta}\mathrm{tr}_{\sf{g}}(\sf{h})\bigg\}\lambda+\mathcal{O}(\lambda^{2})\\
		\text{\emph{(ii)}}\quad &\mathrm{Scal}(\hat{\sf{g}}(\lambda))=\mathrm{Scal}(\sf{g})+\bigg\{-\nabla^{\rho}\nabla_{\rho}\mathrm{tr}_{\sf{g}}(\sf{h})+\nabla^{\alpha}\nabla^{\beta}\sf{h}_{\alpha\beta}-\mathrm{Ric}(\sf{g})^{\alpha\beta}\sf{h}_{\alpha\beta}\bigg\}\lambda+\mathcal{O}(\lambda^{2})\\
		\text{\emph{(iii)}}\quad &\mathrm{Ein}_{\Lambda}(\hat{\sf{g}}(\lambda))_{\alpha\beta}=\mathrm{Ein}_{\Lambda}(\sf{g})_{\alpha\beta}+\frac{1}{2}\bigg\{-\nabla^{\rho}\nabla_{\rho}\sf{h}_{\alpha\beta}+2\nabla_{\rho}\nabla_{(\alpha}\sf{h}_{\beta)}^{\rho}-\nabla_{\alpha}\nabla_{\beta}\mathrm{tr}_{\sf{g}}(\sf{h})\\&\hspace*{0.6cm}+\sf{g}_{\alpha\beta}\bigg(\nabla^{\rho}\nabla_{\rho}\mathrm{tr}_{\sf{g}}(\sf{h})-\nabla_{\mu}\nabla_{\nu}\sf{h}^{\mu\nu}+\mathrm{Ric}(\sf{g})_{\mu\nu}\sf{h}^{\mu\nu}\bigg)+(2\Lambda-\mathrm{Scal}(\sf{g}))\sf{h}_{\alpha\beta}\bigg\}\lambda+\mathcal{O}(\lambda^{2})\,,
	\end{aligned}\end{align*}
	where $\nabla$ and $\square_{i}$ are the operators with respect to the background metric $\sf{g}$ and where we use the background metric to raise and lower indices.
\end{lemma}

\begin{proof}
	As a first step, let us choose a local coordinate chart $(\mathcal{U},\varphi=(x^{\alpha})_{\alpha=0,\dots,d})$ on $\sf{M}$. Then, $\hat{\sf{g}}(\lambda)_{\alpha\beta}=\sf{g}_{\alpha\beta}+\lambda \sf{h}_{\alpha\beta}+\mathcal{O}(\lambda^{2})$. Now, observe that
	\begin{align*}
		0=\frac{\d}{\d\lambda}\bigg\vert_{\lambda=0}\underbrace{\hat{\sf{g}}(\lambda)_{\alpha\beta}\hat{\sf{g}}(\lambda)^{\beta\gamma}}_{=\delta_{\alpha}^{\gamma}}=\sf{g}_{\alpha\beta}\frac{\d}{\d\lambda}\bigg\vert_{\lambda=0}\hat{\sf{g}}(\lambda)^{\beta\gamma}+\sf{h}_{\alpha\beta}\sf{g}^{\beta\gamma}\quad\Rightarrow\quad \frac{\d}{\d\lambda}\bigg\vert_{\lambda=0}\hat{\sf{g}}(\lambda)^{\alpha\beta}=-\sf{g}^{\alpha\gamma}\sf{g}^{\beta\delta}\sf{h}_{\gamma\delta}\, ,
	\end{align*}
	which, using the convention that indices are raised/lowered via the background metric $\sf{g}$, yields
	\begin{align*}
		\hat{\sf{g}}(\lambda)^{\alpha\beta}=\sf{g}^{\alpha\beta}-\lambda \sf{h}^{\alpha\beta}+\mathcal{O}(\lambda^{2})\, .
	\end{align*}
	Next, let us denote the Christoffel symbols of $\hat{\sf{g}}(\lambda)$ by $\Gamma(\lambda)^{\alpha}_{\beta\gamma}$ and $\Gamma^{\alpha}_{\beta\gamma}:=\Gamma(0)^{\alpha}_{\beta\gamma}$ for the background metric. Then,
	\begin{align*}
		\frac{\d}{\d\lambda}&\bigg\vert_{\lambda=0}\Gamma(\lambda)^{\alpha}_{\beta\gamma}=\frac{1}{2}\frac{\d}{\d\lambda}\bigg\vert_{\lambda=0}(\hat{\sf{g}}(\lambda)^{\alpha\delta}(\partial_{\beta}\hat{\sf{g}}(\lambda)_{\gamma\delta}+\partial_{\gamma}\hat{\sf{g}}(\lambda)_{\beta\delta}-\partial_{\delta}\hat{\sf{g}}(\lambda)_{\beta\gamma})=\\&=\frac{1}{2}\sf{g}^{\alpha\delta}(\partial_{\beta}\sf{h}_{\gamma\delta}+\partial_{\gamma}\sf{h}_{\beta\delta}-\partial_{\delta}\sf{h}_{\beta\gamma})-\frac{1}{2}\sf{h}^{\alpha\delta}(\partial_{\beta}\sf{g}_{\gamma\delta}+\partial_{\gamma}\sf{g}_{\beta\delta}-\partial_{\delta}\sf{g}_{\beta\gamma})=\frac{1}{2}(2\nabla_{(\beta}\sf{h}_{\gamma)}^{\alpha}-\nabla^{\alpha}\sf{h}_{\beta\gamma})\, ,
	\end{align*}
	where the last equality can easily be seen by writing out the right-hand side explicitly. To sum up, we found the following expansion of the Christoffel symbols:
	\begin{align}\label{eq:ChriPert}
		\Gamma(\lambda)^{\alpha}_{\beta\gamma}=\Gamma^{\alpha}_{\beta\gamma}+\frac{1}{2}\bigg\{2\nabla_{(\beta}\sf{h}_{\gamma)}^{\alpha}-\nabla^{\alpha}\sf{h}_{\beta\gamma}\bigg\}\lambda+\mathcal{O}(\lambda^{2})\, .
	\end{align}
	With this expansion, we have all the ingredients to derive the expansion of the Ricci tensor:
	\begin{align*}
        \frac{\mathrm{d}}{\mathrm{d}\lambda}\bigg\vert_{\lambda=0}&\mathrm{Ric}(\hat{\sf{g}}(\lambda))_{\alpha\beta}=\frac{\mathrm{d}}{\mathrm{d}\lambda}\bigg\vert_{\lambda=0}\bigg(\partial_{\rho}\Gamma(\lambda)^{\rho}_{\alpha\beta}-\partial_{\alpha}\Gamma(\lambda)^{\rho}_{\beta\rho}+\Gamma(\lambda)^{\rho}_{\alpha\beta}\Gamma(\lambda)^{\mu}_{\rho\mu}-\Gamma(\lambda)^{\rho}_{\alpha\mu}\Gamma(\lambda)^{\mu}_{\beta\rho}\bigg)=\\=&\frac{1}{2}(\partial_{\rho}\nabla_{\alpha}\sf{h}_{\beta}^{\rho}+\partial_{\rho}\nabla_{\beta}\sf{h}_{\alpha}^{\rho}-\partial_{\rho}\nabla^{\rho}\sf{h}_{\alpha\beta}-\partial_{\alpha}\nabla_{\beta}\sf{h}_{\rho}^{\rho}-\cancel{\partial_{\alpha}\nabla_{\rho}\sf{h}_{\beta}^{\rho}}+\cancel{\partial_{\alpha}\nabla^{\rho}\sf{h}_{\beta\rho}})\\&+\frac{1}{2}(\Gamma^{\mu}_{\rho\mu}\nabla_{\alpha}\sf{h}_{\beta}^{\rho}+\Gamma^{\mu}_{\rho\mu}\nabla_{\beta}\sf{h}_{\alpha}^{\rho}-\Gamma^{\mu}_{\rho\mu}\nabla^{\rho}\sf{h}_{\alpha\beta}+\Gamma^{\rho}_{\alpha\beta}\nabla_{\rho}\sf{h}+\cancel{\Gamma^{\rho}_{\alpha\beta}\nabla_{\mu}\sf{h}_{\rho}^{\mu}}-\cancel{\Gamma^{\rho}_{\alpha\beta}\nabla^{\mu}\sf{h}_{\rho\mu}})\\&-\frac{1}{2}(\Gamma^{\mu}_{\beta\rho}\nabla_{\alpha}\sf{h}_{\mu}^{\rho}+\Gamma^{\mu}_{\beta\rho}\nabla_{\mu}\sf{h}_{\alpha}^{\rho}-\Gamma^{\mu}_{\beta\rho}\nabla^{\rho}\sf{h}_{\alpha\mu}+\Gamma^{\rho}_{\alpha\mu}\nabla_{\beta}\sf{h}_{\rho}^{\mu}+\Gamma^{\rho}_{\alpha\mu}\nabla_{\rho}\sf{h}_{\beta}^{\mu}-\Gamma^{\rho}_{\alpha\mu}\nabla^{\mu}\sf{h}_{\beta\rho})=\\=&-\frac{1}{2}\nabla^{\rho}\nabla_{\rho} \sf{h}_{\alpha\beta}+\frac{1}{2}(2\nabla_{\rho}\nabla_{(\alpha}\sf{h}^{\rho}_{\beta)}-\nabla_{\alpha}\nabla_{\beta}\mathrm{tr}_{\sf{g}}(\sf{h}))\, .
	\end{align*}
	This shows (i). For (ii), we use the expansion of inverse metric and the Ricci tensor to obtain
	\begin{align*}
        \frac{\mathrm{d}}{\mathrm{d}\lambda}\bigg\vert_{\lambda=0}\mathrm{Scal}(\hat{\sf{g}}(\lambda))&=\frac{\mathrm{d}}{\mathrm{d}\lambda}\bigg\vert_{\lambda=0}\hat{\sf{g}}(\lambda)^{\alpha\beta}\mathrm{Ric}(\hat{\sf{g}}(\lambda))_{\alpha\beta}=\\&=\frac{1}{2}g^{\alpha\beta}(2\nabla_{\rho}\nabla_{(\alpha}\sf{h}^{\rho}_{\beta)}-\nabla_{\alpha}\nabla_{\beta}\mathrm{tr}_{\sf{g}}(\sf{h})-\nabla^{\rho}\nabla_{\rho} \sf{h}_{\alpha\beta})-\sf{h}^{\alpha\beta}\mathrm{Ric}(\sf{g})_{\alpha\beta}=\\&=-\nabla^{\rho}\nabla_{\rho}\mathrm{tr}_{\sf{g}}(\sf{h})+\nabla^{\alpha}\nabla^{\beta}\sf{h}_{\alpha\beta}-\mathrm{Ric}(\sf{g})^{\alpha\beta}\sf{h}_{\alpha\beta}
    \end{align*}
    using the Leibniz rule. The linearisation of the Einstein tensor follows now from the linearisation of the Ricci and scalar curvature by an easy computation using the Leibniz rule, i.e.
    \begin{align*}
        \frac{\mathrm{d}}{\mathrm{d}\lambda}\bigg\vert_{\lambda=0}\mathrm{Ein}_{\Lambda}(\hat{\sf{g}}(\lambda))_{\alpha\beta}=&\frac{\mathrm{d}}{\mathrm{d}\lambda}\bigg\vert_{\lambda=0}\bigg(\mathrm{Ric}(\hat{\sf{g}}(\lambda))_{\alpha\beta}-\frac{1}{2}\mathrm{Scal}(\hat{\sf{g}}(\lambda))\hat{\sf{g}}(\lambda)_{\alpha\beta}+\Lambda \hat{\sf{g}}(\lambda)\bigg)=\\=&-\frac{1}{2}\nabla^{\rho}\nabla_{\rho} \sf{h}_{\alpha\beta}+\frac{1}{2}(2\nabla_{\rho}\nabla_{(\alpha}\sf{h}^{\rho}_{\beta)}-\nabla_{\alpha}\nabla_{\beta}\mathrm{tr}_{\sf{g}}(\sf{h}))-\frac{1}{2}\mathrm{Scal}(\sf{g})\sf{h}_{\alpha\beta}\\&-\frac{1}{2}\sf{g}_{\alpha\beta}\bigg(-\nabla^{\rho}\nabla_{\rho}\mathrm{tr}_{\sf{g}}(\sf{h})+\nabla^{\mu}\nabla^{\nu}\sf{h}_{\mu\nu}-\mathrm{Ric}(\sf{g})^{\mu\nu}\sf{h}_{\mu\nu}\bigg)+\Lambda\sf{h}_{\alpha\beta}\,,
    \end{align*}
    which concludes the proof. 
\end{proof}

For the following discussion, it will be useful to establish some notation for symmetric tensor fields, which will allow us to perform computations in a coordinate-free manner. First of all, in what follows, $(\sf{M},\sf{g})$ will be a globally hyperbolic spacetime of dimension $d=1+3$. On this manifold, we denote the bundles whose sections of symmetric $(0,k)$-tensor fields equipped with their natural bundle metrics by
\begin{align}\label{eq:BundleSymTens}
	\sf{S}_{k}:=\sf{T}^{\ast}\sf{M}^{\otimes_{s}k},\quad\qquad \langle\sf{h},\sf{k}\rangle_{\sf{S}_{k}}:=k!(\sf{g}^{\sharp})^{\otimes k}(\sf{h},\sf{k})=(k!)\,\sf{h}^{\alpha_{1}\dots\alpha_{k}}\sf{k}_{\alpha_{1}\dots\alpha_{k}}\, .
\end{align}
By definition, note that $\Gamma^{\infty}(\sf{S}_{1})=\Omega^{1}(\sf{M})=\mathfrak{X}^{\ast}(\sf{M})$. As usual, we denote the formal adjoint of a linear differential operator $\sf{A}\in\mathrm{DO}(\sf{S}_{k},\sf{S}_{l})$ with respect to the induced bilinear forms $(\cdot,\cdot)_{\sf{S}_{\bullet}}$ on the level of sections by $\sf{A}^{\ast}$. On these bundles, we introduce the symmetrised covariant derivative and divergence operator by
\begin{align*}
	\d_{s}\:&\Gamma^{\infty}(\sf{S}_{k})\to\Gamma^{\infty}(\sf{S}_{k+1}),\qquad (\d_{s} u)_{\alpha_{1}\dots\alpha_{k+1}}:=\nabla_{(\alpha_{1}}u_{\alpha_{2}\dots\alpha_{k})}\qquad &\textit{symmetrised gradient}\\
\delta_{s}\:&\Gamma^{\infty}(\sf{S}_{k+1})\to\Gamma^{\infty}(\sf{S}_{k}),\qquad (\delta_{s} u)_{\alpha_{1}\dots\alpha_{k}}:=-(k+1)\nabla^{\lambda}u_{\lambda\alpha_{1}\dots\alpha_{k}}\qquad &\textit{divergence}
\end{align*}
where we use the subscript ``$s$'' to distinguish these operators from the antisymmetric analogue for differential forms. The normalisation conventions chosen are such that $\d_{s}$ and $\delta_{s}$ are formal adjoints with respect to $(\cdot,\cdot)_{\sf{S}_{\bullet}}$, i.e.~$\delta_{s}=\d_{s}^{\ast}$. Note that clearly $\d_{s}^{2}\neq 0$ and $\delta_{s}^{2}\neq 0$ as opposed to the exterior derivative and codifferential on antisymmetric tensors. Let us further introduce the following operators on $\Gamma^{\infty}(\sf{S}_{k})$, which will be useful in the following discussion:
\begin{align*}
	\I\:&\Gamma^{\infty}(\sf{S}_{2})\to\Gamma^{\infty}(\sf{S}_{2}),\qquad (\I\sf{h})_{\alpha\beta}:=\sf{h}_{\alpha\beta}-\frac{1}{2}\sf{g}_{\alpha\beta}\mathrm{tr}_{\sf{g}}(\sf{h})\qquad &\textit{trace-reversal}\\
\mathrm{Ric}_{\sf{g}}\:&\Gamma^{\infty}(\sf{S}_{1})\to\Gamma^{\infty}(\sf{S}_{1}),\qquad \mathrm{Ric}_{\sf{g}}(\omega)_{\alpha}:=\mathrm{Ric}(\sf{g})_{\alpha}^{\lambda}\omega_{\lambda}\qquad &\textit{Ricci operator on $\sf{S}_{1}$}\\
\mathrm{Ric}_{\sf{g}}\:&\Gamma^{\infty}(\sf{S}_{2})\to\Gamma^{\infty}(\sf{S}_{2}),\qquad \mathrm{Ric}_{\sf{g}}(\sf{h})_{\alpha\beta}:=\mathrm{Ric}(\sf{g})_{(\alpha}^{\lambda}\sf{h}_{\beta)\lambda}\qquad &\textit{Ricci operator on $\sf{S}_{2}$}\\
\mathrm{Riem}_{\sf{g}}\:&\Gamma^{\infty}(\sf{S}_{2})\to\Gamma^{\infty}(\sf{S}_{2}),\qquad \mathrm{Riem}_{\sf{g}}(\sf{h})_{\alpha\beta}:=\tensor{\mathrm{Riem}(\sf{g})}{^\rho_\alpha_\beta^\sigma}\sf{h}_{\rho\sigma}\qquad &\textit{Riemann operator}
\end{align*}
Note that $\mathrm{I}$ is an involution, i.e.~$\mathrm{I}^{2}=\mathrm{id}$, and satisfies $\mathrm{tr}_{\sf{g}}\circ\mathrm{I}=-\mathrm{tr}_{\sf{g}}$, which is the origin of the name \textit{trace-reversal}. Furthermore,  we recall that we denote by
\begin{align*}
	\square\:\Gamma^{\infty}(\sf{T}^{\ast}\sf{M}^{\otimes k})\to\Gamma^{\infty}(\sf{T}^{\ast}\sf{M}^{\otimes k}),\qquad \square:=g^{\alpha\beta}\nabla_{\alpha}\nabla_{\beta}
\end{align*}
the connection d'Alembertian, as usual. A more natural wave-type operator for symmetric tensor fields is $-\delta_{s}\d_{s}+\d_{s}\delta_{s}$, in analogy to the de Rham-Hodge operator for differential forms, and a straightforward computation in coordinates shows that it is related to $\square$ by a curvature term of order zero by means of the Weitzenböck-type identity
\begin{align*}
		\square&=-\delta_{s}\d_{s}+\d_{s}\delta_{s}-k(k-1)\tensor{\mathrm{Riem}(\sf{g})}{^\lambda_{(\alpha_{1}}_{\alpha_{2}}^{\rho}}\omega_{\alpha_{3}\dots\alpha_{k})\rho}-k\mathrm{Ric}(\sf{g})^{\lambda}_{(\alpha_{1}}\omega_{\alpha_{2}\dots\alpha_{k})\lambda}\quad\text{on}\quad\Gamma^{\infty}(\sf{S}_{k})\, .
\end{align*}
In particular, it holds that $\square=-\delta_{s}\d_{s}$ on $\Gamma^{\infty}(\sf{S}_{0})$ and $\square=-\delta_{s}\d_{s}+\d_{s}\delta_{s}-\mathrm{Ric}_{\sf{g}}$ on $\Gamma^{\infty}(\sf{S}_{1})$. Last but not least, we define the natural bilinear operation
\begin{align*}
	\times_{s}\:\Gamma^{\infty}(\sf{S}_{2})\times\Gamma^{\infty}(\sf{S}_{2})\to\Gamma^{\infty}(\sf{S}_{2}),\qquad (\sf{h}\times_{s}\sf{k})_{\alpha\beta}:=\sf{h}_{(\alpha}^{\lambda}\sf{k}_{\beta)\lambda}\, .
\end{align*}
We remark that $(\Gamma^{\infty}(\sf{S}_{2}),\times_{s})$ is a commutative unital algebra with multiplicative identity element $\sf{g}$. Furthermore, the bilinear map $\langle\cdot,\cdot\rangle_{\sf{S}_{2}}$ is invariant under the action of $\times_{s}$, i.e.
\begin{align}\label{eq:SymTenInv}
	\langle \sf{h}\times_{s}\sf{k},\sf{m}\rangle_{\sf{S}_{2}}=\langle \sf{k},\sf{h}\times_{s}\sf{m}\rangle_{\sf{S}_{2}}
\end{align}
for all $\sf{h},\sf{k},\sf{m}\in\Gamma^{\infty}(\sf{S}_{2})$. Note also that $\mathrm{Ric}_{\sf{g}}(\sf{h})=\mathrm{Ric}(\sf{g})\times_{s}\sf{h}$ for all $\sf{h}\in\Gamma^{\infty}(\sf{S}_{2})$.

In the following discussion, we will call a Lorentzian metric \textit{$\Lambda$-Einstein}, if it satisfies the non-linear Einstein equations in vacuum, i.e.~if $\mathrm{Ein}_{\Lambda}(\sf{g})=0$. Let us record the following properties for later convenience (cf.~\cite[Lemma 4.1-4.3]{GerardMurroWrochna}\footnote{Note that there is a typo in the statement of Lemma 4.2(iii) in \cite{GerardMurroWrochna}; the right-hand side should be with a minus, as written correctly in the last line of their proof.}, \cite[Lemma~2.2]{BeniniMurro}).

\begin{lemma}\label{Lemma:PropLinGrav} Let $(\sf{M},\sf{g})$ be a $(1+3)$-dimensional Lorentzian manifold.
	\begin{itemize}
		\item[\emph{(i)}]$\mathrm{Riem}_{\sf{g}}^{\ast}=\mathrm{Riem}_{\sf{g}}$, \quad $\mathrm{Ric}_{\sf{g}}^{\ast}=\mathrm{Ric}_{\sf{g}}$, \quad $\mathrm{I}^{\ast}=\mathrm{I}$,\quad $\square^{\ast}=\square$ \quad and \quad $\d_{s}^{\ast}=\delta_{s}$. 
		\item[\emph{(ii)}]$\square\circ \I=\I\circ\square$ on $\Gamma^{\infty}(\sf{S}_{2})$.
		\item[\emph{(iii)}]If $\sf{g}$ is $\Lambda$-Einstein, i.e.~$\mathrm{Ein}_{\Lambda}(\sf{g})=0$, then the following holds:
		\begin{itemize}
			\item[\emph{(a)}]\quad $(-\square+2\mathrm{Riem}_{\sf{g}})\circ \d_{s}=\d_{s}\circ (-\square-\Lambda)$ on $\Gamma^{\infty}(\sf{S}_{1})$.
			\item[\emph{(b)}]\quad $\delta_{s}\circ (-\square+2\mathrm{Riem}_{\sf{g}})=(-\square-\Lambda)\circ\delta_{s}$ on $\Gamma^{\infty}(\sf{S}_{2})$.
			\item[\emph{(c)}]\quad $\delta_{s}\circ\mathrm{I}\circ\d_{s}=-\square-\Lambda$ on $\Gamma^{\infty}(\sf{S}_{1})$.
			\item[\emph{(d)}]\quad $\mathrm{Riem}_{\sf{g}}\circ\mathrm{I}=\mathrm{I}\circ\mathrm{Riem}_{\sf{g}}$.
		\end{itemize}
	\end{itemize}
\end{lemma}

\begin{proof}
	The claims (i) and (ii) can easily be seen. For (iii), we will only show (a) and (c), since (b) follows from (a) by taking the formal adjoints and (d) is a one-line computation. Now, for (a), let $\omega\in\Gamma^{\infty}(\sf{S}_{1})$. Then,
	\begin{align*}
			(-\square\circ\d_{s})\omega_{\alpha\beta}=&-\frac{1}{2}\nabla^{\lambda}\nabla_{\lambda}(\nabla_{\alpha}\omega_{\beta}+\nabla_{\beta}\omega_{\alpha})\\=&-\frac{1}{2}\bigg(\nabla^{\lambda}\nabla_{\alpha}\nabla_{\lambda}\omega_{\beta}+\nabla^{\lambda}\nabla_{\beta}\nabla_{\lambda}\omega_{\alpha}+\nabla^{\lambda}(\tensor{\mathrm{Riem}(\sf{g})}{_\lambda_\alpha_\beta^\rho}\omega_{\rho}+\tensor{\mathrm{Riem}(\sf{g})}{_\lambda_\beta_\alpha^\rho}\omega_{\rho})\bigg)\\=&-\frac{1}{2}\bigg(\nabla_{\alpha}\nabla^{\lambda}\nabla_{\lambda}\omega_{\beta}+\nabla_{\beta}\nabla^{\lambda}\nabla_{\lambda}\omega_{\alpha}+\Lambda(\nabla_{\alpha}\omega_{\beta}+\nabla_{\beta}\omega_{\alpha})+4\mathrm{Riem}_{\sf{g}}(\d_{s}\omega)\bigg)\\=&\d_{s}\circ (-\square-\Lambda)\omega_{\alpha\beta}-2\mathrm{Riem}_{\sf{g}}(\d_{s}\omega)_{\alpha\beta}\, ,
	\end{align*}
	where we used Equation~\eqref{eq:Identity} of the convention and notation section, the fact that $\mathrm{Ein}_{\Lambda}(\sf{g})=0$ and hence $\mathrm{Ric}(\sf{g})=\Lambda\sf{g}$ as well as the Bianchi identity to conclude that
	\begin{align*}
		\nabla^{\lambda}\tensor{\mathrm{Riem}(\sf{g})}{_\lambda_\alpha_\beta^\rho}=\nabla^{\lambda}\tensor{\mathrm{Riem}(\sf{g})}{_\beta^\rho_\lambda_\alpha}=\nabla_{\beta}\mathrm{Ric}(\sf{g})^{\rho}_{\alpha}-\nabla^{\rho}\mathrm{Ric}(\sf{g})_{\alpha\beta}=0\, .
	\end{align*}
	For (c), we take again $\omega\in\Gamma^{\infty}(\sf{S}_{1})$ and compute
	\begin{align*}
		(\delta_{s}\mathrm{I}\d_{s}\omega)_{\lambda}=-\nabla^{\rho}(\nabla_{\rho}\omega_{\lambda}+\nabla_{\lambda}\omega_{\rho})+\nabla_{\lambda}\nabla^{\rho}\omega_{\rho}=-(\square\omega)_{\lambda}-\mathrm{Ric}(\sf{g})_{\lambda}^{\rho}\omega_{\rho}=[(-\square-\Lambda)\omega]_{\lambda}\, ,
	\end{align*}
	which concludes the proof.
\end{proof}

Now, coming back to the linearised Einstein equations, we note that the linearisation of the Ricci tensor $\mathrm{Ric}(\sf{g})$ and Einstein tensor $\mathrm{Ein}_{\Lambda}(\sf{g})$ in Lemma~\ref{Lem:LinRic} can be written in terms of the operators introduced above as follows.

\begin{proposition}\label{Prop:LinEinRicOp} \emph{(Linearised Ricci and Einstein Operators)}\newline
	Let $\sf{M}$ be a four-dimensional manifold and $\hat{\sf{g}}(\lambda)$ be a family of Lorentzian metrics on $\sf{M}$ such that $\hat{\sf{g}}(\lambda)=\sf{g}+\lambda \sf{h}+\mathcal{O}(\lambda^{2})$ with $\sf{g}:=\sf{g}(0)$ and $\sf{h}:=\frac{\d}{\d\lambda}\big\vert_{\lambda=0}\sf{g}(\lambda)\in\Gamma^{\infty}(\sf{T}^{\ast}\sf{M}^{\otimes_{s}2})$. Then,
	\begin{align*}\begin{aligned}
		&\mathrm{Ric}(\hat{\sf{g}}(\lambda))=\mathrm{Ric}(\sf{g})+(\sf{R}\sf{h})\lambda+\mathcal{O}(\lambda^{2})\\
		&\mathrm{Ein}_{\Lambda}(\hat{\sf{g}}(\lambda))=\mathrm{Ein}_{\Lambda}(\sf{g})+(\sf{L}\sf{h})\lambda+\mathcal{O}(\lambda^{2})\, ,
	\end{aligned}\end{align*}
	where $\sf{R},\sf{L}\in\mathrm{DO}^{2}(\sf{S}_{2})$ are the second-order linear differential operators given by
	\begin{align*}
		\sf{R}&:=\frac{1}{2}\bigg(-\square-\d_{s}\delta_{s}\mathrm{I}+2\mathrm{Riem}_{\sf{g}}+2\mathrm{Ric}_{\sf{g}}\bigg)\,,\\
		\sf{L}&:=\frac{1}{2}\bigg(-\square-\I\d_{s}\delta_{s}+2\mathrm{Riem}_{\sf{g}}+2(\mathrm{Ein}_{\Lambda}(\sf{g})\times_{s}\cdot)\bigg)\I+\frac{1}{4}\sf{g}\langle\mathrm{Ein}_{\Lambda}(\sf{g}),\cdot\rangle_{\sf{S}_{2}}\, .
	\end{align*}
	In particular, if the background metric $\sf{g}$ is $\Lambda$-Einstein, this reduces to
	\begin{align*}
		\sf{R}=\frac{1}{2}\bigg(-\square-\d_{s}\delta_{s}\mathrm{I}+2\mathrm{Riem}_{\sf{g}}+2\Lambda\bigg)\,,\quad\sf{L}=\frac{1}{2}\bigg(-\square-\I\d_{s}\delta_{s}+2\mathrm{Riem}_{\sf{g}}\bigg)\I\, .
	\end{align*}
\end{proposition}

\begin{proof}
	Let $\sf{h}\in\Gamma^{\infty}(\sf{S}_{2})$. As a first step, we write the $(0,2)$-tensors $\mathrm{I}\d_{s}\delta_{s}\sf{h}$ and $\d_{s}\delta_{s}\mathrm{I}\sf{h}$ explicitly in local coordinates, which yields
	\begin{align*}
		(\I\d_{s}\delta_{s}\sf{h})_{\alpha\beta}&=-2\nabla_{(\alpha}\nabla^{\lambda}\sf{h}_{\beta)\lambda}+\sf{g}_{\alpha\beta}\nabla^{\rho}\nabla^{\sigma}\sf{h}_{\rho\sigma}\\
		(\d_{s}\delta_{s}\I\sf{h})_{\alpha\beta}&=-2\nabla_{(\alpha}\nabla^{\lambda}\sf{h}_{\beta)\lambda}+\nabla_{\alpha}\nabla_{\beta}\mathrm{tr}_{\sf{g}}(\sf{h})
	\end{align*}
	Comparing the expression of $\d_{s}\delta_{s}\I\sf{h}$ with the linearisation of the Ricci tensor in Lemma~\ref{Lem:LinRic}, we see that it almost agrees with the last two terms up to the order of covariant derivatives in $\nabla_{(\alpha}\nabla^{\lambda}\sf{h}_{\beta)\lambda}$. Hence, we compute
	\begin{align*}
		\nabla_{(\alpha}\nabla^{\lambda}\sf{h}_{\beta)\lambda}&=\nabla^{\lambda}\nabla_{(\alpha}\sf{h}_{\beta)\lambda}+\tensor{\mathrm{Riem}(\sf{g})}{_{(\alpha}^\lambda_{\beta)}^\rho}\sf{h}_{\rho\lambda}-\mathrm{Ric}(\sf{g})_{(\alpha}^{\rho}\sf{h}_{\beta)\rho}\\&=\nabla^{\lambda}\nabla_{(\alpha}\sf{h}_{\beta)\lambda}-\mathrm{Riem}_{\sf{g}}(\sf{h})_{\alpha\beta}-\mathrm{Ric}_{\sf{g}}(\sf{h})\, ,
	\end{align*}
	where we used Equation~\eqref{eq:Identity} of the convention and notation section. As a consequence, we can write the operator $\sf{R}$ introduced above in local coordinates as
	\begin{align*}
		(\sf{R}\sf{h})_{\alpha\beta}&=\frac{1}{2}\big\{\big(-\square-\d_{s}\delta_{s}\I+2\mathrm{Riem}_{\sf{g}}+2\mathrm{Ric}_{\sf{g}}\big)\sf{h}\big\}_{\alpha\beta}\\&=\frac{1}{2}\bigg(-(\square\sf{h})_{\alpha\beta}+2\nabla_{\lambda}\nabla_{(\alpha}\sf{h}_{\beta)}^{\lambda}-\nabla_{\alpha}\nabla_{\beta}\mathrm{tr}_{\sf{g}}(\sf{h})\bigg)\, ,
	\end{align*}
	which is exactly the linearised expression of the Ricci tensor derived in Lemma~\ref{Lem:LinRic}. For the Einstein tensor, we first observe replace $\sf{h}\mapsto\mathrm{I}\sf{h}$ in the above equation for $\I\d_{s}\delta_{s}\sf{h}$ to obtain
	\begin{align*}
		(\I\d_{s}&\delta_{s}\I\sf{h})_{\alpha\beta}=-2\nabla_{(\alpha}\nabla^{\lambda}\sf{h}_{\beta)\lambda}+\sf{g}_{\alpha\beta}\nabla^{\rho}\nabla^{\sigma}\sf{h}_{\rho\sigma}+\nabla_{\alpha}\nabla_{\beta}\mathrm{tr}_{\sf{g}}(\sf{h})-\frac{1}{2}\sf{g}_{\alpha\beta}\square\mathrm{tr}_{\sf{g}}(\sf{h})\\&=-2\nabla^{\lambda}\nabla_{(\alpha}\sf{h}_{\beta)\lambda}+\sf{g}_{\alpha\beta}\nabla^{\rho}\nabla^{\sigma}\sf{h}_{\rho\sigma}+\nabla_{\alpha}\nabla_{\beta}\mathrm{tr}_{\sf{g}}(\sf{h})-\frac{1}{2}\sf{g}_{\alpha\beta}\square\mathrm{tr}_{\sf{g}}(\sf{h})+2\mathrm{Riem}_{\sf{g}}(\sf{h})+2\mathrm{Ric}_{\sf{g}}(\sf{h})\, .
	\end{align*}
	Comparing this with the linearisation of the Einstein tensor in Lemma~\ref{Lem:LinRic} we see that the this expression, together with $\square\I\sf{h}$, recovers all the second-order terms in the linearisation expression. It is a straightforward computation to see that the remaining zeroth-order terms can be written as in the statement of this proposition.
\end{proof}

So far, we have linearised the Einstein equations for general background metrics $\sf{g}$. As a next step, we explore the gauge-invariance of the obtained linearised equation. The gauge symmetry of the full, non-linear, Einstein equations~\eqref{eq:EinEq} is given by \textit{diffeomorphism invariance}, i.e.
\begin{align*}
	\Phi^{\ast}\mathrm{Ein}_{\Lambda}(\sf{g})=\mathrm{Ein}_{\Lambda}(\Phi^{\ast}\sf{g})
\end{align*}
for an arbitrary diffeomorphism $\Phi\in\mathrm{Diff}^{\infty}(\sf{M})$. In other words, the gauge group of the Einstein equations is $\mathrm{Diff}^{\infty}(\sf{M})$ and parametrised by $\sf{g}\mapsto\Phi^{\ast}\sf{g}$. The linearised gauge transformations should hence correspond to transformations of the form $\sf{h}\mapsto\sf{h}+\mathcal{L}_{\sf{X}}g$, where $\mathcal{L}_{\sf{X}}$ denotes the \textit{Lie derivative} with respect to a vector field $\sf{X}\in\mathfrak{X}(\sf{M})$. 

\begin{proposition}\label{Prop:GaugeLinGrav} \emph{(Gauge-Invariance of the Linearised Operators)}\newline
	Let $\sf{L},\sf{R}\in\mathrm{DO}^{2}(\sf{S}_{2})$ be the Einstein and Ricci operators linearised around $\sf{g}$. Then,
	\begin{align*}
		\sf{L}\circ\d_{s}=\frac{1}{2}\mathcal{L}_{\cdot^{\sharp}}\mathrm{Ein}_{\Lambda}(\sf{g})\qquad\text{and}\qquad\sf{R}\circ\d_{s}=\frac{1}{2}\mathcal{L}_{\cdot^{\sharp}}\mathrm{Ric}_{\Lambda}(\sf{g})\, .
	\end{align*}
	Hence, $\sf{L}\circ\d_{s}=0$ and $\sf{R}\circ\d_{s}=0$ if and only if the background $\sf{g}$ is a $\Lambda$-Einstein metric.
\end{proposition}

\begin{proof}
	Instead of computing the composition $\sf{L}\circ \d$ directly by a cumbersome computation, we prove the claim in a more abstract way. In fact, the statement becomes almost obvious when we spell out the definitions: choose $\omega\in\Gamma^{\infty}(\sf{E}_{1})$ and denote by $\Phi\:(-\varepsilon,\varepsilon)\times\sf{M}\to\sf{M}$ with suitable $\varepsilon>0$ the flow of the corresponding vector field $\sf{X}:=\omega^{\sharp}\in\mathfrak{X}(\sf{M})$. Then, $\hat{\sf{g}}(\lambda):=\Phi_{\lambda}\sf{g}$ defines a one-parameter family of metrics such that $\hat{\sf{g}}(\lambda)=\sf{g}+\lambda\sf{h}+\mathcal{O}(\lambda^{2})$ with linear part
	\begin{align*}
		\sf{h}:=\frac{\d}{\d\lambda}\bigg\vert_{\lambda=0}\hat{\sf{g}}(\lambda)=\frac{\d}{\d\lambda}\bigg\vert_{\lambda=0}\Phi_{\lambda}^{\ast}\sf{g}=\mathcal{L}_{\sf{X}}\sf{g}=2\d_{s}\omega\, ,
	\end{align*}	
	where we used the well-known formula $(\mathcal{L}_{\sf{X}}g)_{\alpha\beta}=2\nabla_{(\alpha}\sf{X}_{\beta)}	$ in the last step. Now, by definition of the linearised Einstein operator, $\sf{L}\sf{h}$ is the linearisation of $\mathrm{Ein}_{\Lambda}(\hat{\sf{g}}(\lambda))$, which means that
	\begin{align*}
		\sf{L}\sf{h}=\frac{\d}{\d\lambda}\bigg\vert_{\lambda=0}\mathrm{Ein}_{\Lambda}(\hat{\sf{g}}(\lambda))=\frac{\d}{\d\lambda}\bigg\vert_{\lambda=0}\mathrm{Ein}_{\Lambda}(\Phi_{\lambda}^{\ast}\sf{g})=\frac{\d}{\d\lambda}\bigg\vert_{\lambda=0}\Phi_{\lambda}^{\ast}\mathrm{Ein}_{\Lambda}(\sf{g})=\mathcal{L}_{\sf{X}}\mathrm{Ein}_{\Lambda}(\sf{g})\, ,
	\end{align*}
	where we used that $\mathrm{Ein}_{\Lambda}(\sf{g})$ is diffeomorphism invariant, i.e.~$\mathrm{Ein}_{\Lambda}(\Phi^{\ast}_{\lambda}\sf{g})=\Phi^{\ast}_{\lambda}\mathrm{Ein}_{\Lambda}(\sf{g})$. Combining this with the fact that $\sf{h}=2\d_{s}\omega$, as derived above, gives the claimed result. The claim for $\sf{R}$ follows from similar arguments and the fact that also $\mathrm{Ric}(\Phi^{\ast}_{\lambda}\sf{g})=\Phi^{\ast}_{\lambda}\mathrm{Ric}(\sf{g})$.
\end{proof}

\begin{remark}
	If the background metric is $\Lambda$-Einstein, we can prove $\sf{L}\d_{s}=0$ and $\sf{R}\d_{s}=0$ also more directly. For example,
	\begin{align*}
		2\sf{L}\d_{s}&=(-\square-\I\d_{s}\delta_{s}+2\mathrm{Riem}_{\sf{g}})\I\d_{s}\\&=\I(-\square+2\mathrm{Riem}_{\sf{g}})\d_{s}-\I\d_{s}\delta_{s}\I\d_{s}\\&=\I(-\square+2\mathrm{Riem}_{\sf{g}})\d_{s}-\I\d_{s}(-\square-\Lambda)\\&=\I(-\square+2\mathrm{Riem}_{\sf{g}})\d_{s}-\I(-\square+2\mathrm{Riem}_{\sf{g}})\d_{s}=0\, ,
	\end{align*}
	where we used the relations explained in Lemma~\ref{Lemma:PropLinGrav}. 
\end{remark}

As a next step, let us discuss in which cases the linearised operator $\sf{L}$ introduced in the Proposition~\ref{Prop:LinEinRicOp} is formally self-adjoint with respect to $(\cdot,\cdot)_{\sf{S}_{2}}$.

\begin{proposition}\label{Prop:FormAdjLinGrav} \emph{(Formally Self-Adjointness of Linear Einstein Operator)}\newline
	Let $(\sf{M},g)$ be an arbitrary background Lorentzian manifold. Then,
	\begin{align*}
		\sf{L}-\sf{L}^{\ast}=\frac{1}{4}\bigg(\sf{g}\langle\mathrm{Ein}_{\Lambda}(\sf{g}),\cdot\rangle_{\sf{S}_{2}}-2\mathrm{Ein}_{\Lambda}(\sf{g})\mathrm{tr}_{\sf{g}}\bigg)
	\end{align*}
	Furthermore, $\sf{L}=\sf{L}^{\ast}$ if and only if the background metric is $\Lambda$-Einstein.
\end{proposition}

\begin{proof}
	First of all, let us write the linearised operator $\sf{L}$ as
	\begin{align*}
		\sf{L}=&\frac{1}{2}\bigg(-\square-\I\d_{s}\delta_{s}+2\mathrm{Riem}_{\sf{g}}+2(\mathrm{Ein}_{\Lambda}(\sf{g})\times_{s}\cdot)\bigg)\I+\frac{1}{4}\sf{g}\langle\mathrm{Ein}_{\Lambda}(\sf{g}),\cdot\rangle_{\sf{S}_{2}}=\\=&\frac{1}{2}\bigg(-\square-\I\d_{s}\delta_{s}+2\mathrm{Riem}_{\sf{g}}+2\mathrm{Ric}_{\sf{g}}-\mathrm{Scal}_{\sf{g}}+2\Lambda\bigg)\I+\frac{1}{4}\sf{g}\langle\mathrm{Ein}_{\Lambda}(\sf{g}),\cdot\rangle_{\sf{S}_{2}}
	\end{align*}
	Now, a straightforward computation in local coordinates shows that
	\begin{align*}
		(\mathrm{Riem}_{\sf{g}}+\mathrm{Ric}_{\sf{g}})\circ \mathrm{I}=\mathrm{Riem}_{\sf{g}}+\frac{1}{2}\cancel{\mathrm{Ric}(\sf{g})\mathrm{tr}_{\sf{g}}}+\mathrm{Ric}_{\sf{g}}-\frac{1}{2}\cancel{\mathrm{Ric}(\sf{g})\mathrm{tr}_{\sf{g}}}=\mathrm{Riem}_{\sf{g}}+\mathrm{Ric}_{\sf{g}}
	\end{align*}
	Hence, $(\mathrm{Riem}_{\sf{g}}+\mathrm{Ric}_{\sf{g}})\circ\mathrm{I}$ is formally self-adjoint by Lemma~\ref{Lemma:PropLinGrav}(i) and the only term that is not self-adjoint is the zeroth-order operator $\sf{Z}:=\sf{g}\langle\mathrm{Ein}_{\Lambda}(\sf{g}),\cdot\rangle_{\sf{S}_{2}}$. Let $\sf{h},\sf{k}\in\Gamma^{\infty}(\sf{S}_{2})$. Then,
	\begin{align*}
		\langle \sf{Z}\sf{h},\sf{k}\rangle_{\sf{S}_{2}}=4\mathrm{Ein}_{\Lambda}(\sf{g})^{\alpha\beta}\sf{h}_{\alpha\beta}\sf{g}^{\rho\sigma}\sf{k}_{\rho\sigma}=4\mathrm{Ein}_{\Lambda}(\sf{g})^{\alpha\beta}\sf{h}_{\alpha\beta}\mathrm{tr}_{\sf{g}}(\sf{k})=2\langle \sf{h},\mathrm{Ein}_{\Lambda}(\sf{g})\mathrm{tr}_{\sf{g}}(\sf{k})\rangle_{\sf{S}_{2}}\, .
	\end{align*}
	Hence, we conclude that $\sf{Z}^{\ast}=2\mathrm{Ein}_{\Lambda}(\sf{g})\mathrm{tr}_{\sf{g}}$, which proves the claim.
	
	Now, if $\mathrm{Ein}_{\Lambda}(\sf{g})=0$, then clearly $\sf{L}=\sf{L}^{\ast}$. On the other hand, if $\sf{L}=\sf{L}^{\ast}$, we know that 
	\begin{align*}
		\sf{g}\langle\mathrm{Ein}_{\Lambda}(\sf{g}),\cdot\rangle_{\sf{S}_{2}}-2\mathrm{Ein}_{\Lambda}(\sf{g})\mathrm{tr}_{\sf{g}}=0\, .
	\end{align*}
	Applying this operator to an arbitrary $\sf{h}\in\Gamma^{\infty}(\sf{S}_{2})$ and taking the trace implies
	\begin{align*}
		0=4\langle\mathrm{Ric}(\sf{g}),\sf{h}\rangle_{\sf{S}_{2}}-2\mathrm{Scal}(\sf{g})\mathrm{tr}_{\sf{g}}(\sf{h})=\langle 4\mathrm{Ric}(\sf{g})-\mathrm{Scal}(\sf{g})\sf{g},\sf{h}\rangle_{\sf{S}_{2}}
	\end{align*}
	and by non-degeneracy of $\langle\cdot,\cdot\rangle_{\sf{S}_{2}}$, we conclude that $\mathrm{Ric}(\sf{g})=\frac{1}{4}\mathrm{Scal}(\sf{g})\sf{g}$. Now, according to \cite[Prop.~7.19]{LeeRiemann}, if $\mathrm{Ric}(\sf{g})=f\sf{g}$ for some $f\in C^{\infty}(\sf{M})$, then $f$ is necessarily constant as a consequence of the Bianchi identity. Hence, there exists a constant $\Lambda\in\bb{R}$ such that $\mathrm{Scal}(\sf{g})=4\Lambda$ and hence $\mathrm{Ric}(\sf{g})=\Lambda\sf{g}$, which shows that $\sf{g}$ is $\Lambda$-Einstein.
\end{proof}

After this general discussion of linearised gravity, let us now explain how to fit the theory in the framework of linear gauge theories discussed in Section~\ref{Sec:HackSchenkel}. Let $(\sf{M},\sf{g})$ be a four-dimensional $\Lambda$-Einstein metric. The previous discussion, in particular Proposition~\ref{Prop:FormAdjLinGrav} and Proposition~\ref{Prop:GaugeLinGrav}, suggests to consider the operators
\begin{align*}
	\sf{L}:=\frac{1}{2}\bigg(-\square+2\mathrm{Riem}_{\sf{g}}-\mathrm{I}\d_{s}\delta_{s}\bigg)\I\in\mathrm{DO}^{2}(\sf{S}_{2})\,,\qquad\sf{K}_{\sf{L}}:=\d_{s}\in\mathrm{DO}^{1}(\sf{S}_{1},\sf{S}_{2})\, .
\end{align*}
As discussed, it holds that $\sf{L}^{\ast}=\sf{L}$ and $\sf{L}\circ\sf{K}_{\sf{L}}=0$. However, the quadruple $(\sf{S}_{1},\sf{S}_{2},\sf{L},\sf{K}_{\sf{L}})$ does not directly fit into the framework of Hack-Schenkel in Definition~\ref{Def:LinGaugeTh}: first, the corresponding operator $\sf{D}_{1}:=\sf{K}_{\sf{L}}^{\ast}\sf{K}_{\sf{L}}=\delta_{s}\d_{s}=-\square+\d_{s}\delta_{s}-\Lambda$ is not Green hyperbolic due to the non-hyperbolic term $\d_{s}\delta_{s}$. Note also that this problem is still there when considering the slightly more general definition from the original article of Hack-Schenkel as recalled in Remark~\ref{Rem:HackSchenkel}. Secondly, also the operator $\sf{D}_{2}:=\sf{L}+\sf{K}_{\sf{L}}\sf{K}_{\sf{L}}^{\ast}$ is not hyperbolic, since the operator $\sf{K}_{\sf{L}}\sf{K}_{\sf{L}}^{\ast}=\d_{s}\delta_{s}$ does not cancel the non-hyperbolic term $\I\d_{s}\delta_{s}\I$ contained in $\sf{L}$. In fact, looking at the operator $\sf{L}$, a more natural gauge condition to consider is $\delta_{s}\I\sf{h}=0$ rather than $\sf{K}_{\sf{L}}^{\ast}\sf{h}=\delta_{s}\sf{h}=0$. This obstacle is well known in the physics literature and the solution is provided by performing a \textit{field redefinition}, namely the replacement $\sf{h}\mapsto\I\sf{h}$. In other words, instead of considering the linearised Einstein operator $\sf{L}$, we consider the operator
\begin{align*}
	\sf{P}:=2(\sf{L}\circ\I)=-\square+2\mathrm{Riem}_{\sf{g}}-\mathrm{I}\d_{s}\delta_{s}
\end{align*}
instead, where we have put the additional factor of $2$ for convenience. Since $\I$ is an involution, the solution space described by this operator is equivalent to the one of $\sf{L}$ and hence yields an equivalent description of the linearised gravitational field. Of course, this operator is no longer formally self-adjoint with respect to $(\cdot,\cdot)_{\sf{S}_{2}}$, but we have to consider the modified Hermitian bundle $(\sf{S}_{2,\mathrm{I}},\langle\cdot,\cdot\rangle_{\sf{S}_{2,\mathrm{I}}})$ defined by
\begin{align*}
	\sf{S}_{2,\mathrm{I}}:=\sf{S}_{2}\,,\qquad\qquad\langle\cdot,\cdot\rangle_{\sf{S}_{2,\mathrm{I}}}:=\langle\mathrm{I}\cdot,\cdot\rangle_{\sf{S}_{2}}
\end{align*}
instead. Let us denote the formal adjoint of a linear differential operator $\sf{A}$ between $\sf{S}_{1}$ and $\sf{S}_{2,\mathrm{I}}$ or from $\sf{S}_{2,\mathrm{I}}$ to $\sf{S}_{2,\mathrm{I}}$ by $\sf{A}^{\dagger}$. Clearly, it holds that
\begin{equation*}
\begin{aligned}
	\sf{A}^{\dagger}&=\I\circ\sf{A}^{\ast}\circ\I\,\qquad &\text{for}\quad\sf{A}\:\Gamma^{\infty}(\sf{S}_{2})\to\Gamma^{\infty}(\sf{S}_{2})\, ,\\
	\sf{A}^{\dagger}&=\I\circ\sf{A}^{\ast}\,\qquad &\text{for}\quad\sf{A}\:\Gamma^{\infty}(\sf{S}_{2})\to\Gamma^{\infty}(\sf{S}_{1})\, ,\\
	\sf{A}^{\dagger}&=\sf{A}^{\ast}\circ\I\,\qquad &\text{for}\quad\sf{A}\:\Gamma^{\infty}(\sf{S}_{1})\to\Gamma^{\infty}(\sf{S}_{2})\, .
	\end{aligned}
\end{equation*}
When considering the operator $\sf{P}$ instead of $\sf{L}$, we also have to modify the gauge transformations  accordingly by defining
\begin{align*}
	\sf{K}:=\I\circ\d_{s}\in\mathrm{DO}^{1}(\sf{S}_{1},\sf{S}_{2})\, .
\end{align*}
The corresponding adjoint is given by $\sf{K}^{\dagger}=\delta_{s}$ and the subsidiary gauge condition associated to this operator, 
\begin{align*}
	(\sf{K}^{\dagger}\sf{h})_{\alpha}=(\delta_{s}\sf{h})_{\alpha}=-2\nabla^{\lambda}\sf{h}_{\lambda\alpha}=0\, ,
\end{align*}
is usually called the \textit{de Donder}, \textit{harmonic}, or \textit{Bianchi gauge}. To sum up:

\begin{proposition}\label{eq:EinsteinHS} \emph{(Linearised Gravity as a Linear Gauge Theory)}\newline
	Let $(\sf{M},\sf{g})$ be a globally hyperbolic $\Lambda$-Einstein manifold. Then, $(\sf{S}_{1},\sf{S}_{2,\mathrm{I}},\sf{P},\sf{K})$ with 
	\begin{align*}
		\sf{P}:=-\square+2\mathrm{Riem}_{\sf{g}}-\mathrm{I}\d_{s}\delta_{s}\in\mathrm{DO}^{2}(\sf{S}_{2,\mathrm{I}})\,, \qquad \sf{K}:=\mathrm{I}\d_{s}\in\mathrm{DO}^{1}(\sf{S}_{1},\sf{S}_{2,\mathrm{I}})
	\end{align*}
	is a linear gauge theory in the sense of Definition~\ref{Def:LinGaugeTh}. The corresponding operators $\sf{D}_{1}:=\sf{K}^{\dagger}\sf{K}$ and $\sf{D}_{2}:=\sf{P}+\sf{K}\sf{K}^{\dagger}$ are the normally hyperbolic operators
	\begin{align*}
		\sf{D}_{1}=-\square-\Lambda\,,\qquad\sf{D}_{2}=-\square+2\mathrm{Riem}_{\sf{g}}\, .
	\end{align*}
\end{proposition}

\begin{proof}
	By definition, $\sf{P}:=2\sf{L}\mathrm{I}$ and hence, by Proposition~\ref{Prop:GaugeLinGrav}, $\sf{P}\sf{K}=2\sf{L}\mathrm{I}^{2}\d_{s}=2\sf{L}\d_{s}=0$, since the background metric is assumed to be $\Lambda$-Einstein. Furthermore, $\sf{L}$ is formally self-adjoint with respect to $(\cdot,\cdot)_{\sf{S}_{2}}$ by Proposition~\ref{Prop:FormAdjLinGrav} and hence, it is clear that $\sf{P}=2\sf{L}\I$ is formally self-adjoint with respect to $(\cdot,\cdot)_{\sf{S}_{2},\I}$, since
	\begin{align*}
		\sf{P}^{\dagger}=2(\sf{L}\mathrm{I})^{\dagger}=2\mathrm{I}^{\dagger}\sf{L}^{\dagger}=2\mathrm{I}^{2}\sf{L}^{\ast}\mathrm{I}=2\mathrm{I}\sf{L}\mathrm{I}=\sf{P}\, .
	\end{align*}
	The operator $\sf{D}_{2}=\sf{P}+\sf{K}\sf{K}^{\dagger}=-\square+2\mathrm{Riem}_{\sf{g}}$ is clearly normally hyperbolic and hence in particular Green hyperbolic. Furthermore, by Lemma~\ref{Lemma:PropLinGrav}(iiic), we obtain
	\begin{align*}
		\sf{D}_{1}=\sf{K}^{\dagger}\sf{K}=\delta_{s}\I\d_{s}=-\square-\Lambda\, ,
	\end{align*}
	which is a normally hyperbolic operator as well.
\end{proof}

The example above has been considered in a modified form in the original article of Hack-Schenkel \cite[Example~3.8]{HackSchenkel}. Linearised gravity in the sense above and in the context of the Hack-Schenkel formalism has also been considered in the works related to the construction of Hadamard states for linearised gravity by Benini-Murro-Dappiaggi \cite{BeniniMurro}, Gérard-Murro-Wrochna \cite{GerardMurroWrochna}, Gérard \cite{Gerard} and Gérard-Wrochna \cite{GerardWrochna2}.

There is yet another possibility of fitting linearised gravity in the Hack-Schenkel framework for linear gauge theories, which has been not yet discussed in this context. It is well known that the four-dimensional Einstein equations~\eqref{eq:EinEq} are equivalent to its trace-reversed form, i.e.
\begin{align*}
	\mathrm{Ein}_{\Lambda}(\sf{g})=0\qquad \Leftrightarrow\qquad\mathrm{Ric}(\sf{g})-\Lambda\sf{g}=0\, .
\end{align*}
Hence, instead of considering the linearised Einstein operator, we may consider the linearisation of $\mathrm{Ric}(\sf{g})-\Lambda\sf{g}$ instead, i.e.~the operator $\sf{R}-\Lambda$, where $\sf{R}$ is the linearised Ricci operator as introduced in Proposition~\ref{Prop:LinEinRicOp}. If the background metric is $\Lambda$-Einstein, it holds that 
\begin{align*}
	\sf{L}=\I\circ (\sf{R}-\Lambda)\,,
\end{align*}
which corresponds to the same relation on the non-linear level. In other words, the diagram
\begin{equation*}
	\begin{tikzcd}
		\mathrm{Ein}_{\Lambda}(\sf{g})=0 \arrow[d,swap,"\text{linearise}"]\arrow[r,"\I"] & \mathrm{Ric}(\sf{g})-\Lambda\sf{g}=0\arrow[d,"\text{linearise}"]\\ \sf{L}\sf{h}=0 \arrow[r,"\I"] & (\sf{R}-\Lambda)\sf{h}=0
	\end{tikzcd}
\end{equation*}
commutes whenever the background solves the nonlinear Einstein equations.

Following up on this observation, we obtain the following result: 

\begin{proposition}\label{Prop:EinsteinHS2} \emph{(Linearised Gravity as a Linear Gauge Theory II)}\newline
	Let $(\sf{M},\sf{g})$ be a globally hyperbolic $\Lambda$-Einstein manifold. Then, $(\sf{S}_{1},\sf{S}_{2,\I},\sf{P}_{\sf{R}},\sf{K}_{\sf{R}})$ with 
	\begin{align*}
		\sf{P}_{\sf{R}}:=2(\sf{R}-\Lambda)=-\square+2\mathrm{Riem}_{\sf{g}}+\d_{s}\delta_{s}\I\,, \qquad \sf{K}_{\sf{R}}:=\d_{s}
	\end{align*}
	is a linear gauge theory in the sense of Definition~\ref{Def:LinGaugeTh}. The corresponding operators $\sf{D}_{1}:=\sf{K}_{\sf{R}}^{\dagger}\sf{K}_{\sf{R}}$ and $\sf{D}_{2}:=\sf{P}_{\sf{R}}+\sf{K}_{\sf{R}}\sf{K}_{\sf{R}}^{\dagger}$ are as before given by
	\begin{align*}
		\sf{D}_{1}=-\square-\Lambda,\qquad\sf{D}_{2}=-\square+2\mathrm{Riem}_{\sf{g}}\, .
	\end{align*}
\end{proposition}

\begin{proof}
	The proof is similar to the one of Proposition~\ref{eq:EinsteinHS}. It is clear that $\sf{R}$ is formally self-adjoint with respect to $(\cdot,\cdot)_{\sf{S}_{2,\I}}$. Gauge-invariance, i.e.~$\sf{P}_{\sf{R}}\circ\sf{K}_{\sf{R}}=0$, follows from Prop.~\ref{Prop:GaugeLinGrav}. The remaining claims follow from $\sf{K}_{\sf{R}}^{\dagger}=\I\circ\d_{s}^{\ast}=\I\circ\delta_{s}$.
\end{proof}

The main difference of this formulation compared to the one of Proposition~\ref{eq:EinsteinHS} is that the gauge transformations are parametrised by the operator $\d_{s}$ rather than $\I\d_{s}$. On the other hand, the harmonic gauge condition gets an extra trace-reversal and reads
\begin{align*}
	(\sf{K}_{\sf{R}}^{\ast}\sf{h})_{\alpha}=-2\bigg(\nabla^{\lambda}\sf{h}_{\lambda\alpha}-\frac{1}{2}\nabla_{\alpha}\mathrm{tr}_{\sf{g}}(\sf{h})\bigg)=0\, .
\end{align*}

\begin{remark} (The TT-Gauge)\newline
	Many works on linearised gravity employ the so-called TT-gauge (\textit{transverse-taceless gauge}), in which one combines the de Donder gauge condition with the additional condition that the trace of the perturbation vanishes, i.e.
	\begin{align*}
		\sf{T}\sf{T}\text{-gauge}:\qquad 
			\begin{cases}
				\delta_{s}\sf{h}&=0\\
			 	\mathrm{tr}_{\sf{g}}(\sf{h})&=0
			 \end{cases}\, .
	\end{align*}	 
	The main advantage of this gauge choice is that we are left with exactly two degrees of freedom, corresponding to the two polarisations of gravitational waves. It was first observed in Fewster-Hunt \cite[Sec.~2.2, Appendix.~A.2]{FewsterHunt} that the TT-gauge cannot always be achieved and is subject to topological obstructions on the background manifold. One has to distinguish two cases:
	\begin{itemize}
		\item[$\bullet$]If $\Lambda\neq 0$, the TT-gauge can always be achieved on-shell: let $\sf{h}\in\mathrm{ker}(\sf{P}\vert_{\Gamma^{\infty}_{\mathrm{sc}}})$ and suppose that we have already achieved the de Donder gauge, i.e.~$\delta_{s}\sf{h}=0$. We want to find another gauge transformation $\sf{h}\mapsto\sf{h}^{\prime}:=\sf{h}+\sf{K}\omega$ with $\omega\in\Gamma^{\infty}_{\mathrm{sc}}(\sf{S}_{2})$ preserving the de Donder gauge condition, i.e.~$\delta_{s}\sf{h}^{\prime}=0$, and such that $\mathrm{tr}_{\sf{g}}(\sf{h}^{\prime})=0$. This leads to the system
		\begin{align}\label{eq:System}
			\begin{cases}
				\sf{D}_{1}\omega&=0\\
				\delta_{s}\omega&=-\mathrm{tr}_{\sf{g}}(\sf{h})
			\end{cases}
		\end{align}
	where we used that $\mathrm{tr}_{\sf{g}}\circ\sf{K}=\delta_{s}$. Now, we claim that $\omega:=-\frac{1}{2\Lambda}\d_{s}\mathrm{tr}_{\sf{g}}(\sf{h})$ is the required solution. For this, we first observe that $\sf{D}_{1}\circ\d_{s}=\d_{s}\circ\sf{D}_{0}$ and $\mathrm{tr}_{\sf{g}}\circ \sf{D}_{2}=\sf{D}_{0}\circ\mathrm{tr}_{\sf{g}}$, where $\sf{D}_{0}$ is the normally hyperbolic operator $\sf{D}_{0}:=-\square-2\Lambda$, which follows from a straightforward computation similar to the one in Lemma~\ref{Lemma:PropLinGrav}(iiia). Then,
	\begin{align*}
		\begin{cases}
		\sf{D}_{1}\omega&=-\frac{1}{2\Lambda}\sf{D}_{1}\d_{s}\mathrm{tr}_{\sf{g}}(\sf{h})=-\frac{1}{2\Lambda}\d_{s}\sf{D}_{0}\mathrm{tr}_{\sf{g}}(\sf{h})=-\frac{1}{2\Lambda}\d_{s}\mathrm{tr}_{\sf{g}}(\sf{D}_{2}\sf{h})=0\\
		\delta_{s}\omega &=-\frac{1}{2\Lambda}\delta_{s}\d_{s}\mathrm{tr}_{\sf{g}}(\sf{h})=-\frac{1}{2\Lambda}(\sf{D}_{0}+2\Lambda)\mathrm{tr}_{\sf{g}}(\sf{h})=-\frac{1}{2\Lambda}\mathrm{tr}_{\sf{g}}(\sf{D}_{2}\sf{h})-\mathrm{tr}_{\sf{g}}(\sf{h})=-\mathrm{tr}_{\sf{g}}(\sf{h})
		\end{cases}
	\end{align*}		
	In particular, we obtain the following isomorphisms of phase spaces in this case
	\begin{align*}
		\frac{\mathrm{ker}(\sf{P}\vert_{\Gamma^{\infty}_{\mathrm{sc}}})}{\mathrm{ran}(\sf{K}\vert_{\Gamma^{\infty}_{\mathrm{sc}}})}\cong \frac{\mathrm{ker}(\sf{D}_{2}\vert_{\Gamma^{\infty}_{\mathrm{sc}}})\cap \mathrm{ker}(\sf{K}^{\ast}\vert_{\Gamma^{\infty}_{\mathrm{sc}}})}{\sf{K}(\mathrm{ker}(\sf{D}_{1}\vert_{\Gamma^{\infty}_{\mathrm{sc}}}))}\cong \frac{\mathrm{ker}(\sf{D}_{2}\vert_{\Gamma^{\infty}_{\mathrm{sc}}})\cap \mathrm{ker}(\sf{K}^{\ast}\vert_{\Gamma^{\infty}_{\mathrm{sc}}})\cap \mathrm{ker}(\mathrm{tr}_{\sf{g}}\vert_{\Gamma^{\infty}_{\mathrm{sc}}})}{\sf{K}(\mathrm{ker}(\sf{D}_{1}\vert_{\Gamma^{\infty}_{\mathrm{sc}}})\cap\mathrm{ker}(\delta_{s}\vert_{\Gamma^{\infty}_{\mathrm{sc}}}))}\, .
	\end{align*}
	Note that there is in general still a remaining gauge-freedom in the space on the right.
	\item[$\bullet$]If $\Lambda=0$, the previous Ansatz does not work. As before, let $\sf{h}\in\mathrm{ker}(\sf{P}\vert_{\Gamma^{\infty}_{\mathrm{sc}}})$ be such that $\delta_{s}\sf{h}=0$ and suppose that there exists a $\omega\in\Gamma^{\infty}_{\mathrm{sc}}(\sf{S}_{1})$ solving~\eqref{eq:System}. Let us denote the causal propagators of $\sf{D}_{i}$ by $\sf{G}_{i}$ for $i=0,1,2$. Then, since $\sf{D}_{2}\sf{h}=0$, there exists a $\sf{k}\in\Gamma^{\infty}_{\mathrm{c}}(\sf{S}_{2})$ such that $\sf{h}=\sf{G}_{2}\sf{k}$. Now, since $\mathrm{tr}_{\sf{g}}\circ\sf{D}_{2}=\sf{D}_{0}\circ\mathrm{tr}_{\sf{g}}$, it follows that $\mathrm{tr}_{\sf{g}}(\sf{h})=\sf{G}_{0}\mathrm{tr}_{\sf{g}}(\sf{k})$. On the other hand, since $\sf{D}_{1}\omega=0$, we can write $\omega=\sf{G}_{1}\eta$ for some $\eta\in\Gamma^{\infty}_{\mathrm{c}}(\sf{S}_{1})$ and hence
	\begin{align*}
		\delta_{s}\omega=\delta_{s}\sf{G}_{1}\eta=\sf{G}_{0}\delta_{s}\eta\stackrel{!}{=}-\mathrm{tr}_{\sf{g}}(\sf{h})=-\sf{G}_{0}\mathrm{tr}_{\sf{g}}(\sf{k})\, .
	\end{align*}
	In other words, $\sf{G}_{0}(\mathrm{tr}_{\sf{g}}(\sf{k})+\delta_{s}\eta)=0$, which means that there is a $f\in C^{\infty}_{\mathrm{c}}(\sf{M})$ such that $\mathrm{tr}_{\sf{g}}(\sf{k})+\delta_{s}\eta=\sf{D}_{0}f$. Now, since we assume $\Lambda=0$, we have that $\sf{D}_{0}=-\square=\square_{\mathrm{dRH}}$, where $\square_{\mathrm{dRH}}=\delta\d$ is the de Rham-Hodge d'Alembertian on zero-forms. We conclude that $\mathrm{tr}_{\sf{g}}(\sf{k})=\delta\lambda$ for $\lambda:=\d f-\eta$, where we used that $\delta_{s}=\delta$ on $1$-forms. We have shown that if a solution $\omega\in\Gamma^{\infty}_{\mathrm{sc}}(\sf{S}_{1})$ to~\eqref{eq:System} exists, then necessarily $\mathrm{tr}_{\sf{g}}(\sf{k})\in\delta\Omega_{\mathrm{c}}^{1}(\sf{M})$. In fact, also the other direction is true: if we can write $\mathrm{tr}_{\sf{g}}(\sf{k})=\delta\lambda$ for some $\lambda\in\Omega^{1}_{\mathrm{c}}(\sf{M})$, then clearly $\omega=-\sf{G}_{1}\lambda$ is a solution to~\eqref{eq:System}. To sum up, in the case $\Lambda=0$:
	\begin{align*}
		\exists\omega\in\Gamma^{\infty}_{\mathrm{sc}}(\sf{S}_{1})\text{ solving }~\eqref{eq:System}\quad\Leftrightarrow\quad\sf{k}\text{ with }\sf{h}=\sf{G}_{2}\sf{k}\text{ can be chosen s.t. }\mathrm{tr}_{\sf{g}}(\sf{k})\in\delta\Omega^{1}_{\mathrm{c}}(\sf{M})\, .
	\end{align*}
	 Now, via the Hodge-star operator, this is equivalent to require that $\ast\mathrm{tr}_{\sf{g}}(\sf{k})\in\d\Omega^{3}_{\mathrm{c}}(\sf{M})$. This is in particular true generically when the de Rham homology group $\sf{H}^{4}_{\mathrm{c}}(\sf{M})$ is trivial. However, this is never true, since $\sf{H}^{4}_{\mathrm{c}}(\sf{M})\cong\sf{H}^{3}_{\mathrm{c}}(\Sigma)\cong\mathbb{R}$. Hence, the TT-gauge can in general not be achieved for \textit{all} $\sf{h}\in\Gamma^{\infty}_{\mathrm{sc}}(\sf{S}_{2})$, but only for a subspace of it. If we relax the condition on the supports and we consider fields $\sf{h}\in\Gamma^{\infty}(\sf{S}_{2})$ and $\omega\in\Gamma^{\infty}(\sf{S}_{2})$, then repeating the same arguments yields the requirement $\sf{H}_{\mathrm{tc}}^{4}(\sf{M})\stackrel{!}{=}0$. It holds that $\sf{H}_{\mathrm{tc}}^{4}(\sf{M})\cong\sf{H}^{3}(\Sigma)$, see~\cite{Benini,Khavkine}, which implies that this condition is always fulfilled if $\Sigma$ is non-compact. To sum up: if $\Lambda=0$, then the TT-gauge can in general only be achieved for a subspace of $\sf{h}\in\mathrm{ker}(\sf{P}\vert_{\Gamma^{\infty}_{\mathrm{sc}}})$. In the non-compact case, when instead working in the larger spaces $\Gamma^{\infty}(\sf{S}_{i})$, then the TT-gauge can again always be achieved. This is in particular true for Minkowski spacetime.
	\end{itemize}
\end{remark}

\subsection{Linearised Einstein-Klein Gordon System}
As a final example of a linear gauge theory, we consider linearised gravity in the presence of a non-trivial energy-momentum tensor. The simplest case arises from a (real) \textit{scalar field theory}, whose energy-momentum tensor can be written as  
\begin{align}\label{eq:EMT}
    \sf{T}_{\sf{V}}(g,\phi) := \d_{s}\phi \otimes \d_{s}\phi - \bigg(\frac{1}{2} \Vert\d_{s}\phi\Vert^{2}_{\sf{S}_{1}} + \sf{V}(\phi) \bigg) \sf{g} \in \Gamma^{\infty}(\sf{S}_{2})  
\end{align}  
for a given metric $\sf{g}$, a scalar field $\phi \in C^{\infty}(\sf{M})$, and an arbitrary potential $\sf{V} \in C^{\infty}(\bb{R})$, where $\langle\cdot,\cdot\rangle_{\sf{S}_{k}}$ are the bundle metrics on $\sf{S}_{k}:=\sf{T}^{\ast}\sf{M}^{\otimes_{s}k}$ as defined in~\eqref{eq:BundleSymTens}. In local coordinates, the components of this tensor, defined as $\sf{T}_{\mu\nu} := \sf{T}_{\sf{V}}(g,\phi)(\partial_{\mu},\partial_{\nu})$, take the form  
\begin{align*}  
    \sf{T}_{\mu\nu} = \nabla_{\mu} \phi \nabla_{\nu} \phi - \bigg(\frac{1}{2} \nabla^{\lambda} \phi \nabla_{\lambda} \phi + V(\phi) \bigg) \sf{g}_{\mu\nu}.  
\end{align*}  
To ensure consistency with the Einstein equations~\eqref{eq:EinEq}, we need to impose $\delta_{s} \sf{T}_{\sf{V}}(g,\phi) = 0$, which in local coordinates translates to $\nabla^{\lambda} \sf{T}_{\lambda\alpha}=0$. This condition corresponds to the energy-momentum conservation of the system and leads to the following equation for $\phi$:  
\begin{align*}  
    \delta_{s} \sf{T}_{\sf{V}}(g,\phi) = 2[-\square \phi + \sf{V}^{\prime}(\phi)]\d_{s}\phi \stackrel{!}{=} 0.  
\end{align*}  
In other words, energy-momentum conservation enforces the scalar field $\phi$ to satisfy the \textit{Klein-Gordon equation} $-\square\phi+\sf{V}^{\prime}(\phi)=0$. Thus, in the following discussion, we consider the coupled system of nonlinear equations\footnote{Note that the Einstein equations $\mathrm{Ein}_{\Lambda}(\sf{g})=\sf{T}_{\sf{V}}(g,\phi)$ are equivalent to $\mathrm{Ein}_{\Lambda=0}=\sf{T}_{\sf{V}^{\prime}:=\sf{V}-\Lambda}(g,\phi)$. So, without loss of generality, we could shift the cosmological constant into the potential and set $\Lambda=0$.}
\begin{align}\label{eq:EinKG}  
    0=\mathrm{EKG}_{\Lambda,\sf{V}}(\sf{g},\phi):=\begin{cases}  
        \mathrm{Ein}_{\Lambda}(\sf{g})-\sf{T}_{\sf{V}}(g,\phi)\\  
        -\square\phi + \sf{V}^{\prime}(\phi)
    \end{cases}  
\end{align}  
known as the \textit{Einstein-Klein-Gordon equations}. For later convenience, we record that the first line, Einstein's equation $\mathrm{Ein}_{\Lambda}(\sf{g})=\sf{T}_{\sf{V}}(g,\phi)$, is equivalent to the trace-reversed equation
\begin{align}\label{eq:EinTR}
	\mathrm{Ric}(\sf{g})=\d_{s}\phi\otimes\d_{s}\phi+(\Lambda+\sf{V}(\phi))\sf{g}\, .
\end{align}
Well-posedness of Eq.~\eqref{eq:EinKG} can be studied similarly as in the vacuum case, see for instance the monograph \cite{Ringstrom}. The corresponding \textit{linearised} system, viewed as a linear gauge theory, has been studied by Hack in \cite{Hack} (see also \cite[Sec.~3.3]{HackBook}). Our exposition differs from that of \cite{Hack} in two respects. First, while \cite{Hack} employs a more general framework for linear gauge theories, as briefly recalled in Remark~\ref{Rem:HackSchenkel}, we show that the system under consideration also fits naturally into the refined framework of Definition~\ref{Def:LinGaugeTh}. Second, our presentation is formulated in a global and coordinate-free language, in contrast to the local, coordinate-based approach in \cite{Hack}, which allows us to derive more geometrically transparent expressions.
 
The linearised Einstein-Klein-Gordon equations and their quantisation play a crucial role in cosmology, especially in the study of cosmological perturbations. These perturbations are fundamental to modern theoretical cosmology, providing the framework for understanding the origin and evolution of structure in the universe. One of the central questions in this field is how primordial inhomogeneities, which later evolve into the anisotropies observed in the \textit{cosmic microwave background}, eventually lead to the large-scale distribution of galaxies we observe today. The inflationary paradigm offers an explanation for these perturbations. According to this theory, quantum fluctuations in a scalar field, the \textit{inflaton}, serve as the seeds for cosmological structure. Initially, these fluctuations exist on Planckian scales, but during the rapid inflationary expansion of space postulated in the very early universe, they are ``stretched'' to macroscopic distances. This process hence creates a direct link between high-energy microphysics and the large-scale structure of the universe, making inflation a key testbed for fundamental physics. We refer to \cite{Mukhanov1,Mukhanov2} for more details on the physics side; see also \cite{AQFTCos} for a discussion from the angle of algebraic quantum field theory.

We consider all the operators on the bundles $\sf{S}_{k}:=\sf{T}^{\ast}\sf{M}^{\otimes_{k}s}$ as defined in the previous section. Now, consider a one-parameter family of pairs $(\hat{\sf{g}}(\lambda),\hat{\phi}(\lambda))$ consisting of a Lorentzian metric $\sf{g}(\lambda)$ and scalar field $\phi(\lambda)$ depending smoothly on the parameter $\lambda\in\bb{R}$. We then formally expand
\begin{align*}
	(\hat{\sf{g}}(\lambda),\hat{\phi}(\lambda))=(\sf{g},\phi)+\lambda (\sf{h},\varphi)+\mathcal{O}(\lambda^{2}),\qquad (\sf{h},\varphi):=\frac{\d}{\d\lambda}\bigg\vert_{\lambda=0}(\hat{\sf{g}}(\lambda),\hat{\phi}(\lambda))\, ,
\end{align*}
where $\sf{g}:=\hat{\sf{g}}(\lambda)$, $\phi:=\hat{\phi}(0)$ are the background objects and $\sf{h}\in\Gamma^{\infty}(\sf{S}_{2})$, $\varphi\in\Gamma^{\infty}(\sf{S}_{0})\cong C^{\infty}(\sf{M})$ the perturbations. We note that the linearised fields live in the Whitney sum bundle 
\begin{align*}
	\mathcal{S}_{2}:=\sf{S}_{2}\oplus\sf{S}_{0},\qquad \Gamma^{\infty}(\mathcal{S}_{2})\cong \Gamma^{\infty}(\sf{S}_{2})\oplus C^{\infty}(\sf{M})\, ,
\end{align*}
which we shall below equip with a suitable bundle metric. As in the previous section, all the differential operators appearing in the following are the one associated to the background metric $\sf{g}$, if not explicitly stated otherwise. 

Now, before deriving the linearised equations explicitly, let us work on a more abstract setting. We define the \textit{linearised Einstein-Klein-Gordon operator} to be the linear differential operator $\mathscr{L}\in\mathrm{DO}^{2}(\mathcal{S}_{2})$ defined by
\begin{align}\label{eq:LinEGG}
	\mathscr{L}(\sf{h},\varphi):=\frac{\d}{\d\lambda}\bigg\vert_{\lambda=0}\mathrm{EKG}_{\Lambda,\sf{V}}(\hat{\sf{g}}(\lambda),\hat{\phi}(\lambda))\,,
\end{align}
i.e.~$\mathrm{EKG}_{\Lambda,\sf{V}}(\hat{\sf{g}},\hat{\phi})=\mathrm{EKG}_{\Lambda,\sf{V}}(\sf{g},\phi)+\lambda\mathscr{L}(\sf{h},\varphi)+\mathcal{O}(\lambda^{2})$. Now, let us discuss the gauge-invariance in this context. First, we recall that the non-linear equations~\eqref{eq:EinKG} are invariant under the action of diffeomorphisms, i.e.~it holds that
\begin{align*}
	\mathrm{EKG}_{\Lambda,\sf{V}}(\sf{g},\phi)=\mathrm{EKG}_{\Lambda,\sf{V}}(\Phi^{\ast}\sf{g},\Phi^{\ast}\phi)
\end{align*}
for all $\Phi\in\mathrm{Diff}^{\infty}(\sf{M},\sf{M})$, where $\Phi^{\ast}\phi:=\phi\circ\Phi$. Generalising Proposition~\ref{Prop:GaugeLinGrav} for the case of vacuum gravity, we obtain the following:

\begin{proposition} \emph{(Gauge-Invariance of Linearised Einstein-Klein-Gordon Operator)}\newline
Consider the linearised operator $\mathscr{L}\in\mathrm{DO}^{2}(\mathcal{S}_{2})$ as defined in~\eqref{eq:LinEGG}. Then,
\begin{align*}
	\mathscr{L}\mathcal{L}_{X}(\sf{g},\phi)=\mathcal{L}_{X}\mathrm{EKG}_{\Lambda,\sf{V}}(\sf{g},\phi)
\end{align*}
for all $X\in\mathfrak{X}(\sf{M})$, where $\mathcal{L}_{X}$ acts on $(\sf{h},\varphi)\in\Gamma^{\infty}(\mathcal{S}_{2})$ component-wise. Hence, $\mathscr{L}\circ\mathcal{L}_{X}=0$ if and only if the background solves the non-linear equations, i.e.~$\mathrm{EKG}_{\Lambda,\sf{V}}(\sf{g},\phi)=0$.
\end{proposition}

\begin{proof}
	The proof is essentially the same as the one for Proposition~\ref{Prop:GaugeLinGrav}.
\end{proof}

Now, recall that the Lie derivative $\mathcal{L}_{X}$ for some vector field $X\in\mathfrak{X}(\sf{M})$ acts on a smooth function $f\in C^{\infty}(\sf{M})$ as 
\begin{align*}
	\mathcal{L}_{X}f=X^{\lambda}\nabla_{\lambda}f=\langle X^{\flat},\d_{s}f\rangle_{\sf{S}_{1}}\,,
\end{align*}
while $\mathcal{L}_{X}\sf{g}=2\d_{s}\sf{X}^{\flat}$. In particular, considering a pair of background fields $(\sf{g},\phi)$ satisfying the Einstein-Klein-Gordon equations~\eqref{eq:EinKG}, gauge transformation parameters are sections of $\sf{S}_{1}$ and the corresponding transformations are parametrised by the operator
\begin{align}\label{eq:GaugeTrafoEKG}
	\begin{pmatrix}
		\d_{s}\\
		\frac{1}{2}\langle\d_{s}\phi,\cdot\rangle_{\sf{S}_{1}}
	\end{pmatrix}\:\Gamma^{\infty}(\sf{S}_{1})\to\Gamma^{\infty}(\mathcal{S}_{2})\, .
\end{align}

Assuming that the background fields satisfy the non-linear Einstein-Klein-Gordon equations, the linearised operator $\mathscr{L}$ is explicitly given as follows:

\begin{proposition}\label{Prop:GaugeInvEKG} \emph{(Linearised Einstein-Klein-Gordon Operator)}\newline
	Let $\sf{M}$ be a four-dimensional manifold admitting Lorentzian metrics and $(\sf{g},\phi)$ be a solution to $\mathrm{EKG}_{\Lambda,\sf{V}}(\sf{g},\phi)=0$. Then, the linearised operator $\mathscr{L}\in\mathrm{DO}^{2}(\mathcal{S}_{2})$ takes the form
	\begin{align*}
		\mathscr{L}=
		\begin{pmatrix}
			\mathscr{L}_{11} &\mathscr{L}_{12}\\
			\mathscr{L}_{21} &\mathscr{L}_{22}
		\end{pmatrix},\qquad
		\begin{cases}
			\mathscr{L}_{11}:=\frac{1}{2}\big(-\square-\I\d_{s}\delta_{s}+2\mathrm{Riem}_{\sf{g}}+2[(\d_{s}\phi\otimes\d_{s}\phi)\times_{s}\cdot]\big)\I \\
			\mathscr{L}_{12}:=-2\I(\d_{s}\phi\otimes_{s}\d_{s}\cdot)+\sf{g}\sf{V}^{\prime}(\phi)\\
			\mathscr{L}_{21}:=\frac{1}{2}\big[\langle\d_{s}^{2}\phi,\cdot\rangle_{\sf{S}_{2}}-\langle\d_{s}\phi,\delta_{s}\I\cdot\rangle_{\sf{S}_{1}}\big]\\
			\mathscr{L}_{22}:=-\square+\sf{V}^{\prime\prime}(\phi)
		\end{cases}
	\end{align*}
\end{proposition}

\begin{proof}
	We start by linearising the Klein-Gordon equation $-\square\phi+\sf{V}^{\prime}(\phi)=0$. Consider a one-parameter family $(\hat{\sf{g}}(\lambda),\hat{\phi}(\lambda))$ consisting of a $\lambda$-depended Lorentzian metric $\hat{\sf{g}}(\lambda)$ and scalar field $\hat{\phi}(\lambda)$ with $(\sf{g},\phi)=(\hat{\sf{g}}(0),\hat{\phi}(0))$ and $(\sf{h},\varphi):=\frac{\d}{\d\lambda}\big\vert_{\lambda=0}(\hat{\sf{g}}(\lambda),\hat{\phi}(\lambda))$. Then, denoting the Levi-Civita connection of $\hat{\sf{g}}$ by $\hat{\nabla}$, a straightforward computation yields
	\begin{align*}
	\frac{\mathrm{d}}{\mathrm{d}\lambda}\bigg\vert_{\lambda=0}&(-\square\hat{\phi}+\sf{V}^{\prime}(\hat{\phi}))(\lambda)=\frac{\mathrm{d}}{\mathrm{d}\lambda}\bigg\vert_{\lambda=0}(-\hat{\sf{g}}^{\alpha\beta}\hat{\nabla}_{\alpha}\partial_{\beta}\hat{\phi}+\sf{V}^{\prime}(\hat{\phi}))(\lambda)=\\&=\sf{h}^{\alpha\beta}\nabla_{\alpha}\nabla_{\beta}\phi-\square\varphi+\frac{1}{2}(2\nabla^{\rho}\sf{h}_{\rho}^{\nu}-\nabla^{\nu}\mathrm{tr}_{\sf{g}}(\sf{h}))\nabla_{\nu}\phi+\sf{V}^{\prime\prime}(\phi)\varphi=\\&=[-\square+\sf{V}^{\prime\prime}(\phi)]\varphi+\frac{1}{2}\langle \d_{s}^{2}\phi,\sf{h}\rangle_{\sf{S}_{2}}-\frac{1}{2}\langle\d_{s}\phi,\delta_{s}\I \sf{h}\rangle_{\sf{S}_{1}}\, ,
\end{align*}
where we used the linearisation of the Christoffel symbols as derived in~\eqref{eq:ChriPert} and we note that $(\d_{s}^{2}\phi)_{\alpha\beta}=\nabla_{\alpha}\nabla_{\beta}\phi$. In particular, this allows us to determine the operators $\mathscr{L}_{21}\in\mathrm{DO}^{2}(\sf{S}_{2},\sf{S}_{0})$ and $\mathscr{L}_{22}\in\mathrm{DO}^{2}(\sf{S}_{0})$ and we obtain
\begin{align*}
	\mathscr{L}_{21}:=\frac{1}{2}\big[\langle\d_{s}^{2}\phi,\cdot\rangle_{\sf{S}_{2}}-\langle\d_{s}\phi,\delta_{s}\I\cdot\rangle_{\sf{S}_{1}}\big],\qquad\text{and}\qquad \mathscr{L}_{22}:=-\square+\sf{V}^{\prime\prime}(\phi)\, .
\end{align*}
Now, the linearised Einstein operator has already been derived in Proposition~\ref{Prop:LinEinRicOp} for the non-vacuum case, which leaves us with deriving the linearisation of the energy-momentum tensor. A straightforward computation in coordinates shows that
\begin{align*}
	\frac{\mathrm{d}}{\mathrm{d}\lambda}\bigg\vert_{\lambda=0}\sf{T}_{\sf{V}}(\hat{\sf{g}},\hat{\phi})&=\frac{\mathrm{d}}{\mathrm{d}\lambda}\bigg\vert_{\lambda=0}\bigg(\d_{s}\hat{\phi} \otimes \d_{s}\hat{\phi} - \bigg(\frac{1}{2} \Vert\d_{s}\hat{\phi}\Vert^{2}_{\sf{S}_{1}} + \sf{V}(\hat{\phi}) \bigg) \hat{\sf{g}}\bigg)\\&=2\I(\d_{s}\phi\otimes_{s}\d_{s}\varphi)-\bigg[\frac{1}{2}\Vert\d_{s}\phi\Vert^{2}_{\sf{S}_{1}}+\sf{V}(\phi)\bigg]\sf{h}+\bigg[\frac{1}{4}\langle\d_{s}\phi\otimes\d_{s}\phi,h\rangle_{\sf{S}_{2}}-\sf{V}^{\prime}(\phi)\varphi\bigg]\sf{g}\,,
\end{align*}
where $(\d_{s}\phi\otimes_{s}\d_{s}\varphi)_{\alpha\beta}=\nabla_{(\alpha}\phi\nabla_{\beta)}\varphi$. Using the non-linear Einstein equations of the background, i.e.~$\mathrm{Ein}_{\Lambda}(\sf{g})=\sf{T}_{\sf{V}}(\sf{g},\phi)$, the linearisation of the energy-momentum just derived and the linearisation of the Einstein tensor as written in Proposition~\ref{Prop:LinEinRicOp}, we obtain
\begin{align*}
	\frac{\mathrm{d}}{\mathrm{d}\lambda}\bigg\vert_{\lambda=0}&\bigg(\mathrm{Ein}_{\Lambda}(\hat{\sf{g}})-\sf{T}_{\sf{V}}(\hat{\sf{g}},\hat{\phi})\bigg)=\\=&\frac{1}{2}\bigg(-\square-\I\d_{s}\delta_{s}+2\mathrm{Riem}_{\sf{g}}+2(\sf{T}_{\sf{V}}(\sf{g},\phi)\times_{s}\cdot)\bigg)\I\sf{h}+\frac{1}{4}\sf{g}\langle\sf{T}_{\sf{V}}(\sf{g},\phi),\sf{h}\rangle_{\sf{S}_{2}}\\&-2\I(\d_{s}\phi\otimes_{s}\d_{s}\varphi)+\bigg[\frac{1}{2}\Vert\d_{s}\phi\Vert^{2}_{\sf{S}_{1}}+\sf{V}(\phi)\bigg]\sf{h}-\bigg[\frac{1}{4}\langle\d_{s}\phi\otimes\d_{s}\phi,h\rangle_{\sf{S}_{2}}-\sf{V}^{\prime}(\phi)\varphi\bigg]\sf{g}=\\=&\frac{1}{2}\bigg(-\square-\I\d_{s}\delta_{s}+2\mathrm{Riem}_{\sf{g}}+2[(\d_{s}\phi\otimes\d_{s}\phi)\times_{s}\cdot]\bigg)\I\sf{h}-2\I(\d_{s}\phi\otimes_{s}\d_{s}\varphi)+\sf{V}^{\prime}(\phi)\varphi\sf{g}
\end{align*}
The last line of this computation allows us to determine the operators $\mathscr{L}_{11}$ and $\mathscr{L}_{12}$ and we find
\begin{align*}
	\mathscr{L}_{11}:=\frac{1}{2}\bigg(-\square-\I\d_{s}\delta_{s}+2\mathrm{Riem}_{\sf{g}}+2[(\d_{s}\phi\otimes\d_{s}\phi)\times_{s}\cdot]\bigg)\I,\quad\mathscr{L}_{12}:=-2\I(\d_{s}\phi\otimes_{s}\d_{s}\cdot)+\sf{V}^{\prime}(\phi)\sf{g}\, ,
\end{align*}
	as claimed in the proposition.
\end{proof}

\begin{lemma}\label{Lemma:FSAEKG} \emph{(Formally Self-Adjointness)}\newline
	Let $\sf{M}$ be a four-dimensional manifold admitting Lorentzian metrics and $(\sf{g},\phi)$ be a solution to $\mathrm{EKG}_{\Lambda,\sf{V}}(\sf{g},\phi)=0$. Then, the linearised operator $\mathscr{L}\in\mathrm{DO}^{2}(\mathcal{S}_{2})$ is formally self-adjoint with respect to the bundle metric
	\begin{align*}
		\langle\cdot,\cdot\rangle_{\mathcal{S}_{2}}:=\langle\cdot,\cdot\rangle_{\sf{S}_{2}}\oplus 4\langle\cdot,\cdot\rangle_{\sf{S}_{0}}\, .
	\end{align*}
\end{lemma}

\begin{proof}
	Within this proof, let us denote the formal adjoint with respect to $(\cdot,\cdot)_{\sf{S}_{k}}$ by $\ast$. Now, clearly, it holds that $\mathcal{L}_{22}^{\ast}=\mathcal{L}_{22}$. For formally self-adjointness of $\mathcal{L}_{11}$, the only terms that are not formally self-adjoint individually are the operators $\mathrm{Riem}_{\sf{g}}\circ\I$ and $(\d_{s}\phi\otimes\d_{s}\phi)\times_{s}\I(\cdot)$. A straightforward computation in coordinated shows that
	\begin{align*}
		&\text{(i)}\quad\mathrm{Riem}_{\sf{g}}\circ\I-\I\circ\mathrm{Riem}_{\sf{g}}=\frac{1}{2}\bigg(\mathrm{Ric}(\sf{g})\mathrm{tr}_{\sf{g}}-\frac{1}{2}\langle\mathrm{Ric}(\sf{g}),\cdot\rangle_{\sf{S}_{2}}\sf{g}\bigg)\\
		&\text{(ii)}\quad (\d_{s}\phi\otimes\d_{s}\phi)\times_{s}\I(\cdot)-\I(\d_{s}\phi\otimes\d_{s}\phi)\times_{s}\cdot)=\frac{1}{2}\bigg(\frac{1}{2}\langle\d_{s}\phi\otimes\d_{s}\phi,\sf{h}\rangle_{\sf{S}_{2}}\sf{g}-(\d_{s}\phi\otimes\d_{s}\phi)\mathrm{tr}_{\sf{g}}\bigg)\,,
	\end{align*}
	where the first line generalises Lemma~\ref{Lemma:PropLinGrav}(iii)(d) to the case of non-Einstein backgrounds. Now, let $\sf{h},\sf{k}\in\Gamma^{\infty}(\sf{S}_{2})$. Then, using~\eqref{eq:SymTenInv} as well as Lemma~\ref{Lemma:PropLinGrav}(i), we obtain
	\begin{align*}
		\langle \mathrm{Riem}_{\sf{g}}(\I\sf{h})+(\d_{s}\phi &\otimes\d_{s}\phi)\times_{s}\I\sf{h},\sf{k}\rangle_{\sf{S}_{2}}-\langle\sf{h},\mathrm{Riem}_{\sf{g}}(\I\sf{k})+(\d_{s}\phi \otimes\d_{s}\phi)\times_{s}\I\sf{k}\rangle_{\sf{S}_{2}}=\\&=\langle(\mathrm{Riem}_{\sf{g}}\circ\I-\I\circ\mathrm{Riem}_{\sf{g}})\sf{h}+(\d_{s}\phi\otimes\d_{s}\phi)\times_{s}\I\sf{h}-\I(\d_{s}\otimes\d_{s}\phi)\sf{h}),\sf{k}\rangle_{\sf{S}_{2}}\\&=\frac{1}{2}\bigg(\mathrm{tr}_{\sf{g}}(\sf{h})\langle\mathrm{Ric}(\sf{g})-\d_{s}\phi\otimes\d_{s}\phi,\sf{k}\rangle_{\sf{S}_{2}}-\langle\mathrm{Ric}(\sf{g})-\d_{s}\phi\otimes\d_{s}\phi,\sf{h}\rangle_{\sf{S}_{2}}\mathrm{tr}_{\sf{g}}(\sf{k})\bigg)\\&=\frac{1}{2}\bigg(\mathrm{tr}_{\sf{g}}(\sf{h})\langle\mathrm{Ein}_{\Lambda}(\sf{g})-\sf{T}_{\sf{V}}(\sf{g},\phi),\sf{k}\rangle_{\sf{S}_{2}}-\langle\mathrm{Ein}_{\Lambda}(\sf{g})-\sf{T}_{\sf{V}}(\sf{g},\phi),\sf{h}\rangle_{\sf{S}_{2}}\mathrm{tr}_{\sf{g}}(\sf{k})\bigg)\, .
	\end{align*}
	Hence, the operator $(\mathrm{Riem}_{\sf{g}}+[(\d_{s}\phi\otimes\d_{s}\phi)\times_{s}\cdot])\I$ is formally self-adjoint whenever the background satisfies the Einstein equations $\mathrm{Ein}_{\Lambda}(\sf{g})=\sf{T}_{\sf{V}}(\sf{g},\phi)$. This shows $\mathcal{L}_{11}^{\ast}=\mathcal{L}_{11}$. Now, let us turn to the operator $\mathscr{L}_{12}$. First of all, note that
	\begin{align*}
		\langle \sf{g}\sf{V}^{\prime}(\phi)\varphi,\sf{h}\rangle_{\sf{S}_{2}}=2\sf{V}^{\prime}(\phi)\varphi\sf{g}^{\alpha\beta}\sf{h}_{\alpha\beta}=\langle\varphi, 2\sf{V}^{\prime}(\phi)\mathrm{tr}_{\sf{g}}(\sf{h})\rangle_{\sf{S}_{0}}
	\end{align*}
	for all $\varphi\in\Gamma^{\infty}(\sf{S}_{0})$ and $\sf{h}\in\Gamma^{\infty}(\sf{S}_{2})$ and hence $(\sf{g}\sf{V}^{\prime}(\phi)\mathrm{id})^{\ast}=2\sf{V}^{\prime}(\phi)\mathrm{tr}_{\sf{g}}$ with respect to $(\cdot,\cdot)_{\sf{S}_{k}}$. Next, we consider the operator $\d\phi\otimes_{s}\d_{s}\cdot\in\mathrm{DO}^{1}(\sf{S}_{0},\sf{S}_{2})$. A computation in coordinates shows
	\begin{align*}
		\langle\d_{s}\phi\otimes_{s}\d_{s}\varphi,\sf{h}\rangle_{\sf{S}_{2}}&=2\sf{h}_{\alpha\beta}\nabla^{\alpha}\phi\nabla^{\beta}\varphi =2 (\sf{h}_{\alpha\beta}\nabla^{\alpha}\phi)\nabla^{\beta}\varphi\\&=-2\varphi (\nabla^{\alpha}\phi\nabla^{\beta}\sf{h}_{\alpha\beta}+\sf{h}_{\alpha\beta}\nabla^{\alpha}\nabla^{\beta}\phi)\quad\text{mod}\quad\,\mathrm{ran}(\delta_{s})\\&=\langle\varphi,\langle\d_{s}\phi,\delta_{s} \sf{h}\rangle_{\sf{S}_{1}}-\langle\d_{s}^{2}\phi,\sf{h}\rangle_{\sf{S}_{2}}\rangle_{\sf{S}_{0}}\quad\text{mod}\quad\,\mathrm{ran}(\delta_{s})
\end{align*}
Integrating over $\sf{M}$ and using Stokes' theorem, we conclude that the formal adjoint of $\d\phi\otimes_{s}\d_{s}\cdot$ with respect to $(\cdot,\cdot)_{\sf{S}_{k}}$ is given by $\langle\d_{s}\phi,\delta_{s}\cdot\rangle_{\sf{S}_{1}}-\langle\d_{s}^{2}\phi,\cdot\rangle_{\sf{S}_{2}}$. Combining everything, we get
\begin{align*}
	\mathscr{L}_{12}^{\ast}=2\langle\d_{s}^{2}\phi,\I\cdot\rangle_{\sf{S}_{2}}-2\langle\d_{s}\phi,\delta_{s}\I\cdot\rangle_{\sf{S}_{1}}+2\sf{V}^{\prime}(\phi)\mathrm{tr}_{\sf{g}}=4\mathscr{L}_{21}+2(-\square\phi+\sf{V}^{\prime}(\phi))\mathrm{tr}_{\sf{g}}\, .
\end{align*}
Hence, $\mathscr{L}_{12}^{\ast}=4\mathscr{L}_{21}$ if the background satisfies the Klein-Gordon equation $-\square\phi+\sf{V}^{\prime}(\phi)=0$.
\end{proof}

We now have all the necessary ingredients to demonstrate how the linearised Einstein-Klein-Gordon theory fits into the framework introduced in Definition~\ref{Def:LinGaugeTh}. As in the case of vacuum gravity, we must work with trace-reversed perturbations of the gravitational field. This requires composing the operators $\mathcal{L}_{11}$ and $\mathcal{L}_{21}$, which act on $\sf{h}\in \Gamma^{\infty}(\sf{S}{2})$, with the trace-reversal operator $\I$. To formalise this, we first define \begin{align*} 
	\mathcal{I} := \begin{pmatrix} 2\I & 0 \\ 0 & \mathrm{id}_{\Gamma^{\infty}(\sf{S}_{0})} 
\end{pmatrix} \in \mathrm{DO}^{0}(\mathcal{S}_{2}) \qquad \mathcal{I}^{-1}=\begin{pmatrix} \frac{1}{2}\I & 0\\ 0 & \mathrm{id}_{\Gamma^{\infty}(\sf{S}_{0})}\end{pmatrix}\, , \end{align*} 
where the factor of $2$ in the first entry will turn out to be convenient. With this definition, rather than working directly with the operator $\mathscr{L}$, we equivalently describe the linearised Einstein-Klein-Gordon system via the transformed operator
\begin{align*}
	\mathscr{P}:=\mathscr{L}\circ\mathcal{I}=
	\begin{pmatrix}
		-\square-\I\d_{s}\delta_{s}+2\mathrm{Riem}_{\sf{g}}+2[(\d_{s}\phi\otimes\d_{s}\phi)\times_{s}\cdot] &
		-2\I(\d_{s}\phi\otimes_{s}\d_{s}\cdot)+\sf{g}\sf{V}^{\prime}(\phi)\\
		\langle\d_{s}^{2}\phi,\I\cdot\rangle_{\sf{V}_{2}}-\langle\d_{s}\phi,\delta_{s}\cdot\rangle_{\sf{V}_{1}}&
		-\square+\sf{V}^{\prime\prime}(\phi)	
	\end{pmatrix}
\end{align*}
When considering the operator $\mathscr{P}$ instead of $\mathscr{L}$, we have to modify the bundle metrics and gauge transformations accordingly. First of all, we define the Hermitian bundle
\begin{align}\label{eq:EKG:BundMet1}
	\mathcal{S}_{2,\mathcal{I}}:=\mathcal{S}_{2}=\sf{S}_{2}\oplus\sf{S}_{0},\qquad \langle\cdot,\cdot\rangle_{\mathcal{S}_{2},\mathcal{I}}:=\langle\mathcal{I}\cdot,\cdot\rangle_{\mathcal{S}_{2}}=2\langle\I\cdot,\cdot\rangle_{\sf{S}_{2}}+4\langle\cdot,\cdot\rangle_{\sf{S}_{0}}\, ,
\end{align}
where $\langle\cdot,\cdot\rangle_{\mathcal{S}_{2}}$ is as defined in \ref{Lemma:FSAEKG}. As previously explained, gauge transformations are parametrised by $(0,1)$-tensor fields, i.e.~sections of $\sf{S}_{1}$ and hence, to unify the notation, we set
\begin{align}\label{eq:EKG:BundMet2}
	\mathcal{S}_{1}:=\sf{S}_{1},\qquad\langle\cdot,\cdot\rangle_{\mathcal{S}_{1}}:=\langle\cdot,\cdot\rangle_{\sf{S}_{1}}, .
\end{align}
We shall denote the formal adjoints of operators acting between the bundles $\mathcal{S}_{1}$ and $\mathcal{S}_{2,\mathcal{I}}$ by $\dagger$. Now, when considering the operator $\mathcal{P}$ and $\mathcal{L}$, we have to compose the operator~\eqref{eq:GaugeTrafoEKG} with $\mathcal{I}^{-1}$ from the left. We set 
\begin{align*}
	\mathscr{K}:=
	c\begin{pmatrix}
		\I\d_{s}\\
		\langle \d_{s}\phi,\cdot\rangle_{\sf{S}_{1}}
	\end{pmatrix}\in\mathrm{DO}^{1}(\mathcal{S}_{1},\mathcal{S}_{2,\mathcal{I}})
\end{align*}
for some constant $c\in\bb{R}$ chosen below. A straightforward computation in coordinates, shows that the formal adjoint of $\mathscr{K}$ with respect to $\langle\cdot,\cdot\rangle_{\mathcal{S}_{1}}$ and $\langle\cdot,\cdot\rangle_{\mathcal{S}_{2},\I}$ is given by
\begin{align*}
	\mathscr{K}^{\dagger}=2c\begin{pmatrix}
	\delta_{s} & 2(\d_{s}\phi)
	\end{pmatrix}\in\mathrm{DO}^{1}(\mathcal{S}_{2,\mathcal{I}},\mathcal{S}_{1})
\end{align*}
Now, it is easy to see that the $11$-component of the composed operator $\mathscr{K}\mathscr{K}^{\dagger}$ is given by $2c^{2}\I\d_{s}\delta_{s}$. Hence, in order to match the non-hyperbolic term in $\mathscr{P}_{11}$ above, we choose $c=1/\sqrt{2}$. Summing up everything discussed in this section, we obtain the following result:

\begin{proposition}\label{Prop:GaugeEinKG} \emph{(Linearised Einstein-Klein-Gordon Theory)}\newline
	Let $(\sf{M},\sf{g})$ be a globally hyperbolic manifold and $\phi\in C^{\infty}(\sf{M})$ a scalar field such that $\mathrm{EKG}_{\Lambda,\sf{V}}(\sf{g},\phi)=0$. Then, $(\mathcal{S}_{1},\mathcal{S}_{2,\mathcal{I}},\mathscr{P},\mathscr{K})$, where the Hermitian bundles $\mathcal{S}_{1}$ and $\mathcal{S}_{2,\mathcal{I}}$ are as defined in~\eqref{eq:EKG:BundMet1} and \eqref{eq:EKG:BundMet2}, with
	\begin{align*}
		\mathscr{P}&:=\begin{pmatrix}
		-\square-\I\d_{s}\delta_{s}+2\mathrm{Riem}_{\sf{g}}+2[(\d_{s}\phi\otimes\d_{s}\phi)\times_{s}\cdot] &
		-2\I(\d_{s}\phi\otimes_{s}\d_{s}\cdot)+\sf{g}\sf{V}^{\prime}(\phi)\\
		\langle\d_{s}^{2}\phi,\I\cdot\rangle_{\sf{S}_{2}}-\langle\d_{s}\phi,\delta_{s}\cdot\rangle_{\sf{S}_{1}}&
		-\square+\sf{V}^{\prime\prime}(\phi)	
	\end{pmatrix}\\
	\mathscr{K}&:=\frac{1}{\sqrt{2}}\begin{pmatrix}
		\I\d_{s}\\
		\langle \d_{s}\phi,\cdot\rangle_{\sf{S}_{1}}
	\end{pmatrix}
	\end{align*}
	is a linear gauge theory in the sense Definition~\ref{Def:LinGaugeTh}. Furthermore, the corresponding operators $\mathscr{D}_{1}:=\mathscr{K}^{\dagger}\mathscr{K}$ and $\mathscr{D}_{2}:=\mathscr{P}+\mathscr{K}\mathscr{K}^{\dagger}$ are the normally hyperbolic operators
	\begin{align*}
		\mathscr{D}&=-\square-\Lambda-\sf{V}(\phi)+(\d_{s}\phi)\langle\d_{s}\phi,\cdot\rangle_{\sf{S}_{1}}\\
		\mathscr{D}_{2}&=\begin{pmatrix}
		-\square+2\mathrm{Riem}_{\sf{g}}+2[(\d_{s}\phi\otimes\d_{s}\phi)\times_{s}\cdot] &
		2(\d_{s}^{2}\phi)\\
		\langle\d_{s}^{2}\phi,\I\cdot\rangle_{\sf{S}_{2}} &
		-\square+\sf{V}^{\prime\prime}(\phi)+2\Vert\d_{s}\phi\Vert_{\sf{S}_{1}}^{2}
	\end{pmatrix}\, .
	\end{align*}
\end{proposition}

\begin{proof}
	Formally self-adjointness of $\mathscr{P}$ with respect to $\langle\cdot,\cdot\rangle_{\mathcal{S}_{2},\mathcal{I}}$ and gauge-invariance, i.e.~the fact that $\mathscr{P}\circ\mathscr{K}=0$, follows from Lemma~\ref{Lemma:FSAEKG} and Proposition~\ref{Prop:GaugeInvEKG} as well as the discussion of above. It remains to derive the explicit expressions for $\mathscr{D}_{1}$ and $\mathscr{D}_{2}$. The adjoint of $\mathscr{K}$ is given by $\mathscr{K}^{\dagger}=\sqrt{2}\big(\delta_{s} \,\,\, 2(\d_{s}\phi)\big)$ and a straightforward computation in coordinates shows that
	\begin{align*}
		\mathscr{K}\mathscr{K}^{\dagger}&=\begin{pmatrix}
			\I\d_{s}\delta_{s} & 2\I(\d_{s}\phi\otimes_{s}\d_{s}\cdot)+2\I(\d_{s}^{2}\phi)\\
			\langle\d_{s}\phi,\delta_{s}\cdot\rangle_{\sf{S}_{1}} & 2\Vert\d_{s}\phi\Vert_{\sf{S}_{1}}
		\end{pmatrix}=\\&=\begin{pmatrix}
			\I\d_{s}\delta_{s} & 2\I(\d_{s}\phi\otimes_{s}\d_{s}\cdot)+2(\d_{s}^{2}\phi)-\sf{g}\sf{V}^{\prime}(\phi)\\
			\langle\d_{s}\phi,\delta_{s}\cdot\rangle_{\sf{S}_{1}} & 2\Vert\d_{s}\phi\Vert_{\sf{S}_{1}}^{2}
		\end{pmatrix}
	\end{align*}
	where we used the Klein-Gordon equations of the background to write $\square\phi=\sf{V}^{\prime}(\phi)$. Adding this to $\mathscr{P}$ gives the claimed result for $\mathscr{D}_{2}$. The operator $\mathscr{D}_{2}$ is clearly normally hyperbolic and hence in particular Green hyperbolic. For $\mathscr{D}_{1}$, we compute
	\begin{align}\label{eq:ProofEKG1}
		\mathscr{D}_{1}=\mathscr{K}^{\dagger}\mathscr{K}=\delta_{s}\I\d_{s}+2(\d_{s}\phi)\langle\d_{s}\phi,\cdot\rangle_{\sf{S}_{1}}
	\end{align}
	Now, it is easy to see that Lemma~\ref{Lemma:PropLinGrav}(c) in the case of a non-Einstein manifold generalises to
	\begin{align}\label{eq:ProofEKG2}
		\delta_{s}\I\d_{s}=-\square-\mathrm{Ric}_{\sf{g}}\, ,
	\end{align}
	where $\mathrm{Ric}_{\sf{g}}\:\Gamma^{\infty}(\sf{S}_{1})\to\Gamma^{\infty}(\sf{S}_{1})$ denotes the Ricci operator as introduced in the previous section. Using the background equation~\eqref{eq:EinTR} together with Eq.~\eqref{eq:ProofEKG1} and Eq.~\eqref{eq:ProofEKG2}, we obtain
	\begin{align*}
		\mathscr{D}_{1}=-\square-\Lambda-\sf{V}(\phi)+(\d_{s}\phi)\langle\d_{s}\phi,\cdot\rangle_{\sf{S}_{1}}\, ,
	\end{align*}
	which is clearly normally hyperbolic.
\end{proof}

\section{Quantisation of Linear Gauge Theories}\label{Sec.AQFT}
Having discussed the Cauchy problem and the phase space of \textit{classical} linear field theories with a gauge symmetry in depth, together with a detailed analysis of various examples with direct relevance to physical applications, we now turn our attention to the next crucial step, their \textit{quantisation}. We will follow the \textit{algebraic approach} to quantum field theory, which offers a mathematically rigorous and conceptually compelling quantisation scheme, ideally suited for field theories defined on \textit{curved} Lorentzian backgrounds. After establishing the foundational features of algebraic quantum field theory, we proceed by discussing the concept of \textit{Hadamard states}, which play a central role in ensuring the physical viability of quantum field theories on curved spacetimes. In this context, we review the \emph{Gérard-Wrochna formalism} of quantising linear gauge theories and Hadamard states developed in \cite{GerardWrochna} (see also \cite{Gerard,GerardWrochna2,MurroSchmid}). We conclude this section by providing a detailed bibliography of various existence results and explicit constructions of Hadamard states for a wide range of different field theories.

\subsection{A Glance of Algebraic Quantum Field Theory}
As mentioned above, we will follow the so-called \textit{algebraic approach to quantum field theory} (short:~AQFT). AQFT roots back to the pioneering works of Haag, Kastler \cite{HaagLille,HaagKastler} and Araki \cite{Araki1,Araki2} in the early 1960s (see also the foundational monographs \cite{HaagBook,ArakiBook} as well as early applications to scattering theory in \cite{HaagScattering,Ruelle}), with earlier works on \textit{algebraic quantum mechanics} by Jordan-von Neumann-Wigner \cite{Neumann} and Segal \cite{Segal1,Segal2}, and is nowadays a well-established and rigorous formalism for quantum field theory. Instead of working with quantum fields in the sense of operator-valued distributions, as in the pioneering program on mathematical and axiomatic quantum field theory initiated by Gårding-Wightman \cite{WightmanGarding,Wightman} in the 1950s (see also the detailed treatment in \cite[Chap.~IX, Sec.~8]{ReedSimonII}), the point of view in AQFT is shifted to local observables and their algebraic relations. Quantum fields are then merely tools that emerge as a way of labelling observables. In this framework, the quantisation is interpreted as a two step procedure. First, one assigns a \textit{$\ast$-algebra of observables}, which encodes structural
properties such as causality, dynamics and canonical commutation relations, to a physical bosonic system described by a classical phase space. The second step calls for the identification of a physical \textit{state}, i.e.~a positive, linear and normalised functional on the algebra of observables. 

To start with, we recall some basic notions of linear algebra: 
\begin{itemize}
	\item[$\bullet$]A \textit{$\bb{C}$-algebra} is a pair $(\mathcal{A},\cdot)$ consisting of a $\bb{C}$-vector space $\mathcal{A}$ and a $\mathbb{C}$-bilinear product $\cdot\:\mathcal{A}\times\mathcal{A}\to\mathcal{A}$. It is called \textit{unital}, if there exists a two-sided multiplicative unit, i.e.~an element $\mathds{1}_{\mathcal{A}}$ such that $\mathds{1}_{\mathcal{A}}\cdot a=a\cdot\mathds{1}_{\mathcal{A}}=a$ for all $a\in\mathcal{A}$. It is easy to see that the unit is unique whenever it exists. Last but not least, a \textit{$\ast$-algebra} is a $\bb{C}$-algebra $(\mathcal{A},\cdot)$ together with an antilinear involution $\ast\:\mathcal{A}\to\mathcal{A}$ satisfying $(a\cdot b)^{\ast}=b^{\ast}\cdot a^{\ast}$.
	\item[$\bullet$]If $(\mathcal{A},\cdot_{\mathcal{A}})$ and $(\mathcal{B},\cdot_{\mathcal{B}})$ are two $\bb{C}$-algebra, an \textit{algebra homomorphism} is a $\bb{C}$-linear map $\varphi\:\mathcal{A}\to\mathcal{B}$ preserving the products, i.e.~$\varphi(a\cdot_{\mathcal{A}}b)=\varphi(a)\cdot_{\mathcal{B}}\varphi(b)$ for all $a,b\in\mathcal{A}$. If $\varphi$ is bijective, it is called an \textit{algebra isomorphism}. If both $\mathcal{A}$ and $\mathcal{B}$ are unital, an algebra homomorphism is called \textit{unital} if it preserves the unit, i.e.~$\varphi(\mathds{1}_{\mathcal{A}})=\mathds{1}_{\mathcal{B}}$. If both $\mathcal{A}$ and $\mathcal{B}$ are $\ast$-algebras, an algebra homomorphism is called a \textit{$\ast$-homomorphism} if preserves the involutions, i.e.~$\varphi(a^{\ast_{\mathcal{A}}})=\varphi(a)^{\ast_{\mathcal{B}}}$ for all $a\in\mathcal{A}$.
	\item[$\bullet$]Given a $\bb{C}$-algebra $(\mathcal{A},\cdot)$, a linear subspace $\mathcal{I}\subset\mathcal{A}$ is called a \textit{two-sided ideal}, if $a\cdot b,b\cdot a\in\mathcal{I}$ for all $a\in\mathcal{I}$ and $b\in\mathcal{A}$. If $\mathcal{A}$ is a $\ast$-algebra, then a \textit{two-sided $\ast$-ideal} is a two-sided ideal that is in addition closed under the $\ast$-involution, i.e.~$a^{\ast}\in\mathcal{I}$ for all $a\in\mathcal{I}$.
\end{itemize}

Now, suppose we are given an arbitrary set $\mathcal{G}$. Then, we wish to define the ``smallest'' unital $\ast$-algebra $\mathcal{A}_{\mathcal{G}}$ such that $\mathcal{G}\subset\mathcal{A}_{\mathcal{G}}$. As usual, those kind of constructions can abstractly be defined by means of a universal property: given an arbitrary (not necessarily countable) set $\mathcal{G}$, a \textit{free unital $\ast$-algebra generated by $\mathcal{G}$} is a unital $\ast$-algebra $\mathcal{A}_{\mathcal{G}}$ together with a map $\alpha\:\mathcal{G}\to\mathcal{A}_{\mathcal{G}}$ such that for every other pair $(\mathcal{B},\beta)$ consisting of a unital $\ast$-algebra and map $\beta\:\mathcal{G}\to\mathcal{B}$, there exists a unique unital $\ast$-homomorphism $\varphi\:\mathcal{A}_{\mathcal{G}}\to\mathcal{B}$ such that the following diagram commutes: 
\begin{equation*}
        \begin{tikzcd}
           \mathcal{G} \arrow[rd,swap,"\beta"] \arrow[r,"\alpha"] & \mathcal{A}_{\mathcal{G}}\arrow[d,"\varphi"] \\& \mathcal{B}
        \end{tikzcd}
    \end{equation*}
As usual, the universal property implies that the pair $(\mathcal{A}_{\mathcal{G}},\alpha)$ is unique up to canonical isomorphism and hence it makes sense to speak of \textit{the} free unital $\ast$-algebra $\mathcal{A}_{\mathcal{G}}$ generated by $\mathcal{G}$. Existence can be shown by an explicit construction: let $\mathcal{S}$ be the set of finite, ordered tuples of elements in $\mathcal{G}$, i.e.~$\mathcal{S}:=\{(g_{1},\dots,g_{k})\mid k\in\bb{N}_{0}\,,g_{i}\in\mathcal{G}\}$ and consider a $\bb{C}$-vector space $\mathcal{A}_{\mathcal{G}}$ spanned by a basis labelled by $\mathcal{S}$, i.e.~$(e_{S})_{S\in\mathcal{S}}$. We set $\mathds{1}:=e_{()}$, where $()$ denotes the empty tuple, and define a multiplication $\cdot\:\mathcal{A}_{\mathcal{G}}\times\mathcal{A}_{\mathcal{G}}\to\mathcal{A}_{\mathcal{G}}$ by $\mathds{1}\cdot e_{S}=e_{S}\cdot\mathds{1}=e_{S}$ and $e_{S}\cdot e_{T}:=e_{S\cdot T}$, where $S\cdot T:=(g_{1},\dots,g_{k}h_{1},\dots, h_{l})$ for $S=(g_{1},\dots,g_{k})$ and $T=(h_{1},\dots,h_{l})$. Furthermore, let $\ast\:\mathcal{G}\to\mathcal{G}$  be any involution and define $\ast\:\mathcal{A}_{\mathcal{G}}\to\mathcal{A}_{\mathcal{G}}$ by $\mathds{1}^{\ast}=\mathds{1}$ and $(e_{S})^{\ast}=e_{S^{\ast}}$, where $S^{\ast}=(\ast g_{k},\dots,\ast g_{1})$ for $S=(g_{1},\dots,g_{k})$. With this definition, it is easy to see that $\mathcal{A}_{\mathcal{G}}$ together with the map $\gamma(g):=e_{(g)}$ satisfies the above universal property.

In many applications, one often seeks to construct the smallest algebra that contains a certain given sets while also satisfying specific relations among its generators. For instance, given a set of generators $\mathcal{G}$, we  asks for the smallest unital $\ast$-algebra containing $\mathcal{G}$ such that $g^{\ast}=g$ for all $g\in\mathcal{G}$. To construct this algebra, we consider the two-sided $\ast$-ideal spanned by the relation we want to impose, for instance, $g-g^{\ast}$ for all $g\in\mathcal{G}$ in the provided example. Then, we set $\mathcal{A}_{\mathcal{G},\mathcal{I}}:=\mathcal{A}_{\mathcal{G}}/\mathcal{I}$. If there is more than one relation one want to impose, one considers the corresponding intersection of ideals. We remark that the so-constructed algebra $\mathcal{A}_{\mathcal{G},\mathcal{I}}$ can again be characterised by means of a universal property, see e.g.~\cite[Sec.~5.1.2]{MorettiKhavkine}.

After this preliminary discussion, we return to algebraic quantisation. As mentioned above, the first step consists of assigning a unital $\ast$-algebra to a given classical phase space. 

\begin{definition}\label{Def:CCR} (CCR-Algebra)\newline
	Let $(\mathcal{V},\sigma)$ be a pair consisting of a $\bb{C}$-vector space together with a (possibly degenerate) Hermitian sesquilinear form $\sigma$ (for instance, the classical phase space of a linear gauge theory). Then, the \textit{CCR-algebra} $\mathrm{CCR}(\mathcal{V},\sigma)$ is the unital $\ast$-algebra generated by an identity $\mathds{1}$ and elements $\{\Phi(v),\Phi^{\ast}(v)\}_{v\in\mathcal{V}}$ subject to the following relations:
	\begin{align*}
	\text{(i)}&\hspace*{1cm}\Phi(v)^{\ast}=\Phi^{\ast}(v)\\
	\text{(ii)}&\hspace*{1cm}v\mapsto\Phi(v)\text{ is $\bb{C}$-antilinear and }v\mapsto\Phi^{\ast}(v)\text{ is $\bb{C}$-linear}\\
	\text{(iii)}&\hspace*{1cm}[\Phi(v),\Phi(w)]=[\Phi^{\ast}(v),\Phi^{\ast}(w)]=0\quad\text{and}\quad[\Phi(v),\Phi^{\ast}(w)]=\sigma(v,w)\mathds{1}
\end{align*}
The relation $[\Phi(v),\Phi^{\ast}(w)]=\sigma(v,w)\mathds{1}$ are called \textit{canonical commutation relations}.
\end{definition}

\begin{remark}
	We restrict our analysis to \textit{bosonic} field theories. For fermionic theories, such as the Dirac or Rarita-Schwinger field theories, one needs to consider the \textit{canonical anti-computation relations}, i.e.~replacing the commutators in (iii) above by anti-commutators.
\end{remark}

Having defined a non-commutative algebra of observables, the second step consists of the choice of a \textit{state}. From a physical perspective, an \textit{observable} is a quantity which we want to measure.  Now, it is essential to assume that experiments can be reliably repeated, enabling a consistent measurement of a given observable under identical conditions, which is a fundamental requirement for an empirical science. Such conditions, the preparation of the system, are described by the choice of a \textit{state}, that assigns to a given observable the expectation value of the relevant measurement in the system. To this end, we make the following definition.

\begin{definition}\label{Def:States} (Algebraic States)\newline
	Let $(\mathcal{A},\cdot,\mathds{1},\ast)$ be a complex unital $\ast$-algebra. A $\bb{C}$-linear functional $\omega\:\mathcal{A}\to\bb{C}$ is called a \textit{state} if the following two conditions hold:
	\begin{itemize}
		\item[(i)]$\omega(\mathds{1})=1$\hfill (``\textit{normalisation}'')
		\item[(ii)]$\omega(a^{\ast}a)\geq 0$ for all $a\in\mathcal{A}$. \hfill (``\textit{positivity}'')
	\end{itemize}
\end{definition}

\begin{remark}\label{Remark:States}
	Note that the definition implies $\omega(a^{\ast})=\overline{\omega(a)}$ for all $a\in\mathcal{A}$. Indeed, 
	\begin{align*}
		0\leq \omega((a+\lambda\mathds{1})^{\ast}(a+\lambda\mathds{1}))=\omega(a^{\ast}a)+\vert\lambda\vert^{2}+\overline{\lambda}\omega(a)+\lambda\omega(a^{\ast})\leq \overline{\lambda}\omega(a)+\lambda\omega(a^{\ast})
	\end{align*}
	for all $\lambda\in\bb{C}$ and $a\in\mathcal{A}$, which implies $\overline{\lambda}\omega(a)+\lambda\omega(a^{\ast})\in\bb{R}$. Choosing $\lambda=i$, this implies $\mathrm{Re}(\omega(a))=\mathrm{Re}(\omega(a^{\ast}))$ and choosing $\lambda=1$, we obtain $\mathrm{Im}(\omega(a))=-\mathrm{Im}(\omega(a^{\ast}))$.
\end{remark}

\begin{remark} (The GNS-Construction)\newline
Let us briefly explain how algebraic quantum mechanics and field theory are related to the traditional Hilbert space approach. Let $\mathcal{A}$ be a unital $\ast$-algebra and $\omega\:\mathcal{A}\to\bb{C}$ be a state. Then, there exists a quadruple $(\mathcal{H}_{\omega},\mathcal{D}_{\omega},\Omega_{\omega},\pi_{\omega})$ consisting of a Hilbert space $\mathcal{H}_{\omega}$, a dense subspace $\mathcal{D}_{\omega}\subset\mathcal{H}_{\omega}$, a vector $\Omega_{\omega}\in\mathcal{H}_{\omega}$ and a unital $\ast$-representation\footnote{A \textit{representation} of a unital $\ast$-algebra $\mathcal{A}$ on a fixed dense subspace $\mathcal{D}$ of some Hilbert space $\mathcal{H}$ is an algebra homomorphism $\pi\:\mathcal{A}\to L(\mathcal{D},\mathcal{H})$, where $L(\mathcal{D},\mathcal{H})$ is the algebra of linear operators $T\:\mathcal{D}\to\mathcal{H}$ that are closable and that satisfy $\mathrm{ran}(T)\subset\mathcal{D}$. It is a \textit{unital $\ast$-representation}, if $\pi(\mathds{1})=\mathrm{id}_{\mathcal{D}}$, if $\mathcal{D}\subset\mathcal{D}(\pi(a)^{\ast})$ for all $a\in\mathcal{A}$, where $\mathcal{D}(\pi(a)^{\ast})\subset\mathcal{H}$ denotes the domain of $\pi(a)^{\ast}$, and if $\pi(a^{\ast})=\pi(a)^{\ast}\vert_{\mathcal{D}}$ for all $a\in\mathcal{A}$, see e.g.~\cite[Chap.~8]{Schmudgen}.} $\pi_{\omega}\:\mathcal{A}_{\omega}\to L(\mathcal{D}_{\omega},\mathcal{H}_{\omega})$, such that $\pi_{\omega}(\mathcal{A})\Omega_{\omega}$ is dense in $\mathcal{H}_{\omega}$ and such that
\begin{align*}
	\omega(a)=\langle\Omega_{\omega},\pi_{\omega}(a)\Omega_{\omega}\rangle_{\mathcal{H}_{\omega}}
\end{align*}
for all $a\in\mathcal{A}$. Furthermore, this quadruple is unique up to \textit{unitary equivalence}, i.e.~for any other quadruple $(\mathcal{H}_{\omega}^{\prime},\mathcal{D}_{\omega}^{\prime},\Omega_{\omega}^{\prime},\pi_{\omega}^{\prime})$ with the same properties, there exists a unitary operator $\sf{U}\:\mathcal{H}_{\omega}\to\mathcal{H}_{\omega}^{\prime}$, such that $\sf{U}\Omega_{\omega}=\Omega^{\prime}_{\omega}$, $\sf{U}\mathcal{D}_{\omega}=\mathcal{D}_{\omega}^{\prime}$ and such that
\begin{align*}
	\sf{U}\pi_{\omega}(\cdot)\sf{U}^{-1}=\pi_{\omega}^{\prime}(\cdot)
\end{align*}
This result first appeared implicitly in the work of Gelfand-Naimark \cite{GNS} on what is now known as the \textit{Gelfand-Naimark theorem}, i.e.~the fact that every $C^{\ast}$-algebra is isometrically $\ast$-isomorphic to a $C^{\ast}$-subalgebra of the $\ast$-algebra of bounded operators. Segal later formulated the above construction explicitly in \cite{Segal1}. Henceforth, the construction $(\mathcal{A},\omega)\mapsto (\mathcal{H}_{\omega},\mathcal{D}_{\omega},\Omega_{\omega},\pi_{\omega})$ is known as the \textit{Gelfand-Naimark-Segal construction}. The idea of the proof is in fact relatively simple: set $\mathcal{N}_{\omega}:=\{a\in\mathcal{A}\mid\omega(a^{\ast}a)=0\}$, which can easily be seen to be a linear subspace of $\mathcal{A}$. Then, consider the vector space $\mathcal{D}_{\omega}:=\mathcal{A}\backslash\mathcal{N}_{\omega}$ equipped with the Hermitian sesquilinear form $\langle [a],[b]\rangle_{\mathcal{H}_{\omega}}:=\omega(a^{\ast}b)$, which can easily be seen to be well-defined and positive-definite. Now, denote the completion of $\mathcal{D}_{\omega}$ with respect to $\langle\cdot,\cdot\rangle_{\mathcal{H}_{\omega}}$ by $(\mathcal{H}_{\omega},\langle\cdot,\cdot\rangle_{\mathcal{H}_{\omega}})$. Then, a straightforward computation shows that $(\mathcal{H}_{\omega},\mathcal{D}_{\omega},\Omega_{\omega},\pi_{\omega})$ with $\Omega_{\omega}:=[\mathds{1}]$ and $\pi_{\omega}(a)[b]:=[ab]$ provides the required quadruple. For more details, we refer to \cite[Thm.~5.1.13]{MorettiKhavkine} or \cite[Thm.~10]{RejznerFewster}.
\end{remark}

\subsection{Hadamard States}
The definition of states given in Definition~\ref{Def:States} is fairly general and just includes the minimal physical requirements. Among the plethora of all possible states, however, most of them have pathological physical behaviour and we need to find a selecting criterion to distinguish those which are physical ``acceptable''. While in Minkowski spacetime there is always a preferred state\footnote{More generally, there always exists a preferred state on a stationary spacetime, that is, on a globally hyperbolic spacetime for which there exists a complete and future-directed timelike Killing vector field $\xi$, namely the \textit{ground state} that is invariant under the isometries induced by $\xi$ and that has \textit{minimal energy} in a suitable sense, see~\cite[Chap.~9]{Gerard} and \cite[Sec.~5.2]{MorettiKhavkine} for details.}, namely the \textit{vacuum}, characterised by its invariance and stability properties with respect to the Poincaré symmetry group, it turns out this is a highly non-trivial task on general Lorentzian backgrounds.

Let $\mathrm{CCR}(\mathcal{V},\sigma)$ be a CCR-algebra and $\omega$ a state on it. By definition, every $a\in\mathrm{CCR}(\mathcal{V},\sigma)$ can be written as a finite linear combination of products of the generators $\mathds{1},\Phi(v_{i}),\Phi^{\ast}(w_{j})$. Hence, the state $\omega$ is completely determined by its \textit{correlation functions},
\begin{align*}
	\omega_{m,n}\:\mathcal{V}_{\mathrm{c}}^{m+n}\to\bb{C}\,,\qquad \omega_{m,n}(v_{1},\dots,v_{m},w_{1},\dots,w_{n}):=\omega(\Phi(v_{1})\dots\Phi(v_{n})\Phi^{\ast}(w_{1})\dots\Phi^{\ast}(w_{n}))\,.
\end{align*}
With this notation, we will only consider states contained in the following class.

\begin{definition} (Quasifree States)\newline
	A state $\omega$ on $\mathrm{CCR}(\mathcal{V},\sigma)$ is called \textit{quasi-free} or \textit{Gaussian}, if its purely determined by its \emph{$2$-point function} $\omega_{1,1}(\cdot,\cdot)=\omega(\Phi(\cdot)\Phi^{\ast}(\cdot))$ in the sense that
	\begin{align*}
		\omega_{m,n}(v_{1},\dots,v_{m},w_{1},\dots,w_{n})=\delta_{mn}\sum_{\sigma\in\mathfrak{S}_{m}}\prod_{i=1}^{m}\omega_{1,1}(v_{i},w_{\sigma(i)})
	\end{align*}
\end{definition}

In other words, for a quasi-free state, only the correlations functions with the same number of $\Phi$- and $\Phi^{\ast}$-fields are non-zero and in this case, they are given by the the sum over all \textit{pairwise disjoint contractions} of $\Phi$ and $\Phi^{\ast}$ fields. This generalises the well-known \textit{Wick theorem} for the free quantum scalar field and the \textit{vacuum state} in the standard approach to quantum field theory in Minkowski spacetime (after \cite{Wick}; see \cite[Sec.~4.3]{Peskin} for a textbook treatment from a physics perspective). This is also related to the \textit{Isserlis theorem} \cite{Isserlis1,Isserlis2} of probability theory, which allows to express higher-order momenta of (zero-mean) random Gaussian variables in terms of their covariances in a similar manner. In other words, quasi-free states mimic Gaussian probability distributions, which is the origin of the name \textit{Gaussian}.

Now, if $\omega$ is a quasifree state on $\mathrm{CCR}(\mathcal{V},\sigma)$, we define its \textit{$2$-point functions} to be the sesquilinear forms $\Lambda^{\pm}\:\mathcal{V}\times\mathcal{V}\to\bb{C}$ defined by
\begin{align*}
	\Lambda^{+}(v,w):=\omega_{1,1}(v,w):=\omega(\Phi(v)\Phi^{\ast}(w))\,,\qquad\Lambda^{-}(v,w):=\omega(\Phi(w)^{\ast}\Phi(v))\, .
\end{align*}
Following Remark~\ref{Remark:States}, it is easy to see that $\Lambda^{\pm}$ are Hermitian. Of course, $\Lambda^{+}$ and $\Lambda^{-}$ are not independent, but rather satisfy $\Lambda^{+}-\Lambda^{-}=\sigma$, as a consequence of the canonical commutation relation $[\Phi(v),\Phi^{\ast}(w)]=\sigma(v,w)\mathds{1}$. Furthermore, $\Lambda^{\pm}$ are positive in the sense that $\Lambda^{\pm}(v,v)\geq 0$ for all $v\in\mathcal{V}$. By definition, $\omega$ is completely determined by $\Lambda^{\pm}$. On the other hand, every pair of Hermitian sesquilinear forms $\Lambda^{\pm}$ on $\mathcal{V}$ satisfying 
\begin{align*}
	\text{(i)}\qquad & \Lambda^{+}-\Lambda^{-}=\sigma\\
	\text{(ii)}\qquad & \Lambda^{\pm}(v,v)\geq 0\qquad\forall v\in\mathcal{V}
\end{align*}
gives rise to a unique quasi-free state on $\mathrm{CCR}(\mathcal{V},\sigma)$. 

Now, after this abstract discussion, let us turn to linear gauge theories. For practical purposes, we will only consider quasifree states coming from a pair of \textit{pseudo-covariances}\footnote{The name ``\textit{pseudo}''-covariance was chosen in the original article~\cite{GerardWrochna} since condition (iv) only needs to be satisfied on the subspace $\mathrm{ker}(\sf{K}^{\ast}\vert_{\Gamma_{\mathrm{c}}^{\infty}})$ as opposed to ordinary field theories such as Klein-Gordon fields.} (following \cite[Sec.~3.2.1]{GerardWrochna} and \cite[Lemma 2.9]{GerardMurroWrochna} for linear gauge theories; see also \cite[Prop.~6.1.3]{GerardBook} for a similar definition in the context of Klein-Gordon fields).

\begin{proposition} \emph{(Spacetime Pseudo-Covariances)}\newline
	Let $(\sf{E}_{1},\sf{E}_{2},\sf{P},\sf{K})$ be a linear gauge theory with corresponding CCR-algebra $\mathrm{CCR}(\mathcal{V}_{\mathrm{c}},\sigma)$. A pair of linear and continuous operators $\lambda^{\pm}\:\Gamma_{\mathrm{c}}^{\infty}(\sf{E}_{2})\to\Gamma^{\infty}(\sf{E}_{2})$ give rise to a quasifree state $\omega$ on $\mathrm{CCR}(\mathcal{V}_{\mathrm{c}},\sigma)$ by setting 
	\begin{align*}\Lambda^{\pm}([\psi],[\varphi]):=(\psi,\lambda^{\pm}\varphi)_{\sf{E}_{2}}
	\end{align*}
	for all $f,g\in\mathrm{ker}(\sf{K}^{\ast}\vert_{\Gamma^{\infty}_{\mathrm{c}}})$, if the following conditions are satisfied
	\begin{align*}
		\text{\emph{(i)}}\quad &\sf{D}_{2}\circ\lambda^{\pm}=\lambda^{\pm}\circ\sf{D}_{2}=\mathrm{id}\quad\text{on}\quad\Gamma^{\infty}_{\mathrm{c}}(\sf{E}_{2})\qquad\text{and}\qquad (\lambda^{\pm})^{\ast}=\lambda^{\pm}\quad\text{w.r.t.}\quad (\cdot,\cdot)_{\sf{E}_{2}}\\
		\text{\emph{(ii)}}\quad & \lambda^{\pm}(\mathrm{ran}(\sf{K}\vert_{\Gamma^{\infty}_{\mathrm{c}}}))\subset\mathrm{ran}(\sf{K}\vert_{\Gamma^{\infty}})\\
		\text{\emph{(iii)}}\quad & \lambda^{+}-\lambda^{-}=\i\sf{G}_{2}\\
		\text{\emph{(iv)}}\quad & \langle\psi,\lambda^{\pm}\psi\rangle_{\sf{E}_{2}}\geq 0\qquad\forall\psi\in\mathrm{ker}(\sf{K}^{\ast}\vert_{\Gamma^{\infty}_{\mathrm{c}}})\, .
	\end{align*}
	The operators $\lambda^{\pm}$ are called the \emph{spacetime pseudo-covariances} of the state $\omega$.
\end{proposition}

\begin{proof}
	Set $\Lambda^{\pm}([\psi],[\varphi]):=(\psi,\lambda^{\pm}\varphi)_{\sf{E}_{2}}$ for all $f,g\in\mathrm{ker}(\sf{K}^{\ast}\vert_{\Gamma^{\infty}_{\mathrm{c}}})$. For this to be well-defined, this needs to be independent of the chosen representatives, which is guaranteed by (ii) and the first part of (i), since
	\begin{align*}
		(\psi,\lambda^{\pm}\sf{P}\varphi)_{\sf{E}_{2}}=(\psi,\lambda^{\pm}(\sf{D}_{2}-\sf{K}\sf{K}^{\ast})\varphi)_{\sf{E}_{2}}=-(\psi,\sf{K}\omega)_{\sf{E}_{2}}=-(\sf{K}^{\ast}\psi,\omega)_{\sf{E}_{2}}=0
	\end{align*}
	for all $\psi\in\mathrm{ker}(\sf{K}^{\ast}\vert_{\Gamma^{\infty}_{\mathrm{c}}})$ and $\psi\in\Gamma^{\infty}_{\mathrm{c}}(\sf{E}_{2})$, where $\omega\in\Gamma^{\infty}(\sf{E}_{2})$ is such that $\lambda^{\pm}\sf{K}\sf{K}^{\ast}\varphi=\sf{K}\omega$. The second condition in (i), formal self-adjointness, shows that $\Lambda^{\pm}$ is Hermitian. Condition (iii) implies the CCR-relations $\Lambda^{+}-\Lambda^{-}=\sigma$ and (iv) the positivity $\Lambda^{\pm}(v,v)\geq 0$ for all $v\in\mathcal{V}_{\mathrm{c}}$.
\end{proof}

\begin{remark}
Note that the previous proposition is not an ``if-and-only-if statement'' and one could consider more general conditions. For instance, instead of (ii), it would be enough to require $\lambda^{\pm}(\mathrm{ran}(\sf{K}\sf{K}^{\ast}\vert_{\Gamma^{\infty}_{\mathrm{c}}}))\subset\mathrm{ran}(\sf{K}\vert_{\Gamma^{\infty}})$, cf.~\cite[Lem.~2.8]{GerardWrochna2}, as one can easily see from the proof. However, for our purposes, it will not be necessary to consider a more general construction.
\end{remark}

Now, recall from Proposition~\ref{Prop:EqPS} that there is an equivalent phase space in terms of initial data, denoted by $(\mathcal{V}_{\Sigma},\sigma_{\Sigma,2})$, which is unitarily isomorphic to $(\mathcal{V}_{\mathrm{c}},\sigma)$ via 
\begin{align*}
\bigg(\mathcal{V}_{\mathrm{c}}=\cfrac{\mathrm{ker}(\sf{K}^{\ast}\vert_{\Gamma^{\infty}_{\mathrm{c}}})}{\mathrm{ran}(\sf{P}\vert_{\Gamma^{\infty}_{\mathrm{c}}})} ,\sigma\bigg)\xrightarrow{[\rho_{2}\sf{G}_{2}]}\bigg(\mathcal{V}_{\Sigma}:=\cfrac{\mathrm{ker}(\sf{K}_{\Sigma}^{\dagger}\vert_{\Gamma_{\mathrm{c}}^{\infty}})}{\mathrm{ran}(\sf{K}_{\Sigma}\vert_{\Gamma_{\mathrm{c}}^{\infty}})},\sigma_{\Sigma,2}\bigg)\, .
\end{align*}
On $\mathcal{V}_{\Sigma}$, quasifree states can be obtained as follows (see~\cite[Sec.~3.3]{GerardWrochna} and \cite[Prop.~2.11]{GerardMurroWrochna}).

\begin{proposition}\label{Prop:CauchyCov} \emph{(Cauchy Pseudo-Covariances)}\newline
	Let $(\sf{E}_{1},\sf{E}_{2},\sf{P},\sf{K})$ be a linear gauge theory with corresponding CCR-algebra $\mathrm{CCR}(\mathcal{V}_{\mathrm{c}},\sigma)$. Furthermore, let $\rho_{i}\:\Gamma^{\infty}_{\mathrm{sc}}(\sf{E}_{i})\to\Gamma^{\infty}_{\mathrm{c}}(\sf{E}_{\rho_{i}})$ be initial data maps of $\sf{D}_{i}$ and $(\mathcal{V}_{\Sigma},\sigma_{\Sigma,2})$ the equivalent phase space in terms of initial data as in Proposition~\ref{Prop:EqPS}. A pair of linear and continuous operators $c^{\pm}\:\Gamma_{\mathrm{c}}^{\infty}(\sf{E}_{\rho_{2}})\to\Gamma^{\infty}(\sf{E}_{\rho_{2}})$ define spacetime pseudo-covariances 
	\begin{align*}\lambda^{\pm}:=(\rho_{2}\sf{G}_{2})^{\ast}(\pm i\sf{G}_{\Sigma,2}c^{\pm})(\rho_{2}\sf{G}_{2})=\pm i\mathcal{U}_{2} c^{\pm}(\rho_{2}\sf{G}_{2})
	\end{align*}
	of a state $\omega$ on $\mathrm{CCR}(\mathcal{V}_{\mathrm{c}},\sigma)$, if the following conditions are satisfied
	\begin{align*}
		\emph{(i)}\quad &(c^{\pm})^{\dagger}=c^{\pm}\quad\text{w.r.t.}\quad \sigma_{\Sigma,2}\\
		\emph{(ii)}\quad & c^{\pm}(\mathrm{ran}(\sf{K}_{\Sigma}\vert_{\Gamma^{\infty}_{\mathrm{c}}}))\subset\mathrm{ran}(\sf{K}_{\Sigma}\vert_{\Gamma^{\infty}})\\
		\emph{(iii)}\quad & c^{+}+c^{-}=\mathrm{id}\\
		\emph{(iv)}\quad & \pm\sigma_{\Sigma,2}(\mathfrak{f},c^{\pm}\mathfrak{f})\geq 0\qquad\forall\mathfrak{f}\in\mathrm{ker}(\sf{K}_{\Sigma}^{\dagger}\vert_{\Gamma^{\infty}_{\mathrm{c}}})\, .
	\end{align*}
	The operators $c^{\pm}$ are called the \emph{Cauchy pseudo-covariances} of the state $\omega$.
\end{proposition}

\begin{proof}
	Recall the defining relation $\sf{G}_{2}=(\rho_{2}\sf{G}_{2})\sf{G}_{\Sigma,2}(\rho_{2}\sf{G}_{2})$. The claim then follows easily from the definition of $\sigma_{\Sigma,2}$ and the relations in Proposition~\ref{Lem:PropKSigma} and Proposition~\ref{Prop:PropGSigma}.
\end{proof}

\begin{remark}
	Instead of considering $c^{\pm}$, one could also consider the pair of operators $\lambda^{\pm}_{\Sigma}:=\pm i\sf{G}_{\Sigma,2}c^{\pm}$. In this case, the characterising conditions from the proposition above become
	\begin{align*}
		\text{(i)}\quad &(\lambda^{\pm}_{\Sigma})^{\ast}=\lambda^{\pm}_{\Sigma}\quad\text{w.r.t.}\quad \langle\cdot,\cdot\rangle_{\sf{E}_{\rho_{2}}}\\
		\text{(ii)}\quad & \lambda_{\Sigma}^{\pm}(\mathrm{ran}(\sf{K}_{\Sigma}\vert_{\Gamma^{\infty}_{\mathrm{c}}}))\subset\mathrm{ran}(\sf{K}_{\Sigma}\vert_{\Gamma^{\infty}})\\
		\text{(iii)}\quad & \lambda^{+}_{\Sigma}-\lambda^{-}_{\Sigma}=i\sf{G}_{\Sigma,2}\\
		\text{(iv)}\quad & \langle \mathfrak{f},\lambda^{\pm}_{\Sigma}\mathfrak{f}\rangle_{\sf{E}_{\rho_{2}}}\geq 0\qquad\forall\mathfrak{f}\in\mathrm{ker}(\sf{K}_{\Sigma}^{\dagger}\vert_{\Gamma^{\infty}_{\mathrm{c}}})\, .
	\end{align*}
	 The advantage of using $c^{\pm}$ is that $\sigma_{\Sigma,2}$ are \emph{physical}, i.e.~uniquely determined by the linear gauge theory, while the bundle metrics $(\cdot,\cdot)_{\sf{E}_{\rho_{i}}}$ can be chosen arbitrarily, cf.~Remark~\ref{Remark:PhysicalCharge}.
\end{remark}

Focusing on quasi-free states coming from pseudo-covariances already provides a restriction on the class of all states on the CCR-algebra $\mathrm{CCR}(\mathcal{V}_{\mathrm{c}},\sigma)$. However, there is still a plethora of states and we need a further selection criterion. A main motivation, we consider yet another essential property of the \textit{Minkowski vacuum state}. So far we have only considered \textit{linear} observables. However, from a physical perspective, it is clear that also non-linear observables play an essential role. For instance, consider the energy-momentum tensor of a massive scalar field (see Eq.~\eqref{eq:EMT}) with potential $\sf{V}(\phi):=m^{2}\phi^{2}$ for some $m\in\bb{R}_{>0}$, which contains also quadratic expressions of the fields such as $\phi^{2}$ and $\Vert\d_{s}\phi\Vert_{\sf{S}_{1}}^{2}=\sf{g}^{\alpha\beta}(\nabla_{\alpha}\phi)(\nabla_{\beta}\phi)$. 

Now, in the following, we briefly sketch how this problem can be overcome in the context of the real Klein-Gordon field in Minkowski spacetime, without dwelling into technical details. We refer to the extensive discussions in \cite{MorettiKhavkine} and \cite[Chap.~8]{GerardBook}. Let us consider a state $\omega$ on the CCR-algebra of the Klein-Gordon theory on Minkowski spacetime $\sf{M}:=\bb{M}^{1,k}$ and consider the corresponding GNS-construction $(\mathcal{H}_{\omega},\mathcal{D}_{\omega},\Omega_{\omega},\pi_{\omega})$. Now, let us write $\hat{\Phi}_{\omega}(f):=\pi_{\omega}(\Phi([f]))$ for all $f\in C^{\infty}_{\mathrm{c}}(\sf{M})$, which we view as an \textit{operator-valued distribution}. Now, in general, it is not possible to define $\hat{\Phi}_{\omega}^{2}$ directly, since the $2$-point function $\omega_{2}(\cdot,\cdot):=\omega(\Phi([\cdot])\Phi([\cdot]))\in\mathcal{D}^{\prime}(\sf{M}\times\sf{M})$ is usually divergent on the diagonal. This is for instance the case for the \textit{Minkowski vacuum state}. The solution to this problem is well known and consists of \textit{Wick ordering}. The idea is to fist define the Wick-ordered product
\begin{align*}
	\: \hat{\Phi}_{\omega}(f)\hat{\Phi}_{\omega}(g)\: :=\hat{\Phi}_{\omega}(f)\hat{\Phi}_{\omega}(g)-\langle\Omega_{\omega},\hat{\Phi}_{\omega}(f)\hat{\Phi}_{\omega}(g)\Omega_{\omega}\rangle_{\mathcal{H}_{\omega}}\mathds{1} \, .
\end{align*}
As it turns out, the pull-back of this distribution on the diagonal is well-defined due to Hörmander's theorem of multiplication of distributions and the singularity structure of the 2-point function of the Minkowski vacuum. Hence, we obtain a well-defined observable $\: \hat{\Phi}^{2}_{\omega}(f)\:$ in this case, called a \textit{Wick polynomial}. In other words, the idea of the Wick ordering procedure is to subtract the vacuum contribution to obtain a well-defined object. We also note that the previous discussion of Wick polynomials is equivalent in Minkowski spacetime to the popular procedure in the physics literature based on the re-ordering of creation and annihilation operators (see for instance \cite[Sec.~4.3]{Peskin} for a standard treatment of the latter approach).

Hence, while on general curved (and non-static) spacetimes, there is in general no way to define a ``preferred vacuum state'', the idea is to consider those states whose 2-point function mimics the ultraviolet structure of the Minkowski vacuum state, in order to be able to define physical important observables such as the energy-momentum tensor. The corresponding condition can be formulated conveniently using the language of \textit{microlocal analysis}.

\begin{definition}\label{Def.Hadamard} (Hadamard State)\newline
	A quasi-free state $\omega$ on the CCR-algebra $\mathrm{CCR}(\mathcal{V}_{\mathrm{c}},\sigma)$ of some linear gauge theory $(\sf{E}_{1},\sf{E}_{2},\sf{P},\sf{K})$ coming from a pair of pseudo-covariances $\lambda^{\pm}\:\Gamma^{\infty}_{\mathrm{c}}(\sf{E}_{2})\to\Gamma^{\infty}(\sf{E}_{2})$ is called \textit{Hadamard}, if
	\begin{align*}
		\mathrm{WF}^{\prime}(\lambda^{\pm})\subset\mathcal{N}^{\pm}\times\mathcal{N}^{\pm}\,,
	\end{align*}
	where we identify $\lambda^{\pm}$ with its Schwartz kernel $\lambda^{\pm}\in\mathcal{D}^{\prime}(\sf{M}\times\sf{M},\sf{E}\boxtimes\sf{E}^{\ast})$ and where $\mathcal{N}^{\pm}$ are the \textit{positive/negative energy cones}, i.e. 
	\begin{align*}
		\mathcal{N}^{\pm}:=\{\xi\in\sf{T}^{\ast}\sf{M}\mid \sf{g}^{\sharp}(\xi,\xi)=0\text{ and }\pm \xi(v)>0 \text{ for all future-directed timelike } v\in\sf{T}\sf{M}\}\, .
	\end{align*}
\end{definition}

The \emph{wavefront set} $\mathrm{WF}(\sf{T})\in\sf{T}^{\ast}\sf{M}\backslash\{\textbf{0}\}$ of a distribution $\sf{T}\in\mathcal{D}^{\prime}(\sf{M},\sf{E})$ is a refinement of the singular support of $\sf{T}$ that includes additional informations on the singular direction in Fourier space. In particular, the above definition makes precise the idea of \emph{states with the same singular ultraviolet structure of the Minkowski vacuum}. For details on distributions, the Schwartz kernel theorem and microlocal analysis, we refer to Appendix~\ref{Appendix:Micro}.

\begin{remarks}
\begin{itemize}
	\item[]
	\item[(i)]With the general definition of Hadamard states on curved spacetimes, \textit{Wick polynomials} can then also be defined as abstract algebraic objects. More precisely, one can define an \textit{extended algebra} including Wick polynomials, see e.g.~the work of Brunetti-Fredenhagen-Kohler \cite{BrunettiFredenhagenKohler}, Hollands-Wald \cite{HollandsWald1,HollandsWald2} as well as the review by Khavkine-Moretti \cite{MorettiKhavkine}.
	\item[(ii)]The concept of Hadamard states emerged out of the the question of how to rigorously define a regularised quantum energy-momentum tensor. This quantity is, of course, interesting on its own right, as it described the energy and momentum content of a theory, but it is also of central relevance in \textit{semiclassical gravity}, in which one studies the \textit{backreactions} of a \textit{quantum} field $\phi$ on a \textit{classical} gravitational background $\sf{g}$ by means of the \textit{semiclassical Einstein equations} $\mathrm{Ein}_{\Lambda}(\sf{g})=\omega_{\sf{g}}(\mathclose{:}\sf{T}(\sf{g},\phi)\mathclose{:})$, where $\omega_{\sf{g}}$ is a state on the corresponding algebra of observables, see the introduction to Section~\ref{Sec:SHSNonLoc}. In particular, it has been observed in the 1970s by Wald \cite{Wald1977,Wald1978}, Christensen \cite{Christensen} and Adler-Lieberman-Ng \cite{AdlerLieberman1,AdlerLieberman2} that the right-hand side of the above equation can be consistently defined if $\omega_{\sf{g}}$ has the same short-distance behaviour as the Minkowski vacuum. The notion of \textit{Hadamard states} was then further developed and clarified by Fulling-Sweeny-Wald \cite{FullingSweenyWald}, Fulling-Narcowich-Wald \cite{FullingNarcowichWald} and Kay-Wald \cite{KayWald}, among others, in which also the viewpoint has emerged that only states satisfying the Hadamard condition should be considered as ``physically acceptable''. In those early papers, the Hadamard condition is usually formulated by specifying the singularities of the 2-point function explicitly in terms of the so-called \textit{Hadamard parametrix}, a local approximate solution (a ``parametrix'') of the wave equation written in terms of geometric quantity developed by Hadamard \cite{Hadamard} in the 1920s (see also \cite{DeWitt} and \cite[Sec.~17.4]{HormanderIII}). In particular, Kay-Wald described in the aforementioned paper \cite{KayWald} of 1991 a \textit{global} description of the Hadamard parametrix and provided the first definition of Hadamard states (see \cite{MorettiHadamard} for a recent technical correction to the original definition). It was then later observed in the seminal works of Radzikoswki \cite{Radzikowski1,Radzikowski2} that the Hadamard condition written in terms of the global Hadamard parametrix can be related to the theory of distinguished parametrices \cite{DuistermaatHormander} and microlocal analysis. In particular, it has been shown that the definition of Kay-Wald can equivalently be characterised by a condition of the wavefront set of the $2$-point function (see \cite[Sec.~8.4]{GerardBook} for a modern proof). The condition in Definition~\ref{Def.Hadamard} is slightly different to the one given by Radzikowski and its equivalence can be found in \cite[Sec.~8.4]{Gerard} and \cite{WrochnaPhD}. All the previous sources consider \emph{scalar fields}; for fields taking values in a vector bundle, the Hadamard condition has been worked out by Sahlmann-Verch \cite{SahlmannVerch} (with earlier works for \emph{Dirac fields}, e.g.~\cite{HollandsDirac,Kohler,Verch}).
	\item[(iii)]It has been argued in \cite{FewsterVerch} in the specific case of ultrastatic spacetimes with compact Cauchy surfaces that the Hadamard condition is in fact \textit{necessary} for the Wick squares of all time derivatives a quantised Klein-Gordon field to have finite fluctuations.
\end{itemize}
\end{remarks}

For applications it is useful to consider an equivalent condition for the Cauchy pseudo-covariances of a quasi-free Hadamard states. The prove of the following condition is essentially an application of the well-known \textit{propagation of singularities} theorem (see Section~\ref{Subsec:PropSing}) and can be found in \cite[Prop.~2.11]{GerardMurroWrochna} (cf.~also \cite[Prop.~11.1.1]{Gerard})).

\begin{proposition}\label{Prop:HadCauchyCov} Let $\omega$ be a quasi-free state on the CCR-algebra $\mathrm{CCR}(\mathcal{V}_{\mathrm{c}},\sigma)$ of a linear gauge theory $(\sf{E}_{1},\sf{E}_{2},\sf{P},\sf{K})$ coming from a pair of Cauchy pseudo-covariances $c^{\pm}\:\Gamma^{\infty}_{\mathrm{c}}(\sf{E}_{2})\to\Gamma^{\infty}(\sf{E}_{2})$. If $\sf{D}_{2}$ is normally hyperbolic, then $\omega$ is a Hadamard state, if
	\begin{align*}
		\exists\,\mathcal{U}\subset\sf{M}\text{ neighbourhood of }\Sigma:~\mathrm{WF}^{\prime}(\mathcal{U}_{2}\circ c^{\pm})\subset(\mathcal{N}^{\pm}\times\mathcal{F})\times\sf{T}^{\ast}\Sigma\text{ over }\mathcal{U}\times\sf{M}\,,
	\end{align*}
	where $\mathcal{F}\subset\sf{T}^{\ast}\sf{M}$ is a conic set with $\mathcal{F}\cap\mathcal{N}^{\pm}=\emptyset$.
\end{proposition}
 
\subsection{Existence of Hadamard States and Construction: An Overview}\label{Sec:BiblioStates}
In this final section, we provide a bibliography and literature review on results concerning the existence and explicit construction of Hadamard states for various field theories. We stress that this survey is by no means complete, given the vast amount of literature on Hadamard states, but rather, reflects a selection of research directions that are particularly relevant in the context of this thesis through some representative and influential works. Hence, the focus naturally emphasises certain approaches and contributions, while other important developments and articles may not be covered.

\paragraph{Klein-Gordon Theory:} The general existence of Hadamard states for Klein-Gordon theory on general globally hyperbolic spacetimes has been established by Fulling-Narcowich-Wald \cite{FullingNarcowichWald} (see~\cite[Thm.~8.9.1]{Gerard} or \cite[Sec.~4.10.2]{GerardHafnerBook} for a modern proof). In their proof, Hadamard states are first explicitly constructed on ultrastatic spacetimes, generalising the definition of the vacuum state from Minkowski spacetime. We briefly review the central construction and argument following the aforementioned references: let $(\Sigma,\sf{h})$ be geodesically complete Riemannian manifold with Levi-Civita connection $\nabla^{\sf{h}}$ and let $\sf{M}=\bb{R}\times\Sigma$ with $\sf{g}=-\d t\otimes \d t+\sf{h}$ be the corresponding ultrastatic spacetime. Furthermore, consider the Klein-Gordon operator
\begin{align*}
	\sf{P}=-\square+m^{2}=\partial_{t}^{2}+\varepsilon^{2}=(\partial_{t}+i\varepsilon)(\partial_{t}-i\varepsilon)\,\qquad\varepsilon^{2}:=-\Delta+m^{2}\,,
\end{align*}
acting on complex scalar fields $C^{\infty}(\sf{M},\bb{C})$, for a mass $m>0$, where $\Delta:=\sf{h}^{ij}\nabla^{\sf{h}}_{i}\nabla^{\sf{h}}_{j}$ is the Laplace-Beltrami operator of $(\Sigma,\sf{h})$. Now, since $(\Sigma,\sf{h})$ is geodesically complete, $\varepsilon^{2}$ is essentially self-adjoint on $\sf{L}^{2}(\Sigma,\d\mu_{\sf{h}})$ with spectrum $\sigma(\varepsilon)\subset [m^{2},\infty)$. In particular, using spectral calculus, it follows that the square root $\varepsilon$ of its unique self-adjoint extension is well-defined, self-adjoint and invertible on its respective domain. Now, with this notation, it is not too hard to see that
\begin{align*}
	c^{\pm}:=\frac{1}{2}\begin{pmatrix}
	\mathrm{id} & \pm \varepsilon^{-1}\\ \pm\varepsilon^{\pm} & \mathrm{id}
	\end{pmatrix}\qquad\text{on}\qquad C_{\mathrm{c}}^{\infty}(\Sigma,\bb{C})\oplus C^{\infty}_{\mathrm{c}}(\Sigma,\bb{C})
\end{align*}
are the Cauchy pseudo-covariances of a quasifree Hadamard state on the CCR-algebra of the complex Klein-Gordon theory defined by $\sf{P}$. In fact, if we view the Klein-Gordon system equivalently as the evolutionary problem
\begin{align*}
	\sf{S}\Psi:=(\partial_{t}-i\sf{H})\Psi\,,\qquad\Psi:=\begin{pmatrix}
		\psi \\ i^{-1}\partial_{t}\psi
	\end{pmatrix}\,,\qquad\sf{H}:=\begin{pmatrix}
		0 & \mathrm{id}\\ \varepsilon^{2} & 0
	\end{pmatrix}
\end{align*}
with \textit{Hamiltonian} $\sf{H}$, a straightforward computation shows that $\mathcal{U}_{\sf{S}}(t)c^{\pm}\mathfrak{f}=e^{\pm i\varepsilon t}c^{\pm}\mathfrak{f}$ for all $t\in\bb{R}$ and initial datum $\mathfrak{f}\in C^{\infty}_{\mathrm{c}}(\Sigma,\bb{C})\oplus C_{\mathrm{c}}^{\infty}(\Sigma,\bb{C})$, where $\mathcal{U}_{\sf{S}}(t):=e^{i\sf{H}t}$ denotes the Cauchy evolution operator of $\sf{S}$. In particular, $c^{+}$ (resp.~$c^{-}$) are the projectors onto the spaces of initial data that evolve into \textit{positive} (resp.~\textit{negative}) \textit{energy solutions}. Hence, this construction is a generalisation of the well-known \textit{frequency splitting} procedure used to define the Minkowski vacuum state.

Now, in order to obtain the existence of Hadamard states on general globally hyperbolic spacetimes, the idea of Fulling-Narcowich-Wald is to employ a \textit{deformation argument}: consider two globally hyperbolic metrics $\sf{g}_{1}$ and $\sf{g}_{2}$ on the same manifold $\sf{M}$ such that there exists a Cauchy surface $\Sigma$ of both $(\sf{M},\sf{g}_{1})$ and $(\sf{M},\sf{g}_{2})$ with the property that $\sf{g}_{1}=\sf{g}_{2}$ on a (causally compatible) neighbourhood $\mathcal{U}$ of $\Sigma$ in $\sf{M}$. Now, by construction, the causal propagators $\sf{G}_{1}$ and $\sf{G}_{2}$ of $\sf{P}_{\sf{g}_{1}}$ and $\sf{P}_{\sf{g}_{2}}$, respectively, agree on $\mathcal{U}\times\mathcal{U}$ and hence, if $c^{\pm}_{1}$ are the Cauchy pseudo-covariances of a Hadamard state on $(\sf{M},\sf{g}_{1})$, we see that the corresponding state defined on $(\sf{M},\sf{g}_{2})$ via $\sf{G}_{2}$ will also be Hadamard on $\mathcal{U}\times\mathcal{U}$. The propagation of singularity theorem (see Section~\ref{Subsec:PropSing}) then implies that the state on $(\sf{M},\sf{g}_{2})$ is Hadamard everywhere. In order to apply this fact, the idea is to choose an \textit{interpolating metric} between an arbitrary globally hyperbolic metric $\sf{g}$ on $\sf{M}$ and an ultrastatic metric $\sf{g}_{\mathrm{u}}$ on $\sf{M}$ and applying this argument twice.

While the construction of Fulling-Narcowich-Wald works in full generality, it has the disadvantage of being very indirect. A general construction of Hadamard states for Klein-Gordon fields without a deformation argument using techniques from pseudodifferential calculus has been achieved by Gérard-Wrochna \cite{GerardWrochnaPseudo} and Gérard-Oulghazi-Wrochna \cite{GerardOulghaziWrochna} for globally hyperbolic manifolds of \textit{bounded geometry}\footnote{Riemannian (and Lorentzian) manifolds of \textit{bounded geometry} are a class of manifolds that behave ``nicely'' at infinity and allow for a well-defined pseudodifferential calculus. We will define this concept in more detail in the next chapter. For completeness, we just mention that virtually all globally hyperbolic spacetimes of physical interest are included in that class.}, also allowing for general smooth potentials, (a detailed treatment of the same proof can also be found in the monograph \cite[Chap.~11]{Gerard}), following earlier works of Junker-Schrohe \cite{Junker1,Junker2,JunkerSchrohe}. The proof is in fact similar in spirit of the construction for ultrastatic manifolds reviewed above, i.e., by factorising the Cauchy evolution operator accordingly. However, due to the time-dependence of the metric $\sf{h}$, the splitting becomes more complicated and requires methods of pseudodifferential calculus and microlocal analysis. We will review this method in more details lateron in this thesis.

Using these pseudodifferential techniques, one in fact constructs an entire family of states. For concrete applications, however, the question remains as to which state should be selected. In many cases, particularly when working with spacetimes of direct physical relevance, this choice is often guided by suitable invariance properties with respect to the isometries of the background. Beyond the general existence proofs and constructions, there is hence also a substantial body of literature devoted to the explicit construction of Hadamard states on specific spacetimes, often emphasising physical properties of direct importance for applications. An important line of research is the construction of Hadamard state for conformally coupled scalar fields from asymptotic data for asymptotically flat spacetimes, initiated by Dappiaggi-Pinamonti-Moretti \cite{Moretti1,Moretti2,DMP1} (see also the monograph \cite{DMPBook}). In this work, it is shown that for asymptotically flat spacetimes at future null infinity $\mathscr{I}^{+}$ (see~\cite[Chap.~11]{Wald}) naturally supports a bosonic quantum field theory on its boundary. Within this framework, a distinguished quasifree state can be constructed using the celebrated \textit{BMS-group} \cite{BMS1,BMS2}, which encodes the asymptotic symmetries of $\mathscr{I}^{+}$. Moreover, there exists an injective $\ast$-homomorphism from the algebra of observables in the bulk to that at $\mathscr{I}^{+}$, allowing one to pull back the boundary state to the bulk,  which has been shown to satisfy the Hadamard condition and to be invariant under all isometries of the bulk.

The previous results have also been expanded to other situations, namely for cosmological spacetimes in~\cite{DMP2,DMP3} and to the construction of the so-called \textit{Unruh state} on Schwarzschild spacetime \cite{DMP4}. As a side comment, it should be mentioned that the latter construction is more complicated than the construction of states for asymptotically flat spacetimes at future null infinity, since in the case of Schwarzschild spacetime one needs suitable decay bounds of solutions to the wave equations at $i^{\pm}$ in order to construct the aforementioned injective $\ast$-algebra homomorphism. In particular, the result in \cite{DMP4} depends on the highly non-trivial bounds derived by Dafermos-Rodnianski in \cite{DafermosRodnianski}. Using asymptotic/boundary data, Hadamard states have also been constructed for various different types of spacetimes and physical situations using various different techniques and estimates, e.g.~\cite{BrumJoras,SandersHad,Klein,VasyWrochna,HollandsHad,GerardWrochnaBH1,GerardWrochnaBH2}.

\paragraph{Dirac Theory:} An explicit construction of Hadamard states for Dirac fields on general spatially compact globally hyperbolic manifolds has been performed by Capoferri-Murro in \cite{CapoferriMurro} by writing the Cauchy evolution operator as a sum of to \textit{global} Fourier integral operators with complex-valued phase functions, based on the \textit{global parametrix} techniques developed in \cite{Laptev,Capoferri1,Capoferri2,Capoferri3}, and relating it to the Feynman propagators. An alternative construction using methods from pseudodifferential calculus, as mentioned above, has been applied to globally hyperbolic spacetimes of bounded geometry by Gérard-Stoskopf \cite{GerardStoskopf}. The existence of Hadamard states for Dirac fields can also be proven by means of a deformation argument, as shown by Murro-Volpe \cite{MurroVolpe}, using \textit{intertwining operators} (with an extension to globally hyperbolic manifolds with timelike boundaries in \cite{MurroGinouxDrago}). Last but not least, an alternative construction of Hadamard states from Feynman propagators has  been discussed by Islam-Strohmaier \cite{IslamStrohmaier} and Fewster-Strohmaier \cite{FewsterStrohmaier}, generalising the classic approach based on \emph{distinguished parametrices} by Duistermaat-Hörmander \cite{DuistermaatHormander} to the case of normally hyperbolic operators on Hermitian vector bundles with positive definite bundle metrics.

Earlier works by Sahlmann-Verch \cite{SahlmannVerch2} have shown that the ground and thermal states for Dirac fields on stationary spacetimes are Hadamard. Furthermore, employing asymptotic techniques, Dappiaggi-Hack-Pinamonti \cite{DHP} have constructed Hadamard states for Dirac fields on a large class of FLRW-spacetimes. Based on the scattering calculus developed by Häfner-Nicolas \cite{HafnerNicolas}, the Unruh state for massless Dirac fields on (very slowly rotating) Kerr spacetime has been constructed and its Hadamard property proven by Gérard-Häfner-Wrochna \cite{GerardHafner} (see also Häfner-Klein \cite{HafnerKlein} for a recent extension of this result). A different line of research is based on the functional analytic construction of the \textit{fermionic projectors} developed in \cite{FinsterFer1,FinsterFer2}, which has been applied to construct Hadamard states for (massive) Dirac fields in Minkowski \cite{Fer1}, Rindler \cite{Fer2} and de Sitter spacetime \cite{Fer3}.

\paragraph{Proca Theory:} The existence of Hadamard states for Proca fields has been established for general globally hyperbolic spacetimes by Moretti-Murro-Volpe in \cite{MurroProca}. Proca field theory is somehow a special situation, because despite being not a gauge theory, it shares some difficulties with them. The Proca equation is given by 
\begin{align*}
	\sf{P}=\delta\d+m^{2}\:\Omega^{1}(\sf{M})\to\Omega^{1}(\sf{M})
\end{align*}
for some mass $m>0$. It is not normally hyperbolic, but rather pre-normally hyperbolic, see Example~\ref{Ex:PreNor}. The bundle metric of the theory is given by the de Rham-Hodge sequilinear form on $\Omega^{1}(\sf{M})$ and hence in particular not positive-definite. Furthermore, it is easy to show that the Proca equation $\sf{P}A=0$ is equivalent to the hyperbolic equation $(\square_{\mathrm{dRH}}+m^{2})\sf{A}=0$ with the additional constraint $\delta\sf{A}=0$.

Now, it has been argued that a general vector Klein-Gordon operator with a non-positive fibre metric does not admit a Hadamard state under some mild analytic assumptions, see~\cite[Sec.~6.3]{GerardWrochna} and \cite{FewsterStrohmaier} as well as the comment after Prop.~5.6 in \cite{SahlmannVerch} for an earlier mentioning of this fact. Nevertheless, does not affect the Proca case. Indeed, it has been shown that the state is positive on divergence-free solutions to the Klein-Gordon equation. The construction of states in~\cite{MurroProca} is explicitly performed on ultrastatic manifolds of bounded geometry and the existence on general globally hyperbolic spacetimes by means of deformation argument based on suitable Møller operators as defined in \cite{Paracausal}.

\paragraph{Maxwell and Linearised Yang-Mills Theory:}
The existence of Hadamard states for Maxwell's theory was  investigated under various assumptions on the Cauchy hypersurface $\Sigma$ of the spacetime: Fewster-Pfenning~\cite{FewsterPfenning} (generalising results of Furlani~\cite{FurlaniMaxwell} in the case of static spacetimes) assumed $\Sigma$ to be compact and with vanishing first cohomology group, Finster-Strohmaier~\cite{FinsterStrohmaier}, extending the so-called \textit{Gupta–Bleuler formalism} \cite{Gupta,Bleuler}, considered Cauchy surfaces $\Sigma$ subjected to an ``absence of zero resonance'' condition for the Hodge Laplacian on 1-forms, while Dappiaggi-Siemssen~\cite{DappiaggiSiemssen} worked out the construction on asymptotically flat spacetimes with similar methods mentioned above. The construction of Hadamard states for Maxwell's theory on general globally hyperbolic manifolds has been established by Murro-Schmid~in\cite{MurroSchmid} and this paper will be the content of Chapter~\ref{Chap:Maxwell} of this thesis.

The construction of Hadamard states for linearised Yang-Mills theory was first studied by Hollands~\cite{Hollands} (within the BRST formalism), which considered the theory linearised around the zero solution on globally hyperbolic spacetimes with compact Cauchy surface and
vanishing first cohomology group. More recently, Gérard-Wrochna~\cite{GerardWrochna} constructed Hadamard states with techniques from pseudodifferential calculus for Yang-Mills theory linearised around non-zero solutions on globally hyperbolic spacetimes of bounded geometry in the case in which $\Sigma$ is compact and parallelisable and the case $\Sigma=\bb{R}^{n}$ with a Riemannian metric $\sf{h}$ satisfying 
\begin{align*}
c^{-1}  \leq [\sf{h}_{ij}(x)] \leq c \, \quad \text{for } c > 0  \qquad \text{ and } \qquad |\partial_x^\alpha \sf{h}_{ij} (x)| \leq C_\alpha \quad \forall \alpha \in \mathbb{N}^n \quad x \in \bb{R}^n\, .
\end{align*}

\paragraph{Linearised Gravity:} For linearised gravity, the existence of Hadamard states has been established for  Cauchy compact globally hyperbolic spacetimes of bounded geometry by Gérard in \cite{Gerard} via an approach based on complete gauge fixing. Furthermore, Hadamard stated for linearised gravity for asymptotically flat spacetimes, using asymptotic data as in the aforementioned works by Dappiaggi-Pinamonti-Moretti and Dappiaggi-Siemssen, have been constructed by Benini-Dappiaggi-Murro \cite{BeniniMurro}, however, the quantisation turns out to be limited to a subspace of observables due to divergences at null infinity. An open question in this direction is whether there exists a gauge condition that propagates to $\mathcal{I}^{+}$. 

An earlier result about existence of Hadamard states for linearised gravity has appeared in the PhD thesis of Hunt \cite{HuntPhD}, in which it has been shown that the Fock vacuum state of linearised gravity in Minkowski spacetime is Hadamard. Furthermore, in the preprint of \cite[App.~D of v3 on arXiv]{RejznerGravity}, it has been shown that there exists a Hadamard state for linearised gravity for every ultrastatic and Cauchy compact globally hyperbolic manifold, employing the BRST formalisms and using functional analytic methods. The existence of Hadamard states for gravity linearised around a \textit{general} globally hyperbolic spacetime with \textit{non-compact} Cauchy surfaces remains an important open question. A possible approach is discussed in the outlook section of this thesis.

An approach for globally hyperbolic manifold of bounded geometry that are analytic in Gaußian time based on Wick rotation has been discussed by Gérard-Murro-Wrochna in \cite{GerardMurroWrochna}. In this approach, the relevant normally hyperbolic operators are transformed into elliptic operators on the Wick rotated manifold, which, however, are only defined in some strip in imaginary Gaussian time. Then, one can define a quasi-free state for the Lorentzian theory from the \textit{Calderón
projectors} of the elliptic theory, a well-known tool in elliptic boundary value problems. In the context of Hadamard states, the use of Calderón projectors to construct Hadamard states has been first discussed for Klein-Gordon fields by Gérard-Wrochna in \cite{GerardWrochnaCalderon}. However, for linear gauge theories, this approach inherits a number of difficulties: the Wick rotated elliptic operators should be invertible and for this, one needs to impose proper boundary conditions. In~\cite{GerardMurroWrochna}, Dirichlet conditions for all the metric components are used, which has the advantage of easily giving invertibility of the Wick rotated operators and a modified positivity property. However, they are not gauge invariant. Hence, the states constructed in~\cite{GerardMurroWrochna} are only gauge-invariant (and also positive) up to a smoothing contribution. Motivated by this problem, the spectral properties of more general gauge-invariant boundary conditions satisfying the \textit{Shapiro-Lopatinski conditions} (which guarantee a well-posed elliptic boundary value problem in suitable Sobolev spaces \cite{SL1,SL2}, see e.g.~\cite[Chap.~XX]{HormanderIII} and \cite{SL} for a modern discussion) for linearised Euclidean gravity have been discussed by Capoferri-Murro-Schmid~in \cite{SchmidCapoferriMurro}, generalising boundary conditions considered by Anderson \cite{Anderson,Anderson2,Anderson3} (cf.~also the review in \cite{Witten}) in the linearised setting, however, none of them gives invertibility of the relevant elliptic operators on \textit{all} spacetimes. Nevertheless, building up on \cite{GerardMurroWrochna}, a Hadamard states for linearised gravity on de Sitter space have been constructed by Gérard-Wrochna in \cite{GerardWrochna2}, strongly relying on the fact that the Wick rotation of de Sitter spacetime is, in fact, a \textit{compact} Riemannian manifold.

In the specific case of Kerr spacetime, another line of research it to circumvent the difficulties arising from the gauge freedom of linearised gravity by expressing the theory equivalently in terms of the \textit{Teukolsky scalar fields} \cite{Teu1,Teu2} and to construct Hadamard states in the scalar setting, see e.g.~the recent articles~\cite{KleinTeu,IulianoTeu}.
\chapter{Quantisation of Maxwell's Theory via Complete Gauge Fixing}
\label{Chap:Maxwell}
Looking at the overview of existence theorems and explicit constructions of Hadamard states for various different linear field theories provided in~Section~\ref{Sec:BiblioStates}, it becomes evident that much less is known for theories admitting a (non-trivial) gauge symmetry. The reason for this is that there are several additional technical difficulties present in this case. First of all, if $(\sf{E}_{1},\sf{E}_{2},\sf{P},\sf{K})$ is a linear gauge theory in the sense of Definition~\ref{Def:LinGaugeTh}, then, following our discussion of the Cauchy problem for linear gauge theories in Section~\ref{Sec:CauchyGauge}, we know that the solution space of the theory can be characterised as
\begin{equation*}
        \begin{tikzcd}
        		\mathrm{Sol}_{\mathrm{sc}}:=\cfrac{\mathrm{ker}(\sf{P}\vert_{\Gamma^{\infty}_{\mathrm{sc}}})}{\mathrm{ran}(\sf{K}\vert_{\Gamma^{\infty}_{\mathrm{sc}}})}&& \cfrac{\mathrm{ker}(\sf{D}_{2}\vert_{\Gamma^{\infty}_{\mathrm{sc}}})\cap\mathrm{ker}(\sf{K}^{\ast}\vert_{\Gamma^{\infty}_{\mathrm{sc}}})}{\sf{K}(\mathrm{ker}(\sf{D}_{1}\vert_{\Gamma^{\infty}_{\mathrm{sc}}}))}\arrow[ll,hook',"\cong"]\,,
        \end{tikzcd}
    \end{equation*}
where $\sf{D}_{2}:=\sf{P}+\sf{K}\sf{K}^{\ast}$ is the hyperbolic gauge-fixed operator with respect to the canonical (or \textit{subsidiary}) condition $\sf{K}^{\ast}\psi=0$ and where $\sf{D}_{1}:=\sf{K}^{\ast}\sf{K}$ is a hyperbolic operator parametrising the residual gauge freedom present. Now, naively, one idea would be to construct a state for the hyperbolic theory defined by $\sf{D}_{2}$, subsequently restrict it to the relevant subspace, and pull it back to the phase space of gauge-invariant observables. However, as one can see from our discussion of Section~\ref{Sec:Exam}, essentially all the examples of linear gauge theories of physical interest admit a bundle metric that is \emph{not} positive-definite, but merely non-degenerate. For example, in the case of Maxwell's theory of electromagnetism, the bundle metric is given by the Lorentzian metric $\sf{g}^{\sharp}$ on $\sf{T}^{\ast}\sf{M}$. But, as already mentioned previously, it is known that normally hyperbolic operators with a non-positive fibre metric do \emph{not} admit a Hadamard state, but only \emph{pseudo-states}, see e.g.~the discussion in~\cite[Sec.~6.3]{GerardWrochna}. Hence, the positivity condition of the state can only hoped to be achievable on the level of the quotient space. In other words, the non-positivity of the bundle metric is in tension with the positivity of the state to be constructed.

On the other hand, linear gauge theories are usually plagued by \textit{infrared divergencies}: following similar steps as explained in Section~\ref{Sec:BiblioStates}, states on ultrastatic spacetimes are usually constructed by considering projection operators onto the subspace of positive/negative-frequency solutions. In the case of Klein–Gordon theory, the definition of the corresponding Cauchy covariances $c^{\pm}$ requires defining the square root of the ``spatial part'' of the Klein-Gordon operator, i.e.~the operator $ \varepsilon:=(-\Delta + m^{2})^{1/2} $, which is well defined provided $m>0$ by means of spectral calculus. For massless theories, including gauge theories, however, this issue becomes more delicate. In the case of massless Klein–Gordon theory, the problem can be overcome by using tools from \emph{pseudodifferential calculus} (see e.g.~\cite[Chap.~11]{GerardBook}, \cite[Sec.~5]{GerardMurroWrochna}): the idea is essentially to find a suitable \emph{smoothing operator} $r_{-\infty}$ such that $\varepsilon^{2}:=-\Delta + r_{-\infty}$ admits a well-defined and invertible square root. Since $r_{-\infty}$ is only a smoothing contribution, it does not affect the Hadamard property and hence allows one to construct a state in a similar manner. For gauge theories, however, one must additionally respect \emph{gauge invariance}, encoded in the condition $c^{\pm}\mathsf{K}_{\Sigma}=\mathsf{K}_{\Sigma}\widetilde{c}^{\pm}$ for some suitable $\widetilde{c}^{\pm}$, see~Proposition~\ref{Prop:CauchyCov}. This requirement is, of course, spoiled by the introduction of $r_{-\infty}$. In other words, the pseudodifferential techniques that successfully resolve infrared issues in the scalar case turn out to be inherent conflict with the gauge symmetry. This tension creates a significant obstruction for the construction of Hadamard states and calls for some more refined techniques and approaches to be developed.

Several approaches to this problem have been proposed in the literature, as discussed in the bibliography in Section~\ref{Sec:BiblioStates}. The goal of this chapter, based on the article \cite{MurroSchmid}, is to prove the existence of Hadamard states for \emph{Maxwell’s theory} on general globally hyperbolic spacetimes. Our strategy relies on a \emph{complete gauge fixing} procedure. Specifically, we impose a gauge condition at the level of the initial data, which eliminates all the unphysical degrees of freedom. Moreover, the condition is chosen in such a way that it removes the negative contributions in the bundle metric at the initial-data stage, thereby enabling the construction of positive Cauchy covariances for the fully gauge-fixed theory. States for Maxwell’s theory are then defined as the pull-back along the gauge-fixing projector, and we show that this construction preserves the Hadamard property. A central ingredient in establishing this new gauge condition is a novel Hodge decomposition for differential forms in Sobolev spaces on complete Riemannian manifolds and a dedicated section is devoted to the development of this decomposition and the needed functional analytic methods. The construction of states will then explicitly be performed on \emph{ultrastatic spacetimes}, while the existence on general globally hyperbolic spacetimes follows from a \emph{deformation argument}. We emphasise that, although our construction of states relies on a complete gauge-fixing procedure, the underlying algebra of observables remains that of \emph{gauge-invariant} observables and the role of gauge fixing is purely auxiliary.

As it turns out, the construction of states via complete gauge fixing at the level of initial data is both natural and admits several technical advantages. Besides Maxwell's theory, a strategy based on complete gauge fixing has recently also independently been worked out for linearised gravity on spatially compact globally hyperbolic spacetimes by Gérard in~\cite{Gerard}.

\section{Maxwell's Theory on Globally Hyperbolic Spacetimes}\label{Sec:MaxTheLorMa}
We have already discussed Maxwell’s theory briefly in Section~\ref{Sec:Maxwell}, where we also saw that it naturally fits into the framework of linear gauge theories of Hack–Schenkel (Definition~\ref{Def:LinGaugeTh}). In this section, we provide a more detailed analysis of Maxwell's theory to prepare the ground for the subsequent chapters. First of all, we discuss differential forms on globally hyperbolic spacetimes and their decomposition into temporal and spatial parts adapted to the geometry of a chosen foliation by spacelike Cauchy hypersurfaces. We then provide explicit expressions for the all the operators acting on the space of initial data introduced in Section~\ref{Sec:CauchyGauge} and Section~\ref{Sec:PhaseGauge}, such as $\sf{K}_{\Sigma},\sf{K}_{\Sigma}^{\dagger}$ and $\sf{G}_{\Sigma,i}$. Last but not least, we examine various different gauge conditions for Maxwell's theory. In particular, we will define the \emph{Cauchy radiation gauge}, which will play a central role in the subsequent construction of Hadamard states.

The purpose of this section is twofold. On the one hand, it prepares the ground for the construction of Hadamard states of Maxwell's theory. On the other hand, it provides a detailed and self-contained treatment of both the Cauchy problem for Maxwell's equations on general globally hyperbolic spacetimes and the corresponding gauge problem. Although the construction of Hadamard states will be carried out specifically in \emph{ultrastatic spacetimes}, with the existence of such states on general spacetimes following from a \emph{deformation argument}, we hence keep the discussion here at a general level. This broader perspective is included both for its intrinsic interest and to illustrate how the ultrastatic case simplifies within the general framework.

\subsection{Differential Forms on Globally Hyperbolic Spacetimes}\label{Sec.DiffFormsGlobHyp}
Throughout this section, let $n\in\bb{N}$ and consider a $(1+n)$-dimensional globally hyperbolic spacetime $(\sf{M},\sf{g})$. Furthermore, we \emph{fix} a Cauchy temporal function $t\in C^{\infty}(\sf{M},\bb{R})$ and identify
\begin{align*}
	\sf{M}=\bb{R}\times\Sigma\,,\qquad \sf{g}=-\beta^{2}\d t\otimes\d t+\sf{h}_{t}\,,
\end{align*}
where $\Sigma$ is a smooth spacelike Cauchy hypersurface, $\beta\in C^{\infty}(\sf{M},(0,\infty))$ a lapse function and $(\sf{h}_{t})_{t\in\bb{R}}$ a one-parameter family of Riemannian metrics on $\Sigma$ depending smoothly on $t$, in accordance with the Bernal-Sánchez theorem (see Theorem~\ref{Thm:BernalSanchez}). The corresponding foliation of $\sf{M}$ by spacelike Cauchy surfaces is denoted by $(\Sigma_{t}:=\{t\}\times\Sigma)_{t\in\bb{R}}$, as usual.

In the following pages, we discuss the notation and terminology used for differential forms on $\sf{M}$, (time-dependent) forms on $\Sigma$ and for operators related to the geometry of the foliation $(\Sigma_{t})_{t\in\bb{R}}$. Furthermore, we introduce a natural a decomposition of differential forms on $\sf{M}$ into a \emph{temporal} and \emph{spatial part}. We refer to~pp.~\pageref{DiffForms}ff.~for the conventions used for differential forms.

\paragraph{Differential Forms on $\sf{M}$:} As opposed to the presentation of Maxwell's theory in Section~\ref{Sec:Maxwell}, we will work with \textit{complex-valued} fields, for convenience. With a slight abuse of notation, we use the same notation for the \textit{complex-valued} exterior bundles as we did for their real-valued analogue. More precisely, denoting by $\underline{\bb{C}}_{\sf{M}}$ the \emph{trivial $\bb{C}$-line bundle} over $\sf{M}$, we consider the Hermitian vector bundles
\begin{align*}
	\sf{A}_{k}:=\underline{\bb{C}}_{\sf{M}}\otimes_{\bb{R}}\bigwedge^{k}\sf{T}^{\ast}\sf{M}\,,\qquad\qquad \langle\alpha,\beta\rangle_{\sf{A}_{k}}:=\frac{1}{k!}(\sf{g}^{\sharp})^{\otimes k}(\overline{\alpha},\beta)=\frac{1}{k!}\overline{\alpha}^{\mu_{1}\dots\mu_{k}}\beta_{\mu_{1}\dots\mu_{k}}\, ,
\end{align*}
whose sections are smooth \textit{complex-valued} differential $k$-forms, i.e.~$\Omega^{k}(\sf{M},\bb{C}):=\Gamma^{\infty}(\sf{A}_{k})$. The corresponding non-degenerate Hermitian sesquilinear form on the level of sections is given by
\begin{align}\label{eq:HodgeSF}
	(\alpha,\beta)_{\sf{A}_{k}}:=\int_{\sf{M}}\,\langle\alpha,\beta\rangle_{\sf{A}_{k}}\,\d\mu_{\sf{g}}=\int_{\sf{M}}\overline{\alpha}\wedge\ast\beta
\end{align}
for all $\alpha,\beta\in\Omega^{k}(\sf{M},\bb{C})$ with $\mathrm{supp}(\alpha)\cap\mathrm{supp}(\beta)$ compact, where $\ast\:\Omega^{k}(\sf{M},\bb{C})\to\Omega^{d-k}(\sf{M},\bb{C})$ denotes the Hodge $\ast$-operator, as usual. Furthermore, following the notation section of this thesis (see~\pageref{DiffForms}ff.), we denote the exterior derivative and its formal adjoint with respect to $(\cdot,\cdot)_{\sf{A}_{\bullet}}$, the codifferential, by
\begin{align*}
	\d\:\Omega^{k}(\sf{M},\bb{C})\to\Omega^{k+1}(\sf{M},\bb{C})\,,\qquad\delta:=(-1)^{k+1}\ast^{-1}\d\ast\:\Omega^{k+1}(\sf{M},\bb{C})\to\Omega^{k}(\sf{M},\bb{C})\, .
\end{align*}
In particular, if $(x^{\mu})_{\mu=0,\dots,n}$ are local coordinates defined on some open subset $\mathcal{U}\subset\sf{M}$, the operator $\d$ and $\delta$ have the local expression
\begin{align*}
	(\d\omega)_{\alpha_{1}\dots\alpha_{k+1}}:=(1+k)\nabla_{[\alpha_{1}}\omega_{\alpha_{1}\dots\alpha_{k+1}]}\,,\quad(\delta\omega)_{\alpha_{1}\dots\alpha_{k-1}}:=-\sf{g}^{\mu\nu}\nabla_{\mu}\omega_{\nu\alpha_{1}\dots\alpha_{k-1}}\quad\forall\omega\in\Omega^{k}(\sf{M},\sf{g})
\end{align*}
on $\mathcal{U}$, where we denote by $\nabla\:\Gamma^{\infty}(\sf{T}^{p,q}\sf{M})\to\Gamma^{\infty}(\sf{T}^{p,q+1}\sf{M})$ the Levi-Civita connection of $(\sf{M},\sf{g})$. Last but not least, we will consider the \textit{de Rham-Hodge d'Alembertian} as defined in Example~\ref{Example:HodgeLapl}. Since we will mainly work with differential forms throughout this chapter, we will use the simplified notation
\begin{align*}
	\square:=\square_{\mathrm{dRH}}:=\d\delta+\delta\d\:\Omega^{k}(\sf{M},\bb{C})\to\Omega^{k}(\sf{M},\bb{C})\,,
\end{align*}
which should not be confused with the \textit{connection d'Alembertian} $\sf{g}^{\alpha\beta}\nabla_{\alpha}\nabla_{\beta}$ (see Example~\ref{Ex:ConDA}) for which we previously have used the notation $\square$ and which differs from the de Rham-Hodge d'Alembertian by the \textit{Weitzenböck identity}, i.e.
\begin{align*}
	\square=-\sf{g}^{\alpha\beta}\nabla_{\alpha}\nabla_{\beta}+\mathcal{R}_{k}\qquad\text{on}\quad\Omega^{k}(\sf{M},\bb{C})
\end{align*} 
for some zeroth-order operator $\mathcal{R}_{k}\in\mathrm{DO}^{0}(\sf{A}_{k})$ constructed out of the curvature of $(\sf{M},\sf{g})$. Important examples are $\square=-\sf{g}^{\alpha\beta}\nabla_{\alpha}\nabla_{\beta}$ on $0$-forms as well as $\square=-\sf{g}^{\alpha\beta}\nabla_{\alpha}\nabla_{\beta}+\mathrm{Ric}_{\sf{g}}$ with zeroth-order operator $\mathcal{R}_{1}:=\mathrm{Ric}_{\sf{g}}(\omega):=\mathrm{Ric}(\sf{g})_{\alpha}^{\beta}\omega_{\beta}$ for $1$-forms.

\paragraph{Differential Forms on $\Sigma$:} Let us fix a time $t\in\bb{R}$ for now. Denoting by $\underline{\bb{C}}_{\Sigma}$ the trivial $\bb{C}$-line bundle over $\Sigma$, we consider the Hermitian vector bundles
\begin{align*}
	\sf{A}_{\Sigma,k}:=\underline{\bb{C}}_{\Sigma}\otimes_{\bb{R}}\bigwedge^{k}\sf{T}^{\ast}\Sigma\,,\qquad\qquad \langle\alpha,\beta\rangle^{t}_{\sf{A}_{\Sigma,k}}:=\frac{1}{k!}(\sf{h}_{t}^{\sharp})^{\otimes k}(\overline{\alpha},\beta)=\frac{1}{k!}\overline{\alpha}^{i_{1}\dots i_{k}}\beta_{i_{1}\dots i_{k}}\, ,
\end{align*}
where indices are raised/lowered with $\sf{h}_{t}$, whose sections are smooth \textit{complex-valued} differential $k$-forms on $\Sigma$, i.e.~$\Omega^{k}(\Sigma,\bb{C}):=\Gamma^{\infty}(\sf{A}_{\Sigma,k})$ and where the bundle metric explicitly depends on time. As above, we denote the corresponding Hermitian sesquilinear form on sections by
\begin{align*}
	(\alpha,\beta)^{t}_{\sf{A}_{\Sigma,k}}:=\int_{\sf{M}}\,\langle\alpha,\beta\rangle^{t}_{\sf{A}_{\Sigma,k}}\,\d\mu_{\sf{h}_{t}}=\int_{\sf{M}}\overline{\alpha}\wedge\ast_{t}\beta\, ,
\end{align*}
where $\ast_{t}\:\Omega^{k}(\Sigma,\bb{C})\to\Omega^{d-k}(\Sigma,\bb{C})$ is the Hodge $\ast$-operator of $(\Sigma,\sf{h}_{t})$. $(\cdot,\cdot)^{t}_{\sf{A}_{\Sigma,k}}$ is positive-definite and defines an inner product on $\Omega_{\mathrm{c}}^{k}(\Sigma,\bb{C})$ for each $t\in\bb{R}$, which we will refer to as the \textit{de Rham-Hodge inner product}. The exterior derivative and codifferential on $(\Sigma,\sf{h}_{t})$ are given by
\begin{align*}
	\d_{\Sigma}\:\Omega^{k}(\Sigma,\bb{C})\to\Omega^{k+1}(\Sigma,\bb{C})\,,\qquad\delta_{\Sigma}^{t}:=(-1)^{k+1}\ast_{t}^{-1}\d\ast_{t}\:\Omega^{k+1}(\Sigma,\bb{C})\to\Omega^{k}(\Sigma,\bb{C})\, ,
\end{align*}
respectively. Now, let $(x^{i})_{i=1,\dots,n}$ be local coordinates of $\Sigma$ defined on some open subset $\mathcal{U}\subset\Sigma$ and let us denote the Levi-Civita connection of $(\Sigma,\sf{h}_{t})$ by $\nabla^{\Sigma_{t}}$. Then, locally 
\begin{align*}
	(\d_{\Sigma}\omega)_{i_{1}\dots i_{k+1}}:=(1+k)\nabla^{\Sigma_{t}}_{[i_{1}}\omega_{i_{1}\dots i_{k+1}]}\,,\quad(\delta_{\Sigma}^{t}\omega)_{i_{1}\dots i_{k-1}}:=-\sf{h}_{t}^{kl}\nabla_{k}^{\Sigma_{t}}\omega_{li_{1}\dots i_{k-1}}\quad\forall\omega\in\Omega^{k}(\sf{M},\sf{g})\, ,
\end{align*}
where we stress that $\d_{\Sigma}$ is independent of time and we could replace $\nabla^{\Sigma_{t}}$ in the formula above by any torsion-free connection. Last but not least, we denote the \textit{de Rham-Hodge Laplacian} of the Riemannian manifold $(\Sigma,\sf{h}_{t})$ by
\begin{align*}
	\Delta^{t}:=\d_{\Sigma}\delta_{\Sigma}^{t}+\delta_{\Sigma}^{t}\d_{\Sigma}\:\Omega^{k}(\Sigma,\bb{C})\to\Omega^{k}(\Sigma,\bb{C})\, .
\end{align*}
Similarly to the situations for the de Rham-Hodge d'Alembertian, $\Delta_{t}$ is related to the \textit{connection} or \textit{rough Laplacian} $\sf{h}_{t}^{ij}\nabla^{\Sigma_{t}}_{i}\nabla^{\Sigma_{t}}_{j}$ by means of a \textit{Weitzenböck identity}, i.e.~
\begin{align*}
	\Delta^{t}=-\sf{h}_{t}^{ij}\nabla^{\Sigma_{t}}_{i}\nabla^{\Sigma_{t}}_{j}+\mathcal{R}_{\Sigma,k}^{t}\qquad\text{on}\quad\Omega^{k}(\Sigma,\bb{C})
\end{align*} 
for some zeroth-order operator $\mathcal{R}_{\Sigma,k}^{t}\in\mathrm{DO}^{0}(\sf{A}_{\Sigma,k})$ constructed out of the curvature of $(\Sigma,\sf{h}_{t})$. 

\paragraph{Time-Dependent Differential Forms on $\Sigma$}
Now, if we denote by $\pi_{2}\:\sf{M}\to\Sigma$ the natural projection onto the second factor, let us also consider the complex vector bundles
\begin{align*}
	\pi_{2}^{\ast}(\sf{A}_{\Sigma,k})\cong\underline{\bb{C}}_{\sf{M}}\otimes_{\bb{R}}\bigwedge^{k}\pi^{\ast}_{2}(\sf{T}^{\ast}\Sigma)\, .
\end{align*}
By definition, sections of this bundle can be viewed as a \textit{time-dependent} differential $k$-forms on $\Sigma$ and we will use the suggestive notation
\begin{align*}
	C^{\infty}(\bb{R},\Omega^{k}(\Sigma,\bb{C})):=\Gamma^{\infty}(\pi_{2}^{\ast}(\sf{A}_{\Sigma,k}))\, .
\end{align*}
The space $C^{\infty}(\bb{R},\Omega^{k}(\Sigma,\bb{C}))$ can be naturally viewed as a $C^{\infty}(\sf{M})$-submodule of $\Omega^{k}(\sf{M},\bb{C})$. Furthermore, if $(x^{i})_{i=1,\dots,n}$ are local coordinates of $\Sigma$ defined on some open set $\mathcal{U}\subset\Sigma$ and we extend them to local coordinates $(x^{\mu})_{\mu=0,\dots,n}$ of $\sf{M}$ by adding $x^{0}:=t$, then any element $\omega\in C^{\infty}(\bb{R},\Omega^{k}(\Sigma,\bb{C}))$ can locally be written in the form $\omega=\omega_{i}(t,\vec{x})\d x^{i}$ for coefficient functions $\omega_{i}\in C^{\infty}(\bb{R}\times\mathcal{U},\bb{C})$. For elements of this space, we define the metric-like pairing $\langle\cdot,\cdot\rangle_{\sf{A}_{\Sigma,k}}$ and the operators $\nabla^{\Sigma},\d_{\Sigma},\delta_{\Sigma}$ and $\Delta_{\Sigma}$ acting pointwise in time. More precisely, we set
\begin{align*}
	\langle\alpha,\beta\rangle_{\sf{A}_{\Sigma,k}}&:=\frac{1}{k!}(\sf{h}^{\sharp})^{\otimes k}(\overline{\alpha},\beta)\in C^{\infty}(\sf{M},\bb{C})\,,\quad\text{i.e.}\quad (\langle\alpha,\beta\rangle_{\sf{A}_{\Sigma,k}})(t):=\langle\alpha(t),\beta(t)\rangle_{\sf{A}_{\Sigma,k}}^{t}
\end{align*}
for all $\alpha,\beta\in C^{\infty}(\bb{R},\Omega^{k}(\Sigma,\bb{C}))$, $(\nabla^{\Sigma}\alpha)(t):=\nabla^{\Sigma_{t}}(\alpha(t))$ for all $\alpha\in C^{\infty}(\bb{R},\Omega^{k}(\Sigma,\bb{C}))$ and
\begin{align*}
	&\d_{\Sigma}\:C^{\infty}(\bb{R},\Omega^{k}(\Sigma))\to C^{\infty}(\bb{R},\Omega^{k+1}(\Sigma))\,,\qquad (\d_{\Sigma}\alpha)(t):=\d_{\Sigma}(\alpha(t))\\
	&\delta_{\Sigma}\:C^{\infty}(\bb{R},\Omega^{k+1}(\Sigma))\to C^{\infty}(\bb{R},\Omega^{k}(\Sigma))\,,\qquad (\delta_{\Sigma}\alpha)(t):=\delta_{\Sigma}^{t}(\alpha(t))\\
	&\Delta\:C^{\infty}(\bb{R},\Omega^{k}(\Sigma))\to C^{\infty}(\bb{R},\Omega^{k}(\Sigma))\,,\qquad\quad\,\, (\Delta\alpha)(t):=\Delta^{t}(\alpha(t))\, .
\end{align*}

Last but not least, for a given time-dependent $k$-form $\omega\in C^{\infty}(\bb{R},\Omega^{k}(\Sigma,\bb{C}))$, it makes sense to define its \textit{(partial) time-derivative}. There are several, equivalent, ways to make this precise: first of all, we can define $\partial_{t}\omega\in C^{\infty}(\bb{R},\Omega^{k}(\Sigma,\bb{C}))$ to be the time-dependent $k$-form, which in local coordinates $(x^{i})_{i=1,\dots,n}$ on $\mathcal{U}\subset\Sigma$ takes the form $(\partial_{t}\omega)_{i_{1}\dots i_{k}}=\partial_{t}\omega_{i_{1}\dots i_{k}}$, where $\omega_{i_{1}\dots i_{k}}\in C^{\infty}(\bb{R}\times\mathcal{U},\bb{C})$ are the local components of $\omega$. It is clear that this independent of the chosen coordinates on $\Sigma$ and hence well-defined, however, we stress that it clearly depends on the choice of Cauchy temporal function $t\in C^{\infty}(\sf{M},\bb{R})$ (and hence on the foliation chosen), which here is assumed to be fixed from the beginning. From a differential geometric point of view, a more natural (and equivalent) definition is provided by setting
\begin{align*}
	\partial_{t}\omega:=\mathcal{L}_{\partial_{t}}\omega\in C^{\infty}(\bb{R},\Omega^{k}(\Sigma,\bb{C}))\,,
\end{align*}
where $\mathcal{L}_{\partial_{t}}\:\Omega^{k}(\sf{M},\bb{C})\to\Omega^{k}(\sf{M},\bb{C})$ denotes the Lie-derivative along $\partial_{t}$. The fact that this is indeed equivalent follows easily from the definition of $\mathcal{L}_{\partial_{t}}$.

\begin{remark}\label{Rem:FrechetSmooth}
	As explained above, we use the notation $C^\infty(\mathbb{R},\Omega^k(\Sigma,\mathbb{C}))$ for smooth sections of the bundle $\pi_2^*(\bigwedge^k \sf{T}^{\ast}\Sigma)$ and we interpret elements of this module as time-dependent $k$-forms on $\Sigma$, i.e.~$t\mapsto\omega(t,\cdot)\in\Omega^{k}(\Sigma,\bb{C})$, where smoothness has to be understood in the sense that $\omega$ is a smooth section of $\pi_2^*(\bigwedge^k \sf{T}^{\ast}\Sigma)$ on $\sf{M}$. This is, of course, the natural differential-geometric definition. From an analytic point of view, we note that this is equivalent to define  $C^\infty(\mathbb{R},\Omega^k(\Sigma,\mathbb{C}))$ as the space of assignments $t\mapsto \omega(t,\cdot) \in \Omega^k(\Sigma,\mathbb{C})$ that are smooth with respect to the \emph{Fréchet space structure} on $\Omega^k(\Sigma,\mathbb{C})=\Gamma^{\infty}(\underline{\bb{C}}_{\Sigma}\otimes\bigwedge^{k}\sf{T}^{\ast}\Sigma)$. In the special case $k=0$, this is known as the \emph{exponential law} and states that $C^{\infty}(\bb{R}\times\Sigma)=C^{\infty}(\bb{R},C^{\infty}(\Sigma))$. More generally, the exponential law states that there is a (topological) isomorphism $C^{\infty}(\sf{M}\times\sf{N},\mathcal{F})\cong C^{\infty}(\sf{M},C^{\infty}(\sf{N},\mathcal{F}))$ (in fact, even a diffeomorphism in a suitable sense), where $\sf{M}$, $\sf{N}$ are smooth manifolds and $\mathcal{F}$ a Fréchet space. We refer to \cite[Sec.~I.3, Thm.~3.12 and Cor.~3.13]{KrieglMichor} for details.
\end{remark}

\paragraph{Decomposition of Differential Forms:}
Now, using the notation introduced above, we note that every $\omega\in\Omega^{k}(\sf{M},\bb{C})$ can be decomposed into a \emph{temporal} and \emph{spatial} part as 
\begin{align*}
	\omega=\eta\wedge\omega_{\mathrm{T}}+\omega_{\Sigma}\quad\text{with}\quad\omega_{\mathrm{T}}:=\nu\lrcorner\omega\quad\text{and}\quad\omega_{\Sigma}:=\omega-\eta\wedge\omega_{t}\, ,
\end{align*}
where $\nu=\beta^{-1}\partial_{t}\in\mathfrak{X}(\sf{M})$ denotes the future-directed timelike unit normal vector field and $\eta:=-\nu^{\flat}=\beta\d t\in\mathfrak{X}^{\ast}(\sf{M})$ the corresponding future-directed timelike unit normal covector field of the foliation $(\Sigma_{t})_{t\in\bb{R}}$. Clearly $\omega_{\mathrm{T}}(\cdot,\dots,\nu,\dots,\cdot)=0$ and $\omega_{\Sigma}(\cdot,\dots,\nu,\dots,\cdot)=0$ and hence, $\omega_{\mathrm{T}}$ and $\omega_{\Sigma}$ are well-defined sections of $\pi^{\ast}_{2}(\sf{A}_{\Sigma,k-1})$ and $\pi^{\ast}_{2}(\sf{A}_{\Sigma,k})$, respectively. In local coordinates $(x^{0}=t,x^{i})_{i=1,\dots,k}$, as above,  the above decomposition of $\omega$ reads
\begin{align*}
	\omega=\frac{1}{k!}\omega_{\mu_{1}\dots\mu_{k}}&\d x^{\mu_{1}}\wedge\dots\wedge\d x^{\mu_{k}}=\\&=\beta\d t\wedge\bigg(\frac{1}{(k-1)!}\frac{1}{\beta}\omega_{0i_{2}\dots i_{k}}\d x^{i_{2}}\wedge\dots\wedge\d x^{i_{k}}\bigg)+\frac{1}{k!}\omega_{i_{1}\dots i_{k}}\d x^{i_{1}}\wedge\dots\wedge\d x^{i_{k}}\, .
\end{align*}
In other words, if the components of $\omega$ in local coordinates are given by $\omega_{\mu_{1}\dots\mu_{k}}$, then $\omega_{\mathrm{T}}$ is the $(k-1)$-form on $\Sigma$ with time-dependent components $\frac{1}{\beta}\omega_{0i_{2}\dots i_{k}}$ and $\omega_{\Sigma}$ the time-dependent $k$-form on $\Sigma$ with components $\omega_{i_{1}\dots i_{k}}$. To sum up, we obtain the canonical decomposition
\begin{equation}\label{eq:DecompDiffForms}
\begin{aligned}
	\Omega^{k}(\sf{M},\bb{C})&\xrightarrow{\cong} C^{\infty}(\bb{R},\Omega^{k-1}(\Sigma,\bb{C}))\oplus C^{\infty}(\bb{R},\Omega^{k}(\Sigma,\bb{C}))\\
	\omega &\mapsto (\hspace*{1.35cm}\omega_{\mathrm{T}}\hspace*{1.55cm},\hspace*{1.15cm}\omega_{\Sigma}\hspace*{1.2cm})\,.
\end{aligned}
\end{equation}
We stress that this splitting depends, of course, on the choice of Cauchy temporal function $t\in C^{\infty}(\sf{M},\bb{R})$, which here is assumed to be fixed from the beginning. An important special case for our purposes is the case of $\Omega^{1}(\sf{M},\bb{C})$. In this case, any $\sf{A}\in\Omega^{1}(\sf{M},\bb{C})$ can be written as $\sf{A}=\sf{A}_{\mathrm{T}}\eta+\sf{A}_{\Sigma}$ with $\sf{A}_{\mathrm{T}}:=\sf{A}(\nu)=\beta^{-1}\sf{A}(\partial_{t})\in C^{\infty}(\sf{M},\bb{C})$ and $\sf{A}_{\Sigma}=\sf{A}_{i}\d x^{i}\in C^{\infty}(\bb{R},\Omega^{1}(\Sigma,\bb{C}))$.

\paragraph{Second Fundamental Form and Geometry of Foliation.} Denoting the space of time-dependent symmetric $(0,k)$-tensor fields by $C^{\infty}(\bb{R},\Gamma^{\infty}(\sf{T}^{\ast}\Sigma^{\otimes_{s}k})):=\Gamma^{\infty}(\pi_{2}^{\ast}(\sf{T}^{\ast}\Sigma^{\otimes_{s}k}))$, in analogy to the notation used above, we define the time-dependent symmetric $(0,2)$-tensor field
\begin{align*}
	\sf{k}:=-\frac{1}{2\beta}\partial_{t}\sf{h}\in C^{\infty}(\bb{R},\Gamma^{\infty}(\sf{T}^{\ast}\Sigma^{\otimes_{s}2}))\, ,
\end{align*}
which is the \textit{second fundamental form} of the foliation $(\Sigma_{t})_{t\in\bb{R}}$. More precisely, the future-directed timelike normal vector of $\Sigma_{t}$ in $\sf{M}$ is given by $\nu_{t}=\beta^{-1}\partial_{t}\vert_{\Sigma_{t}}$. With this definition, the second fundamental form\footnote{More precisely, this is the \textit{scalar} 2nd fundamental form: if $(\sf{M},\sf{g})$ is a pseudo-Riemannian manifold with pseudo-Riemannian embedded submanifold $(i\:\Sigma\hookrightarrow\sf{M},\sf{h}:=i^{\ast}\sf{g})$, the 2nd fundamental form is the $\sf{N}\Sigma$-valued symmetric $(0,2)$-tensor field $\mathrm{II}(X,Y):=(\nabla^{\sf{M}}_{X^{\prime}}Y^{\prime})^{\perp}$ for $X,Y\in\mathfrak{X}(\Sigma)$ and extensions $X^{\prime},Y^{\prime}$ to $\sf{M}$, where $\sf{N}\Sigma$ is the normal bundle and $\cdot^{\perp}\:\sf{T}\sf{M}\vert_{\Sigma}\to\sf{N}\Sigma$ the projection. If $\Sigma$ is a \textit{hyper}surface (i.e.~codimension $1$), one defines the \emph{scalar} 2nd fundamental form (up to a sign) by $\sf{k}(X,Y):=\sf{g}(\nu,\mathrm{II}(X,Y))$, where $\nu$ is a unit normal vector.} of $\Sigma_{t}$ in $\sf{M}$ is defined as
\begin{align*}
	\sf{k}_{t}(X,Y):=\sf{g}(\nu_{t},\nabla_{X}Y)=-\sf{g}(\nabla_{X}\nu_{t},Y)\,,
\end{align*}
for all $X,Y\in\mathfrak{X}(\Sigma_{t})$, where we have used the same convention as in ~\cite[Chap.~8, Eq.~(8.9)]{LeeRiemann}. In local coordinates, we hence obtain
\begin{align*}
	(\sf{k}_{t})_{ij}=\sf{k}_{t}(\partial_{i},\partial_{j})=-\nabla_{i}(\nu_{t})_{j}=-\beta\Gamma^{0}_{ij}\vert_{\Sigma_{t}}=-\frac{1}{2\beta_{t}}\partial_{t}(\sf{h}_{t})_{ij}\, ,
\end{align*}
where $\beta_{t}(\cdot):=\beta(t,\cdot)$ and where we used that $(\nu_{t})_{\alpha}=\sf{g}_{\alpha\beta}(\nu_{t})^{\beta}=-\beta_{t}\delta_{\alpha}^{0}$. Collecting all of these tensor fields for different times, we obtain the tensor field $\sf{k}=-\frac{1}{2\beta}\partial_{t}\sf{h}$, as defined above.

Furthermore, at a given time $t\in\bb{R}$, we define the \textit{Weingarten map}, or \textit{shape operator}, $\sf{W}_{\sf{h}_{t}}\:\Omega^{1}(\Sigma,\bb{C})\to\Omega^{1}(\Sigma,\bb{C})$ by $\sf{W}_{\sf{h}_{t}}(\omega)_{i}:=(\sf{k}_{t})_{i}^{k}\omega_{k}$. Collecting these operators for all times, we obtain the corresponding Weingarten map of the foliation $(\Sigma_{t})_{t\in\bb{R}}$, i.e.
\begin{align*}
\mathrm{W}_{\sf{h}}\:C^{\infty}(\bb{R},\Omega^{1}(\Sigma,\bb{C}))\to C^{\infty}(\bb{R},\Omega^{1}(\Sigma,\bb{C}))\,,\qquad \mathrm{W}_{\sf{h}}(\omega)_{i}:=\sf{k}^{j}_{i}\omega_{j}=-\frac{1}{2\beta}\sf{h}^{jl}(\partial_{t}\sf{h}_{ij})\omega_{l}\, ,
\end{align*}
where we stress that indices of time-dependent tensor fields on $\Sigma$ are raised and lowered with respect to the time-dependent metric $\sf{h}_{\bullet}$ on $\Sigma$. Last but not least, we define the smooth function $\mathrm{tr}_{\sf{h}}(\sf{k}):=\sf{h}^{ij}\sf{k}_{ij}\in C^{\infty}(\sf{M})$, whose restriction to $\Sigma_{t}$ is (a multiple of) the \textit{mean curvature} of $(\Sigma,\sf{h}_{t})$.

\subsection{Maxwell's Theory Revisited and (1+n)-Decomposition}
Following Section~\ref{Sec:Maxwell}, Maxwell's theory, viewed as a linear gauge theory in the sense of Definition~\ref{Def:LinGaugeTh}, is defined as the quadruple $(\sf{A}_{0},\sf{A}_{1},\sf{P},\sf{K})$, where the two operators $\sf{P}\in\mathrm{DO}^{2}(\sf{A}_{1})$ and $\sf{K}\in\mathrm{DO}^{1}(\sf{A}_{0},\sf{A}_{1})$ are given by
\begin{align*}
	\sf{P}:=\delta\d\:\Omega^{1}(\sf{M},\bb{C})\to\Omega^{1}(\sf{M},\bb{C})\,,\qquad\sf{K}:=\d\:C^{\infty}(\sf{M},\bb{C})\to\Omega^{1}(\sf{M},\bb{C})\, .
\end{align*}
By the previous discussion, we note that $\sf{P}$ is not hyperbolic and can also be be written as
\begin{align*}
	\sf{P}=\delta\d=\square-\d\delta=-\sf{g}^{\alpha\beta}\nabla_{\alpha}\nabla_{\beta}-\d\delta+\mathrm{Ric}_{\sf{g}}\, .
\end{align*}
Clearly, it holds that $\sf{P}\circ\sf{K}=0$, which encodes the gauge-invariance of the theory. The canonical gauge condition in this case is given by $\sf{K}^{\ast}\sf{A}=\delta\sf{A}=0$ for $\sf{A}\in\Omega^{1}(\sf{M},\bb{C})$, where we denote by ``$\ast$'' the formal adjoints with respect to the Hodge sesquilinear form~\eqref{eq:HodgeSF}, and usually called the \textit{Lorenz gauge}. Furthermore, the two operators $\sf{D}_{1}:=\sf{K}^{\ast}\sf{K}$ and $\sf{D}_{2}:=\sf{P}+\sf{K}\sf{K}^{\ast}$ are given by
\begin{align*}
	\sf{D}_{i}=\square\:\Omega^{i-1}(\sf{M},\bb{C})\to\Omega^{i-1}(\sf{M},\bb{C})\, .
\end{align*}

For later convenience, we shall now derive a $(1+n)$-decomposition of the Maxwell equations, i.e.~a decomposition into the temporal and spatial parts $\sf{A}_{\mathrm{T}}$ and $\sf{A}_{\Sigma}$ for a given Maxwell field $\sf{A}\in\Omega^{1}(\sf{M},\bb{C})$. This can be seen as an analogue to the way one derives the \textit{ADM decomposition} of general relativity (see the review \cite{ADM} by the original authors and Fischer-Marsden \cite{FischerMarsden00,FischerMarsden1,FischerMarsden2} for a mathematically oriented discussion). We start with the following technical lemma.

\begin{lemma}\label{Lemma:Dec} Let $(\sf{M},\sf{g})$ be a globally hyperbolic spacetime with Cauchy temporal function $t\in C^{\infty}(\sf{M},\bb{R})$ and recall the decomposition $\Omega^{1}(\sf{M},\bb{C})\ni\omega=\beta\d t\wedge\omega_{\mathrm{T}}+\omega_{\Sigma}$ with $\omega_{\mathrm{T}}:=\nu\lrcorner\omega\in C^{\infty}(\bb{R},\Omega^{k-1}(\sf{M},\bb{C}))$ and $\omega_{\Sigma}\in C^{\infty}(\bb{R},\Omega^{k}(\sf{M},\bb{C}))$ as introduced in~Eq.~\eqref{eq:DecompDiffForms}. Then:
\begin{align*}
	\text{\emph{(i)}}&\quad (\d f)_{\mathrm{T}}=\frac{1}{\beta}\partial_{t}f\qquad\text{and}\qquad (\d f)_{\Sigma}=\d_{\Sigma}f\,,\hspace*{4.2cm}  f\in C^{\infty}(\sf{M},\bb{C})\\
	&\quad (\d\omega)_{\mathrm{T}}=\frac{1}{\beta}\big(\partial_{t}\omega_{\Sigma}-\d_{\Sigma}\beta\wedge\omega_{\mathrm{T}}\big)-\d_{\Sigma}\omega_{\mathrm{T}}\quad\text{and}\quad (\d\omega)_{\Sigma}=\d_{\Sigma}\omega_{\Sigma}\,,\hspace*{0.65cm} \omega\in\Omega^{k\geq 1}(\sf{M},\bb{C})\\
	\text{\emph{(ii)}}&\quad \delta\sf{A}=\frac{1}{\beta}\partial_{t}\sf{A}_{\mathrm{T}}-\mathrm{tr}_{\sf{h}}(\sf{k})\sf{A}_{\mathrm{T}}+\delta_{\Sigma}\sf{A}_{\Sigma}-\frac{1}{\beta}\sf{h}^{\sharp}(\d_{\Sigma}\beta,\sf{A}_{\Sigma})\,,\hspace*{2.75cm} \sf{A}\in\Omega^{1}(\sf{M},\bb{C})\\
	\text{\emph{(iii)}}&\quad (\delta\sf{F})_{\mathrm{T}}=-\delta_{\Sigma}\sf{F}_{\mathrm{T}}\,,\hspace*{8.35cm}\sf{F}\in\Omega^{2}(\sf{M},\bb{C})\\&\quad (\delta\sf{F})_{\Sigma}=\frac{1}{\beta}\partial_{t}\sf{F}_{\mathrm{T}}+2\sf{W}_{\sf{h}}(\sf{F}_{\mathrm{T}})-\mathrm{tr}_{\sf{h}}(\sf{k})\sf{F}_{\mathrm{T}}+\delta_{\Sigma}\sf{F}_{\Sigma}-\frac{1}{\beta}(\d_{\Sigma}\beta)^{\sharp}\lrcorner\sf{F}_{\Sigma}\,.
\end{align*}
\end{lemma}

\begin{proof}
	As usual, we choose local coordinates $(x^{i})_{i=1,\dots,n}$ of $\Sigma$ and extent them to local coordinates $(x^{\mu})_{\mu=0,\dots,n}$ of $\sf{M}$ by adding $x^{0}=t$. Then, a straightforward computation in coordinates shows that the Christoffel symbols $\Gamma_{\alpha\beta}^{\gamma}$ of $(\sf{M},\sf{g})$ are given by
	\begin{align*}
    \Gamma^{0}_{00}&=\frac{1}{\beta}\partial_{t}\beta, \qquad\qquad \Gamma^{0}_{i0}=\frac{1}{\beta}(\d_{\Sigma}\beta)_{i}=\frac{1}{\beta}\partial_{i}\beta,\qquad\qquad \Gamma^{k}_{00}=\beta ((\d_{\Sigma}\beta)^{\sharp})^{k}=\beta \sf{h}^{kl}\partial_{l}\beta\,,\\
    \Gamma^{0}_{ij}&=\frac{1}{2\beta^{2}}\partial_{t}\sf{h}_{ij}=-\frac{1}{\beta}\sf{k}_{ij},\qquad\Gamma^{k}_{i0}=\frac{1}{2}\sf{h}^{kl}\partial_{t}\sf{h}_{lj}=-\beta \sf{k}^{k}_{i}, \qquad \Gamma^{k}_{ij}=(\Gamma_{\Sigma})_{ij}^{k}\,,
\end{align*}
where $\Gamma_{\Sigma}$ denote the (time-dependent) Christoffel symbols of $(\Sigma,\sf{h}_{\bullet})$. Claim (i) for $f\in C^{\infty}(\sf{M},\bb{C})$ follows from $\d f=(\partial_{\mu}f)\d x^{\mu}$. Let $\omega\in\Omega^{k}(\sf{M},\bb{C})$ be arbitrary with local components $\omega_{\mu_{1}\dots\mu_{k}}$. Then, $\omega_{\mathrm{T}}$ is the time-dependent $(k-1)$-form with components $\beta^{-1}\omega_{0i_{1}\dots i_{k-1}}$ and $\omega_{\Sigma}$ the time-dependent $k$-form with components $\omega_{i_{1}\dots i_{k}}$. Now, it is easy to see that $(\d\omega)_{\Sigma}=\d_{\Sigma}\omega_{\Sigma}$, since the exterior derivative is independent of the choice of (torsion-free) connection, i.e.
\begin{align*}
	((\d\omega)_{\Sigma})_{i_{1}\dots i_{k+1}}&=(k+1)\nabla_{[i_{1}}\omega_{i_{2}\dots i_{k+1}]}=(k+1)\partial_{[i_{1}}\omega_{i_{2}\dots i_{k+1}]}=\\&=(k+1)\nabla^{\Sigma}_{[i_{1}}(\omega_{\Sigma})_{i_{2}\dots i_{k+1}]}=(\d_{\Sigma}\omega_{\Sigma})_{i_{1}\dots i_{k+1}}\, .
\end{align*}
A straightforward computation in coordinates shows that the temporal part of $\d\omega$ is given by
\begin{align*}
	((\d\omega)_{\mathrm{T}})_{i_{1}\dots i_{k}}&=\beta^{-1}(\d\omega)_{0i_{1}\dots i_{k}}=(k+1)\beta^{-1}\partial_{[0}\omega_{i_{1}\dots i_{k}]}=\beta^{-1}\partial_{0}\omega_{i_{1}\dots i_{k}}-k\beta^{-1}\partial_{[i_{1}}\omega_{|0|i_{2}\dots i_{k}]}=\\&=\beta^{-1}(\partial_{t}\omega_{\Sigma})_{i_{1}\dots i_{k}}-k\beta^{-1}\nabla^{\Sigma}_{[i_{1}}(\beta\omega_{\mathrm{T}})_{i_{2}\dots i_{k}]}=\beta^{-1}(\partial_{t}\omega_{\Sigma})_{i_{1}\dots i_{k}}-\beta^{-1}(\d_{\Sigma}(\beta\omega_{\mathrm{T}}))_{i_{1}\dots i_{k}}\\&=\beta^{-1}(\partial_{t}\omega_{\Sigma})_{i_{1}\dots i_{k}}-(\d_{\Sigma}\omega_{\mathrm{T}})_{i_{1}\dots i_{k}}-(\beta^{-1}\d_{\Sigma}\beta\wedge\omega_{\mathrm{T}})_{i_{1}\dots i_{k}}
\end{align*}
and hence $(\d\omega)_{\mathrm{T}}=\beta^{-1}(\partial_{t}\omega_{\Sigma}-\d_{\Sigma}\beta\wedge\omega_{\mathrm{T}})-\d_{\Sigma}\omega_{\mathrm{T}}$, where $[\dots|\alpha|\dots]$ for some index $\alpha$ means that it is not contained in the antisymmetrisation $[\dots]$. Now, for a given $1$-form $\sf{A}\in\Omega^{1}(\sf{M},\bb{C})$ decomposed as $\sf{A}=\beta\sf{A}_{\mathrm{T}}\d t+\sf{A}_{\Sigma}$, a straightforward computation in coordinates shows that
\begin{align*}
	\delta\sf{A}&=-\sf{g}^{\alpha\beta}\nabla_{\alpha}\sf{A}_{\beta}=\frac{1}{\beta^{2}}\nabla_{0}\sf{A}_{0}-\sf{h}^{ij}\nabla_{i}\sf{A}_{j}=\\&=\frac{1}{\beta^{2}}(\partial_{0}\sf{A}_{0}-\Gamma_{00}^{0}\sf{A}_{0}-\Gamma^{k}_{00}\sf{A}_{k})-\sf{h}^{ij}(\partial_{i}\sf{A}_{j}-\Gamma^{0}_{ij}\sf{A}_{0}-\Gamma^{k}_{ij}\sf{A}_{k})\\&=\frac{1}{\beta^{2}}\bigg(\partial_{0}\sf{A}_{0}-\frac{1}{\beta}(\partial_{t}\beta)\sf{A}_{0}-\beta\sf{h}^{kl}(\partial_{k}\beta)\sf{A}_{l}\bigg)-\sf{h}^{ij}\bigg(\nabla^{\Sigma}_{i}(\sf{A}_{\Sigma})_{j}+\frac{1}{\beta}\sf{k}_{ij}\sf{A}_{0}\bigg)=\\&=\frac{1}{\beta^{2}}\partial_{t}(\beta\sf{A}_{\mathrm{T}})+\delta_{\Sigma}\sf{A}_{\Sigma}-\bigg(\frac{\partial_{t}\beta}{\beta^{2}}+\mathrm{tr}_{\sf{h}}(\sf{k})\bigg)\sf{A}_{\mathrm{T}}-\frac{1}{\beta}\sf{h}^{\sharp}(\d_{\Sigma}\beta,\sf{A}_{\Sigma})=\\&=\frac{1}{\beta}\partial_{t}\sf{A}_{\mathrm{T}}+\delta_{\Sigma}\sf{A}_{\Sigma}-\mathrm{tr}_{\sf{h}}(\sf{k})\sf{A}_{\mathrm{T}}-\frac{1}{\beta}\sf{h}^{\sharp}(\d_{\Sigma}\beta,\sf{A}_{\Sigma})\, .
\end{align*}
A similar computation yields the temporal and spatial components of the divergence of a given $2$-form $\sf{F}\in\Omega^{2}(\sf{M},\bb{C})$. For the temporal component, we compute
\begin{align*}
	(\delta\sf{F})_{\mathrm{T}}&=\frac{1}{\beta}(\delta\sf{F})_{0}=-\frac{1}{\beta}\sf{g}^{\alpha\beta}\nabla_{\alpha}\sf{F}_{\beta 0}=-\frac{1}{\beta}\sf{h}^{ij}\nabla_{i}\sf{F}_{j0}=\\&=-\frac{1}{\beta}\sf{h}^{ij}(\partial_{i}\sf{F}_{j0}-\Gamma^{0}_{ij}\sf{F}_{00}-\Gamma^{k}_{ij}\sf{F}_{k0}-\Gamma^{0}_{i0}\sf{F}_{j0}-\Gamma^{k}_{i0}\sf{F}_{j k})\\&=\frac{1}{\beta}\sf{h}^{ij}\nabla^{\Sigma}_{i}(\beta\sf{F}_{\mathrm{T}})_{j}-\frac{1}{\beta}\sf{h}^{ij}(\partial_{i}\beta)(\sf{F}_{\mathrm{T}})_{j}-\sf{k}^{ij}\sf{F}_{ij}=-\delta_{\Sigma}\sf{F}_{\mathrm{T}}
\end{align*}
and hence $(\delta\sf{F})_{\mathrm{T}}=-\delta_{\Sigma}\sf{F}_{\mathrm{T}}$. A similar computation shows that the spatial part is given by
\begin{align*}
	(\delta\sf{F})_{i}&=-\sf{g}^{\alpha\beta}\nabla_{\alpha}\sf{F}_{\beta i}=\frac{1}{\beta^{2}}\nabla_{0}\sf{F}_{0i}-\sf{h}^{kl}\nabla_{k}\sf{F}_{li}=\\&=\frac{1}{\beta^{2}}\big(\partial_{0}\sf{F}_{0i}-\Gamma^{0}_{00}\sf{F}_{0i}-\Gamma^{k}_{00}\sf{F}_{ki}-\Gamma^{0}_{0i}\sf{F}_{00}-\Gamma^{k}_{0i}\sf{F}_{0k}\big)-\sf{h}^{kl}\big(\nabla^{\Sigma}_{k}(\sf{F}_{\Sigma})_{li}-\Gamma^{0}_{kl}\sf{F}_{0i}-\Gamma^{0}_{ki}\sf{F}_{l0}\big)=\\&=\frac{1}{\beta^{2}}\partial_{t}(\beta\sf{F}_{\mathrm{T}})_{i}+(\delta_{\Sigma}\sf{F}_{\Sigma})_{i}-\bigg(\frac{\partial_{t}\beta}{\beta^{2}}+\mathrm{tr}_{\sf{h}}(\sf{k})\bigg)(\sf{F}_{\mathrm{T}})_{i}+2\sf{k}^{k}_{i}(\sf{F}_{\mathrm{T}})_{k}-\frac{1}{\beta}\sf{h}^{kl}(\partial_{k}\beta)(\sf{F}_{\Sigma})_{li}=\\&=\frac{1}{\beta}\partial_{t}(\sf{F}_{\mathrm{T}})_{i}+(\delta_{\Sigma}\sf{F}_{\Sigma})_{i}-\mathrm{tr}_{\sf{h}}(\sf{k})(\sf{F}_{\mathrm{T}})_{i}+2\sf{k}^{k}_{i}(\sf{F}_{\mathrm{T}})_{k}-\frac{1}{\beta}\sf{h}^{kl}(\partial_{k}\beta)(\sf{F}_{\Sigma})_{li}\,,
\end{align*}
which gives the claimed equation when written in coordinate-free notation.
\end{proof}

As a direct consequence of the previous lemma, we obtain a decomposition of Maxwell's equations $\sf{P}\sf{A}=0$ in terms of the temporal and spatial parts of $\sf{A}$.

\begin{proposition}\label{Prop:DecoMax} \emph{(Decomposition of the Maxwell Equations)}\newline
Let $(\sf{M},\sf{g})$ be a globally hyperbolic spacetime and fix a Cauchy temporal function $t\in C^{\infty}(\sf{M},\bb{R})$. Then, decomposing $\sf{A}\in\Omega^{1}(\sf{M},\bb{C})$ as $\sf{A}=\sf{A}_{\sf{T}}\eta+\sf{A}_{\Sigma}$ with $\sf{A}_{\mathrm{T}}:=\nu\lrcorner\sf{A}$, where $\nu=\beta^{-1}\partial_{t}$ and $\eta=\beta\d t$, the Maxwell equations $\sf{P}\sf{A}=0$ are equivalent to the coupled system
\begin{align*}
0=\begin{cases}
\Delta\sf{A}_{\mathrm{T}}-\frac{1}{\beta}\delta_{\Sigma}(\partial_{t}\sf{A}_{\Sigma}-\sf{A}_{\mathrm{T}}\d_{\Sigma}\beta)-\frac{1}{\beta^{2}}\sf{h}^{\sharp}(\d_{\Sigma}\beta,\partial_{t}\sf{A}_{\Sigma}-\sf{A}_{\mathrm{T}}\d_{\Sigma}\beta)\\\frac{1}{\beta^{2}}\partial_{t}^{2}\sf{A}_{\Sigma}+\delta_{\Sigma}\d_{\Sigma}\sf{A}_{\Sigma}-\frac{1}{\beta}\d_{\Sigma}(\partial_{t}\sf{A}_{\mathrm{T}})-\frac{1}{\beta^{3}}(\partial_{t}\beta)(\partial_{t}\sf{A}_{\Sigma}-\sf{A}_{\mathrm{T}}\d_{\Sigma}\beta)\\\hspace*{0.5cm}-\frac{1}{\beta^{2}}\partial_{t}(\sf{A}_{\mathrm{T}}\d_{\Sigma}\beta)-\frac{1}{\beta}(\d_{\Sigma}\beta)^{\sharp}\lrcorner\d_{\Sigma}\sf{A}_{\Sigma}+\frac{1}{\beta}(2\sf{W}_{\sf{h}}-\mathrm{tr}_{\sf{h}}(\sf{k})\mathrm{id})(\partial_{t}\sf{A}_{\Sigma}-\sf{A}_{\mathrm{T}}\d_{\Sigma}\beta-\beta\d_{\Sigma}\sf{A}_{\mathrm{T}})
\end{cases}
\end{align*}
\end{proposition}

\begin{proof}
	Set $\sf{F}^{\sf{A}}:=\d\sf{A}$. Then, by Lemma~\ref{Lemma:Dec}(i), $\sf{F}^{\sf{A}}_{\mathrm{T}}=\beta^{-1}(\partial_{t}\sf{A}_{\Sigma}-\sf{A}_{\mathrm{T}}\d_{\Sigma}\beta)-\d_{\Sigma}\sf{A}_{\mathrm{T}}$ and $\sf{F}^{\sf{A}}_{\Sigma}=\d_{\Sigma}\sf{A}_{\Sigma}$. Using this, Lemma~\ref{Lemma:Dec}(iii) implies 
	\begin{align*}
		\begin{cases}
			(\sf{P}\sf{A})_{\sf{T}}=&-\delta_{\Sigma}(\beta^{-1}(\partial_{t}\sf{A}_{\Sigma}-\sf{A}_{\mathrm{T}}\d_{\Sigma}\beta)-\d_{\Sigma}\sf{A}_{\mathrm{T}})\\
			(\sf{P}\sf{A})_{\Sigma}=&\frac{1}{\beta}\partial_{t}(\beta^{-1}(\partial_{t}\sf{A}_{\Sigma}-\sf{A}_{\mathrm{T}}\d_{\Sigma}\beta)-\d_{\Sigma}\sf{A}_{\mathrm{T}})+2\sf{W}_{\sf{h}}(\beta^{-1}(\partial_{t}\sf{A}_{\Sigma}-\sf{A}_{\mathrm{T}}\d_{\Sigma}\beta)-\d_{\Sigma}\sf{A}_{\mathrm{T}})\\&\hspace{2cm}-\mathrm{tr}_{\sf{h}}(\sf{k})(\beta^{-1}(\partial_{t}\sf{A}_{\Sigma}-\sf{A}_{\mathrm{T}}\d_{\Sigma}\beta)-\d_{\Sigma}\sf{A}_{\mathrm{T}})+\delta_{\Sigma}\d_{\Sigma}\sf{A}_{\Sigma}-\frac{1}{\beta}(\d_{\Sigma}\beta)^{\sharp}\lrcorner\d_{\Sigma}\sf{A}_{\Sigma}
		\end{cases}
	\end{align*}
	Now, for $f\in C^{\infty}(\sf{M})$ and $\omega\in C^{\infty}(\bb{R},\Omega^{1}(\Sigma,\bb{C}))$, there is the Leibniz rule $\delta_{\Sigma}(f\omega)=f\delta_{\Sigma}\omega-\sf{h}^{\sharp}(\d_{\Sigma}f,\omega)$, which yields the claimed result.
\end{proof}

\begin{remark} (Electric and Magnetic Fields)\newline
	We briefly recall how the decomposition into temporal and spatial part in Proposition~\ref{Prop:DecoMax} is related to the Maxwell equations in terms of the electric magnetic fields. For this, we consider a $(1+3)$-dimensional ultrastatic globally hyperbolic spacetime $(\sf{M}=\bb{R}\times\Sigma,\sf{g}=-\d t\otimes\d t+\sf{h})$ (cf.~Example~\ref{Examples:GlobHyp}(i)) and recall the Maxwell equations from Example~\ref{Ex:Maxwell} in terms of the electric and magnetic fields $\mathscr{E},\mathscr{B}\in C^{\infty}(\bb{R},\mathfrak{X}(\Sigma))$, which are time-dependent vector fields on $\Sigma$, i.e.
	\begin{align}\label{eq:MaxwellSy}
		\begin{cases}
			\partial_{t}\mathscr{E}-\mathrm{curl}_{\Sigma}(\mathscr{B})&=0\\
			\partial_{t}\mathscr{B}+\mathrm{curl}_{\Sigma}(\mathscr{E})&=0
		\end{cases}\qquad\text{and}\qquad\begin{cases}\mathrm{div}_{\Sigma}(\mathscr{E})&=0\\\mathrm{div}_{\Sigma}(\mathscr{B})&=0\end{cases}\,,
	\end{align}
	where $\mathrm{curl}_{\Sigma}(\mathscr{E}):=(\ast\d_{\Sigma}\mathscr{E}^{\flat})^{\sharp}$ and $\mathrm{div}_{\Sigma}(\mathscr{E}):=-\delta_{\Sigma}\mathscr{E}^{\flat}$ denote the curl and divergence operators on $(\Sigma,\sf{h})$. Now, like in the classical theory of electrodynamics on Minkowski spacetime, we consider an \textit{electric potential} $\phi\in C^{\infty}(\sf{M})$ and a \textit{magnetic potential} $\mathscr{A}\in C^{\infty}(\bb{R},\mathfrak{X}(\Sigma))$ and set
	\begin{align*}
		\begin{cases}
			\mathscr{E}&:=-\d_{\Sigma}\phi-\partial_{t}\mathscr{A}\\
			\mathscr{B}&:=\mathrm{curl}_{\Sigma}(\mathscr{A})
		\end{cases}\, .
	\end{align*}
	The pair $(\phi,\mathscr{A})$ is not uniquely determined by $(\mathscr{E},\mathscr{B})$, but subject to a gauge symmetry, i.e.
	\begin{align}\label{eq:GaugeTrafoMax}
		(\phi,\mathscr{A})\mapsto (\phi-\partial_{t}f,\mathscr{A}+\mathrm{grad}_{\Sigma}(f))
	\end{align}
	for all $f\in C^{\infty}(\sf{M})$ leaves $(\mathscr{E},\mathscr{B})$ invariant, where $\mathrm{grad}_{\Sigma}(f):=(\d_{\Sigma}f)^{\sharp}$. Now, consider the $1$-form $\sf{A}:=\sf{A}_{\mathrm{T}}\d t+\sf{A}_{\Sigma}$ with $\sf{A}_{\mathrm{T}}:=-\phi$ and $\sf{A}_{\Sigma}:=\mathscr{A}^{\flat}$, where the additional minus sign in $\sf{A}_{\mathrm{T}}$ comes from the fact that we lower the indices of the vector field $\phi\partial_{t}+\mathscr{A}$ with the Lorentzian metric $\sf{g}$. Note that the gauge transformations~\eqref{eq:GaugeTrafoMax} exactly correspond to $\sf{A}\mapsto\sf{A}+\d f$, since $(\d f)_{\mathrm{T}}=\partial_{t}f$ and $(\d f)_{\Sigma}=(\d_{\Sigma}f)=(\mathrm{grad}_{\Sigma}(f))^{\flat}$. Furthermore, the system~\eqref{eq:MaxwellSy} yields
	\begin{align*}
		0=&\begin{cases}
			\partial_{t}\mathscr{E}^{\flat}-\mathrm{curl}_{\Sigma}(\mathscr{B})^{\flat}=\partial_{t}\d_{\Sigma}\sf{A}_{\mathrm{T}}-\partial_{t}^{2}\sf{A}_{\Sigma}-\ast\d_{\Sigma}\ast\d_{\Sigma}\sf{A}_{\Sigma}=-\partial_{t}^{2}\sf{A}_{\Sigma}-\delta_{\Sigma}\d_{\Sigma}\sf{A}_{\Sigma}+\d_{\Sigma}\partial_{t}\sf{A}_{\mathrm{T}}\\
			\partial_{t}\mathscr{B}^{\flat}+\mathrm{curl}_{\Sigma}(\mathscr{E})^{\flat}=\partial_{t}\ast\d_{\Sigma}\sf{A}_{\Sigma}-\ast\d_{\Sigma}\partial_{t}\sf{A}_{\Sigma}=\ast\d_{\Sigma}\partial_{t}\sf{A}_{\Sigma}-\ast\d_{\Sigma}\partial_{t}\sf{A}_{\Sigma}
		\end{cases}\\
		0=&\begin{cases}
			\mathrm{div}_{\Sigma}(\mathscr{E})=-\delta_{\Sigma}\mathscr{E}^{\flat}=-\delta_{\Sigma}\d_{\Sigma}\sf{A}_{\mathrm{T}}+\partial_{t}\delta_{\Sigma}\sf{A}_{\Sigma}=-\Delta\sf{A}_{\mathrm{T}}+\partial_{t}\delta_{\Sigma}\sf{A}_{\Sigma}\\
			\mathrm{div}_{\Sigma}(\mathscr{B})=-\delta_{\Sigma}\mathscr{B}^{\flat}=-\delta_{\Sigma}\ast\d_{\Sigma}\sf{A}_{\Sigma}=\delta_{\Sigma}\delta_{\Sigma}\ast\sf{A}_{\Sigma}
		\end{cases}
	\end{align*}
	where $\ast$ denotes the Hodge $\ast$-operator on $(\Sigma,\sf{h})$. Now, it is easy to see that the second line of each pairs of equations are trivially fulfilled, which is an artefact of the \emph{Bianchi identity} $\d\sf{F}^{\sf{A}}=0$ for $\sf{F}^{\sf{A}}:=\d\sf{A}$. The first equation of both pairs yield the system
	\begin{align*}
0=\begin{cases}
\Delta\sf{A}_{\mathrm{T}}-\partial_{t}\delta_{\Sigma}\sf{A}_{\Sigma}\\\partial^{2}_{t}\sf{A}_{\Sigma}+\delta_{\Sigma}\d_{\Sigma}\sf{A}_{\Sigma}-\d_{\Sigma}\partial_{t}\sf{A}_{\mathrm{T}}
\end{cases}
\end{align*}
which is exactly the system in Proposition~\ref{Prop:DecoMax} in the ultrastatic case ($\beta=1$ and $\sf{k}=0$). 
\end{remark}

Note that the equations in Proposition~\ref{Prop:DecoMax} simplify a lot in the case $\beta=1$. As it turns out, at least in the case of $(1+3)$-dimensional spacetimes, we can always reduce ourselves to this situation. The following proposition is well known, see e.g.~\cite[App.~D]{Wald} or \cite[Sec.~2.2]{DappiaggiSiemssen}, but we prove it here for completeness and the reader's convenience.

\begin{proposition}\label{Prop:MaxConf} \emph{(Conformal Invariance in (1+3)-Dimensions)}\newline
Let $(\sf{M},\sf{g})$ be a $(1+3)$-dimensional spacetime, consider a conformal transformation $\Phi\in C^{\infty}(\sf{M})$, i.e.~$\sf{g}^{\prime}:=e^{2\Phi}\sf{g}$ and denote the Maxwell operators by $\sf{P}_{\sf{g}}$ and $\sf{P}_{\sf{g}^{\prime}}$. Then, it holds that $\sf{P}_{\sf{g}^{\prime}}=e^{-2\Phi}\sf{P}_{\sf{g}}$, i.e.~Maxwell's equations in (1+3)-dimensions are conformally invariant.
\end{proposition}

\begin{proof}
	First of all, denoting the Christoffel symbols of $\sf{g}$ by $\Gamma^{\gamma}_{\alpha\beta}$ and the ones of $\sf{g}^{\prime}$ by $(\Gamma^{\prime})^{\gamma}_{\alpha\beta}$, we obtain the following transformation law:
	\begin{align*}
	\Gamma_{\alpha\beta}^{\gamma}\quad\xrightarrow{\sf{g}\mapsto \sf{g}^{\prime}=e^{2\Phi}\sf{g}}\quad (\Gamma^{\prime})^{\gamma}_{\alpha\beta}=\Gamma_{\alpha\beta}^{\gamma}+2\delta_{(\alpha}^{\gamma}\nabla_{\beta)}\Phi-\sf{g}_{\alpha\beta}\sf{g}^{\gamma\delta}\nabla_{\delta}\Phi\, 
\end{align*}
Now, since $\d\sf{A}$ for given $\sf{A}\in\Omega^{1}(\sf{M},\bb{C})$ is independent of the metric, it stays trivially invariant under conformal transformations, i.e.~$(\d\sf{A})_{\alpha\beta}=2\nabla_{[\alpha}\sf{A}_{\beta]}=2\partial_{[\alpha}\sf{A}_{\beta]}=2\nabla^{\prime}_{[\alpha}\sf{A}_{\beta]}$, where we denote by $\nabla$ the Levi-Civita connection of $(\sf{M},\sf{g})$ and by $\nabla^{\prime}$ the one of $(\sf{M},\sf{g}^{\prime})$. Let us set $\sf{F}^{\sf{A}}:=\d\sf{A}$, in accordance to the notation used in Section~\ref{Sec:LinYM}. Then, using the transformation law for the Christoffel symbols derived above, a straightforward computation in local coordinates shows
\begin{align*}
		(\sf{P}_{\sf{g}^{\prime}}\sf{A})_{\lambda}&=-(\sf{g}^{\prime})^{\alpha\beta}\nabla^{\prime}_{\alpha}\sf{F}^{\sf{A}}_{\beta\lambda}=-e^{-2\Phi}\sf{g}^{\alpha\beta}\nabla^{\prime}_{\alpha}\sf{F}^{\sf{A}}_{\beta\lambda}\\&=-e^{-2\Phi}\sf{g}^{\alpha\beta}\big(\nabla_{\alpha}\sf{F}^{\sf{A}}_{\beta\lambda}-(2\delta_{(\alpha}^{\rho}\nabla_{\beta)}\Phi-\sf{g}_{\alpha\beta}\sf{g}^{\rho\sigma}\nabla_{\sigma}\Phi)\sf{F}^{\sf{A}}_{\rho\lambda}-(2\delta_{(\alpha}^{\rho}\nabla_{\lambda)}\Phi-\sf{g}_{\alpha\lambda}\sf{g}^{\rho\sigma}\nabla_{\sigma}\Phi)\sf{F}^{\sf{A}}_{\beta\rho}\big)\\&=-e^{-2\Phi}\sf{g}^{\alpha\beta}\nabla_{\alpha}\sf{F}^{\sf{A}}_{\beta\lambda}+2e^{-2\Phi}(\nabla^{\rho}\Phi)\sf{F}^{\sf{A}}_{\rho\lambda}-e^{-2\Phi}(\sf{g}^{\rho\beta}\nabla_{\lambda}\Phi+\delta_{\lambda}^{\rho}\nabla^{\beta}\Phi-\delta_{\lambda}^{\beta}\nabla^{\rho}\Phi)\sf{F}^{\sf{A}}_{\beta\rho}\\&=-e^{-2\Phi}\sf{g}^{\alpha\beta}\nabla_{\alpha}\sf{F}^{\sf{A}}_{\beta\lambda}+\cancel{2e^{-2\Phi}(\nabla^{\rho}\Phi)\sf{F}^{\sf{A}}_{\rho\lambda}}-\cancel{2e^{-2\Phi}(\nabla^{\rho}\Phi)\sf{F}^{\sf{A}}_{\rho\lambda}}\\&=-e^{-2\Phi}\sf{g}^{\alpha\beta}\nabla_{\alpha}\sf{F}^{\sf{A}}_{\beta\lambda}=e^{-2\Phi}(\sf{P}_{\sf{g}}\sf{A})_{\lambda}\,.
	\end{align*}
	Hence, Maxwell's equations are conformally invariant (of conformal weight zero) in (1+3)-dimensional spacetimes.
\end{proof}

Since conformal transformations preserve the causal structure of a Lorentzian manifold, we note that global hyperbolicity is preserved as well. Hence, when working on a $(1+3)$-dimensional globally hyperbolic spacetime, we can perform the conformal transformation $\sf{g}\mapsto\beta^{-2}\sf{g}=:\sf{g}^{\prime}$, which yields a new globally hyperbolic metric $\sf{g}^{\prime}$ with lapse function $\beta^{\prime}=1$. By Proposition~\ref{Prop:MaxConf}, it follows that the solution spaces of the Maxwell operator on $(\sf{M},\sf{g})$ and $(\sf{M},\sf{g}^{\prime})$ are equivalent.

\subsection{Cauchy Problem for Maxwell's Equations and Phase Space}\label{Subsec:CauchyMax}
As a next step, we discuss the Cauchy problem for Maxwell's equations, following the abstract discussion in Section~\ref{Sec:CauchyGauge}. Moreover, we examine the various formulations of the phase space of Maxwell’s theory, following Section~\ref{Sec:PhaseGauge}. To start with, we need to set up the notation for the spaces of initial data. Recall the decomposition
\begin{align}\label{eq:DecoID}
	\Gamma^{\infty}(\sf{A}_{k})=\Omega^{k}(\sf{M},\bb{C})\cong C^{\infty}(\bb{R},\Omega^{k-1}(\Sigma,\bb{C}))\oplus C^{\infty}(\bb{R},\Omega^{k}(\Sigma,\bb{C}))\, 
\end{align}
defined by $\omega=\eta\wedge \omega_{\mathrm{T}}+\omega_{\Sigma}$ for $\omega_{\mathrm{T}}:=\nu\lrcorner\omega$ with $\nu=\beta^{-1}\partial_{t}$ and $\eta=\beta\d t$, cf.~Eq.~\eqref{eq:DecompDiffForms}. Throughout the following, we assign initial data on the slice $\Sigma_{0}=\{0\}\times\Sigma$. Following the notation introduced in Section~\ref{Sec:GreenCauchyHyp}, we define the $\bb{C}$-vector bundles $\sf{A}_{\rho_{i}}:=\sf{A}_{i}\vert_{\Sigma_{0}}\oplus\sf{A}_{i}\vert_{\Sigma_{0}}$ over $\Sigma_{0}\cong\Sigma$ and identify their $C^{\infty}(\Sigma,\bb{C})$-modules of sections in accordance to the decomposition~\eqref{eq:DecoID} as
\begin{align*}
	\Gamma^{\infty}(\sf{A}_{\rho_{0}})&\cong C^{\infty}(\Sigma,\bb{C})\oplus C^{\infty}(\Sigma,\bb{C})\,,\\  \Gamma^{\infty}(\sf{A}_{\rho_{1}})&\cong C^{\infty}(\Sigma,\bb{C})\oplus C^{\infty}(\Sigma,\bb{C})\oplus\Omega^{1}(\Sigma,\bb{C})\oplus\Omega^{1}(\Sigma,\bb{C})\,.
\end{align*}
Furthermore, we define the corresponding initial data maps $\rho_{i}\:\Gamma^{\infty}(\sf{A}_{i})\to\Gamma^{\infty}(\sf{A}_{\rho_{i}})$ by
\begin{align}\label{eq:CauchyRhoMax}
	\rho_{0}(f):=\bigg(f,\,\frac{1}{i\beta}\partial_{t}f\bigg)\bigg\vert_{t=0}\,,\qquad \rho_{1}(\sf{A}):=\bigg(\sf{A}_{\mathrm{T}},\,\frac{1}{i\beta}\partial_{t}\sf{A}_{\mathrm{T}},\,\sf{A}_{\Sigma},\,\frac{1}{i\beta}\partial_{t}\sf{A}_{\Sigma}\bigg)\bigg\vert_{t=0}\,,
\end{align}
where $\sf{A}=\sf{A}_{\mathrm{T}}\eta+\sf{A}_{\Sigma}$ with $\sf{A}_{\mathrm{T}}:=\sf{A}(\nu)$. The choices of $\rho_{i}$ in~\eqref{eq:CauchyRhoMax} are not unique and we could also consider different initial data maps, which might have some advantages in different applications. In the case of $\rho_{0}$, the definition in Eq.~\eqref{eq:CauchyRhoMax} agrees with the general definition of initial data maps for normally hyperbolic operators as defined in Section~\ref{Sec:GreenCauchyHyp} (see Eq.~\eqref{eq:IDM.NH}), i.e.~$\rho_{0}(f)=(f,\frac{1}{i}\nabla_{\nu}f)\vert_{t=0}$, where we put an additional factor of $i$ for convenience. In the case of $\rho_{1}$, the definition in~\eqref{eq:CauchyRhoMax} is slightly different and the discussion in Section~\ref{Sec:GreenCauchyHyp} rather suggests 
\begin{align}\label{eq:CauchyRhoMax2}
	\widetilde{\rho}_{1}\:\Gamma^{\infty}(\sf{A}_{1})\to\Gamma^{\infty}(\sf{A}_{\rho_{1}})\,,\qquad\widetilde{\rho}_{1}(\sf{A}):=\bigg(\sf{A}_{\mathrm{T}},\,\frac{1}{i}(\nabla_{\nu}\sf{A})_{\mathrm{T}},\,\sf{A}_{\Sigma},\,\frac{1}{i}(\nabla_{\nu}\sf{A})_{\Sigma}\bigg)\bigg\vert_{t=0}\, .
\end{align}
To see how $\rho_{1}$ and $\widetilde{\rho}_{1}$ are related, we first observe that the normal derivative $\nabla_{\nu}\sf{A}\in\Omega^{1}(\sf{M},\bb{C})$ of $\sf{A}\in\Omega^{1}(\sf{M},\bb{C})$, decomposed as $\nabla_{\nu}\sf{A}=(\nabla_{\nu}\sf{A})_{\mathrm{T}}\eta+(\nabla_{\nu}\sf{A})_{\Sigma}$, is given by
\begin{align}\label{eq:NormalDer}
	(\nabla_{\nu}\sf{A})_{\mathrm{T}}=\frac{1}{\beta}\partial_{t}\sf{A}_{\mathrm{T}}-\frac{1}{\beta}\sf{h}^{\sharp}(\d_{\Sigma}\beta,\sf{A}_{\Sigma})\,,\qquad (\nabla_{\nu}\sf{A})_{\Sigma}=\frac{1}{\beta}\partial_{t}\sf{A}_{\Sigma}-\frac{1}{\beta}(\d_{\Sigma}\beta)\sf{A}_{\mathrm{T}}+\sf{W}_{\sf{h}}(\sf{A}_{\Sigma})\, ,
\end{align}
following computations analogous to those in Lemma~\ref{Lemma:Dec}. Hence, the initial data map for $1$-forms in~\eqref{eq:CauchyRhoMax} is related to the one in~\eqref{eq:CauchyRhoMax2} via $\widetilde{\rho}_{1}=\sf{S}\circ\rho_{1}$, where $\sf{S}\:\Gamma^{\infty}(\sf{A}_{\rho_{1}})\to\Gamma^{\infty}(\sf{A}_{\rho_{1}})$ is the linear isomorphism
\begin{align*}
	\sf{S}:=
	\begin{pmatrix}
		\mathrm{id} & 0 & 0 & 0\\
		0 & \mathrm{id} & \frac{i}{\beta_{0}}\sf{h}_{0}^{\sharp}(\d_{\Sigma}\beta_{0},\cdot) & 0\\
		0 & 0 & \mathrm{id} & 0\\
		\frac{i}{\beta_{0}}(\d_{\Sigma}\beta_{0})\mathrm{id} & 0 & -i\sf{W}_{\sf{h}_{0}} &\mathrm{id}
	\end{pmatrix},\,\,\, \sf{S}^{-1}=
	\begin{pmatrix}
		\mathrm{id} & 0 & 0 & 0\\
		0 & \mathrm{id} & -\frac{i}{\beta_{0}}\sf{h}_{0}^{\sharp}(\d_{\Sigma}\beta_{0},\cdot) & 0\\
		0 & 0 & \mathrm{id} & 0\\
		-\frac{i}{\beta_{0}}(\d_{\Sigma}\beta_{0})\mathrm{id} & 0 & i\sf{W}_{\sf{h}_{0}} &\mathrm{id}
	\end{pmatrix}
\end{align*}
with $\beta_{0}:=\beta\vert_{t=0}\in C^{\infty}(\Sigma,(0,\infty))$. This also implies that our choice of initial data map $\rho_{1}$ in~\eqref{eq:CauchyRhoMax} is well-defined in the sense that it leads to a \emph{well-posed Cauchy problem}. In other words, $\rho_{i}$ are invertible when viewed as a map of the form $\rho_{i}\:\mathrm{ker}(\square\vert_{\Gamma^{\infty}_{\mathrm{sc}}})\to\Gamma^{\infty}_{\mathrm{c}}(\sf{A}_{\rho_{i}})$ and we denote the corresponding inverse, the \emph{Cauchy evolution operator}, by 
\begin{align*}
	\mathcal{U}_{i}:=(\rho\vert_{\mathrm{ker}(\square\vert_{\Gamma^{\infty}_{\mathrm{sc}}})})^{-1}\:\Gamma^{\infty}_{\mathrm{c}}(\sf{A}_{\rho_{i}})\to\Gamma^{\infty}_{\mathrm{sc}}(\sf{A}_{i})\, .
\end{align*}
In view of the decomposition $\sf{A}=\sf{A}_{\sf{T}}\eta+\sf{A}_{\Sigma}$, the choice of $\rho_{1}$ in~\eqref{eq:CauchyRhoMax} is more natural compared to $\widetilde{\rho}_{1}$ employing normal covariant derivatives, since the latter mixes the $\sf{A}_{\mathrm{T}}$ and $\sf{A}_{\Sigma}$ components in a non-trivial way. Of course, in the case of ultrastatic spacetimes with adapted Cauchy temporal function ($\beta=1$ and $\sf{k}=0$), we have $\sf{S}=\mathrm{id}$ and the two initial data maps are equivalent.

In the specific case of Maxwell's theory, there is yet another choice of initial data map, namely the one introduced by Furlani in \cite{FurlaniCauchy} (see also the discussion in \cite{Pfenning,GerardWrochna}), which is more adapted to differential forms and Maxwell's theory. In our notation\footnote{It is worth noting that the original paper of Furlani, which appeared before the work of Bernal-Sánchez, used a slightly different notation: fix a Cauchy surface $\Sigma$ with embedding $i\:\Sigma\hookrightarrow\sf{M}$ and let $\ast_{\sf{h}}$ and $\ast_{\sf{g}}$ be the Hodge $\ast$-operators of $(\sf{M},\sf{g})$ and $(\Sigma,\sf{h}:=i^{\ast}\sf{g})$. Then, Furlani considered (up to signs) the Cauchy data 
\begin{align*}
	\sf{A}_{(0)}:=i^{\ast}\sf{A}\,,\qquad \sf{A}_{(\d)}:=\ast_{\sf{h}}i^{\ast}\ast_{\sf{g}}\d\sf{A}\,,\qquad \sf{A}_{(\delta)}:=i^{\ast}\delta\sf{A}\,,\qquad \sf{A}_{(n)}:=\ast_{\sf{h}}i^{\ast}\ast_{\sf{g}}\,.
\end{align*}
Now, if we choose a Cauchy temporal function $t\in C^{\infty}(\sf{M})$ such that $\Sigma=t^{-1}(0)$, it is easy to see that the operator $\ast_{\sf{h}}i^{\ast}\ast_{\sf{g}}$ is exactly the operator that maps a $k$-form $\omega$ (up to a possible sign) to $\omega_{\mathrm{T}}\vert_{t=0}$, while the pull-back $i^{\ast}$ gives $\omega_{\Sigma}\vert_{t=0}$. In particular, the Cauchy data $\sf{A}_{(0)}$ and $\sf{A}_{(n)}$ are exactly $\sf{A}_{\Sigma}\vert_{t=0}$ and $\sf{A}_{\mathrm{T}}\vert_{t=0}$, respectively, while $\sf{A}_{(\d)}$ and $\sf{A}_{(\delta)}$ correspond to $(\d\sf{A})_{\mathrm{T}}\vert_{t=0}$ and $\delta\sf{A}\vert_{t=0}$.}, it is given by
\begin{align}\label{eq:CauchyRhoMaxFur}
	\rho^{\sf{F}}_{1}\:\Gamma^{\infty}(\sf{A}_{1})\to\Gamma^{\infty}(\sf{A}_{\rho_{1}})\,,\qquad\rho_{1}^{\sf{F}}(\sf{A}):=\bigg(\sf{A}_{\mathrm{T}},\,\frac{1}{i}\delta\sf{A},\,\sf{A}_{\Sigma},\,\frac{1}{i}(\d\sf{A})_{\mathrm{T}}\bigg)\bigg\vert_{t=0}\, .
\end{align}
The relation to our choice of initial data map $\rho_{1}$ in~\eqref{eq:CauchyRhoMax} can be seen by recalling the formulae 
\begin{align*}
	\delta\sf{A}=\frac{1}{\beta}\partial_{t}\sf{A}_{\mathrm{T}}-\mathrm{tr}_{\sf{h}}(\sf{k})\sf{A}_{\mathrm{T}}+\delta_{\Sigma}\sf{A}_{\Sigma}-\frac{1}{\beta}\sf{h}^{\sharp}(\d_{\Sigma}\beta,\sf{A}_{\Sigma})\quad\text{and}\quad (\d\sf{A})_{\mathrm{T}}=\frac{1}{\beta}(\partial_{t}\sf{A}_{\Sigma}-\sf{A}_{\mathrm{T}}\d_{\Sigma}\beta)-\d_{\Sigma}\sf{A}_{\mathrm{T}}
\end{align*}
from Lemma~\ref{Lemma:Dec}(i) and (ii). In particular, the initial datum $\delta\sf{A}\vert_{t=0}$ is a replacement for $\partial_{t}\sf{A}_{\mathrm{T}}\vert_{t=0}$, while $(\d\sf{A})_{\mathrm{T}}\vert_{t=0}$ is a replacement for $\partial_{t}\sf{A}_{\Sigma}\vert_{t=0}$. Denoting by 
\begin{align*}
	\mathcal{U}_{1}^{\sf{F}}\:\Gamma^{\infty}_{\mathrm{c}}(\sf{A}_{\rho_{1}})\to\Gamma^{\infty}_{\mathrm{sc}}(\sf{A}_{1})
\end{align*}
the corresponding Cauchy evolution operator, i.e.~the inverse of $\rho_{1}^{\sf{F}}$ restricted to $\mathrm{ker}(\square\vert_{\Gamma^{\infty}_{\mathrm{sc}}})$, the relation between the two maps $\rho_{1}$ and $\rho_{1}^{\sf{F}}$ is provided by a linear isomorphism $\sf{S}_{\sf{F}}\:\Gamma^{\infty}(\sf{A}_{\rho_{1}})\to\Gamma^{\infty}(\sf{A}_{\rho_{1}})$ satisfying
\begin{align*}
	\rho^{\sf{F}}_{1}=\sf{S}_{\sf{F}}\circ\rho_{1}\qquad\text{and}\qquad \mathcal{U}_{1}^{\sf{F}}=\mathcal{U}_{1}\circ\sf{S}_{\sf{F}}^{-1}\, .
\end{align*}
Using the explicit formulas of $\delta\sf{A}$ and $(\d\sf{A})_{\mathrm{T}}$ in terms of $\sf{A}_{\mathrm{T}},\sf{A}_{\Sigma}$ and $\partial_{t}\sf{A}_{\mathrm{T}},\partial_{t}\sf{A}_{\Sigma}$ from above, recalling the notation $\delta_{\Sigma}^{0}=\delta_{\Sigma}\vert_{t=0}$ and writing $\beta_{0}:=\beta\vert_{t=0}\in C^{\infty}(\Sigma,(0,\infty))$ as well as $\sf{k}_{0}:=\sf{k}\vert_{t=0}=-\frac{1}{2\beta_{0}}(\partial_{t}\sf{h})\vert_{t=0}\in\Gamma^{\infty}(\sf{T}^{\ast}\Sigma^{\otimes_{s}2})$, we obtain the explicit expressions
\begin{subequations}\label{eq:IsoCauchy}
\begin{align}
	\sf{S}_{\sf{F}}&=
	\begin{pmatrix}
		\mathrm{id} & 0 & 0 & 0\\
		i\mathrm{tr}_{\sf{h}_{0}}(\sf{k}_{0})\mathrm{id} & \mathrm{id} & \cfrac{i}{\beta}\sf{h}_{0}^{\sharp}(\d_{\Sigma}\beta_{0},\cdot)-i\delta^{0}_{\Sigma} & 0 \\
		0 & 0 & \mathrm{id} & 0 \\
		i\d_{\Sigma}+\frac{i}{\beta_{0}}(\d_{\Sigma}\beta_{0})\mathrm{id} & 0 & 0 & \mathrm{id}
	\end{pmatrix}\\&\text{and}\hspace*{2cm}\sf{S}_{\sf{F}}^{-1}=
	\begin{pmatrix}
		\mathrm{id} & 0 & 0 & 0\\
		-i\mathrm{tr}_{\sf{h}_{0}}(\sf{k}_{0})\mathrm{id} & \mathrm{id} & i\delta_{\Sigma}^{0}-\frac{i}{\beta_{0}}\sf{h}_{0}^{\sharp}(\d_{\Sigma}\beta_{0},\cdot) & 0\\
		0 & 0 & \mathrm{id} & 0 \\
		-i\d_{\Sigma}-\frac{i}{\beta_{0}}(\d_{\Sigma}\beta_{0})\mathrm{id} & 0 & 0 & \mathrm{id}
	\end{pmatrix}\, .
\end{align}
\end{subequations}

As a next step, we derive explicit expressions for the differential operators $\sf{K}_{\Sigma}$ and $\sf{K}_{\Sigma}^{\dagger}$ (see~Definition~\ref{Def:KSigma}), which encode the gauge transformations and the Lorenz gauge condition at the level of initial data. It turns out to be more convenient to compute these operators via the Furlani initial data map~\eqref{eq:CauchyRhoMaxFur}, and then compose them with the isomorphism $\sf{S}_{\sf{F}}$ to obtain the corresponding operators for the choice of initial 
data maps in~\eqref{eq:CauchyRhoMax}, which we will use lateron.

\begin{proposition}\label{Prop:KSigma} \emph{(The Operators $\sf{K}_{\Sigma}$ and $\sf{K}_{\Sigma}^{\dagger}$)}\newline
	Let $(\sf{M},\sf{g})$ be a globally hyperbolic manifold and fix a Cauchy temporal function $t\in C^{\infty}(\sf{M},\bb{R})$. Using $\rho_{0},\rho_{1}^{\sf{F}}$, the operators $\sf{K}_{\Sigma}^{\sf{F}}:=\rho_{1}^{\sf{F}}\sf{K}\mathcal{U}_{0}$ and $(\sf{K}^{\sf{F}}_{\Sigma})^{\dagger}:=\rho_{0}\sf{K}^{\ast}\mathcal{U}_{1}^{\sf{F}}$ are given by
	\begin{align*}
		\sf{K}_{\Sigma}^{\sf{F}}=\begin{pmatrix}0 & i\cdot\mathrm{id}\\ 0 & 0 \\ \d_{\Sigma} & 0 \\ 0 & 0 \end{pmatrix}\,,\qquad(\sf{K}_{\Sigma}^{\sf{F}})^{\dagger}=\begin{pmatrix} 0 & i\cdot\mathrm{id} & 0 & 0 \\ 0 & 0 & 0 & \delta^{0}_{\Sigma}\end{pmatrix}\, .
	\end{align*}
	Using the initial data map $\rho_{1}$ instead of $\rho_{1}^{\sf{F}}$, $\sf{K}_{\Sigma}:=\rho_{1}\sf{K}\mathcal{U}_{0}$ and $\sf{K}^{\dagger}_{\Sigma}:=\rho_{0}\sf{K}^{\ast}\mathcal{U}_{1}$ are given by
	\begin{align*}
		\sf{K}_{\Sigma}&=\begin{pmatrix} 0 & i\cdot\mathrm{id}\\ i\Delta^{0}-\frac{i}{\beta_{0}}\sf{h}^{\sharp}_{0}(\d_{\Sigma}\beta_{0},\d_{\Sigma}\cdot)& \mathrm{tr}_{\sf{h}_{0}}(\sf{k}_{0})\mathrm{id}\\ \d_{\Sigma} & 0 \\ 0 & \d_{\Sigma}+\frac{1}{\beta_{0}}(\d_{\Sigma}\beta_{0})\mathrm{id}\end{pmatrix}\,,\\[1em] \sf{K}_{\Sigma}^{\dagger}&=
		\begin{pmatrix}
		-\mathrm{tr}_{\sf{h}_{0}}(\sf{k}_{0})\mathrm{id} & i\cdot\mathrm{id} & \delta^{0}_{\Sigma}-\frac{1}{\beta_{0}}\sf{h}_{0}^{\sharp}(\d_{\Sigma}\beta_{0},\cdot) & 0 \\ i\Delta^{0}+i\delta^{0}_{\Sigma}\big(\frac{1}{\beta_{0}}(\d_{\Sigma}\beta_{0})\cdot\big) & 0 & 0 &\delta^{0}_{\Sigma}
		\end{pmatrix}\,.
	\end{align*}
\end{proposition}

\begin{proof}
	Consider an initial datum $\mathfrak{c}_{0}:=(\mathfrak{a},\pi)\in\Gamma^{\infty}(\sf{A}_{\rho_{0}})$ and set $f:=\mathcal{U}_{0}\mathfrak{c}\in C^{\infty}(\sf{M},\bb{C})$, i.e.~$f$ is the unique solution to $\square f=0$ with initial data $f\vert_{t=0}=\mathfrak{a}$ and $\partial_{t}f\vert_{t=0}=i\beta\pi$. Then,
	\begin{align*}
		\sf{K}_{\Sigma}^{\sf{F}}\mathfrak{c}_{0}=\rho^{\sf{F}}_{1}\sf{K}f=
		\begin{pmatrix}
			(\d f)_{\mathrm{T}}\\
			\frac{1}{i}\delta\d f\\
			(\d f)_{\Sigma}\\
			\frac{1}{i}(\d\d f)_{\mathrm{T}}
		\end{pmatrix}\bigg\vert_{t=0}=
		\begin{pmatrix}
			\frac{1}{\beta}\partial_{t}f\\
			0\\
			\d_{\Sigma}f\\
			0
		\end{pmatrix}\bigg\vert_{t=0}=
		\begin{pmatrix}
			i\pi\\ 0\\ \d_{\Sigma}\mathfrak{a}\\ 0
		\end{pmatrix}\,,
	\end{align*}
	where we used Lemma~\ref{Lemma:Dec}(i) and $\sf{D}_{1} f=\square f=\delta\d f=0$. Analogously, we choose an initial datum $\mathfrak{c}_{1}:=(\mathfrak{a}_{\mathrm{T}},\pi_{\mathrm{T}},\mathfrak{a}_{\Sigma},\pi_{\Sigma})\in\Gamma^{\infty}(\sf{A}_{\rho_{1}})$ and set $\sf{A}:=\mathcal{U}_{1}^{\sf{F}}\mathfrak{c}_{1}\in \Omega^{1}(\sf{M},\bb{C})$, i.e.~$\sf{A}$ is the unique solution to the hyperbolic equation $\square\sf{A}=0$ with Furlani initial data $\sf{A}_{\mathrm{T}}\vert_{t=0}=\mathfrak{a}_{\mathrm{T}}$, $\delta\sf{A}\vert_{t=0}=i\pi_{\mathrm{T}}$, $\sf{A}_{\Sigma}\vert_{t=0}=\mathfrak{a}_{\Sigma}$ and $(\d\sf{A})_{\mathrm{T}}\vert_{t=0}=i\pi_{\Sigma}$. Then,
	\begin{align*}
		(\sf{K}^{\sf{F}}_{\Sigma})^{\dagger}\mathfrak{c}_{1}=\rho_{0}\sf{K}^{\ast}\sf{A}=
		\begin{pmatrix}
			\delta\sf{A}\\
			\frac{1}{i\beta}\partial_{t}\delta\sf{A}
		\end{pmatrix}\bigg\vert_{t=0}=
		\begin{pmatrix}
			i\pi_{\mathrm{T}}\\
			\delta^{0}_{\Sigma}\pi_{\Sigma}		
		\end{pmatrix}
	\end{align*}
	where in the last equation, we used the fact that $\sf{D}_{2}\sf{A}=\square\sf{A}=(\delta\d+\d\delta)\sf{A}=0$ as well as Lemma~\ref{Lemma:Dec}(i) and (iii) to obtain
	\begin{align*}
		\frac{1}{i\beta}\partial_{t}\delta\sf{A}=\frac{1}{i}(\d\delta\sf{A})_{\mathrm{T}}=-\frac{1}{i}(\delta\d\sf{A})_{\sf{T}}=\frac{1}{i}\delta_{\Sigma}(\d\sf{A})_{\mathrm{T}}\stackrel{t=0}{=}\delta_{\Sigma}\pi_{\Sigma}\, .
	\end{align*}
	Now, for $\sf{K}_{\Sigma}$ and $\sf{K}_{\Sigma}^{\dagger}$, we could in principle follow the same strategy. However, as in the case above, we would have to use the equations $\square f=0$ and $\square\sf{A}=0$ at some point and their decomposition into temporal and spatial part are quite complicated. Hence, we follow a simpler path: consider the isomorphism $\sf{S}_{\sf{F}}$ defined above that satisfies $\rho_{1}^{\sf{F}}=\sf{S}_{\sf{F}}\circ\rho_{1}$ and $\mathcal{U}_{1}^{\sf{F}}=\mathcal{U}_{1}\circ\sf{S}_{\sf{F}}^{-1}$. Then, it follows that 
	\begin{align*}
		\sf{K}_{\Sigma}=\sf{S}_{\sf{F}}^{-1}\circ\sf{K}_{\Sigma}^{\sf{F}}\qquad\text{and}\qquad\sf{K}_{\Sigma}^{\dagger}=(\sf{K}_{\Sigma}^{\sf{F}})^{\dagger}\circ\sf{S}_{\sf{F}}\, .
	\end{align*}
	Using the explicit expressions for $\sf{S}_{\sf{T}}$ and $\sf{S}_{\sf{T}}^{-1}$ from Eq.~\eqref{eq:IsoCauchy}, a straightforward computation yields the claimed results for $\sf{K}_{\Sigma}$ and $\sf{K}_{\Sigma}^{\dagger}$.
\end{proof}

Using the isomorphism $\sf{S}$, we could also derive explicit expressions for $\sf{K}_{\Sigma}$ and $\sf{K}_{\Sigma}^{\dagger}$ using the initial data map $\widetilde{\rho}_{1}$ defined in~\eqref{eq:CauchyRhoMax2}, however, we shall no need them in the following discussion. The expressions for $\sf{K}_{\Sigma}^{\sf{F}}$ and $(\sf{K}_{\Sigma}^{\sf{F}})^{\dagger}$ are extremely simple, which is one of the main advantages of working with the Furlani initial data. This is also not too surprising, since $\rho_{\sf{F}}$ is perfectly adapted to differential forms and the Maxwell operator. However, for our purposes, it will be more convenient to work with $\sf{K}_{\Sigma}$ and $\sf{K}_{\Sigma}^{\dagger}$, for reasons that will become clear later. While the explicit expressions in full generality look quite complicated, we stress that we will mainly work in spacetimes with $\beta=1$ (which in $1+3$ dimensions is no loss of generality, cf.~Proposition~\ref{Prop:MaxConf}), in which case these operators simplify to
\begin{align*}
	\sf{K}_{\Sigma}=\begin{pmatrix} 0 & i\cdot\mathrm{id}\\ i\Delta^{0} & \mathrm{tr}_{\sf{h}_{0}}(\sf{k}_{0})\mathrm{id}\\ \d_{\Sigma} & 0 \\ 0 & \d_{\Sigma}\end{pmatrix}\,,\qquad\sf{K}_{\Sigma}^{\dagger}=
		\begin{pmatrix}
		-\mathrm{tr}_{\sf{h}_{0}}(\sf{k}_{0})\mathrm{id} & i\cdot\mathrm{id} & \delta_{\Sigma}^{0} & 0 \\ i\Delta^{0} & 0 & 0 &\delta_{\Sigma}^{0}
		\end{pmatrix}\, .
\end{align*}

Now, let us denote the advanced and retarded Green operators of the normally hyperbolic operators $\square\:\Omega^{i}(\sf{M},\bb{C})\to\Omega^{i}(\sf{M},\bb{C})$ by $\sf{G}_{i}^{\pm}\:\Omega^{i}_{\mathrm{c}}(\sf{M},\bb{C})\to\Omega^{i}_{\mathrm{sc}}(\sf{M},\bb{C})$ and the corresponding causal propagators, as defined in Definition~\ref{Def.CausProp}, by
\begin{align*}
	\sf{G}_{i}\:\Omega^{i}_{\mathrm{c}}(\sf{M},\bb{C})\to\Omega^{i}_{\mathrm{sc}}(\sf{M},\bb{C})\, .
\end{align*}
Then, following the discussion of general linear gauge theories in Section~\ref{Sec:PhaseGauge}, and in particular the discussion of the Cauchy problem in Proposition~\ref{Prop:CauchyGauge}, there are isomorphisms
\begin{align}\label{eq:PSMaxwell}
\mathcal{V}_{\mathrm{c}}=\cfrac{\mathrm{ker}(\sf{K}^{\ast}\vert_{\Gamma^{\infty}_{\mathrm{c}}})}{\mathrm{ran}(\sf{P}\vert_{\Gamma^{\infty}_{\mathrm{c}}})}\quad\xrightarrow[\cong]{\quad[\sf{G}_{1}]\quad}\quad \mathrm{Sol}_{\mathrm{sc}}=\cfrac{\mathrm{ker}(\sf{P}\vert_{\Gamma^{\infty}_{\mathrm{sc}}})}{\mathrm{ran}(\sf{K}\vert_{\Gamma^{\infty}_{\mathrm{sc}}})}\quad\xrightarrow[\cong]{\quad[\rho_{1}]\quad}\quad\mathcal{V}_{\Sigma}:=\cfrac{\mathrm{ker}(\sf{K}_{\Sigma}^{\dagger}\vert_{\Gamma_{\mathrm{c}}^{\infty}})}{\mathrm{ran}(\sf{K}_{\Sigma}\vert_{\Gamma_{\mathrm{c}}^{\infty}})}\,,
\end{align}
where $\mathcal{V}_{\mathrm{c}}$ is the space of  \emph{classical observables} of Maxwell's theory and $\mathrm{Sol}_{\mathrm{sc}}$ its \emph{solution space}. Furthermore, let us define the following Hermitian sesquilinear form on $\Omega_{\mathrm{c}}^{1}(\sf{M},\bb{C})$:
\begin{align*}
	\sigma&\:\Omega^{1}_{\mathrm{c}}(\sf{M},\bb{C})\times\Omega^{1}_{\mathrm{c}}(\sf{M},\bb{C})\to\bb{C}\,,\qquad\sigma(\sf{A},\sf{B}):=i(\sf{A},\sf{G}_{1}\sf{B})_{\sf{A}_{1}}=i\int_{\sf{M}}\sf{g}^{\sharp}(\overline{\sf{A}},\sf{G}_{1}\sf{B})\,\d\mu_{\sf{g}}\, .
\end{align*}
As shown in Proposition~\ref{Prop:PSSym}, $\sigma$ gives rise to a well-defined Hermitian sesquilinear form on $\mathcal{V}_{\mathrm{c}}$, i.e.~$\sigma\:\mathcal{V}_{\mathrm{c}}\times\mathcal{V}_{\mathrm{c}}\to\bb{C}$, and as already noted in Remark~\ref{Rem:Deg}, we stress that this form is in general \emph{degenerate}. The pair $(\mathcal{V}_{\mathrm{c}},\sigma)$ is the \emph{classical phase space} of Maxwell's theory and lies in the heart of the algebraic quantisation procedure outlined in Section~\ref{Sec.AQFT}.

As discussed in Section~\ref{Sec:PhaseGauge} for general linear gauge theories, the Hermitian form $\sigma$ induces a corresponding Hermitian structure on $\mathcal{V}_{\Sigma}$, thereby defining the \emph{phase space} at the level of initial data. Let us now discuss the construction of this space in the case of Maxwell's theory. To start with, we equip the vector bundle $\sf{A}_{\rho_{i}}\cong\sf{A}_{i}\vert_{\Sigma_{0}}\oplus\sf{A}_{i}\vert_{\Sigma_{0}}$ with suitable bundle metrics \begin{align*}
	\langle\cdot,\cdot\rangle_{\sf{A}_{\rho_{i}}}\in\Gamma^{\infty}(\overline{\sf{A}}_{\rho_{i}}^{\ast}\otimes\sf{A}_{\rho_{i}}^{\ast})\,.
\end{align*}
Following Remark~\ref{Remark:PhysicalCharge}, the choice of these bundle metrics is arbitrary and only serves an auxiliary purpose. For simplicity, we choose them to be \emph{positive definite}:
\begin{align*}
		\bigg\langle
		\begin{pmatrix}
			\mathfrak{a}\\\pi
		\end{pmatrix},
		\begin{pmatrix}
			\mathfrak{b}\\\rho
		\end{pmatrix}\bigg\rangle_{\sf{A}_{\rho_{0}}}:&=\langle\mathfrak{a},\mathfrak{b}\rangle_{\sf{A}_{\Sigma,0}}^{0}+\langle\pi,\rho\rangle_{\sf{A}_{\Sigma,0}}^{0}=\overline{\mathfrak{a}}\mathfrak{b}+\overline{\pi}\rho\\
		\bigg\langle
		\begin{pmatrix}
			\mathfrak{a}_{\mathrm{T}}\\\pi_{\mathrm{T}}\\\mathfrak{a}_{\Sigma}\\\pi_{\Sigma}
		\end{pmatrix},
		\begin{pmatrix}
			\mathfrak{b}_{\mathrm{T}}\\\rho_{\mathrm{T}}\\\mathfrak{b}_{\Sigma}\\\rho_{\Sigma}
		\end{pmatrix}\bigg\rangle_{\sf{A}_{\rho_{1}}}:&=\langle\mathfrak{a}_{\mathrm{T}},\mathfrak{b}_{\mathrm{T}}\rangle_{\sf{A}_{\Sigma,0}}^{0}+\langle\pi_{\mathrm{T}},\rho_{\mathrm{T}}\rangle_{\sf{A}_{\Sigma,0}}^{0}+\langle\mathfrak{a}_{\Sigma},\mathfrak{b}_{\Sigma}\rangle_{\sf{A}_{\Sigma,1}}^{0}+\langle\pi_{\Sigma},\rho_{\Sigma}\rangle_{\sf{A}_{\Sigma,1}}^{0}\\&=\overline{\mathfrak{a}_{\mathrm{T}}}\mathfrak{b}_{\mathrm{T}}+\overline{\pi_{\mathrm{T}}}\rho_{\mathrm{T}}+\sf{h}_{0}^{\sharp}(\overline{\mathfrak{a}_{\Sigma}},\mathfrak{b}_{\Sigma})+\sf{h}^{\sharp}_{0}(\overline{\pi_{\Sigma}},\rho_{\Sigma})
\end{align*}
for all $(\mathfrak{a},\pi),(\mathfrak{b},\rho)\in\Gamma^{\infty}(\sf{A}_{\rho_{0}})$ and $(\mathfrak{a}_{\mathrm{T}},\pi_{\mathrm{T}},\mathfrak{a}_{\Sigma},\pi_{\Sigma}),(\mathfrak{b}_{\mathrm{T}},\rho_{\mathrm{T}},\mathfrak{b}_{\Sigma},\rho_{\Sigma})\in\Gamma^{\infty}(\sf{A}_{\rho_{1}})$, where we recall that $\langle\cdot,\cdot\rangle_{\sf{A}_{\Sigma,k}}^{t}:=\frac{1}{k!}\sf{h}_{t}^{\sharp}(\overline{\cdot},\cdot)$ are bundle metrics on $\sf{A}_{\Sigma,k}$, see Section~\ref{Sec.DiffFormsGlobHyp}. As usual, we denote the corresponding positive-definite inner product on the level of sections by
\begin{align*}
	(\mathfrak{f},\mathfrak{g})_{\sf{A}_{\rho_{i}}}:=\int_{\Sigma}\langle\mathfrak{f},\mathfrak{g}\rangle_{\sf{A}_{\rho_{i}}}\,\d\mu_{\sf{h}_{0}}
\end{align*} 
for all $\mathfrak{f},\mathfrak{g}\in\Gamma^{\infty}(\sf{A}_{\rho_{i}})$ with compactly overlapping supports. We shall denote the formal adjoint of operators acting on and between the bundle $\sf{A}_{i}$ and $\sf{A}_{\rho_{i}}$ with respect to $(\cdot,\cdot)_{\sf{A}_{i}}$ and $(\cdot,\cdot)_{\sf{A}_{\rho_{i}}}$ by ``$\ast$''. Moreover, following the general discussion of Section~\ref{Sec:PhaseGauge} (see Corollary~\ref{Cor:KSigmaAdj}), we define the (degenerate) Hermitian sesquilinear forms 
\begin{align*}
	\sigma_{\Sigma,0}&\:\Gamma^{\infty}_{\mathrm{c}}(\sf{A}_{\rho_{0}})\times\Gamma^{\infty}_{\mathrm{c}}(\sf{A}_{\rho_{0}})\to\bb{C}\,,\qquad &&\sigma_{\Sigma,0}(\mathfrak{f},\mathfrak{g}):=i(\mathfrak{f},\sf{G}_{\Sigma,0}\mathfrak{g})_{\sf{A}_{\rho_{0}}}\\
	\sigma_{\Sigma,1}&\:\Gamma^{\infty}_{\mathrm{c}}(\sf{A}_{\rho_{1}})\times\Gamma^{\infty}_{\mathrm{c}}(\sf{A}_{\rho_{1}})\to\bb{C}\,,\qquad &&\sigma_{\Sigma,1}(\mathfrak{f},\mathfrak{g}):=i(\mathfrak{f},\sf{G}_{\Sigma,1}\mathfrak{g})_{\sf{A}_{\rho_{1}}}\,.
\end{align*}
where the linear differential operators $\sf{G}_{\Sigma,i}\:\Gamma^{\infty}(\sf{A}_{\rho_{i}})\to\Gamma^{\infty}(\sf{A}_{\rho_{i}})$ are defined as in Proposition~\ref{Prop:PropGSigma}, i.e.
\begin{align*}
	\sf{G}_{i}=(\rho_{i}\sf{G}_{i})^{\ast}\sf{G}_{\Sigma,i}(\rho_{i}\sf{G}_{i})\,,
\end{align*}
We recall that the operators $\sf{G}_{\Sigma,i}$ explicitly depend on the choice of bundle metrics $\langle\cdot,\cdot\rangle_{\sf{A}_{\rho_{i}}}$ on $\sf{A}_{\rho_{i}}$, however, the Hermitian forms $\sigma_{\Sigma,i}$ are independent of this choice (cf.~Remark~\ref{Remark:PhysicalCharge}). Following Proposition~\ref{Prop:EqPS}, the operator $\rho_{1}\sf{G}_{1}$ induces a unitary isomorphism
\begin{align}\label{eq:sdfsafangagasge}
\bigg(\mathcal{V}_{\mathrm{c}}:=\cfrac{\mathrm{ker}(\sf{K}^{\ast}\vert_{\Gamma^{\infty}_{\mathrm{c}}})}{\mathrm{ran}(\sf{P}\vert_{\Gamma^{\infty}_{\mathrm{c}}})} ,\sigma\bigg)\xrightarrow[\cong]{\quad[\rho_{1}\sf{G}_{1}]\quad}\bigg(\mathcal{V}_{\Sigma}:=\cfrac{\mathrm{ker}(\sf{K}_{\Sigma}^{\dagger}\vert_{\Gamma_{\mathrm{c}}^{\infty}})}{\mathrm{ran}(\sf{K}_{\Sigma}\vert_{\Gamma_{\mathrm{c}}^{\infty}})},\sigma_{\Sigma,1}\bigg)\,,
\end{align}
where $\sigma_{\Sigma,1}$ is well-defined on the quotient space $\mathcal{V}_{\Sigma}$ on account of Proposition~\ref{Prop:EqPS}.

It remains to derive explicit expressions of the operators $\sf{G}_{\Sigma,i}\:\Gamma^{\infty}(\sf{A}_{\rho_{i}})\to\Gamma^{\infty}(\sf{A}_{\rho_{i}})$. To start with, we first need to establish a suitable \emph{Green's identity}. 

\begin{proposition}\label{Prop:Green} \emph{(Green's Identity for Normally Hyperbolic Operators)}\newline
	Let $(\sf{E},\langle\cdot,\cdot\rangle_{\sf{E}})$ be a Hermitian vector bundle over a globally hyperbolic spacetime $(\sf{M},\sf{g})$ and $\sf{N}\in\mathrm{DO}^{2}(\sf{E})$ be a formally self-adjoint normally hyperbolic operator with causal propagator $\sf{G}\:\Gamma^{\infty}_{\mathrm{c}}(\sf{E})\to\Gamma^{\infty}_{\mathrm{sc}}(\sf{E})$. Furthermore, we fix a Cauchy temporal function $t\in C^{\infty}(\sf{M})$ such that $\sf{M}=\bb{R}\times\Sigma$ and $\sf{g}=-\beta^{2}\d t\otimes\d t+\sf{h}_{t}$, as usual. Then, for all $\psi,\varphi\in\Gamma^{\infty}_{\mathrm{c}}(\sf{E})$,
	\begin{align*}
		(\psi,\sf{G}\varphi)_{\sf{E}}=\int_{\Sigma_{0}}\bigg(\langle\nabla_{\nu}\mathfrak{G}_{\psi},\mathfrak{G}_{\varphi}\rangle_{\sf{E}}-\langle\mathfrak{G}_{\psi},\nabla_{\nu}\mathfrak{G}_{\varphi}\rangle_{\sf{E}}\bigg)\bigg\vert_{t=0}\,\d\mu_{\sf{h}_{0}}
	\end{align*}
	where $\mathfrak{G}_{\psi}:=\sf{G}\psi$ and $\mathfrak{G}_{\varphi}:=\sf{G}\varphi$ and $\nu=\beta^{-1}\partial_{t}$ is the future-directed timelike unit normal.
\end{proposition}

\begin{proof}
	The proposition and its proof are similar to the proof of Lemma~\ref{Lemma:AdjointSHS}, in which we derived the Green's identity for symmetric hyperbolic systems. Following Lemma~\ref{Lem.NH}, we can find a connection $\nabla^{\sf{E}}\:\Gamma^{\infty}(\sf{E})\to\Gamma^{\infty}(\sf{E}\otimes\sf{T}^{\ast}\sf{M})$ on $\sf{E}$ and a zeroth-order operator $\sf{N}_{0}$ such that $\sf{N}=-\square^{\nabla^{\sf{E}}}+\sf{N}_{0}$. Since $\sf{N}$ is formally self-adjoint, $\nabla^{\sf{E}}$ is compatible with $\langle\cdot,\cdot\rangle_{\sf{E}}$ and $\sf{N}_{0}^{\ast}=\sf{N}_{0}$ (see Remark~\ref{Rem:ExBM}). Now, for fixed $\psi,\varphi\in\Gamma^{\infty}_{\mathrm{c}}(\sf{E})$, consider the $n$-form $\omega\in\Omega^{n}(\sf{M},\bb{C})$ defined by
	\begin{align*}
		\omega_{\psi,\varphi}:=(\chi_{\psi,\varphi}^{\sharp})\lrcorner\d\mu_{\sf{g}}\,\qquad\text{with}\qquad\chi_{\psi,\varphi}:=\big(\langle\nabla_{\partial_{\alpha}}^{\sf{E}}\psi,\varphi\rangle_{\sf{E}}-\langle\psi,\nabla_{\partial_{\alpha}}^{\sf{E}}\varphi\rangle_{\sf{E}}\big)\d x^{\alpha}\in\Omega^{1}(\sf{M},\bb{C})\,,
	\end{align*}
	where we chose local coordinates $(x^{\alpha})_{\alpha=0,\dots,n}$. Now, using Cartan's magic formula, we find
	\begin{align*}
		\d\omega_{\psi,\varphi}&=\mathrm{div}_{\sf{g}}(\chi_{\psi,\varphi}^{\sharp})\d\mu_{\sf{g}}=\sf{g}^{\alpha\beta}(\partial_{\alpha}(\chi_{\psi,\varphi})_{\beta}-\Gamma_{\alpha\beta}^{\gamma}(\chi_{\psi,\varphi})_{\gamma})\d\mu_{\sf{g}}=\\&=\bigg(\langle \square^{\nabla^{\sf{E}}}\psi,\varphi\rangle_{\sf{E}}+\cancel{\sf{g}^{\alpha\beta}\langle\nabla_{\partial_{\beta}}^{\sf{E}}\psi,\nabla^{\sf{E}}_{\partial_{\alpha}}\varphi\rangle_{\sf{E}}}-\langle\psi,\square^{\nabla^{\sf{E}}}\varphi\rangle_{\sf{E}}-\cancel{\sf{g}^{\alpha\beta}\langle\nabla_{\partial_{\alpha}}^{\sf{E}}\psi,\nabla_{\partial_{\beta}}^{\sf{E}}\varphi\rangle_{\sf{E}}}\bigg)\d\mu_{\sf{g}}=\\&=\big(-\langle\sf{N}\psi,\varphi\rangle_{\sf{E}}+\langle\psi,\sf{N}\varphi\rangle_{\sf{E}}\big)\d\mu_{\sf{g}}\,,
	\end{align*}
	where we used $\square^{\nabla^{\sf{E}}}=\sf{g}^{\alpha\beta}(\nabla^{\sf{E}}_{\partial_{\alpha}}\nabla^{\sf{E}}_{\partial_{\beta}}-\Gamma^{\gamma}_{\alpha\beta}\nabla^{\sf{E}}_{\partial_{\gamma}})$ (see Example~\ref{Ex:ConDA}) and the fact that $\nabla^{\sf{E}}$ is compatible with $\langle\cdot,\cdot\rangle_{\sf{E}}$. Furthermore, we added and subtracted $\langle\sf{N}_{0}\psi,\varphi\rangle_{\sf{E}}=\langle\psi,\sf{N}_{0}\varphi\rangle_{\sf{E}}$ in the last step, using the fact that $\sf{N}_{0}$ is an operator of order zero, which implies that the adjoint relation holds pointwise. Now, writing $\psi^{\pm}:=\sf{G}^{\pm}\psi$, Stoke's theorem and the decomposition $\sf{M}=\mathcal{J}^{+}(\Sigma_{0})\cup\mathcal{J}^{-}(\Sigma_{0})$ imply
\begin{align*}
	(\psi,\sf{G}\varphi)_{\sf{E}}&=\int_{\sf{M}}\langle\psi,\mathfrak{G}_{\varphi}\rangle_{\sf{E}}\,\d\mu_{\sf{g}}=\int_{\mathcal{J}^{+}(\Sigma_{0})}\langle\sf{N}\psi^{-},\mathfrak{G}_{\varphi}\rangle_{\sf{E}}\,\d\mu_{\sf{g}}+\int_{\mathcal{J}^{-}(\Sigma_{0})}\langle\sf{N}\psi^{+},\mathfrak{G}_{\varphi}\rangle_{\sf{E}}\,\d\mu_{\sf{g}}=\\&=-\int_{\mathcal{J}^{+}(\Sigma_{0})}\d\omega_{\psi^{-},\mathfrak{G}_{\varphi}}-\int_{\mathcal{J}^{-}(\Sigma_{0})}\d\omega_{\psi^{+},\mathfrak{G}_{\varphi}}=\int_{\Sigma_{0}}i_{0}^{\ast}(\omega_{\psi^{-},\mathfrak{G}_{\varphi}})-\int_{\Sigma_{0}}i_{0}^{\ast}(\omega_{\psi^{+},\mathfrak{G}_{\varphi}})\,,
\end{align*}	
where we used $\sf{N}\mathfrak{G}_{\varphi}=\sf{N}\sf{G}\varphi=0$ and  $\sf{N}\psi^{\pm}=\sf{N}\sf{G}^{\pm}\psi=\psi$ and where $i_{0}:\Sigma_{0}\hookrightarrow\sf{M}$ denotes the embedding. Note also the sign change in the last step, reflecting the fact that the induced orientation on $\partial\mathcal{J}^{+}(\Sigma_{0})=\Sigma_{0}$ is the opposite then the usual one on $\Sigma_{0}$, since the \textit{outwards pointing normal} in this case is the \textit{past-directed} timelike unit normal $-\beta^{-1}\partial_{t}$. Now, let us denote by $\nu=\beta^{-1}\partial_{t}$ the future-directed normal vector. Then, we can decompose any $\sf{X}\in\mathfrak{X}(\sf{M})$ into a normal component $\sf{X}^{\perp}=-\sf{g}(\sf{X},\nu)\nu$ and tangential component $\sf{X}^{\top}=\sf{X}-\sf{X}^{\perp}$. In particular, 
\begin{align*}
	i^{\ast}_{0}(\omega_{\psi,\varphi})&=i^{\ast}_{0}((\chi_{\psi,\varphi}^{\sharp})^{\perp}\lrcorner\d\mu_{\sf{g}})=-\sf{g}(\chi_{\psi,\varphi}^{\sharp},\nu)\vert_{t=0}i_{0}^{\ast}(\nu\lrcorner\d\mu_{\sf{g}})=-\sf{g}(\chi_{\psi,\varphi}^{\sharp},\nu)\vert_{t=0}\d\mu_{\sf{h}_{0}}=\\&=\bigg(-\langle\nabla_{\nu}^{\sf{E}}\psi,\varphi\rangle_{\sf{E}}+\langle\psi,\nabla_{\nu}^{\sf{E}}\varphi\rangle_{\sf{E}}\bigg)\bigg\vert_{t=0}\d\mu_{\sf{h}_{0}}\, .
\end{align*}
Plugging this equation into the formula obtained above, we obtain the claimed result
\begin{align*}
	(\psi,\sf{G}\varphi)_{\sf{E}}&=\int_{\Sigma_{0}}\bigg(-\langle\nabla_{\nu}^{\sf{E}}\psi^{-},\mathfrak{G}_{\varphi}\rangle_{\sf{E}}+\langle\psi^{-},\nabla_{\nu}^{\sf{E}}\mathfrak{G}_{\varphi}\rangle_{\sf{E}}+\langle\nabla_{\nu}^{\sf{E}}\psi^{+},\mathfrak{G}_{\varphi}\rangle_{\sf{E}}-\langle\psi^{+},\nabla_{\nu}^{\sf{E}}\mathfrak{G}_{\varphi}\rangle_{\sf{E}}\bigg)\bigg\vert_{t=0}\d\mu_{\sf{h}_{0}}\\&=\int_{\Sigma_{0}}\bigg(\langle\nabla_{\nu}\mathfrak{G}_{\psi},\mathfrak{G}_{\varphi}\rangle_{\sf{E}}-\langle\mathfrak{G}_{\psi},\nabla_{\nu}\mathfrak{G}_{\varphi}\rangle_{\sf{E}}\bigg)\bigg\vert_{t=0}\,\d\mu_{\sf{h}_{0}}\,,
\end{align*}	
	where we used that $\psi^{+}-\psi^{-}=\sf{G}^{+}\psi-\sf{G}^{-}\psi=\sf{G}\psi=\mathfrak{G}_{\psi}$ and similarly $\varphi^{+}-\varphi^{-}=\mathfrak{G}_{\varphi}$.
\end{proof}

Now, using Green's identity, it is straightforward to derive explicit expressions of the linear differential operators $\sf{G}_{\Sigma,i}$. By definition, they depend on the choice of initial data map. For $i=0$, we recall from~\eqref{eq:CauchyRhoMax} that $\rho_{0}$ is defined by
\begin{align*}
	\Gamma^{\infty}(\sf{A}_{0})\ni f\quad\mapsto\quad \rho_{0}(f)=\bigg(f,\,\frac{1}{i}\nabla_{\nu}f\bigg)\bigg\vert_{t=0}=\bigg(f,\,\frac{1}{i\beta}\partial_{t}f\bigg)\bigg\vert_{t=0}\in\Gamma^{\infty}(\sf{A}_{\rho_{0}})\,,	
\end{align*}
while for $1$-forms, we have considered three different choices, namely~\eqref{eq:CauchyRhoMax}, \eqref{eq:CauchyRhoMax2} and \eqref{eq:CauchyRhoMaxFur}, which for convenience of the reader we display here again:
\begin{align*}
	\Gamma^{\infty}(\sf{A}_{1})\ni \sf{A}\quad\mapsto\quad\rho_{1}(\sf{A})&:=\bigg(\sf{A}_{\mathrm{T}},\,\frac{1}{i\beta}\partial_{t}\sf{A}_{\mathrm{T}},\,\sf{A}_{\Sigma},\,\frac{1}{i\beta}\partial_{t}\sf{A}_{\Sigma}\bigg)\bigg\vert_{t=0}\in\Gamma^{\infty}(\sf{A}_{\rho_{1}})\\
\Gamma^{\infty}(\sf{A}_{1})\ni \sf{A}\quad\mapsto\quad\widetilde{\rho}_{1}(\sf{A})&:=\bigg(\sf{A}_{\mathrm{T}},\,\frac{1}{i}(\nabla_{\nu}\sf{A})_{\mathrm{T}},\,\sf{A}_{\Sigma},\,\frac{1}{i}(\nabla_{\nu}\sf{A})_{\Sigma}\bigg)\bigg\vert_{t=0}\in\Gamma^{\infty}(\sf{A}_{\rho_{1}})\\
	\Gamma^{\infty}(\sf{A}_{1})\ni \sf{A}\quad\mapsto\quad\rho_{1}^{\sf{F}}(\sf{A})&:=\bigg(\sf{A}_{\mathrm{T}},\,\frac{1}{i}\delta\sf{A},\,\sf{A}_{\Sigma},\,\frac{1}{i}(\d\sf{A})_{\mathrm{T}}\bigg)\bigg\vert_{t=0}\in\Gamma^{\infty}(\sf{A}_{\rho_{1}})
\end{align*}

\begin{proposition}\label{Prop:GSigma} \emph{(The Operators $\sf{G}_{\Sigma,i}$)}\newline
	Let $(\sf{M},\sf{g})$ be a globally hyperbolic manifold and fix a Cauchy temporal function $t\in C^{\infty}(\sf{M},\bb{R})$. The operator $\sf{G}_{\Sigma,0}$ takes the form
	\begin{align*}
		\sf{G}_{\Sigma,0}=\frac{1}{i}
		\begin{pmatrix}
			0 & \mathrm{id}\\
			\mathrm{id} & 0
		\end{pmatrix}\,,
	\end{align*}
	while $\sf{G}_{\Sigma,1},\widetilde{\sf{G}}_{\Sigma,1}$ and $\sf{G}_{\Sigma,1}^{\sf{F}}$ defined using $\rho_{1},\widetilde{\rho}_{1}$ and $\rho_{1}^{\sf{F}}$, respectively, take the forms
	\begin{align*}
		\sf{G}_{\Sigma,1}^{\sf{F}}=\frac{1}{i}
		\begin{pmatrix}
			0 & -\mathrm{id} & 0 & 0\\
			-\mathrm{id} & 0 & 0 & 0\\
			0 & 0 & 0 & \mathrm{id}\\
			0 & 0 & \mathrm{id} & 0
		\end{pmatrix},\, \sf{G}_{\Sigma,1}=\frac{1}{i}
		\begin{pmatrix}
			0 & -\mathrm{id} & -\frac{2i}{\beta_{0}}\sf{h}^{\sharp}(\d_{\Sigma}\beta_{0},\cdot) & 0\\
			-\mathrm{id} & 0 & 0 & 0\\
			\frac{2i}{\beta_{0}}(\d_{\Sigma}\beta_{0})\mathrm{id} & 0 & 0 & \mathrm{id}\\
			0 & 0 & \mathrm{id} & 0
		\end{pmatrix}
	\end{align*}
	and $\widetilde{\sf{G}}_{\Sigma,1}=\sf{G}_{\Sigma,1}^{\sf{F}}$. In particular, we note that $\sf{G}_{\Sigma,1}=\widetilde{\sf{G}}_{\Sigma,1}=\sf{G}_{\Sigma,1}^{\sf{F}}$ in the case $\beta=1$.
\end{proposition}

\begin{proof}
	Take $f,g\in C^{\infty}_{\mathrm{c}}(\sf{M},\bb{C})$ and write $\mathcal{F}:=\sf{G}_{0}f$ and $\mathcal{G}:=\sf{G}_{0}g$. Then, by Proposition~\ref{Prop:Green},
	\begin{align*}
		i(f,\sf{G}_{0}g)_{\sf{A}_{0}}&=i\int_{\Sigma_{0}}\bigg\{\bigg(\frac{1}{\beta}\overline{\partial_{t}\mathcal{F}}\bigg)\mathcal{G}-\overline{\mathcal{F}}\bigg(\frac{1}{\beta}\partial_{t}\mathcal{G}\bigg)\bigg\}\bigg\vert_{t=0}\d\mu_{\sf{h}_{0}}=\\&=i\int_{\Sigma_{0}}\frac{1}{i}\bigg\{\bigg(\overline{\frac{1}{i\beta}\partial_{t}\mathcal{F}}\bigg)\mathcal{G}+\overline{\mathcal{F}}\bigg(\frac{1}{i\beta}\partial_{t}\mathcal{G}\bigg)\bigg\}\bigg\vert_{t=0}\d\mu_{\sf{h}_{0}}=\\&=i\bigg\langle\rho_{0}\mathcal{F},\frac{1}{i}\begin{pmatrix} 0 & \mathrm{id}\\ \mathrm{id}& 0\end{pmatrix}\rho_{0}\mathcal{G}\bigg\rangle_{\sf{A}_{\rho_{0}}}=i\bigg\langle f,(\rho_{0}\sf{G}_{0})^{\ast}\frac{1}{i}\begin{pmatrix}0 & \mathrm{id}\\ \mathrm{id}& 0\end{pmatrix}(\rho_{0}\sf{G}_{0})g\bigg\rangle_{\sf{A}_{0}}\, .
	\end{align*}
	Since $f\in C^{\infty}_{\mathrm{c}}(\sf{M},\bb{C})$ was arbitrary, non-degeneracy of $\langle\cdot,\cdot\rangle_{\sf{A}_{0}}$ and the defining relation $\sf{G}_{0}=(\rho_{0}\sf{G}_{0})^{\ast}\sf{G}_{\Sigma,0}(\rho_{0}\sf{G}_{0})$ yield the claimed result for $\sf{G}_{\Sigma,0}$. Similarly, take $\sf{A},\sf{B}\in\Omega^{1}(\sf{M},\bb{C})$ and write $\mathcal{A}:=\sf{G}_{1}\sf{A}$ and $\mathcal{B}:=\sf{G}_{1}\sf{B}$. Then, using Proposition~\ref{Prop:Green}, a similar computation implies
	\begin{align*}
		i(\sf{A},&\sf{G}_{1}\sf{B})_{\sf{A}_{1}}=i\int_{\Sigma_{0}}\bigg\{\langle\nabla_{\nu}\mathcal{A},\mathcal{B}\rangle_{\sf{A}_{1}}-\langle\mathcal{A},\nabla_{\nu}\mathcal{B}\rangle_{\sf{A}_{1}}\bigg\}\bigg\vert_{t=0}\,\d\mu_{\sf{h}_{0}}=\\&=i\int_{\Sigma_{0}}\bigg\{-\overline{(\nabla_{\nu}\mathcal{A})}_{\mathrm{T}}\mathcal{B}_{\mathrm{T}}+\overline{\mathcal{A}}_{\mathrm{T}}(\nabla_{\nu}\mathcal{B})_{\mathrm{T}}+\sf{h}^{\sharp}(\overline{(\nabla_{\nu}\mathcal{A})}_{\Sigma},\mathcal{B}_{\Sigma})-\sf{h}^{\sharp}(\overline{\mathcal{A}}_{\Sigma},(\nabla_{\nu}\mathcal{B})_{\Sigma})\bigg\}\bigg\vert_{t=0}\d\mu_{\sf{h}_{0}}\\&=i\bigg\langle\sf{A},(\widetilde{\rho}_{1}\sf{G}_{1})^{\ast}\frac{1}{i}
		\begin{pmatrix}
			0 & -\mathrm{id} & 0 & 0\\
			-\mathrm{id} & 0 & 0 & 0\\
			0 & 0 & 0 & \mathrm{id}\\
			0 & 0 & \mathrm{id} & 0
		\end{pmatrix}(\widetilde{\rho}_{1}\sf{G}_{1})\sf{B}\bigg\rangle_{\sf{A}_{1}}\,,
	\end{align*}
	where we used that $\langle\cdot,\cdot\rangle_{\sf{A}_{1}}=\sf{g}^{\sharp}(\overline{\cdot},\cdot)$ and $\sf{g}^{\sharp}(\mathcal{A},\mathcal{B})=-\overline{\mathcal{A}}_{\mathrm{T}}\mathcal{B}_{\mathrm{T}}+\sf{h}^{\sharp}(\overline{\mathcal{A}}_{\Sigma},\mathcal{B}_{\Sigma})$. As before, non-degeneracy of $\langle\cdot,\cdot\rangle_{\sf{A}_{1}}$ yields the claimed result for $\widetilde{\sf{G}}_{\Sigma,1}$. In this context, the Cauchy data map $\widetilde{\rho}_{1}$ appears quite natural, since the normal covariant derivatives also appear in the Green identity from Proposition~\ref{Prop:Green}. Now, to derive $\sf{G}_{\Sigma,1}^{\sf{F}}$, we note that
	\begin{align*}
		(\nabla_{\nu}\sf{C})_{\mathrm{T}}=\delta\sf{C}+\mathrm{tr}_{\sf{h}}(\sf{k})\sf{C}_{\mathrm{T}}-\delta_{\Sigma}\sf{C}_{\Sigma}\,,\qquad (\nabla_{\nu}\sf{C})_{\Sigma}=(\d\sf{C})_{\mathrm{T}}+\d_{\Sigma}\sf{C}_{\mathrm{T}}+\sf{W}_{\sf{h}}(\sf{C}_{\Sigma})\, .
	\end{align*}
	for all $\sf{C}\in\Omega^{1}(\sf{M},\bb{C})$, by Lemma~\ref{Lemma:Dec}(i),(ii) and Eq.~\eqref{eq:NormalDer}. With these equations, it follows that
	\begin{equation}\begin{aligned}\label{eq:ProofGSigma}
		-\overline{(\nabla_{\nu}\mathcal{A})}_{\mathrm{T}}\mathcal{B}_{\mathrm{T}}&+\overline{\mathcal{A}}_{\mathrm{T}}(\nabla_{\nu}\mathcal{B})_{\mathrm{T}}+\sf{h}^{\sharp}(\overline{(\nabla_{\nu}\mathcal{A})}_{\Sigma},\mathcal{B}_{\Sigma})-\sf{h}^{\sharp}(\overline{\mathcal{A}}_{\Sigma},(\nabla_{\nu}\mathcal{B})_{\Sigma})=\\=&-(\overline{\delta\mathcal{A}})\mathcal{B}_{\mathrm{T}}+\overline{\mathcal{A}}_{\mathrm{T}}(\delta\mathcal{B})+\sf{h}^{\sharp}(\overline{(\d\mathcal{A})}_{\mathrm{T}},\mathcal{B}_{\Sigma})-\sf{h}^{\sharp}(\overline{\mathcal{A}}_{\Sigma},(\d\mathcal{B})_{\mathrm{T}})\\&+\overline{(\delta_{\Sigma}\mathcal{A}_{\Sigma})}\mathcal{B}_{\mathrm{T}}-\overline{\mathcal{A}}_{\mathrm{T}}(\delta_{\Sigma}\mathcal{B}_{\Sigma})+\sf{h}^{\sharp}(\overline{\d_{\Sigma}\mathcal{A}}_{\mathrm{T}},\mathcal{B}_{\Sigma})-\sf{h}^{\sharp}(\overline{\mathcal{A}}_{\Sigma},\d_{\Sigma}\mathcal{B}_{\mathrm{T}})\,,
	\end{aligned}
	\end{equation}
	where we used that the zeroth-order operator $\sf{W}_{\sf{h}_{0}}$ is formally self-adjoint with respect to $\sf{h}_{0}$, as one can easily see from its definition, i.e.
	\begin{align*}
		\sf{h}^{\sharp}(\sf{W}_{\sf{h}}(\overline{\omega}),\eta)=\sf{h}^{ij}(\sf{W}_{\sf{h}}(\overline{\omega}))_{i}\eta_{j}=\sf{h}^{ij}\sf{k}^{l}_{i}\overline{\omega}_{l}\eta_{j}=\sf{k}^{ij}\overline{\omega}_{i}\eta_{j}=\sf{h}^{ij}\overline{\omega}_{i}\sf{k}^{l}_{j}\eta_{l}=\sf{h}^{ij}\overline{\omega}_{i}(\sf{W}_{\sf{h}}(\eta))_{j}=\sf{h}^{\sharp}(\overline{\omega},\sf{W}_{\sf{h}}(\eta))
	\end{align*}
	for all $\omega,\eta\in C^{\infty}(\bb{R},\Omega^{1}(\Sigma,\bb{C}))$. Now, after integrating over $\Sigma_{0}$, we see that the second line of the right-hand side in Eq.~\eqref{eq:ProofGSigma} vanishes, since $\d_{\Sigma}^{\ast}=\delta_{\Sigma}$. Hence, the computation for deriving $\widetilde{\sf{G}}_{\Sigma,1}$ above stays virtually the same with $(\nabla_{\nu}\mathcal{A})_{\mathrm{T}}$ replaced by $\delta\mathcal{A}$ and $(\nabla_{\nu}\mathcal{A})_{\Sigma}$ replaced by $(\d\mathcal{A})_{\mathrm{T}}$. In particular, we conclude that $\sf{G}_{\Sigma,1}^{\sf{F}}=\widetilde{\sf{G}}_{\Sigma,1}$. Analogously, to compute $\sf{G}_{\Sigma,1}$, we recall that
	\begin{align*}
		(\nabla_{\nu}\sf{C})_{\mathrm{T}}=\frac{1}{\beta}\partial_{t}\sf{C}_{\mathrm{T}}-\frac{1}{\beta}\sf{h}^{\sharp}(\d_{\Sigma}\beta,\sf{C}_{\Sigma})\,,\qquad (\nabla_{\nu}\sf{C})_{\Sigma}=\frac{1}{\beta}\partial_{t}\sf{C}_{\Sigma}-\frac{1}{\beta}(\d_{\Sigma}\beta)\sf{C}_{\mathrm{T}}+\sf{W}_{\sf{h}}(\sf{C}_{\Sigma})\, ,
\end{align*}
for all $\sf{C}\in\Omega^{1}(\sf{M},\bb{C})$, by Eq.~\eqref{eq:NormalDer}. Using these relations, we find
\begin{align*}
		-\overline{(\nabla_{\nu}\mathcal{A})}_{\mathrm{T}}\mathcal{B}_{\mathrm{T}}&+\overline{\mathcal{A}}_{\mathrm{T}}(\nabla_{\nu}\mathcal{B})_{\mathrm{T}}+\sf{h}^{\sharp}(\overline{(\nabla_{\nu}\mathcal{A})}_{\Sigma},\mathcal{B}_{\Sigma})-\sf{h}^{\sharp}(\overline{\mathcal{A}}_{\Sigma},(\nabla_{\nu}\mathcal{B})_{\Sigma})=\\=-&\overline{\bigg(\frac{1}{\beta}\partial_{t}\mathcal{A}_{\mathrm{T}}\bigg)}\mathcal{B}_{\mathrm{T}}+\overline{\mathcal{A}}_{\mathrm{T}}\bigg(\frac{1}{\beta}\partial_{t}\mathcal{B}_{\mathrm{T}}\bigg)+\sf{h}^{\sharp}\bigg(\overline{\bigg(\frac{1}{\beta}\partial_{t}\mathcal{A}_{\Sigma}\bigg)},\mathcal{B}_{\Sigma}\bigg)-\sf{h}^{\sharp}\bigg(\overline{\mathcal{A}}_{\Sigma},\bigg(\frac{1}{\beta}\partial_{t}\mathcal{B}_{\Sigma}\bigg)\bigg)\\&+\frac{1}{\beta}\sf{h}^{\sharp}(\d_{\Sigma}\beta,\overline{\mathcal{A}}_{\Sigma})\mathcal{B}_{\mathrm{T}}-\frac{1}{\beta}\overline{\mathcal{A}}_{\mathrm{T}}\sf{h}^{\sharp}(\d_{\Sigma}\beta,\mathcal{B}_{\Sigma})-\frac{1}{\beta}\overline{\mathcal{A}}_{\mathrm{T}}\sf{h}^{\sharp}(\d_{\Sigma}\beta,\mathcal{B}_{\Sigma})+\frac{1}{\beta}\sf{h}^{\sharp}(\overline{\mathcal{A}}_{\Sigma},\d_{\Sigma}\beta)\mathcal{B}_{\mathrm{T}}\, .
	\end{align*}
	Now, in this case, we get two extra terms, namely $\frac{2}{\beta}\sf{h}^{\sharp}(\overline{\mathcal{A}}_{\Sigma},\d_{\Sigma}\beta)\mathcal{B}_{\mathrm{T}}$ and $-\frac{2}{\beta}\overline{\mathcal{A}}_{\mathrm{T}}\sf{h}^{\sharp}(\d_{\Sigma}\beta,\mathcal{B}_{\Sigma})$, which gives the claimed result for $\sf{G}_{\Sigma,1}$.
\end{proof}

\begin{remark}
	As discussed in Corollary~\ref{Cor:KSigmaAdj}, $\sf{K}_{\Sigma}^{\dagger}$ is the formal adjoint of $\sf{K}_{\Sigma}$ with respect to $\sigma_{\Sigma,i}$. Indeed, using Proposition~\ref{Prop:KSigma} and Proposition~\ref{Prop:GSigma}, it is straightforward to verify that 
	\begin{align*}
		(\sf{K}^{\sf{F}}_{\Sigma})^{\dagger}=\sf{G}_{\Sigma,0}^{-1}\circ(\sf{K}_{\Sigma}^{\sf{F}})^{\ast}\circ\sf{G}_{\Sigma,1}^{\sf{F}}\qquad\text{and}\qquad \sf{K}_{\Sigma}^{\dagger}=\sf{G}_{\Sigma,0}^{-1}\circ\sf{K}_{\Sigma}^{\ast}\circ\sf{G}_{\Sigma,1}\,,
	\end{align*}
	where $(\sf{K}^{\sf{F}})_{\Sigma}^{\ast}$ and $\sf{K}_{\Sigma}^{\ast}$ are the formal adjoints of $\sf{K}_{\Sigma}^{\sf{F}}$ and $\sf{K}_{\Sigma}$ with respect to $(\cdot,\cdot)_{\sf{A}_{i}}$ and $(\cdot,\cdot)_{\sf{A}_{\rho_{i}}}$.
\end{remark}

\subsection{Gauge Conditions for Maxwell's Theory and Complete Gauge Fixing}\label{Subsec:GCMax}
In the previous section, we provided a detailed description of the \emph{Cauchy problem} of Maxwell's theory and introduced all the relevant operators. As the next step in our analysis of Maxwell's theory on globally hyperbolic spacetimes, we now turn to a more detailed discussion of the \emph{gauge problem}. This section is based on \cite[Sec.~2]{MurroSchmid}.

To start with, for a given Maxwell field $\sf{A}=\sf{A}_{\mathrm{T}}\eta+\sf{A}_{\Sigma}\in\Omega^{1}(\sf{M},\bb{C})$ with $\eta=\beta\d t$ and $\sf{A}_{\mathrm{T}}:=\beta^{-1}\sf{A}(\partial_{t})$, as usual, we consider the following gauge conditions.
\begin{align*}
 &(\sf{T}) \qquad && 0 = \sf{A}_t \quad &&& \textit{temporal (or Weyl) gauge} \\
& (\sf{L}) \qquad && 0 = \delta \sf{A} \quad &&& \textit{Lorenz gauge} \\
& (\sf{C}) \qquad && 0 = \delta_\Sigma \sf{A}_\Sigma \quad &&& \textit{Coulomb (or transverse) gauge}\\
& (\sf{R})=(\sf{L})+(\sf{T})\qquad && 0=\delta\sf{A}=\sf{A}_{\mathrm{T}} \quad &&& \textit{radiation gauge}
\end{align*}
We stress that different authors might use different names or different definitions for those kind of conditions. Furthermore, we stress that the gauge conditions $(\sf{T})$, $(\sf{C})$ and $(\sf{R})$ depend on the choice of Cauchy temporal function $t\in C^{\infty}(\sf{M})$, which here is fixed from the start. Only the Lorenz gauge condition $(\sf{L})$ is fully covariant.

A comment is needed regarding the terminology. Consider $(1+n)$-dimensional Minkowski spacetime $\bb{M}^{1,n}:=(\bb{R}^{1+n},\eta)$. Then, for a given $4$-potential $\sf{A}=\sf{A}_{\mu}\d x^{\mu}\in\Omega^{1}(\mathbb{R}^{1+n})$, we note that the Lorenz gauge condition decomposes as $0=\partial^{\mu}\sf{A}_{\mu}=-\partial_{t}\sf{A}_{0}+\partial^{i}\sf{A}_{i}$. Hence, we conclude that $(\sf{T})+(\sf{L})=(\sf{T})+(\sf{C})$. This condition is usually called the \emph{radiation gauge}, owing to the fact that it isolates the \textit{propagating, transverse} degrees of freedom. Indeed, the Maxwell equations in this case reduce to the wave equation $\partial_{t}^{2}\sf{A}_{i}-\partial^{k}\partial_{k}\sf{A}_{i}=0$ for the transverse part $\sf{A}_{\Sigma}=\sf{A}_{i}\d x^{i}$. In fact, at least on-shell, it turns out that the radiation gauge in Minkowski space is also equivalent to the Coulomb gauge when choosing the function spaces properly: if $\sf{A}$ satisfies the Coulomb gauge condition, i.e.~$\partial^{i}\sf{A}_{i}=0$, then the Maxwell equations imply $\Delta\sf{A}_{\mathrm{T}}=0$ and hence, when considering fields that are compactly supported in space, or at least vanishing at spatial infinity, the only solution to this elliptic equation is $\sf{A}_{\mathrm{T}}=0$. For these reasons, the terms \emph{Coulomb} and \emph{radiation gauge} are often used interchangeably in the physics literature, as are the conditions $(\sf{T})+(\sf{L})$ and $(\sf{T})+(\sf{C})$ (see, for instance, \cite[Chap.~6]{Jackson}). In fact, a similar statement is true, more generally, for \textit{ultrastatic} globally hyperbolic manifolds:

\begin{proposition}\label{Prop:EqiGauges} Let $(\sf{M}=\bb{R}\times\Sigma,\sf{g}=-\d t\otimes\d t+\sf{h})$ be an ultrastatic globally hyperbolic spacetime and let $\sf{A}\in\Omega^{1}(\sf{M},\bb{C})$. Then:
\begin{itemize}
\item[\emph{(i)}]$\sf{A}$ satisfies $(\sf{R})=(\sf{L})+(\sf{T})$ if and only if it satisfies $(\sf{C})+(\sf{T})$.
\item[\emph{(ii)}]If $\Sigma$ is non-compact, then, for $\sf{A}\in\mathrm{ker}(\sf{P}\vert_{\Omega^{1}_{\mathrm{sc}}})$, conditions $(\sf{R})$ and $(\sf{C})$ are equivalent. In this case, the Maxwell equations reduce to the hyperbolic equation $(\partial_{t}^{2}+\Delta)\sf{A}_{\Sigma}=0$.
\end{itemize}
\end{proposition}

\begin{proof}
	Claim (i) follows directly from Lemma~\ref{Lemma:Dec}(ii), which in the case $\beta=1$ and $\sf{k}=-\frac{1}{2\beta}\partial_{t}\sf{h}=0$ reduces to $\delta\sf{A}=\partial_{t}\sf{A}_{\mathrm{T}}+\delta_{\Sigma}\sf{A}_{\Sigma}$ and hence implies $\delta\sf{A}=\delta_{\Sigma}\sf{A}_{\Sigma}$ provided $\sf{A}_{\mathrm{T}}=0$. For (ii), we note that the Maxwell equations in the case $\beta=1$ and $\sf{k}=0$ are equivalent to the system
\begin{align*}
0=\begin{cases}
\Delta\sf{A}_{\mathrm{T}}-\partial_{t}\delta_{\Sigma}\sf{A}_{\Sigma}\\\partial^{2}_{t}\sf{A}_{\Sigma}+\delta_{\Sigma}\d_{\Sigma}\sf{A}_{\Sigma}-\d_{\Sigma}\partial_{t}\sf{A}_{\mathrm{T}}=(\partial_{t}^{2}+\Delta)\sf{A}_{\Sigma}-\d_{\Sigma}(\partial_{t}\sf{A}_{\mathrm{T}}+\delta_{\Sigma}\sf{A}_{\Sigma})
\end{cases}
\end{align*}	
by Proposition~\ref{Prop:DecoMax}. In particular, the Coulomb gauge $\delta_{\Sigma}\sf{A}_{\Sigma}$ implies $\Delta\sf{A}_{\mathrm{T}}=0$. If $\Sigma$ is non-compact, there are no non-trivial compactly-supported harmonic $1$-forms, which implies $\sf{A}_{\mathrm{T}}\vert_{t=s}=0$ for all $s\in\bb{R}$ and hence in particular $\sf{A}_{\mathrm{T}}=0$.
\end{proof}

\begin{remark}
	Proposition~\ref{Prop:EqiGauges}(ii) can also be generalised: consider the case in which the volume of $(\Sigma,\sf{h})$ is infinite and $\sf{A}\in\mathrm{ker}(\sf{P})$ is such that $\sf{A}_{\Sigma}\vert_{t=s}\in\sf{L}_{1}^{2}(\Sigma)$ for all $s\in\bb{R}$, where $\sf{L}_{1}^{2}(\Sigma)$ is the natural $\sf{L}^{2}$-space of $1$-forms on $(\Sigma,\sf{h})$, i.e.~the completion of $\Omega^{1}_{\mathrm{c}}(\Sigma,\bb{C})$ with respect to the de Rham-Hodge inner product on $(\Sigma,\sf{h})$. Then, since $(\sf{M},\sf{g})$ is ultrastatic, $(\Sigma,\sf{h})$ is complete (cf.~Example~\ref{Examples:GlobHyp}(i)), which implies that any harmonic $k$-form is closed and coclosed. In particular, the Maxwell equation $\Delta\sf{A}_{\mathrm{T}}=0$ implies that $\sf{A}_{\mathrm{T}}\vert_{t=s}=C_{s}$ for some constant $C_{s}\in\bb{R}$ depending on time $s\in\bb{R}$. But if $\mathrm{vol}_{\sf{h}}(\Sigma)=\infty$, this constant must necessarily be zero, which shows again that $\sf{A}_{\mathrm{T}}=0$. 
\end{remark}

Now, the situation is different when considering more general, non-ultrastatic, globally hyperbolic spacetimes: a natural generalisation of the Coulomb gauge condition in Minkowski spacetime to arbitrary globally hyperbolic spacetimes is the transversality condition $\delta_{\Sigma}\sf{A}_{\Sigma}=0$, as defined above. However, in this case, it is no longer true that $(\sf{T})+(\sf{L})=(\sf{T})+(\sf{C})$, at least in general. Indeed, assuming $\sf{A}_{\mathrm{T}}=0$, i.e.~condition $(\sf{T})$, Lemma~\ref{Lemma:Dec}(ii) implies
\begin{align*}
	\delta\sf{A}=\delta_{\Sigma}\sf{A}_{\Sigma}-\beta^{-1}\sf{h}^{\sharp}(\d_{\Sigma}\beta,\sf{A}_{\Sigma})\,.
\end{align*}
Hence, $(\sf{T})+(\sf{L})=(\sf{T})+(\sf{C})$ only holds true if the Cauchy temporal function is chosen such that $\beta=\mathrm{const}$. At least in $(1+3)$-dimensional spacetimes we can always reduce ourselves to this situation, since we can always perform a conformal transformation $\sf{g}\mapsto\beta^{-2}\sf{g}$ to normalise $\sf{g}$ in time to $\sf{g}^{\prime}=-\d t\otimes\d t+\sf{h}_{t}^{\prime}$, by conformal invariance of Maxwell's theory, cf.~Proposition~\ref{Prop:MaxConf}. For more general spacetimes, however, this is not the case.

Furthermore, since the Maxwell's equations in general curved spacetimes are the strongly coupled system in Proposition~\ref{Prop:DecoMax}, the two conditions $(\sf{T})+(\sf{L})$ and $(\sf{T})+(\sf{C})$ both lead to different equations on $\sf{A}_{\Sigma}$, both of which would deserve to be called \textit{radiation gauge}. Here we decide to use the former condition, as it contains the coordinate-independent and hence geometrically more natural Lorenz gauge condition. In general, not much has been written about different gauge conditions for Maxwell's theory on general globally hyperbolic spacetimes, but we mention that our concept of radiation gauge is somewhat close in spirit to the one considered by Tolksdorf in \cite{Tolksdorf}, in which he calls \textit{radiation gauge} the combination of the Lorenz gauge and the condition $\xi\lrcorner\sf{A}=0$ for some timelike Killing vector field $\xi$. Last but not least, we remind the reader that both our temporal and Coulomb gauge depend on the choice of Cauchy temporal function $t\:\sf{M}\to\bb{R}$. In principle, also in Minkowski spacetime one could choose a different, non-trivial, foliation in which the same subtleties would appear.

\begin{remark}\label{Rem:WeightedCoulomb}
	There is a way of defining a modified Coulomb gauge condition such that the equivalence $(\sf{L})+(\sf{T})=(\sf{L})+(\sf{C})$ in Proposition~\ref{Prop:EqiGauges}(i) holds true on general globally hyperbolic spacetimes and with arbitrary Cauchy temporal functions. Let $p\in\bb{R}$. Then, under the conformal transformation $\sf{h}\mapsto \sf{h}^{\prime}:=\beta^{p}\sf{h}$, denoting the codifferential on $(\Sigma,\sf{h}^{\prime})$ by $\delta_{\Sigma}^{\prime}$, a computation similar to the one in Proposition~\ref{Prop:MaxConf} yields
	\begin{align*}
		\delta^{\prime}_{\Sigma}\sf{A}_{\Sigma}=\beta^{-p}\bigg(\delta_{\Sigma}\sf{A}_{\Sigma}-\frac{(2-n)p}{2}\beta^{-1}\sf{h}^{\sharp}(\d_{\Sigma}\beta,\sf{A}_{\Sigma})\bigg)
	\end{align*}
	for all $\sf{A}_{\Sigma}\in C^{\infty}(\bb{R},\Omega^{1}(\Sigma,\bb{C}))$, where we recall that $n=\mathrm{dim}(\Sigma)$. Now, set $p_{0}:=2/(2-n)$ and $\sf{h}_{\beta}:=\beta^{p_{0}}\sf{h}$ and denote the corresponding codifferential by $\delta_{\Sigma}^{\beta}$. Then, the above formula implies
	\begin{align*}
		\delta_{\Sigma}^{\beta}\sf{A}_{\Sigma}=\beta^{-p_{0}}\big(\delta_{\Sigma}\sf{A}_{\Sigma}-\beta^{-1}\sf{h}^{\sharp}(\d_{\Sigma}\beta,\sf{A}_{\Sigma})\big)\, .
	\end{align*}
	Therefore, one could define a \emph{$\beta$-weighted Coulomb gauge condition} by the requirement
	\begin{align*}
& (\sf{C}_{\beta}) \qquad && 0 = \delta^{\beta}_\Sigma \sf{A}_\Sigma \quad &&& \textit{$\beta$-weighted Coulomb gauge}
\end{align*} 
	With this definition, Lemma~\ref{Lemma:Dec}(ii) implies $(\sf{L})+(\sf{T})=(\sf{L})+(\sf{C}_{\beta})$ on general globally hyperbolic spacetimes. In the ultrastatic case, one clearly recovers $(\sf{C}_{\beta}) = (\sf{C})$.
	
	Now, in the case of $1+3$ spacetime dimensions, we find $p_{0}=-2$. In this case, $\beta^{-2}$ corresponds precisely to the conformal factor normalising the globally hyperbolic metric $\sf{g} = -\beta^{2}\mathrm{d}t \otimes \mathrm{d}t + \sf{h}_{t}$ in time to $\sf{g}^{\prime}=-\d t\otimes\d t+(\sf{h}_{\beta})_{t}$, matching the discussion from above. On more general spacetimes, however, there seems to be no clear physical motivation for considering this $\beta$-weighted condition other than that it generalises the Coulomb gauge condition from Minkowski spacetime.
\end{remark}

Choosing the radiation gauge $(\sf{R})$ has several noteworthy implications, which, particularly in the context of constructing Hadamard states, make it an especially attractive choice: besides the Lorenz gauge condition $\delta\sf{A}=0$, which allows to turn Maxwell's equations $\sf{P}\sf{A}=\delta\d\sf{A}=0$ into the hyperbolic problem $\sf{D}_{2}\sf{A}=\square\sf{A}=0$ and is central for all the analysis of Maxwell's theory considered so far, it includes in addition the condition $\sf{A}_{\mathrm{T}}=0$, which exactly sets the \emph{negative contribution} in the bundle metric $\langle\sf{A},\sf{B}\rangle_{\sf{A}_{1}}=\sf{g}^{\sharp}(\overline{\sf{A}},\sf{B})=-\overline{\sf{A}}_{\mathrm{T}}\sf{B}_{\mathrm{T}}+\sf{h}^{\sharp}(\overline{\sf{A}}_{\Sigma},\sf{B}_{\Sigma})$ to zero. As mentioned in the introduction of this chapter, the non positivity of the bundle metric is one of the main complications for constructing Hadamard states for linear gauge theories and hence, the radiation gauge provides a promising gauge fixing condition in this context. Unfortunately, on general globally hyperbolic spacetimes, it is not clear whether the radiation gauge condition can be \emph{achieved}, i.e.~whether it can consistently be applied.

At this stage, let us clarify the terminology: when looking at the solution space of Maxwell's theory, or more generally, of an arbitrary linear gauge theory, i.e.~the space
\begin{align*}
	\mathrm{Sol}_{\mathrm{sc}}=\cfrac{\mathrm{ker}(\sf{P}\vert_{\Gamma^{\infty}_{\mathrm{sc}}})}{\mathrm{ran}(\sf{K}\vert_{\Gamma^{\infty}_{\mathrm{sc}}})}\,,
\end{align*}
a gauge condition $(\sf{G})$ is \emph{achievable}, if in any equivalence class $[\sf{A}]\in \mathrm{Sol}_{\mathrm{sc}}$, there is \emph{at least one} representative satisfying condition $(\sf{G})$. Equivalently, for every Maxwell field $\sf{A}\in\Omega^{1}_{\mathrm{sc}}(\sf{M},\bb{C})$, we can find a $f\in C^{\infty}_{\mathrm{sc}}(\sf{M},\bb{C})$ such that $\sf{A}^{\prime}:=\sf{A}-\d f$ satisfies condition $(\sf{G})$, which in turn typically amounts to solving a PDE. In other words, the achievability of a certain gauge condition is usually equivalent to the existence of solutions to some differential equations. In particular, we stress that this concept crucially depends on the function spaces we are considering. For instance, instead of working with forms having spatially compact support, one could work with a bigger class, such as forms that are $\sf{L}^{2}$ or $\sf{H}^{s}$ for some $s\in\bb{R}$ in space, in which a gauge condition that is not achievable for spatially compact fields, becomes achievable.

In the case of the radiation gauge condition, achievability is equivalent to the question whether the system
\begin{align*}
\begin{cases}
	\square f&=\delta\sf{A}\\
	\partial_{t}f &=\sf{A}_{\mathrm{T}}
\end{cases}
\end{align*}
admits a solution $f\in C^{\infty}_{\mathrm{sc}}(\sf{M},\bb{C})$ for every source $\sf{A}\in\Omega_{\mathrm{sc}}^{1}(\sf{M},\bb{C})$. Besides some specific and highly symmetric spacetimes (see, for instance, the discussion in \cite{Tolksdorf}), this system is highly non-trivial to solve and it is a priori not clear if solutions can be found.

Motivated from this problem, we hence look for a gauge condition that serves similar purposes but turns out to be easier to achieve. More precisely, since we may equivalently work on the level of initial data using the equivalent characterisation of the phase space in~\eqref{eq:PSMaxwell}, the idea is to define an analogues gauge condition on the level of initial data by combining the Lorenz gauge with an additional condition that sets the negative contributions to the fibre metric on initial data to zero. To this end, we make the following definition.

\begin{definition} (Cauchy Radiation Gauge)\newline
Let $(\sf{M},\sf{g})$ be a globally hyperbolic spacetime and fix a Cauchy temporal function $t\in C^{\infty}(\sf{M})$. Then, for $\sf{A}\in\Omega^{1}(\sf{M},\bb{C})$ decomposed as $\sf{A}=\sf{A}_{\mathrm{T}}\eta+\sf{A}_{\Sigma}$ with $\eta=\beta\d t$ and $\sf{A}_{\mathrm{T}}:=\beta^{-1}\sf{A}(\partial_{t})$, as usual, we define the \emph{Cauchy radiation gauge} by
\begin{align*}
	(\sf{R}_{\mathrm{Cauchy}})\qquad\qquad \delta\sf{A}=0\,\qquad\text{and}\qquad \sf{A}_{\mathrm{T}}\vert_{t=0}=\partial_{t}\sf{A}_{\mathrm{T}}\vert_{t=0}=0\, .
\end{align*}
\end{definition}

We stress that the Cauchy radiation gauge $(\sf{R}_{\mathrm{Cauchy}})$ depends both on the choice of Cauchy temporal gauge, as the temporal gauge $(\sf{T})$, Coulomb gauge $(\sf{C})$ and radiation gauge $(\sf{R})$, but, at least in general, also on the choice of \emph{initial slice}, which here is simply chosen to be $\Sigma_{t=0}$.

\begin{remark}
	The definition of the Cauchy radiation gauge is slightly different from the one considered in the original article \cite[Def.~2.1]{MurroSchmid}, in which the condition $\partial_{t}\sf{A}_{\mathrm{T}}\vert_{t=0}=0$ is replaced by $(\nabla_{\nu}\sf{A})_{\mathrm{T}}\vert_{t=0}=0$. In fact, by Eq.~\eqref{eq:NormalDer}, it holds that
	\begin{align*}
		(\nabla_{\nu}\sf{A})_{\mathrm{T}}\vert_{t=0}=\frac{1}{\beta_{0}}\partial_{t}\sf{A}_{\mathrm{T}}\vert_{t=0}-\frac{1}{\beta_{0}}\sf{h}^{\sharp}_{0}(\d_{\Sigma}\beta_{0},\sf{A}_{\Sigma}\vert_{t=0})
	\end{align*}
	with $\beta_{0}:=\beta\vert_{t=0}\in C^{\infty}(\Sigma,(0,\infty))$. Hence, whenever $\beta$ is not constant in space, these two conditions are not equivalent. However, for all practical purposes, this will not make any difference, besides some minor modification in the proof of Proposition~\ref{Prop:AchievCRG} below.
\end{remark}

To set this new gauge condition $(\sf{R}_{\mathrm{Cauchy}})$ into context, we first note that, at least on-shell, it is equivalent to the spacetime radiation gauge condition $(\sf{R})$ on ultrastatic spacetimes.

\begin{proposition}\label{Prop:EquivRad}
	Let $(\sf{M}=\bb{R}\times\Sigma,\,\sf{g}=-\d t\otimes\d t+\sf{h})$ be an ultrastatic globally hyperbolic spacetime and let $\sf{A}\in\Omega^{1}(\sf{M},\bb{C})$. If $\sf{A}\in\mathrm{ker}(\sf{P})$, then $\sf{A}$ satisfies the radiation gauge $(\sf{R})$ if and only if it satisfies $(\sf{R}_{\mathrm{Cauchy}})$.
\end{proposition}

\begin{proof}
	The direction ``$(\sf{R})\Rightarrow(\sf{R}_{\mathrm{Cauchy}})$'' is obvious. For the other direction, we first note that, in the ultrastatic case, the hyperbolic equation $\square\sf{A}$ decouples into two separate wave equations for the temporal and spatial component, i.e.~$(\square\sf{A})_{\mathrm{T}}=(\partial_{t}+\Delta)\sf{A}_{\mathrm{T}}=\square(\sf{A}_{\mathrm{T}})$ and $(\square\sf{A})_{\Sigma}=(\partial_{t}+\Delta)\sf{A}_{\Sigma}$, where $\Delta$ denotes the de Rham-Hodge Laplacian acting on $0$- and $1$-forms, respectively. In other words, the following systems of equations are equivalent:
	\begin{align*}
		\begin{cases}
			\sf{P}\sf{A} &=0\\
			\delta\sf{A}	&=0
		\end{cases} \qquad\Longleftrightarrow\qquad 			\begin{cases}
			\square\sf{A}&=0\\
			\delta \sf{A}&=0
		\end{cases} \qquad\Longleftrightarrow\qquad 			\begin{cases}
			(\partial_{t}^{2}+\Delta)\sf{A}_{\mathrm{T}}=0\\
			(\partial_{t}^{2}+\Delta)\sf{A}_{\Sigma} =0\\
			\delta\sf{A}=0
		\end{cases} 
	\end{align*}
	Hence, the equations for $\sf{A}_{\mathrm{T}}$ and $\sf{A}_{\Sigma}$ are decoupled in this case and if the initial data $\sf{A}_{\mathrm{T}}\vert_{t=0}$ and $\partial_{t}\sf{A}_{\mathrm{T}}\vert_{t=0}$ are zero, we conclude that also $\sf{A}_{\mathrm{T}}=0$ (cf.~Theorem~\ref{Thm:NHCauchy}).
\end{proof}

To sum up, combining Proposition~\ref{Prop:EqiGauges} and Proposition~\ref{Prop:EquivRad}, we conclude that on \emph{ultrastatic} globally hyperbolic spacetimes, there is the following equivalence of gauge conditions:
\begin{align*}
	(\sf{R})=(\sf{L})+(\sf{T})\xLeftrightarrow{} (\sf{C})+(\sf{T})\xLeftrightarrow[\text{on}\,\mathrm{ker}(\sf{P})]{} (\sf{R}_{\mathrm{Cauchy}})\xLeftrightarrow[\text{if }\Sigma\text{ is non-compact; on}\,\mathrm{ker}(\sf{P}\vert_{\Omega_{\mathrm{sc}}^{1}})]{}(\sf{C})
\end{align*}

Now, before addressing the achievability of the Cauchy radiation gauge $(\sf{R}_{\mathrm{Cauchy}})$, let us introduce another important concept, one that is closely related to \emph{achievability} of a gauge condition: we call an (achievable) gauge condition $(\sf{G})$ \emph{complete}, if any equivalence class $[\sf{A}]\in \mathrm{Sol}_{\mathrm{sc}}$ has \emph{exactly one} representative satisfying $(\sf{G})$. This is usually equivalent to the question whether the solution of a given differential equations is unique and again depends crucially one the space of fields one considers. In the context of Maxwell's theory, we consider the following illustrating example:

\begin{example} (The Coulomb Gauge on Spatially Compact Ultrastatic Spacetimes)\newline
Let $(\sf{M}=\bb{R}\times\Sigma,\sf{g}=-\d t\otimes\d t+\sf{h})$ be a globally hyperbolic ultrastatic spacetime such that $\Sigma$ is \emph{compact}, i.e.~$\Gamma^{\infty}_{\mathrm{sc}}(\sf{A}_{k})=\Gamma^{\infty}(\sf{A}_{k})$. Now, let $\sf{A}\in\mathrm{ker}(\sf{P}\vert_{\Gamma^{\infty}})$ and consider the \emph{Coulomb gauge condition} $\delta_{\Sigma}\sf{A}_{\Sigma}$. The goal is to find a gauge transformation $f\in C^{\infty}(\sf{M},\bb{C})$ such that $\sf{A}^{\prime}:=\sf{A}-\d f$ satisfies the Coulomb gauge, i.e.~$\delta_{\Sigma}\sf{A}_{\Sigma}^{\prime}=0$. A short computation shows that this is equivalent to solve the \emph{Poisson equation}
\begin{align*}
	\Delta f=\delta_{\Sigma}\sf{A}_{\Sigma}\,,
\end{align*}
where $\Delta=\delta_{\Sigma}\d_{\Sigma}$ denotes the de Rham-Hodge Laplacian on $(\Sigma,\sf{h})$, as usual. Now, for compact manifolds, this equation can always be solved and the solution is unique up to a constant. Hence, we conclude that the Coulomb gauge is achievable. Furthermore, it is in fact a \emph{complete} gauge fixing: if $\sf{A}\in\mathrm{ker}(\sf{P}\vert_{\Gamma^{\infty}})$ satisfies $\delta_{\Sigma}\sf{A}_{\Sigma}=0$, then any other gauge transformation $\sf{A}\mapsto \sf{A}+\d f$ leaving this condition untouched results into the Laplace equation $\Delta f=0$. The only solutions to this equation are the constant functions and hence $\d f=0$. In other words, every equivalent class of the solution space $\mathrm{Sol}_{\mathrm{sc}}$ has a \emph{unique} representative satisfying the Coulomb gauge condition.

Now, this discussion is specific to the case of \emph{spatially compact} spacetimes. If one considers the more general case in which $\Sigma$ is non-compact, the previous statements are no longer true: it is not clear in general that one can solve the Poisson equation $\Delta f=\delta_{\Sigma}\sf{A}_{\Sigma}$ within the space of \emph{compactly supported functions} for a \emph{compactly supported source}. Instead, it becomes more natural to consider a different space of fields. For instance, in Minkowski spacetime, one usually works with fields $\sf{A}$ and $f$ that vanish at (spatial) infinity (see e.g.~\cite{Jackson}).
\end{example}

As it turns out, the Cauchy radiation gauge can always be achieved on \emph{spatially compact} spacetimes, since its achievability reduces to a Poisson-type equation for the initial data, analogous to the one discussed in the example above.

\begin{proposition}\label{Prop:AchievCRG} \emph{(Cauchy Radiation Gauge on Spatially Compact Spacetimes)}\newline
Let $(\sf{M}=\bb{R}\times\Sigma,\sf{g}=-\beta^{2}\d t\otimes\d t+\sf{h}_{t})$ be a globally hyperbolic spacetime with $\Sigma$ compact. Then, the Cauchy radiation gauge $(\sf{R}_{\mathrm{Cauchy}})$ is an achievable and complete gauge-fixing in $\mathrm{Sol}_{\mathrm{sc}}$, i.e.~for any $\sf{A}\in\mathrm{ker}(\sf{P}\vert_{\Gamma^{\infty}})$ there exists a unique $f\in C^{\infty}(\sf{M},\bb{C})$ (up to constant) such that $\sf{A}^{\prime}:=\sf{A}-\sf{K}f$ satisfies the gauge condition $(\sf{R}_{\mathrm{Cauchy}})$.
\end{proposition}

\begin{proof}
	Let $\sf{A}\in\mathrm{ker}(\sf{P}\vert_{\Gamma^{\infty}})$ be arbitrary. First of all, we observe that the gauge condition $(\sf{R}_{\mathrm{Cauchy}})$ is equivalent to the conditions
	\begin{align*}
		\delta\sf{A}=0\,,\qquad \sf{A}_{\mathrm{T}}\vert_{t=0}=\bigg(\delta_{\Sigma}\sf{A}_{\Sigma}-\frac{1}{\beta}\sf{h}^{\sharp}(\d_{\Sigma}\beta,\sf{A}_{\Sigma})\bigg)\bigg\vert_{t=0}=0
	\end{align*}
	on account of $\delta\sf{A}=\frac{1}{\beta}\partial_{t}\sf{A}_{\mathrm{T}}-\mathrm{tr}_{\sf{h}}(\sf{k})\sf{A}_{\mathrm{T}}+\delta_{\Sigma}\sf{A}_{\Sigma}-\frac{1}{\beta}\sf{h}^{\sharp}(\d_{\Sigma}\beta,\sf{A}_{\Sigma})$, see Lemma~\ref{Lemma:Dec}(ii), and the fact that $\beta>0$. In particular, if we use the notation from Remark~\ref{Rem:WeightedCoulomb} and write $\sf{h}_{\beta}:=\beta^{p}\sf{h}$ for $p=2/(2-\mathrm{dim}(\Sigma))$, then the latter condition can be written as $\sf{A}_{\mathrm{T}}\vert_{t=0}=\delta_{\Sigma}^{\beta}\sf{A}_{\Sigma}\vert_{t=0}=0$, where $\delta_{\Sigma}^{\beta}$ denotes the codifferential on the conformally equivalent (time-dependent) Riemannian manifold $(\Sigma,\sf{h}_{\beta})$. Now, we try to find a function $f\in C^{\infty}(\sf{M},\bb{C})$ such that $\sf{A}^{\prime}=\sf{A}-\sf{K}f$ satisfies the Cauchy radiation gauge, which hence is equivalent to finding a solution to the system
	\begin{align*}
		\begin{cases}
			\square f &=\delta\sf{A}\\
			\partial_{t}f\vert_{t=t_{0}}&=\beta\sf{A}_{\mathrm{T}}\vert_{t=t_{0}}\\
			\Delta^{\beta}f\vert_{t=0}&=\delta^{\beta}_{\Sigma}\sf{A}_{\Sigma}\vert_{t=0}
		\end{cases}\,,
	\end{align*}
	where $\Delta=\delta_{\Sigma}^{\beta}\d_{\Sigma}$ denotes the de Rham-Hodge Laplacian on $(\Sigma,\sf{h}_{\beta})$. Now, consider the Poisson equation $\Delta^{\beta}f\vert_{t=0}=\delta^{\beta}_{\Sigma}\sf{A}_{\Sigma}\vert_{t=0}$. Since $(\Sigma,\sf{h}_{\beta})$ is a compact Riemannian manifold, the \emph{Hodge decomposition} applied to $(\Sigma,(\sf{h}_{\beta})_{0})$, which states that there is a $\sf{L}^{2}$-orthogonal decomposition
\begin{align*}
	\Omega^{1}(\Sigma,\bb{C})\cong \mathrm{ran}(\d_{\Sigma}\vert_{C^{\infty}})\oplus\mathrm{ran}(\delta^{\beta,0}_{\Sigma}\vert_{\Omega^{2}})\oplus\mathrm{ker}(\Delta\vert_{\Omega^{1}})\,,
\end{align*}
	where $\delta^{\beta,0}_{\Sigma}$ denotes the codifferential of $(\Sigma,(\sf{h}_{\beta})_{0})$, allows us to decompose $\sf{A}_{\Sigma}\vert_{t=0}\in\Omega^{1}(\Sigma,\bb{C})$ uniquely as $\sf{A}_{\Sigma}\vert_{t=0}=\d_{\Sigma}\mathfrak{f}+\omega$ for some $\mathfrak{f}\in C^{\infty}(\Sigma,\bb{C})$ (unique up to constant) and $\omega\in\mathrm{ker}(\delta_{\Sigma}^{\beta,0})$. In particular, $\mathfrak{f}$ is the unique (up to constant, since $\mathrm{ker}(\d_{\Sigma}\vert_{C^{\infty}})\cong\mathbb{R}$ on connected manifolds) solution to $\Delta^{\beta}f\vert_{t=0}=\delta^{\beta}_{\Sigma}\sf{A}_{\Sigma}\vert_{t=0}$ and the above system is equivalent to the Cauchy problem
	\begin{align*}
		\begin{cases}
			\square f &=\delta\sf{A}\\
			\partial_{t}f\vert_{t=t_{0}}&=\beta\sf{A}_{\mathrm{T}}\vert_{t=t_{0}}\\
			f\vert_{t=0}&=\mathfrak{f}\quad\text{mod}\quad\text{constant}
		\end{cases}\, .
	\end{align*}
	We conclude that there is a unique $f\in C^{\infty}(\sf{M},\bb{C})$ (up to constant) such that $\sf{A}^{\prime}=\sf{A}-\sf{K}f$ satisfies the Cauchy radiation gauge, on account of Theorem~\ref{Thm:NHCauchy}.
\end{proof}

\begin{remark}
	Alternatively, we may deduce the existence and uniqueness of solutions of the Poisson equation $\Delta^{\beta}f\vert_{t=0}=\delta^{\beta}_{\Sigma}\sf{A}_{\Sigma}\vert_{t=0}$ on $(\Sigma,\sf{h}_{0})$ also from the \emph{Fredholm alternative}: let $\sf{D}\in\mathrm{DO}(\sf{E})$ be an elliptic differential operator on a compact Riemannian manifold $(\sf{M},\sf{g})$ with $\sf{E}$ denoting a vector bundle equipped with a positive-definite bundle metric $\langle\cdot,\cdot\rangle_{\sf{E}}$. Then, in the smooth setting, the equation $\sf{D}\psi=\varphi$ has a smooth solution $\psi\in\Gamma^{\infty}(\sf{E})$ if and only if the source $\varphi\in\Gamma^{\infty}(\sf{E})$ is $\sf{L}^{2}$-orthogonal to $\mathrm{ker}(\sf{D}^{\ast})$. Now, in the case $\sf{D}=\Delta^{\beta}$ for $(\sf{M}=\Sigma,\sf{g}=(\sf{h}_{\beta})_{0})$, this implies the claim, since clearly $\Delta^{\beta}=(\Delta^{\beta})^{\ast}$ and $\mathrm{ker}(\delta^{\beta}_{\Sigma})\perp\mathrm{ker}(\Delta^{\beta})$,
\end{remark}

Now, if $(\sf{M},\sf{g})$ is a globally hyperbolic Lorentzian manifold, consider the linear zeroth-order operator defined by
\begin{align*}
	\sf{R}_{\Sigma}\:\Gamma^{\infty}(\sf{A}_{\rho_{1}})\to\Gamma^{\infty}(\sf{A}_{\rho_{1}})\,,\qquad \sf{R}_{\Sigma}=
		\begin{pmatrix}
			\mathrm{id} & 0 & 0 & 0\\
			0 & \mathrm{id} & 0 & 0\\
			0 & 0 & 0 & 0\\
			0 & 0 & 0 & 0
		\end{pmatrix}\, .
\end{align*}
By definition, $\sf{A}\in\Omega^{1}(\sf{M},\bb{C})$ satisfies the Cauchy radiation gauge, if and only if $\delta\sf{A}=0$ and $\rho_{1}(\sf{A})\in\mathrm{ker}(\sf{R}_{\Sigma})$. With this notation, Proposition~\ref{Prop:AchievCRG} implies that, in the case in which $\Sigma$ is compact, there exists an operator $\sf{T}_{\Sigma}\:\Gamma^{\infty}(\sf{A}_{\rho_{1}})\to\Gamma^{\infty}(\sf{A}_{\rho_{1}})$ satisfying $\sf{T}_{\Sigma}\circ\sf{K}_{\Sigma}=0$, which induces a linear isomorphism of the form 
\begin{align}\label{eq:PhaseSpaceCRG}
	\mathcal{V}_{\Sigma}:=\cfrac{\mathrm{ker}(\sf{K}_{\Sigma}^{\dagger}\vert_{\Gamma^{\infty}})}{\mathrm{ran}(\sf{K}_{\Sigma}\vert_{\Gamma^{\infty}})}\qquad \xrightarrow[\cong]{\quad [\sf{T}_{\Sigma}]\quad}\qquad \mathcal{V}_{\sf{R}}:=\mathrm{ker}(\sf{K}_{\Sigma}^{\dagger}\vert_{\Gamma^{\infty}})\cap\mathrm{ker}(\sf{R}_{\Sigma}\vert_{\Gamma^{\infty}})\, .
\end{align}
More precisely, $\sf{T}_{\Sigma}$ is the operator that maps the initial data of a given Maxwell field $\sf{A}\in\Omega^{1}(\sf{M},\bb{C})$ to the initial data of the unique representative of $[\sf{A}]\in\mathcal{V}_{\mathrm{c}}$ satisfying the Cauchy radiation gauge. An explicit expression of this operator in the ultrastatic case can be found in Remark~\ref{Rem:GaugeInvData} below. A more detailed discussion of the projector $\sf{T}_{\Sigma}$ will follow in Section~\ref{Sec:ConsHadMax}.

Using the explicit formula of $\sf{K}_{\Sigma}^{\dagger}$ derived in Proposition~\ref{Prop:KSigma}, it is easy to see that the space $\mathcal{V}_{\sf{R}}$ is explicitly given by
\begin{align*}
	\mathcal{V}_{\sf{R}}=\bigg\{(\mathfrak{a}_{\mathrm{T}},\pi_{\mathrm{T}},\mathfrak{a}_{\Sigma},\pi_{\Sigma})\in\Gamma^{\infty}(\sf{A}_{\rho_{1}})\mid \mathfrak{a}_{\mathrm{T}}=\pi_{\mathrm{T}}=\delta_{\Sigma}^{\beta,0}\mathfrak{a}_{\Sigma}=\delta_{\Sigma}^{0}\pi_{\Sigma}=0\}\,,
\end{align*}
where we recall the notation $\delta_{\Sigma}^{\beta,0}=\beta^{-p_{0}}(\delta_{\Sigma}-\beta_{0}^{-1}\sf{h}_{0}^{\sharp}(\d_{\Sigma}\beta_{0},\cdot))$ with $p_{0}=2/(2-\mathrm{dim}(\Sigma))$ and where $\beta_{0}:=\beta\vert_{t=0}\in C^{\infty}(\Sigma,(0,\infty))$.

\begin{remark}\label{Rem:GaugeInvData} (Cauchy Radiation Gauge and Gauge-Invariant Initial Data)\newline
	At this stage, it is quite interesting to note that the Cauchy radiation gauge has a natural interpretation from the physical point of view, namely in terms of \emph{gauge-invariant initial data}. For simplicity, we assume that $(\sf{M},\sf{g})$ is ultrastatic and spatially compact, i.e. $\sf{M}=\bb{R}\times\Sigma$ with $\Sigma$ compact and $\sf{g}=-\d t\otimes\d t+\sf{h}$. The following arguments can easily be generalised to the non-ultrastatic case by inserting factors of $\sf{k}$ and replacing $\delta_{\Sigma}$ with $\delta_{\Sigma}^{\beta}$ in the relevant places.
	
	Now, the Hodge decomposition theorem for $(\Sigma,\sf{h})$ implies that any $1$-form $\omega\in\Omega^{1}(\Sigma,\bb{C})$ can be decomposed as $\omega=\d\varphi+\eta$ for $\varphi\in C^{\infty}(\Sigma,\bb{C})$, unique up to constant, and a unique $\eta\in\mathrm{ker}(\delta_{\Sigma})$. In order to avoid issues concerning the ``up to constant'' definition of $\varphi$, we consider the linear subspace 
	\begin{align*}
		C^{\infty}_{\mathrm{v}}(\Sigma,\bb{C}):=\bigg\{\varphi\in C^{\infty}(\Sigma,\bb{C})\mid\int_{\Sigma}\varphi\,\d\mu_{\sf{h}}=0\bigg\}\, .
	\end{align*}
	Using this notation, we note that every $1$-form can \emph{uniquely} be decomposed according to
	\begin{equation}
	\begin{aligned}\label{eq:GI1}
		\Omega^{1}(\Sigma,\bb{C})&\xrightarrow{\cong}C^{\infty}_{\mathrm{v}}(\Sigma,\bb{C})\oplus\mathrm{ker}(\delta_{\Sigma}\vert_{\Omega^{1}})\\
		\omega=\d_{\Sigma}\varphi+\eta&\mapsto  (\varphi,\eta)
	\end{aligned}
	\end{equation}
	Now, let us denote the natural projection of $C^{\infty}(\Sigma,\bb{C})$ onto its subspace $C^{\infty}_{\mathrm{v}}(\Sigma,\bb{C})$ by
\begin{align*}
		\mathcal{I}:=\mathrm{id}-\frac{1}{\mathrm{vol}_{\sf{h}}(\Sigma)}\int_{\Sigma}\cdot\,\d\mu_{\sf{h}}\:C^{\infty}(\Sigma,\bb{C})\to C^{\infty}_{\mathrm{v}}(\Sigma,\bb{C})\, .
	\end{align*}	
	The de Rham-Hodge Laplacian $\Delta$ viewed as an operator $\Delta\: C^{\infty}_{\mathrm{v}}(\Sigma,\bb{C})\to\mathrm{ran}(\delta_{\Sigma}\vert_{\Omega^{1}})$ is bijective. Indeed, injectivity is clear, since any harmonic function on a compact manifold is necessarily constant, and for surjectivity, we recall that the equation $\Delta\varphi=\delta_{\Sigma}\omega$ for given $\omega\in\Omega^{1}(\Sigma,\bb{C})$ has a unique solution up to constant, as consequence of the Hodge decomposition (cf.~the proof of Proposition~\ref{Prop:AchievCRG}). We note that $\Delta\Delta^{-1}=\mathrm{id}$ on $\mathrm{ran}(\delta_{\Sigma}\vert_{\Omega^{1}})$, while $\Delta^{-1}\Delta=\mathcal{I}$ on $C^{\infty}(\Sigma,\bb{C})$. With this notation, we can write the projectors of the decomposition~\eqref{eq:GI1} as
	\begin{align*}
		\Delta^{-1}\delta_{\Sigma}\:\Omega^{1}(\Sigma,\bb{C})\to C^{\infty}_{\mathrm{v}}(\Sigma,\bb{C})\,,\qquad \mathrm{id}-\d_{\Sigma}\Delta^{-1}\delta_{\Sigma}\:\Omega^{1}(\Sigma,\bb{C})\to\mathrm{ker}(\delta_{\Sigma}\vert_{\Omega^{1}})\,,
	\end{align*}
	as one can easily verify via an explicit computation.
	
	Having fixed the relevant notation, let $(\mathfrak{a}_{\mathrm{T}},\pi_{\mathrm{T}},\mathfrak{a}_{\Sigma},\pi_{\Sigma})\in\Gamma^{\infty}(\sf{A}_{\rho_{1}})$ be an arbitrary $1$-form initial datum. Using the Hodge decomposition in the more specific form of Eq.~\eqref{eq:GI1}, we decompose $\mathfrak{a}_{\Sigma},\pi_{\Sigma}\in\Omega^{1}(\Sigma,\bb{C})$ as 
	\begin{align*}
		\mathfrak{a}_{\Sigma}=\d_{\Sigma}\varphi_{\mathfrak{a}}+\eta_{\mathfrak{a}}\,,\qquad\pi_{\Sigma}=\d_{\Sigma}\varphi_{\pi}+\eta_{\pi}
	\end{align*}
	for uniquely determined $\varphi_{\mathfrak{a}},\varphi_{\pi}\in C^{\infty}_{\mathrm{v}}(\Sigma,\bb{C})$ and $\eta_{\mathfrak{a}},\eta_{\pi}\in\mathrm{ker}(\delta_{\Sigma})$. Now, instead of considering $\mathfrak{a}_{\Sigma}$ and $\pi_{\Sigma}$ as independent variables, let us consider $\varphi_{\mathfrak{a}},\varphi_{\pi}$ and $\eta_{\mathfrak{a}},\eta_{\pi}$. In other words, we consider the bijective map
	\begin{align*}
		\Gamma^{\infty}(\sf{A}_{\rho_{1}})&\xrightarrow{\cong} C^{\infty}(\Sigma,\bb{C})^{\oplus 2}\oplus C^{\infty}_{\mathrm{v}}(\Sigma,\bb{C})^{\oplus 2}\oplus\mathrm{ker}(\delta_{\Sigma}\vert_{\Omega^{1}})^{\oplus 2}\\
		(\mathfrak{a}_{\mathrm{T}},\pi_{\mathrm{T}},\mathfrak{a}_{\Sigma},\pi_{\Sigma}) &\mapsto (\mathfrak{a}_{\mathrm{T}},\pi_{\mathrm{T}},\varphi_{\mathfrak{a}},\varphi_{\pi},\eta_{\mathfrak{a}},\eta_{\pi})\,.
	\end{align*}
	Let us analyse how the gauge transformations act on those fields individually. To start with, we observe that on the level of initial data, gauge transformations act as
	\begin{align*}
		\begin{pmatrix}
			\mathfrak{a}_{\mathrm{T}}\\
			\pi_{\mathrm{T}}\\
			\mathfrak{a}_{\Sigma}\\
			\pi_{\Sigma}
		\end{pmatrix}+\sf{K}_{\Sigma}\begin{pmatrix}
			\mathfrak{f}\\\mathfrak{g}
		\end{pmatrix}=
		\begin{pmatrix}
			\mathfrak{a}_{\mathrm{T}}+i\mathfrak{g}\\
			\pi_{\mathrm{T}}+i\Delta\mathfrak{f}\\
			\mathfrak{a}_{\Sigma}+\d_{\Sigma}\mathfrak{f}\\
			\pi_{\Sigma}+\d_{\Sigma}\mathfrak{g}
		\end{pmatrix}
	\end{align*}
	for all $(\mathfrak{f},\mathfrak{g})\in\Gamma^{\infty}(\sf{A}_{\rho_{0}})$, where we used Proposition~\ref{Prop:KSigma} for the explicit expression of $\sf{K}_{\Sigma}$. On the individual variables introduced above, the gauge transformation hence act as
	\begin{align*}
		\begin{cases}
			\mathfrak{a}_{\mathrm{T}}\mapsto \mathfrak{a}_{\mathrm{T}}+i\mathfrak{g}\\
			\pi_{\mathrm{T}}\mapsto  \pi_{\mathrm{T}}+i\Delta\mathfrak{f}
		\end{cases}\qquad
		\begin{cases}
			\varphi_{\mathfrak{a}}\mapsto\varphi_{\mathfrak{a}}+\mathcal{I}\mathfrak{f}\\
			\varphi_{\pi}\mapsto\varphi_{\pi}+\mathcal{I}\mathfrak{g}
		\end{cases}\qquad
		\begin{cases}
			\eta_{\mathfrak{a}}\mapsto\eta_{\mathfrak{a}}\\
			 \eta_{\pi}\mapsto\eta_{\pi}
		\end{cases}\,,
	\end{align*}
	where we used that $\d_{\Sigma}\circ\mathcal{I}=\d_{\Sigma}$ and the fact that $\Omega^{1}(\Sigma)\cong\d C^{\infty}_{\mathrm{v}}(\Sigma)\oplus \mathrm{ker}(\delta_{\Sigma}^{0})$ is a direct sum decomposition. Now, at this point, we see that the divergence-free parts $\eta_{\mathfrak{a}}$ and $\eta_{\pi}$ are already \emph{gauge-invariant}. Furthermore, we can find linear combinations of the four scalar initial data to obtain two \textit{gauge-invariant} scalar initial data, namely
	\begin{align*}
		\Phi_{\mathfrak{a}}:=\mathcal{I}\mathrm{a}_{\mathrm{T}}-i\varphi_{\pi}\in C^{\infty}_{\mathrm{v}}(\Sigma,\bb{C})\,,\qquad \Phi_{\pi}:=\pi_{\mathrm{T}}-i\Delta\varphi_{\mathfrak{a}}\in C^{\infty}(\Sigma,\bb{C})\, .
	\end{align*}
	To sum up, instead of considering the initial datum $(\mathfrak{a}_{\mathrm{T}},\pi_{\mathrm{T}},\mathfrak{a}_{\Sigma},\pi_{\Sigma})$ we could equivalently consider $(\Phi_{\mathfrak{a}},\Phi_{\pi},\eta_{\mathfrak{a}},\eta_{\pi})$, which contains the gauge-invariant degrees of freedom on the level of initial data. The projector that maps a given initial datum $(\mathfrak{a}_{\mathrm{T}},\pi_{\mathrm{T}},\mathfrak{a}_{\Sigma},\pi_{\Sigma})$ into its gauge-invariant components $(\Phi_{\mathfrak{a}},\Phi_{\pi},\eta_{\mathfrak{a}},\eta_{\pi})$ can be written as
	\begin{align*}
		\Pi:=
		\begin{pmatrix}
			\mathcal{I} & 0 & 0 & -i\Delta^{-1}\delta_{\Sigma} \\
			0 & \mathrm{id} & -i\delta_{\Sigma} & 0\\
			0 & 0 & \mathrm{id}-\d_{\Sigma}\Delta^{-1}\delta_{\Sigma} & 0\\
			0 & 0& 0 &\mathrm{id}-\d_{\Sigma}\Delta^{-1}\delta_{\Sigma}
		\end{pmatrix}\:\Gamma^{\infty}(\sf{A}_{\rho_{1}})\to\Gamma^{\infty}(\sf{A}_{\rho_{1}})\,,\quad
		\begin{pmatrix}
			\mathfrak{a}_{\mathrm{T}}\\
			\pi_{\mathrm{T}}\\
			\mathfrak{a}_{\Sigma}\\
			\pi_{\Sigma}
		\end{pmatrix}\mapsto 
		\begin{pmatrix}
			\Phi_{\mathfrak{a}}\\
			\Phi_{\pi}\\
			\eta_{\mathfrak{a}}\\
			\eta_{\pi}
		\end{pmatrix}\, .
	\end{align*}
	By construction, the operator $\Pi$ is a projector, i.e.~$\Pi^{2}=\Pi$ that has the following additional properties, which one can easily verify:
	\begin{itemize}
		\item[(i)]$\Pi\sf{K}_{\Sigma}=0$ (gauge invariance), $\mathrm{ker}(\Pi)=\mathrm{ran}(\sf{K}_{\Sigma})$ and $\sf{K}_{\Sigma}^{\dagger}\Pi=\sf{K}_{\Sigma}^{\dagger}$.
		\item[(ii)]When restricted to $\mathrm{ker}(\sf{K}_{\Sigma}^{\dagger})$, the operator $\Pi$ takes the simple form
		\begin{align*}
			\Pi\vert_{\mathrm{ker}(\sf{K}_{\Sigma}^{\dagger})}=\begin{pmatrix}
				0 & 0& 0 & 0\\
				0&0&0&0\\0 & 0 & \mathrm{id}-\d_{\Sigma}\Delta^{-1}\delta_{\Sigma} & 0\\
			0 & 0& 0 &\mathrm{id}-\d_{\Sigma}\Delta^{-1}\delta_{\Sigma}
		\end{pmatrix}\, .
		\end{align*}
		\item[(iii)]$\mathrm{ran}(\Pi\vert_{\mathrm{ker}(\sf{K}_{\Sigma}^{\dagger})})=\mathcal{V}_{\sf{R}}$, i.e.~$\Pi$ restricted to $\mathrm{ker}(\sf{K}_{\Sigma}^{\dagger})$ is exactly the projector $\sf{T}_{\Sigma}$ from Eq.~\eqref{eq:PhaseSpaceCRG} that projects onto the space of initial datum satisfying the Cauchy radiation gauge.
		\item[(iv)]The operator $\Pi\:\Gamma^{\infty}(\sf{A}_{\rho_{1}})\to\Gamma^{\infty}(\sf{A}_{\rho_{1}})$ allows us to find a right-inverse of $\sf{K}_{\Sigma}$ in the following sense: consider the operator $\sf{L}\:\Gamma^{\infty}(\sf{A}_{\rho_{1}})\to\Gamma^{\infty}(\sf{A}_{\rho_{0}})$ defined by
		\begin{align*}
			\sf{L}:=
			\begin{pmatrix}
				0 & 0 & \Delta^{-1}\delta_{\Sigma} & 0\\
				-i(\mathrm{id}-\mathcal{I}) & 0 & 0 &\Delta^{-1}\delta_{\Sigma}
			\end{pmatrix}\, .
		\end{align*}
		Then, it holds that $\sf{K}_{\Sigma}\sf{L}=\mathrm{id}-\Pi$ on $\Gamma^{\infty}(\sf{A}_{\rho_{1}})$ and hence $\sf{K}_{\Sigma}\Pi=0$ on $\mathrm{ker}(\Pi)=\mathrm{ran}(\sf{K}_{\Sigma})$. On the other hand, it holds that $\sf{L}\sf{K}_{\Sigma}=0$ on $C^{\infty}_{\mathrm{v}}(\Sigma,\bb{C})^{\oplus 2}\subset\Gamma^{\infty}(\sf{A}_{\rho_{0}})$. 
	\end{itemize}
	In summary, the Cauchy radiation gauge is related to \emph{gauge-invariant} initial data.\footnote{The derivation of the gauge invariant degrees of freedom is similar to the definition of the  \textit{Bardeen potentials} \cite{Bardeen} in \emph{cosmological perturbation theory}. In the latter, one considers linearised gravity in cosmological spacetimes and decomposes the linearised gravitational field $\sf{h}\in\Gamma^{\infty}(\sf{T}^{\ast}\sf{M}^{\otimes_{s}2})$ into scalar, vectorial and tensorial degrees of freedom. The transverse-traceless tensorial parts are already gauge-invariant, while the scalar and vectorial degrees of freedom are replaced by linear combinations thereof that are gauge invariant, see e.g.~\cite{Mukhanov2,Hack}.}
\end{remark}

The space $\mathcal{V}_{\sf{R}}$ can be equipped with a Hermitian sesquilinear form, defined in such way that the isomorphism $[\sf{T}_{\Sigma}]$ in \eqref{eq:PhaseSpaceCRG} becomes unitary. The idea for constructing Hadamard states is then to construct the states on the phase space $\mathcal{V}_{\sf{R}}$, on which the bundle metric is positive definite, and then to use the projector $\sf{T}_{\Sigma}$ to obtain a state on the original phase space $\mathcal{V}_{\Sigma}$. One of the key parts of this construction is, of course, to make sure that all the relevant properties such as positivity and the Hadamard property are preserved. 

Before addressing the explicit construction, we need to extend Proposition~\ref{Prop:AchievCRG} also to spacetimes for which $\Sigma$ is non-compact. This requires solving the Poisson equation $\Delta f\vert_{t=0}=\delta_{\Sigma}\sf{A}_{\Sigma}\vert_{t=0}$ also on non-compact manifolds. Clearly, working with spatially compactly supported fields is not appropriate in the setting of elliptic equations. Therefore, we enlarge our functional framework by employing suitable Sobolev spaces. In particular, the goal is to derive an analogue of the Hodge decomposition theorem for Sobolev spaces of differential forms, which will enable us to achieve the Cauchy radiation gauge in appropriate function spaces.

\section{Hodge Decomposition in Sobolev Spaces}
\label{Sec:HodgeDecomp}
Motivated by the previous discussion, our goal is to derive a Hodge decomposition for differential forms in suitable Sobolev spaces. While this is explicitly needed for our approach of  construction Hadamard states, such a decomposition is of course also interesting also on its own right.  We will start by defining suitable Sobolev spaces and the relevant functional analytic preliminaries. Afterwards, we establish the Hodge decomposition on those spaces and discuss the solvability of the Poisson equation in this setting. This section is based on \cite[Sec.~3]{MurroSchmid}.

\subsection{Sobolev Spaces of Differential Forms}
There are several ways to define (bundle-valued) Sobolev spaces on manifolds, each motivated by one of the equivalent definitions of Sobolev spaces in $\bb{R}^{d}$, which, however, turn out to be \emph{not} equivalent in the general setting of non-compact Riemannian manifolds. Here, we follow an approach based on \emph{spectral calculus} for the de Rham-Hodge Laplacian, which is perfectly adapted for later applications and the derivation of a Hodge decomposition. To start with, let $(\Sigma,\sf{h})$ be an oriented, connected and \emph{complete} $n$-dimensional Riemannian manifold. Furthermore, we denote the $\sf{L}^{2}$-space of $k$-forms by
\begin{align*}
	\sf{L}^{2}_{k}(\Sigma):=\overline{\Omega^{k}_{\mathrm{c}}(\Sigma,\bb{C})}^{\Vert\cdot\Vert_{\sf{L}^{2}}}\,,\qquad \langle\alpha,\beta\rangle_{\sf{L}^{2}}:=\int_{\Sigma}\,\overline{\alpha}\wedge\ast\beta=\int_{\Sigma}(\sf{h}^{\sharp})^{\otimes k}(\overline{\alpha},\beta)\,\d\mu_{\sf{h}}\,.
\end{align*} 
In this section, we restrict our attention to Riemannian manifolds without assuming the existence of a globally hyperbolic \emph{ambient} manifold. Ultimately, however, $(\Sigma,\sf{h})$ should be regarded as a Cauchy hypersurface of a globally hyperbolic spacetime, in which case the inner product $\langle\cdot,\cdot\rangle_{\sf{L}^{2}}$ corresponds to $(\cdot,\cdot)_{\sf{A}_{\Sigma,k}}$ in the notation introduced in the previous section.

As usual, let $\d_{\Sigma}\:\Omega^{k}(\Sigma,\bb{C})\to\Omega^{k+1}(\Sigma,\bb{C})$ and $\delta_{\Sigma}\:\Omega^{k+1}(\Sigma,\bb{C})\to\Omega^{k}(\Sigma,\bb{C})$ be the exterior derivative and codifferential and let us view the de Rham-Hodge Laplacian $\Delta:=\d_{\Sigma}\delta_{\Sigma}+\delta_{\Sigma}\d_{\Sigma}$ as a densely-defined operator of the form
\begin{align*}
	\Delta=\d_{\Sigma}\delta_{\Sigma}+\delta_{\Sigma}\d_{\Sigma}\:\mathrm{dom}(\Delta)\to \sf{L}^{2}_{k}(\Sigma)
\end{align*}
with $\mathrm{dom}(\Delta):=\Omega_{\mathrm{c}}^{k}(\Sigma,\bb{C})$. The main reason for assuming \emph{completeness} is the following fact:

\begin{theorem}\label{Thm:EsLap} The de Rham-Hodge Laplacian $\Delta=\d_{\Sigma}\delta_{\Sigma}+\delta_{\Sigma}\d_{\Sigma}\:\mathrm{dom}(\Delta)\to \sf{L}^{2}_{k}(\Sigma)$ with $\mathrm{dom}(\Delta):=\Omega_{\mathrm{c}}^{k}(\Sigma,\bb{C})$ is essentially self-adjoint on a complete Riemannian manifold $(\Sigma,\sf{h})$. 
\end{theorem}

The previous theorem is classical and well-documented in the literature. One of the earliest proofs can probably be attributed to Gaffney\footnote{Its worth mentioning, however, that Gaffney considered a larger domain. For $\d_{\Sigma}$ (resp.~$\delta_{\Sigma}$), he considers the set of all $C^{1}$-forms such that both $\alpha$ and $\d_{\Sigma}\alpha$ (resp.~$\delta_{\Sigma}\alpha$) are in $\sf{L}^{2}$ and for $\Delta=\d_{\Sigma}\delta_{\Sigma}+\delta_{\Sigma}\d_{\Sigma}$ the obvious domain induced thereof. Essentially self-adjointness for $\mathrm{dom}(\Delta)=\Omega^{k}_{\mathrm{c}}(\Sigma,\bb{C})$ is proven, for instance, in \cite{Chernoff,Strichartz}.}~\cite{Gaffney1951,GaffneyPhD}, who used mollification techniques to show that the minimal and maximal extensions (see Section~\ref{Sec:LinOp}) of $\Delta$ coincide. Shortly after, a different proof was obtained independently by Roelcke \cite{Roelcke}. Since then, several alternative proofs have been given. Among them, notable examples are those by Chernoff \cite{Chernoff} and Strichartz~\cite{Strichartz}. The latter one of Strichartz is particularly simple and only requires basic tools from functional analysis. For completeness, we sketch the main ideas of the proof:

\begin{proof}[Proof (sketch) of Theorem~\ref{Thm:EsLap}.]
	Let $\sf{A}\:\mathrm{dom}(\sf{A})\to\mathcal{H}$ be a densely-defined, closed and symmetric operator in a Hilbert space $\mathcal{H}$ that is \emph{non-negative}, i.e.~$\langle\sf{A}\psi,\psi\rangle_{\mathcal{H}}\geq 0$ for all $\psi\in\mathrm{dom}(\sf{A})$. Then, $\sf{A}$ is self-adjoint if and only if $\mathrm{ker}(\sf{A}^{\dagger}-\lambda\mathrm{id})=\{0\}$ for all $\lambda<0$, i.e.~if its adjoint $\sf{A}^{\dagger}$ does \emph{not} admit \emph{eigenvectors with negative eigenvalue}. Indeed, ``$\Rightarrow$'' is obvious and for ``$\Leftarrow$'', we observe that $\mathrm{dim}(\mathrm{ker}(\lambda\mathrm{id}-\sf{A}^{\dagger}))$ is constant on $\lambda\in\bb{C}\backslash [0,\infty)$ in this case (see e.g.~\cite[p.~137]{ReedSimonII}). In particular, combining this with the assumption, we conclude that the deficiency indices $\mathcal{N}^{\pm}=\mathrm{dim}(\mathrm{ker}(i\mp\sf{A}^{\dagger}))$ are zero, which implies that $\sf{A}$ is self-adjoint. 
	
	Using this fact, self-adjointness of $\overline{\Delta}$ is equivalent to show that $\overline{\Delta}^{\dagger}\omega=\lambda\omega$ implies $\omega=0$ for $\lambda<0$ and $\omega\in\mathrm{dom}(\overline{\Delta})$. In fact, we can simplify this even a bit further. First, note that $\overline{\Delta}^{\dagger}=\Delta^{\dagger}=\Delta_{\mathrm{max}}$, where $\Delta_{\mathrm{max}}$ denotes the \emph{maximal closed extension} (see Section~\ref{Sec:LinOp}) of $\Delta$. Furthermore, by elliptic regularity, we know that $\omega$ is smooth whenever $\Delta_{\mathrm{max}}\omega$ is smooth and hence, self-adjointness of $\overline{\Delta}$ is equivalent to
	\begin{align*}
		\Delta\omega=\lambda\omega\quad\Rightarrow\quad\omega=0
		\qquad\qquad\forall\omega\in\Omega^{k}(\Sigma,\bb{C})\cap\sf{L}^{2}_{k}(\Sigma)\,,\quad\lambda<0\, .
	\end{align*}
	Now, let $\Phi$ be any function on $\Sigma$ and assume that $\omega\in\Omega^{k}(\Sigma,\bb{C})\cap\sf{L}^{2}_{k}(\Sigma)$ is such that $\Delta\omega=\lambda\omega$ for some $\lambda<0$. Then, a straightforward computation using the Leibniz rule shows
	\begin{align*}
		0\geq \langle\Phi^{2}\omega,\Delta\omega\rangle_{\sf{L}^{2}}=\Vert\Phi\d_{\Sigma}\omega\Vert_{\sf{L}^{2}}^{2}+\Vert\Phi\delta_{\Sigma}\omega\Vert_{\sf{L}^{2}}^{2}+2\langle\Phi\d_{\Sigma}\Phi\wedge\omega,\d_{\Sigma}\omega\rangle_{\sf{L}^{2}}-2\langle\omega,\Phi\d_{\Sigma}\Phi\wedge\delta_{\Sigma}\omega\rangle_{\sf{L}^{2}}
	\end{align*}		
	In particular, we find $\Vert\Phi\d_{\Sigma}\omega\Vert_{\sf{L}^{2}}+\Vert\Phi\delta_{\Sigma}\omega\Vert_{\sf{L}^{2}}\leq 4\Vert\d_{\Sigma}\Phi\Vert_{\infty}\Vert\omega\Vert_{\sf{L}^{2}}$. The idea is now to construct a suitable \emph{approximation of unity}, i.e.~a family $\Phi_{r,s}$ of Lipschitz continuous functions with $0\leq\Phi_{r,s}\leq 1$ such that $\Phi_{r,s}=1$ on a ball $\mathcal{B}_{r}(p)$ of radius $r>0$ around a fixed (but arbitrary) point $p\in\sf{M}$, such that $\Phi_{r,s}=0$ on $\Sigma\backslash\mathcal{B}_{s}(p)$ for $s>r$ and such that $\Vert\d_{\Sigma}\Phi_{r,s}\Vert_{\infty}\leq C(s-r)^{-1}$ for some constant $C>0$. The construction of such a family for \emph{complete} manifolds is straightforward and was established by Strichartz in \cite[Lemma~2.2]{Strichartz} (based on earlier ideas of Yau in \cite{Yau}). With such a family, choose $\Phi=\Phi_{r,s}$ in the above estimate and let $s\to\infty$ for $r$ fixed. It follows that $\d_{\Sigma}\omega=0$ and $\delta_{\Sigma}\omega=0$ and hence $\omega=\lambda^{-1}\Delta\omega=0$, which concludes the proof.
\end{proof}

\begin{remark}
	It is important to stress that completeness of $(\Sigma,\mathsf{h})$ is a sufficient but clearly not a necessary condition for essential self-adjointness of $\Delta\:\Omega^{k}_{\mathrm{c}}(\Sigma,\bb{C})\to\sf{L}^{2}_{k}(\Sigma)$. In fact, the question of essential self-adjointness of $\Delta$ is closely related to whether an $\mathsf{L}^{2}$-version of \emph{Stokes' theorem} holds. In his original paper \cite{Gaffney1951}, Gaffney considers manifolds for which such a theorem holds and calls them \emph{manifolds of negligible boundary}. In a subsequent work, he then shows that complete manifolds fall into this class \cite{Gaffney1954} (see also the generalisation by Yau in \cite{Yau}). However, there are many more examples. For instance, following \cite{Gaffney1951,Gaffney1954}, any manifold obtained from removing a $(n-2)$-dimensional surface from $\bb{R}^{n}$ has negligible boundary.
	
	While there are many situations in which $\Delta$ is essentially self-adjoint, there are some restrictions. For instance, as argued in \cite[Lem.~3.3]{BruningLesch}, essential self-adjointness of $\Delta$ implies that the minimal and maximal closed extensions of $\d_{\Sigma}$ on $\mathrm{dom}(\d_{\Sigma})=\Omega^{\bullet}_{\mathrm{c}}(\Sigma,\bb{C})$ coincide.
\end{remark}

Now, since $\Delta\:\Omega^{k}_{\mathrm{c}}(\Sigma,\bb{C})\to\sf{L}^{2}_{k}(\Sigma)$ is essentially self-adjoint, it has a unique self-adjoint extension, namely its closure, which we denote by 
\begin{align*}
	\overline{\Delta}\:\mathrm{dom}(\overline{\Delta})\to\sf{L}^{2}_{k}(\Sigma)\,.
\end{align*}
By definition, $\omega\in\mathrm{dom}(\overline{\Delta})$ if and only if there exists a sequence $(\omega_{n})_{n\in\bb{N}}$ in $\Omega^{k}_{\mathrm{c}}(\Sigma,\bb{C})=\mathrm{dom}(\Delta)$ such that $\lim_{n\to\infty}\omega_{n}=\omega$ and such that $(\Delta\omega_{n})_{n}$ converges in $\sf{L}^{2}_{k}(\Sigma)$. Then, $\overline{\Delta}\omega:=\lim_{n\to\infty}\Delta\omega_{n}$. 

It is well-known that the de Rham-Hodge Laplacian is a \emph{non-negative} operator:

\begin{lemma}\label{Lemma:SpecLap} The spectrum\footnote{We denote by $\sigma(\sf{A})$ the spectrum of some operator $\sf{A}$ in a Hilbert space, see Appendix~\ref{App:FuncAna} for details.} of $\overline{\Delta}$ satisfies $\sigma(\overline{\Delta})\subset [0,\infty)$.
\end{lemma}

\begin{proof}
	To start with, let $\omega\in\Omega^{k}_{\mathrm{c}}(\Sigma,\bb{C})$. Then, $\langle\Delta\omega,\omega\rangle_{\sf{L}^{2}}=\Vert\d_{\Sigma}\omega\Vert_{\sf{L}^{2}}^{2}+\Vert\delta_{\Sigma}\omega\Vert_{\sf{L}^{2}}^{2}\geq 0$ and by continuity of the inner product and definition of the closure, we conclude that also $\langle\overline{\Delta}\omega,\omega\rangle_{\sf{L}^{2}}\geq 0$ for all $\omega\in\mathrm{dom}(\overline{\Delta})$. Now, since $\overline{\Delta}$ is self-adjoint, we know that $\sigma(\overline{\Delta})\subset\bb{R}$. Take $\lambda\in\bb{R}$ such that $\lambda<0$. Then, using $\langle\overline{\Delta}\omega,\omega\rangle_{\sf{L}^{2}}\geq 0$, the Cauchy-Schwartz inequality implies 
	\begin{align}\label{eq:EstimateSpec}
		\Vert(\overline{\Delta}-\lambda\mathrm{id})\omega\Vert_{\sf{L}^{2}}\geq\vert\lambda\vert\cdot\Vert\omega\Vert_{\sf{L}^{2}}
		\end{align}
for all $\omega\in\mathrm{dom}(\overline{\Delta})\backslash\{0\}$ and hence $\mathrm{ker}(\overline{\Delta}-\lambda\mathrm{id})=\{0\}$. Next, let $\eta\in\overline{\mathrm{ran}(\overline{\Delta}-\lambda\mathrm{id})}$ and consider a sequence $\omega_{n}\in\mathrm{dom}(\overline{\Delta})$ such that $ (\overline{\Delta}-\lambda\mathrm{id})\omega_{n}\to\eta$ as $n\to\infty$. Then, since
	\begin{align*}
		\Vert\omega_{n}\Vert_{\sf{L}^{2}}\leq\frac{1}{\vert\lambda\vert}\Vert (\overline{\Delta}-\lambda\mathrm{id})\omega_{n}\Vert_{\sf{L}^{2}}\xrightarrow{n\to\infty}\frac{1}{\vert\lambda\vert}\Vert\eta\Vert_{\sf{L}^{2}}\,,
	\end{align*}
	we conclude that also $(\omega_{n})_{n\in\mathbb{N}}$ is convergent. By definition of the closure, it is clear that its limit $\omega\in\sf{L}^{2}_{k}(\Sigma)$ is contained in $\mathrm{dom}(\overline{\Delta})$ and satisfies $(\overline{\Delta}-\lambda\mathrm{id})\omega=\eta$. Hence, $\eta\in\mathrm{ran}(\overline{\Delta}-\lambda\mathrm{id})$, which shows that $\mathrm{ran}(\overline{\Delta}-\lambda\mathrm{id})$ is closed. Combining everything, we find
	\begin{align*}
		\mathrm{ran}(\overline{\Delta}-\lambda\mathrm{id})=\overline{\mathrm{ran}(\overline{\Delta}-\lambda\mathrm{id})}=\mathrm{ker}(\overline{\Delta}-\lambda\mathrm{id})^{\perp}=\{0\}^{\perp}=\sf{L}^{2}_{k}(\Sigma)\,.
	\end{align*}
	To sum up, we have shown that $(\overline{\Delta}-\lambda\mathrm{id})\:\mathrm{dom}(\overline{\Delta})\to\sf{L}^{2}_{k}(\Sigma)$ is bijective and from the Estimate~\eqref{eq:EstimateSpec}, it follows that its inverse is bounded. Hence, $\lambda\notin\sigma(\overline{\Delta})$.
\end{proof}

\begin{definition} We define $\sf{E}:=\overline{\mathrm{id}+\Delta}=\mathrm{id}+\overline{\Delta}\:\mathrm{dom}(\overline{\Delta})\to\sf{L}^{2}_{k}(\Sigma)$. Then, for $s\in\bb{R}$, we define $\sf{E}^{s}\:\mathrm{dom}(\sf{E}^{s})\to\sf{L}^{2}_{k}(\Sigma)$ by means of spectral calculus, i.e.
	\begin{align*}
		\mathrm{dom}(\sf{E}^{s}):=\{\omega\in\sf{L}^{2}_{k}(\Sigma)\mid (\lambda\mapsto\lambda^{s})\in\sf{L}^{2}(\sigma(\sf{E}),\d\mu^{\sf{E}}_{\omega})\}\,,\qquad\sf{E}^{s}\omega:=\int_{\sigma(\sf{E})}\lambda^{s}\,\d\mu_{\omega}^{\sf{E}}(\lambda)\,,
	\end{align*}
	where $\mu^{\sf{E}}$ denotes the spectral measure of the self-adjoint operator $\sf{E}$ and $\mu_{\omega}^{\sf{E}}$ for $\omega\in\sf{L}^{2}_{k}(\Sigma)$ the corresponding complex measure $\mu^{\sf{E}}_{\omega}(\cdot):=\langle\mu^{\sf{E}}(\cdot)\omega,\omega\rangle_{\sf{L}^{2}}$.
\end{definition}

Note that $\sf{E}^{s}$ is well-defined for all $s \in \mathbb{R}$, since $\sigma(\sf{E}) \subset [1,\infty)$ by Lemma~\ref{Lemma:SpecLap}. Furthermore, we remark that $\sf{E}^{-\frac{\alpha}{2}}$ is commonly referred to as the \emph{Bessel potential of order $\alpha$} for $\mathrm{Re}(\alpha) \geq 0$. For the spectral theorem, we refer to Theorem~\ref{Thm:SpecThm} in the appendix.

There exist several alternative (and equivalent) approaches for defining the operators $\sf{E}^{s}$ without resorting to spectral calculus. For example, Strichartz~\cite{Strichartz} provides a construction employing the heat kernel and semigroup techniques. One of the main advantages of the spectral-theoretic definition, however, is that many useful properties follow directly from the general framework of spectral calculus:

\begin{proposition} \label{Prop:SobProp}
	The self-adjoint operators $\sf{E}^{s}$ have the following properties:
	\begin{itemize}
		\item[\emph{(i)}]$\sigma(\sf{E}^{s})\subset [1,\infty)$ for $s\geq 0$, while $\sigma(\sf{E}^{s})\subset [0,1]$ for $s<0$. 
		\item[\emph{(ii)}]The operators $\sf{E}^{s}$ for $s<0$ are bounded with $\mathrm{dom}(\sf{E}^{s})=\sf{L}^{2}_{k}(\Sigma)$ and $\Vert\sf{E}^{s}\Vert_{\mathrm{op}}=1$.
		\item[\emph{(iii)}]$\sf{E}^{s}\:\mathcal{D}(\sf{E}^{s})\to\sf{L}^{2}_{k}(\Sigma)$ for $s>0$ is bijective with bounded inverse $\sf{E}^{-s}$. 
		\item[\emph{(iv)}]$\sf{E}^{m}=\overline{(\mathrm{id}+\Delta)^{m}}$ for all $m\in\bb{N}$, where $\mathrm{dom}((\mathrm{id}+\Delta)^{m})=\Omega^{k}_{\mathrm{c}}(\Sigma,\bb{C})$. In other words, taking the closure and the $m^{\mathrm{th}}$ power are commuting operations for the operator $\mathrm{id}+\Delta$ defined on $\Omega^{k}_{\mathrm{c}}(\sf{M},\bb{C})$.
		\item[\emph{(v)}]$\Omega^{k}_{\mathrm{c}}(\Sigma,\bb{C})\subset\mathrm{dom}(\sf{E}^{s})\subset\mathrm{dom}(\sf{E}^{r})\subset\sf{L}^{2}_{k}(\Sigma)$ for all $s,r\in\bb{R}$ with $r\leq s$.
		\item[\emph{(vi)}]$\sf{E}^{r+s}=\sf{E}^{r}\sf{E}^{s}=\sf{E}^{s}\sf{E}^{r}$ on $\mathrm{dom}(\sf{E}^{\gamma})$ with $\gamma:=\max\{r,s,r+s\}$ for all $r,s\in\bb{R}$.
	\end{itemize}
\end{proposition}

\begin{proof}
	Define $f_{s}\in C(\sigma(\sf{E}),\bb{R})$ by $f(\lambda):=\lambda^{s}$. By definition of the spectral calculus, it holds that $\sf{E}^{s}:=f_{s}(\sf{E})$. Most of the claims then follow from the general properties of the functional calculus (see Appendix~\ref{App:FuncAna} for details): by continuity of $f_{s}$, it holds that 
	\begin{align*}
		\sigma(f_{s}(\sf{E}))=\overline{f_{s}(\sigma(\sf{E}))}\subset\overline{\{\lambda^{s}\mid\lambda\in [0,\infty)\}}
	\end{align*}
	and hence $\sigma(\sf{E}^{s})\subset [1,\infty)$ for $s\geq 0$ as well as $\sigma(\sf{E}^{s})\subset [0,1]$ for $s<0$. This shows (i). For (ii), we note that $f_{s}$ is a bounded function with $\Vert f_{s}\Vert_{\infty}=1$ on $\sigma(\sf{E})\subset [1,\infty)$ for $s<0$ and for (iii), since $f_{s}(\lambda)\neq 0$ on $\sigma(\sf{E})$, we conclude that $\sf{E}=f_{s}(\sf{E})$ is bijective with inverse $(1/f_{s})(\sf{E})=\sf{E}^{-s}$. For (iv), we note that in fact all the powers of the de Rham-Hodge Laplacian are essentially self-adjoint, see~\cite{Cordes,Chernoff}. Hence, the claim follows from uniqueness of the self-adjoint extension. For (v), let us first take $r,s\in\bb{R}$ such that $r\leq s$. Then, for $\omega\in\mathrm{dom}(\sf{E}^{s})$, it follows that
	\begin{align*}
		\Vert\sf{E}^{r}\omega\Vert^{2}_{\sf{L}^{2}}=\int_{\sigma(\sf{E})}\lambda^{2r}\d\mu_{\omega}^{\sf{E}}(\lambda)\leq \int_{\sigma(\sf{E})}\lambda^{2s}\d\mu_{\omega}^{\sf{E}}(\lambda)=\Vert\sf{E}^{s}\omega\Vert^{2}_{\sf{L}^{2}}
	\end{align*}
	and hence $\omega\in\mathrm{dom}(\sf{E}^{r})$. Furthermore, by (iv), we know that $\Omega^{k}_{\mathrm{c}}(\Sigma,\bb{C})\subset\mathrm{dom}(\sf{E}^{m})$ for all $m\in\bb{N}$, which proves the claim. Last but not least, for $r,s\in\bb{N}$ arbitrary, it holds that $\sf{E}^{r+s}=\overline{\sf{E}^{s}\sf{E}^{r}}=\overline{\sf{E}^{r}\sf{E}^{s}}$ as well as $\mathrm{dom}(\sf{E}^{r}\sf{E}^{s})=\mathrm{dom}(\sf{E}^{s})\cap\mathrm{dom}(\sf{E}^{r+s})$ and $\mathrm{dom}(\sf{E}^{s}\sf{E}^{r})=\mathrm{dom}(\sf{E}^{r})\cap\mathrm{dom}(\sf{E}^{r+s})$. In particular, we see that the operators $\sf{E}^{r+s}$, $\sf{E}^{r}\sf{E}^{s}$ and $\sf{E}^{s}\sf{E}^{r}$ agree on the common domain $\mathrm{dom}(\sf{E}^{\gamma})$ for $\gamma=\max\{r,s,r+s\}$ by (v).
\end{proof}

With this definition, we define the Sobolev spaces of differential forms on complete Riemannian manifolds with \emph{positive degree} as follows.

\begin{definition}\label{Def.Sobolev} (Sobolev Spaces of Differential Forms)\newline
	Let $s\in\bb{R}$ with $s\geq 0$. Then, we define the Hilbert space
	\begin{align*}
		\sf{H}^{s}_{k}(\Sigma):=\mathrm{dom}(\sf{E}^{\frac{s}{2}})\,,\qquad\langle\cdot,\cdot\rangle_{\sf{H}^{s}}:=\langle\sf{E}^{\frac{s}{2}}\cdot,\sf{E}^{\frac{s}{2}}\cdot\rangle_{\sf{L}^{2}}\,,
	\end{align*}
	called the \emph{$\sf{L}^{2}$-Sobolev space of order $s$}. 
\end{definition}

\begin{remark}
	By definition, it holds that $\sf{H}^{2}_{k}(\Sigma)=\mathrm{dom}(\sf{E})=\mathrm{dom}(\mathrm{id}+\overline{\Delta})=\mathrm{dom}(\overline{\Delta})$. In other words, $\overline{\Delta}\:\sf{H}^{2}_{k}(\Sigma)\to\sf{L}^{2}_{k}(\Sigma)$ is the closure of $\Delta\:\Omega^{k}_{\mathrm{c}}(\Sigma,\bb{C})\to\sf{L}^{2}_{k}(\Sigma)$, as expected.
\end{remark}

The fact that $\langle\cdot,\cdot\rangle_{\sf{H}^{s}}$ is a well-defined \emph{positive-definite inner product} on the vector space $\sf{H}^{s}_{k}(\Sigma)$ follows easily from the fact that $\sf{E}^{\frac{s}{2}}$ is invertible. Similarly, it is clear that  $(\sf{H}^{s}_{k}(\Sigma),\langle\cdot,\cdot\rangle_{\sf{H}^{s}})$ is \emph{complete} and hence in particular a Hilbert space.

Note that we restrict ourselves to Sobolev spaces of \emph{non-negative} order. In fact, the above definition is clearly only sensible for $s\geq 0$, since $\mathrm{dom}(\sf{E}^{\frac{s}{2}})=\sf{L}^{2}_{k}(\Sigma)$ for all $s<0$ by Proposition~\ref{Prop:SobProp}(ii). One may define the spaces $\sf{H}^{s}_{k}(\Sigma)$ for $s<0$ as the topological dual space of $\sf{H}^{-s}_{k}(\Sigma)$, however, for our purposes, it will be enough to work with $\sf{H}^{s}_{k}(\Sigma)$ for $s\geq 0$. 

\begin{remark}\label{Rem:Sobolev}
	As already indicated at the beginning of this section, there are several non-equivalent definitions of Sobolev spaces on complete Riemannian manifolds. As we will show below, our functional-analytic definition is equivalent to define $\sf{H}^{s}_{k}(\Sigma)$ as the completion of $\Omega^{k}_{\mathrm{c}}(\Sigma,\bb{C})$ with respect to $\Vert\cdot\Vert_{\sf{H}^{s}}$. Another definition is provided by considering the completion of $\{\omega\in\Omega^{k}(\Sigma,\bb{C})\mid\Vert\omega\Vert_{\sf{H}^{s}}<\infty\}$ with respect to $\Vert\cdot\Vert_{\sf{H}^{s}}$ instead. Furthermore, one could generalise the standard definition on $\bb{R}^{d}$ by defining $\sf{H}^{m}_{k}(\Sigma)$ for $m\in\bb{N}_{0}$ using \emph{weak/distributional covariant derivatives} (see e.g.~\cite[Def.~10.2.33]{Nicoleascu}).
	
It is known, however, that these three definitions agree in the special case of a \emph{Riemannian manifold of bounded geometry} (see Section~\ref{Subsec:BoundGeom} in the appendix), which are Riemannian manifolds with non-zero injectivity radius and with the property that the Riemann curvature tensor and all its derivatives are uniformly bounded. Examples include the Euclidean space $(\bb{R}^{d},\delta)$ and compact Riemannian manifolds. For a proof, we refer to \cite{Eichhorn,Aubin}. In the specific case of differential forms, a short proof can also be found in \cite{Dodziuk}. For further details on Sobolev spaces on non-compact manifolds, we refer to the monographs \cite{EichhornBook,AubinBook,Hebey}.

Furthermore, we mention that also \emph{Sobolev embedding theorems} do not hold in general in the non-compact setting, they do, however, hold in the case of bounded geometry, i.e.~for a Riemannian manifold of bounded geometry there is a continuous embedding $\sf{H}^{s}_{k}(\Sigma)\hookrightarrow\Omega^{l}(\Sigma,\bb{C})$ for $k,l\in\mathbb{N}_{0}$ such that $k>l+\mathrm{dim}(\Sigma)/2$.
\end{remark}

The Sobolev spaces $\sf{H}^{s}_{k}(\Sigma)$ as defined in Definition~\ref{Def.Sobolev} have the following properties.

\begin{proposition}\label{Prop:Sob} The Sobolev spaces $\sf{H}^{s}_{k}(\Sigma)$ for $s\geq 0$ have the following properties:
\begin{itemize}
	\item[\emph{(i)}]The operator $\sf{E}^{\frac{s}{2}}\:\sf{H}^{s}_{k}(\Sigma)\to\sf{L}^{2}_{k}(\Sigma)$ is unitary and $\sf{H}^{s}_{k}(\Sigma)=\mathrm{ran}(\sf{E}^{-\frac{s}{2}}\:\sf{L}^{2}_{k}(\Sigma)\to\sf{L}^{2}_{k}(\Sigma))$.
	\item[\emph{(ii)}]$\sf{H}^{s}_{k}(\Sigma)\subset\sf{H}_{k}^{r}(\Sigma)$ for $0\leq r\leq s$. and the inclusion map $i\:\sf{H}^{s}_{k}(\Sigma)\hookrightarrow\sf{H}^{r}_{k}(\Sigma)$ is continuous with $\Vert i\Vert_{\mathrm{op}}\leq 1$, i.e.~$\Vert\omega\Vert_{\sf{L}^{2}}\leq\Vert\omega\Vert_{\sf{H}^{r}}\leq\Vert\omega\Vert_{\sf{H}^{s}}$ for all $\omega\in\sf{H}^{s}_{k}(\Sigma)$.
\end{itemize}
\end{proposition}

\begin{proof}
	Claim (i) is obvious and follows directly from the definition and Proposition~\ref{Prop:SobProp}(iii). For (ii), let $s,r\in\bb{R}$ with $0\leq r\leq s$. By Proposition~\ref{Prop:SobProp}(v), we conclude that $\sf{H}^{s}_{k}(\Sigma)\subset\sf{H}_{k}^{r}(\Sigma)$. Furthermore, it holds that
	\begin{align*}
		\Vert\omega\Vert_{\sf{H}^{s}}=\Vert\sf{E}^{\frac{s}{2}}\omega\Vert_{\sf{L}^{2}}=\Vert\sf{E}^{\frac{s-r}{2}}\sf{E}^{\frac{r}{2}}\omega\Vert_{\sf{L}^{2}}\leq\Vert\sf{E}^{\frac{r}{2}}\omega\Vert_{\sf{L}^{2}}=\Vert\omega\Vert_{\sf{H}^{r}}\, ,
	\end{align*}
	where we used Proposition~\ref{Prop:SobProp}(vi) and where we used the fact that $\sf{E}^{\frac{s-r}{2}}$ is a bounded operator on $\sf{L}^{2}_{\bullet}(\Sigma)$ with operator norm $\Vert\sf{E}^{\frac{s-r}{2}}\Vert_{\mathrm{op}}=1$ by Proposition~\ref{Prop:SobProp}(ii).
\end{proof}

By Proposition~\ref{Prop:Sob}(ii), it makes sense to define the \emph{Sobolev space of order $\infty$} by
\begin{align*}
	\sf{H}^{\infty}_{k}(\Sigma):=\bigcap_{s\geq 0}\sf{H}^{s}_{k}(\Sigma)\, .
\end{align*}
We shall equip the space $\sf{H}^{\infty}_{k}(\Sigma)$ with the natural \emph{projective limit topology}, i.e.~the coarsest topology on $\sf{H}^{\infty}_{k}(\Sigma)$ such that all the inclusions $i_{s}\:\sf{H}^{\infty}_{k}(\Sigma)\to\sf{H}^{s}_{k}(\Sigma)$ are continuous. In other words, the space $\sf{H}^{\infty}_{k}(\Sigma)$ is the locally convex vector space whose topology is induced by the family of norms $\Vert\cdot\Vert_{\sf{H}^{s}}$. It is easy to see that $\sf{H}^{\infty}_{k}(\Sigma)$ is in fact metrisable, Hausdorff and complete and hence in particular a Fréchet space. By definition, a sequence $(\omega_{n})_{n\in\bb{N}}$ in $\sf{H}^{\infty}_{k}(\Sigma)$ converges to some $\omega\in\sf{H}^{\infty}_{k}(\Sigma)$ in $\sf{H}^{\infty}_{k}(\Sigma)$ if and only if $\omega_{n}\to\omega$ as $n\to\infty$ in $\sf{H}^{s}_{k}(\Sigma)$ for all $s\in [0,\infty)$.

\begin{remark}
	If $\Sigma$ is compact, then $\sf{H}^{\infty}_{k}(\Sigma)=\Omega^{k}(\Sigma,\bb{C})$. Inclusion ``$\supset$'' is obvious and ``$\subset$'' follows from the Sobolev embedding theorems. On general complete Riemannian manifolds, however, we stress that neither of these inclusions is true in general (cf.~Remark~\ref{Rem:Sobolev}).
\end{remark}

It turns out that the inclusion maps $i\:\sf{H}^{s}_{k}(\Sigma)\hookrightarrow\sf{H}^{r}_{k}(\Sigma)$ for $s\geq r$ are not only continuous, but in fact, also have dense images. A fast way to prove this is by employing suitable \emph{mollifiers}. Adapting the definitions employed in the lecture notes \cite[Proof of Thm.~3.7.7]{BaerLecture} to the non-compact (but complete) setting, we make the following definition.

\begin{lemma} \label{Lemma:Mollifiers} \emph{(Mollifiers)}\newline
	Consider the self-adjoint and bounded operator $\mathcal{J}_{\varepsilon}:= e^{-\varepsilon\sf{E}}\:\sf{L}^{2}_{k}(\Sigma)\to\sf{L}^{2}_{k}(\Sigma)$ defined using spectral calculus for all $k\in\bb{N}_{0}$, depending on a real parameter $\varepsilon>0$. Then,
	\begin{itemize}
		\item[\emph{(i)}]$\mathcal{J}_{\varepsilon}$ commutes with $\sf{E}^{s}$ on $\sf{H}^{2s}_{k}(\Sigma)$ for $s\in [0,\infty)$. In particular, $\mathcal{J}_{\varepsilon}$ is self-adjoint in $\sf{H}^{s}_{\bullet}(\Sigma)$ and $\mathcal{J}_{\varepsilon}\:\sf{H}^{s}_{\bullet}(\Sigma)\to\sf{H}^{r}_{\bullet}(\Sigma)$ is well-defined and bounded for all $r,s\in [0,\infty)$.
		\item[\emph{(ii)}]$\mathcal{J}_{\varepsilon}$ is regularising, i.e.~$\mathrm{ran}(\mathcal{J}_{\varepsilon})\subset\Omega^{k}(\Sigma,\bb{C})$.
		\item[\emph{(iii)}]$\mathcal{J}_{\varepsilon}\:\mathsf{H}^{s}_{k}(\Sigma)\to\sf{H}^{s}_{k}(\Sigma)$ converges strongly to $\mathrm{id}_{\sf{H}^{s}_{k}(\Sigma)}$ as $\varepsilon\to 0^{+}$. 
	\end{itemize}
\end{lemma}

\begin{proof}
	To start with, we note that $\mathcal{J}_{\varepsilon}$ is indeed well-defined and bounded in $\sf{L}^{2}_{k}(\Sigma)$, since $g_{\varepsilon}(\lambda):=e^{-\varepsilon\lambda}$ is a bounded and continuous function on $\sigma(\sf{E})\subset [1,\infty)$. For (i), we recall that for two measurable functions $f,g\:\sigma(\sf{E})\to\bb{C}$, it holds that $f(\sf{E})g(\sf{E})=g(\sf{E})f(\sf{E})$ on $\mathcal{D}(f(\sf{E}))\cap\mathcal{D}(g(\sf{E}))\cap\mathcal{D}((f\cdot g)(\sf{E}))$, where $f\cdot g$ denotes the pointwise product. Taking $f_{s}(\lambda):=\lambda^{\frac{s}{2}}$ and $g_{\varepsilon}(\lambda)=e^{-\varepsilon\lambda}$, it follows that $\mathcal{J}_{\varepsilon}\sf{E}^{\frac{s}{2}}=\sf{E}^{\frac{s}{2}}\mathcal{J}_{\varepsilon}$ on $\sf{H}^{s}_{\bullet}(\Sigma)$. Secondly, we compute 
	\begin{align*}
		\Vert\mathcal{J}_{\varepsilon}\omega\Vert_{\sf{H}^{s}}=\Vert \sf{E}^{\frac{s}{2}}\mathcal{J}_{\varepsilon}\omega\Vert_{\sf{L}^{2}}=\Vert \sf{E}^{\frac{s-r}{2}}\mathcal{J}_{\varepsilon}\sf{E}^{\frac{r}{2}}\omega\Vert_{\sf{L}^{2}}\leq C\cdot\Vert\sf{E}^{\frac{r}{2}}\omega\Vert_{\sf{L}^{2}}=C\cdot\Vert\omega\Vert_{\sf{H}^{r}}
	\end{align*}
	with $C>0$ for all $\omega\in\sf{H}^{s}_{k}(\Sigma)$, where we used that $\sf{E}^{\frac{s-r}{2}}\mathcal{J}_{\varepsilon}$ is a bounded operator on $\sf{L}^{2}_{\bullet}(\Sigma)$. For (ii) we observe that $\mathcal{J}_{\varepsilon}\omega$ satisfies the hypoelliptic heat-type equation $(\partial_{\varepsilon}+\sf{E})\mathcal{J}_{\varepsilon}\omega=0$ for any $\omega\in\sf{L}^{2}_{\bullet}(\Sigma)$. Statement (iii) is a consequence of the dominant convergence theorem and the fact that $g_{\varepsilon}(\lambda)=e^{-\varepsilon\lambda}$ converges to the constant function $\lambda\mapsto 1$ as $\varepsilon\to 0^{+}$. 
\end{proof}

The mollifiers $\mathcal{J}_{\varepsilon}$ provide a powerful tool that will be crucial in the subsequent sections. As a first application, they allow us to show that $\sf{H}^{s}_{k}(\Sigma)$ is dense in $\sf{H}^{r}_{k}(\Sigma)$ for all $r\leq s$. Furthermore, we can show that also $\Omega^{k}_{\mathrm{c}}(\Sigma,\bb{C})$ is dense in $\sf{H}^{s}_{k}(\Sigma)$ for all $s\in [0,\infty)$, generalising the analogues fact for the standard Sobolev spaces on $\bb{R}^{d}$, which is one of the main reasons for our choice of Sobolev spaces compared to the other definitions briefly mentioned in Remark~\ref{Rem:Sobolev}. 

\begin{proposition}\label{Prop:Sob2} The Sobolev spaces $\sf{H}^{s}_{k}(\Sigma)$ for $s\geq 0$ have the following properties:
\begin{itemize}
	\item[\emph{(i)}]The inclusion map $i\:\sf{H}^{s}_{k}(\Sigma)\hookrightarrow\sf{H}^{r}_{k}(\Sigma)$ for $0\leq r\leq s$ has dense image.
	\item[\emph{(ii)}]$\Omega^{k}_{\mathrm{c}}(\Sigma,\bb{C})$ is dense in $\sf{H}^{s}_{k}(\Sigma)$ for all $s\in [0,\infty)$.
\end{itemize}
\end{proposition}

\begin{proof}
	Claim (i) follows immediately from Lemma~\ref{Lemma:Mollifiers}: take $\omega\in\sf{H}^{r}_{k}(\Sigma)$ and consider the one-parameter family $(\mathcal{J}_{\varepsilon}\omega)_{\varepsilon\geq 0}$. By Lemma~\ref{Lemma:Mollifiers}(i), $\mathcal{J}_{\varepsilon}\omega\in\sf{H}^{s}_{k}(\Sigma)$ and by Lemma~\ref{Lemma:Mollifiers}(ii), $(\mathcal{J}_{\varepsilon}\omega)_{\varepsilon\geq 0}\to\omega$ in $\sf{H}^{r}_{k}(\Sigma)$ as $\varepsilon\downarrow 0$. 
	
	For (ii), we first show the claim for $s=2m$ with $m\in\bb{N}_{0}$. In this case, we know by Proposition \ref{Prop:SobProp}(iv) that $\sf{H}^{2m}_{k}(\Sigma)=\mathrm{dom}(\overline{\sf{T}})$, where $\sf{T}:=(\mathrm{id}+\Delta)^{m}$ is equipped with the domain $\mathrm{dom}(\sf{T})=\Omega^{k}_{\mathrm{c}}(\Sigma,\bb{C})$. Now, by definition of the closure, $\mathrm{dom}(\overline{\sf{T}})$ is the completion of $\Omega^{k}_{\mathrm{c}}(\Sigma,\bb{C})$ with respect to the graph norm $\Vert\cdot\Vert:=(\Vert\cdot\Vert_{\sf{L}^{2}}+\Vert\sf{T}\cdot\Vert_{\sf{L}^{2}})^{\frac{1}{2}}$. It is easy to see that $\Vert\cdot\Vert$ is equivalent to $\Vert\cdot\Vert_{\sf{H}^{2m}}=\Vert\sf{T}\cdot\Vert_{\sf{L}^{2}}$ and hence, we conclude that $\Omega^{k}_{\mathrm{c}}(\Sigma,\bb{C})$ is dense in $\sf{H}^{2m}_{k}(\Sigma)$. Now, let $s\in [0,\infty)$ by arbitrary, take $\omega\in\sf{H}^{s}_{k}(\Sigma)$ and $m\in\bb{N}_{0}$ large enough such that $2m\geq s$. By (i), $\sf{H}^{2m}_{k}(\Sigma)$ is dense in $\sf{H}^{s}_{k}(\Sigma)$ and hence we can find a sequence $(\omega_{n})_{n\in\bb{N}}$ in $\sf{H}^{2m}_{k}(\Sigma)$ such that $\Vert\omega_{n}-\omega\Vert_{\sf{H}^{s}}\to 0$ as $n\to\infty$. But now, by density of $\Omega^{k}_{\mathrm{c}}(\Sigma,\bb{C})$ in $\sf{H}^{2m}_{k}(\Sigma)$, we can find a $\eta_{m}\in\Omega_{\mathrm{c}}^{k}(\Sigma,\bb{C})$ for all $m\in\bb{N}$ such that $\Vert\omega_{m}-\eta_{m}\Vert_{\sf{H}^{2m}}\leq\frac{1}{m}$. It follows that
	\begin{align*}
		\Vert\eta_{m}-\omega\Vert_{\sf{H}^{s}}&\leq \Vert\eta_{m}-\omega_{m}\Vert_{\sf{H}^{s}}+\Vert\omega_{m}-\omega\Vert_{\sf{H}^{s}}\leq \Vert\eta_{m}-\omega_{m}\Vert_{\sf{H}^{2m}}+\Vert\omega_{m}-\omega\Vert_{\sf{H}^{s}}\leq\\&\leq \frac{1}{m}+\Vert\omega_{m}-\omega\Vert_{\sf{H}^{s}}\xrightarrow{m\to\infty} 0
\end{align*}	 
which shows that $(\eta_{m})_{m\in\bb{N}}\in\Omega^{k}_{\mathrm{c}}(\Sigma,\bb{C})^{\bb{N}}$ converges to $\omega$ in $\sf{H}^{s}_{k}(\Sigma)$.
\end{proof}

\subsection{Hodge Decomposition and Poisson Equation}\label{Sec:HodgeDec}
As in the previous section, let $(\Sigma,\sf{h})$ be a complete, connected and oriented Riemannian manifold and consider the Sobolev spaces $\sf{H}^{s}_{k}(\Sigma)$ of order $s\in [0,\infty]$ with $k\in\bb{N}_{0}$. We shall denote the Hilbert space adjoint of a densely-defined linear operators $\sf{A}\:\mathrm{dom}(\sf{A})\to\sf{H}^{s}_{k}(\Sigma)$ by $\sf{A}^{\dagger}\:\mathrm{dom}(\sf{A}^{\dagger})\to\sf{H}^{s}_{k}(\Sigma)$. Furthermore, we  introduce the following notation:

\begin{definition} For $s\in [0,\infty]$ and $k\in\bb{N}$, we denote the linear subspace of \emph{smooth} $k$-forms in the Sobolev space $\sf{H}^{s}_{k}(\Sigma)$ by 
\begin{align*}
	\Omega^{k}_{s}(\Sigma):=\sf{H}^{s}_{k}(\Sigma)\cap\Omega^{k}(\Sigma,\bb{C})\, .
\end{align*}
\end{definition}

By Proposition~\ref{Prop:Sob2}(ii), we observe that the subspace $\Omega^{k}_{s}(\Sigma)$ is dense in the Sobolev space $\sf{H}^{k}_{s}(\Sigma)$, since $\Omega^{k}_{\mathrm{c}}(\Sigma,\bb{C})\subset\Omega^{k}_{s}(\Sigma)\subset\sf{H}^{k}_{s}(\Sigma)$. Furthermore, following the same logic, we observe that the space $\Omega_{\infty}^{k}(\Sigma)$ is dense in $\sf{H}^{k}_{s}(\Sigma)$ for all $s\in [0,\infty]$.

Before we are able to establish a (weak) Hodge-type decomposition for the space $\sf{H}^{s}_{k}(\Sigma)$ (and its linear subspace $\Omega^{k}_{s}(\Sigma)$), we first need show the following proposition:

\begin{proposition}\label{Prop:dinSob} \emph{(Exterior Derivative and Codifferential in Sobolev Spaces)}\newline
	The operators $\d_{\Sigma}$ and codifferential and $\delta_{\Sigma}$ are well-defined and closable as linear operators
	\begin{align*}
		\d_{\Sigma}\:\mathrm{dom}(\d_{\Sigma})\to\sf{H}^{s}_{\bullet+1}(\Sigma)\,,\qquad \delta_{\Sigma}\:\mathrm{dom}(\delta_{\Sigma})\to\sf{H}^{s}_{\bullet-1}(\Sigma)
	\end{align*}
	with dense domain $\mathrm{dom}(\d_{\Sigma})=\mathrm{dom}(\delta_{\Sigma})=\Omega^{\bullet}_{\infty}(\Sigma)$ and their closures, which we denote by 
	\begin{align*}
		\overline{\d}_{\Sigma}\:\mathrm{dom}_{s}(\overline{\d}_{\Sigma})\to\sf{H}^{s}_{\bullet+1}(\Sigma)\,,\qquad \delta_{\Sigma}\:\mathrm{dom}_{s}(\overline{\delta}_{\Sigma})\to\sf{H}^{s}_{\bullet-1}(\Sigma)\,,
	\end{align*}
	are $\sf{H}^{s}$-adjoints of each other, i.e.~$\overline{\d}_{\Sigma}^{\dagger}=\overline{\delta}_{\Sigma}$.
\end{proposition}

In order to prove this proposition, we first need to establish some technical preliminary results. First of all, we show that $\d_{\Sigma}$ and $\delta_{\Sigma}$ are not only formal $\sf{L}^{2}$-adjoints, but in fact also formal adjoints in $\sf{H}^{s}_{k}(\Sigma)$ in a suitable sense.

\begin{lemma}\label{Lemma:FormalAdjointSobolev}
	Let $s\in [0,\infty)$. For every $\alpha\in\Omega^{\bullet}_{s}(\Sigma)$ and $\beta\in\Omega^{\bullet+1}_{s}(\Sigma)$ with the property that $\d_{\Sigma}\alpha\in\Omega^{\bullet+1}_{s}(\Sigma)$ and $\delta_{\Sigma}\beta\in\Omega^{\bullet}_{s}(\Sigma)$, it holds that
 \begin{align*}
		\langle\d_{\Sigma}\alpha,\beta\rangle_{\sf{H}^{s}}=\langle\alpha,\delta_{\Sigma}\beta\rangle_{\sf{H}^{s}}\, .
	\end{align*}
\end{lemma}

\begin{proof}
	We recall some general facts about $\d_{\Sigma}$ and $\delta_{\Sigma}$ viewed as operators in $\sf{L}^{2}_{\bullet}(\Sigma)$.
	\begin{itemize}
		\item[(i)]Let us view $\d_{\Sigma}$ and $\delta_{\Sigma}$ as densely-defined operators of the form $\d_{\Sigma}\:\Omega^{\bullet}_{\mathrm{c}}(\Sigma,\bb{C})\to\sf{L}^{2}_{\bullet+1}(\Sigma)$ and  $\delta_{\Sigma}\:\Omega^{\bullet+1}_{\mathrm{c}}(\Sigma,\bb{C})\to\sf{L}^{2}_{\bullet}(\Sigma)$. Now, these operators are, of course, formal adjoints of each other in $\sf{L}^{2}_{\bullet}(\Sigma)$. Now, as it turns out, their closures, which we shall denote for the moment by $\widetilde{\d}_{\Sigma}\:\mathrm{dom}(\widetilde{\d}_{\Sigma})\to\sf{L}^{2}_{\bullet+1}(\Sigma)$ and $\widetilde{\delta}_{\Sigma}\:\mathrm{dom}(\widetilde{\delta}_{\Sigma})\to\sf{L}^{2}_{\bullet}(\Sigma)$ to distinguish them from the $\sf{H}^{s}$-closures $\overline{\d}_{\Sigma}$ and $\overline{\delta}_{\Sigma}$ considered in Proposition~\ref{Prop:dinSob} above, are $\sf{L}^{2}$-adjoints of each other, i.e.~$\widetilde{\d}_{\Sigma}^{\dagger}=\widetilde{\delta}_{\Sigma}$ (see e.g.~\cite{Gaffney1951} or \cite[Lemma~3.3]{BruningLesch}), assuming $(\Sigma,\sf{h})$ to be complete. Now, the fact that $\widetilde{\d}_{\Sigma}^{\dagger}=\widetilde{\delta}_{\Sigma}$ is equivalent to saying that the minimal and maximal closed extension (cf.~Section~\ref{Sec:LinOp}) of $\d_{\Sigma}\:\Omega^{\bullet}_{\mathrm{c}}(\Sigma,\bb{C})\to\sf{L}^{2}_{\bullet+1}(\Sigma)$ agree. In particular, we note that
		\begin{align*}
			\mathrm{dom}(\widetilde{\d}_{\Sigma})=\{\omega\in\sf{L}^{2}_{k}(\Sigma)\mid \d_{\Sigma}\omega\in\sf{L}^{2}_{k+1}(\Sigma)\text{ in the distributional sense }\}\, .
		\end{align*}
		As a consequence, it follows that $\Omega^{k}(\Sigma,\bb{C})\cap\sf{L}^{2}_{k}(\Sigma)\subset \mathrm{dom}(\widetilde{\d}_{\Sigma})$. A similar statement is true for $\widetilde{\delta}_{\Sigma}$, i.e.~$\Omega^{k}(\Sigma,\bb{C})\cap\sf{L}^{2}_{k}(\Sigma)\subset \mathrm{dom}(\widetilde{\delta}_{\Sigma})$.
		\item[(ii)]On smooth forms, it is well known that $\d_{\Sigma}\Delta=\Delta\d_{\Sigma}$ and $\delta_{\Sigma}\Delta=\Delta\delta_{\Sigma}$. Now, one can ask the question whether this is true also for the corresponding closures $\widetilde{\d}_{\Sigma}$, $\widetilde{\delta}_{\Sigma}$ and $\overline{\Delta}$ (recall that $\overline{\Delta}$ is the $\sf{L}^{2}$-closure of $\Delta=\Omega^{k}_{\mathrm{c}}(\Sigma,\bb{C})\to\sf{L}^{2}_{\bullet}(\Sigma)$). This is of course, purely a question on the domains and using techniques from spectral calculus, one can show that 
		\begin{align*}
		\sf{E}^{s}\widetilde{\d}_{\Sigma}&=\widetilde{\d}_{\Sigma}\sf{E}^{s},\qquad \text{on }\quad\mathcal{D}(\sf{E}^{s})\cap\mathcal{D}(\sf{E}^{s}\widetilde{\d}_{\Sigma})\subset\mathcal{D}(\widetilde{\d}_{\Sigma}\sf{E}^{s})\\
		\sf{E}^{s}\widetilde{\delta}_{\Sigma}&=\widetilde{\delta}_{\Sigma}\sf{E}^{s},\qquad \text{on }\quad\mathcal{D}(\sf{E}^{s})\cap\mathcal{D}(\sf{E}^{s}\widetilde{\delta}_{\Sigma})\subset\mathcal{D}(\widetilde{\delta}_{\Sigma}\sf{E}^{s})\,.
	\end{align*}
	We refer to \cite[Lemma 5.9.]{MurroProca} for details on this fact.
	\end{itemize}
	With the previous two facts, the claim follows from a straightforward computation: let $\alpha,\beta$ be as in the formulation of the lemma. Then,
	\begin{align*}
		\langle \alpha,\delta_{\Sigma}\beta\rangle_{\sf{H}^{s}}&=\langle\sf{E}^{\frac{s}{2}}\alpha,\sf{E}^{\frac{s}{2}}\delta_{\Sigma}\beta\rangle_{\sf{L}^{2}}=\langle\sf{E}^{\frac{s}{2}}\alpha,\widetilde{\delta}_{\Sigma}\sf{E}^{\frac{s}{2}}\beta\rangle_{\sf{L}^{2}}=\\&=\langle\widetilde{\d}_{\Sigma}\sf{E}^{\frac{s}{2}}\alpha,\sf{E}^{\frac{s}{2}}\beta\rangle_{\sf{L}^{2}}=\langle\sf{E}^{\frac{s}{2}}\d_{\Sigma}\alpha,\sf{E}^{\frac{s}{2}}\beta\rangle_{\sf{L}^{2}}=\langle \d_{\Sigma}\alpha,\beta\rangle_{\sf{H}^{s}}\,,
	\end{align*}
	which concludes the proof.	
\end{proof}

As a next step, we need to show that $\d_{\Sigma}$ and $\delta_{\Sigma}$ are well-defined in $\sf{H}^{s}_{\bullet}(\Sigma)$ when using the domain $\Omega^{\bullet}_{\infty}(\Sigma)$. Since $\d_{\Sigma}$ and $\delta_{\Sigma}$ are first-order differential operators one expects, of course, that they map any Sobolev space of order $s\geq 1$ into the Sobolev space of order $s-1$. The following lemma shows that this is indeed true for our choice of Sobolev spaces:

\begin{lemma}\label{Lemma:Tec2}
	Let $s\in [1,\infty)$. Then, $\d_{\Sigma}$ and $\delta_{\Sigma}$ are well-defined as operators 
	\begin{align*}
		\d_{\Sigma}\:\Omega^{\bullet}_{s}(\Sigma)\to\Omega^{\bullet+1}_{s-1}(\Sigma)\,,\qquad
		\delta_{\Sigma}\:\Omega^{\bullet+1}_{s}(\Sigma)\to\Omega^{\bullet}_{s-1}(\Sigma)\, .
	\end{align*}
\end{lemma}

\begin{proof}
	We show the claim only for $\d_{\Sigma}$, since the the proof for $\delta_{\Sigma}$ can be done analogously. First of all, we claim that the following inequality holds true for all $\varphi\in\Omega^{\bullet}_{\mathrm{c}}(\Sigma,\bb{C})$:
	\begin{align}\label{eq:Ineq}
		\Vert\d_{\Sigma}\varphi\Vert_{\sf{H}^{s-1}}\leq\Vert\varphi\Vert_{\sf{H}^{s}}\, .
	\end{align}
	Indeed, using Lemma~\ref{Lemma:FormalAdjointSobolev}, Proposition~\ref{Prop:SobProp}(iv) and the fact that $\sf{E}$ is self-adjoint in $\sf{L}^{2}_{\bullet}(\Sigma)$, a straightforward computation shows
	\begin{align*}
		\Vert\varphi\Vert_{\sf{H}^{s}}^{2}&=\langle\sf{E}^{\frac{s}{2}}\varphi,\sf{E}^{\frac{s}{2}}\varphi\rangle_{\sf{L}^{2}}=\langle\sf{E}^{\frac{s-1}{2}}\varphi,\sf{E}^{\frac{s-1}{2}}\sf{E}\varphi\rangle_{\sf{L}^{2}}=\langle\varphi,\sf{E}\varphi\rangle_{\sf{H}^{s-1}}=\Vert\varphi\Vert_{\sf{H}^{s-1}}^{2}+\langle\varphi,\Delta\varphi\rangle_{\sf{H}^{s}}\\&=\Vert\varphi\Vert_{\sf{H}^{s-1}}^{2}+\Vert\d_{\Sigma}\varphi\Vert_{\sf{H}^{s-1}}^{2}+\Vert\delta_{\Sigma}\varphi\Vert_{\sf{H}^{s-1}}^{2}\geq\Vert\d_{\Sigma}\varphi\Vert_{\sf{H}^{s-1}}^{2}\,.
	\end{align*}
	Now, let $\omega\in\Omega^{k}_{s}(\Sigma)$. Since $\Omega^{k}_{\mathrm{c}}(\Sigma,\bb{C})$ is dense in $\sf{H}^{s}_{k}(\Sigma)$ (cf.~Proposition~\ref{Prop:Sob2}(ii)), we can find a sequence $(\omega_{n})_{n\in\bb{N}}$ in $\Omega^{k}_{\mathrm{c}}(\Sigma,\bb{C})$ such that $\omega=\sf{H}^{s}\text{-}\lim_{n\to\infty}\omega_{n}$. By~\eqref{eq:Ineq}, it follows that $(\d_{\Sigma}\omega_{n})_{n\in\bb{N}}$ converges in $\sf{H}^{s-1}_{k+1}(\Sigma)$, i.e.~there exists $\eta\in\sf{H}^{s-1}_{k+1}(\Sigma)$ such that $\eta=\sf{H}^{s-1}\text{-}\lim_{n\to\infty}\d_{\Sigma}\omega_{n}$. We claim that $\eta=\d_{\Sigma}\omega$. First of all, Proposition~\ref{Prop:Sob}(ii), we know that also $\omega=\sf{L}^{2}\text{-}\lim_{n\to\infty}\omega_{n}$ and $\eta=\sf{L}^{2}\text{-}\lim_{n\to\infty}\d_{\Sigma}\omega_{n}$. Hence, for all $\varphi\in\Omega^{k+1}_{\mathrm{c}}(\Sigma,\bb{C})$ it holds that 
\begin{align*}
		&\langle\d_{\Sigma}\omega_{n},\varphi\rangle_{\sf{L}^{2}}\xrightarrow{n\to\infty}\langle\eta,\varphi\rangle_{\sf{L}^{2}}\\
		&\hspace*{0.8cm}\verteq\\
		&\langle\omega_{n},\delta_{\Sigma}\varphi\rangle_{\sf{L}^{2}}\xrightarrow{n\to\infty}\langle\omega,\delta\varphi\rangle_{\sf{L}^{2}}=\langle\d_{\Sigma}\omega,\varphi\rangle_{\sf{L}^{2}}
	\end{align*}
and hence $\langle\eta-\d_{\Sigma}\omega,\varphi\rangle_{\sf{L}^{2}}=0$. We conclude that $\eta=\d_{\Sigma}\omega$, by non-degeneracy of $\langle\cdot,\cdot\rangle_{\sf{L}^{2}}$ and the fact that the space of test sections $\Omega^{k+1}_{\mathrm{c}}(\Sigma,\bb{C})$ is dense in $\sf{L}^{2}_{k+1}(\Sigma)$.
\end{proof}

Before being able to prove Proposition~\ref{Prop:dinSob} we need one last ingredient. Recall the mollifiers $\mathcal{J}_{\varepsilon}:= e^{-\varepsilon\sf{E}}\:\sf{L}^{2}_{k}(\Sigma)\to\sf{L}^{2}_{k}(\Sigma)$ for $\varepsilon>0$ as defined in Lemma~\ref{Lemma:Mollifiers}. Then, the following holds true:

\begin{lemma}\label{Lemma:TecMol}
	The mollifiers $\mathcal{J}_{\varepsilon}:= e^{-\varepsilon\sf{E}}\:\sf{L}^{2}_{\bullet}(\Sigma)\to\sf{L}^{2}_{\bullet}(\Sigma)$ satisfy
	\begin{align*}
		\mathcal{J}_{\varepsilon}\d_{\Sigma}=\d_{\Sigma}\mathcal{J}_{\varepsilon}
	\end{align*}
	on the space $\{\omega\in\sf{L}^{2}_{\bullet}(\Sigma)\cap\Omega^{k}(\Sigma,\bb{C})\mid\d_{\Sigma}\omega\in\sf{L}^{2}_{\bullet}(\Sigma)\}$ and similarly for $\delta_{\Sigma}$.
\end{lemma}

\begin{proof}
	First of all, we use the following trick: observe that $\mathcal{J}_{\varepsilon}$ can equivalently be defined by spectral calculus of the \emph{bounded} operator $\sf{E}^{-1}$ by 
	\begin{align*}
		\mathcal{J}_{\varepsilon}=g_{\varepsilon}(\sf{E}^{-1}),\hspace*{1cm} g_{\varepsilon}(\lambda):= \begin{cases} e^{-\frac{\varepsilon}{\lambda}}, &\lambda> 0\\
		0, &\text{else}\end{cases}\, .
	\end{align*}
	Note that $g_{\varepsilon}$ is bounded and continuous on $\sigma(\sf{E}^{-1})\subset [0,1]$. The fact that this indeed yields the same operator can easily be seen from uniqueness of the spectral calculus. Now, if $\sf{A}\:\mathrm{dom}(\sf{A})\to\mathcal{H}$ is any closed and densely-defined operator in some Hilbert space $\mathcal{H}$ and $\sf{B}\:\mathcal{H}\to\mathcal{H}$ bounded and self-adjoint, then $\sf{B}\sf{A}=\sf{A}\sf{B}$ on $\mathrm{dom}(\sf{A})$ implies $f(\sf{B})\sf{A}=\sf{A} f(\sf{B})$ on $\mathrm{dom}(\sf{A})\cap\mathrm{dom}(\sf{A}f(\sf{B}))$ for any bounded and Borel-measurable function $f$ defined on $\sigma(\sf{B})$. Now, in our case, it holds that $\sf{E}^{-1}\widetilde{\d}_{\Sigma}=\widetilde{\d}_{\Sigma}\sf{E}^{-1}$ on $\mathrm{dom}(\widetilde{\d}_{\Sigma})$, as explained in the proof of Lemma~\ref{Lemma:FormalAdjointSobolev}, where $\sf{E}^{-1}$ is a bounded operator (cf.~Proposition~\ref{Prop:SobProp}(ii)) and where $\widetilde{\d}_{\Sigma}$ denotes the $\sf{L}^{2}$-closure of $\d_{\Sigma}\:\Omega^{\bullet}_{\mathrm{c}}(\Sigma,\bb{C})\to\sf{L}^{2}_{\bullet+1}(\Sigma)$, as in the proof of Lemma~\ref{Lemma:FormalAdjointSobolev}. In particular, we conclude that $\mathcal{J}_{\varepsilon}\widetilde{\d}_{\Sigma}=\widetilde{\d}_{\Sigma}\mathcal{J}_{\varepsilon}$ on $\mathrm{dom}(\widetilde{\d}_{\Sigma})\cap\mathrm{dom}(\widetilde{\d}_{\Sigma}\mathcal{J}_{\varepsilon})$. The claim then follows from the fact that any $\omega\in\sf{L}^{2}_{\bullet}(\Sigma)\cap\Omega^{k}(\Sigma,\bb{C})$ with $\d_{\Sigma}\omega\in\sf{L}^{2}_{\bullet}(\Sigma)$ is contained in $\mathrm{dom}(\widetilde{\d}_{\Sigma})$ and Lemma~\ref{Lemma:Mollifiers}(ii).
\end{proof}

After this preliminary discussion, we are now able to prove Proposition~\ref{Prop:dinSob}.

\begin{proof}[Proof of Proposition~\ref{Prop:dinSob}.]
	First of all, by Lemma~\ref{Lemma:Tec2}, we now that $\d_{\Sigma}$ and $\delta_{\Sigma}$ are well-defined as operators 
	\begin{align*}
		\d_{\Sigma}\:\mathrm{dom}(\d_{\Sigma})\to\sf{H}^{s}_{\bullet+1}(\Sigma)\,,\qquad \d_{\Sigma}\:\mathrm{dom}(\d_{\Sigma})\to\sf{H}^{s}_{\bullet-1}(\Sigma)\, .
	\end{align*}
	Furthermore, Lemma~~\ref{Lemma:FormalAdjointSobolev} implies that $\d_{\Sigma}$ and $\delta_{\Sigma}$ with these domains are \emph{closable}, since $\mathrm{dom}(\delta_{\Sigma})\subset\mathrm{dom}(\overline{\d}_{\Sigma}^{\dagger})$ and $\mathrm{dom}(\d_{\Sigma})\subset\mathrm{dom}(\overline{\delta}_{\Sigma}^{\dagger})$, which in particular implies that the adjoints $\overline{\d}_{\Sigma}^{\dagger}$ and $\overline{\delta}_{\Sigma}^{\dagger}$ have a dense domain.
	
	Let us now turn to the operator equality $\overline{\d}_{\Sigma}^{\dagger}=\overline{\delta}_{\Sigma}$. First, note that $\overline{\delta}_{\Sigma}^{\dagger}=\overline{\d}_{\Sigma}$ is equivalent to $\delta_{\Sigma}^{\dagger}=\overline{\d}_{\Sigma}$, since $(\overline{\sf{A}})^{\dagger}=\overline{\sf{A}^{\dagger}}=\sf{A}^{\dagger}$ for any densely-defined closable operator $\sf{A}$ on a Hilbert space. To show $\overline{\d}_{\Sigma}\subset\delta_{\Sigma}^{\dagger}$, let $\omega\in\mathrm{dom}_{s}(\overline{\d}_{\Sigma})$. By definition of the closure, there exists a sequence $\{\omega_{n}\}_{n\in\bb{N}}\in\Omega^{\bullet}_{\infty}(\Sigma)$ such that 
	\begin{align*}
		\omega=\sf{H}^{s}\text{-}\lim_{n\to\infty}\omega_{n}\hspace*{1cm}\text{and}\hspace*{1cm}\overline{\d}_{\Sigma}\omega=\sf{H}^{s}\text{-}\lim_{n\to\infty}\d_{\Sigma}\omega_{n}\, .
	\end{align*}
	As a consequence, for all test forms $\varphi\in\Omega^{\bullet}_{\infty}(\Sigma)$, it holds that 
	\begin{align*}
		&\langle\d_{\Sigma}\omega_{n},\varphi\rangle_{\sf{H}^{s}}\xrightarrow{n\to\infty}\langle\overline{\d}_{\Sigma}\omega,\varphi\rangle_{\sf{H}^{s}}\\
		&\hspace*{0.8cm}\verteq\\
		&\langle\omega_{n},\delta_{\Sigma}\varphi\rangle_{\sf{H}^{s}}\xrightarrow{n\to\infty}\langle\omega,\delta_{\Sigma}\varphi\rangle_{\sf{H}^{s}}
	\end{align*}
	where we used Lemma \ref{Lemma:FormalAdjointSobolev}. We conclude that $\langle\omega,\delta_{\Sigma}\varphi\rangle_{\sf{H}^{s}}=\langle\overline{\d}_{\Sigma}\omega,\varphi\rangle_{\sf{H}^{s}}$ and hence $\omega\in\mathrm{dom}(\delta^{\dagger}_{\Sigma})$ as well as $\delta_{\Sigma}^{\dagger}\omega=\overline{\d}_{\Sigma}\omega$. It is left to show $\mathrm{dom}(\delta_{\Sigma}^{\dagger})\subset\mathrm{dom}_{s}(\overline{\d}_{\Sigma})$. Let $\omega\in\mathrm{dom}(\delta_{\Sigma}^{\dagger})$. To show that $\omega\in\mathrm{dom}_{s}(\overline{\d}_{\Sigma})$, it is enough to show that there exists a sequence $\{\omega_{n}\}_{n\in\bb{N}}\in\Omega^{\bullet}_{\infty}(\Sigma)$ such that 
	\begin{align*}
		\omega=\sf{H}^{s}\text{-}\lim_{n\to\infty}\omega_{n}\hspace*{1cm}\text{and}\hspace*{1cm}(\d_{\Sigma}\omega_{n})_{n}\text{ is convergent in }\sf{H}^{s}_{\bullet}(\Sigma)\, ,
	\end{align*}
	because then it follows that $\omega\in\mathrm{dom}_{s}(\overline{\d}_{\Sigma})$ as well as $\sf{H}^{s}\text{-}\lim_{n\to\infty}\d_{\Sigma}\omega_{n}=\overline{\d}_{\Sigma}\omega=\delta^{\dagger}_{\Sigma}\omega$. Consider the class of mollifiers introduces in Lemma \ref{Lemma:Mollifiers}. Now, we argue that for $\omega\in\mathrm{dom}(\delta^{\dagger}_{\Sigma})$, it holds that $\d_{\Sigma}\mathcal{J}_{\varepsilon}\omega=\mathcal{J}_{\varepsilon}\delta_{\Sigma}^{\dagger}\omega$. By Lemma~\ref{Lemma:Mollifiers}(i) and (ii), it holds that $\mathcal{J}_{\varepsilon}\omega\in\Omega^{\bullet}_{\infty}(\Sigma)$. Now, let $\varphi\in\Omega^{\bullet}_{\infty}(\Sigma)$ be an arbitrary test form. Then
	  \begin{align*}
	  	\langle\d_{\Sigma}\mathcal{J}_{\varepsilon}\omega,\varphi\rangle_{\sf{H}^{s}}=\langle\mathcal{J}_{\varepsilon}\omega,\delta_{\Sigma}\varphi\rangle_{\sf{H}^{s}}=\langle\omega,\mathcal{J}_{\varepsilon}\delta_{\Sigma}\varphi\rangle_{\sf{H}^{s}}=\langle\omega,\delta_{\Sigma}\mathcal{J}_{\varepsilon}\varphi\rangle_{\sf{H}^{s}}=\langle\delta_{\Sigma}^{\dagger}\omega,\mathcal{J}_{\varepsilon}\varphi\rangle_{\sf{H}^{s}}=\langle\mathcal{J}_{\varepsilon}\delta^{\dagger}_{\Sigma}\omega,\varphi\rangle_{\sf{H}^{s}}\, ,
	  \end{align*}
	where we used Lemma~\ref{Lemma:FormalAdjointSobolev}, the fact that $\mathcal{J}_{\varepsilon}$ is self-adjoint in $\sf{L}^{2}$ (cf.~Lemma~\ref{Lemma:Mollifiers}(i)) as well as Lemma~\ref{Lemma:TecMol}. Now, by non-degeneracy of $\langle\cdot,\cdot\rangle_{\sf{H}^{s}}$ and the fact that the space of test forms $\Omega^{s}_{\infty}(\Sigma)$ is dense in $\sf{H}^{s}_{\bullet}(\Sigma)$, it follows that $\d_{\Sigma}\mathcal{J}_{\varepsilon}\omega=\mathcal{J}_{\varepsilon}\delta^{\dagger}_{\Sigma}\omega$ as claimed and hence $\d_{\Sigma}\mathcal{J}_{\varepsilon}\omega\to\delta_{\Sigma}^{\dagger}\omega$ as $\varepsilon\to 0^{+}$ by Lemma~\ref{Lemma:Mollifiers}(iii).	
\end{proof}

As a next step, we prove that there exists a (weak) Hodge-type decomposition for the Sobolev spaces $\sf{H}^{s}_{\bullet}(\Sigma)$ of differential forms. We start be recalling the following general result on  Hodge-type decompositions established by Brüning-Lesch in \cite{BruningLesch}.

\begin{lemma}\label{Lemma:WeakHodge} \emph{(Weak Hodge Decomposition \cite[Lem.~2.1]{BruningLesch})}\newline
Let $(\mathcal{H}_{i},\langle\cdot, \cdot\rangle_{i})_{i=0,\dots,n}$ be a family of Hilbert spaces and set $\mathcal{H}_{n+1}:=\{0\}$. Furthermore, consider a family $(\sf{D}_{i}\:\mathrm{dom}(\sf{D}_{i})\to\mathcal{H}_{i+1})_{i=0,\dots,n}$ of closed operators with dense domain $\mathrm{dom}(\sf{D}_{i})\subset\mathcal{H}_{i}$. If they form a cochain complex
\begin{align*}
	0\xrightarrow{}\mathrm{dom}(\sf{D}_{0})\xrightarrow{\sf{D}_{0}}\mathrm{dom}(\sf{D}_{1})\xrightarrow{\sf{D}_{1}}\mathrm{dom}(\sf{D}_{2})\xrightarrow{\sf{D}_{2}}\dots\xrightarrow{\sf{D}_{n-1}}\mathrm{dom}(\sf{D}_{n})\xrightarrow{}0\,,
\end{align*}
i.e.~$\mathrm{ran}(\sf{D}_{i})\subset\mathrm{dom}(\sf{D}_{i+1})$ and $\sf{D}_{i+1}\circ\sf{D}_{i}=0$, then there is a $\mathcal{H}_{i}$-orthogonal decomposition
\begin{align*}
	\mathcal{H}_{i}\cong\overline{\mathrm{ran}(\sf{D}_{i-1})}\oplus\overline{\mathrm{ran}(\sf{D}^{\dagger}_{i+1})}\oplus\mathrm{Har}_{i}\qquad\text{with}\qquad\mathrm{Har}_{i}:=\mathrm{ker}(\sf{D}_{i})\cap\mathrm{ker}(\sf{D}^{\dagger}_{i})\, ,
\end{align*}
where $\sf{D}_{i+1}^{\dagger}\:\mathrm{dom}(\sf{D}^{\dagger}_{i+1})\to\mathcal{H}_{i}$ denotes the adjoint of $\sf{D}_{i}$ with respect to $\mathcal{H}_{i}$ and $\mathcal{H}_{i+1}$.
\end{lemma}

\begin{proof}
	Since $\sf{D}_{i}$ is a closed operator, its kernel $\mathrm{ker}(\sf{D}_{i})$ is a closed subset of $\mathcal{H}_{i}$ and by the projection theorem of Hilbert spaces we obtain a $\mathcal{H}_{i}$-orthogonal decomposition
	\begin{align*}
		\mathcal{H}_{i}\cong\mathrm{ker}(\sf{D}_{i})^{\perp}\oplus\mathrm{ker}(\sf{D}_{i})\, .
	\end{align*}
	Furthermore, since $\mathrm{ker}(\sf{D}_{i})$ is closed, it is a Hilbert space on its own right and we can choose the closed subset $\overline{\mathrm{ran}(\sf{D}_{i-1})}$ of $\mathrm{ker}(\sf{D}_{i})$ to further decompose
	\begin{align*}
		\mathcal{H}_{i}&\cong\mathrm{ker}(\sf{D}_{i})^{\perp}\oplus\overline{\mathrm{ran}(\sf{D}_{i-1})}\oplus \overline{\mathrm{ran}(\sf{D}_{i-1})}^{\perp_{\mathrm{ker}}}\\&\cong\mathrm{ker}(\sf{D}_{i})^{\perp}\oplus\overline{\mathrm{ran}(\sf{D}_{i-1})}\oplus (\overline{\mathrm{ran}(\sf{D}_{i-1})}^{\perp}\cap\mathrm{ker}(\sf{D}_{i}))\, ,
	\end{align*}
	where by $\perp_{\mathrm{ker}}$ we denote the orthogonal complement in the Hilbert space $\mathrm{ker}(\sf{D}_{i})$. The claim then follows from the fact that $\mathrm{ker}(\sf{A})^{\perp}=\overline{\mathrm{ran}(\sf{A}^{\dagger})}$ for any closed densely-defined operator $\sf{A}$.
\end{proof}

\begin{remark}\label{Remark:HodgeTerminology}
	We should briefly clarify the terminology: a Hodge decomposition of the type in Lemma~\ref{Lemma:WeakHodge} is referred to as \emph{weak} if it contains the closures of the spaces $\mathrm{ran}(\sf{D}_{i-1})$ and $\mathrm{ran}(\sf{D}^{\dagger}_{i+1})$. If the decomposition of $\mathcal{H}_{i}$ can also be written without the closures, the corresponding decomposition is called a \emph{strong} Hodge decomposition. 
\end{remark}

We now show that the operators $\overline{\d}_{\Sigma}$ form a Hilbert complex, which then allows us to derive a weak Hodge decomposition as a special case of the previous lemma:

\begin{lemma}\label{Lemma:HilbertComplexSob}
	The sequence
	\begin{align*}
		0\xrightarrow{}\mathrm{dom}_{s}(\overline{\d}_{\Sigma})\subset\sf{H}_{0}^{s}(\Sigma)\xrightarrow{\overline{\d}_{\Sigma}}\mathrm{dom}_{s}(\overline{\d}_{\Sigma})\subset\sf{H}_{1}^{s}(\Sigma)\xrightarrow{\overline{\d}_{\Sigma}}\dots\xrightarrow{\overline{\d}_{\Sigma}}\mathrm{dom}_{s}(\overline{\d}_{\Sigma})\subset\sf{H}_{n}^{s}(\Sigma)\xrightarrow{\overline{\d}_{\Sigma}}0
	\end{align*}
	is a well-defined cochain complex, i.e.~$\mathrm{ran}(\overline{\d}_{\Sigma})\subset\mathrm{dom}_{s}(\overline{\d}_{\Sigma})$ as well as $\overline{\d}_{\Sigma}\circ\overline{\d}_{\Sigma}=0$. 
\end{lemma}

\begin{proof}
	Consider $\omega\in\mathrm{dom}_{s}(\overline{\d}_{\Sigma})\subset\sf{H}^{s}_{\bullet}(\Sigma)$. By definition, this means that there is a sequence $(\omega_{n})_{n\in\bb{N}}$ in $\Omega^{\bullet}_{\infty}(\Sigma)$ such that $\sf{H}^{s}\text{-}\lim_{n\to\infty}\omega_{n}=\omega$ and such that $(\d_{\Sigma}\omega_{n})_{n\in\bb{N}}$ is convergent in $\sf{H}^{s}_{\bullet+1}(\Sigma)$. Moreover, $\overline{\d}_{\Sigma}\omega=\sf{H}^{s}\text{-}\lim_{n\to\infty}\d_{\Sigma}\omega_{n}$. Now, consider the sequence $(\d_{\Sigma}\omega_{n})_{n}$. Clearly, also $(\d_{\Sigma}\d_{\Sigma}\omega_{n}=0)_{n}$ is converging, since it just the constant zero-sequence. Hence, $\overline{\d}_{\Sigma}\omega\in\mathrm{dom}_{s}(\overline{\d}_{\Sigma})$. Furthermore
	\begin{align*}
		\overline{\d}_{\Sigma}\overline{\d}_{\Sigma}\omega=\sf{H}^{s}\text{-}\lim_{n\to\infty}\d_{\Sigma}\d_{\Sigma}\omega_{n}=0\, ,
	\end{align*}
	by definition of the closure, which concludes the proof.
\end{proof}

\begin{theorem}\label{Thm:HodgeDecom} \emph{(Weak Hodge Decomposition in Sobolev Spaces)}\newline
	The Sobolev space $\sf{H}^{s}_{k}(\Sigma)$ admits the following $\sf{H}^{s}$-orthogonal decompositions:
	\begin{align*}
		\sf{H}^{s}_{k}(\Sigma)\cong \overline{\d_{\Sigma}\Omega^{k-1}_{\infty}(\Sigma)}^{\sf{H}^{s}}\oplus\overline{\delta_{\Sigma}\Omega^{k+1}_{\infty}(\Sigma)}^{\sf{H}^{s}}\oplus\mathrm{Har}^{s}_{k}(\Sigma)
	\end{align*}
	for all $s\in [0,\infty)$ and $k\in\bb{N}_{0}$, where $\mathrm{Har}^{s}_{k}(\Sigma):= \mathrm{ker}(\overline{\d}_{\Sigma})\cap\mathrm{ker}(\overline{\delta}_{\Sigma})=\mathrm{ker}(\Delta\vert_{\Omega^{k}_{\infty}})$.
\end{theorem}

\begin{proof}
	By Lemma~\ref{Lemma:HilbertComplexSob}, we know that $(\sf{H}^{s}_{\bullet}(\Sigma),\overline{\d}_{\Sigma})$ is a Hilbert complex. Furthermore, by Proposition~\ref{Prop:dinSob}, we know that the $\sf{H}^{s}$-adjoint of $\overline{\d}_{\Sigma}$ is given by $\overline{\delta}_{\Sigma}$. Hence, by Lemma~\ref{Lemma:WeakHodge}, there is a $\sf{H}^{s}$-orthogonal decomposition
	\begin{align*}
		\sf{H}^{s}_{k}(\Sigma)\cong \overline{\mathrm{ran}(\overline{\d}_{\Sigma})}^{\sf{H}^{s}}\oplus\overline{\mathrm{ran}(\overline{\delta}_{\Sigma})}^{\sf{H}^{s}}\oplus\mathrm{Har}^{s}_{k}(\Sigma)
	\end{align*}
	Now, if $\sf{A}\:\mathrm{dom}(\sf{A})\to\mathcal{H}$ is any densely-defined and closable operator in some Hilbert space $\mathcal{H}$, we claim that 
	\begin{align*}
		\overline{\mathrm{ran}(\sf{A})}=\overline{\mathrm{ran}(\overline{\sf{A}})}\, .
	\end{align*}
	The inclusion ``$\subset$'' follows trivially from the fact that $\overline{\sf{A}}$ is an extension of $\sf{A}$, which implies $\mathrm{ran}(\sf{A})=\mathrm{ran}(\overline{\sf{A}})$. For ``$\supset$'', we observe that $\mathrm{ran}(\overline{\sf{A}})\subset\overline{\mathrm{ran}(\sf{A})}$. Indeed, by definition of the closure, we recall that $\omega\in\mathrm{dom}(\overline{\sf{A}})$ if and only if there exists a sequence $(\omega_{n})_{n\in\bb{N}}$ such that $\omega=\lim_{n\to\infty}\omega_{n}$ and $\overline{\sf{A}}\omega=\lim_{n\to\infty}\sf{A}\omega_{n}$, which implies that $\overline{\sf{A}}\omega\in\overline{\mathrm{ran}(\sf{A})}$. Applying this fact to our operators, we obtain
	\begin{align*}
		\overline{\mathrm{ran}(\overline{\d}_{\Sigma})}^{\sf{H}^{s}}=\,\,\overline{\d_{\Sigma}\Omega^{\bullet-1}_{\infty}(\Sigma)}^{\sf{H}^{s}}\,\qquad \overline{\mathrm{ran}(\overline{\delta}_{\Sigma})}^{\sf{H}^{s}}=\,\,\overline{\delta_{\Sigma}\Omega^{\bullet+1}_{\infty}(\Sigma)}^{\sf{H}^{s}}\,,
	\end{align*}
	
	It remains to show $\mathrm{Har}^{s}_{k}(\sf{M})=\mathrm{ker}(\Delta\vert_{\Omega^{k}_{\infty}})$. For ``$\subset$'', let $\omega\in\mathrm{Har}_{k}^{s}(\Sigma)$, which means that $\omega\in\mathrm{dom}_{s}(\overline{\d}_{\Sigma})\cap\mathrm{dom}_{s}(\overline{\delta}_{\Sigma})$ and $\overline{\d}_{\Sigma}\omega=\overline{\delta}_{\Sigma}\omega=0$. We first show that $\omega$ is smooth: by assumption, there exists a sequence $(\omega_{n})_{n\in\bb{N}}$ in $\Omega^{k}_{\infty}(\Sigma)$ such that $\omega=\sf{H}^{s}\text{-}\lim_{n\to\infty}\omega_{n}$ and $\overline{\d}_{\Sigma}\omega=\sf{H}^{s}\text{-}\lim_{n\to\infty}\d_{\Sigma}\omega_{n}=0$. By Lemma~\ref{Prop:Sob}(i), the same convergence holds in the $\sf{L}^{2}$-topology, which implies
\begin{align*}
		&\langle\omega_{n},\delta_{\Sigma}\varphi\rangle_{\sf{L}^{2}}\xrightarrow{n\to\infty}\langle\omega,\delta_{\Sigma}\varphi\rangle_{\sf{L}^{2}}\\
		&\hspace*{0.8cm}\verteq\\
		&\langle\d_{\Sigma}\omega_{n},\varphi\rangle_{\sf{L}^{2}}\xrightarrow{n\to\infty}\langle\overline{\d}_{\Sigma}\omega,\varphi\rangle_{\sf{L}^{2}}=0
	\end{align*}
and hence $\langle\omega,\delta_{\Sigma}\varphi\rangle_{\sf{L}^{2}}=0$ for all test forms $\varphi\in\Omega^{k+1}_{\mathrm{c}}(\Sigma,\bb{C})$. Similarly, we show $\langle\omega,\d_{\Sigma}\varphi\rangle_{\sf{L}^{2}}=0$ for all $\varphi\in\Omega^{k-1}_{\mathrm{c}}(\Sigma,\bb{C})$. We conclude that $(\d_{\Sigma}+\delta_{\Sigma})\omega=0$ in the \emph{distributional sense} and hence $\omega\in\Omega^{k}(\Sigma,\bb{C})$ be elliptic regularity of the operator $\d_{\Sigma}+\delta_{\Sigma}$. It follows that $\omega\in\mathrm{ker}(\Delta)\cap\Omega^{k}_{s}(\Sigma)$. For ``$\supset$'', let $\omega\in\Omega^{k}_{s}(\Sigma)$ be such that $\Delta\omega=0$. Now, it is a well-known fact that $\Delta\eta=0$ for some $\eta\in\sf{L}^{2}_{k}(\Sigma)$ if and only if $\d_{\Sigma}\eta=\delta_{\Sigma}\eta=0$ on complete manifolds (see e.g.~\cite{Gaffney1954,Yau}), which is a direct consequence of the fact that the $\sf{L}^{2}$-closures of $\d_{\Sigma}$ and $\delta_{\Sigma}$ are $\sf{L}^{2}$-adjoints of each other (cf.~the proof of Lemma~\ref{Lemma:FormalAdjointSobolev}). We conclude that $\d_{\Sigma}\omega=0$ and $\delta_{\Sigma}\omega=0$. Hence, we have shown that $\ker(\Delta\vert_{\Omega^{k}_{s}})\subset\ker(\d_{\Sigma}\vert_{\Omega^{k}_{s}})\cap\ker(\delta_{\Sigma}\vert_{\Omega^{k}_{s}})$. It is left to show that 
\begin{align*}
	\ker(\d_{\Sigma}\vert_{\Omega^{k}_{s}})\cap\ker(\delta_{\Sigma}\vert_{\Omega^{k}_{s}})\subset\ker(\overline{\d}_{\Sigma})\cap\ker(\overline{\delta}_{\Sigma})\,.
\end{align*}
To show $\mathrm{ker}(\d_{\Sigma})\cap\Omega^{k}_{s}(\Sigma)\subset\ker(\overline{\d}_{\Sigma})$, we take $\omega\in\Omega^{k}_{s}(\Sigma)$ such that $\d_{\Sigma}\omega=0$ and we have to construct a sequence $(\omega_{n})_{n}$ in $\Omega_{\infty}^{k}(\Sigma)$ such that $\omega=\sf{H}^{s}\text{-}\lim_{n\to\infty}\omega_{n}$ and $\sf{H}^{s}\text{-}\lim_{n\to\infty}\d_{\Sigma}\omega_{n}=0$. We consider the mollifiers $\mathcal{J}_{\varepsilon}$ introduced in Lemma \ref{Lemma:Mollifiers} and notice that $\d_{\Sigma}\mathcal{J}_{\varepsilon}\omega=\mathcal{J}_{\varepsilon}\d_{\Sigma}\omega=0$ by Lemma~\ref{Lemma:TecMol} and $\mathcal{J}_{\varepsilon}\omega\in\Omega^{k}_{\infty}(\Sigma)$ by Lemma~\ref{Lemma:Mollifiers}(ii). Hence $\mathcal{J}_{\varepsilon}\omega$ gives the required sequence. Similarly, we show that $\mathrm{ker}(\delta_{\Sigma})\cap\Omega^{k}_{s}(\Sigma)\subset\ker(\overline{\delta}_{\Sigma})$.
\end{proof} 

\begin{remark} (The Hodge-Kodaira Decomposition)\newline
Consider the special case $s=0$, i.e.~$\sf{H}^{0}_{k}(\Sigma)=\sf{L}^{2}_{k}(\Sigma)$. Now, it is easy to show that in fact
\begin{align*}
	\overline{\d_{\Sigma}\Omega^{k-1}_{\infty}(\Sigma)}^{\Vert\cdot\Vert_{\sf{L}^{2}}}=\,\,\overline{\d_{\Sigma}\Omega^{k-1}_{\mathrm{c}}(\Sigma,\bb{C})}^{\Vert\cdot\Vert_{\sf{L}^{2}}}
\end{align*}
in this case, and similarly for $\delta_{\Sigma}$, e.g.~by using similar arguments as in the lecture notes \cite[Lem.~1.5]{Carron}. In particular, we obtain the $\sf{L}^{2}$-orthogonal decomposition
\begin{align*}
	\sf{L}^{2}_{k}(\Sigma)=\overline{\d_{\Sigma}\Omega^{k-1}_{\mathrm{c}}(\Sigma,\bb{C})}^{\Vert\cdot\Vert_{\sf{L}^{2}}}\oplus\overline{\delta_{\Sigma}\Omega^{k+1}_{\mathrm{c}}(\Sigma,\bb{C})}^{\Vert\cdot\Vert_{\sf{L}^{2}}}\oplus\mathrm{ker}(\Delta\vert_{\sf{L}^{2}_{k}\cap\Omega^{k}})
\end{align*}
as a special case of Theorem~\ref{Thm:HodgeDecom}, which is the well-known \emph{Hodge-Kodaira decomposition}, introduced by Kodaira in \cite{Kodaira} (see also the monograph \cite[Thm.~24]{deRhamBook} for a proof).
\end{remark}

\begin{remark}\label{Rem:StrongHodge} (Strong Hodge Decomposition)\newline
	Following the terminology briefly mentioned in Remark~\ref{Remark:HodgeTerminology}, we note that the Hodge decomposition in Theorem~\ref{Thm:HodgeDecom} is a \emph{weak Hodge decomposition}. One might ask the question under which assumptions on the manifold one obtains a \emph{strong} Hodge decomposition, which would be more similar in spirit to the Hodge decomposition on compact manifolds. An important result in this direction has been established by Gromov~\cite{Gromov}, in which a strong Hodge decomposition of the form
	\begin{align*}
		\sf{L}^{2}_{k}(\Sigma)=\d_{\Sigma}\sf{H}^{1}_{k-1}(\Sigma)\oplus\delta_{\Sigma}\sf{H}^{1}_{k+1}(\Sigma)\oplus\mathrm{ker}(\Delta\vert_{\sf{L}_{k}^{2}\cap\Omega^{k}})
	\end{align*}
	has been established under the (rather strong) assumption that all de Rham-Hodge Laplacians $\Delta$ have a \emph{spectral gap} (see also \cite{Li,Scott} for some extensions of this result to the $\sf{L}^{p}$-setting). Further results in this direction have been obtained by Cantor \cite{Cantor} on asymptotically Euclidean manifolds and by Lockhart \cite{Lockhart} on noncompact manifolds with finitely many ends and asymptotically translational invariant metrics, both for suitable \emph{weighted Sobolev spaces}. A strong Hodge decomposition in weighted $\sf{L}^{p}$-Sobolev spaces on submanifolds of $\bb{R}^{d}$ with compact boundaries has been established by Schwarz in \cite{Schwarz}. Last but not least, let us mention the recent preprint \cite{Czubak}, in which a strong Hodge-decomposition of $\sf{H}^{1}_{k}(\Sigma)$ is established for noncompact manifolds of nonpositive constant sectional curvature.
\end{remark}

As a direct consequence of elliptic regularity, we obtain a smooth version of the Hodge decomposition stated in Theorem~\ref{Thm:HodgeDecom}. Moreover, although this decomposition does not constitute a \emph{strong} Hodge decomposition as in Remark~\ref{Rem:StrongHodge}, we can still express its summands in terms of \emph{exact} and \emph{coexact forms} in a suitable sense. Before proving these two results, we require an additional tool: the \emph{Poincaré duality}, or more precisely, a corollary thereof. For a given oriented and connected Riemannian $n$-manifold $(\Sigma,\sf{h})$, let us denote the \emph{de Rham cohomology} its global and compactly supported versions, respectively, by
\begin{align*}
	\sf{H}^{k}(\Sigma):=\cfrac{\mathrm{ker}(\d_{\Sigma}\vert_{\Omega^{k}})}{\mathrm{ran}(\d_{\Sigma}\vert_{\Omega^{k-1}})}\,,\qquad\qquad\sf{H}_{\mathrm{c}}^{k}(\Sigma):=\cfrac{\mathrm{ker}(\d_{\Sigma}\vert_{\Omega^{k}_{\mathrm{c}}})}{\mathrm{ran}(\d_{\Sigma}\vert_{\Omega^{k-1}_{\mathrm{c}}})}\,,
\end{align*}
which represent the cohomology groups of the cochain complexes $(\Omega^{\bullet}(\Sigma),\d_{\Sigma})$ and $(\Omega^{\bullet}_{\mathrm{c}}(\Sigma),\d_{\Sigma})$, respectively. Now, if $\Sigma$ is compact, then $\sf{H}^{k}(\Sigma)=\sf{H}^{k}_{\mathrm{c}}(\Sigma)$ are finite-dimensional vector spaces (see e.g.~\cite[Thm.~10.17]{LeeManifolds}) and the Poincaré duality in its differential geometric formulation states that the sesquilinear pairing
\begin{align*}
	\langle \cdot,\cdot\rangle\:\sf{H}^{k}(\Sigma)\times\sf{H}^{n-k}(\Sigma)\to\bb{R}\,,\qquad \langle[\alpha],[\beta]\rangle:=\int_{\Sigma}\alpha\wedge\beta
\end{align*}
is well-defined and \emph{non-degenerate}, i.e.~the induced map $\mathrm{PD}_{k}\:\sf{H}^{k}(\Sigma)\to(\sf{H}^{n-k}(\Sigma))^{\ast}$ defined by $\mathrm{PD}_{k}([\alpha]):=([\beta]\mapsto\langle[\alpha],[\beta]\rangle)$ is injective (see e.g.~\cite[Thm.~9.57]{LeeManifolds}). In particular, finite dimensionality implies that there are linear isomorphisms $\sf{H}^{k}(\Sigma)\cong(\sf{H}^{n-k}(\Sigma))^{\ast}\cong\sf{H}^{n-k}(\Sigma)$.

Now, if $\Sigma$ is non-compact, the situation is slightly different. First of all, the natural generalisation of the bilinear pairing $\langle\cdot,\cdot\rangle$ in the non-compact setting is given by
\begin{align*}
	\langle \cdot,\cdot\rangle\:\sf{H}^{k}(\Sigma)\times\sf{H}_{\mathrm{c}}^{n-k}(\Sigma)\to\bb{R}\,,\qquad \langle[\alpha],[\beta]\rangle:=\int_{\Sigma}\alpha\wedge\beta\, .
\end{align*}
In this case, one often considers the case in which $\Sigma$ has a \emph{finite good cover}, i.e.~an open cover $(\mathcal{U}_{i})_{i\in\mathrm{I}}$ such that any intersection of arbitrary but finitely many members of $(\mathcal{U}_{i})_{i\in\mathrm{I}}$ is diffeomorphic to $\bb{R}^{n}$. In this case, both $\sf{H}^{k}(\Sigma)$ and $\sf{H}^{k}_{\mathrm{c}}(\Sigma)$ are finite-dimensional (see e.g.~\cite[Prop.~5.3.1 and 5.3.2]{Bott}) and the Poincaré duality in this case states once again that $\langle \cdot,\cdot\rangle$ is non-degenerate and hence $\sf{H}^{k}(\Sigma)\cong(\sf{H}^{n-k}_{\mathrm{c}}(\Sigma))^{\ast}\cong\sf{H}_{\mathrm{c}}^{n-k}(\Sigma)$ (see e.g.~\cite[Thm.~10.26]{LeeManifolds}).

While most textbook deal with the case of manifolds admitting a finite good cover, it is well known that the Poincaré duality can also be generalised when using more refined techniques from algebraic topology. In this case, $\sf{H}^{k}(\Sigma)$ and $\sf{H}^{k}_{\mathrm{c}}(\Sigma)$ are not finite-dimensional in general and the Poincaré duality in this setting states that the induced map $\mathrm{PD}_{k}\:\sf{H}^{k}(\Sigma)\mapsto(\sf{H}_{\mathrm{c}}^{n-k}(\Sigma))^{\ast}$ is bijective and hence $\sf{H}^{k}(\Sigma)\cong(\sf{H}^{n-k}_{\mathrm{c}}(\Sigma))^{\ast}$ (see e.g.~\cite[Sec.~5.12]{Greub}). It is important to stress that the dual statement $(\sf{H}^{k}(\Sigma))^{\ast}\cong\sf{H}^{n-k}_{\mathrm{c}}(\Sigma)$ is \emph{not} true in general. Similarly, an infinite-dimensional vector space is in general not isomorphic to its (algebraic) dual and hence, also $\sf{H}^{k}(\Sigma)\cong\sf{H}^{n-k}_{\mathrm{c}}(\Sigma)$ fails for general non-compact manifolds (see \cite[Rem.~5.7]{Bott} for details).

As a direct consequence of the Poincaré duality, we obtain the following lemma.

\begin{lemma}\label{Lemma:PD} Let $(\Sigma,\sf{h})$ be a connected and oriented Riemannian manifold. Then,
\begin{align*}
	\text{\emph{(i)}}\quad&\forall \alpha\in\mathrm{ker}(\d_{\Sigma}\vert_{\Omega^{k}}): \big(\forall\beta\in\mathrm{ker}(\delta_{\Sigma}\vert_{\Omega^{k+1}_{\mathrm{c}}}): \langle\alpha,\beta\rangle_{\sf{L}^{2}}=0\quad\Rightarrow\quad\alpha\in\mathrm{ran}(\d_{\Sigma}\vert_{\Omega^{k-1}})\big)\, .\\
	\text{\emph{(ii)}}\quad& \forall \alpha\in\mathrm{ker}(\delta_{\Sigma}\vert_{\Omega^{k}}): \big(\forall\beta\in\mathrm{ker}(\d_{\Sigma}\vert_{\Omega^{k-1}_{\mathrm{c}}}): \langle\alpha,\beta\rangle_{\sf{L}^{2}}=0\quad\Rightarrow\quad\alpha\in\mathrm{ran}(\delta_{\Sigma}\vert_{\Omega^{k+1}})\big)\, .
\end{align*} 
\end{lemma}

\begin{proof}
	As discussed above, the Poincaré duality states that the linear map $\mathrm{PD}_{k}\:\sf{H}^{k}(\Sigma)\mapsto(\sf{H}_{\mathrm{c}}^{n-k}(\Sigma))^{\ast}$ defined by $\mathrm{PD}_{k}([\alpha]):=([\beta]\mapsto\langle[\alpha],[\beta]\rangle)$ is bijective, which in particular implies that the pairing $\langle \cdot,\cdot\rangle\:\sf{H}^{k}(\Sigma)\times\sf{H}_{\mathrm{c}}^{n-k}(\Sigma)\to\bb{R}$ is weakly non-degenerate in the first entry, i.e.~for every fixed $[\alpha]\in \sf{H}^{k}(\Sigma)$, it holds that 
	\begin{align*}
		\forall [\beta]\in\sf{H}_{\mathrm{c}}^{n-k}(\Sigma): \langle [\alpha],[\beta]\rangle=0\qquad\Rightarrow\qquad [\alpha]=0\, .
	\end{align*}
	The claim in the lemma is essentially a reformulation of this: the pairing $\langle\cdot,\cdot\rangle$ is independent of the chosen representative and satisfies $\langle [\alpha],[\beta]\rangle=\langle\alpha,\ast\beta\rangle_{\sf{L}^{2}}$ up to a sign, by definition. Furthermore, the Hodge $\ast$-operator is an isomorphism $\ast\:\Omega^{n-k}_{\mathrm{c}}(\Sigma)\to\Omega^{k}_{\mathrm{c}}(\Sigma)$ that maps bijectively the kernel (resp.~range) of $\d_{\Sigma}$ into the kernel (resp.~range) of $\delta_{\Sigma}$. In particular, the above statement is equivalent to saying that for every $\alpha\in\mathrm{ker}(\d_{\Sigma}\vert_{\Omega^{k}})$, it holds that
	\begin{align*}
		\forall \beta\in\mathrm{ker}(\delta_{\Sigma}\vert_{\Omega_{\mathrm{c}}^{k}}): \langle \alpha,\beta\rangle_{\sf{L}^{2}}=0\qquad\Rightarrow\qquad \alpha\in\mathrm{ran}(\d_{\Sigma}\vert_{\Omega^{k-1}})\, ,
	\end{align*}
	which proves (i). Claim (ii) follows by duality. We formulated the Poincaré duality for $\bb{R}$-valued forms. The $\bb{C}$-valued case is obtained by decomposing forms into real and imaginary parts.
\end{proof}

Using this lemma, we obtain the following smooth version of Theorem~\ref{Thm:HodgeDecom}:

\begin{theorem}\label{Thm:HodgeDecomSmooth} \emph{(Smooth Hodge Decomposition in Sobolev Spaces)}\newline
	The space $\Omega^{k}_{s}(\Sigma)=\sf{H}^{s}_{k}(\Sigma)\cap\Omega^{k}(\Sigma,\bb{C})$ admits the following $\sf{H}^{s}$-orthogonal decompositions:
	\begin{align*}
		\Omega^{k}_{s}(\Sigma)\cong \Omega^{k}_{s,\d}(\Sigma)\oplus\Omega^{k}_{s,\delta}\oplus\mathrm{ker}(\Delta\vert_{\Omega^{k}_{s}})
	\end{align*}
	for all $s\in [0,\infty)$ and $k\in\bb{N}_{0}$, where $\Omega_{s,\d}^{k}(\Sigma)$ and $\Omega_{s,\delta}^{k}(\Sigma)$ are the subspaces defined by
	\begin{align*}
		\Omega_{s,\d}^{k}(\Sigma):=\Omega^{k}(\Sigma,\bb{C})\cap\overline{\d_{\Sigma}\Omega^{k-1}_{\infty}(\Sigma)}^{\sf{H}^{s}}\,,\qquad \Omega_{s,\delta}^{k}(\Sigma):=\Omega^{k}(\Sigma,\bb{C})\cap\overline{\delta_{\Sigma}\Omega^{k+1}_{\infty}(\Sigma)}^{\sf{H}^{s}}\, .
	\end{align*}
	Furthermore, if we decompose $\omega\in\Omega^{s}_{k}(\Sigma)$ accordingly as $\omega=\alpha+\beta+\gamma$ for (unique) $\alpha\in\Omega^{k}_{s}(\Sigma)$, $\beta\in\Omega^{k}_{s,\delta}(\Sigma)$ and $\gamma\in\mathrm{ker}(\Delta\vert_{\Omega^{s}_{k}})$, the following holds true:
	\begin{itemize}
		\item[\emph{(i)}]$\Omega_{s,\d}^{k}(\Sigma)\subset\mathrm{ran}(\d_{\Sigma}\vert_{\Omega^{k-1}})\cap\Omega^{k}_{s}(\Sigma)$, i.e.~$\alpha$ is exact.
		\item[\emph{(ii)}]$\Omega_{s,\delta}^{k}(\Sigma)\subset\mathrm{ran}(\delta_{\Sigma}\vert_{\Omega^{k+1}})\cap\Omega^{k}_{s}(\Sigma)$, i.e.~$\beta$ is coexact.
		\item[\emph{(iii)}]$\mathrm{ker}(\d_{\Sigma}\vert_{\Omega^{k}_{s}})=\Omega^{k}_{s,\d}(\Sigma)\oplus\mathrm{ker}(\Delta\vert_{\Omega^{k}_{s}})$, i.e.~$\d_{\Sigma}\omega=0$ if and only if $\beta=0$.
		\item[\emph{(iv)}]$\mathrm{ker}(\delta_{\Sigma}\vert_{\Omega^{k}_{s}})=\Omega^{k}_{s,\delta}(\Sigma)\oplus\mathrm{ker}(\Delta\vert_{\Omega^{k}_{s}})$, i.e.~$\delta_{\Sigma}\omega=0$ if and only if $\alpha=0$.
	\end{itemize}
\end{theorem}

\begin{proof}
	Let $\omega\in\Omega_{k}^{s}(\Sigma)$. By Theorem~\ref{Thm:HodgeDecom}, we can decompose it uniquely as $\omega=\alpha+\beta+\gamma$ with 
	\begin{align*}
		\alpha\in\overline{\d_{\Sigma}\Omega^{k-1}_{\infty}(\Sigma)}^{\sf{H}^{s}}\,,\quad\beta\in \overline{\delta_{\Sigma}\Omega^{k+1}_{\infty}(\Sigma)}^{\sf{H}^{s}}\,,\quad \gamma\in\mathrm{ker}(\Delta\vert_{\Omega^{k}_{s}})\, .
	\end{align*}
	We will argue that both $\alpha$ and $\beta$ have to smooth individually in this case. The fact that the corresponding decomposition is a direct sum decomposition is then clear by construction. First of all, we claim that any $\alpha$ is closed in the distributional sense. Indeed, by definition, there exists a sequence $(\alpha_{n})_{n\in\bb{N}}$ in $\Omega^{k-1}_{\infty}(\Sigma)$ such that $\alpha=\sf{H}^{s}\text{-}\lim_{n\to\infty}\d_{\Sigma}\alpha_{n}$. By Proposition~\ref{Prop:Sob}(ii), we know that also $\alpha=\sf{L}^{2}\text{-}\lim_{n\to\infty}\d_{\Sigma}\alpha_{n}$ and therefore, we obtain
	\begin{align*}
		\langle\alpha,\delta_{\Sigma}\varphi\rangle_{\sf{L}_{2}}=\lim_{n\to\infty}\langle\d_{\Sigma}\alpha_{n},\delta_{\Sigma}\varphi\rangle_{\sf{L}_{2}}=\lim_{n\to\infty}\langle\d^{2}_{\Sigma}\alpha_{n},\varphi\rangle_{\sf{L}_{2}}=0
	\end{align*}
	for all test forms $\varphi\in\Omega^{k+1}_{\mathrm{c}}(\Sigma,\bb{C})$. Similarly, we show that $\beta$ is coclosed in the distributional sense, i.e.	 $\langle\beta,\d_{\Sigma}\varphi\rangle_{\sf{L}_{2}}=0$ for all test forms $\varphi\in\Omega^{k-1}_{\mathrm{c}}(\Sigma,\bb{C})$. Now, consider $\Omega(\Sigma,\bb{C}):=\bigoplus_{k=0}^{n}\Omega^{k}(\Sigma,\bb{C})$ equipped with the $\sf{L}^{2}$-inner product $\langle\cdot,\cdot\rangle_{\sf{L}^{2}}$ induced from the $\sf{L}^{2}$-inner products on each summand. Viewing $\omega$ as an element in $\Omega(\Sigma,\bb{C})$ via the inclusion $\Omega^{k}(\Sigma,\bb{C})\hookrightarrow\Omega(\Sigma,\bb{C})$, we find
	\begin{align*}
		\langle\delta_{\Sigma}\omega,\varphi\rangle_{\sf{L}^{2}}=\langle\omega,\d_{\Sigma}\varphi\rangle_{\sf{L}^{2}}=\langle\alpha,\d_{\Sigma}\varphi\rangle_{\sf{L}^{2}}=\langle\alpha,(\d_{\Sigma}+\delta_{\Sigma})\varphi\rangle_{\sf{L}^{2}}
	\end{align*}
	for all $\varphi\in\Omega_{\mathrm{c}}(\Sigma,\bb{C})$. In other words, $(\d_{\Sigma}+\delta_{\Sigma})\alpha=\delta_{\Sigma}\omega$ in the distributional sense and since $\omega$ is smooth, we conclude that also $\alpha$ is smooth by elliptic regularity of the operator $\d_{\Sigma}+\delta_{\Sigma}$. With similar arguments, we show that also $\beta$ is smooth.
	
	Claim (i) is a consequence of Lemma~\ref{Lemma:PD}: as argued above, $\alpha$ is closed in the distributional sense, i.e.~$\langle\alpha,\delta_{\Sigma}\varphi\rangle_{\sf{L}_{2}}=0$ for all $\varphi\in\Omega^{k+1}_{\mathrm{c}}(\Sigma,\bb{C})$ and, as just argued, we also know that $\alpha$ smooth. In particular, $\langle\d_{\Sigma}\alpha,\varphi\rangle_{\sf{L}^{2}}=0$ for all $\varphi\in\Omega^{k+1}_{\mathrm{c}}(\Sigma,\bb{C})$ and hence $\d_{\Sigma}\alpha=0$, by non-degeneracy. Now, by definition, there is a sequence $(\alpha_{n})_{n\in\bb{N}}$ in $\Omega^{k-1}_{\infty}(\Sigma)$ such that $\alpha=\sf{L}^{2}\text{-}\lim_{n\to\infty}\d_{\Sigma}\alpha_{n}$, as above. Then, for every $\varphi\in\mathrm{ker}(\delta_{\Sigma}\vert_{\Omega_{\mathrm{c}}^{k+1}})$, it holds that
	\begin{align*}
		\langle\alpha,\beta\rangle_{\sf{L}^{2}}=\lim_{n\to\infty}\langle\d_{\Sigma}\alpha_{n},\varphi\rangle_{\sf{L}^{2}}=\lim_{n\to\infty}\langle\alpha_{n},\delta_{\Sigma}\varphi\rangle_{\sf{L}^{2}}=0\, .
	\end{align*}
	Lemma~\ref{Lemma:PD} then implies $\alpha=\d_{\Sigma}\psi$ for some $\psi\in\Omega^{k-1}(\Sigma,\bb{C})$. Claim (ii) follows by duality.
	
	For (iii), we only need to show the inclusion ``$\subset$'', since the other inclusion is obvious. Assume that $\d_{\Sigma}\omega=0$. Then, since $\d_{\Sigma}\alpha=\d_{\Sigma}\gamma=0$, we know that $\d_{\Sigma}\beta=0$. Hence, $\beta\in\Omega^{k}_{s}(\Sigma)$ is both closed and coclosed in this case and hence contained both in $\Omega^{k}_{s,\delta}(\Sigma)$ and $\mathrm{ker}(\Delta\vert_{\Omega^{k}_{s}})$. Since the Hodge decomposition is a direct sum decomposition, this implies $\beta=0$.  Claim (iv) follows from similar arguments.
\end{proof}

\begin{remark}
	While every $k$-form $\alpha \in \Omega^{k}_{s,\d}(\Sigma)$ can be expressed as $\alpha =\d_{\Sigma}\psi$ for some $\psi \in \Omega^{k-1}(\Sigma, \bb{C})$, we emphasise that, in general, nothing can be said about the behaviour of $\psi$ at infinity and it may not even belong to $\sf{L}^{2}$. The only information we have is that $\d_{\Sigma}\psi$ can be approximated by a sequence of the form $\d_{\Sigma}\alpha_{n}$ in $\sf{H}^{s}$, where $\alpha_{n} \in \Omega^{k-1}_{\infty}(\Sigma)$. An analogous observation applies to $\Omega^{k}_{s,\delta}(\Sigma)$.
\end{remark}

\begin{remark}
	Consider the case of a \emph{compact} manifold $\Sigma$. Then, $\Omega^{k}_{s,\d}(\Sigma)=\d_{\Sigma}\Omega^{k-1}(\Sigma,\bb{C})$ and $\Omega^{k}_{s,\delta}(\Sigma)=\delta_{\Sigma}\Omega^{k+1}(\Sigma,\bb{C})$ for all $s\in [0,\infty]$. Indeed, the inclusion ``$\subset$'' is the content of Theorem~\ref{Thm:HodgeDecomSmooth}(i) and (ii), which holds true on every (complete) Riemannian manifold, while ``$\supset$'' is trivially true for compact manifolds. Hence, we obtain the standard Hodge decomposition 
	\begin{align*}
		\Omega^{k}(\Sigma,\bb{C})=\d_{\Sigma}\Omega^{k-1}(\Sigma,\bb{C})\oplus\delta_{\Sigma}\Omega^{k+1}(\Sigma,\bb{C})\oplus\mathrm{ker}(\Delta\vert_{\Omega^{k}})
	\end{align*}
	as a special case of Theorem~\ref{Thm:HodgeDecomSmooth} in the compact case.
\end{remark}

To end our discussion on the Hodge decomposition in Sobolev space, we derive a corresponding decomposition in the setting of $\sf{H}^{\infty}$, which we will be used in the context of constructing Hadamard state in the forthcoming discussion. We first observe the following.

\begin{lemma}\label{Lemma:HInfCon}
	If $\sf{M}\subset\sf{H}^{\infty}_{\bullet}(\Sigma)$ is an arbitrary subset, then it holds that
\begin{align*}
	\overline{\sf{M}}^{\sf{H}^{\infty}}=\bigcap_{s\in [0,\infty)}\overline{\sf{M}}^{\sf{H}^{s}}\, ,
\end{align*}
where we recall that $\sf{H}^{\infty}_{\bullet}(\Sigma)$ is equipped with its projective limit topology.
\end{lemma}

\begin{proof}
The inclusion ``$\subset$'' is obvious: recall that $\sf{H}^{\infty}_{\bullet}(\Sigma)$ is a Fréchet space and hence in particular metrisable, which implies that it is first countable. Hence, closed sets can be characterised by means of converging sequences (i.e.~we do not need to talk about \emph{nets} or \emph{filters}). In other words, $x$ is contained in the $\sf{H}^{\infty}$-closure of $\sf{M}$ if and only if there exists a sequence $(x_{n})_{n\in\bb{N}}$ converging to $x$ in the $\sf{H}^{\infty}$-topology. By definition of the projective limit topology, this is equivalent to saying that $(x_{n})_{n\in\bb{N}}$ converges to $x$ in every $\sf{H}^{s}_{\bullet}(\Sigma)$ for $s\geq 0$, which shows that $x$ is contained in the $\sf{H}^{s}$-closure of $\sf{M}$ for all $s\geq 0$.

For ``$\supset$'', we first note that we can replace the index set in the right-hand side without loss of generality by $n\in\bb{N}$, since $\sf{H}_{\bullet}^{r}(\Sigma)\subset\sf{H}^{s}_{\bullet}(\Sigma)$ for all $r\geq s$ by Proposition~\ref{Prop:Sob}(ii). Now, suppose that $x$ is contained in the space on the right-hand side in the formulation of the lemma, i.e.~there is a sequence $(x_{k}^{n})_{k\in\bb{N}}$ converging to $x$ in $\sf{H}^{n}_{\bullet}(\Sigma)$ for each $n\in\bb{N}$. Now, to show that $x$ is contained in the $\sf{H}^{\infty}$-closure of $\sf{M}$, we need to construct a \emph{single} sequence converging to $x$ in \emph{all} Sobolev spaces. We use a \emph{diagonalisation argument}: for every $n\in\bb{N}$, consider $y_{n}:=x^{n}_{k_{n}}$, where $k_{n}\in\bb{N}$ is chosen such that $\Vert x^{n}_{k_{n}}-x\Vert_{\sf{H}^{n}}\leq \frac{1}{n}$. Without loss of generality, we choose $k_{n}$ such that $k_{m}\geq k_{n}$ for $m\geq n$. Then, for each $m\in\bb{N}$, 
\begin{align*}
	\Vert x-y_{n}\Vert_{\sf{H}^{m}}\leq \Vert x-y_{n}\Vert_{\sf{H}^{n}}\leq \frac{1}{n}\qquad\text{for}\qquad \forall n\geq m\,,
\end{align*}
where we used Proposition~\ref{Prop:Sob}(ii). Hence, $\Vert x-y_{n}\Vert_{\sf{H}^{m}}\to 0$ as $n\to\infty$.
\end{proof}

\begin{corollary}\label{Cor:HodgeDecomInf} \emph{(Hodge Decompositions of $\sf{H}^{\infty}$)}\newline
The Fréchet space $\sf{H}^{\infty}_{\bullet}(\Sigma)$ admits the direct sum decomposition
\begin{align*}
	\sf{H}^{\infty}_{\bullet}(\Sigma)\cong\overline{\d_{\Sigma}\Omega^{k-1}_{\infty}(\Sigma)}^{\sf{H}^{\infty}}\oplus\overline{\delta_{\Sigma}\Omega^{k+1}_{\infty}(\Sigma)}^{\sf{H}^{\infty}}\oplus\mathrm{Har}^{\infty}_{k}(\Sigma)\,,
\end{align*}
where $\mathrm{Har}^{\infty}_{k}(\Sigma):=\mathrm{ker}(\Delta\vert_{\Omega^{k}_{\infty}})$, which is orthogonal with respect to $\langle\cdot,\cdot\rangle_{\sf{H}^{s}}$ for every $s\in [0,\infty]$. Furthermore, there is an orthogonal direct sum decomposition
	\begin{align*}
		\Omega^{k}_{\infty}(\Sigma)\cong \Omega^{k}_{\infty,\d}(\Sigma)\oplus\Omega^{k}_{\infty,\delta}(\Sigma)\oplus\mathrm{ker}(\Delta\vert_{\Omega^{k}_{\infty}})\, ,
	\end{align*}
	satisfying conditions (i)-(iv) with $s=\infty$ in Theorem~\ref{Thm:HodgeDecomSmooth}, where we wrote
	\begin{align*}
		\Omega_{\infty,\d}^{k}(\Sigma):=\Omega^{k}(\Sigma,\bb{C})\cap\overline{\d_{\Sigma}\Omega^{k-1}_{\infty}(\Sigma)}^{\sf{H}^{\infty}}\,,\qquad \Omega_{s,\delta}^{k}(\Sigma):=\Omega^{k}(\Sigma,\bb{C})\cap\overline{\delta_{\Sigma}\Omega^{k+1}_{\infty}(\Sigma)}^{\sf{H}^{\infty}}\, .
	\end{align*}
\end{corollary}

\begin{proof}
	Let $\omega\in\sf{H}^{\infty}_{\bullet}(\Sigma)$. Since $\omega\in\sf{L}^{2}_{\bullet}(\Sigma)$, Theorem~\ref{Thm:HodgeDecom} implies that $\omega$ uniquely decomposes as
	\begin{align}\label{eq:InfDecompProof}
		\omega=\alpha_{0}+\beta_{0}+\gamma_{0}\,,\qquad \alpha_{0}\in\overline{\d_{\Sigma}\Omega^{k-1}_{\infty}(\Sigma)}^{\sf{H}^{0}}\,,\quad\beta_{0}\in\overline{\delta_{\Sigma}\Omega^{k+1}_{\infty}(\Sigma)}^{\sf{H}^{0}}\,,\quad\gamma_{0}\in\mathrm{ker}(\Delta\vert_{\Omega^{k}_{0}})\, .
	\end{align}
	We argue that in fact $\alpha_{0}\in\overline{\d_{\Sigma}\Omega^{k-1}_{\infty}(\Sigma)}^{\sf{H}^{\infty}}$, $\beta_{0}\in\overline{\delta_{\Sigma}\Omega^{k+1}_{\infty}(\Sigma)}^{\sf{H}^{\infty}}$ and $\gamma_{0}\in\mathrm{ker}(\Delta\vert_{\Omega^{k}_{\infty}})$. Let $s\in [0,\infty)$ be arbitrary. Since $\omega\in\sf{H}^{s}_{\bullet}(\Sigma)$, Theorem~\ref{Thm:HodgeDecom} implies that $\omega$ uniquely decomposes as
	\begin{align*}
		\omega=\alpha_{s}+\beta_{s}+\gamma_{s}\,,\qquad \alpha_{s}\in\overline{\d_{\Sigma}\Omega^{k-1}_{\infty}(\Sigma)}^{\sf{H}^{s}}\,,\quad\beta_{s}\in\overline{\delta_{\Sigma}\Omega^{k+1}_{\infty}(\Sigma)}^{\sf{H}^{s}}\,,\quad\gamma_{s}\in\mathrm{ker}(\Delta\vert_{\Omega^{k}_{s}})\, .
	\end{align*}
	But now, by Proposition~\ref{Prop:SobProp}(ii), it clearly holds that
	\begin{align*}
		\overline{\d_{\Sigma}\Omega^{k-1}_{\infty}(\Sigma)}^{\sf{H}^{0}}\subset \overline{\d_{\Sigma}\Omega^{k-1}_{\infty}(\Sigma)}^{\sf{H}^{s}}\,,\quad \overline{\delta_{\Sigma}\Omega^{k+1}_{\infty}(\Sigma)}^{\sf{H}^{0}}\subset \overline{\delta_{\Sigma}\Omega^{k-1}_{\infty}(\Sigma)}^{\sf{H}^{s}}\,,\quad \mathrm{ker}(\Delta\vert_{\Omega^{k}_{0}})\subset \mathrm{ker}(\Delta\vert_{\Omega^{k}_{s}})\, .
	\end{align*}
	Hence, by uniqueness of the decompositions in Eq.~\eqref{eq:InfDecompProof}, we conclude that $\alpha_{0}=\alpha_{s}$, $\beta_{0}=\beta_{s}$ and $\gamma_{0}=\gamma_{s}$. Since $s\in[0,\infty)$ was arbitrary, Lemma~\ref{Lemma:HInfCon} proves the claim. The smooth setting follows from Theorem~\ref{Thm:HodgeDecomSmooth} using similar steps.
\end{proof}

As an immediate consequence of Theorem~\ref{Thm:HodgeDecomSmooth}, we get existence of solutions of the Poisson equation, which is closely related to the achievability of the Cauchy radiation gauge.

\begin{proposition}\label{prop:Poisson}
	Let $s\in [0,\infty]$ and $\omega\in\Omega_{s}^{1}(\Sigma)$. Then the Poisson equation
	\begin{align*}
		\Delta f=\delta_{\Sigma}\omega\, ,
	\end{align*}
	has a unique solution (up to constant) in the space $\{f\in C^{\infty}(\Sigma,\bb{C})\mid \d_{\Sigma} f\in\Omega_{s,\d}^{1}(\Sigma)\}$.
\end{proposition}

\begin{proof}
	By Theorem~\ref{Thm:HodgeDecomSmooth}, $\omega$ can be written as $\omega=\alpha+\beta$, where $\alpha\in\Omega_{s,\d}^{1}(\Sigma)$ and $\beta\in\ker(\delta_{\Sigma}\vert_{\Omega^{1}_{s}})$. Furthermore, $\alpha=\d_{\Sigma} f$ for some $f\in C^{\infty}(\Sigma,\bb{C})$. It follows that $\delta_{\Sigma}\omega=\delta_{\Sigma}\alpha=\Delta f$. 
	
	For uniqueness, let $f\in C^{\infty}(\Sigma,\bb{C})$ be such that $\d_{\Sigma} f\in\Omega^{1}_{s,\d}(\Sigma)$ and $\Delta f=0$. Then, the $1$-form $\eta:=\d_{\Sigma} f$ satisfies $\d_{\Sigma} \eta=0$ and $\delta_{\Sigma}\eta=0$ and hence in particular $\eta\in\ker(\Delta\vert_{\Omega^{1}_{s}})$. Since also $\eta\in\Omega^{1}_{s,\d}(\Sigma)$ by assumption, we conclude that $\eta=0$, by the fact that the decomposition in Theorem \ref{Thm:HodgeDecomSmooth} is a direct sum decomposition. It follows that $f$ is closed and hence constant.
\end{proof}

\section{Existence of Hadamard States for Maxwell's Theory}\label{Sec:ConsHadMax}
After this functional-analytic \emph{intermezzo}, we return to the discussion of Maxwell's theory and the Cauchy radiation gauge, leading us to the construction of Hadamard states. The discussion in this section follows \cite[Sec.~4-6]{MurroSchmid}. 

As already mentioned at the beginning of this chapter, we restrict our analysis to \emph{ultrastatic spacetimes}, as this suffices for the purpose of constructing Hadamard states. We note, however, that the first part of this section, namely the Cauchy problem for $\sf{H}^{\infty}$-differential forms, the achievability of the Cauchy radiation gauge as well as the definition and equivalence of various phases spaces for Maxwell's theory, can be extended to more general globally hyperbolic spacetimes, provided suitable decay conditions are imposed on their lapse function and on the second fundamental form of the foliation. We refer to~\cite[Sec.~4 and 5]{MurroSchmid} for details.

Throughout the following, let $(\sf{M},\sf{g})$ be a globally hyperbolic ultrastatic spacetime with
adapted Cauchy temporal function $t\in C^{\infty}(\sf{M})$ such that 
\begin{align*}
	\sf{M}=\bb{R}\times\Sigma\,,\qquad\qquad\sf{g}=-\d t\otimes\d t+\sf{h}\,,
\end{align*}
where $(\Sigma,\sf{h})$ is a \emph{complete} Riemannian manifold (cf.~Example~\ref{Examples:GlobHyp}(i)). Furthermore, we assume that $(\Sigma,\sf{h})$ is of \emph{bounded geometry}, which means that its injectivity radius is non-zero and that its Riemann curvature and all its covariant derivatives are uniformly bounded (see Appendix~\ref{Subsec:BoundGeom}). The reason for this technical assumption is that those kind of manifold allow for a nice calculus of \emph{pseudodifferential operators}, which we will employ lateron in the construction of states. 

Since $(\Sigma,\sf{h})$ is complete, we can apply the analysis of the previous section on differential forms defined on $\Sigma$. Let us recall the following notation:
\begin{align*}
	&(\sf{L}^{2}_{k}(\Sigma),\langle\cdot,\cdot\rangle_{\sf{L}^{2}})\qquad &&\text{space of square-intergable $k$-forms}\\
	&(\sf{H}^{s}_{k}(\Sigma),\langle\cdot,\cdot\rangle_{\sf{H}^{s}})\qquad &&\text{Sobolev space of $k$-forms of degree $s\in [0,\infty]$}\\
	&\Omega^{k}_{s}(\Sigma):=\Omega^{k}(\Sigma,\bb{C})\cap\sf{H}^{s}_{k}(\Sigma)\qquad && \text{subspace of smooth elements of $\sf{H}^{s}_{k}(\Sigma)$}
\end{align*}
In addition, we have considered the following two subspaces of the pre-Hilbert space $\Omega^{k}_{s}(\Sigma)$:
\begin{align*}
		\Omega_{s,\d}^{k}(\Sigma):=\Omega^{k}(\Sigma,\bb{C})\cap\overline{\d_{\Sigma}\Omega^{k-1}_{\infty}(\Sigma)}^{\sf{H}^{s}}\,,\qquad \Omega_{s,\delta}^{k}(\Sigma):=\Omega^{k}(\Sigma,\bb{C})\cap\overline{\delta_{\Sigma}\Omega^{k+1}_{\infty}(\Sigma)}^{\sf{H}^{s}}\, .
	\end{align*}
Recall that the space $\sf{H}^{\infty}_{k}(\Sigma)$ is equipped with its projective limit topology, i.e.~the locally convex vector space whose topology is induced by the family of norms $(\Vert\cdot\Vert_{\sf{H}^{s}})_{s\in [0,\infty)}$.

\subsection{The Cauchy Problem for Differential Forms in Spatial $\sf{H}^{\infty}$}
As discussed at the end of Section~\ref{Subsec:GCMax}, the achievability of the Cauchy radiation gauge is closely linked to the solvability of the Poisson equation $\Delta f = \delta_{\Sigma}\omega$ in a suitable function space, while the uniqueness of solutions in an appropriate space ensures that this gauge fixing is \emph{complete} (cf.~Proposition~\ref{Prop:AchievCRG}). As already emphasised, both of these questions depend crucially on the class of fields under consideration. In Proposition~\ref{prop:Poisson}, we showed that the desired properties hold when working within Sobolev spaces. Consequently, as our class of fields, we are guided to take the space of Maxwell fields $\sf{A}:=\sf{A}_{\mathrm{T}}\d t+\sf{A}_{\Sigma}\in\Omega^{1}(\sf{M},\bb{C})$ with the requirement that \begin{align*}\sf{A}_{\mathrm{T}} \in C^{\infty}(\bb{R},\Omega^{0}_{s}(\Sigma,\bb{C}))\,,\qquad \sf{A}_{\Sigma} \in C^{\infty}(\bb{R},\Omega^{1}_{s}(\Sigma,\bb{C}))\end{align*} for some $s \in \bb{R}$. In other words, at each fixed time, both components $\sf{A}_{\mathrm{T}}$ and $\sf{A}_{\Sigma}$ belong to appropriate Sobolev spaces. In practice, it is convenient to work with a class of fields that is \emph{closed} under differential operators; therefore, we will consider the case $s = \infty$.

The first step in our analyse is to show that the Cauchy problem is well-posed in Sobolev spaces. Let us consider the the space 
\begin{align*}
	C^{m}(\bb{R},\sf{H}^{s}_{k}(\Sigma))\end{align*}
for arbitrary $m\in\bb{N}_{0}$ and $s\in [0,\infty)$, which is defined, as the space of all maps $\bb{R}\ni t\mapsto\omega(t)\in \sf{H}^{s}_{k}(\Sigma)$ which are $m$-times continuously differentiable with respect to the the Fréchet-space structure on $\sf{H}^{s}_{k}(\Sigma)$.

Then, in the case of time-dependent $k$-forms on $\Sigma$, we have the following generalisation of Theorem~\ref{Thm:NHCauchy}.

\begin{proposition}\label{Prop:CauchyProblemSob} \emph{(Cauchy Problem for Time-Dependent Forms)}\newline
	Let $(\Sigma,\sf{h})$ be a (connected, oriented) Riemannian manifold of bounded geometry, $s\in [1,\infty)$ and $m\in\bb{N}_{0}$ be arbitrary such that $s\geq m+1$. Then, the Cauchy problem
	\begin{align*}
		\begin{cases}
			(\partial_{t}^{2}+\Delta)\omega &=\varphi\\
			\omega\vert_{t=0}&=\mathfrak{f}\\
			\partial_{t}\omega\vert_{t=0}&=\mathfrak{g}
		\end{cases}
	\end{align*}
	with Cauchy data $(\varphi,\mathfrak{f},\mathfrak{g})\in C^{m}(\bb{R},\sf{H}^{s-1}_{k}(\Sigma))\times\sf{H}^{s}_{k}(\Sigma)\times\sf{H}^{s-1}_{k}(\Sigma)$ admits a unique solution
	\begin{align*}
		\omega\in \bigcap_{l=0}^{m+1}C^{l}(\bb{R},\sf{H}^{s-l}_{k}(\Sigma))\, .
		\end{align*}
		In the special case $m=\infty$ with $s\to\infty$, we hence obtain a (bijective) solution map 
		\begin{align*}
			C^{\infty}(\bb{R},\sf{H}^{\infty}_{k}(\Sigma))\times\sf{H}^{\infty}_{k}(\Sigma)\times\sf{H}^{\infty}_{k}(\Sigma)\to C^{\infty}(\bb{R},\sf{H}^{\infty}_{k}(\Sigma))\, .
		\end{align*}
\end{proposition}

The idea of the proof is to use suitable energy estimates in combination with Theorem~\ref{Thm:NHCauchy} that tells us that the Cauchy problem is well-posed for arbitrary smooth Cauchy data.

\begin{lemma}\label{Lemma:EnEstHInf} \emph{(Energy Estimates)}\newline
	Let $(\Sigma,\sf{h})$ be a (connected, oriented) Riemannian manifold of bounded geometry and consider the \emph{energy} $\mathcal{E}(\omega,\cdot)$ defined by 
	\begin{align*}
		\mathcal{E}_{s}(\omega,\tau):=\Vert\omega_{\tau}\Vert_{\sf{H}^{s}}^{2}+\Vert(\partial_{t}\omega)_{\tau}\Vert_{\sf{H}^{s-1}}^{2}\in\bb{R}
	\end{align*}
	for all $\tau\in\bb{R}$, where we wrote $\omega_{\tau}:=\omega\vert_{\tau}$. Then, for any time intervals $\mathrm{I}:=[\mathrm{T}_{0},\mathrm{T}_{1}]\subset\bb{R}$ with $\mathrm{T}_{0},\mathrm{T}\in\bb{R}$ and $\mathrm{T}_{0}<\mathrm{T}_{1}$, there exists a constant $C=C(\mathrm{T}_{0},\mathrm{T}_{1})>0$ such that
	\begin{align*}
		\mathcal{E}_{s}(\omega,t)\leq \mathcal{E}_{s}(\omega,t_{0})e^{Ct}+\int_{t_0}^{t}e^{C(t-\tau)}\Vert(\partial_{t}^{2}+\Delta)\omega_{\tau}\Vert_{\sf{H}^{s}}^{2}\,\d\tau\, 
	\end{align*}
	for all $\omega\in C^{\infty}(\bb{R},\Omega_{\mathrm{c}}^{k}(\Sigma,\bb{C}))$ and $t_{0},t\in\mathrm{I}$ with $t_{0}\leq t$.
\end{lemma}

\begin{proof}
	The proof strategy is standard and we follow the exposition provided in \cite{BaerWafo} (see also the discussion in \cite[Sec.~3.7]{BaerLecture} or \cite[Chap.~2]{TaylorI}) adapted to our setting. The discussion in these references, however, differs in two small details: on one hand, our setting is slightly simpler, since we effectively deal with the ultrastatic case in which the metric $\sf{h}$ and hence the volume measure has no time dependence. On the other hand, the above references either work on the setting of \emph{compact} manifolds $\Sigma$ or with local Sobolev spaces, while in our setting we work on \emph{complete} Riemannian manifolds with the Sobolev spaces introduced in the previous section. To start with, we estimate
	\begin{equation}\label{eq:FirstPiece}
	\begin{aligned}
		\frac{\d}{\d t}\Vert\omega_{t}\Vert_{\sf{H}^{s}}^{2}&=  2\mathrm{Re}\langle\partial_{t}\sf{E}^{\frac{s}{2}}\omega_{t},\sf{E}^{\frac{s}{2}}\omega_{t}\rangle_{\sf{L}^{2}}\\
		&\leq 2\mathrm{Re}\langle\partial_{t}\omega_{t},\omega_{t}\rangle_{\sf{H}^{s}}+c_{1}(t)\cdot\Vert\omega_{t}\Vert_{\sf{H}^{s}}^{2}\, ,
	\end{aligned}
	\end{equation}
	for some constant $c=c(t)>0$, where we used that the commutator $[\sf{E}^{\frac{s}{2}},\partial_{t}]$ is effectively a pseudodifferential operator of order $\leq s$, as one can easily see be calculating its principal symbol. On Riemannian manifolds of bounded geometry, there is a nice class of pseudodifferential operators (see Appendix~\ref{Subsec:BoundGeom}), to which this operator belongs, and hence we can bound the commutator in terms of the $\sf{H}^{s}$-norm. Replacing $\omega$ by $\partial_{t}\omega$ in the estimate \eqref{eq:FirstPiece}, we immediately obtain 
	\begin{align}\label{eq:TimeDerivativeTerm}
	\frac{\d}{\d t}\Vert\partial_{t}\omega_{t}\Vert_{\sf{H}^{s-1}}^{2}\leq 2\mathrm{Re}\langle\partial_{t}^{2}\omega_{t},\partial_{t}\omega_{t}\rangle_{\sf{H}^{s-1}}+c_{2}(t)\cdot\Vert\partial_{t}\omega_{t}\Vert_{\sf{H}^{s-1}}^{2}
	\end{align}
for some constant $c_{2}=c_{2}(t)>0$. Now, recall that $\sf{E}=\Delta+\mathrm{id}$ on regular enough forms and hence, using the Cauchy-Schwartz inequality, we can estimate the first term in the right-hand side of Estimate~\eqref{eq:TimeDerivativeTerm} as 
\begin{equation}\label{eq:SecondPiece}
	\begin{aligned}
	2\mathrm{Re}\langle\partial_{t}^{2}\omega_{t},\partial_{t}\omega_{t}\rangle_{\sf{H}^{s-1}} &=2\mathrm{Re}\langle(\partial_{t}^{2}+\Delta)\omega_{t}-\sf{E}\omega_{t}+\omega_{t},\partial_{t}\omega_{t}\rangle_{\sf{H}^{s-1}}\\&\leq -2\mathrm{Re}\langle\sf{E}\omega_{t},\partial_{t}\omega_{t}\rangle_{\sf{H}^{s-1}}+2\big(\Vert (\partial_{t}^{2}+\Delta)\omega_{t}\Vert_{\sf{H}^{s-1}}+\Vert \omega_{t}\Vert_{\sf{H}^{s}}\big)\Vert\partial_{t}\omega_{t}\Vert_{\sf{H}^{s-1}}\\&\leq -2\mathrm{Re}\langle\omega_{t},\partial_{t}\omega_{t}\rangle_{\sf{H}^{s}}+\Vert (\partial_{t}^{2}+\Delta)\omega_{t}\Vert_{\sf{H}^{s-1}}^{2}+\Vert \omega_{t}\Vert_{\sf{H}^{s}}^{2}+\Vert\partial_{t}\omega_{t}\Vert_{\sf{H}^{s-1}}^{2}\,,
\end{aligned}
\end{equation}
where we used the Cauchy-Schwartz inequality as well as the algebraic inequality $2ab\leq a^{2}+b^{2}$ for $a,b\in\bb{R}$. Moreover, we used the self-adjointness of $\sf{E}$ as well as Proposition~\ref{Prop:SobProp}(vi) to obtain
\begin{align*}
		\langle\sf{E}\omega_{t},\partial_{t}\omega_{t}\rangle_{\sf{H}^{s-1}}=\langle\sf{E}^\frac{1}{2}\omega_{t},\sf{E}^\frac{1}{2}\partial_{t}\omega_{t}\rangle_{\sf{H}^{s-1}}=\langle\omega_{t},\partial_{t}\omega_{t}\rangle_{\sf{H}^{s}}\, .
	\end{align*}
	Summing~\eqref{eq:FirstPiece} and~\eqref{eq:SecondPiece}, we see that the terms $2\langle\omega_{t},\partial_{t}\omega_{t}\rangle_{\sf{H}^{s}}$ cancel each other and we are left with the inequality
	\begin{align*}
	\frac{\d}{\d t}\mathcal{E}_{s}(\omega,t)\leq c_{3}(t)\cdot \mathcal{E}_{s}(\omega,t)+\Vert (\partial_{t}^{2}+\Delta)\omega_{t}\Vert^{2}_{\sf{H}^{s-1}}
\end{align*}
for some constant $c_{3}=c_{3}(t)>0$ constructed using $c_{1}$ and $c_{2}$. The claimed result follows by applying Grönwall's lemma \cite{Gronwall} (see also \cite[Lemma~1.5.1]{BaerLecture}) to the compact time interval $\mathrm{I}=[\mathrm{T}_{0},\mathrm{T}_{1}]$ with $\mathrm{T}_{0},\mathrm{T}_{1}\in\bb{R}$ and $\mathrm{T}_{0}<\mathrm{T}_{1}$. By construction, we note that the dependency of the constant $c_{3}$ on time is continuous, since it is constructed out of the norms of smooth objects. Hence, we can simply consider its maximum on $\mathrm{I}$ to obtain a time-independent constant before applying Gronwall's lemma.
\end{proof}

Using the energy estimates derived in Lemma~\ref{Lemma:EnEstHInf} above, we now give a proof of Proposition~\ref{Prop:CauchyProblemSob} on the well-posedness of the Cauchy problem for time-dependent differential forms in Sobolev spaces (cf.~\cite[Thm.~13]{BaerWafo} for a related proof). 

\begin{proof}[Proof of Proposition~\ref{Prop:CauchyProblemSob}.]
	It is sufficient to prove the claim for $m=0$, since the claim for $m>0$ follows by induction. More precisely, suppose that the claim holds true for $m=0$. Then, consider Cauchy data $(\varphi,\mathfrak{f},\mathfrak{g})\in C^{1}(\bb{R},\sf{H}^{s-1}_{k}(\Sigma))\times\sf{H}^{s}_{k}(\Sigma)\times\sf{H}^{s-1}_{k}(\Sigma)$. and denote the unique solution to the Cauchy problem by $\omega\in C^{0}(\bb{R},\sf{H}^{s}_{k}(\Sigma))\cap C^{1}(\bb{R},\sf{H}^{s-1}_{k}(\Sigma))$. Then, we note that $\eta:=\partial_{t}\omega$ is the unique solution to the Cauchy problem
	\begin{align*}
		\begin{cases}
			(\partial_{t}^{2}+\Delta)\eta &=\partial_{t}\varphi\in C^{0}(\bb{R},\sf{H}_{k}^{s-1}(\Sigma))\subset C^{0}(\bb{R},\sf{H}_{k}^{s-2}(\Sigma)) \\
			\eta\vert_{t=0}&=\mathfrak{g}\in \sf{H}^{s-1}_{k}(\Sigma)\\
			\partial_{t}\eta\vert_{t=0}&=\partial_{t}^{2}\omega\vert_{t=0}=(\varphi-\Delta\omega)\vert_{t=0}\in\sf{H}_{k}^{s-2}(\Sigma)
		\end{cases}
	\end{align*}
	where we used that $[\partial_{t},\Delta]=0$ in the ultrastatic case. Hence, we conclude that $\partial_{t}\omega\in C^{0}(\bb{R},\sf{H}^{s-1}_{k}(\Sigma))\cap C^{1}(\bb{R},\sf{H}^{s-2}_{k}(\Sigma))$ and hence $\omega\in C^{1}(\bb{R},\sf{H}^{s-1}_{k}(\Sigma))\cap C^{2}(\bb{R},\sf{H}^{s-2}_{k}(\Sigma))$.
	
	Now, let us prove the proposition for $m=0$. To start with, we equip the space $C^{l}(\bb{R},\sf{H}^{s}_{k}(\Sigma))$ for $l\in\bb{N}_{0}$ with its natural locally convect topology induced by the family of seminorms
	\begin{align*}
	\Vert\omega\Vert_{\mathrm{I},j,s}:=\sup_{t\in\mathrm{I}}\Vert(\partial_{t}^{j}\omega)_{t}\Vert_{\sf{H}^{s}}
	\end{align*}
	labelled by compact intervals $\mathrm{I}=[a,b]\subset\bb{R}$ and $j=0,\dots,l$. Clearly, $C^{l}(\bb{R},\sf{H}^{s}_{k}(\Sigma))$ equipped with this topology is metrisable, Hausdorff and complete and hence in particular a Fréchet space. In addition, let us also consider the locally convex topology vector space $\sf{L}^{2}_{\mathrm{loc}}(\bb{R},\sf{H}^{s}_{k}(\Sigma))$ with family of norms
	\begin{align*}
	\Vert\omega\Vert_{\mathrm{I},s}:=\bigg(\int_{\mathrm{I}}\Vert\omega_{t}\Vert_{\sf{H}^{s}}^{2}\,\d t\bigg)^{\frac{1}{2}}\, ,
	\end{align*}
	which again can easily be seen to be a Fréchet space. Now, consider the globally hyperbolic manifold $\sf{M}:=\bb{R}\times\Sigma$ with $\sf{g}=-\d t\otimes\d t+\sf{h}$. Then, by Theorem~\ref{Thm:NHCauchy} applied to the normally hyperbolic operator $\partial_{t}^{2}+\Delta$ acting on sections of the bundle $\pi_{2}^{\ast}\big(\bigwedge^{k}\sf{T}^{\ast}\Sigma\big)$ with $\pi_{2}\:\sf{M}\to\Sigma$ denoting the projection, we know that there is a well-defined solution map
	\begin{align*}
		\mathcal{S}\:C^{\infty}(\bb{R},\Omega^{k}_{\mathrm{c}}(\Sigma,\bb{C}))\times\Omega^{k}_{\mathrm{c}}(\Sigma,\bb{C})\times\Omega^{k}_{\mathrm{c}}(\Sigma,\bb{C})\to C^{\infty}(\bb{R},\Omega^{k}_{\mathrm{c}}(\Sigma,\bb{C}))\,,
	\end{align*}
	where $C^{\infty}(\bb{R},\Omega_{\mathrm{c}}^{k}(\Sigma,\bb{C}))\subset\Omega_{\mathrm{sc}}^{k}(\sf{M},\bb{C})$ denotes the subset of time-dependent $k$-forms on $\Sigma$, which are spatially compactly supported in spacetime $\sf{M}$. Now, for an arbitrary compact time interval $\mathrm{I}\subset\sf{R}$, Lemma~\ref{Lemma:EnEstHInf} yields estimates
	\begin{subequations}\label{eq:EstimatesCauchyEnergy}
	\begin{equation}
		\Vert\omega\Vert_{\mathrm{I},0,s}\leq C_{0}\big(\Vert \varphi\Vert_{\mathrm{I},s-1}+\Vert\mathfrak{f}\Vert_{\sf{H}^{s}}+\Vert\mathfrak{g}\Vert_{\sf{H}^{s-1}}\big)
		\end{equation}
		\begin{equation}
		\Vert\partial_{t}\omega\Vert_{\mathrm{I},0,s-1}\leq C_{1}\big(\Vert \varphi\Vert_{\mathrm{I},s-1}+\Vert\mathfrak{f}\Vert_{\sf{H}^{s}}+\Vert\mathfrak{g}\Vert_{\sf{H}^{s-1}}\big)
		\end{equation}
	\end{subequations}
	for all $(\varphi,\mathfrak{f},\mathfrak{g})\in C^{\infty}(\bb{R},\Omega^{k}_{\mathrm{c}}(\Sigma,\bb{C}))\times\Omega^{k}_{\mathrm{c}}(\Sigma,\bb{C})\times\Omega^{k}_{\mathrm{c}}(\Sigma,\bb{C})$ with $\omega:=\mathcal{S}(\varphi,\mathfrak{f},\mathfrak{g})$, where $C_{0},C_{1}>0$ are constants depending on the interval $\mathrm{I}$ but not on $\varphi,\mathfrak{f}$ and $\mathfrak{g}$. In particular, for arbitrary initial data $(\varphi,\mathfrak{f},\mathfrak{g})\in C^{0}(\bb{R},\sf{H}^{s-1}_{k}(\Sigma))\times\sf{H}^{s}_{k}(\Sigma)\times\sf{H}^{s-1}_{k}(\Sigma)$, we choose a sequence $(\mathfrak{f}_{n})_{n\in\bb{N}}$ in $\Omega^{k}_{\mathrm{c}}(\Sigma,\bb{C})$ converging to $\mathfrak{f}$ in $\sf{H}^{s}$, a sequence $(\mathfrak{g}_{n})_{n\in\bb{N}}$ in $\Omega^{k}_{\mathrm{c}}(\Sigma,\bb{C})$ converging to $\mathfrak{g}$ in $\sf{H}^{s-1}$ and a sequence $(\varphi_{n})_{n\in\bb{N}}$ in $C^{\infty}(\bb{R},\Omega^{k}_{\mathrm{c}}(\Sigma,\bb{C}))$ converging to $\varphi$ in $\sf{L}^{2}_{\mathrm{loc}}(\bb{R},\sf{H}^{s}_{k}(\Sigma))$, then the Estimates in Eq.~\eqref{eq:EstimatesCauchyEnergy} imply that the sequence $\omega_{n}:=\mathcal{S}(\varphi_{n},\mathfrak{f}_{n},\mathfrak{g}_{n})$ converges in both $C^{0}(\bb{R},\sf{H}^{s}_{k}(\Sigma))$ and $C^{1}(\bb{R},\sf{H}^{s-1}_{k}(\Sigma))$. It is also clear that the limit $\omega$ is independent of the chosen sequences $(\mathfrak{f}_{n})_{n\in\bb{N}}, (\mathfrak{g}_{n})_{n\in\bb{N}}$ and $(\varphi_{n})_{n\in\bb{N}}$, as a consequence of Eq.~\eqref{eq:EstimatesCauchyEnergy}. Hence, we obtain a (continuous) extension of the solution map $\mathcal{S}$ of the form\footnote{The space $C^{0}(\bb{R},\sf{H}^{s}_{k}(\Sigma))\cap C^{1}(\bb{R},\sf{H}^{s-1}_{k}(\Sigma))$ is usually called the space of \emph{finite energy sections}, since it is essentially the biggest space of sections $\omega$ for which the energy $\mathcal{E}_{s}(\omega,\cdot)$ is a well-defined and continuous function.}
	\begin{align*}
		\mathcal{S}\:C^{0}(\bb{R},\sf{H}^{s-1}_{k}(\Sigma))\times\sf{H}^{s}_{k}(\Sigma)\times\sf{H}^{s-1}_{k}(\Sigma)\to C^{0}(\bb{R},\sf{H}^{s}_{k}(\Sigma))\cap C^{1}(\bb{R},\sf{H}^{s-1}_{k}(\Sigma))\, .
	\end{align*}
	Of course, this map is not bijective, since not every $\omega\in C^{0}(\bb{R},\sf{H}^{s}_{k}(\Sigma))\cap C^{1}(\bb{R},\sf{H}^{s-1}_{k}(\Sigma))$ satisfies $(\partial_{t}^{2}+\Delta)\omega\in C^{0}(\bb{R},\sf{H}^{s-1}_{k}(\Sigma))$. To become an honest projection, one needs to consider a bigger space of sources, i.e.~all of $\sf{L}^{2}_{\mathrm{loc}}(\bb{R},\sf{H}^{s-1}_{k}(\Sigma))$, we refer to \cite[Sec.~3]{BaerWafo} for details. On the other hand, if we consider the case $m=\infty$ and $s\to\infty$, we obtain clearly an isomorphism $C^{\infty}(\bb{R},\sf{H}^{\infty}_{k}(\Sigma))\times\sf{H}^{\infty}_{k}(\Sigma)\times\sf{H}^{\infty}_{k}(\Sigma)\to C^{\infty}(\bb{R},\sf{H}^{\infty}_{k}(\Sigma))$.
\end{proof}

Now, for our purposes, we will exclusively work with \emph{smooth fields} that are in $\sf{H}^{\infty}$ at each fixed time. For this, we need to introduce a suitable notation. To start with, recall the notation $C^{\infty}(\bb{R},\Omega^{k}(\Sigma,\bb{C})):=\Gamma^{\infty}(\pi_{2}^{\ast}(\sf{A}_{\Sigma,k}))$, where $\pi_{2}\:\sf{M}\to\Sigma$ denotes the obvious projection, which is the space of time-dependent $k$-forms on $\Sigma$ (cf.~also~Remark~\ref{Rem:FrechetSmooth}). With this notation, following Eq.~\eqref{eq:DecompDiffForms}, we obtain the decomposition
\begin{align*}
	\Gamma^{\infty}(\sf{A}_{k})=\Omega^{k}(\sf{M},\bb{C})\cong C^{\infty}(\bb{R},\Omega^{k-1}(\Sigma,\bb{C}))\oplus C^{\infty}(\bb{R},\Omega^{k}(\Sigma,\bb{C}))
\end{align*}
defined by $\omega\mapsto (\omega_{\mathrm{T}},\omega_{\Sigma})$, where $\omega=\d t\wedge\omega_{\mathrm{T}}+\omega_{\Sigma}$ with $\omega_{\mathrm{T}}:=\partial_{t}\lrcorner\omega$ and $\omega_{\Sigma}:=\omega-\d t\wedge\omega_{\mathrm{T}}$. For initial data, we consider the bundles $\sf{A}_{\rho_{k}}:=\sf{A}_{k}\vert_{\Sigma_{0}}\oplus\sf{A}_{k}\vert_{\Sigma_{0}}$ and identify
\begin{align*}
	\Gamma^{\infty}(\sf{A}_{\rho_{k}})\cong\Omega^{k-1}(\Sigma,\bb{C})\oplus \Omega^{k-1}(\Sigma,\bb{C})\oplus\Omega^{k}(\Sigma,\bb{C})\oplus\Omega^{k}(\Sigma,\bb{C})
\end{align*}
as in Section~\ref{Subsec:CauchyMax} to arbitrary $k$-forms. In the case $k=0$, this should be understood with the convention $\Omega^{-1}=\emptyset$. The corresponding Cauchy data maps are then given by
\begin{align*}
	\rho_{k}\:\Gamma^{\infty}(\sf{A}_{k})\to\Gamma^{\infty}(\sf{A}_{\rho_{k}})\,,\qquad\rho_{k}(\omega):=\bigg(\omega_{\mathrm{T}},\,\frac{1}{i}\partial_{t}\omega_{\mathrm{T}},\,\omega_{\Sigma},\,\frac{1}{i}\partial_{t}\omega_{\Sigma}\bigg)\bigg\vert_{t=0}\,,
\end{align*}
as discussed at length in Section~\ref{Subsec:CauchyMax}. Let us introduce the following notation.

\begin{definition} (Notation: Smooth Spatially $\sf{H}^{\infty}$-Forms)\newline
	We consider the space of smooth spatially $\sf{H}^{\infty}$-sections by
	\begin{align*}
		\Gamma^{\infty}_{\infty}(\sf{A}_{k}):=\Gamma^{\infty}(\sf{A}_{k})\cap \big(C^{\infty}(\bb{R},\sf{H}^{\infty}_{k-1}(\Sigma))\oplus C^{\infty}(\bb{R},\sf{H}^{\infty}_{k}(\Sigma))\big)\, .
	\end{align*}
	Furthermore, we consider the space of smooth $\sf{H}^{\infty}$-initial data
	\begin{align*}
		\Gamma^{\infty}_{\infty}(\sf{A}_{\rho_{k}}):=\begin{cases}\Omega_{\infty}^{0}(\Sigma)\oplus \Omega_{\infty}^{0}(\Sigma) & \quad ,\,k=0\\\Omega_{\infty}^{k-1}(\Sigma)\oplus \Omega_{\infty}^{k-1}(\Sigma)\oplus\Omega_{\infty}^{k}(\Sigma)\oplus\Omega_{\infty}^{k}(\Sigma) & \quad ,\,k\geq 1\end{cases}\, . 
	\end{align*}
\end{definition}

As a consequence of Proposition~\ref{Prop:CauchyProblemSob}, we obtain the following result on the level of forms defined on the spacetime manifold $\sf{M}$:

\begin{theorem}\label{Thm:CauchySob} \emph{(The Cauchy Problem for Smooth Spatial $\sf{H}^{\infty}$-Sections)}\newline
	Let $(\sf{M}=\bb{R}\times\Sigma,\sf{g}=-\d t\otimes\d t+\sf{h})$ be a globally hyperbolic ultrastatic manifold such that $(\Sigma,\sf{h})$ has bounded geometry. Then, the unique solution $\omega\in\Gamma^{\infty}(\sf{A}_{k})$ to the Cauchy problem for the de Rham-Hodge d'Alembertian $\square=\d\delta+\delta\d$
	\begin{align*}
		\begin{cases}
			\square\omega&=\varphi\\
			\rho_{k}(\omega)&=\mathfrak{f}
		\end{cases}
	\end{align*}
	for Cauchy data $(\varphi,\mathfrak{f})\in \Gamma^{\infty}_{\infty}(\sf{A}_{k})\times\Gamma^{\infty}_{\infty}(\sf{A}_{\rho_{k}})$ satisfies $\omega\in\Gamma^{\infty}_{\infty}(\sf{A}_{k})$. 
\end{theorem}

\begin{proof}
	On ultrastatic spacetimes, it is easy to check that $\square\omega\in\Omega^{k}(\sf{M},\bb{C})$ decomposes as $(\square\omega)_{\mathrm{T}}=(\partial_{t}^{2}+\Delta)\omega_{\mathrm{T}}$ and $(\square\omega)_{\Sigma}=(\partial_{t}^{2}+\Delta)\omega_{\Sigma}$, where $\omega=\d t\wedge\omega_{\mathrm{T}}+\omega_{\Sigma}$ and where $\Delta=\d_{\Sigma}\delta_{\Sigma}+\delta_{\Sigma}\d_{\Sigma}$ denotes the de Rham-Hodge Laplacian, following from the fact that the Christoffel symbols $\Gamma^{\gamma}_{\alpha\beta}$ of $(\sf{M},\sf{g})$ are zero whenever at least one of the indices $\gamma,\alpha,\beta$ is temporal. Hence, $\square\omega=\varphi$ decouples into the temporal and spatial components. The claim then follows from Theorem~\ref{Thm:NHCauchy} and Proposition~\ref{Prop:CauchyProblemSob} applied to both components individually. 
	
	More precisely, consider the equation $(\partial_{t}^{2}+\Delta)\omega=\varphi$ for some time-dependent form $\varphi$ that is contained both in $C^{\infty}(\bb{R},\Omega^{k}(\Sigma,\bb{C}))$ and $C^{\infty}(\bb{R},\sf{H}^{\infty}_{k}(\Sigma))$ and initial data $(\mathfrak{f},\mathfrak{g})$ contained in $\Omega^{k}_{\infty}(\Sigma)=\sf{H}^{\infty}_{k}(\Sigma)\cap\Omega^{k}(\Sigma,\bb{C})$. Then, since $\Omega^{k}_{\mathrm{c}}(\Sigma,\bb{C})$ is dense both in $\Omega^{k}(\Sigma,\bb{C})$ and $\sf{H}^{\infty}_{k}(\Sigma)$ and since $C^{\infty}(\bb{R},\Omega^{k}_{\mathrm{c}}(\Sigma,\bb{C}))$ is dense both in $C^{\infty}(\bb{R},\Omega^{k}(\Sigma,\bb{C}))$ and $\sf{L}^{2}_{\mathrm{loc}}(\bb{R},\sf{H}^{\infty}_{k}(\Sigma))$, we can find a sequence $(\varphi_{n},\mathfrak{f}_{n},\mathfrak{g}_{n})$ in $C^{\infty}(\bb{R},\Omega^{k}_{\mathrm{c}}(\Sigma,\bb{C}))\times\Omega^{k}_{\mathrm{c}}(\Sigma,\bb{C})\times\Omega^{k}_{\mathrm{c}}(\Sigma,\bb{C})$ converging to $(\varphi,\mathfrak{f},\mathfrak{g})$ in both topologies. Furthermore, by Theorem~\ref{Thm:NHCauchy} and Proposition~\ref{Prop:CauchyProblemSob}, we know that the solution map is continuous both with respect to the smooth Fréchet topologies and the topologies induced by the Sobolev spaces and hence, we conclude that the unique smooth solution $\omega\in C^{\infty}(\bb{R},\Omega^{k}(\Sigma,\bb{C}))$ is contained in the space $\omega\in C^{\infty}(\bb{R},\sf{H}^{\infty}_{k}(\Sigma))$.
\end{proof}

As a byproduct of Theorem~\ref{Thm:CauchySob}, following the general discussion of \emph{Green hyperbolic operators} in Section~\ref{Sec:GreenCauchyHyp}, we obtain the following result.

\begin{corollary}
	Let $(\sf{M}=\bb{R}\times\Sigma,\sf{g}=-\d t\otimes\d t+\sf{h})$ be a globally hyperbolic ultrastatic manifold such that $(\Sigma,\sf{h})$ has bounded geometry.
	\begin{itemize}
		\item[\emph{(i)}]The Cauchy data map $\rho_{k}\:\Gamma^{\infty}(\sf{A}_{k})\to\Gamma^{\infty}(\sf{A}_{\rho_{k}})$ is well-defined and bijective as a map $\rho_{k}\:\mathrm{ker}(\square\vert_{\Gamma^{\infty}_{\infty}})\to\Gamma_{\infty}^{\infty}(\sf{A}_{\rho_{k}})$ with inverse $\mathcal{U}_{k}\:\Gamma_{\infty}^{\infty}(\sf{A}_{\rho_{k}})\to\mathrm{ker}(\square\vert_{\Gamma^{\infty}_{\infty}})$.
		\item[\emph{(ii)}]The retarded and advances Green operators are well-defined as linear operators of the form $\sf{G}_{k}^{\pm}\:\Gamma_{\mathrm{tc},\infty}^{\infty}(\sf{A}_{k})\to\Gamma_{\infty}^{\infty}(\sf{A}_{k})$ satisfying 
		\begin{align*}
        &\text{\emph{(i)}}\hspace*{1cm}\sf{G}_{k}^{\pm}\circ \square\vert_{\Gamma^{\infty}_{\mathrm{tc},\infty}}=\square\circ \sf{G}_{k}^{\pm}=\mathrm{id}_{\Gamma^{\infty}_{\mathrm{tc},\infty}}\\
        &\text{\emph{(ii)}}\hspace*{0.8cm}\mathrm{supp}(\sf{G}_{k}^{\pm}\psi)\subset \mathcal{J}^{\pm}(\mathrm{supp}(\psi))\hspace*{1cm}\forall \psi\in\Gamma^{\infty}_{\mathrm{tc},\infty}(\sf{A}_{k})\, ,
\end{align*}
where we wrote $\Gamma^{\infty}_{\mathrm{tc},\infty}(\sf{A}_{k}):=\Gamma^{\infty}_{\mathrm{tc}}(\sf{A}_{k})\cap\Gamma^{\infty}_{\infty}(\sf{A}_{k})$. Furthermore, the corresponding causal propagator $\sf{G}_{k}:=\sf{G}_{k}^{+}-\sf{G}_{k}^{-}\:\Gamma_{\mathrm{tc},\infty}^{\infty}(\sf{A}_{k})\to\Gamma^{\infty}_{\infty}(\sf{A}_{k})$ defines an exact sequence
	\begin{align*}
		0\xrightarrow{}\Gamma_{\mathrm{tc},\infty}^{\infty}(\sf{A}_{k})\xrightarrow{\square}\Gamma_{\mathrm{tc},\infty}^{\infty}(\sf{A}_{k})\xrightarrow{\sf{G}_{k}}\Gamma_{\infty}^{\infty}(\sf{A}_{k})\xrightarrow{\square}\Gamma_{\infty}^{\infty}(\sf{A}_{k})\, .
	\end{align*}
	\end{itemize}
\end{corollary}

\begin{proof}
	Claim (i) is a direct consequence of Theorem~\ref{Thm:CauchySob}. The first part of claim (ii) follows from Theorem~\ref{Thm:CauchySob} using similar arguments as in Example~\ref{Example:GreenHyp}, while the exact sequence can be established following the same steps as used in the proof of Proposition~\ref{Prop:ExactSeq}(iv). We refer to \cite[Prop.~4.7]{MurroSchmid} for more details.
\end{proof}
\subsection{Cauchy Radiation Gauge Revisited}
Let us now generalise Proposition~\ref{Prop:AchievCRG}, namely the achievability of the Cauchy radiation gauge, to ultrastatic spacetimes with \emph{non-compact} Cauchy surface. To this end, let us consider the following two subspaces of $\Gamma^{\infty}_{\infty}(\sf{A}_{1})$ and $\Gamma^{\infty}_{\infty}(\sf{A}_{\rho_{1}})$:
\begin{align*}
	\Gamma^{\infty}_{\infty,\d}(\sf{A}_{1})&:=\{\sf{A}\in\Gamma^{\infty}_{\infty}(\sf{A}_{1})\mid \sf{A}_{\Sigma}\vert_{t=0},\partial_{t}\sf{A}_{\Sigma}\vert_{t=0}\in\Omega^{1}_{\infty,\d}(\Sigma)\}\\
	\Gamma^{\infty}_{\infty,\d}(\sf{A}_{\rho_{1}})&:=\Omega_{\infty}^{0}(\Sigma)\oplus \Omega_{\infty}^{0}(\Sigma)\oplus\Omega_{\infty,\d}^{1}(\Sigma)\oplus\Omega_{\infty,\d}^{1}(\Sigma)\, .
\end{align*}
Two comments are on to place: first, the bijective initial data map $\rho_{1}\:\mathrm{ker}(\square\vert_{\Gamma^{\infty}_{\infty}})\to\Gamma^{\infty}_{\infty}(\sf{A}_{1})$ obviously restricts to a bijective map $\rho_{1}\:\mathrm{ker}(\square\vert_{\Gamma^{\infty}_{\infty,\d}})\to\Gamma^{\infty}_{\infty,\d}(\sf{A}_{1})$. Secondly, we stress that the conditions  $\sf{A}_{\Sigma}\vert_{t=0},\partial_{t}\sf{A}_{\Sigma}\vert_{t=0}\in\Omega^{1}_{\infty,\d}(\Sigma)$ are not propagating in general. In other words, if we consider the Cauchy problem $\square\sf{A}=0$ with $\rho_{1}(\sf{A})\in\Gamma^{\infty}_{\infty,\d}(\sf{A}_{1})$, the solution $\sf{A}\in\Gamma^{\infty}_{\infty,\d}(\sf{A}_{1})$ will in general satisfy the conditions $\sf{A}_{\Sigma}\vert_{t},\partial_{t}\sf{A}_{\Sigma}\vert_{t}\in\Omega^{1}_{\infty,\d}(\Sigma)$ just at time $t=0$.

\begin{proposition}\label{Prop:AchievCRG2}\emph{(Achievability of the Cauchy Radiation Gauge)}\newline
	Let $(\sf{M}=\bb{R}\times\Sigma,\sf{g}=-\d t\otimes\d t+\sf{h})$ be a globally hyperbolic ultrastatic spacetime such that $(\Sigma,\sf{h})$ is of bounded geometry. Then, the Cauchy radiation gauge $(\sf{R}_{\mathrm{Cauchy}})$ is an achievable and complete gauge-fixing in the following sense: for any $\sf{A}\in\mathrm{ker}(\sf{P}\vert_{\Gamma^{\infty}_{\infty}})$ there exists a unique $f\in C^{\infty}(\sf{M},\bb{C})$ (up to constant) such that $\sf{K} f\in\Gamma^{\infty}_{\infty,\d}(\sf{A}_{1})$ and such that $\sf{A}^{\prime}:=\sf{A}-\sf{K}f$ satisfies the gauge condition $(\sf{R}_{\mathrm{Cauchy}})$.
\end{proposition}

\begin{proof}
	Let $\sf{A}\in\Gamma^{\infty}_{\infty}(\sf{A}_{1})$ be such that $\sf{P}\sf{A}=0$. As in the proof of Proposition~\ref{Prop:AchievCRG}, we observe that the Cauchy radiation gauge, i.e.~$\delta\sf{A}=0$ and $\sf{A}_{\mathrm{T}}\vert_{t=0}=\partial_{t}\sf{A}_{\mathrm{T}}\vert_{t=0}=0$ is equivalent to the conditions $\delta\sf{A}=0$ and $\sf{A}_{\mathrm{T}}\vert_{t=0}=\delta_{\Sigma}\sf{A}_{\Sigma}\vert_{t=0}=0$ on account of $\delta\sf{A}=\partial_{t}\sf{A}_{\mathrm{T}}+\delta_{\Sigma}\sf{A}_{\Sigma}$, cf.~Lemma~\ref{Lemma:Dec}(ii). Now, the goal is to show that there exists a function $f\in C^{\infty}(\sf{M},\bb{C})$ such that $\sf{A}^{\prime}=\sf{A}-\sf{K} f$ satisfies this gauge condition. This is equivalent to finding a solution to the system
	\begin{align*}
		\begin{cases}
			\square f&=\delta\sf{A}\\
			\partial_{t}f\vert_{t=0}&=\sf{A}_{\mathrm{T}}\vert_{t=0}\\
			\Delta f\vert_{t=0}&=\delta_{\Sigma}\sf{A}_{\Sigma}\vert_{t=0}\, .
		\end{cases}
	\end{align*}
	Since $\sf{A}_{\Sigma}\vert_{t=0}\in\Omega^{1}_{\infty}(\Sigma)$, by assumption, Proposition~\ref{prop:Poisson} implies that there exists a unique solution $\mathfrak{f}\in C^{\infty}(\Sigma,\bb{C})$ (up to constant) satisfying $\d_{\Sigma}\mathfrak{f}\in\Omega^{1}_{\infty,\d}(\Sigma)$ of the equation $\Delta\mathfrak{f}=\delta_{\Sigma}\sf{A}_{\Sigma}\vert_{t=0}$. Hence, the above system is equivalent to
	\begin{align*}
		\begin{cases}
			\square f&=\delta\sf{A}\\
			\partial_{t}f\vert_{t=0}&=\sf{A}_{\mathrm{T}}\vert_{t=0}\\
			f\vert_{t=0}&=\mathfrak{f}\quad\text{mod}\quad\text{constant}\, .
		\end{cases}
	\end{align*}
	Hence, by Theorem~\ref{Thm:NHCauchy}, there exists a unique $f\in C^{\infty}(\Sigma,\bb{C})$ (up to constant) to this equation. Now, by construction, we only know that $\d_{\Sigma}f$ at time $t=0$ lies in $\Omega^{1}_{\infty}(\Sigma)$. It remains to show that in fact $\sf{K}f\in\Gamma^{\infty}_{\infty}(\sf{A}_{1})$, i.e.~that both $\partial_{t}f$ and $d_{\Sigma}f$ are in $\sf{H}^{\infty}$ at any fixed time. For this, we will make use of Theorem~\ref{Thm:CauchySob}. First, we observe that $\d f$ is a solution to the Cauchy problem
	\begin{align*}
		\begin{cases}
			\square\d f &=\d\delta\sf{A}\\
			(\d f)_{\mathrm{T}}\vert_{t=0}&=\partial_{t}f\vert_{t=0}=\sf{A}_{\mathrm{T}}\vert_{t=0}\\
			\partial_{t}(\d f)_{\mathrm{T}}\vert_{t=0}&=(\partial_{t}^{2}f)\vert_{t=0}=(-\Delta f+\delta\sf{A})\vert_{t=0}=\partial_{t}\sf{A}_{\mathrm{T}}\vert_{t=0}\\
			(\d_{\Sigma}f)\vert_{t=0}&=\d_{\Sigma}\mathfrak{f}\\
			\partial_{t}(\d_{\Sigma}f)\vert_{t=0}&=\d_{\Sigma}(\partial_{t}f)\vert_{t=0}=\d_{\Sigma}\sf{A}_{\mathrm{T}}\vert_{t=0}
		\end{cases}	\quad
		\begin{array}{l}
		\in\Gamma^{\infty}_{\infty}(\sf{A}_{1})\\[0.1cm]
		\in\Omega^{0}_{\infty}(\Sigma)\\[0.1cm]
		\in\Omega^{0}_{\infty}(\Sigma)\\[0.1cm]
		\in\Omega^{1}_{\infty,\d}(\Sigma)\\[0.1cm]
		\in\Omega^{1}_{\infty,\d}(\Sigma)
\end{array}		\,,	
	\end{align*}
	where we used that clearly $\d_{\Sigma}\Omega^{k}_{\infty}(\Sigma)\subset\Omega^{k+1}_{\infty,\d}(\Sigma)$ for all $k\in\bb{N}_{0}$. Hence, we conclude that $\d f\in\Gamma^{\infty}_{\infty,\d}(\sf{A}_{1})$, by Theorem~\ref{Thm:CauchySob}.
\end{proof}

Now, throughout the following, let us denote by $\sf{R}_{\Sigma}\:\Gamma^{\infty}(\sf{A}_{\rho_{1}})\to \Gamma^{\infty}(\sf{A}_{\rho_{1}})$ the linear differential operator defined by
\begin{align*}
	\sf{R}_{\Sigma}\:\Gamma^{\infty}(\sf{A}_{\rho_{1}})\to\Gamma^{\infty}(\sf{A}_{\rho_{1}})\,,\qquad \sf{R}_{\Sigma}=
		\begin{pmatrix}
			\mathrm{id} & 0 & 0 & 0\\
			0 & \mathrm{id} & 0 & 0\\
			0 & 0 & 0 & 0\\
			0 & 0 & 0 & 0
		\end{pmatrix}\, .
\end{align*}
By definition, $\sf{R}:=\mathcal{U}_{1}\sf{R}_{\Sigma}\rho_{1}\:\Gamma^{\infty}(\sf{A}_{1})\to\Gamma^{\infty}(\sf{A}_{1})$, is the differential operator that parametrises the \emph{Cauchy temporal gauge}, i.e.~$\sf{A}\in\mathrm{ker}(\sf{R})$ if and only if $\sf{A}_{\mathrm{T}}\vert_{t=0}=\partial_{t}\sf{A}_{\mathrm{T}}\vert_{t=0}=0$. Since the Cauchy radiation gauge can be achieved in suitable Sobolev spaces by Proposition~\ref{Prop:AchievCRG2}, a similar discussion as the one after Proposition~\ref{Prop:AchievCRG} at the end of Section~\ref{Subsec:GCMax}, yields the following:

\begin{corollary}\label{Corollary:PSSob} Let $(\sf{M},\sf{g})$ be as in Proposition~\ref{Prop:AchievCRG2}. Then, there exists a projection operator $\sf{T}_{\Sigma}\:\Gamma^{\infty}_{\infty}(\sf{A}_{1})\to\Gamma^{\infty}_{\infty}(\sf{A}_{1})$ such that the following diagram commutes:
\begin{equation*}
        \begin{tikzcd}
			\mathrm{Sol}_{\infty}:=\cfrac{\mathrm{ker}(\sf{P}\vert_{\Gamma^{\infty}_{\infty}})}{\mathrm{ran}(\sf{K}\vert_{\Gamma^{\infty}})\cap\Gamma^{\infty}_{\infty,\d}(\sf{A}_{1})}\arrow[d,swap,"\cong"]\arrow[d,"{[\rho_{1}]}"]\arrow[rr,swap,hookleftarrow,"{[\mathrm{id}]}"]&&\arrow[ll,swap,"\cong"]\mathrm{ker}(\square\vert_{\Gamma^{\infty}_{\infty}})\cap\mathrm{ker}(\sf{K}^{\ast}\vert_{\Gamma^{\infty}_{\infty}})\cap\mathrm{ker}(\sf{R}\vert_{\Gamma^{\infty}_{\infty}})\arrow[d,"\rho_{1}"]\arrow[d,swap,"\cong"]\\\mathcal{V}^{\infty}_{\Sigma}:=\cfrac{\mathrm{ker}(\sf{K}_{\Sigma}^{\dagger}\vert_{\Gamma_{\infty}^{\infty}})}{\mathrm{ran}(\sf{K}_{\Sigma}\vert_{\Gamma^{\infty}})\cap\Gamma^{\infty}_{\infty,\d}(\sf{A}_{\rho_{1}})} \arrow[rr,"{[\sf{T}_{\Sigma}]}"]\arrow[rr,swap,"\cong"] &&\mathcal{V}_{\sf{R}}^{\infty}:=\mathrm{ker}(\sf{K}_{\Sigma}^{\dagger}\vert_{\Gamma^{\infty}_{\infty}})\cap\mathrm{ker}(\sf{R}_{\Sigma}\vert_{\Gamma^{\infty}_{\infty}})\, .
\end{tikzcd}\end{equation*}
\end{corollary}

\begin{proof}
	The fact that $[\mathrm{id}]$, $[\rho_{1}]$ and $\rho_{1}$ are isomorphisms follows from Theorem~\ref{Thm:CauchySob} and similar arguments as in Proposition~\ref{Prop:CauchyGauge}. The operator $\sf{T}_{\Sigma}$ is defined as the projector that maps arbitrary $1$-form initial data $\mathfrak{c}\in\mathrm{ker}(\sf{K}^{\dagger}_{\Sigma}\vert_{\Gamma^{\infty}_{\infty}})$ into the unique representatives satisfying the Cauchy radiation gauge on the level of initial data.
\end{proof}

The space $\mathcal{V}_{\Sigma}^{\infty}$ can be seen as a space of observables on the level of initial data, similar to $\mathcal{V}_{\Sigma}$. The identity clearly induces a linear map $[\mathrm{id}]\:\mathcal{V}_{\Sigma}\to\mathcal{V}_{\Sigma}^{\infty}$, however, this map is in general neither injective nor surjective. In other words, there is no simple relation between $\mathcal{V}_{\Sigma}^{\infty}$ and 
\begin{align*}
	\mathcal{V}_{\Sigma}=\cfrac{\mathrm{ker}(\sf{K}_{\Sigma}^{\dagger}\vert_{\Gamma_{\infty}^{\mathrm{c}}})}{\mathrm{ran}(\sf{K}_{\Sigma}\vert_{\Gamma_{\mathrm{c}}^{\infty}})}\, .
\end{align*}
The space $\mathcal{V}_{\Sigma}^{\infty}$ can be seen as the class of observables allowing for the Cauchy radiation gauge, however, in the end this space will only play an auxiliary role in our construction and Hadamard states will be defined on $\mathcal{V}_{\Sigma}$. The space $\mathcal{V}^{\infty}_{\sf{R}}$ contains initial data $\mathfrak{c}\in\Gamma^{\infty}_{\infty}(\sf{A}_{\rho_{1}})$ such that $\sf{A}:=\mathcal{U}_{1}\mathfrak{c}\in\Gamma^{\infty}_{\infty}(\sf{A}_{1})$ satisfies the Cauchy radiation gauge. Explicitly, the space $\mathcal{V}^{\infty}_{\sf{R}}$ is given by
\begin{align*}
	\mathcal{V}^{\infty}_{\sf{R}}=\bigg\{(\mathfrak{a}_{\mathrm{T}},\pi_{\mathrm{T}},\mathfrak{a}_{\Sigma},\pi_{\Sigma})\in\Gamma^{\infty}_{\infty}(\sf{A}_{\rho_{1}})\mid \mathfrak{a}_{\mathrm{T}}=\pi_{\mathrm{T}}=\delta_{\Sigma}\mathfrak{a}_{\Sigma}=\delta_{\Sigma}\pi_{\Sigma}=0\}\, .
\end{align*}
The operator $\sf{T}_{\Sigma}\:\Gamma^{\infty}_{\infty}(\sf{A}_{\rho_{1}})\to \Gamma^{\infty}_{\infty}(\sf{A}_{\rho_{1}})$ is the corresponding gauge projector that maps initial data for $1$-form into the unique representative satisfying the Cauchy radiation gauge on the level of initial data. More precisely, for initial data $\mathfrak{c}\in\Gamma^{\infty}_{\infty}(\sf{A}_{1})$, the operator $\sf{T}_{\Sigma}$ acts as $\sf{T}_{\Sigma}\mathfrak{c}:=\mathfrak{c}+\sf{K}_{\Sigma}\mathfrak{f}$, where $\mathfrak{f}\in\Gamma^{\infty}(\sf{A}_{\rho_{0}})$ is the unique element with $\sf{K}_{\Sigma}\mathfrak{f}\in\Gamma^{\infty}_{\infty,\d}(\sf{A}_{\rho_{1}})$ such that $\mathfrak{c}+\sf{K}_{\Sigma}\mathfrak{f}\in\mathrm{ker}(\sf{R}_{\Sigma})$. By definition, $\sf{T}_{\Sigma}$ is clearly a projection operator, i.e.~$\sf{T}_{\Sigma}^{2}=\sf{T}_{\Sigma}$. Moreover, it holds that $\sf{T}_{\Sigma}\circ\sf{K}_{\Sigma}=0$, which reflects the fact that the Cauchy radiation gauge is a \emph{complete} gauge fixing on the chosen function spaces.

As a next step in our analysis, we derive an explicit expression of the operator $\sf{T}_{\Sigma}$. To start with, we recall the Hodge decomposition of smooth $\sf{H}^{\infty}$-forms from Corollary~\ref{Cor:HodgeDecomInf}, which states that there is a direct sum decomposition of the vector space $\Omega^{k}_{\infty}(\Sigma)$ given by
\begin{align}\label{eq:DecompoInf}
	\Omega^{k}_{\infty}(\Sigma)\cong \Omega^{k}_{\infty,\d}(\Sigma)\oplus\mathrm{ker}(\delta_{\Sigma}\vert_{\Omega^{k}_{\infty}})\,,
\end{align}
where the linear subspace $\mathrm{ker}(\delta_{\Sigma}\vert_{\Omega^{k}_{\infty}})$ further decomposes as $\mathrm{ker}(\delta_{\Sigma}\vert_{\Omega^{k}_{\infty}})\cong \Omega^{k}_{\infty,\delta}(\Sigma)\oplus\mathrm{ker}(\Delta_{\Omega^{k}_{\infty}})$. Let us denote the projector of $\Omega^{k}_{\infty}(\Sigma)$ onto $\mathrm{ker}(\delta_{\Sigma}\vert_{\Omega^{k}_{\infty}})$ in the following by $\Pi\:\Omega^{k}_{\infty}(\Sigma)\to\Omega^{k}_{\infty}(\Sigma)$. In other words, $\Pi$ is the unique projection operator defined by Eq.~\eqref{eq:DecompoInf} satisfying
\begin{align*}
	\Pi^{2}=\Pi\,,\qquad \mathrm{ker}(\Pi)=\Omega^{k}_{\infty,\d}(\Sigma)\,,\qquad\mathrm{ran}(\Pi)=\mathrm{ker}(\delta_{\Sigma}\vert_{\Omega^{k}_{\infty}})\, .
\end{align*}
Furthermore, the operator $\Pi$ has the following properties:

\begin{lemma}\label{Lemma:ProjectionsHodgeDecomp}
	Let $(\Sigma,\sf{h})$ be a complete Riemannian manifold. The projector $\Pi\:\Omega^{k}_{\infty}(\Sigma)\to\Omega^{k}_{\infty}(\Sigma)$ has the following properties:
\begin{itemize}
	\item[\emph{(i)}]$\Pi$ is orthogonal w.r.t.~$\langle\cdot,\cdot\rangle_{\sf{H}^{s}}$ for all $s\in [0,\infty)$.
	\item[\emph{(ii)}]$\Pi\Delta=\Delta\Pi$ on $\Omega^{k}_{\infty}(\Sigma)$.
	\item[\emph{(iii)}]For $k=1$, we have $\Pi=\mathrm{id}-\d_{\Sigma}\Delta^{-1}\delta_{\Sigma}$ on $\Omega^{1}_{\infty}(\Sigma)$, in which $\Delta$ is viewed as an operator
	\begin{align*}
		\Delta\:\{f\in C^{\infty}(\Sigma,\bb{C})\mid\d_{\Sigma}f\in\Omega^{1}_{\infty,\d}(\Sigma)\}/_{\sim_{\mathrm{c}}}\to\mathrm{ran}(\delta_{\Sigma}\vert_{\Omega^{1}_{\infty}})
	\end{align*}
	with $\sim_{\mathrm{c}}$ denoting the equivalence relation on $C^{\infty}(\Sigma,\bb{C})$ identifying functions that differ by a constant. By Proposition~\ref{prop:Poisson}, $\Delta$ is bijective with this domain and codomain.
\end{itemize}	 
\end{lemma}

\begin{proof}
	(i) follows from the fact that the decomposition in Eq.~\eqref{eq:DecompoInf} is $\sf{H}^{s}$-orthogonal for all $s\in [0,\infty)$, by Corollary~\ref{Cor:HodgeDecomInf}. 
	
	For (ii), let  $\omega\in\Omega^{k}_{\infty}(\Sigma)$ be arbitrary and  let us decompose it uniquely in accordance to  Eq.~\eqref{eq:DecompoInf} as
	\begin{align*}
		\omega=\alpha+\beta\qquad\text{with}\qquad \alpha\in\Omega^{k}_{\infty,\d}(\Sigma)\,,\quad\beta\in\mathrm{ker}(\delta_{\Sigma}\vert_{\Omega^{1}_{\infty}})\,.\end{align*}
By definition of $\Pi$, it holds that $\Pi\omega=\beta$ and hence $\Delta\Pi\omega=\Delta\beta$. On the other hand, clearly $\Delta\omega=\Delta\alpha+\Delta\beta$ and $\Delta\beta\in\ker(\delta_{\Sigma}\vert_{\Omega^{1}_{\infty}})$ on account of $\delta_{\Sigma}\Delta=\Delta\delta_{\Sigma}$. It remains to show that $\Delta\alpha\in\Omega^{1}_{\infty,\d}(\Sigma)$, since then $\Pi\Delta\omega=\Delta\beta$, which shows that $\Pi\Delta\omega= \Delta\beta=\Delta\Pi\omega$. By assumption, there exists a sequence $(\alpha_{n})_{n\in\mathbb{N}}$ in $\Omega^{k-1}_{\infty}(\Sigma)$ such that $(\d_{\Sigma}\alpha_{n})_{n\in\bb{N}}$ converges to $\alpha$ in $\sf{H}^{s}_{k}(\Sigma)$ for all $s\in [0,\infty)$. But now, it is easy to see that $\Delta$ is bounded as an operator $\Delta\:\Omega^{s+2}_{k}(\Sigma)\to\Omega^{s}_{k}(\Sigma)$, following similar steps as in the proof of Lemma~\ref{Lemma:Tec2}, which implies that the sequence $\Delta\d_{\Sigma}\alpha_{n}=\d_{\Sigma}\Delta\alpha_{n}$ converges as well and $\Delta\alpha=\sf{H}^{s}\text{-}\lim_{n\to\infty}\d_{\Sigma}\Delta\alpha$. We conclude that $\Delta\alpha\in\Omega^{k}_{\infty,\d}(\Sigma)$. 
	
	Claim (iii) follows from a straightforward computation: let $\omega=\alpha+\beta$ be as in the proof of (ii). By Corollary~\ref{Cor:HodgeDecomInf}, we can find a function $f\in C^{\infty}(\Sigma)$ such that $\alpha=\d_{\Sigma}f$. Then,
	\begin{align*}
		(\mathrm{id}-\d_{\Sigma}\Delta^{-1}\delta_{\Sigma})\omega=\omega-\d_{\Sigma}\Delta^{-1}\delta_{\Sigma}\omega=\omega-\d_{\Sigma}f=\omega-\alpha=\beta=\Pi\omega\,,
	\end{align*}
	where we used that $f$ is the unique (up to constant) solution to $\Delta f=\delta_{\Sigma}\omega$ with the property that $\d_{\Sigma}f\in\Omega^{1}_{\infty,\d}(\Sigma)$, by Proposition~\ref{prop:Poisson}.
\end{proof}

Using the projection operator $\Pi$, we can define a projection operator $\sf{T}_{\Sigma}\:\Gamma_{\infty}^{\infty}(\sf{A}_{\rho_{1}})\to \Gamma_{\infty}^{\infty}(\sf{A}_{\rho_{1}})$ that, when restricted to the subspace $\mathrm{ker}(\sf{K}_{\Sigma}^{\dagger}\vert_{\Gamma^{\infty}_{\infty}})$, provides the projection operator implicitly defined in Corollary~\ref{Corollary:PSSob}, i.e.~the operator that realises the Cauchy radiation gauge on the level of initial data.

\begin{proposition} \emph{(The Gauge Projector $\sf{T}_{\Sigma}$)}\label{Prop:ProjOp}\newline
	Let $(\sf{M},\sf{g})$ be an ultrastatic globally hyperbolic spacetime such that $(\Sigma,\sf{h})$ is of bounded geometry. Then, the projection operator $\sf{T}_{\Sigma}\:\Gamma_{\infty}^{\infty}(\sf{A}_{\rho_{1}})\to \Gamma_{\infty}^{\infty}(\sf{A}_{\rho_{1}})$ defined by
\begin{align*}
		\sf{T}_{\Sigma}=\begin{pmatrix}
	0&0&0&0\\0&0&0&0\\0 & 0 & \Pi& 0\\ 0&0 &0&\Pi
	\end{pmatrix}
	\end{align*}
	has the following properties:
	\begin{itemize}
	\item[\emph{(i)}] $\sf{T}_{\Sigma}^{2}=\sf{T}_{\Sigma}$  and $\sf{T}_{\Sigma}\vert_{\mathcal{V}_{\sf{R}}^{\infty}}=\mathrm{id}_{\mathcal{V}_{\sf{R}}^{\infty}}$
	\item[\emph{(ii)}]$\sf{T}_{\Sigma}=\mathrm{id}-\sf{K}_{\Sigma}(\sf{R}_{\Sigma}\sf{K}_{\Sigma})^{-1}\sf{R}_{\Sigma}$ on $\mathrm{ker}(\sf{K}_{\Sigma}^{\dagger}\vert_{\Gamma^{\infty}_{\infty}})$ and it has the following properties:
	\begin{itemize}
	\item[\emph{(iia)}] $\ker(\sf{T}_{\Sigma}\vert_{\mathrm{ker}(\sf{K}_{\Sigma}^{\dagger}\vert_{\Gamma^{\infty}_{\infty}})})=\mathrm{ran}(\sf{K}_{\Sigma})\cap\Gamma^{\infty}_{\infty,\d}(\sf{A}_{\rho_{1}})$
	\item[\emph{(iib)}] $\mathrm{ran}(\sf{T}_{\Sigma}\vert_{\ker(\sf{K}_{\Sigma}^{\dagger}\vert_{\Gamma^{\infty}_{\infty}})})=\mathcal{V}^{\infty}_{\sf{R}}$
	\end{itemize}
	\end{itemize}
	In particular, (ii) implies that $\sf{T}_{\Sigma}$ is well-defined and bijective as a map $\sf{T}_{\Sigma}\:\mathcal{V}_{\Sigma}^{\infty}\to\mathcal{V}^{\infty}_{\sf{R}}$.
\end{proposition}

\begin{proof}
For (i), the claim that $\sf{T}_{\Sigma}^{2}=\sf{T}_{\Sigma}$ follows from $\Pi^{2}=\Pi$, while $\sf{T}_{\Sigma}\vert_{\mathcal{V}_{\sf{R}}^{\infty}}=\mathrm{id}$ follows from the fact that $\Pi=\mathrm{id}$ on $\mathrm{ker}(\delta_{\Sigma}\vert_{\Omega^{1}_{\infty}})$ and 
\begin{align*}
\mathcal{V}^{\infty}_{\sf{R}}=\bigg\{(\mathfrak{a}_{\mathrm{T}},\pi_{\mathrm{T}},\mathfrak{a}_{\Sigma},\pi_{\Sigma})\in\Gamma^{\infty}_{\infty}(\sf{A}_{\rho_{1}})\mid \mathfrak{a}_{\mathrm{T}}=\pi_{\mathrm{T}}=\delta_{\Sigma}\mathfrak{a}_{\Sigma}=\delta_{\Sigma}\pi_{\Sigma}=0\}	\, .
\end{align*}

Let us now turn to (ii): by Proposition~\ref{Prop:KSigma} (in the ultrastatic case, i.e.~$\sf{k}=0$ and $\beta=1$), the operator $\sf{R}_{\Sigma}\sf{K}_{\Sigma}\:\Gamma^{\infty}(\sf{A}_{\rho_{0}})\to\Gamma^{\infty}(\sf{A}_{\rho_{1}})$ is given by 
\begin{align*}
        \sf{R}_{\Sigma}\sf{K}_{\Sigma}=\begin{pmatrix}0& i\cdot\mathrm{id}\\ i\Delta & 0\\ 0&0\\0&0\end{pmatrix}\, .
\end{align*}
Now, following Proposition~\ref{prop:Poisson}, this operator is well-defined and bijective as an operator
\begin{align*}
	\sf{R}_{\Sigma}\sf{K}_{\Sigma}\:\{(\mathfrak{a},\pi)\in\Gamma^{\infty}(\sf{A}_{\rho_{0}})\mid \d_{\Sigma}\mathfrak{a}\in\Omega^{1}_{\infty,\d}(\Sigma)\}/_{\sim_{\mathrm{c}}}\to\{(\mathfrak{a},\pi,0,0)\in\Gamma^{\infty}(\sf{A}_{\rho_{1}})\mid \pi\in\mathrm{ran}(\delta_{\Sigma}\vert_{\Omega^{1}_{\infty}})\}\,,
\end{align*}
where $\sim_{\mathrm{c}}$ denotes the equivalence relation on $\Gamma^{\infty}(\sf{A}_{\rho_{0}})$ identifying $(\mathfrak{a}_{0},\pi_{0})$ and $(\mathfrak{a}_{1},\pi_{1})$ if and only if $\mathfrak{a}_{0}$ and $\mathfrak{a}_{1}$ differ by a constant. It follows that $(\sf{R}_{\Sigma}\sf{K}_{\Sigma})^{-1}\sf{R}_{\Sigma}$ is well-defined on $\mathrm{ker}(\sf{K}_{\Sigma}^{\dagger}\vert_{\Gamma^{\infty}_{\infty}})$. Furthermore, it is straightforward to verify that 
\begin{align*}
		\sf{K}_{\Sigma}(\sf{R}_{\Sigma}\sf{K}_{\Sigma})^{-1}\sf{R}_{\Sigma}\begin{pmatrix}\mathfrak{a}_{\mathrm{T}}\\\pi_{\mathrm{T}}\\\mathfrak{a}_{\Sigma}\\\pi_{\Sigma}\end{pmatrix}=\begin{pmatrix}
		\mathfrak{a}_{\mathrm{T}}\\\pi_{\mathrm{T}}\\-\d_{\Sigma}\Delta^{-1}(i\pi_{\mathrm{T}})\\-i\d_{\Sigma}\mathfrak{a}_{0}
		\end{pmatrix}=\begin{pmatrix}
		\mathfrak{a}_{\mathrm{T}}\\\pi_{\mathrm{T}}\\\d_{\Sigma}\Delta^{-1}(\delta_{\Sigma}a_{\Sigma})\\\d_{\Sigma}\Delta^{-1}(\delta_{\Sigma}\pi_{\Sigma})
		\end{pmatrix}\, ,
	\end{align*}
for all $(\mathfrak{a}_{\mathrm{T}},\pi_{\mathrm{T}},\mathfrak{a}_{\Sigma},\pi_{\Sigma})\in\ker(\sf{K}_{\Sigma}\vert_{\Gamma^{\infty}_{\infty}})$, where $\Delta^{-1}$ is as defined in Lemma~\ref{Lemma:ProjectionsHodgeDecomp}(iii) and where we used Proposition~\ref{Prop:KSigma} to conclude that $i\pi_{\mathrm{T}}=-\delta_{\Sigma}\mathfrak{a}_{\Sigma}$ as well as $i\Delta\mathfrak{a}_{\mathrm{T}}=-\delta_{\Sigma}\pi_{\Sigma}$, which implies $\Delta^{-1}\delta_{\Sigma}\pi_{\Sigma}=-i\mathfrak{a}_{\mathrm{T}}$ where $a_{\mathrm{T}}$ is such that $\d_{\Sigma}\mathfrak{a}_{\mathrm{T}}\in\Omega^{1}_{\infty,\d}(\Sigma)$.  We conclude that $\sf{T}_{\Sigma}=\mathrm{id}-\sf{K}_{\Sigma}(\sf{R}_{\Sigma}\sf{K}_{\Sigma})^{-1}\sf{R}_{\Sigma}$ on $\mathrm{ker}(\sf{K}_{\Sigma}^{\dagger}\vert_{\Gamma^{\infty}_{\infty}})$ by Proposition~\ref{Prop:KSigma}.

It remains to check (iia) and (iib). Both of these equation follow essentially from the fact that $\sf{T}_{\Sigma}=\mathrm{id}-\sf{K}_{\Sigma}(\sf{R}_{\Sigma}\sf{K}_{\Sigma})^{-1}\sf{R}_{\Sigma}$ on $\mathrm{ker}(\sf{K}_{\Sigma}^{\dagger}\vert_{\Gamma^{\infty}_{\infty}})$ and the definition of $\Pi$. However, we can also show them directly: for (iia), we note that $\mathfrak{c}_{1}:=(\mathfrak{a}_{\mathrm{T}},\pi_{\mathrm{T}},\mathfrak{a}_{\Sigma},\pi_{\Sigma})\in\Gamma^{\infty}_{\infty}(\sf{A}_{\rho_{1}})$ satisfies $\sf{T}_{\Sigma}\mathfrak{c}_{1}=\sf{K}_{\Sigma}^{\dagger}\mathfrak{c}_{1}=0$ if and only if $\mathfrak{a}_{\Sigma},\pi_{\Sigma}\in\Omega^{1}_{\infty,\d}(\Sigma)$ and $i\pi_{\mathrm{T}}+\delta_{\Sigma}\mathfrak{a}_{\Sigma}=0$ as well as $i\Delta\mathfrak{a}_{\mathrm{T}}+\delta_{\Sigma}\pi_{\Sigma}=0$. In particular, this implies that $\pi_{\Sigma}=-i\d_{\Sigma}\mathfrak{a}_{\mathrm{T}}$, since $\mathfrak{a}_{\mathrm{T}}\in\Omega^{0}_{\infty}(\Sigma)$ and hence trivially $\d_{\Sigma}\mathfrak{a}_{\mathrm{T}}\in\Omega^{1}_{\infty,\d}(\Sigma)$, as well as $\mathfrak{a}_{\Sigma}=\d_{\Sigma}\mathfrak{f}$ for some $\mathfrak{f}\in C^{\infty}(\Sigma)$ with $\d_{\Sigma}\mathfrak{f}\in\Omega^{1}_{\infty,\d}(\Sigma)$. Clearly, $\mathfrak{c}_{0}:=(\mathfrak{f},-i\mathfrak{a}_{\mathrm{T}})$ satisfies $\sf{K}_{\Sigma}\mathfrak{c}_{0}\in\Gamma^{\infty}_{\infty,\d}(\sf{A}_{\rho_{0}})$ and $\mathfrak{c}_{1}=\sf{K}_{\Sigma}\mathfrak{c}_{0}$. The other direction follows from reversing these computations. Claim (iib) follows immediately from $\sf{T}_{\Sigma}=\mathrm{id}$ on $\mathcal{V}_{\sf{R}}^{\infty}$ and the fact that $\mathrm{ran}(\Pi)=\mathrm{ker}(\delta_{\Sigma}\vert_{\Omega^{1}_{\infty}})$.
\end{proof}

\subsection{Construction of States on Ultrastatic Spacetimes}
As next step, we construct Hadamard states for Maxwell's theory on ultrastatic spacetimes. More precisely, let $(\sf{M}=\bb{R}\times\Sigma,\sf{g}=-\d t\otimes\d t+\sf{h})$ be an ultrastatic globally hyperbolic spacetime, where we assume that $(\Sigma,\sf{h})$ is of bounded geometry, as above.

Now, following the discussion of Section~\ref{Subsec:CauchyMax}, we recall that we have equipped the spaces of initial data, $\Gamma^{\infty}(\sf{A}_{\rho_{i}})$, with the (positive-definite) inner products
\begin{align*}
		\bigg(
		\begin{pmatrix}
			\mathfrak{a}\\\pi
		\end{pmatrix},
		\begin{pmatrix}
			\mathfrak{b}\\\rho
		\end{pmatrix}\bigg)_{\sf{A}_{\rho_{0}}}:=&\int_{\Sigma}\bigg(\overline{\mathfrak{a}}\mathfrak{b}+\overline{\pi}\rho\bigg)\,\d\mu_{\sf{h}}=\langle\mathfrak{a},\mathfrak{b}\rangle_{\sf{L}^{2}}+\langle\pi,\rho\rangle_{\sf{L}^{2}}\\
		\bigg(
		\begin{pmatrix}
			\mathfrak{a}_{\mathrm{T}}\\\pi_{\mathrm{T}}\\\mathfrak{a}_{\Sigma}\\\pi_{\Sigma}
		\end{pmatrix},
		\begin{pmatrix}
			\mathfrak{b}_{\mathrm{T}}\\\rho_{\mathrm{T}}\\\mathfrak{b}_{\Sigma}\\\rho_{\Sigma}
		\end{pmatrix}\bigg)_{\sf{A}_{\rho_{1}}}:=&\int_{\Sigma}\bigg(\overline{\mathfrak{a}_{\mathrm{T}}}\mathfrak{b}_{\mathrm{T}}+\overline{\pi_{\mathrm{T}}}\rho_{\mathrm{T}}+\sf{h}^{\sharp}(\overline{\mathfrak{a}_{\Sigma}},\mathfrak{b}_{\Sigma})+\sf{h}^{\sharp}(\overline{\pi_{\Sigma}},\rho_{\Sigma})\bigg)\,\d\mu_{\sf{h}}=\\=&\langle\mathfrak{a}_{\mathrm{T}},\mathfrak{b}_{\mathrm{T}}\rangle_{\sf{L}^{2}}+\langle\pi_{\mathrm{T}},\rho_{\mathrm{T}}\rangle_{\sf{L}^{2}}+\langle\mathfrak{a}_{\Sigma},\mathfrak{b}_{\Sigma}\rangle_{\sf{L}^{2}}+\langle\pi_{\Sigma},\rho_{\Sigma}\rangle_{\sf{L}^{2}}\, ,
\end{align*}
which are the component-wise $\sf{L}^{2}$-inner products on $(\Sigma,\sf{h})$. With this notation, we have further introduce the corresponding Hermitian sesquilinear forms $\sigma_{\Sigma,i}$ on $\Gamma^{\infty}(\sf{A}_{\rho_{i}})$ defined by
\begin{align*}
	\sigma_{\Sigma,i}(\mathfrak{f},\mathfrak{g}):=i(\mathfrak{f},\sf{G}_{\Sigma,i}\mathfrak{g})_{\sf{A}_{\rho_{i}}}\,,
\end{align*}
where the operators $\sf{G}_{\Sigma,i}\:\Gamma^{\infty}(\sf{A}_{\rho_{i}})\to\Gamma^{\infty}(\sf{A}_{\rho_{i}})$ are zeroth-order operators that take the form
\begin{align*}
	\sf{G}_{\Sigma,0}=\frac{1}{i}
		\begin{pmatrix}
			0 & \mathrm{id}\\
			\mathrm{id} & 0
		\end{pmatrix}\,,\qquad \sf{G}_{\Sigma,1}=\frac{1}{i}
		\begin{pmatrix}
			0 & -\mathrm{id} & 0 & 0\\
			-\mathrm{id} & 0 & 0 & 0\\
			0 & 0 & 0 & \mathrm{id}\\
			0 & 0 & \mathrm{id} & 0
		\end{pmatrix}\,,
\end{align*}
as derived in Proposition~\ref{Prop:GSigma}. Now, consider the the phase spaces $\mathcal{V}_{\Sigma}$ and $\mathcal{V}_{\Sigma}^{\infty}$ defined by
\begin{align*}
	\mathcal{V}_{\Sigma}=\cfrac{\mathrm{ker}(\sf{K}_{\Sigma}^{\dagger}\vert_{\Gamma_{\infty}^{\mathrm{c}}})}{\mathrm{ran}(\sf{K}_{\Sigma}\vert_{\Gamma_{\mathrm{c}}^{\infty}})}\,,\qquad\qquad \mathcal{V}^{\infty}_{\Sigma}:=\cfrac{\mathrm{ker}(\sf{K}_{\Sigma}^{\dagger}\vert_{\Gamma_{\infty}^{\infty}})}{\mathrm{ran}(\sf{K}_{\Sigma}\vert_{\Gamma^{\infty}})\cap\Gamma^{\infty}_{\infty,\d}(\sf{A}_{\rho_{1}})}\, .
\end{align*}
Following Proposition~\ref{Prop:EqPS}, we know that $\sigma_{\Sigma,1}$ gives rise to a well-defined Hermitian sesquilinear form on the space $\mathcal{V}_{\Sigma}$. The space $(\mathcal{V}_{\Sigma},\sigma_{\Sigma,1})$, in turn, is unitary equivalent to the \emph{classical phase space} $(\mathcal{V}_{\mathrm{c}},\sigma)$ on the level of spacetime fields, see Eq.~\eqref{eq:sdfsafangagasge} and the discussion for general linear gauge theories in Proposition~\ref{Prop:EqPS}. Now, it is not too hard to see that $\sigma_{\Sigma,1}$ is also well-defined on the space $\mathcal{V}^{\infty}_{\Sigma}$ defined using $\sf{H}^{\infty}$-initial data.

\begin{lemma} The map $\sigma_{\Sigma,1}$ induces a well-defined Hermitian sesquilinear form on the phase space $\mathcal{V}_{\Sigma}^{\infty}$ introduced in Corollary~\ref{Corollary:PSSob}.
\end{lemma}

\begin{proof}
	We only need to check well-definedness: let $\mathfrak{f}\in\mathrm{ker}(\sf{K}^{\dagger}_{\Sigma}\vert_{\Gamma^{\infty}_{\infty}})$ and $\sf{g}=\sf{K}_{\Sigma}\mathfrak{h}$ for some $\mathfrak{h}\in\Gamma^{\infty}(\sf{A}_{\rho_{0}})$ with $\sf{K}_{\Sigma}\mathfrak{h}\in\Gamma^{\infty}_{\infty}(\sf{A}_{\rho_{1}})$. Now, naively, we would like to perform the computation
	\begin{align*}
		\sigma_{\Sigma,1}(\mathfrak{f},\mathfrak{g})=\sigma_{\Sigma,1}(\mathfrak{f},\sf{K}_{\Sigma}\mathfrak{h})=\sigma_{\Sigma,0}(\sf{K}_{\Sigma}^{\dagger}\mathfrak{f},\mathfrak{h})=0\,,
	\end{align*}
	however, the right-hand side is not necessarily well-defined, since $\mathfrak{h}$ is only smooth and not necessarily square integrable. However, the property $\sf{K}_{\Sigma}\sf{h}\in\Gamma^{\infty}_{\infty}(\sf{A}_{\rho_{1}})$ allows us to use similar arguments. More precisely, let us write $\mathfrak{h}=(\mathfrak{a},\pi)$ and $\mathfrak{f}=(\mathfrak{a}_{\mathrm{T}},\pi_{\mathrm{T}},\mathfrak{a}_{\Sigma},\pi_{\Sigma})$. By assumption, $\mathfrak{a}_{\mathrm{T}},\pi_{\mathrm{T}}\in\Omega^{0}_{\infty}(\Sigma)$, $\mathfrak{a}_{\Sigma},\pi_{\Sigma}\in\Omega^{1}_{\infty}(\Sigma)$ as well as $\mathfrak{a},\pi\in C^{\infty}(\Sigma,\bb{C})$. Moreover, by Proposition~\ref{Prop:KSigma}, we know that $\mathfrak{g}=\sf{K}_{\Sigma}\mathfrak{h}=(i\pi,i\Delta\mathfrak{a},\d_{\Sigma}\mathfrak{a},\d_{\Sigma}\pi)$ and hence, by assumption, $\pi\in\Omega^{0}_{\infty}(\Sigma)$ and $\d_{\Sigma}\mathfrak{a}=\lim_{n\to\infty}\d_{\Sigma}\mathfrak{a}_{n}$ for some sequence $(\mathfrak{a}_{n})_{n\in\bb{N}}$ in $\Omega^{0}_{\infty}(\Sigma)$. It follows that
	\begin{align*}
		\sigma_{\Sigma,1}(\mathfrak{f},\sf{K}_{\Sigma}\mathfrak{h})&=-\frac{1}{i}\langle\mathfrak{a}_{\mathrm{T}},i\Delta\mathfrak{a}\rangle_{\sf{L}^{2}}-\frac{1}{i}\langle\pi_{\mathrm{T}},i\pi\rangle_{\sf{L}^{2}}+\frac{1}{i}\langle\mathfrak{a}_{\Sigma},\d_{\Sigma}\pi\rangle_{\sf{L}^{2}}+\frac{1}{i}\langle\pi_{\Sigma},\d_{\Sigma}\mathfrak{a}\rangle_{\sf{L}^{2}}\\&=\lim_{n\to\infty}\bigg(\frac{1}{i}\langle i\mathfrak{a}_{\mathrm{T}},\Delta\mathfrak{a}_{n}\rangle_{\sf{L}^{2}}+\frac{1}{i}\langle i\pi_{\mathrm{T}},\pi\rangle_{\sf{L}^{2}}+\frac{1}{i}\langle\mathfrak{a}_{\Sigma},\d_{\Sigma}\pi\rangle_{\sf{L}^{2}}+\frac{1}{i}\langle\pi_{\Sigma},\d_{\Sigma}\mathfrak{a}_{n}\rangle_{\sf{L}^{2}}\bigg)\\&=\lim_{n\to\infty}\bigg(\frac{1}{i}\langle i\Delta\mathfrak{a}_{\mathrm{T}}+\delta_{\Sigma}\pi_{\Sigma},\mathfrak{a}_{n}\rangle_{\sf{L}^{2}}+\frac{1}{i}\langle i\pi_{\mathrm{T}}+\delta_{\Sigma}\mathfrak{a}_{\Sigma},\pi\rangle_{\sf{L}^{2}}\bigg)\\&=\lim_{n\to\infty}\sigma_{\Sigma,0}\bigg(\sf{K}_{\Sigma}^{\dagger}\mathfrak{f},\begin{pmatrix}
		\mathfrak{a}_{n}\\\pi
		\end{pmatrix}\bigg)=\lim_{n\to\infty}0=0\,,
	\end{align*}
	which proves that $\sigma_{\Sigma,1}$ is well-defined on the quotient space $\mathcal{V}_{\Sigma}^{\infty}$.
\end{proof}

Now, consider the \emph{gauge projector} $\sf{T}_{\Sigma}\:\Gamma^{\infty}_{\infty}(\sf{A}_{\rho_{1}})\to \Gamma^{\infty}_{\infty}(\sf{A}_{\rho_{1}})$ as defined in Proposition~\ref{Prop:ProjOp}. We equip the space $\mathcal{V}_{\sf{R}}^{\infty}$, as defined in Corollary~\ref{Corollary:PSSob}, i.e.~the space of initial data in the Cauchy radiation gauge, with a suitable sesquilinear Hermitian form $\sigma_{\sf{R}}\:\mathcal{V}^{\infty}_{\sf{R}}\times\mathcal{V}^{\infty}_{\sf{R}}\to\bb{C}$ to obtain a unitary isomorphism 
\begin{align}\label{eq:jjjjj}
\bigg(\mathcal{V}^{\infty}_{\Sigma}:=\cfrac{\mathrm{ker}(\sf{K}_{\Sigma}^{\dagger}\vert_{\Gamma_{\infty}^{\infty}})}{\mathrm{ran}(\sf{K}_{\Sigma}\vert_{\Gamma^{\infty}})\cap\Gamma^{\infty}_{\infty,\d}(\sf{A}_{\rho_{1}})},\,\sigma_{\Sigma,1}\bigg)\xrightarrow{[\sf{T}_{\Sigma}]} \big(\mathcal{V}_{\sf{R}}^{\infty}:=\mathrm{ker}(\sf{K}_{\Sigma}^{\dagger}\vert_{\Gamma^{\infty}_{\infty}})\cap\mathrm{ker}(\sf{R}_{\Sigma}\vert_{\Gamma^{\infty}_{\infty}})\,,\sigma_{\sf{R}}\big)\, .
\end{align}
It is easy to see that the induced sesquilinear form $\sigma_{\sf{R}}$ is simply the restriction to the respective subspace, i.e.~$\sigma_{\sf{R}}(\mathfrak{f},\mathfrak{g})=(\mathfrak{f},i\sf{G}_{\sf{R}}\mathfrak{g})_{\sf{A}_{\rho_{1}}}$, where the operator $\sf{G}_{\sf{R}}\:\Gamma^{\infty}(\sf{A}_{\rho_{1}})\to\Gamma^{\infty}(\sf{A}_{\rho_{1}})$ is given by
\begin{align*}
	\sf{G}_{\sf{R}}:=\frac{1}{i}
		\begin{pmatrix}
			0 & 0 & 0 & 0\\
			0 & 0 & 0 & 0\\
			0 & 0 & 0 & \mathrm{id}\\
			0 & 0 & \mathrm{id} & 0
		\end{pmatrix}\, .
\end{align*}
Indeed, given $\mathfrak{f},\mathfrak{g}\in\mathcal{V}_{\sf{R}}^{\infty}$, we obtain corresponding equivalence classes $[\mathfrak{f}],[\mathfrak{g}]\in\mathcal{V}_{\Sigma}^{\infty}$ and, by definition, $\sf{T}_{\Sigma}[\mathfrak{f}]=\mathfrak{f}$ and $\sf{T}_{\Sigma}[\mathfrak{g}]=\mathfrak{g}$. In particular, using the above definitions, it clearly holds that $\sigma_{1,\Sigma}([\mathfrak{f}],[\mathfrak{g}])=\sigma_{\sf{R}}(\sf{T}_{\Sigma}[\mathfrak{f}],\sf{T}_{\Sigma}[\mathfrak{g}])$.\bigskip

Now, having set up the notation for the relevant phase spaces, let us turn to the construction of Hadamard states on the ultrastatic spacetime $(\sf{M},\sf{g})$. As a first step, we show that the hyperbolic operator $\sf{D}_{2}=\sf{P}+\sf{K}^{\ast}\sf{K}=\square$ admits a \emph{microlocal factorisation}, which corresponds to the splitting into \emph{positive} and \emph{negative energy solutions} briefly sketched in Section~\ref{Sec:BiblioStates}. By assumption, $(\Sigma,\sf{h})$ is a Riemannian manifold of \emph{bounded geometry}. On those class of manifolds, there is a well-behaved class of \textit{pseudodifferential operators}, which we review in Section~\ref{Subsec:BoundGeom} of the appendix in some detail. For convenience of the reader, let us recall the basic set-up and central ideas. We refer to  Section~\ref{Subsec:BoundGeom} for a detailed discussion.

The manifold $(\Sigma,\sf{h})$ is of \emph{bounded geometry} (see Definition~\ref{Definition:BoundGeom}), if its injectivity radius is non-zero and if its Riemannian curvature tensor together with all its covariant derivatives is (uniformly) bounded with respect to the norm induced by the metric $\sf{h}$. An equivalent characterisation, which is useful in practise, is to require that there exists a local chart $\varphi_{p}\:\mathcal{U}_{p}\to \mathcal{B}_{1}(0)$ around every $p\in\sf{M}$, where $\mathcal{B}_{1}(0)\subset\mathbb{R}^{d}$ denotes the open ball around $0$ with radius $1$, such that the family $(\varphi_{p}^{\ast}\sf{h})_{p\in\sf{M}}$ is bounded in the space of (uniformly) bounded $(0,2)$-tensor fields on the Euclidean manifold $(\bb{R}^{d},\delta)$ equipped with its natural Fréchet topology, and such that 
\begin{align*}
	\exists C>0:\qquad C^{-1}\delta\leq \sf{g}_{p}\leq C\delta
\end{align*}
for all $p\in\sf{M}$. In other words, around every point $p$, the metric $\sf{g}$ behaves like the Euclidean metric $\delta$, while being controlled in a uniform way. Now, consider a vector bundle of \emph{bounded geometry}, such as $\sf{A}_{\Sigma,k}=\underline{\bb{C}}_{\Sigma}\otimes\bigwedge^{k}\sf{T}^{\ast}\Sigma$, i.e.~a vector bundle for which there is an atlas of bounded charts whose corresponding transition functions are a bounded family of matrix-valued functions. For such bundles, we can define a well-behaved class of \emph{symbols} $\mathcal{S}_{\mathrm{bd}}^{m}(\sf{M},\sf{E})\subset\Gamma^{\infty}(\sf{T}^{\ast}\sf{M},\pi^{\ast}\mathrm{End}(\sf{E}))$ (see Definition~\ref{Definition:BoundSym}), where $\pi\:\sf{T}^{\ast}\sf{M}\to\sf{M}$ denotes the bundle projection, which are a subset of the usual \emph{classical} (or \emph{polyhomogenous}) $\sf{E}$-valued symbols $\mathcal{S}^{m}_{\mathrm{cl}}(\sf{M},\sf{E})$ on $\sf{M}$ with the additional property that the family $(\varphi_{p}^{\ast}a)_{p\in\sf{M}}$ for some $a\in\mathcal{S}^{m}_{\mathrm{bd}}(\sf{M},\sf{E})$ is bounded in the Fréchet space $\mathcal{S}_{\mathrm{cl}}^{m}(\mathcal{B}_{1}(0),\bb{C}^{\mathrm{N}\times\mathrm{N}})$ with $\mathrm{N}:=\mathrm{rank}_{\bb{C}}(\sf{E})$. Having defined a suitable notion of symbols, we can define a corresponding pseudodifferential calculus via the usual (Kohn-Nirenberg) quantisation procedure, i.e.~we define an operator $\mathrm{Op}_{\mathrm{bd}}(a)\:\Gamma^{\infty}_{\mathrm{c}}(\sf{E})\to\Gamma^{\infty}(\sf{E})$ for a given symbol $a\in\mathcal{S}^{m}_{\mathrm{bd}}(\sf{M},\sf{E})$ by patching together local constructions using a (bounded) partition of unity. One can show that the definition of this \emph{quantisation map} is independent of the relevant choices up to a smoothing corrections and gives rise to the space of \emph{bounded pseudodifferential operators} (see Definition~\ref{Definition:PSIDOBoun}), i.e.
\begin{align*}
		\Psi^{m}_{\mathrm{bd}}(\sf{M},\sf{E}):=\mathrm{Op}_{\mathrm{bd}}(\mathcal{S}_{\mathrm{bd}}^{m}(\sf{M},\sf{E}))+\mathcal{W}^{-\infty}(\sf{M},\sf{E})\,.
	\end{align*}
	where the space $\mathcal{W}^{-\infty}(\sf{M},\sf{E})\subset\Psi^{-\infty}(\sf{M},\sf{E})$ is a suitable ideal of (bounded) smoothing operators (see Eq.~\eqref{eq:BoundedSmoothing}). By definition, every \emph{properly-supported} and \emph{classical} pseudodifferential operator on $\Sigma$ is also bounded and hence an element of $\Psi^{m}_{\mathrm{bd}}(\sf{M},\sf{E})$. However, the space of bounded pseudodifferential operators is strictly larger and intermediate between the spaces of properly-supported classical pseudodifferential operators and classical pseudodifferential operators on $\sf{M}$, while still being ``small enough'' to be closed under composition, taking adjoints and other natural operations on the space of pseudodifferential operators. 	

Coming back to the microlocal factorisation, let us start with the following technical lemma.

\begin{lemma}\label{lem:commuting}
	Let $(\Sigma,h)$ be a Riemannian manifold of bounded geometry, $\Pi$ be the projector defined in Lemma~\ref{Lemma:ProjectionsHodgeDecomp} and let $\Delta$ be the Hodge-Laplacian acting on $i$-forms for $i=0,1$.  Then, there exist operators $\varepsilon_{i}\in \Psi^{1}_{\mathrm{bd}}(\Sigma,\sf{A}_{\Sigma,i})$ and smoothing operators $r_{i,-\infty}\in\Psi^{-\infty}(\Sigma,\sf{A}_{\Sigma,i})$ such that the following holds true:
	\begin{itemize}
		\item[\emph{(i)}]$\varepsilon_{i}^{2}=\Delta+r_{i,-\infty}$ and $\varepsilon_{k}^{\ast}=\varepsilon_{k}$ with respect to the de Rham-Hodge inner product $\langle\cdot,\cdot\rangle_{\sf{L}^{2}}$.
		\item[\emph{(ii)}]The principal symbol of $\varepsilon_{i}$ is given by $\sigma_{\varepsilon_{i}}(\xi)=(\sf{h}^{\sharp}(\xi,\xi))^{\frac{1}{2}}\mathrm{id}_{\sf{A}_{\Sigma,i}}$.
		\item[\emph{(iii)}]Choosing the domain $\sf{H}^{1}_{i}(\Sigma)$, the operator $\varepsilon$ is closed in the Hilbert space $\sf{L}^{2}_{i}(\Sigma)$, elliptic, positive and invertible with inverse $\varepsilon_{i}^{-1}\in\Psi^{-1}_{\mathrm{bd}}(\Sigma,\sf{A}_{\Sigma,i})$.
		\item[\emph{(iv)}]$\varepsilon_{1}$ commutes with $\Pi$ up to smoothing, i.e. $\varepsilon_{1} \Pi-\Pi\varepsilon_{1}\in\Psi^{-\infty}(\Sigma,\sf{A}_{\Sigma,1})$.
\end{itemize}		
\end{lemma}

\begin{proof}[Proof of Lemma~\ref{lem:commuting}.]
	The fact that there exists an operator $\varepsilon_{i}\in \Psi^{1}_{\mathrm{bd}}(\Sigma,\sf{A}_{\Sigma,i})$ satisfying properties (i)-(iii) has been shown in~\cite[Prop.~5.11]{GerardOulghaziWrochna} in the scalar case and in \cite[Lemma~5.1]{GerardMurroWrochna} for the general case for operators taking values in a vector bundle. In fact, the construction on those references is more general and applies to general formally self-adjoint  (and possibly time-dependent) operators with the same principal symbol as the Laplacian. Let us briefly sketch the main arguments. Recall that $\Delta$ is well-defined and self-adjoint as an operator $\Delta\:\sf{H}^{2}_{i}(\Sigma)\to\sf{L}^{2}_{i}(\Sigma)$. Now, fix $\chi\in C^{\infty}_{\mathrm{c}}(\bb{R})$ with $\chi(0)=1$ and set $\chi_{r}(x):=\chi(r^{-1}x)$ for some $r>0$. Ellipticity of $\Delta$ implies that $\chi_{r}(\Delta)$ is a smoothing operator. Now, set $r_{i,-\infty}:=r\chi_{r}(\Delta)$, where we choose $r$ large enough so that $\Delta+r_{i,-\infty}\geq 1$. By spectral calculus, we define its square root and denote it by $\varepsilon:=(\Delta+r_{i,-\infty})^{\frac{1}{2}}$. The fact that this defines an operator in $\Psi_{\mathrm{bd}}^{1}(\Sigma,\sf{A}_{\Sigma,1})$ satisfying properties (i)-(iii) follows from spectral calculus and a routine computation.
	
It remains to prove (iv). First of all, we notice that 
\begin{align*}
	(\varepsilon^2_{1} - r_{1,-\infty})\Pi = \Delta_1\Pi =\Pi\Delta_1  =\Pi(\varepsilon^2_1 -  r_{1,-\infty})\,, 
\end{align*}
by Lemma~\ref{Lemma:ProjectionsHodgeDecomp}(ii). Note that $r_{1,-\infty}:\sf{H}^{s}_{i}(\Sigma)\to\sf{H}_{i}^{\infty}(\Sigma)$ for any $s\geq 0$, since it is a smoothing operators, which shows that the composition of $\Pi$ with $r_{1,-\infty}$ is well-defined. This implies that
\begin{align*}
	\varepsilon^2_{1} \Pi  =\Pi\varepsilon^2_1  +  \bar r_{1,-\infty} 
\end{align*}
for $\bar r_{1,-\infty}:=r_{1,-\infty}\Pi -\Pi \tilde{r}_{1,-\infty}$. Furthermore, $\bar r_{1,-\infty}$ is clearly a smoothing operator. Then, $\varepsilon_1^2\Pi= \Pi \varepsilon_1^2$ up to a smoothing operator, and, iterating this argument, the same holds true for polynomials of $\varepsilon_1^2$, and in particular, $\varepsilon_1^{-2}\Pi= \Pi \varepsilon_1^{-2}$ again up to smoothing. Using the \emph{Stone-Weierstrass theorem} and spectral calculus, we can generalise this fact to continuous function $f(\varepsilon_1^{-2})$ and the generalisation to unbounded functions, such as the square root, can be obtained choosing a sequence of bounded measurable functions $f_n$ that converge pointwise to $f$, as usual.
\end{proof}

Using the operators $\varepsilon_{i}$ constructed in Lemma~\ref{lem:commuting}, we next perform a \emph{microlocal factorisation} of the Cauchy evolution operator $\mathcal{U}_{i}$ of the hyperbolic operators $\sf{D}_{1}$ and $\sf{D}_{2}$, i.e.~the de Rham-Hodge d'Alembertians $\square$ acting on $0$- and $1$-forms. On ultrastatic spacetimes, the equations $\square\sf{A}=0$ for $\sf{A}\in\Gamma^{\infty}(\sf{A}_{1})$ decomposes as 
\begin{align*}
	\square\sf{A}=0\qquad\Leftrightarrow\qquad
	\begin{cases}(\partial_{t}+\Delta)\sf{A}_{\mathrm{T}}=0\\ (\partial_{t}+\Delta)\sf{A}_{\Sigma}=0
	\end{cases}
\end{align*}
for $\sf{A}_{\mathrm{T}}\in C^{\infty}(\bb{R},\Omega^{0}(\Sigma,\bb{C}))$ and $\sf{A}_{\Sigma}\in C^{\infty}(\bb{R},\Omega^{1}(\Sigma,\bb{C}))$ defined by $\sf{A}=\sf{A}_{\mathrm{T}}\d t+\sf{A}_{\Sigma}$, where $\Delta$ denotes the de Rham-Hodge Laplacian on $(\Sigma,\sf{h})$, as usual. Hence, we consider the Cauchy evolution operators $\mathcal{U}:=\mathcal{U}_{\partial_{t}^{2}+\Delta}$ of the hyperbolic operators $\partial_{t}^{2}+\Delta$ acting on $i$-forms. 

Now, as in the discussion of Klein-Gordon theory in Section~\ref{Sec:BiblioStates}, we rewrite the hyperbolic equation $(\partial_{t}+\Delta)\psi=0$ for some $\psi\in C^{\infty}(\bb{R},\Omega^{i}(\Sigma,\bb{C}))$ equivalently as a first-order linear evolutionary problem, i.e.~we write
 \begin{align}\label{cauchy}
	\sf{S}\Psi:=(\partial_{t}-i\sf{H})\Psi=0\,,\qquad\Psi:=\begin{pmatrix}
		\psi \\ i^{-1}\partial_{t}\psi
	\end{pmatrix}\,,\qquad\sf{H}:=\begin{pmatrix}
		0 & \mathrm{id}\\ \Delta & 0
	\end{pmatrix}\, .
\end{align}
We denote the corresponding Cauchy evolution operator by $\mathcal{U}_{\sf{S}}(t)$, which maps a given initial datum $\Psi(0)$ to the solution evaluated at time $t$, i.e.~$\Psi(t)= \mathcal{U}_{\sf{S}}(t)\Psi(0)$. Now, the operator $\partial_{t}^{2}+\Delta$ admits a \emph{microlocal factorisation} by Lemma~\ref{lem:commuting}, i.e.
\begin{align*}
	\partial_{t}^{2}+\Delta=(\partial_{t}+i\varepsilon)(\partial_{t}-i\varepsilon)\quad\text{mod}\quad\Psi^{-\infty}(\sf{M},\pi_{2}^{\ast}(\sf{A}_{\Sigma,i}))
\end{align*}
and hence, also the the Cauchy evolution operator $\mathcal{U}$ of $\partial_{t}^{2}+\Delta$ can be factorised accordingly. To this end, we define the pseudodifferential operators $\sf{T}$ and $\sf{T}^{-1}$, respectively, by  
\begin{align*}
  \sf{T}= \frac{1}{i}\begin{pmatrix}
  	\mathrm{id} &-\mathrm{id}\\\varepsilon_{i} & \varepsilon_{i}
  \end{pmatrix}(2\varepsilon_{i})^{-1}\,,\qquad\sf{T}^{-1}=i\begin{pmatrix}
  \varepsilon_{i} & \mathrm{id}\\-\varepsilon_{i} & \mathrm{id}
  \end{pmatrix}
\end{align*}
It is straightforward to verify that these operators are indeed the inverses of each other. Now, the Cauchy problem~\eqref{cauchy} for the operator $\sf{S}$ acting on $\Psi$ can be rewritten as a Cauchy problem for the section $\Psi^{\prime}:= \sf{T}^{-1}\Psi$, i.e.
\begin{equation}\label{cauchyapp}
\sf{S}^{\prime}\Psi^{\prime}:=(\partial_{t}-i\sf{H}^{\prime})\Psi^{\prime}=0\,,\qquad\text{where}\qquad\sf{H}^{\prime}:=\sf{T}^{-1}\circ\sf{H}\circ\sf{T}=\begin{pmatrix}
		\varepsilon_{i} & 0\\ 0 & -\varepsilon
	\end{pmatrix}+\sf{H}_{-\infty}
\end{equation}
where we used that $\varepsilon_{i}^{2}=\Delta+r_{i,-\infty}$ by Lemma~\ref{lem:commuting}(i) and where $\sf{H}_{-\infty}$ is a smoothing operator constructed using $r_{i,-\infty}$. In other words, we have (microlocally) diagonalised the problem, which in particular implies that the Cauchy evolution operator $\mathcal{U}_{\sf{S}}$ for the Cauchy problem~\eqref{cauchy} can be factorized as 
\begin{align*}
	\mathcal{U}_{\sf{S}}=\sf{T}\circ\mathcal{U}_{\sf{S}^{\prime}}\circ\sf{T}^{-1}\,,
\end{align*}
where $\mathcal{U}_{\sf{S}^{\prime}}$ is the Cauchy evolution operator associated to the operator $\sf{S}^{\prime}$ in the Cauchy problem~\eqref{cauchyapp}. Now, there is one remaining issue that needs to be addressed: the Hermitian sesquilinear form $\sigma_{\Sigma,i}$ is in general not preserved, since it holds that
\begin{align*}
	\sf{T}^{\ast}\begin{pmatrix}
	0& \mathrm{id}\\ \mathrm{id} & 0
\end{pmatrix}\sf{T}=(2\varepsilon_i)^{-1}\begin{pmatrix} \mathrm{id}&0\\0 &-\mathrm{id}\end{pmatrix}\,. 
\end{align*}
Hence, it is convenient to consider the modified operator $\sf{T}_\varepsilon=\sf{T}(2\varepsilon_i)^{1/2}$ instead of $\sf{T}$, which again provides a microlocal factorization of $\mathcal{U}_{\sf{S}}$. 
 
Summarising the previous discussion, we obtain the following result:

\begin{proposition}\label{prop:cauchyevdiag}
 Let $(\sf{M}=\bb{R}\times\Sigma,\sf{g}=-\d t\otimes\d t+\sf{h})$ be an ultrastatic globally hyperbolic spacetime such that $(\Sigma,\sf{h})$ is a Riemannian manifold of bounded geometry. Denote by $\mathcal{U}_{\sf{S}}$ the Cauchy evolution operator of 
\begin{align*}
	\sf{S}:=\partial_{t}+i\sf{H}\qquad\text{with}\qquad\sf{H}:=\begin{pmatrix}
	0 & \mathrm{id}\\ \Delta & 0
	\end{pmatrix}\,,
\end{align*}
acting on $C^{\infty}(\bb{R},\Omega^{i}(\Sigma,\bb{C}))\oplus C^{\infty}(\bb{R},\Omega^{i}(\Sigma,\bb{C}))$ whose Cauchy problem is equivalent to the one of $\partial_{t}^{2}+\Delta$ via the identification $\omega\mapsto (\omega,i^{-1}\partial_{t}\omega)$ for $\omega\in C^{\infty}(\bb{R},\Omega^{i}(\Sigma,\bb{C}))$. Moreover, let $\mathcal{U}_{\sf{S}}$ be the corresponding Cauchy evolution operator and define
\begin{align*}
	\sf{T}_\varepsilon= \frac{1}{i}\begin{pmatrix}
		\mathrm{id} & -\mathrm{id}\\ \varepsilon_{i} & \varepsilon_{i}
	\end{pmatrix}(2 \varepsilon_{i})^{-\frac{1}{2}}\,.
\end{align*}
Then, there exists a smoothing operator $\sf{R}_{-\infty}$ such that $\mathcal{U}_{\sf{S}}$ factorises as 
\begin{align*}
	\mathcal{U}_{\sf{S}}(t)=\sf{T}_{\varepsilon}\begin{pmatrix}
		\mathrm{exp}(i\varepsilon_{i}t) & 0 \\ 0 & \mathrm{exp}(-i\varepsilon_{i}t)
	\end{pmatrix}\sf{T}_{\varepsilon}^{-1}+\sf{R}_{-\infty}\, .
\end{align*}
  \end{proposition}

Now, we are finally in the position to prove the existence of Hadamard states for Maxwell's theory on ultrastatic globally hyperbolic spacetimes. If $(\mathcal{V}_{\mathrm{c}},\sigma)$ is the classical phase space of Maxwell's theory, i.e.
\begin{align*}
		 \mathcal{V}_{\mathrm{c}}:=\cfrac{\mathrm{ker}(\sf{K}^{\ast}\vert_{\Gamma^{\infty}_{c}})}{\mathrm{ran}(\sf{P}\vert_{\Gamma^{\infty}_{c}})}\,,\qquad \sigma([\psi],[\varphi]):=(\psi,i\sf{G}_{1}\varphi)_{\sf{A}_{1}}\, ,
\end{align*}
we recall that $\mathrm{CCR}(\mathcal{V}_{\mathrm{c}},\sigma)$ denotes the corresponding CCR-algebra as defined in Definition~\ref{Def:CCR} for general linear gauge theories. More precisely, $\mathrm{CCR}(\mathcal{V}_{\mathrm{c}},\sigma)$ is the unital $\ast$-algebra generated by elements $\Phi(v),\Phi^{\ast}(v)$ depending antilinearly and linearly on $v\in\mathcal{V}_{\mathrm{c}}$, respectively, such that $\Phi(v)^{\ast}=\Phi^{\ast}(v)$ and such that the \emph{canonical commutation relations} hold, i.e.
\begin{align*}
	[\Phi(v),\Phi(w)]=[\Phi^{\ast}(v),\Phi^{\ast}(w)]=0\quad\text{and}\quad[\Phi(v),\Phi^{\ast}(w)]=\sigma(v,w)\mathds{1}
\end{align*}
for all $v,w\in\mathcal{V}_{\mathrm{c}}$. Then, as discussed in Proposition~\ref{Prop:CauchyCov} and Proposition~\ref{Prop:HadCauchyCov}, constructing a Hadamard state for Maxwell's theory on $(\sf{M},\sf{g})$ (see Definition~\ref{Def.Hadamard}) amounts to construct a pair of Cauchy pseudo-covariances $c^{\pm}\:\Gamma_{\mathrm{c}}^{\infty}(\sf{A}_{\rho_{1}})\to\Gamma^{\infty}(\sf{A}_{\rho_{1}})$.

\begin{proposition}\label{Prop:HadamardUltra} \emph{(Hadamard States for Maxwell's Theory on Ultrastatic Spacetimes)}\newline
Let $(\sf{M}=\bb{R}\times\Sigma,\sf{g}=-\d t\otimes\d t+\sf{h})$ be an ultrastatic globally hyperbolic spacetime such that $(\Sigma,\sf{h})$ is a Riemannian manifold of bounded geometry and consider the gauge projector $\sf{T}_{\Sigma}\:\Gamma^{\infty}_{\infty}(\sf{A}_{\rho_{1}})\to\Gamma^{\infty}_{\infty}(\sf{A}_{\rho_{1}})$ defined in Proposition~\ref{Prop:ProjOp}. Moreover, let $\varepsilon_i\in\Psi_{\mathrm{bd}}^{1}(\Sigma,\sf{A}_{\Sigma,i})$ be approximate square roots of the de Rham Hodge-Laplacians $\Delta$ of $(\Sigma,\sf{h})$ as in Lemma~\ref{lem:commuting} and define operators $\pi^{\pm}\:\Gamma^{\infty}_{\infty}(\sf{A}_{\rho_{1}})\to\Gamma^{\infty}_{\infty}(\sf{A}_{\rho_{1}})$ by
	\begin{align*}
		\pi^{\pm}\defeq\frac{1}{2}\begin{pmatrix}\mathrm{id} & \pm\varepsilon_{0}^{-1} &0&0\\\pm\varepsilon_{0} &\mathrm{id}&0&0\\ 0&0&\mathrm{id} & \pm\varepsilon^{-1}_{1}\\0&0&\pm\varepsilon_{1} &\mathrm{id}\end{pmatrix}\, .
	\end{align*}
 	Then, the operators $c^{\pm}:=\sf{T}_{\Sigma}\pi^{\pm}\sf{T}_{\Sigma}\:\Gamma^{\infty}_{\infty}(\sf{A}_{\rho_{1}})\to\Gamma^{\infty}_{\infty}(\sf{A}_{\rho_{1}})$ have the following properties:
	\begin{itemize}
		\item[\emph{(i)}]$(c^{\pm})^{\dagger}=c^{\pm}$ with respect to $\sigma_{\Sigma,1}$.
		\item[\emph{(ii)}]$c^{\pm}(\mathrm{ran}(\sf{K}_{\Sigma}\vert_{\Gamma^{\infty}})\cap\Gamma^{\infty}_{\infty}(\sf{A}_{\rho_{1}}))\subset\mathrm{ran}(\sf{K}_{\Sigma}\vert_{\Gamma^{\infty}})$.
		\item[\emph{(iii)}]$(c^+ +c^-)\mathfrak{f}=\mathfrak{f}$ modulo $\mathrm{ran}(\sf{K}_{\Sigma}\vert_{\Gamma^{\infty}})$ for any $\mathfrak{f}\in\ker(\sf{K}_{\Sigma}^{\dagger}\vert_{\Gamma^{\infty}_\infty})$;
		\item[\emph{(iv)}]$\pm \sigma_{\Sigma,1}(\mathfrak{f},c^{\pm}\mathfrak{f})\geq 0$ for any $\mathfrak{f}\in\ker(\sf{K}_{\Sigma}^{\dagger}\vert_{\Gamma^{\infty}_\infty})$;
	\item[\emph{(v)}]$\mathrm{WF}^{\prime}(\mathcal{U}_{1} c^{\pm})\subset (\mathcal{N}^{\pm}\cup\mathcal{F})\times\sf{T}^{\ast}\Sigma$ for $\mathcal{F}=\{k=0\}\subset\sf{T}^*\sf{M}$.
	\end{itemize}
	In particular, the restrictions $c^{\pm}\:\Gamma^{\infty}_{\mathrm{c}}(\sf{A}_{\rho_{1}})\to\Gamma^{\infty}(\sf{A}_{\rho_{1}})$ are the Cauchy pseudo-covariances of a Hadamard state on $\mathrm{CCR}(\mathcal{V}_{\mathrm{c}},\sigma)$.
\end{proposition}

\begin{proof}
Since $\varepsilon_{i}$ are formally self-adjoint with respect to the Hodge $\sf{L}^{2}$-inner product on $\Sigma$, it immediately follows that
	\begin{align*}
		(\pi^{\pm})^{\dagger}=\sf{G}_{\Sigma,1}^{-1}(\pi^{\pm})^{\ast}\sf{G}_{\Sigma,1}=\pi^{\pm}\hspace*{1cm}\text{with}\hspace*{1cm}(\pi^{\pm})^{\ast}=\frac{1}{2}\begin{pmatrix}\mathrm{id} & \pm\varepsilon_{0} &0&0\\\pm\varepsilon_{0}^{-1} &\mathrm{id}&0&0\\ 0&0&\mathrm{id} & \pm\varepsilon_{1}\\0&0&\pm\varepsilon_{1}^{-1} &\mathrm{id}\end{pmatrix}\, ,
	\end{align*}
	where $(\pi^{\pm})^{\ast}$ denotes the adjoints with respect to $\langle\cdot,\cdot\rangle_{\sf{V}_{\rho_{1}}}$. In other words, $\pi^{\pm}$ are formally self-adjoint w.r.t.~$\sigma_{1,\Sigma}$. A direct computations shows that $\sf{T}_{\Sigma}$ is formally self-adjoint w.r.t.~$\sigma_{\Sigma,1}$, which implies (i), i.e.~$(c^{\pm})^{\dagger}=c^{\pm}$. 
	
	Claim (ii) follows from the fact that $\sf{T}_{\Sigma}$ are the gauge projections of the Cauchy radiation gauge, which is a complete gauge fixing. Indeed, for a given element $\mathfrak{f}\in\mathrm{ran}(\sf{K}_{\Sigma}\vert_{\Gamma^{\infty}})\cap\Gamma^{\infty}_{\infty}(\sf{A}_{\rho_{1}})$ it holds that $\sf{T}_{\Sigma}\mathfrak{f}=0$ and hence $c^{\pm}\mathfrak{f}=0$.
	
	For claim (iii), we note that clearly $\pi^{+}+\pi^{-}=\mathrm{id}$ on the whole space of initial data $\Gamma^{\infty}_\infty(\sf{A}_{\rho_{1}})$. In particular, this implies that
	\begin{align*}
		(c^{+}+c^{-})\mathfrak{f}=\sf{T}_{\Sigma}^{2}\mathfrak{f}=\sf{T}_{\Sigma}\mathfrak{f}=\mathfrak{f}\quad\text{mod}\quad\mathrm{ran}(\sf{K}_{\Sigma}\vert_{\Gamma^{\infty}})
	\end{align*}
	for all $\mathfrak{f}\in\ker(\sf{K}_{\Sigma}^{\dagger}\vert_{\Gamma^{\infty}_\infty})$, where in the last step we used that for every $\mathfrak{f}\in\ker(\sf{K}_{\Sigma}^{\dagger}\vert_{\Gamma^{\infty}_\infty})$ there exists a (unique) $\mathfrak{g}\in\mathrm{ran}(\sf{K}_{\Sigma}\vert_{\Gamma^{\infty}})\cap\Gamma^{\infty}_{\infty}(\sf{A}_{\rho_{1}})$ such that $\sf{T}_{\Sigma}(\mathfrak{f}+\mathfrak{g})=\mathfrak{f}+\mathfrak{g}$ and hence $\sf{T}_{\Sigma}\mathfrak{f}=\mathfrak{f}+\mathfrak{g}$. 
	
	Claim (iv) follows from a direct computation using the positivity of $\varepsilon^{\pm 1}$ and the fact that $[\sf{T}_{\Sigma}]$ provides a unitary isomorphism from $(\mathcal{V}_{\Sigma}^{\infty},\sigma_{\Sigma,1})$ to $(\mathcal{V}_{\sf{R}}^{\infty},\sigma_{\sf{R}})$, see Eq.~\eqref{eq:jjjjj}. Indeed, 
	\begin{align*}
		\pm\sigma_{\Sigma,1}(\mathfrak{f},c^{\pm}\mathfrak{f})=\pm\sigma_{\Sigma,1}(\mathfrak{f},\sf{T}_{\Sigma}\pi^{\pm}\sf{T}_{\Sigma}\mathfrak{f})=\pm\sigma_{\sf{R}}(\sf{T}_{\Sigma}\mathfrak{f},\pi^{\pm}\sf{T}_{\Sigma}\mathfrak{f})\geq 0\,.
\end{align*}	 
	
It remain to show (iv), i.e.~the Hadamard property. By Lemma~\ref{lem:commuting} (iv), $\pi^\pm$ commutes with $\sf{T}_\Sigma$ modulo a smoothing operator and hence, we only need to check that $\pi^\pm$ satisfies 
\begin{align*}
\mathrm{WF}^{\prime}(\mathcal{U}_{1}\pi^{\pm})\subset (\mathcal{N}^{\pm}\cup \mathcal{F})\times\sf{T}^{*}\Sigma
\end{align*}
for $\mathcal{F}=\{k=0\}\subset \sf{T}^*\sf{M}$. Now, since the operator $\square$ acting on $1$-forms on $\sf{M}$ decompose as a direct sum of operators $\partial_{t}^{2}+\Delta$ acting on time-dependent $0$ and $1$-forms on $\Sigma$, we can look at the Cauchy evolution operator $\mathcal{U}(t)$ of $\partial_{t}^{2}+\Delta$ instead. Hence, it remains to show that $\mathcal{U}(t)c^{\pm}$ has the right wavefront set. To this end, we follows the discussion in \cite[Sec.~6,7]{GerardOulghaziWrochna} and \cite[Sec.~5.2]{GerardMurroWrochna}. To start with, writing accordingly $\pi^{\pm}=\pi_{0}^{\pm}\oplus\pi_{1}^{\pm}$, we observe that 
\begin{align*}
	\pi_{i}^{\pm}=\sf{T}_{\varepsilon}\Pi^{\pm}\sf{T}_{\varepsilon}^{-1}\,\quad \text{with}\quad \pi_{i}^{\pm}=\frac{1}{2}\begin{pmatrix}
	\mathrm{id} & \pm\varepsilon_{i}^{-1}\\ \pm\varepsilon_{i} & \mathrm{id}
	\end{pmatrix}\quad \text{and}\quad\Pi^{+}=\begin{pmatrix}
	\mathrm{id} & 0 \\ 0 & 0 
	\end{pmatrix}\,,\,\Pi^{-}=\begin{pmatrix}
	0 & 0 \\ 0 & \mathrm{id}
	\end{pmatrix}\, ,
\end{align*}
where the operator $\sf{T}_{\varepsilon}$ has been introduced in Proposition~\ref{prop:cauchyevdiag}. By definition and Proposition~\ref{prop:cauchyevdiag}, $\pi^{\pm}$ are exactly the projectors on the space of initial data projection onto the solutions with positive and negative energy, i.e.~$\mathcal{U}(t)\pi_{i}^{\pm}\mathfrak{f}=e^{\pm i\varepsilon_{i} t}c^{\pm}\mathfrak{f}$. Moreover, setting $\mathcal{U}^{\pm}(t):=\mathcal{U}(t)\pi_{i}^{\pm}$ we obtain a corresponding factorisation of the evolution operator, i.e.~$\mathcal{U}(t)=\mathcal{U}^{+}(t)+\mathcal{U}^{-}(t)$ and $\mathrm{WF}^{\prime}(\mathcal{U}^{\pm}(\cdot)\mathfrak{f})\subset\mathcal{N}^{\pm}$. Now, consider the operator $\sf{Q}_{\pm}=\partial_{t} \pm \i \varepsilon_i$ as an operator acting on the first group of variables on $\sf{M}\times\Sigma$ and let us denote the distributional kernel of $\mathcal{U}(t)\pi_{i}^{\pm}$ by $\mathcal{U}(t,x,x^{\prime})$. From Proposition~\ref{prop:cauchyevdiag}, it follows that $\sf{Q}_{\pm} \mathcal{U}$ has a smooth kernel. Now, if $\sf{Q}_{\pm}$ were \emph{classical} pseudodifferential operators on $\sf{M}\times \Sigma$, then we would have that $\mathrm{WF}^{\prime}(\mathcal{U}\pi_{i}^\pm)\subset\mathcal{N}^{\pm}\times \sf{T}^{\ast}\Sigma$ by elliptic regularity. We reduce ourselves to this situation by an argument from \cite[Lemma~6.5.5]{DuistermaatHormander}, see for example \cite[Proposition~11.3.2]{GerardBook} for details.

Noting that clearly $\mathrm{ran}(\sf{K}_{\Sigma}\vert_{\Gamma^{\infty}_{\mathrm{c}}})\subset\mathrm{ran}(\sf{K}_{\Sigma})\cap\Gamma^{\infty}_{\infty,\d}(\sf{A}_{\rho_{1}})$, we conclude that the restriction of $c^{\pm}$ to $\Gamma^{\infty}_{\mathrm{c}}(\sf{A}_{\rho_{1}})$ satisfies the conditions of Proposition~\ref{Prop:CauchyCov} and Proposition~\ref{Prop:HadCauchyCov} and hence gives rise to a Hadamard state. More precisely, setting $\lambda^{\pm}:=(\rho_{1}\sf{G}_{1})^{\ast}(\pm i\sf{G}_{\Sigma,1}c^{\pm})(\rho_{1}\sf{G}_{1})=\pm i\mathcal{U}_{1}c^{\pm}\:\Gamma_{\mathrm{c}}^{\infty}(\sf{A}_{1})\to\Gamma^{\infty}(\sf{A}_{1})$ defines a quasi-free Hadamard state $\omega\:\mathrm{CCR}(\mathcal{V}_{\mathrm{c}},\sigma)\to\bb{C}$ by setting
\begin{align*}
	(\psi,\lambda^{+}\phi)_{\sf{A}_{1}}=:\omega(\Phi([\phi])\Phi^{\ast}([\psi]))\,,\qquad (\psi,\lambda^{-}\phi)_{\sf{A}_{1}}=:\omega(\Phi([\psi])\Phi^{\ast}([\phi]))\, .
\end{align*}
Note that in Proposition~\ref{Prop:CauchyCov} we had the slightly stronger condition that $c^{+}+c^{-}=\mathrm{id}$. However, it is clear that $c^{+}+c^{-}=\mathrm{id}$ modulo an operator mapping to $\mathrm{ran}(\sf{K}_{\Sigma})$ equally gives rise to a well-defined Hadamard state, since the property $\Lambda^{+}-\Lambda^{-}=\sigma$ for $\Lambda^{\pm}\:\mathcal{V}_{\mathrm{c}}\times\mathcal{V}_{\mathrm{c}}\to\bb{C}$ defined by $\Lambda^{\pm}([\cdot],[\cdot]):=(\cdot,\lambda^{\pm}\cdot)_{\sf{A}_{1}}$ is clearly not affected by this additional contribution. 
\end{proof}

\subsection{A Deformation Argument}
In this last section, we prove the main theorem of this chapter, namely the existence of Hadamard states for Maxwell's theory on general globally hyperbolic spacetimes. To achieve this goal, the idea is to employ a \emph{deformation argument}, similar in spirit to the one briefly sketched in Section~\ref{Sec:BiblioStates} for Klein-Gordon's theory, which allows us to reduce the proof of existence of Hadamard states to the ultrastatic setting. In the discussion of Klein-Gordon's theory in Section~\ref{Sec:BiblioStates}, we explained the deformation argument developed by Fulling-Narcowich-Wald \cite{FullingNarcowichWald} and a similar deformation argument following those lines of reasoning for Maxwell's theory, and, more generally, linearised Yang-Mills theory, can be found in \cite[Sec.~3.5]{GerardWrochna}. Here we follow a different approach based on \emph{paracausal deformations}.

If $\sf{M}$ is any smooth manifold, it is easy to see that $\sf{g}:=\lambda\sf{g}_{1}+(1-\lambda)\sf{g}_{2}$ with $\lambda\in [0,1]$ for two \emph{Riemannian} metrics $\sf{g}_{1},\sf{g}_{2}$ on $\sf{M}$ is again a Riemannian metric on $\sf{M}$, since positive-definiteness is preserved by taking convex combinations. The same statement, however, is not true for pseudo-Riemannian (and hence also Lorentzian) metrics, since non-degeneracy is in general lost by taking such combinations. This naturally raises the following question: 

\begin{center}
	\textit{\underline{Question}: Is there a natural operation on the space of (globally hyperbolic) Lorentzian metrics on $\sf{M}$ that preserves both the Lorentzian signature and possibly global hyperbolicity?}
\end{center}

A suitable operation has been proposed by Moretti-Murro-Volpe in \cite{Paracausal}, which we shall briefly review here. Let $\sf{M}$ be a smooth manifold that admits globally hyperbolic metrics. This does, of course, impose some restrictions on the topology of $\sf{M}$. For instance, $\sf{M}$ has necessarily to be non-compact (see Lemma~\ref{Lemma:NonCompactGH}) and, moreover, admit a product structure, i.e.~$\sf{M}\cong\bb{R}\times\Sigma$ for some smooth hypersurface $\Sigma$, as discussed in Theorem~\ref{Thm:BernalSanchez}.

To start with, let $\sf{g}$ and $\sf{g}^{\prime}$ be two time-oriented\footnote{Recall that the existence of \emph{time-oriented} Lorentzian metrics on $\sf{M}$ is equivalent to the existence of Lorentzian metrics on $\sf{M}$, see Remark~\ref{Rem:ExTOLM}.} Lorentzian metrics on $\sf{M}$. Then, we define the preorder, i.e.~reflexive and transitive, relation
\begin{align*}
	\sf{g}\preceq\sf{g}^{\prime}\qquad :\Leftrightarrow\qquad \forall p\in\sf{M}:\,\sf{V}_{p}^{\sf{g}}\subset\sf{V}_{p}^{\sf{g}^{\prime}}\,,
\end{align*}
where $\sf{V}^{\sf{g}}_{p}:=\{v\in\sf{T}_{p}\sf{M}\mid \sf{g}_{p}(v,v)<0\}$ denotes the (open) \emph{light cone} of $p\in\sf{M}$, as in Eq.~\eqref{eq:LightCone}. If the two time-orientations of $\sf{g},\sf{g}^{\prime}$ coincide, i.e.~if they are induced by the same global continuous timelike vector field, then we also have $(\sf{V}_{p}^{\sf{g}})^{\pm}\subset (\sf{V}_{p}^{\sf{g}^{\prime}})^{\pm}$, where $(\sf{V}^{\sf{g}}_{p})^{\pm}$ are two connected components of $\sf{V}^{\sf{g}}_{p}$, i.e.~the \emph{future} and \emph{past lightcones} of $p$, see~Eq.~\eqref{eq:LightCone2}. In this case, $\mathcal{J}_{\sf{g}}^{\pm}(\sf{A})\subset \mathcal{J}^{\pm}_{\sf{g}^{\prime}}(\sf{A})$ for all subsets $\sf{A}\subset\sf{M}$. 

With this terminology, there is the following result:

\begin{proposition}\label{Prop:Paracausal} \emph{(following \cite[Thm.~2.18(1),(5)]{Paracausal})}\newline
Let $\sf{M}$ be a smooth manifold, $\sf{g},\sf{g}^{\prime}$ be two smooth Lorentzian metrics on $\sf{M}$ with $\sf{g}\preceq\sf{g}^{\prime}$ and $\chi\in C^{\infty}(\sf{M},[0,1])$. Then, $\sf{g}_{\chi}:=(1-\chi)\sf{g}+\chi\sf{g}^{\prime}$ defines a Lorentzian metric on $\sf{M}$ satisfying $\sf{g}\preceq\sf{g}_{\chi}\preceq\sf{g}^{\prime}$. If, moreover, $\sf{g}$ is globally hyperbolic and $\sf{g}^{\prime}$ time-oriented, then $\sf{g}_{\chi}$ is globally hyperbolic as well.
\end{proposition}

Now, with this notation, we define the notion of \emph{paracausal deformations}, which encodes the idea of deforming a given Lorentzian metric equipped with a time-orientation into another Lorentzian metric with a corresponding time-orientation in a finite number of steps, each of which is compatible with the respective causal structures. 

\begin{definition} (Paracausal Relation, \cite[Def.~2.19]{Paracausal})\newline
	Let $\sf{M}$ be a smooth manifold and $\sf{g},\sf{g}^{\prime}$ be two globally hyperbolic metrics on $\sf{M}$. We say that \emph{$\sf{g}$ is paracausally related to $\sf{g}^{\prime}$} and write $\sf{g}\sim\sf{g}^{\prime}$, if there exists a finite sequence of globally hyperbolic metrics $(\sf{g}_{i})_{i=1,\dots,\mathrm{N}}$ with $\sf{g}_{1}=\sf{g}$ and $\sf{g}_{\mathrm{N}}=\sf{g}^{\prime}$, such that
	\begin{align*}
		\forall p\in\sf{M}\,,i\in\{1,\dots,\mathrm{N}-1\}:\quad\text{either}\quad (\sf{V}_{p}^{\sf{g}_{i}})^{+}\subset (\sf{V}_{p}^{\sf{g}_{i+1}})^{+}\quad\text{or}\quad (\sf{V}_{p}^{\sf{g}_{i+1}})^{+}\subset (\sf{V}_{p}^{\sf{g}_{i}})^{+}\, .
	\end{align*}
	In other words, for every $i\in\{1,\dots,\mathrm{N}-1\}$, either $\sf{g}_{i}\preceq\sf{g}_{i+1}$ or $\sf{g}_{i+1}\preceq\sf{g}_{i}$ while preserving the respective time orientations.
\end{definition}

Clearly, $\sim$ defines an equivalence relation on the space of globally hyperbolic metrics on $\sf{M}$. Furthermore, we stress that this definition explicitly depends on the choice of time-orientations. 

Now, the most important result for our purposes is the fact that any globally hyperbolic metric can paracausally be deformed into a ultrastatic one.

\begin{theorem}\label{Thm:Paracausal} Every globally hyperbolic metric $\sf{g}$ on some smooth manifold $\sf{M}=\bb{R}\times\Sigma$ is paracausally related to a globally hyperbolic ultrastatic metric $\sf{g}_{\mathrm{u}}=-\d t\otimes\d t+\sf{h}$ such that $(\Sigma,\sf{h})$ is of bounded geometry.
\end{theorem}

\begin{proof}[Proof (sketch).]
The previous theorem appeared in \cite[Sec.~5]{MurroProca} and is a combination of several results established in \cite{Paracausal}. First of all, in \cite[Prop.~2.22]{Paracausal}, it is shown that if $(\sf{V}_{p}^{\sf{g}})^{+}\cap (\sf{V}_{p}^{\sf{g}^{\prime}})^{+}\neq\emptyset$ for all $p\in\sf{M}$ and for two globally hyperbolic metrics $\sf{g}$ and $\sf{g}^{\prime}$ on a manifold $\sf{M}$, then necessarily $\sf{g}\sim\sf{g}^{\prime}$. The idea of the proof of this claim is to construct a globally hyperbolic metric $\sf{h}$ on $\sf{M}$ such that $\sf{h}\preceq\sf{g}$ and $\sf{h}\preceq\sf{g}^{\prime}$ and such that their time orientations agree. We omit the technical proof, which can be found in \cite{Paracausal}, and refer to Figure~\ref{Fig:Par} for the intuitive idea. 

\vspace*{-0.4cm}
\begin{figure}[H]
\centering
\includegraphics[scale=1.8]{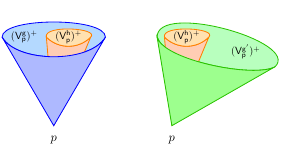}\vspace*{-0.2cm}
\caption{Lorentzian metrics $\sf{g}$, $\sf{g}^{\prime}$ and $\sf{h}$ such that $\sf{h}\preceq\sf{g}$ and $\sf{h}\preceq\sf{g}^{\prime}$.\label{Fig:Par}}
\end{figure}
\vspace*{-0.2cm}

Having established this result, it is straightforward to show that any globally hyperbolic metric $\sf{g}$ is paracausally related to an ultrastatic one, see \cite[Cor.~2.23]{Paracausal}. We identify $\sf{M}=\bb{R}\times\Sigma$ and $\sf{g}=-\d t\otimes\d t+\sf{h}_{t}$ for some time-dependent Riemannian metric $\sf{h}_{t}$. This is not a loss of generality, since we can always normalise a globally hyperbolic in time to obtain a lapse function $\beta=1$, e.g.~by performing a conformal transformation, where we recall that such a transformation preserves the causal structure and hence global hyperbolicity. Now, take any complete Riemannian metric $\sf{h}$ on $\Sigma$ and set $\sf{g}^{\prime}:=-\d t\otimes\d t+\sf{h}$, which, as explained in Example~\ref{Examples:GlobHyp}(i), is globally hyperbolic. Moreover, the vector field $\partial_{t}$ is clearly contained in the intersection $(\sf{V}_{p}^{\sf{g}})^{+}\cap (\sf{V}_{p}^{\sf{g}^{\prime}})^{+}$ at each point, from which the claim follows. Last but not least, it is well-known that every smooth manifold admits a Riemannian metric of bounded geometry, see~e.g.~\cite{Greene}, which shows that $\sf{h}$ may be chosen such that $(\Sigma,\sf{h})$ is of bounded geometry.
\end{proof}

We are finally in the position to prove the main theorem of this chapter, namely the existence of Hadamard states for Maxwell's theory on general globally hyperbolic spacetimes.

\begin{theorem}\label{Thm:Hadamard} \emph{(Existence of Hadamard States for Maxwell's Theory)}\newline
	Let $(\sf{M},\sf{g})$ be a globally hyperbolic spacetime and consider Maxwell's theory 
	\begin{align*}
		(\sf{A}_{0}=\underline{\bb{C}}_{\sf{M}},\,\sf{A}_{1}=\underline{\bb{C}}_{\sf{M}}\otimes\sf{T}^{\ast}\sf{M},\,\sf{P}=\delta\d,\,\sf{K}=\d)
	\end{align*}
	as defined in Proposition~\ref{Prop:MaxGauge}. Moreover, let $\mathrm{CCR}(\mathcal{V}_{\mathrm{c}},\sigma)$ be the CCR-algebra as defined in Definition~\ref{Def:CCR} associated to the classical phase space $(\mathcal{V}_{\mathrm{c}},\sigma)$ as defined in Definition~\ref{def.phasspace}. Then, there exists a Hadamard state on $\mathrm{CCR}(\mathcal{V}_{\mathrm{c}},\sigma)$.
\end{theorem}

\begin{proof} The deformation argument uses the notion of paracausal deformations, as discussed in \cite{Paracausal,MurroProca}, and techniques using \emph{Møller operators} discussed in \cite{MurroGinouxDrago}. \bigskip

\textit{Step 1:} Consider two globally hyperbolic metrics $\sf{g}_0$ and $\sf{g}_1$ on $\sf{M}$ such that $\sf{g}_0\preceq\sf{g}_1$, i.e.~such that the open lightcone $\sf{V}_p^{\sf{g}_0}$ is contained in the open lightcone $\sf{V}_p^{\sf{g}_1}$ for all $p\in\sf{M}$. Now, let $\chi\in C^\infty(\sf{M},[0,1])$  and define $\sf{g}_\chi \defeq (1-\chi) \sf{g}_0 + \chi  \sf{g}_1$. Following Proposition~\ref{Prop:Paracausal}, $\sf{g}_\chi$ is is globally hyperbolic and satisfies $\sf{g}_0\preceq \sf{g}_\chi \preceq \sf{g}_1$. Let us further consider two Cauchy hypersurfaces $\Sigma_0,\Sigma_1\subset \sf{M}$ with respect to the metric $\sf{g}_\chi$ such that
$\Sigma_1 \subset \mathcal{J}_{\sf{g}_\chi}^+(\Sigma_0)$  and such that
\begin{align*}
	\chi_{|_{\mathcal{J}_{\sf{g}_\chi}^+(\Sigma_1)}}=1\qquad\text{and}\qquad \chi_{|_{\mathcal{J}_{\sf{g}_\chi}^-(\Sigma_0)}}=0\, .
\end{align*}
In other words, $\Sigma_{i}$ is a Cauchy surface of $(\sf{M},\sf{g}_{i})$. Now, consider the corresponding phase spaces $(\mathcal{V}_{\Sigma_i},\sigma_{\Sigma_{i},1})$ defined by
\begin{align*}
		 \mathcal{V}_{\Sigma_i}:=\cfrac{\mathrm{ker}(\sf{K}^{\dagger}_{\Sigma_{i}}\vert_{\Gamma^{\infty}_{c}})}{\mathrm{ran}(\sf{K}_{\Sigma_{i}}\vert_{\Gamma^{\infty}_{c}})}\,,\qquad\sigma_{\Sigma_{i},1}([\mathfrak{f}],[\mathfrak{g}]):=(\mathfrak{f},i\sf{G}_{\Sigma_{i},1}\mathfrak{g})_{\sf{A}_{\rho_{1}}}\, .
	\end{align*}
	for $i=0,1$, where $\sf{K}_{\Sigma_{i}},\sf{K}^{\dagger}_{\Sigma_i}$ and $\sf{G}_{\Sigma_{i},1}$ are the operators defined in Proposition~\ref{Prop:KSigma} and Proposition~\ref{Prop:GSigma} on the Cauchy surface $\Sigma_{0}$ and $\Sigma_{i}$, respectively. Now, consider the operator $\sf{R}\defeq\rho_{\Sigma_{1}}\mathcal{U}^{\chi}_{\Sigma_{0}}$, where $\rho_{\Sigma_{1}}\:\Gamma^{\infty}_{\mathrm{sc}}(\sf{M},\sf{A}_{1})\to\Gamma^{\infty}_{\mathrm{c}}(\Sigma_{1},\sf{A}_{\rho_{1}})$ is the initial data map for $1$-forms and initial data assigned at the Cauchy surface $\Sigma_{1}$, while $\mathcal{U}^{\chi}_{\Sigma_{0}}\:\to\Gamma^{\infty}_{\mathrm{c}}(\Sigma_{0},\sf{A}_{\rho_{1}})\to\Gamma^{\infty}_{\mathrm{sc}}(\sf{M},\sf{A}_{1})$ is the Cauchy evolution operator of initial data assigned at $\Sigma_{0}$ with respect to the de Rham-Hodge d'Alembertian $\square_{\chi}$ acting on $1$-forms of $(\sf{M},\sf{g}_{\chi})$. The operator $\sf{R}$ implements an unitary isomorphism between the phase spaces $(\mathcal{V}_{\Sigma_0},\sigma_{\Sigma_{0},1})$ and $(\mathcal{V}_{\Sigma_1},\sigma_{\Sigma_{1},1})$. Therefore, given a pair of pseudo-covariances $c_0^\pm$ of Hadamard states for $(\mathcal{V}_{\Sigma_0},\sigma_{\Sigma_{0},1})$ (see Proposition~\ref{Prop:CauchyCov} and Proposition~\ref{Prop:HadCauchyCov}), we can define the operators $c_1^\pm\defeq(\sf{R}^{-1})^\dagger c^\pm_0 \sf{R}^{-1}$. A routine computations shows that if $c^\pm_0$ satisfy properties (i)-(iv) of Proposition~\ref{Prop:CauchyCov}, then so does $c^\pm_1$. Indeed, conditions (i), (iii) and (iv) follow from the fact that $\sf{R}$ is unitary, while the gauge-invariance, condition (ii), follows from the fact that $\sf{R}(\mathrm{ran}(\sf{K}_{\Sigma_{0}}))\subset\mathrm{ran}(\sf{K}_{\Sigma_{1}})$. Moreover, if $c_{0}^{\pm}$ satisfies the Hadamard property from Proposition~\ref{Prop:HadCauchyCov}, then so does $c^{\pm}_{1}$, which follows easily from similar arguments as in the proof of Proposition~\ref{Prop:HadamardUltra}(iv). Similarly, given two linear and continuous operators $c_1^\pm$ defined on $(\mathcal{V}_{\Sigma_1},\sigma_{\Sigma_{1},1})$, we can define the operators $c_0^\pm\defeq \sf{R}^\dagger c^\pm_1 \sf{R}$, which satisfy properties (i)-(iv) of  Proposition~\ref{Prop:CauchyCov} and the Hadamard condition of Proposition~\ref{Prop:HadCauchyCov} provided $c_{1}^{\pm}$ does. To sum up, if we have (Cauchy) pseudo-covariances of a Hadamard state for Maxwell's theory on $(\sf{M},\sf{g}_{1})$, we obtain the pseudo-covariances for a Hadamard state on $(\sf{M},\sf{g}_{2})$.\bigskip
	
\textit{Step 2:} Now, let $(\sf{M},\sf{g})$ be an arbitrary globally hyperbolic spacetime. By Theorem~\ref{Thm:Paracausal}, we can find a family of metric $\{g_i\}_{i=0,\dots,\mathrm{N}}$ for some $\mathrm{N}\in\bb{N}$ such that $\sf{g}_{0}=\sf{g}$ and $\sf{g}_{\mathrm{N}}=\sf{g}_{\mathrm{u}}$, where $\sf{g}_{\mathrm{u}}$ is a globally hyperbolic ultrastatic spacetime whose spacelike Cauchy surface is of bounded geometry, and such that either $\sf{g}_i\preceq\sf{g}_{i+1}$ or $\sf{g}_{i+1}\preceq \sf{g}_{i}$. Therefore, the claim follows by iterating the arguments from Step 1 and the existence of Hadamard states in the ultrastatic case, i.e.~Proposition~\ref{Prop:HadamardUltra}.
	\end{proof}
\phantomsection
\chapter*{Summary and Outlook}
\addtocontents{toc}{\vspace*{2ex}}
\addcontentsline{toc}{section}{\hspace{-16pt}\textbf{Summary and Outlook}}
\markboth{Summary and Outlook}{Summary and Outlook}

At the beginning of this thesis, we performed a detailed analysis of the Cauchy problem for linear hyperbolic differential operators on globally hyperbolic Lorentzian manifolds. In particular, we showed that the Cauchy problem is well posed for both symmetric hyperbolic systems and normally hyperbolic operators. We next studied two departures from the classical theory of hyperbolic PDEs. First, we examined what happens when we give up on locality, at least in the lower-order terms, by adding nonlocal potentials to an otherwise local symmetric hyperbolic system. In full generality, one cannot expect well-posedness, essentially due to an inherent conflict with the classical energy methods. Nonetheless, we analysed the Cauchy problem for two specific classes of nonlocal potentials, namely retarded potentials and potentials with a small time range, and we established existence of solutions under reasonable assumptions.

Our second deviation concerned linear gauge theories, which form the central part of this thesis. Such theories are inherently non-hyperbolic. However, thanks to the gauge symmetry present, they can be transformed into hyperbolic problems supplemented by gauge constraints, when we consider gauge equivalent classes of solution. After discussing numerous examples of direct physical interest in many details, we moved on to the quantum theory by presenting the general and abstract framework for quantising linear gauge theories. Finally, in the last chapter we carried out a detailed analysis of the quantisation of a specific linear gauge theory, namely Maxwell’s theory of electromagnetism. In particular, we proved the existence of Hadamard states for this theory on globally hyperbolic spacetimes.

In what follows, we provide a brief discussion and outlook on the construction of Hadamard states for yet another linear gauge theory, namely linearised gravity (see Section~\ref{Sec:LinGrav}), which remains an important open problem.

\section*{Gauge-Invariant Initial Data and Linearised Gravity}
As highlighted in Section~\ref{Sec:BiblioStates}, the existence of Hadamard states for linearised gravity is an important open question and so far, constructing such states has only be achieved for spatially compact globally hyperbolic manifolds of bounded geometry \cite{Gerard} and, in the case of manifolds with a non-compact Cauchy surface, for Minkowski spacetime~\cite{HuntPhD}, de-Sitter spacetime~\cite{GerardWrochna2} and, for a suitable subspace of observables, for asymptotically flat spacetimes \cite{BeniniMurro}.

One of the main complications in linearised gravity compared to other linear gauge theories is the fact that a \emph{deformation argument} cannot consistently be applied. As discussed in Section~\ref{Sec:LinGrav}, the linearised Einstein equations only admit a gauge symmetry if the background metric around which we linearise is a solution to the non-linear Einstein equations (see Proposition~\ref{Prop:GaugeLinGrav}) and, as a consequence of the non-linearity of the Einstein equations, it is not clear how to deform a given metric such that we stay within the class of Einstein metrics. Therefore, for showing the existence of Hadamard states for linearised gravity, one needs to find an explicit construction on general globally hyperbolic spacetimes, which is a highly non-trivial task. The approach in \cite{Gerard} for spatially compact spacetimes is based on a complete gauge fixing approach, however, crucially depends on the compactness assumption. 

Now, as explained in Remark~\ref{Rem:GaugeInvData}, the Cauchy radiation gauge and the general idea of complete gauge fixing on the level of initial data admits a nice interpretation in terms of \emph{gauge-invariant} (``physical'') \emph{initial data}. In the following, we give a short outlook on how such ideas can be applied in the context of linearised gravity. We only consider the case of a \emph{spatially-compact} and \emph{ultrastatic} globally hyperbolic manifold of the form $(\sf{M}=\bb{R}\times\Sigma,\sf{g}=-\d t+\d t+\sf{h})$, where $(\Sigma,\sf{h})$ denotes a compact Riemannian manifolds. We will comment below on the problems that must be addressed in order to generalise this approach to more general spacetimes.

To start with, let us assume that the ultrastatic metric $\sf{g}$ is a solution to the vacuum Einstein equations, i.e.
\begin{align*}
	\mathrm{Ein}_{\Lambda}(\sf{g})=\mathrm{Ric}(\sf{g})-\frac{1}{2}\mathrm{Scal}(\sf{g})\sf{g}+\Lambda\sf{g}=0\, .
\end{align*}
Now, let us choose local coordinates $(x^{0}=t,x^{i})$ on $\sf{M}$, where $(x^{i})_{i=1,\dots,n}$ are local coordinates on $\Sigma$. In the ultrastatic case, we have $\mathrm{Riem}(\sf{g})_{\alpha\beta\gamma\delta}=0$ whenever at least one index is zero, while $\mathrm{Riem}(\sf{g})_{ijkl}=\mathrm{Riem}(\sf{h})_{ijkl}$ for spatial indices. In particular, we have $\mathrm{Ric}(\sf{g})_{\alpha\beta}=0$ if either $\alpha=0$ or $\beta=0$ and $\mathrm{Ric}(\sf{g})_{ij}=\mathrm{Ric}(\sf{h})_{ij}$ as well as $\mathrm{Scal}(\sf{g})=\mathrm{Scal}(\sf{h})$. Therefore, the Einstein equations for $\sf{g}=-\d t\otimes\d t+\sf{h}$ decompose as
\begin{align*}
	0=\mathrm{Ein}_{\Lambda}(\sf{g})=
	\begin{cases}
		\mathrm{Scal}(\sf{h})-2\Lambda \\
		\mathrm{Ric}(\sf{h})-\frac{1}{2}\mathrm{Scal}(\sf{h})\sf{h}+\Lambda\sf{h}
	\end{cases}
\end{align*}
Taking the trace of the last equations gives $(2-n)\mathrm{Scal}(\sf{h})+2n\Lambda=0$ and hence, for this system of equations to admit a solution, we necessarily need to require $\Lambda=0$. To sum up, for an ultrastatic metric $\sf{g}$ to be an Einstein metric, we necessarily have to require $\Lambda=0$ and in this case, being a solution to the Einstein equations is equivalent to require that $(\Sigma,\sf{h})$ is Ricci flat.

We will sue the same notation for linearised gravity as in Section~\ref{Sec:LinGrav}. We consider the Hermitian vector bundles $\sf{S}_{k}:=\sf{T}^{\ast}\sf{M}^{\otimes_{s}k}$ equipped with the bundle metrics
\begin{align*}
	\langle\cdot,\cdot\rangle_{\sf{S}_{1}}:=\sf{g}^{\sharp}(\cdot,\cdot)\,,\qquad\qquad \langle\cdot,\cdot\rangle_{\sf{S}_{2,\mathrm{I}}}:=2(\sf{g}^{\sharp})^{\otimes 2}(\mathrm{I}\cdot,\cdot)\,,
\end{align*}
where $\mathrm{I}$ denotes the trace-reversal. By Proposition~\ref{eq:EinsteinHS}, Einstein gravity linearised around $(\sf{M},\sf{g})$ as a linear gauge theory according to Definition~\ref{Def:LinGaugeTh} is described by the operators
\begin{align*}
	\sf{P}:=-\square+2\mathrm{Riem}_{\sf{g}}-\I\d_{s}\delta_{s}\,,\qquad \sf{K}=\mathrm{I}\d_{s}
\end{align*}
and the corresponding hyperbolic operators $\sf{D}_{1}=\sf{K}^{\dagger}\sf{K}$ and $\sf{D}_{2}=\sf{P}+\sf{K}\sf{K}^{\dagger}$ are given by
\begin{align*}
	\sf{D}_{1}=-\square-\Lambda\,,\qquad \sf{D}_{2}=-\square+2\mathrm{Riem}_{\sf{g}}\, ,
\end{align*}
where $\d_{s}$ and $\delta_{s}$ denote the symmetrised gradient and divergence, $\square:=\sf{g}^{\alpha\beta}\nabla^{\sf{g}}_{\alpha}\nabla^{\sf{g}}_{\beta}$ the connection d'Alembertian of $(\sf{M},\sf{g})$ and $\mathrm{Riem}_{\sf{g}}$ the Riemann operator.

Next, let us fix some notation for initial data. To start with, we define the space of \emph{time-dependent symmetric $(0,k)$-tensor fields on $\Sigma$} by $C^{\infty}(\bb{R},\Gamma^{\infty}(\sf{S}_{\Sigma,k})):=\Gamma^{\infty}(\pi_{2}^{\ast}(\sf{S}_{\Sigma,k}))$, where $\sf{S}_{\Sigma,k}:=\sf{T}^{\ast}\Sigma^{\otimes_{s}k}$ and where $\pi_{2}:=\sf{M}\to\Sigma$ denotes the natural projection. In analogy to the decomposition of differential forms in Eq.~\eqref{eq:DecompDiffForms}, we obtain identifications
\begin{align*}
	\Gamma^{\infty}(\sf{S}_{1})&\xrightarrow{\cong} C^{\infty}(\bb{R},\Gamma^{\infty}(\sf{S}_{\Sigma,0}))\oplus C^{\infty}(\bb{R},\Gamma^{\infty}(\sf{S}_{\Sigma,1}))\,,\quad \omega\mapsto (\omega_{\mathrm{T}},\omega_{\Sigma})\\
	\Gamma^{\infty}(\sf{S}_{2})&\xrightarrow{\cong} C^{\infty}(\bb{R},\Gamma^{\infty}(\sf{S}_{\Sigma,0}))\oplus C^{\infty}(\bb{R},\Gamma^{\infty}(\sf{S}_{\Sigma,1}))\oplus C^{\infty}(\bb{R},\Gamma^{\infty}(\sf{S}_{\Sigma,2}))\,,\quad u\mapsto (u_{\mathrm{TT}},u_{\mathrm{T}\Sigma},u_{\Sigma\Sigma})\\
	\Gamma^{\infty}(\sf{S}_{k})&\xrightarrow{\cong} \bigoplus_{i=0}^{k}C^{\infty}(\bb{R},\Gamma^{\infty}(\sf{S}_{\Sigma,i}))\,,\quad t\mapsto (t_{\mathrm{T}\dots\mathrm{T}},t_{\mathrm{T}\dots\mathrm{T}\Sigma},\dots,t_{\Sigma\dots\Sigma})
\end{align*}
which are defined in terms of the obvious decompositions, i.e.
\begin{align*}
	\omega=\omega_{\mathrm{T}}\d t+\omega_{\Sigma}\,,\qquad\quad u=u_{\mathrm{TT}}\d t\otimes\d t+(u_{\mathrm{T}\Sigma}\otimes\d t+\d t\otimes u_{\mathrm{T}\Sigma}) +u_{\Sigma\Sigma}
\end{align*}
and similarly for higher order tensor fields, where $\omega_{\mathrm{T}}:=\omega(\partial_{t})$ as well as $u_{\mathrm{TT}}:=u(\partial_{t},\partial_{t})$ and $u_{\mathrm{T}\Sigma}:=u(\partial_{t},\cdot)-u_{\mathrm{T}\mathrm{T}}\d t$. Now, let us introduce the bundles of initial data by
\begin{align*}
	\sf{S}_{\rho_{k}}:=\bigoplus_{i=0}^{k}(\sf{S}_{\Sigma,i}\oplus\sf{S}_{\Sigma,i})\, .
\end{align*}
The obvious choices of initial data maps for the hyperbolic operators $\sf{D}_{k}$ in view of these decompositions are the maps $\rho_{k}\:\Gamma^{\infty}(\sf{S}_{k})\to\Gamma^{\infty}(\sf{S}_{\rho_{k}})$ defined by
\begin{align*}
	\rho_{1}(\omega):=&(\omega_{\mathrm{T}},\,\partial_{t}\omega_{\mathrm{T}}\,,\omega_{\Sigma},\,\partial_{t}\omega_{\Sigma})\vert_{t=0}\,,\\ \rho_{2}(u):=&(u_{\mathrm{TT}},\partial_{t}u_{\mathrm{TT}},u_{\mathrm{T}\Sigma},\partial_{t}u_{\mathrm{T}\Sigma},u_{\Sigma\Sigma},\partial_{t}u_{\Sigma\Sigma})\vert_{t=0}\, .
\end{align*}
However, just as Maxwell’s theory admits Furlani-type initial data maps that are more naturally suited to the language of differential forms, one can likewise introduce \emph{adapted initial data maps} $\rho^{\sf{A}}_{k}\:\Gamma^{\infty}(\sf{S}_{k})\to\Gamma^{\infty}(\sf{S}_{\rho_{k}})$ for linearised gravity, an approach that, to the best of my knowledge, has not yet been explored in the literature:
\begin{align*}
	\rho^{\sf{A}}_{1}(\omega):=&(\omega_{\mathrm{T}},\,\delta_{s}\omega\,,\omega_{\Sigma},\,2(\sf{K}\omega)_{\mathrm{T}\Sigma})\vert_{t=0}\,,\\ \rho^{\sf{A}}_{2}(u):=&\bigg(u_{\mathrm{TT}},\frac{1}{2}(\delta_{s}u)_{\mathrm{T}},u_{\mathrm{T}\Sigma},\frac{1}{2}(\delta_{s}u)_{\Sigma},u_{\Sigma\Sigma},3(\d_{s}u)_{\mathrm{T}\Sigma\Sigma}\bigg)\bigg\vert_{t=0}\, .
\end{align*}
As usual, we denote the inverses of $\rho_{i}$ restricted to $\mathrm{ker}(\sf{D}_{i}\vert_{\Gamma^{\infty}_{\mathrm{sc}}})$, the Cauchy evolution operators, by $\mathcal{U}_{i}$ and similarly by $\mathcal{U}_{i}^{\sf{A}}$ for the adapted Cauchy data maps $\rho_{i}^{\sf{A}}$. Now, computing the operator $\sf{K}_{\Sigma}$ for the adapted Cauchy data maps is straightforward and one finds
\begin{align*}
\sf{K}_{\Sigma}^{\sf{A}}:=\rho_{2}^{\sf{A}}\sf{K}\mathcal{U}_{1}^{\sf{A}}
=\begin{pmatrix}
0 & \frac{1}{2}\mathrm{id} & -\delta_{s}^{\Sigma} & 0\\
0 & 0 & 0 & 0\\
0 & 0 & 0 & \frac{1}{2}\mathrm{id}\\
0 & 0 & 0 & 0\\
0 & \frac{1}{2}\sf{h} & \d_{s}^{\Sigma} & 0\\ -\d_{s}^{\Sigma}\d_{s}^{\Sigma}+\sf{h}\Delta & 0& 0& 2\d_{s}^{\Sigma}+\frac{1}{2}\sf{h}\delta_{s}^{\Sigma}
\end{pmatrix}\,,
\end{align*}
where $\d_{s}^{\Sigma}\:\Gamma^{\infty}(\sf{S}_{\Sigma,k})\to\Gamma^{\infty}(\sf{S}_{\Sigma,k+1})$ and $\delta_{s}^{\Sigma}\:\Gamma^{\infty}(\sf{S}_{\Sigma,k+1})\to\Gamma^{\infty}(\sf{S}_{\Sigma,k})$ are the symmetrised gradient and divergence on $(\Sigma,\sf{h})$, defined analogously to there counterparts on $(\sf{M},\sf{g})$, and $\Delta:=\sf{h}^{ij}\nabla^{\sf{h}}_{i}\nabla^{\sf{h}}_{j}$ the Laplace-Beltrami operator of $(\Sigma,\sf{h})$. In order to obtain the expression of $\sf{K}_{\Sigma}$ with respect to the original initial data maps, we note that $\sf{T}_{i}\circ\rho_{i}=\rho_{i}^{\sf{A}}$, where $\sf{T}_{i}\:\Gamma^{\infty}(\sf{S}_{\rho_{i}})\to\Gamma^{\infty}(\sf{S}_{\rho_{i}})$ are the invertible operators given by
\begin{align*}
	\quad\sf{T}_{1}=\begin{pmatrix}
		\mathrm{id} & 0 & 0 & 0 \\
		0& \mathrm{id} & \delta_{s}^{\Sigma} & 0 \\
		0 & 0 & \mathrm{id} & 0 \\
		\d_{s}^{\Sigma} & 0 & 0 & \mathrm{id}
		\end{pmatrix},\,\qquad \sf{T}_{2}=\begin{pmatrix}
		\mathrm{id} & 0 & 0 & 0 & 0 & 0 \\
		0 & \mathrm{id} & \delta_{s}^{\Sigma} & 0 & 0 & 0\\
		0 & 0 &\mathrm{id}& 0 & 0 & 0 \\
		0 & 0 & 0 & \mathrm{id} & \frac{1}{2}\delta_{s}^{\Sigma} & 0\\
		0 & 0 & 0 & 0 &\mathrm{id} & 0 \\
		0 & 0 & 2\d_{s}^{\Sigma} & 0 & 0 & \mathrm{id}
		\end{pmatrix}\, .
\end{align*}
With this observation, it is straightforward to verify that $\sf{K}_{\Sigma}=\rho_{2}\sf{K}\mathcal{U}_{1}$ is given by
\begin{align*}
	\sf{K}_{\Sigma}:=\sf{T}_{2}^{-1}\sf{K}_{\Sigma}^{\sf{A}}\sf{T}_{1}=\begin{pmatrix}
		0 & \frac{1}{2}\mathrm{id} & -\frac{1}{2}\delta_{s}^{\Sigma} & 0\\
		\frac{1}{2}\Delta & 0 &0 & -\frac{1}{2}\delta_{s}^{\Sigma}\\
		\frac{1}{2}\d_{s}^{\Sigma} & 0 & 0 & \frac{1}{2}\mathrm{id}\\
		0 & \frac{1}{2}\d_{s}^{\Sigma} & \frac{1}{2}\Delta_{1} & 0 \\
		0 & \frac{1}{2}\sf{h} & \d_{s}^{\Sigma}+\frac{1}{2}\sf{h}\delta_{s}^{\Sigma} & 0\\
		\frac{1}{2}\sf{h}\Delta & 0 & 0 & \d_{s}^{\Sigma}+\frac{1}{2}\sf{h}\delta_{\Sigma}
		\end{pmatrix}\, .
\end{align*}

Now, the goal of the following discussion is to follow similar steps as in Remark~\ref{Rem:GaugeInvData}, i.e.~by decomposing initial data for a linearised gravitational field $u\in\Gamma^{\infty}(\sf{S}_{2})$ suitably and by defining gauge-invariant linear combinations thereof. To start with, we need a suitable decomposition of symmetric $(0,2)$-tensor fields on $\Sigma$. To this end, we have the following result:

Now, for let $(\Sigma,\sf{h})$ be a compact, connected and Ricci-flat $3$-manifold. Then, we claim that there is a direct sum decomposition
	\begin{align*}
	\Gamma(\sf{S}_{\Sigma,2})\cong\mathrm{ran}(\d^{s}_{\Sigma}\d^{s}_{\Sigma})\oplus\mathrm{ran}(\d^{s}_{\Sigma}\vert_{\mathrm{ker}(\delta^{s}_{\Sigma})})\oplus \sf{h}C^{\infty}(\Sigma)\oplus (\mathrm{ker}(\delta_{s}^{\Sigma})\cap\mathrm{ker}(\mathrm{tr}_{\sf{h}}))\, .
\end{align*}

\begin{proof}[Proof (sketch).]
	We define $\sf{R}\:\Gamma^{\infty}(\sf{S}_{\Sigma,2})\to\Gamma^{\infty}(\sf{S}_{\Sigma,2})$ by $\sf{R}=\mathrm{id}-\frac{1}{3}\sf{h}\mathrm{tr}_{\sf{h}}$, which is the projector that maps any symmetric $(0,2)$-tensor field into its trace-free part. Furthermore, we set
\begin{align*}
	&\sf{L}\:\Gamma(\sf{S}_{\Sigma,1})\to\Gamma(\sf{S}_{\Sigma,2}),\qquad &&\sf{L}:=\sf{R}\d^{\Sigma}_{s}=\d^{\Sigma}_{s}+\frac{1}{3}\sf{h}\delta^{\Sigma}_{s}\\
&\mathcal{L}\:\Gamma(\sf{S}_{\Sigma,2})\to\Gamma(\sf{S}_{\Sigma,1}), &&\mathcal{L}:=\delta^{\Sigma}_{s}\sf{R}=\delta^{\Sigma}_{s}+\frac{2}{3}\d_{s}\mathrm{tr}_{\sf{h}}\\
&\Delta_{\sf{L}}\:\Gamma(\sf{S}_{\Sigma,1})\to\Gamma(\sf{S}_{\Sigma,1}), &&\Delta_{\sf{L}}:=\mathcal{L}\sf{L}=\delta^{\Sigma}_{s}\d^{\Sigma}_{s}-\frac{2}{3}\d^{\Sigma}_{s}\delta^{\Sigma}_{s}
\end{align*}
The operator $\sf{L}$ is the \textit{conformal Killing operator}\footnote{$\mathrm{ker}(\sf{L})$ is essentially the space of conformal Killing vector fields, i.e.~$\sf{X}^{\flat}\in\mathrm{ker}(\sf{L})$ iff $\mathcal{L}_{\sf{X}}\sf{g}=\lambda\sf{g}$ for some $\lambda$. This space is known to be finite-dimensional, even in the case in which $\Sigma$ is non-compact, see~\cite{Semmelmann}.} and $\mathcal{L}$ is its formal adjoint w.r.t.~$\langle\cdot,\cdot\rangle_{\sf{S}_{\Sigma,k}}:=(k!)(\sf{h}^{\sharp})^{\otimes k}(\cdot,\cdot)$. The operator $\Delta_{\sf{L}}$ is usually referred to as the \textit{York operator}. Clearly, $\Delta_{\sf{L}}$ is formally self-adjoint. Furthermore, a straightforward computation shows that
\begin{align*}
			\Delta_{\sf{L}}=-\Delta+\frac{1}{3}\d^{\Sigma}_{s}\delta^{\Sigma}_{s}-\mathrm{Ric}_{\sf{h}}\, ,
		\end{align*}
		where $\mathrm{Ric}_{\sf{h}}(\omega)_{i}:=\mathrm{Ric}(\sf{h})_{i}^{j}\omega_{j}$ denotes the Ricci operator. Its principal symbol is given by
		\begin{align*}
		\sigma_{\Delta_{\sf{L}}}(x,\xi)\:\sf{T}^{\ast}\Sigma\to\sf{T}^{\ast}\Sigma\,,\qquad \omega\mapsto -\sf{h}^{\sharp}(\xi,\xi)\omega-\frac{1}{3}\xi \sf{h}^{\sharp}(\xi,\omega)
	\end{align*}
	for all $(x,\xi)\in\sf{T}^{\ast}\Sigma$. Now, we claim that this map is injective for fixed $\xi\neq 0$. Suppose that $-\sf{h}^{\sharp}(\xi,\xi)\omega-\frac{1}{3}\xi \sf{h}^{\sharp}(\xi,\omega)=0$. Applying $\sf{h}^{\sharp}(\xi,\cdot)$ to this expression yields $\sf{h}^{\sharp}(\xi,\omega)=0$. Since $\sf{h}^{\sharp}(\xi,\xi)\neq 0$, we conclude that $\omega=0$. This shows injectivity and hence ellipticity of $\Delta_{\sf{L}}$.
	
	Clearly $\mathrm{ker}(\Delta_{\sf{L}})=\mathrm{ker}(\sf{L})$. Moreover, there is the $\sf{L}^{2}$-orthogonal \emph{York decomposition} \cite{York1,York2} (see also \cite{CantorYork1,CantorYork2} for a generalisation to asymptotically Euclidean manifolds), i.e.
	\begin{align*}
		\Gamma(\sf{S}_{\Sigma,2})\cong \mathrm{ran}(\sf{L})\oplus \sf{h}C^{\infty}(\Sigma)\oplus (\mathrm{ker}(\delta_{s}^{\Sigma})\cap\mathrm{ker}(\mathrm{tr}_{\sf{h}}))\,.
\end{align*}	
Indeed, for $u\in\Gamma^{\infty}(\sf{S}_{\Sigma,2})$, this is equivalent to finding a solution $\omega\in\Gamma^{\infty}(\sf{S}_{\Sigma,1})$ to the elliptic equation $\Delta_{\sf{L}}\omega=\mathcal{L}u$. It is easy to check that $\mathcal{L}u$ is $\sf{L}^{2}$-orthogonal to $\mathrm{ker}(\Delta_{\sf{L}})$ and hence, the Fredholm alternative theorem implies that there exists a solution to this elliptic equation. Furthermore, this solution is unique up to $\mathrm{ker}(\Delta_{\sf{L}})=\mathrm{ker}(\sf{L})$.

Now, let $u\in \Gamma^{\infty}(\sf{S}_{\Sigma,2})$. We decompose it according to the York decomposition as
\begin{align*}
	u=\sf{L}\omega+\sf{h}\varphi+u^{\sf{T}\sf{T}}\, ,
\end{align*}
where $\omega$ is uniquely determined up to $\mathrm{ker}(\sf{L})$, i.e.~up to conformal Killing vector fields. Now, $\omega\in\Gamma^{\infty}(\sf{S}_{\Sigma,1})$ can be decomposed using the Hodge-decomposition theorem as $\omega=\d^{\Sigma}_{s}\psi+\eta$ where $\eta\in\mathrm{ker}(\delta^{\Sigma}_{s})$ and where we used that $\d^{\Sigma}_{s}=\d_{\Sigma}$ on zero-forms and $\delta_{s}^{\Sigma}=\delta_{\Sigma}$ on $1$-forms. Combining these two compositions, we obtain
\begin{align*}
	u&=\d^{\Sigma}_{s}\d^{\Sigma}_{s}\psi+\d_{s}^{\Sigma}\eta-\frac{1}{3}\sf{h}\Delta\psi+\sf{h}\varphi+u^{\sf{T}\sf{T}}=\d^{\Sigma}_{s}\d^{\Sigma}_{s}\psi+\d^{\Sigma}_{s}\eta+\sf{h}\bigg(\varphi-\frac{1}{3}\Delta\psi\bigg)+u^{\sf{T}\sf{T}}
\end{align*}
which is a decomposition of the required form. It is straightforward to verify that this is a direct sum decomposition provided $\mathrm{Ric}(\sf{h})=0$, which, however, is not $\sf{L}^{2}$-orthogonal in general.
\end{proof}

Now, let us fix a $(0,2)$-tensor initial data $\mathfrak{c}_{2}\in\Gamma^{\infty}(\sf{S}_{\rho_{2}})$. Using the Hodge and symmetric $(0,2)$-tensor decompositions from above, respectively, we decompose $\mathfrak{c}$ as follows:
\begin{align*}
	\mathfrak{c}_{2}=\begin{pmatrix}
		\mathfrak{a}\\
		\pi\\
		\d^{\Sigma}_{s}\varphi_{1}+\omega_{1}\\
		\d^{\Sigma}_{s}\varphi_{2}+\omega_{2}\\
		\d^{\Sigma}_{s}\d^{\Sigma}_{s}\psi_{1}+\d^{\Sigma}_{s}\eta_{1}+\sf{h}\chi_{1}+\sf{T}_{1}\\
		 		\d^{\Sigma}_{s}\d^{\Sigma}_{s}\psi_{2}+\d^{\Sigma}_{s}\eta_{2}+\sf{h}\chi_{2}+\sf{T}_{2}
	\end{pmatrix}\quad\text{where}\quad	
	\begin{cases} a,\pi,\varphi_{i},\psi_{i},\chi_{i}\in C^{\infty}(\Sigma)\\\omega_{i},\eta_{i}\in\mathrm{ker}(\delta_{s}^{\Sigma})\\ \sf{T}_{i}\in\mathrm{ker}(\delta_{s}^{\Sigma})\cap\mathrm{ker}(\mathrm{tr}_{\sf{h}})\end{cases}
\end{align*}

Now, as we did for Maxwell's theory in Remark~\ref{Rem:GaugeInvData}, we want to equivalently view $\mathfrak{a}$, $\pi$, $\varphi_{i}$, $\omega_{i}$, $\psi_{i}$, $\eta_{i}$, $\chi_{i}$, $\sf{T}_{i}$ as independent degrees of freedom. Of course, we should stress that $\varphi_{i}$ and $\psi_{i}$ are only determined up to constant, while $\eta_{i}$ up to harmonic $1$-forms (and hence uniquely determined if $\Sigma $ is assumed to be simply connected). However, we will ignore this issue within this outlook, while we should keep in mind that one has to either fix these redundancies accordingly, or work in suitable quotient spaces.

Now, we would like to see how the individual components $\mathfrak{a},\pi,\varphi_{i},\omega_{i},\psi_{i},\eta_{i},\chi_{i},\sf{T}_{i}$ transform under gauge transformations. For this, let us take a $(0,1)$-tensor initial data $\mathfrak{c}_{1}\in\Gamma^{\infty}(\sf{S}_{\rho_{1}})$ and decompose it as
\begin{align*}
	\mathfrak{c}_{1}=\begin{pmatrix}
	\mathfrak{f}_{1}\\\mathfrak{f}_{2}\\ \d^{\Sigma}_{s}\mathfrak{g}_{1}+\mathfrak{V}_{1}\\\d_{s}^{\Sigma}\mathfrak{g}_{2}+\mathfrak{V}_{2}
	\end{pmatrix}\quad\text{where}\quad	
	\begin{cases}
		\mathfrak{f}_{i},\mathfrak{g}_{i}\in C^{\infty}(\Sigma)\\
		\mathfrak{V}_{i}\in\mathrm{ker}(\delta_{s}^{\Sigma})
	\end{cases}\, .
\end{align*}
Again, we stress that $\mathfrak{g}_{1}$ and $\mathfrak{g}_{2}$ are only uniquely determined up to constant. Using the explicit expression of $\sf{K}_{\Sigma}$ as derived before and the fact that $\Delta=-\delta^{\Sigma}_{s}\d^{\Sigma}_{s}$, we hence obtain 
\begin{align*}
	\sf{K}_{\Sigma}\mathfrak{c}_{1}=\frac{1}{2}\begin{pmatrix}
		\mathfrak{f}_{2}+\Delta\mathfrak{g}_{1}\\
		\Delta(\mathfrak{f}_{1}+\mathfrak{g}_{2})\\
		\d^{\Sigma}_{s}(\mathfrak{f}_{1}+\mathfrak{V}_{2})+\mathfrak{V}_{2}\\
		\d^{\Sigma}_{s}(\mathfrak{f}_{2}+\Delta \mathfrak{g}_{1})+\Delta\mathfrak{V}_{1}\\
		2\d^{\Sigma}_{s}\d^{\Sigma}_{s}\mathfrak{g}_{1}+2\d^{\Sigma}_{s}\mathfrak{V}_{1}+\sf{h}(\mathfrak{f}_{2}-\Delta\mathfrak{g}_{1})\\
		2\d^{\Sigma}_{s}\d^{\Sigma}_{s}\mathfrak{g}_{2}+2\d^{\Sigma}_{s}\mathfrak{V}_{2}+\sf{h}\Delta(\mathfrak{f}_{1}-\mathfrak{g}_{2})
	\end{pmatrix}\, ,
\end{align*}
where we used that $\Delta\d^{\Sigma}_{s}=\d^{\Sigma}_{s}\Delta$ on functions and $\Delta\delta^{\Sigma}_{s}=\delta^{\Sigma}_{s}\Delta$ on $(0,1)$-tensors in the Ricci-flat case. Row 5 and 6 are $(0,2)$-tensors decomposed according to the decomposition from the above proposition with the transverse-traceless part being identically zero. Similarly, we note that row (3) and (4) are decomposed according to the Hodge decomposition. Hence, by uniqueness (directness) of these decompositions, we obtain the following transformations laws for individual components
\begin{align*}
	\begin{cases}
		\mathfrak{a}\mapsto \mathfrak{a} + \frac{1}{2}(\mathfrak{f}_{2}+\Delta \mathfrak{g}_{1})\\
		\pi\mapsto\pi +\frac{1}{2}\Delta (\mathfrak{f}_{1}+\mathfrak{g}_{2})\\
		\varphi_{1}\mapsto\varphi_{1}+\frac{1}{2}(\mathfrak{f}_{1}+\mathfrak{g}_{2})\\
		\varphi_{2}\mapsto \varphi_{2} + \frac{1}{2}(\mathfrak{f}_{2}+\Delta \mathfrak{g}_{1})\\
		\psi_{1}\mapsto\psi_{1}+\mathfrak{g}_{1}\\
		\psi_{2}\mapsto\psi_{2}+\mathfrak{g}_{2}\\
		\chi_{1}\mapsto \chi_{1}+ \frac{1}{2}(\mathfrak{f}_{2}-\Delta\mathfrak{g}_{1})\\
		\chi_{2}\mapsto \chi_{2}+ \frac{1}{2}\Delta(\mathfrak{f}_{1}-\mathfrak{g}_{2})
	\end{cases}\,,\qquad
	\begin{cases}
		\omega_{1}\mapsto\omega_{1}+\frac{1}{2}\mathfrak{V}_{2}\\
		\omega_{2}\mapsto\omega_{2}+\frac{1}{2}\Delta \mathfrak{V}_{1}\\
		\eta_{1}\mapsto\eta_{1}+\mathfrak{V}_{1}\\
		\eta_{2}\mapsto\eta_{2}+\mathfrak{V}_{2}\\
	\end{cases}\,,\qquad
	\begin{cases}
		\sf{T}_{1}\mapsto\sf{T}_{1}\\
		\sf{T}_{2}\mapsto\sf{T}_{2}\, .
	\end{cases}
\end{align*}
The transverse-traceless components of the tensorial degrees of freedom are already gauge-invariant, as one would naively expect. For the scalar and vectorial degrees of freedom, we can find suitable linear combinations to obtain gauge-invariant objects, i.e.
\begin{align*}
	\Pi\:\begin{pmatrix}
		\mathfrak{a}\\
		\pi\\
		\d^{\Sigma}_{s}\varphi_{1}+\omega_{1}\\
		\d^{\Sigma}_{s}\varphi_{2}+\omega_{2}\\
		\d^{\Sigma}_{s}\d^{\Sigma}_{s}\psi_{1}+\d^{\Sigma}_{s}\eta_{1}+\sf{h}\chi_{1}+\sf{T}_{1}\\
		 		\d^{\Sigma}_{s}\d^{\Sigma}_{s}\psi_{2}+\d^{\Sigma}_{s}\eta_{2}+\sf{h}\chi_{2}+\sf{T}_{2}
	\end{pmatrix}\mapsto 
	\begin{pmatrix}
		\Phi_{1}\\
		\Phi_{2}\\
		\sf{V}_{1}\\
		\sf{V}_{2}\\
		\sf{T}_{1}\\
		\sf{T}_{2}
	\end{pmatrix}\,\qquad\text{where}\qquad
	\begin{cases}
		\Phi_{1}=\mathfrak{a}+\varphi_{2}-2(\chi_{1}+\Delta\psi_{1})\\
		\Phi_{2}=\pi+\Delta\varphi_{1}-2(\chi_{2}+\Delta\psi_{2})\\
		\sf{V}_{1}:=\omega_{1}-\frac{1}{2}\eta_{2}\\
		\sf{V}_{2}:=\omega_{2}-\frac{1}{2}\Delta\eta_{1}
	\end{cases}\, .
\end{align*}
Once again, we stress that some components are only defined up to constant and hence, to make this precise, one either needs to fix those redundancies or work with suitable quotient spaces. By construction, $\Pi\circ\sf{K}_{\Sigma}=0$ and, similarly to the Maxwell case, it is expected that this projection has similar properties and that it provides an isomorphism
\begin{align*}
	\Pi:=\cfrac{\mathrm{ker}(\sf{K}_{\Sigma}^{\dagger}\vert_{\Gamma^{\infty}})}{\mathrm{ran}(\sf{K}_{\Sigma}\vert_{\Gamma^{\infty}})}\qquad \xrightarrow[\cong]{\quad [\Pi]\quad}\qquad \mathrm{ran}(\Pi\vert_{\mathrm{ker}(\sf{K}_{\Sigma}^{\dagger}\vert_{\Gamma^{\infty}})})\, ,
\end{align*}
where the space on the right-hand side has the interpretation of the space of gauge-invariant, i.e.~``physical'', initial data. Starting from there, one can approach to construct Hadamard states for linearised gravity, either by following similar a strategy as presented in this thesis for Maxwell's theory, or by using an approach based on \emph{modified Cauchy covariances} as studied in \cite{GerardWrochna}.

Now, in the previous approach, we restricted our analysis to \emph{spatially compact} and \emph{ultrastatic} globally hyperbolic spacetimes. To follow a similar strategy for more general globally hyperbolic background spacetimes, the following two aspects need to be considered:
\begin{itemize}
	\item[(i)]First, one needs to find a suitable generalisation of the tensor field decompositions presented above to \emph{non-compact} Riemannian manifolds. Achieving decompositions such as the York decomposition on Riemannian manifolds usually requires to solve elliptic equations and one might hope that the relevant decompositions can be achieved under suitable assumptions on the geometry and topology.
	\item[(ii)]In the non-ultrastatic case, the operators $\sf{K}_{\Sigma}$ and $\sf{K}_{\Sigma}^{\dagger}$ take a more complicated form. However, when working with the adapted initial data maps, it is expected that one can derive suitably simply expressions, at least on specific spacetimes such as static ones.
\end{itemize}

Tackling the previous problems and making the procedure outlined above more precise is subject to future work.

\renewcommand{\theHchapter}{A\arabic{chapter}}
\appendix
\chapter{Appendix}\label{Chap:Appendix}
\fancyhead[RE]{\nouppercase{Appendix}}

\section{Microlocal Analysis I: Distributions and Wavefront Set}\label{Appendix:Micro}
In the first section of the Appendix, we review some basics of the theory of distributions. In particular, we discuss distributions on manifolds, the celebrated \textit{Schwartz kernel theorem} as well as some basic notions of \textit{microlocal analysis}. 

\subsection{Basics of Distribution Theory}
To begin with, we provide a short introduction to distribution theory, starting with the \textit{local} theory in $\bb{R}^{d}$. Afterwards, we review the \textit{global} theory on manifold. We omit citations as the results are well known. We refer to the textbooks \cite{Treves,HormanderI,Friedlander,DuistermaatKolk} for details.

\paragraph{Distributions, the Local Theory.} Let $\bb{K}\subset\{\bb{R},\bb{C}\}$, $d\in\bb{N}$ and consider an arbitrary open subset $\mathcal{U}\subset\bb{R}^{d}$. Throughout the following, we will use the notation $C^{\infty}(\mathcal{U}):=C^{\infty}(\mathcal{U},\bb{K})$. Now, we consider the space $C^{\infty}(\mathcal{U})$ as a \emph{locally convex topological vector space} (short LCTVS), whose topology is induced by the family of seminorms (i.e.~the \emph{initial topology}\footnote{Let $(\sf{X}_{i},\cal{T}_{i})_{i\in \mathrm{I}}$ be a family of topological spaces, $\sf{X}$ be an arbitrary set and $\{f_{i}\:\sf{X}_{i}\to\sf{X}\}_{i\in\mathrm{I}}$, $\{g_{i}\:\sf{X}\to\sf{X}_{i}\}_{i\in\mathrm{I}}$ be two families of functions labelled by an arbitrary index set $\mathcal{I}$. The finest (resp.~coarsest) topology on $\sf{X}$ such that all $f_{i}$ (resp.~$g_{i}$) are continuous is called \textit{final topology} (resp.~\textit{initial topology}) on $\sf{X}$.} with respect to)
\begin{align*}
	p_{n,\sf{K}}(\varphi):=\max_{\alpha\in\bb{N}_{0}^{d},\,\vert\alpha\vert\leq n}\bigg(\sup_{x\in\sf{K}}\vert\partial^{\alpha}\varphi(x)\vert\bigg)
\end{align*}
labelled by $n\in\bb{N}_{0}$ and compact $\sf{K}\subset\mathcal{U}$, called the \emph{$C^{\infty}$-topology}, or \emph{topology of uniform convergence on compact sets}. Taking a \textit{countable exhaustion} of $\mathcal{U}$ by compact sets, we may consider a countable subset of seminorms, which shows that $C^{\infty}(\mathcal{U})$ is metrisable. Moreover, $C^{\infty}(\mathcal{U})$ is in fact Hausdorff and complete. In particular, $C^{\infty}(\mathcal{U})$ is a Fréchet space. Now, consider
\begin{align*}
	C^{\infty}_{\sf{K}}(\mathcal{U}):=\{\varphi\in C^{\infty}(\mathcal{U})\mid\mathrm{supp}(\varphi)\subset\sf{K}\}\subset C^{\infty}(\mathcal{U})\, ,
\end{align*}
for fixed $\sf{K}\subset\mathcal{U}$ compact, equipped with the subspace topology inherited from $C^{\infty}(\mathcal{U})$. The space $C^{\infty}_{\sf{K}}(\mathcal{U})$ is clearly a closed subset of $C^{\infty}(\mathcal{U})$ and hence by itself a Fréchet space. With this notation, we note that 
\begin{align*}
	C^{\infty}_{\mathrm{c}}(\mathcal{U})=\bigcup_{\sf{K}\subset\mathcal{U}\,\text{ compact}}C^{\infty}_{\sf{K}}(\mathcal{U})
\end{align*}
and we equip this space with the \emph{final topology} induced by the inclusions $i_{\sf{K}}\:C^{\infty}_{\sf{K}}(\mathcal{U})\to C^{\infty}_{\mathrm{c}}(\mathcal{U})$. This topology is usually called the \textit{canonical LF-topology}, where ``LF'' stands for ``limit of Fréchet spaces'', owing to the fact that notion of \textit{LF-spaces} can be made more systematic using the language of \emph{direct limits}. Note that the canonical LF-topology is strictly finer than the subspace topology on $C^{\infty}_{\mathrm{c}}(\mathcal{U})$ inherited from the $C^{\infty}$-topology on $C^{\infty}(\mathcal{U})$.

It turns out that the topology of the canonical LF-topology is very rich: first of all, $C^{\infty}_{\mathrm{c}}(\mathcal{U})$ is a complete Hausdorff LCTVS. It is also \emph{nuclear}, which implies that the projective and injective tensor products of $C^{\infty}_{\mathrm{c}}(\mathcal{U})$ with any other LCTVS coincide. Furthermore, it is a so-called \textit{Montel space}, which in particular implies that it is \emph{reflexive}, that a version of the \emph{Banach-Steinhaus theorem} holds and that it has the \emph{Heine-Borel property}, i.e.~that subsets are compact if and only if they are closed and bounded. However, it is important to stress that $C^{\infty}_{\mathrm{c}}(\mathcal{U})$ is \emph{not} metrisable and hence in particular \emph{not} a Fréchet space.

\begin{definition} (Distributions)\newline
	Let $\mathcal{U}\subset\bb{R}^{d}$ be open and consider the space $C^{\infty}_{\mathrm{c}}(\mathcal{U})$ equipped with its canonical LF-space topology. We denote its topological dual space by
	\begin{align*}
		\mathcal{D}^{\prime}(\mathcal{U}):=(C^{\infty}_{\mathrm{c}}(\mathcal{U}))^{\prime}=\{\varphi\:C^{\infty}_{\mathrm{c}}(\mathcal{U})\to\bb{K}\mid\text{ linear and continuous }\}\, .
	\end{align*}
	$\sf{T}\in\mathcal{D}^{\prime}(\mathcal{U})$ is called a \emph{(Schwartz) distribution on $\mathcal{U}$} and we introduce the bilinear pairing
	\begin{align*}
		\langle\cdot,\cdot\rangle_{\mathcal{U}}\:\mathcal{D}^{\prime}(\mathcal{U})\times C^{\infty}_{\mathrm{c}}(\mathcal{U})\to\bb{K}\,,\qquad \langle\sf{T},\varphi\rangle_{\mathcal{U}}:=\sf{T}(\varphi)\, .
	\end{align*}
\end{definition}

\begin{remark}
	Let $(\sf{X},\mathcal{T}_{\sf{X}})$ and $(\sf{Y},\mathcal{T}_{\sf{Y}})$ be two topological spaces and $f\:\sf{X}\to\sf{Y}$ be arbitrary. Then, if $f$ is \emph{continuous}, it is also \emph{sequentially continuous}, i.e.~$f(x_{n})\to f(x)$ for any sequence $(x_{n})_{n\in\mathbb{N}}$ converging to $x\in\sf{X}$. The reverse, however, is not true in general. In fact, it is true if $(\sf{X},\mathcal{T}_{\sf{X}})$ is \emph{first-countable}, however, this is only a sufficient condition.\footnote{Intuitively, this is because a sequence is labelled by the \emph{countable} set $\bb{N}$, which might be ``too small'' to detect continuity in a given (non first-countable) topology. If one considers \emph{nets} (also called \emph{Moore-Smith sequences} \cite{Moore}) instead, i.e.~$(x_{i})_{i\in\mathcal{I}}$ labelled by arbitrary directed index sets $(\mathcal{I},\leq)$, the statement becomes true again, i.e.~a function is continuous if any only if it is \emph{net continuous}.} Now, the space of test functions $C^{\infty}_{\mathrm{c}}(\mathcal{U})$ is \emph{not} first-countable, nevertheless, it turns out that a linear functional $\sf{T}\:C^{\infty}_{\mathrm{c}}(\mathcal{U})\to\bb{K}$ is still continuous if and only if it sequentially continuous. In fact, this is a consequence of the limit structure of the topology on $C^{\infty}_{\mathrm{c}}(\mathcal{U})$: let $\sf{T}\:C^{\infty}_{\mathrm{c}}(\mathcal{U})\to\bb{K}$ be sequentially continuous. Then, by definition, $\varphi\vert_{C^{\infty}_{\sf{K}}}$ is continuous for every $\sf{K}\subset\mathcal{U}$ compact. But $C^{\infty}_{\sf{K}}(\mathcal{U})$ is metrisable and hence in particular first countable, which shows continuity of $\sf{T}\vert_{C^{\infty}_{\sf{K}}(\mathcal{U})}$. Since $\sf{K}$ was arbitrary, we conclude that $\sf{T}$ is continuous. To sum up, a linear map $\sf{T}\:C^{\infty}_{\mathrm{c}}(\mathcal{U})\to\bb{K}$ is in $\mathcal{D}^{\prime}(\mathcal{U})$ if and only if $\sf{T}(\varphi_{n})\to\sf{T}(\varphi)$ for all $(\varphi_{n})_{n\in\bb{N}}$ in $C^{\infty}_{\mathrm{c}}(\mathcal{U})$ converging to $\varphi\in C^{\infty}_{\mathrm{c}}(\mathcal{U})$.
\end{remark}

In practice, the space of distributions $\mathcal{D}^{\prime}(\mathcal{U})$ is typically endowed with one of two topologies, each of which is widely used depending on the application: consider a topological vector space $\sf{X}$ with topological dual space $\sf{X}^{\prime}$ and consider the bilinear pairing $\langle\cdot,\cdot\rangle\:\sf{X}^{\prime}\times\sf{X}\to\bb{R}$ defined by $\langle x^{\prime},x\rangle:=x^{\prime}(x)$. Furthermore, for a given bounded subset $\sf{B}\subset\sf{X}$, consider the seminorm $p_{\sf{B}}(\cdot):=\sup_{x\in\sf{B}}\vert\langle\cdot,x\rangle\vert$. Then, we define the following two topologies on $\sf{X}^{\prime}$:
 \begin{itemize}
\item[(i)]The \emph{weak (dual) topology} on $\sf{X}^{\prime}$ is the coarsest topology on $\sf{X}^{\prime}$ such that that $\langle\cdot,x\rangle\:\sf{X}^{\prime}\to\bb{R}$ is continuous for all $x\in\bb{R}$. Equivalently, $\sf{X}^{\prime}$ equipped with the weak topology is exactly the LCTVS whose topology is induced by the family of seminorms $(p_{\sf{F}})_{\sf{F}\in\mathcal{F}}$, where $\mathcal{F}$ is the collection of all \textit{finite} subsets of $\sf{X}$. 
\item[(ii)]The \emph{strong (dual) topology} on $\sf{X}^{\prime}$ is the locally convex topology induced by the family $(p_{\sf{B}})_{\sf{B}\in\mathcal{B}}$, where $\mathcal{B}$ is the collection of \textit{bounded} subsets of $\sf{X}$.
\end{itemize}
By definition, the strong dual topology is strictly finer than the weak dual topology. Now, we can equip the space $\mathcal{D}^{\prime}(\mathcal{U})$ both with the strong and the weak dual topology induced from the canonical LF-topology on $C^{\infty}_{\mathrm{c}}(\mathcal{U})$. We note that both of them are Hausdorff, but neither of them is metrisable. The space $\mathcal{D}^{\prime}(\mathcal{U})$ equipped with the strong dual topology inherits many properties from $C^{\infty}_{\mathrm{c}}(\mathcal{U})$. In particular, it is a complete, nuclear Montel space. Last but not least, since $C^{\infty}_{\mathrm{c}}(\mathcal{U})$ is a Montel space, it actually turns out that a sequence in $\mathcal{D}^{\prime}(\mathcal{U})$ converges in the strong topology if and only if it converges in the weak topology. 

\begin{examples}\label{Ex:Distributions} (Examples of Distributions)
\begin{itemize}
	\item[(i)]The distribution $\delta\:C^{\infty}_{\mathrm{c}}(\mathcal{U})\to\bb{K}$, $\delta(\varphi):=\varphi(0)$ is called the \emph{(Dirac) $\delta$-distribution}.
	\item[(ii)]For every $\psi\in C^{\infty}(\mathcal{U})$, we define a distribution $\sf{T}_{\psi}\in\mathcal{D}^{\prime}(\mathcal{U})$ by
	\begin{align*}
		\langle\sf{T}_{\psi},\varphi\rangle_{\mathcal{U}}:=\int_{\mathcal{U}}\psi(x)\varphi(x)\,\d^{d}x\,.
	\end{align*}
	It is not too hard to see that the linear assignment $C^{\infty}(\mathcal{U})\to\mathcal{D}^{\prime}(\mathcal{U})$ is in fact a topological embedding if $\mathcal{D}^{\prime}(\mathcal{U})$ is equipped with its strong dual topology. Therefore, it makes sense to view  $C^{\infty}(\mathcal{U})\subset\mathcal{D}^{\prime}(\mathcal{U})$ as a subset. In fact, more generally, we can embed $\sf{L}^{1}_{\mathrm{loc}}(\mathcal{U})$ into $\mathcal{D}^{\prime}(\mathcal{U})$. Last but not least, a distribution $\sf{T}\in\mathcal{D}^{\prime}(\mathcal{U})$ is called \emph{regular}, if there exists a smooth function $\psi\in C^{\infty}(\mathcal{U})$ such that $\sf{T}=\sf{T}_{\psi}$. 
\end{itemize}
\end{examples}

Instead of $C^{\infty}_{\mathrm{c}}(\mathcal{U})$, one might also consider different spaces of test functions. An important example is provided by the Fréchet space $C^{\infty}(\mathcal{U})$. We denote its topological dual by\footnote{The origin of the notation $\mathcal{D}^{\prime}$ and $\mathcal{E}^{\prime}$ stems from the fact that L.~Schwartz used the symbols $\mathcal{D}$ to denote $C^{\infty}_{\mathrm{c}}$ and $\mathcal{E}$ for $C^{\infty}$, see~\cite{Schwartz}.}
\begin{align*}
	\mathcal{E}^{\prime}(\mathcal{U}):=(C^{\infty}(\mathcal{U}))^{\prime}=\{\varphi\:C^{\infty}_{\mathrm{c}}(\mathcal{U})\to\bb{R}\mid\text{ linear and continuous }\}\, .
\end{align*}
Now, consider the inclusion map $i\:C^{\infty}_{\mathrm{c}}(\mathcal{U})\to C^{\infty}(\mathcal{U})$. This map is clearly continuous with respect to the canonical LF-topology on $C^{\infty}_{\mathrm{c}}(\mathcal{U})$ and the Fréchet topology on $C^{\infty}(\mathcal{U})$, since, as mentioned above, the former is strictly finer than the subspace topology induced from $C^{\infty}(\mathcal{U})$. Furthermore, the inclusion map can be shown to be an injection with dense image, which in particular implies that the corresponding transpose map $i^{t}\:\mathcal{E}^{\prime}(\mathcal{U})\to\mathcal{D}^{\prime}(\mathcal{U})$, where both spaces are  equipped with their strong dual topologies, is also a continuous injection. In other words, $\mathcal{E}^{\prime}(\mathcal{U})$ can naturally be viewed as a subspace of $\mathcal{D}^{\prime}(\mathcal{U})$. 

In fact, a distribution $\sf{T}\in\mathcal{D}^{\prime}(\mathcal{U})$ is in $\mathcal{E}(\mathcal{U})$ exactly when it is also continuous in the coarser subspace topology on $C^{\infty}_{\mathrm{c}}(\mathcal{U})$ induced from $C^{\infty}(\mathcal{U})$, in which case it extends to a linear functional on all of $C^{\infty}(\mathcal{U})$, by density. Looking at a \emph{regular} distribution $\sf{T}_{\psi}$, i.e.~Example~\ref{Ex:Distributions}(ii), this is the case when the labelling function $\psi$ is compactly supported. More precisely, distributions in $\mathcal{E}^{\prime}(\mathcal{U})$ can be characterised as follows: first of all, if $\mathcal{V}\subset\mathcal{U}\subset\bb{R}^{d}$ are two open sets, we can extend any function $\varphi\in C^{\infty}_{\mathrm{c}}(\mathcal{V})$ trivially by zero to view it as an element of $C^{\infty}_{\mathrm{c}}(\mathcal{U})$. This defines the \emph{extension map} $e\:C^{\infty}_{\mathrm{c}}(\mathcal{V})\to C^{\infty}_{\mathrm{c}}(\mathcal{U})$, which is a linear and continuous injection.\footnote{However, we stress that this map is not dense. Furthermore, the inverse defined on the image of this map is \emph{not} continuous. In other words, $C^{\infty}(\mathcal{V})$ is a linear subspace of $C^{\infty}(\mathcal{U})$, however, not a topological one.} Hence, its transpose provides a linear continuous map $e^{t}\:\mathcal{D}^{\prime}(\mathcal{U})\to\mathcal{D}^{\prime}(\mathcal{V})$, which is called the \emph{restriction operator} and denoted by $\sf{T}\vert_{\mathcal{V}}:=e^{t}(\sf{T})$ for all $\sf{T}\in \mathcal{D}^{\prime}(\mathcal{U})$. With this terminology, we define the \emph{support} of a distribution by
\begin{align*}
	\mathrm{supp}(\sf{T}):=\mathcal{U}\backslash\{\Omega\subset\mathcal{U}\text{ open }\mid \sf{T}\vert_{\Omega}=0\}\, .
\end{align*}
By definition, this is a closed subset of $\mathcal{U}$. Now, as it turns out, the space $\mathcal{E}^{\prime}(\mathcal{U})$ is exactly the space of \emph{compactly supported distributions}, i.e.
\begin{align*}
	\mathcal{E}^{\prime}(\mathcal{U})=\{\sf{T}\in\mathcal{D}^{\prime}(\mathcal{U})\mid \mathrm{supp}(\sf{T})\text{ compact }\}\, .
\end{align*}
Note that for a regular distribution $\sf{T}_{\psi}$ with $\psi\in C^{\infty}(\mathcal{U})$, it holds that $\mathrm{supp}(\sf{T}_{\psi})=\mathrm{supp}(\psi)$.

\newpage
Another important function space is the space of \emph{Schwartz functions}, which is defined by
\begin{align*}
	\mathcal{S}(\bb{R}^{d}):=\{\varphi\in C^{\infty}(\bb{R}^{d})\mid \Vert\varphi\Vert_{\alpha,\beta}<\infty\}\,,\qquad\Vert\varphi\Vert_{\alpha,\beta}:=\sup_{x\in\bb{R}^{d}}\vert x^{\alpha}\partial^{\beta}\varphi(x)\vert\, .
\end{align*}
It can be shown that $\mathcal{S}(\bb{R}^{d})$ is in fact a Fréchet and nuclear Montel space, whose topology is strictly finer than the subspace topology inherited from $C^{\infty}(\bb{R}^{d})$. We denote its topological dual space, the \emph{space of tempered distributions}, by
\begin{align*}
	\mathcal{S}^{\prime}(\bb{R}^{d}):=(\mathcal{S}(\bb{R}^{d}))^{\prime}=\{\varphi\:\mathcal{S}(\bb{R}^{d})\to\bb{K}\mid\text{ linear and continuous }\}\, .
\end{align*}
The inclusions $C^{\infty}_{\mathrm{c}}(\bb{R}^{d})\to\mathcal{S}(\bb{R}^{d})\to C^{\infty}(\bb{R}^{d})$ are continuous with dense image and hence, by duality, we conclude that $\mathcal{E}^{\prime}(\bb{R}^{d})\subset\mathcal{S}^{\prime}(\bb{R}^{d})\subset\mathcal{D}^{\prime}(\bb{R}^{d})$. One of the most important properties of $\mathcal{S}^{\prime}(\bb{R}^{d})$ is that it allows for a notion of \emph{Fourier transform}. More precisely, the Fourier transform is a well-defined linear continuous operator of the form
\begin{align*}
	\mathcal{F}\:\mathcal{S}(\bb{R}^{d})\to\mathcal{S}(\bb{R}^{d})\,,\qquad \mathcal{F}(\varphi):=\int_{\bb{R}^{d}}e^{-i\xi\cdot x}\varphi(x)\,\d^{d}x\, .
\end{align*}
By duality, this extends to a map $\mathcal{F}\:\mathcal{S}^{\prime}(\bb{R}^{d})\to\mathcal{S}^{\prime}(\bb{R}^{d})$, i.e.~by setting $\langle\mathcal{F}(\sf{T}),\varphi\rangle_{\bb{R}^{d}}=\langle\sf{T},\mathcal{F}(\varphi)\rangle_{\bb{R}^{d}}$.

\paragraph{Distributions, the Global Theory.} Now, let $\sf{M}$ be a smooth manifold and $\sf{E}\xrightarrow{\pi}\sf{M}$ be a finite-rank smooth $\bb{K}$-vector bundle over $\sf{M}$. First of all, the goal is to equip the spaces $\Gamma^{\infty}(\sf{E})$ and $\Gamma^{\infty}_{\mathrm{c}}(\sf{E})$ with similar topologies as in the local theory. For this, we choose a Riemannian metric $\sf{g}$ on $\sf{M}$ with Levi-Civita connection $\nabla$, a connection $\nabla^{\sf{E}}$ on $\sf{E}$ as well as a positive-definite bundle metric $\langle\cdot,\cdot\rangle_{\sf{E}}$ on $\sf{E}$. Now, let us set $\sf{E}_{k}:=\sf{E}\otimes\sf{T}^{\ast}\sf{M}^{\otimes k}$, equipped with the obvious bundle metric $\langle\cdot,\cdot\rangle_{\sf{E}_{k}}$ induced by $\langle\cdot,\cdot\rangle_{\sf{E}}$ and $\sf{g}$, and define the \emph{$k$th covariant derivative} on $\sf{E}$ by
\begin{align*}
	\nabla^{\sf{E},k}\:\Gamma^{\infty}(\sf{E})\xrightarrow{\nabla^{\sf{E}}}\Gamma^{\infty}(\sf{E}_{1})\xrightarrow{\nabla^{\sf{E}_{1}}}\dots\Gamma^{\infty}(\sf{E}_{2})\xrightarrow{\nabla^{\sf{E}_{2}}}\dots\xrightarrow{\nabla^{\sf{E}_{k-1}}}\Gamma^{\infty}(\sf{E}_{k})\,,
\end{align*}
where $\nabla^{\sf{E}_{k}}$ denotes the natural connection on $\sf{E}_{k}$ induced by $\nabla^{\sf{E}}$ and $\nabla$. Then, we consider $\Gamma^{\infty}(\sf{E})$ as the LCTVS whose topology is induced by the family of seminorms
\begin{align*}
	p_{k,\sf{K}}(\varphi):=\max_{0\leq j\leq k}\bigg(\sup_{p\in\sf{K}}\Vert \nabla^{\sf{E},j}\psi(p)\Vert_{\sf{E}_{j}}\bigg)\,,
\end{align*}
labelled by $k\in\bb{N}_{0}$ and compact sets $\sf{K}\subset\sf{M}$. It is not too hard to see that this definition is, in fact, independent of the choice of metrics $\sf{g}$, $\langle\cdot,\cdot\rangle_{\sf{E}}$ and connection $\nabla^{\sf{E}}$, since different choices lead to equivalent seminorms. As in the case of functions, $\Gamma^{\infty}(\sf{E})$ turns out to be a Fréchet space. Moreover, we equip the space $\Gamma^{\infty}_{\mathrm{c}}(\sf{E})$ with the canonical LF-space topology induced from the subspaces $\Gamma^{\infty}_{\sf{K}}(\sf{E}):=\{\varphi\in\Gamma^{\infty}(\sf{E})\mid\mathrm{supp}(\varphi)\subset\sf{K}\}$, as in the local theory. It can be shown that $\Gamma^{\infty}_{\mathrm{c}}(\sf{E})$ has similar topological properties as its scalar counterpart, i.e.~$\Gamma^{\infty}_{\mathrm{c}}(\sf{E})$ is a complete, Hausdorff (but not metrisable) LCTVS that is also a nuclear Montel space.

\begin{definition} (Distributions on Manifolds)\newline
	Let $\sf{M}$ be a smooth manifold and equip $\Gamma^{\infty}_{\mathrm{c}}(\sf{E}^{\ast}),\Gamma^{\infty}(\sf{E}^{\ast})$ with their canonical LF-space and Fréchet space topologies, respectively. Then, we denote their topological dual spaces by
	\begin{align*}
		\mathcal{D}^{\prime}(\sf{E}):=\mathcal{D}^{\prime}(\sf{M},\sf{E}):=(\Gamma_{\mathrm{c}}^{\infty}(\sf{M},\sf{E}^{\ast}))^{\prime}\,,\qquad \mathcal{E}^{\prime}(\sf{E}):=\mathcal{E}^{\prime}(\sf{M},\sf{E}):=(\Gamma^{\infty}(\sf{M},\sf{E}^{\ast}))^{\prime}\, .
	\end{align*}
	The distributional pairing is denoted by $\langle\cdot,\cdot\rangle_{\sf{M}}\:\mathcal{D}^{\prime}(\sf{E})\times\Gamma_{\mathrm{c}}^{\infty}(\sf{E}^{\ast})\to\bb{K}$.
\end{definition}

As in the scalar case, we can equip $\mathcal{E}^{\prime}(\sf{E})$ and $\mathcal{D}^{\prime}(\sf{E})$ with either their weak or strong dual topologies. Furthermore, $\mathcal{E}^{\prime}(\sf{E})$ is exactly the subspace of $\mathcal{D}^{\prime}(\sf{E})$ consisting of distributions with compact support.

\newpage
Now, let us equip the manifold $\sf{M}$ with a pseudo-Riemannian metric $\sf{g}$. Then, the Fréchet space of smooth sections can be embedded into the space of distributions via
\begin{align}\label{eq:RegDist}
	\sf{T}_{\bullet}\:\Gamma^{\infty}(\sf{E})\hookrightarrow\mathcal{D}^{\prime}(\sf{E})\,,\qquad \langle\sf{T}_{\psi},\varphi\rangle_{\sf{M}}:=\int_{\sf{M}}\varphi_{x}(\psi_{x})\,\d\mu_{\sf{g}}(x)\,,
\end{align}
generalising Example~\ref{Ex:Distributions}(ii). As before, we call a distribution in the image of this map \textit{regular}. Note that this embedding explicitly depends on the chosen metric $\sf{g}$ on $\sf{M}$ and we stress that some authors use a slightly different definition (e.g.~\cite[Sec.~1.1.8]{Tarkhanov}): consider the \textit{density bundle} $\mathcal{D}\sf{M}$, i.e.~the line bundle whose fibre $\mathcal{D}_{p}\sf{M}$ at some point $p\in\sf{M}$ is the vector space spanned by $\vert\omega\vert$ for $\omega\in\Omega^{\mathrm{dim}(\sf{M})}(\sf{M})$, see e.g.~\cite[Chap.~16]{LeeSmooth}. Then, they define $\mathcal{D}^{\prime}(\sf{M},\sf{E})$ as the (topological) dual of the space of \emph{densitised test sections} $\Gamma^{\infty}(\sf{M},\sf{E}^{\ast}\otimes\mathcal{D}\sf{M})$. In this way, one obtains a natural embedding $\Gamma^{\infty}(\sf{E})\hookrightarrow\mathcal{D}^{\prime}(\sf{E})$ without the need of specifying a metric or volume measure beforehand. For our purposes, however, we typically work on a pseudo-Riemannian manifold, and it is therefore natural to define the embedding using the associated volume form.

\subsection{The Kernel Theorem of Schwartz}
Having set up the terminology and notation for distributions both in the local and global setting, we now briefly discuss the celebrated \emph{kernel theorem of Schwartz}, which allows us to represent a linear and continuous operator between spaces of test sections and distributions in terms of a bidistributional \emph{kernel}. It was originally proven by Schwartz \cite{SchwartzKernel}; for a modern proof, we refer to \cite[Thm.~51.7]{Treves} and \cite[Thm.~1.5.1]{Tarkhanov}.

For the formulation of the kernel theorem, we recall that $\sf{E}\boxtimes\sf{F}$ for two finite-rank smooth $\bb{K}$-vector bundles $\sf{E},\sf{F}\xrightarrow{\pi}\sf{M}$ over the same manifold $\sf{M}$ denotes the \emph{exterior} (or \textit{outer}) \textit{tensor product}, i.e.~the vector bundle over $\sf{M}\times\sf{M}$ whose fibre at the point $(p,q)\in\sf{M}\times\sf{M}$ is given by $\sf{E}_{p}\otimes\sf{F}_{q}$. In other words, $\sf{E}\boxtimes\sf{F}=(\mathrm{pr}^{\ast}_{1}\sf{E})\otimes(\mathrm{pr}^{\ast}_{2}\sf{F})$, where $\mathrm{pr}_{i}\:\sf{M}\times\sf{M}\to\sf{M}$ denote the obvious projections for $i=1,2$ and $\otimes$ the usual tensor product of vector bundles over $\sf{M}\times\sf{M}$.

\begin{theorem}\label{Thm:Kernel} \emph{(Kernel Theorem of Schwartz)}\newline
	Let $\sf{M}$ be a smooth manifold and $\sf{E},\sf{F}\xrightarrow{\pi}\sf{M}$ be two finite-rank $\bb{K}$-bundle over $\sf{M}$. A linear operator $\sf{A}\:\Gamma^{\infty}_{\mathrm{c}}(\sf{E})\to\mathcal{D}^{\prime}(\sf{F})$ is continuous, where $\mathcal{D}^{\prime}(\sf{F})$ is equipped with the strong topology, if and only if there exists a distribution $k_{\sf{A}}\in\mathcal{D}^{\prime}(\sf{M}\times\sf{M},\sf{F}\boxtimes\sf{E}^{\ast})$ such that
	\begin{align*}
		\langle\sf{A}\psi,\varphi\rangle_{\sf{M}}=\langle k_{\sf{A}},\varphi\boxtimes\psi\rangle_{\sf{M}\times\sf{M}}\,\qquad \forall\psi\in\Gamma^{\infty}_{\mathrm{c}}(\sf{E})\,,\,\,\varphi\in\Gamma^{\infty}(\sf{F}^{\ast})\,.
	\end{align*}
\end{theorem}

\begin{remark}\label{Rem:Schwartz}
	In a more functional-analytic formulation, the kernel theorem of Schwartz asserts that there are linear isomorphisms
	\begin{align}\label{eq:SchwartzKernel}
		\mathcal{D}^{\prime}(\sf{M}\times\sf{M},\sf{F}\boxtimes\sf{E}^{\ast})\cong \mathcal{D}^{\prime}(\sf{F})\hat{\otimes} \mathcal{D}^{\prime}(\sf{E}^{\ast})\cong \mathcal{L}(\Gamma^{\infty}_{\mathrm{c}}(\sf{E}),\mathcal{D}^{\prime}(\sf{F}))\,,
	\end{align}
	where $\mathcal{L}(\cdot,\cdot)$ denotes the space of linear and continuous maps between two LCTVSs, and where the tensor product $\hat{\otimes}$ denotes either the injective or projective tensor product (the choice does not matter, since $\mathcal{D}^{\prime}$ equipped with its strong topology is nuclear). The above isomorphisms are in fact even \emph{topological} isomorphisms, i.e.~homeomorphisms, if the space $\mathcal{L}(\Gamma^{\infty}_{\mathrm{c}}(\sf{E}),\mathcal{D}^{\prime}(\sf{F}))$ is equipped with the \emph{topology of uniform convergence on bounded sets}.\footnote{Let $\mathcal{E},\mathcal{F}$ be two LCTVSs whose topologies are induced by the family of seminorms $(p_{i})_{i\in\mathrm{I}}$ and $(q_{j})_{j\in\mathrm{J}}$. The \emph{topology of uniform convergence on bounded sets} on $\mathcal{L}(\mathcal{E},\mathcal{F})$ is the topology defined by the family of seminorms $r_{\mathcal{B},j}(f):=\sup_{x\in\mathcal{B}}q_{j}(f(x))$ labelled by bounded sets $\mathcal{B}\subset\mathcal{E}$ and $\mathrm{J}$. In the special case $\mathcal{F}:=\bb{K}$, we have $\mathcal{E}^{\prime}=\mathcal{L}(\mathcal{E},\bb{K})$ and the topology coincides with the strong dual topology.}
	
	The kernel theorem also holds for other distribution spaces, i.e.~\eqref{eq:SchwartzKernel} remains valid with $\mathcal{D}^{\prime}$ replaced by $\mathcal{E}^{\prime}$ and, on $\bb{R}^{d}$, also with $\mathcal{S}^{\prime}$. The key property for the kernel theorem is the \emph{nuclearity} of $\mathcal{D}^{\prime},\mathcal{E}^{\prime},\mathcal{S}^{\prime}$ and there exists a generalisation of the kernel theorem for general nuclear spaces (see e.g.~\cite{Schaefer}). In fact, Schwartz's kernel theorem was the main motivation for developing the theory of nuclear spaces by Grothendieck \cite{Grothendieck} in the first place.
	
	Looking at the second isomorphism in Eq.~\eqref{eq:SchwartzKernel}, we note that the kernel theorem of Schwartz can be seen as a functional-analytic analogue of the well-known fact from linear algebra that $\sf{W}\otimes_{\bb{F}}\sf{V}^{\ast}\cong\mathrm{Hom}_{\bb{F}}(\sf{V},\sf{W})$ for finite-dimensional vector spaces $\sf{V}$, $\sf{W}$ over a field $\bb{F}$.
\end{remark}

If a kernel $k_{\sf{A}}$ is \emph{regular}, i.e.~$k_{\sf{A}}\in\Gamma^{\infty}(\sf{F}\boxtimes\sf{E}^{\ast})$, we can view it is a family of linear maps $k_{\sf{A}}(x,y)\:\sf{E}_{y}\to\sf{F}_{x}$ depending smoothly on $x,y\in\sf{M}$. If we further choose a bundle metric $\langle\cdot,\cdot\rangle_{\sf{F}}$ on $\sf{F}$, allowing us to identify $\Gamma^{\infty}(\sf{F})\cong\Gamma^{\infty}(\sf{F}^{\ast})$ via $\psi\mapsto\langle\psi,\cdot\rangle_{\sf{F}}$, the defining relation reads
\begin{align*}
	\int_{\sf{M}}\langle\varphi,\sf{A}\psi\rangle_{\sf{F}}\,\d\mu_{\sf{g}}=\int_{\sf{M}}\int_{\sf{M}}\langle\varphi(x),k_{\sf{A}}(x,y)\psi(y)\rangle_{\sf{F}_{x}}\,\d\mu_{\sf{g}}(x)\,\d\mu_{\sf{g}}(y)
\end{align*}
for all test sections $\psi\in\Gamma^{\infty}_{\mathrm{c}}(\sf{E})$ and $\varphi\in\Gamma^{\infty}_{\mathrm{c}}(\sf{F})$. Motivated by this fact, one often writes
\begin{align*}
	(\sf{A}\psi)(x)=\int_{\sf{M}}k_{\sf{A}}(x,y)\psi(y)\,\d\mu_{\sf{g}}(y)\,,
\end{align*}
even in the case in which $k_{\sf{A}}$ is a (non-regular) distribution.

As explained in Remark~\ref{Rem:Schwartz}, the space of distributional kernels is homeomorphic to the tensor product $\mathcal{D}^{\prime}(\sf{F})\hat{\otimes} \mathcal{D}^{\prime}(\sf{E}^{\ast})$. We consider the following important subspaces:
\begin{itemize}\setlength\itemsep{0em}
	\item[(i)]A kernel $k_{\sf{A}}$ is called \emph{semiregular}, if $k_{\sf{A}}\in\Gamma^{\infty}(\sf{F})\hat{\otimes} \mathcal{D}^{\prime}(\sf{E}^{\ast})$.
	\item[(ii)]A kernel $k_{\sf{A}}$ is called \emph{continuable}, if $k_{\sf{A}}\in\mathcal{D}^{\prime}(\sf{F})\hat{\otimes}\Gamma^{\infty}(\sf{E}^{\ast})$.
	\item[(iii)]A kernel $k_{\sf{A}}$ is called \emph{biregular} if it is both semiregular and continuable.
	\item[(iv)]A kernel $k_{\sf{A}}$ is called \emph{properly supported}, if the projections $\mathrm{pr}_{1,2}\:\mathrm{supp}(k_{\sf{A}})\to\sf{M}$ onto the first and second factor are \emph{proper maps} (i.e.~preimages of compact sets are compact).
\end{itemize}

Formally, a kernel is semiregular (resp.~continuable), if $k_{\sf{A}}(x,y)$ is smooth in the variable $x$ (resp.~$y$). Let us also stress that biregularity is a weaker property than regularity. For instance, the kernel of the identity operator $\mathrm{id}\:C^{\infty}_{\mathrm{c}}(\bb{R})\to C^{\infty}_{\mathrm{c}}(\bb{R})$ is given by $k_{\mathrm{id}}(x,y)=\delta(x-y)$, which can easily seen to be biregular. However, $k_{\mathrm{id}}$ is clearly not regular.

\begin{proposition}\label{Prop:Kernels} \emph{(Mapping Properties of Distributions)}
\begin{itemize}
	\item[\emph{(i)}]$k_{\sf{A}}$ is semiregular if and only if $\sf{A}$ is continuous as an operator $\sf{A}\:\Gamma^{\infty}_{\mathrm{c}}(\sf{E})\to\Gamma^{\infty}(\sf{F})$.
	\item[\emph{(ii)}]$k_{\sf{A}}$ is continuable if and only if $\sf{A}$ is extends continuously to a map $\sf{A}\:\mathcal{E}^{\prime}(\sf{E})\to\mathcal{D}^{\prime}(\sf{F})$.
	\item[\emph{(iii)}]If $k_{\sf{A}}$ is semiregular and properly supported, then $\sf{A}$ is continuous as an operator $\sf{A}\:\Gamma^{\infty}_{\mathrm{c}}(\sf{E})\to\Gamma_{\mathrm{c}}^{\infty}(\sf{F})$, which extends continuously to an operator $\sf{A}\:\Gamma^{\infty}(\sf{E})\to\Gamma^{\infty}(\sf{F})$. Furthermore, if $k_{\sf{A}}$ is continuable and properly supported, then $\sf{A}$ extends continuously to a map $\sf{A}\:\mathcal{E}^{\prime}(\sf{E})\to\mathcal{E}^{\prime}(\sf{F})$, which in turn continuously extends to $\sf{A}\:\mathcal{D}^{\prime}(\sf{E})\to\mathcal{D}^{\prime}(\sf{F})$.
\end{itemize}
\end{proposition}

\begin{proof}
Claim (i) is obvious. For (ii), we first recall that the transpose of $\sf{A}\:\Gamma^{\infty}_{\mathrm{c}}(\sf{E})\to\mathcal{D}^{\prime}(\sf{F})$ is the operator $\sf{A}^{t}\:\Gamma^{\infty}_{\mathrm{c}}(\sf{F}^{\ast})\to\mathcal{D}^{\prime}(\sf{E}^{\ast})$ defined by $\langle\sf{A}\psi,\varphi\rangle_{\sf{M}}=\langle\sf{A}^{t}\varphi,\psi\rangle_{\sf{M}}$ for all $\psi\in\Gamma^{\infty}_{\mathrm{c}}(\sf{E})$ and $\varphi\in\Gamma^{\infty}_{\mathrm{c}}(\sf{F}^{\ast})$. The corresponding kernels are related via $k_{\sf{A}}(x,y)=(k_{\sf{A}}(y,x))^{t}$. Now, with this definition, it is clear that $\sf{A}$ is semiregular (resp.~continuable) if and only if $\sf{A}^{t}$ continuable (resp.~semiregular). Hence, if $\sf{A}$ is continuable, we know that $\sf{A}^{t}$ is continuous as an operator $\sf{A}^{t}\:\Gamma^{\infty}_{\mathrm{c}}(\sf{F}^{\ast})\to\Gamma^{\infty}(\sf{E}^{\ast})$ and we can define an extension $\sf{A}^{\prime}\:\mathcal{E}^{\prime}(\sf{E})\to\mathcal{D}^{\prime}(\sf{F})$ of $\sf{A}$ by 
\begin{align*}
	\langle \sf{A}^{\prime}\sf{T},\varphi\rangle_{\sf{M}}:=\langle\sf{T},\sf{A}^{t}\varphi\rangle_{\sf{M}}
\end{align*}
for all $\sf{T}\in\mathcal{E}^{\prime}(\sf{E})$ and $\varphi\in\Gamma^{\infty}_{\mathrm{c}}(\sf{F}^{\ast})$. For (iii), we first observe that
\begin{align*}
	\mathrm{supp}(\sf{A}\psi)\subset \mathrm{supp}(k_{\sf{A}})\circ\mathrm{supp}(\psi):=\{x\in\sf{M}\mid \exists y\in\mathrm{supp}(\psi): (x,y)\in\mathrm{supp}(k_{\sf{A}})\}
\end{align*}
for all $\psi\in\Gamma^{\infty}_{\mathrm{c}}(\sf{E})$, from which the claims easily follow from (i) and (ii).
\end{proof}

\subsection{Singular Support and Wavefront Set}\label{Sec:Wavefront}
As a next step, we take a closer look at the singularities of distributions by introducing the concept of the wavefront set, a central notion of microlocal analysis. Recall that a distribution $\sf{T} \in \mathcal{D}'(\sf{E})$ is called \emph{regular} if it lies in the image of the map~\eqref{eq:RegDist}, or equivalently, if it can (globally) be identified with a smooth section. For a general distribution $\sf{T} \in \mathcal{D}'(\sf{E})$, we define the singular support as the set of points around which the distribution fails to be regular, i.e.,
\begin{align*}
	\mathrm{sing}\,\mathrm{supp}(\sf{T}):=\sf{M}\backslash\{\Omega\subset\sf{M}\text{ open }\mid \sf{T}\vert_{\Omega}\text{ regular }\}\subset\mathrm{supp}(\sf{T})\, .
\end{align*}

In general, the product of two distributions is not well-defined, since already the concept of a ``value at a point'' lacks meaning in full generality. Nevertheless, motivated by physical applications, one may still ask whether the multiplication of distributions can be defined in a consistent way, at least under suitable restrictions. To start with, if we consider two distributions $\sf{T},\sf{S}\in\mathcal{D}^{\prime}(\mathcal{U})$ for $\mathcal{U}\subset\bb{R}^{d}$ open such that $\mathrm{sing}\,\mathrm{supp}(\sf{T})\cap \mathrm{sing}\,\mathrm{supp}(\sf{S})=\emptyset$, there is an obvious way to define their \emph{product}, e.g.~by choosing an open cover $(\mathcal{U}_{i})_{i\in\mathrm{I}}$ of $\mathcal{U}$ such that for every $i\in\mathrm{I}$, we have either $\sf{T}\vert_{\mathcal{U}_{i}}\in C^{\infty}(\mathcal{U}_{i})$ or $\sf{S}\vert_{\mathcal{U}_{i}}\in C^{\infty}(\mathcal{U}_{i})$, and by defining $\sf{T}\cdot\sf{S}\in\mathcal{D}^{\prime}(\mathcal{U})$ via
\begin{align*}
	(\sf{T}\cdot\sf{S})\psi:=\begin{cases} \sf{T}(\sf{S}\psi) & \text{if }\sf{S}\vert_{\mathcal{U}_{i}}\in C^{\infty}(\mathcal{U}_{i})\\
	\sf{S}(\sf{T}\psi) & \text{if }\sf{T}\vert_{\mathcal{U}_{i}}\in C^{\infty}(\mathcal{U}_{i})
	\end{cases}
\end{align*}
for all $\psi\in C^{\infty}_{\mathrm{c}}(\mathcal{U}_{i})$. It is easy to see that this is well-defined, since on overlaps, on which both $\sf{S}$ and $\sf{T}$ are smooth, both definition agree with the pointwise multiplication. More generally, if we consider two distributions whose singular supports are \emph{not} disjoint, there are still some situations in which a product can meaningfully be defined. For instance, we can always consider the pointwise product of continuous functions, even though their singular supports (which only detects the failure of being \emph{smooth}) might not be disjoint. Furthermore, if we consider the Heaviside function $\theta$ with $\mathrm{sing}\,\mathrm{supp}(\theta)=\{0\}$, it intuitively makes sense to set $\theta^{2}=\theta$. 

By the previous discussion, it is clear that we need a more refined description of the singularities of a distribution, which leads us to the notion of the wavefront set. The wavefront set of a distribution was originally introduced by Sato\footnote{More precisely, the definition of Sato is equivalent to what is nowadays known as the \emph{analytic wavefront set}.} \cite{Sato1,Sato2} and Hörmander \cite{HormanderWF,HormanderFourier} in the early 1970s. We refer to \cite{Brouder} for a gentle introduction and \cite{HormanderI} for a detailed discussion. The central idea is to analyse distributions in Fourier space. As a consequence of the \emph{Paley–Wiener–Schwartz theorem}, a distribution $u \in \mathcal{E}^{\prime}(\mathbb{R}^{d})$ is regular, that is, $u \in C^{\infty}_{\mathrm{c}}(\mathbb{R}^{d})$, if and only if for every $m \in \mathbb{N}$ there exists a constant $C_{m} > 0$ such that
\begin{align*}
	\vert\mathcal{F}(u)(\xi)\vert\leq C_{m}(1+\vert\xi\vert)^{-m}
\end{align*}
for all $\xi\in\mathbb{R}^{d}$, where the left-hand side is well-defined, since $\mathcal{F}(u)\in C^{\infty}(\bb{R}^{d})$ for all $u\in\mathcal{E}^{\prime}(\mathbb{R}^{d})$. 

\begin{definition} (Wavefront Set)\newline
Let $\sf{T}\in\mathcal{D}^{\prime}(\mathcal{U})$. We call a pair $(p,\xi)\in \mathcal{U}\times (\bb{R}^{d}\backslash\{0\})$ \emph{regular directed}, if there exists $\chi\in C_{\mathrm{c}}^{\infty}(\mathcal{U})$ with $\chi(p)=1$ and a closed conic neighbourhood\footnote{i.e.~for all $\lambda>0$ it holds that $\lambda\xi\in\mathcal{V}$.} $\mathcal{V}$ of $\xi$ such that for all $m\in\bb{N}$ there exists a $C_{m}>0$ with 
\begin{align*}
	\vert\mathcal{F}(\chi \sf{T})(\xi)\vert\leq C_{m}(1+\vert\xi\vert)^{-m}\, .
\end{align*}
The \emph{wavefront set} $\mathrm{WF}(\sf{T})$ is the set of all $(p,\xi)\in\mathcal{U}\times (\bb{R}^{d}\backslash\{0\})$ that are \emph{not} regular directed.
\end{definition}

By definition, the projection of $\mathrm{WF}(\sf{T})$ onto the first factor is exactly $\mathrm{sing}\,\mathrm{supp}(\sf{T})$. In other words, $\mathrm{WF}(\sf{T})$ can be understood as a refinement of $\mathrm{sing}\,\mathrm{supp}(\sf{T})$ that does not only include informations on \emph{where} a distribution is singular, but also informations on the \emph{singular direction} in Fourier space. Furthermore, to come back to the central motivation, there is a well-defined notion of a \emph{product} of two distributions $\sf{T},\sf{S}\in\mathcal{D}^{\prime}(\mathcal{U})$ if their wavefront sets satisfy $\mathrm{WF}(\sf{T})\cap (-\mathrm{WF}(\sf{S}))=\emptyset$. In this case, the product $\sf{S}\sf{T}$ is defined as the pullback of their tensor product along the diagonal map $\delta_{\mathcal{U}}\:\mathcal{U}\to\mathcal{U}\times\mathcal{U}$ (see e.g.~\cite[8.2.10]{HormanderI} for details).

\begin{example}\label{Ex:WFdelta}
	Consider the $\delta$-distribution $\delta\in\mathcal{S}^{\prime}(\bb{R}^{d})$ defined by $\delta(\varphi):=\varphi(0)$. Clearly, it holds that $\mathrm{sing}\,\mathrm{supp}(\delta)=\{0\}$. Now, as a distribution, we have $\mathcal{F}(\delta)=1$. Hence, for arbitrary $\chi\in C^{\infty}(\mathcal{U})$ with $\chi(0)=1$, it holds that
	\begin{align*}
		\vert\mathcal{F}(\chi\delta)(\xi)\vert=1
	\end{align*}
	for all $\xi\in\mathcal{V}$. This shows $\mathrm{WF}(\delta)=\{(0,\xi)\mid\xi\in\bb{R}^{d}\backslash\{0\}\}$.
\end{example}

To define the wavefront set on manifolds, we first observe that it behaves covariantly under the change of coordinates. More precisely, if $\varphi\:\mathcal{U}\to\mathcal{V}$ is a diffeomorphism between two sets $\mathcal{U},\mathcal{V}\subset\bb{R}^{d}$, then $\mathrm{WF}(\varphi^{\ast}\sf{T})=\{(x,(\d\varphi_{x})^{t}\xi)\mid (\varphi(x),\xi)\in\mathrm{WF}(\sf{T})\}$ for all $\sf{T}\in\mathcal{D}^{\prime}(\mathcal{V})$, where $\varphi^{\ast}\sf{T}$ is defined by duality. In particular, on a smooth manifold $\sf{M}$, there is a unique way of defining 
\begin{align*}
	\mathrm{WF}(\sf{T})\subset\sf{T}^{\ast}\sf{M}\backslash\{\textbf{0}\}
\end{align*}
for all $\sf{T}\in\mathcal{D}^{\prime}(\sf{M})$, where $\textbf{0}$ denotes the zero-section. More generally, if $\sf{E}$ is any smooth and finite-rank vector bundle over $\sf{M}$, we define $\mathrm{WF}(\sf{T})\subset\sf{T}^{\ast}\sf{M}\backslash\{\textbf{0}\}$ for $\sf{T}\in\mathcal{D}^{\prime}(\sf{M},\sf{E})$ as the union of the wavefront set of its components in a fixed (but arbitrary) local trivialisation.

\begin{remark} (The Primed Wavefront Set)\newline
	Consider the identity operator $\mathrm{id}\:C^{\infty}_{\mathrm{c}}(\bb{R}^{d})\to	C^{\infty}_{\mathrm{c}}(\bb{R}^{d})$ whose Schwartz kernel $k_{\mathrm{id}}\in\mathcal{D}^{\prime}(\bb{R}^{d}\times\bb{R}^{d})$ is given by 
	\begin{align*}
		k_{\mathrm{id}}(\varphi)=\int_{\mathcal{U}}\varphi(x,x)\,\d^{d} x
	\end{align*}
	for all $\varphi\in C^{\infty}_{\mathrm{c}}(\bb{R}^{d}\times\bb{R}^{d})$, or, using the (formal) $\delta$-function notation, $k_{\sf{A}}(x,y)=\delta(x-y)$. Using similar arguments as in Example~\ref{Ex:WFdelta}, one finds
	\begin{align*}
		\mathrm{WF}(\delta)=\{((x,\xi),(x,-\xi))\mid (x,\xi)\in \bb{R}^{d}\times\bb{R}^{d}\backslash\{0\}\}\, .
	\end{align*}
	Now, due to the antisymmetry in this expression, it is customary to define the \emph{primed wavefront set} for all $\sf{T}\in\mathcal{D}^{\prime}(\sf{M}\times\sf{M})$ by
	\begin{align*}
		\mathrm{WF}^{\prime}(\sf{T}):=\{((x,\xi),(y,\eta))\in\sf{T}^{\ast}(\sf{M}\times\sf{M})\backslash\{\textbf{0}\}\mid((x,\xi),(y,-\eta))\in\mathrm{WF}(\sf{T})\}\, .
	\end{align*}
	With this definition, we obtain $\mathrm{WF}^{\prime}(\delta)=\Delta\backslash\{\textbf{0}\}$, where $\Delta:=\{((x,\xi),(x,\xi))\mid (x,\xi)\in\sf{T}^{\ast}\sf{M}\}\}\subset\sf{T}^{\ast}(\sf{M}\times\sf{M})$ denotes the diagonal.
\end{remark}

\section{Microlocal Analysis II: Pseudodifferential Calculus}\label{App:PSIDO}
We ended the discussion of the previous chapter with one of the central notions of \textit{microlocal analysis}, namely the wavefront set of distributions. In this chapter, we continue the discussion of microlocal analysis with the theory of \textit{pseudodifferential operators} (short $\Psi$DOs). The theory of $\Psi$DOs and more generally, of \textit{Fourier integral operators} (FIOs), has been developed in the 1960s by Bokobza-Unterberger \cite{BokobzaUnterberger}, Kohn-Nirenberg \cite{KohnNirenberg}, Hörmander \cite{Hormander,Hormander2,HormanderFourier} and others, generalising earlier work on \textit{singular integral operators} (see e.g.~\cite{Singular,Mikhlin,Calderon1,Calderon2}), as a tool to study (elliptic) partial differential equations. Since then, pseudodifferential calculus has become an indispensable tool for studying linear PDEs. For instance, already the simplest operations one might want to perform on elliptic differential operators (e.g.~taking the inverse, square root, etc.) typically yield operators outside the class of differential operators. $\Psi$DOs provide a generalisation of differential operators, and under some reasonable assumptions on the class of symbols, the operations mentioned above can be performed within this class. Furthermore, the correspondence between operators and their symbols allows many computations and constructions to be carried out purely at the algebraic level of symbols. We refer to \cite{Shubin,Sjorstrand,HormanderIII,Hintz} for a detailed discussion of that subject. 

\subsection{Motivation for Pseudodifferential Calculus}
Let $\varphi\in\mathcal{S}(\mathbb{R}^{d})$ be a Schwartz function. Using the well-known identity $\mathcal{F}(\mathrm{D}^{\alpha}\varphi)=\xi^{\alpha}\mathcal{F}(\varphi)$ with $\mathrm{D}^{\alpha}:=(-i)^{\vert\alpha\vert}\partial^{\alpha}$, we can write $\mathrm{D}^{\alpha}\varphi$ via the inverse Fourier transformation as 
\begin{align*}
    \mathrm{D}^{\alpha}\varphi(x)=\frac{1}{(2\pi)^{d}}\int_{\mathbb{R}^{d}}e^{i\xi\cdot x}\xi^{\alpha}\mathcal{F}(\varphi)(\xi)\,\mathrm{d}^{d}\xi.
\end{align*}
More generally, let $\mathcal{U}\subset\mathbb{R}^{d}$ be open and consider a linear differential operator $A:C^{\infty}(\mathcal{U})\to C^{\infty}(\mathcal{U})$ of order $k\in\mathbb{N}_{0}$ given by
\begin{align*}
    \sf{A}:=\sum_{\alpha\in\mathbb{N}^{d},\,\vert\alpha\vert\leq k}a_{\alpha}(x)\mathrm{D}^{\alpha}
\end{align*}
for smooth coefficients $a_{\alpha}\in C^{\infty}(\mathcal{U})$, as introduced in Eq.~\eqref{eq:LinDiffOp}. Then, we can write
\begin{align}\label{eq:LinDifFou}
    \sf{A}\varphi(x)=\frac{1}{(2\pi)^{d}}\int_{\mathbb{R}^{d}}e^{ix\cdot\xi}a(x,\xi)\mathcal{F}(\varphi)(\xi)\,\mathrm{d}^{d}\xi=\frac{1}{(2\pi)^{d}}\int_{\mathbb{R}^{d}}\int_{\bb{R}^{d}}e^{i(x-y)\xi}a(x,\xi)\varphi(y)\,\mathrm{d}^{d}y\,\mathrm{d}^{d}\xi
\end{align}
for all $\varphi\in C^{\infty}_{\mathrm{c}}(\mathcal{U})$, where $a(\cdot,\xi):=\sum_{\vert\alpha\vert\leq k}a_{\alpha}(\cdot)\xi^{\alpha}$ denotes the \textit{(total) symbol of $\sf{A}$}.

In many cases, it is useful to consider operators of the form~\eqref{eq:LinDifFou} with more general \textit{symbols}~$a$ that are not necessarily polynomials in~$\xi$. This leads naturally to the concept of \textit{pseudodifferential operators} (or~$\Psi$DOs), and, if the term~$(x-y)$ in the exponential is replaced by a more general \emph{phase function}~$\Phi(x, y, \xi)$, to that of \textit{Fourier integral operators} (FIOs). As a simple motivational example, we consider the elliptic equation 
\begin{align*}
    (\mathrm{id}-\Delta)\varphi=f\,,\qquad f\in C_{\mathrm{c}}^{\infty}(\mathbb{R}^{d})
\end{align*}
for $d\geq 2$, where $\Delta:=\sum_{i=1}^{d}\partial_{x^{i}}^{2}$ denotes the Laplace operator on $\mathbb{R}^{d}$. It is well known that a solution of this equation can be obtained via the Fourier transform and reads
\begin{align*}
    \varphi(x)=\frac{1}{(2\pi)^{d}}\int_{\mathbb{R}^{d}}e^{i x\cdot\xi}\frac{\mathcal{F}(f)(\xi)}{1+\vert\xi\vert^{2}}\,\mathrm{d}^{d}\xi.
\end{align*}
In other words, instead of considering $(\mathrm{id}-\Delta)$, one may consider the ``inverse'', which is an operator of the type \eqref{eq:LinDifFou} with \textit{symbol} $a(x,\xi):=1/(1+\vert\xi\vert^{2})$. More generally, let $\sf{A}$ be any linear \emph{elliptic} differential operator of order $k$ with \textit{constant} coefficients $a_{\alpha}\in\mathbb{C}$ of the form
\begin{align*}
    \sf{A}:=\sum_{\alpha\in\mathbb{N}^{d},\,\vert\alpha\vert= k}a_{\alpha}\mathrm{D}^{\alpha}:C^{\infty}(\mathcal{U})\to C^{\infty}(\mathcal{U}).
\end{align*}
By ellipticity, the principal symbol, which in this case coincides with the total symbol $a(\xi)=\sum_{\vert\alpha\vert=k}\alpha_{\alpha}\xi^{\alpha}$, is invertible for $\xi\neq 0$. Then, provided $a$ is nice enough, we may define a linear operator $\sf{P}$ via
\begin{align*}
    \sf{P}f(x):=\frac{1}{(2\pi)^{d}}\int_{\mathbb{R}^{d}}e^{ix\cdot\xi}\frac{1}{a(\xi)}\mathcal{F}(f)(\xi)\,\mathrm{d}^{d}\xi\, .
\end{align*}
It is not too hard to check that $\sf{P}$ is indeed the inverse of $\sf{A}$, i.e.~$\sf{P}\sf{A}\varphi=\sf{A}\sf{P}\varphi=\varphi$ for suitable choices of $\varphi$. Now, if we take an elliptic operator $\sf{A}$ with \textit{non-constant} coefficients, i.e.~$a_{\alpha}\to a_{\alpha}(x)$ and hence $a(\xi)\to a(x,\xi)$, one may try to use a similar strategy and naively set
\begin{align*}
    \sf{P}\varphi(x):=\frac{1}{(2\pi)^{d}}\int_{\mathbb{R}^{d}}e^{ix\cdot\xi}\frac{1}{a(x,\xi)}\mathcal{F}(\varphi)(\xi)\,\mathrm{d}^{d}\xi\, .
\end{align*}
However, in this case, $\sf{P}$ is no longer the inverse of $\sf{P}$, since $a(x,\xi)\mathcal{F}(\varphi)(\xi)$ is in general not the Fourier transform of $\sf{A}\varphi(x)$. However, a straightforward calculation using the Leibniz rule shows 
\begin{align*}
    \sf{A}\sf{P}\varphi(x)=\frac{1}{(2\pi)^{d}}\int_{\mathbb{R}^{d}}\sf{A}\bigg(e^{ix\cdot\xi}\frac{1}{a(x,\xi)}\bigg)\mathcal{F}(\varphi)(\xi)\,\mathrm{d}^{d}\xi=\varphi(x)+\frac{1}{(2\pi)^{d}}\int_{\mathbb{R}^{d}}e^{ix\cdot\xi}r(x,\xi)\mathcal{F}(\varphi)(\xi)\,\mathrm{d}^{d}\xi\, ,
\end{align*}
where we used that $\sf{A}(e^{ix\cdot\xi})=a(x,\xi)$ and where $r(x,\xi)$ consists of terms involving at least first-order derivatives of $1/a(x,\xi)$ with respect to $x$. In other words, we can write
\begin{align*}
    \sf{A}\sf{P}=\mathrm{id}+\sf{R}\,,
\end{align*}
i.e.~$\sf{P}$ is the inverse of $\sf{A}$ up to some (lower, in fact ``negative'', order!) correction operator $\sf{R}$ with symbol $r(x,\xi)$. The operator $\sf{P}$ is called the \textit{parametrix of the elliptic operator $\sf{A}$} and both $\sf{P}$ and $\sf{R}$ provide examples of pseudodifferential operators, which in general are not differential operators, i.e.~operators of the type \eqref{eq:LinDifFou} with non-polynomial ``symbol'' $a(x,\xi)$. 

The goal of pseudodifferential calculus is to make the above considerations more precise. One of the main advantages of this approach is that many calculations can be reduced to purely algebraic considerations on the underlying symbols.
\subsection{Symbol Classes and Oscillatory Integrals}
Let $\mathcal{U}\subset\mathbb{R}^{d}$ be open. Motivated from the previous discussion, our first goal is to make sense out of integrals of the type
\begin{align}\label{OscInt}
   \mathrm{I}_{\Phi,a}(x):=\int_{\mathbb{R}^{d}}\,e^{i\Phi(x,\xi)}a(x,\xi)\,\mathrm{d}^{d}\xi
\end{align}
for a suitable choice of \textit{symbol} $a\in C^{\infty}(\mathcal{U}\times\mathbb{R}^{d})$ and \textit{phase function} $\Phi\in C^{\infty}(\mathcal{U}\times\mathbb{R}^{d})$. In full generality, the integral \eqref{OscInt} is \textit{not} convergent, however, for suitable choices of $a$ and $\Phi$ one can make sense out of them in terms of distributions by means of an appropriate regularisation procedure. Let us start by introducing a suitable class of symbols.

\begin{definition} (Symbol Classes)\newline
    Let $\rho,\sigma\in [0,1]$ and $m\in\mathbb{R}$. We define $\mathcal{S}^{m}_{\rho,\sigma}(\mathcal{U})$ to be the set of functions $a\in C^{\infty}(\mathcal{U}\times\mathbb{R}^{d})$ such that for every compact $\sf{K}\subset\mathcal{U}$ and $\alpha,\beta\in\mathbb{N}^{d}$, there is a constant $C_{\sf{K},\alpha,\beta}>0$, such that
    \begin{align*}
        \vert\partial_{x}^{\alpha}\partial_{\xi}^{\beta} a(x,\xi)\vert\leq C_{\sf{K},\alpha,\beta} (1+\vert\xi\vert)^{m-\rho\vert\beta\vert+\sigma\vert\alpha\vert}\hspace*{1cm}\forall (x,\xi)\in \sf{K}\times\mathbb{R}^{d}.
    \end{align*}
    The set $\mathcal{S}^{m}_{\rho,\sigma}(\mathcal{U})$ is called \textit{symbol class of order $m$ and type $(\rho,\sigma)$}.
\end{definition}

\begin{remark}
	In the above definition, once can replace $(1+\vert\xi\vert)$ also by the \textit{Japanese bracket} $\langle\xi\rangle:=\sqrt{1+\vert\xi\vert^{2}}$, since
	\begin{align*}\frac{1}{\sqrt{2}}(1+\vert c\vert)\leq\sqrt{1+c^{2}}\leq 1+\vert c\vert\end{align*} for any $c\in\mathbb{R}$, as one can easily show using binomial formulas.
\end{remark}

One should think of a symbol $a\in\mathcal{S}^{m}_{\rho,\sigma}(\mathcal{U})$ as a smooth functions on the cotangent bundle $\sf{T}^{\ast}\mathcal{U}\cong \mathcal{U}\times\mathbb{R}^{d}$ with certain decay properties. For many applications, the case $\rho=1,\sigma=0$ is the most important one and we will hence introduce the special notation
\begin{align*}
	\mathcal{S}^{m}(\mathcal{U}):=\mathcal{S}^{m}_{1,0}(\mathcal{U})\, .
\end{align*}
Symbols in this class are also the most intuitive ones: by definition, this class exactly includes those smooth functions $a(x,\xi)$, which asymptotically behave like polynomials in the $\xi$-variable, hence generalising the principal symbols of linear partial differential operators. The symbol class $\mathcal{S}^{m}(\mathcal{U})$ has been introduced in the original work on pseudodifferential operators by Kohn-Nirenberg \cite{KohnNirenberg}, whereas symbol classes with more general $\rho,\sigma$, which also take into account a prescribed asymptotic behaviour in the $x$-variable, have been introduced by Hörmander \cite{Hormander2} in his study of hypoelliptic operators. 

By definition, it holds that $\mathcal{S}^{m_{1}}_{\rho_{1},\sigma_{1}}(\mathcal{U})\subset\mathcal{S}^{m_{2}}_{\rho_{2},\sigma_{2}}(\mathcal{U})$ for $m_{1}\leq m_{2}$ as well as $\sigma_{1}\leq\sigma_{2}$ and $\rho_{1}\geq\rho_{2}$. In particular, the following definition makes sense: 
 
\begin{definition} (The symbol classes $\mathcal{S}^{\infty}$ and $\mathcal{S}^{-\infty}$)\newline
Let $\rho,\sigma\in [0,1]$ be arbitrary. We define
\begin{align*}
    \mathcal{S}^{\infty}(\mathcal{U}):=\bigcup_{m\in\mathbb{R}}\mathcal{S}^{m}_{\rho,\sigma}(\mathcal{U})\hspace*{1cm}\text{and}\hspace*{1cm}\mathcal{S}^{-\infty}(\mathcal{U}):=\bigcap_{m\in\mathbb{R}}\mathcal{S}^{m}_{\rho,\sigma}(\mathcal{U})\,.
\end{align*}
\end{definition}

One can easily check that the definitions of $\mathcal{S}^{\infty}$ and $\mathcal{S}^{-\infty}$ are independent of the choice of $\rho$ and $\sigma$. Symbols in the set $\mathcal{S}^{-\infty}(\mathcal{U})$ are usually called \textit{smoothing symbols}, or \textit{residual symbols}, for reasons that will become clear later.

\begin{examples}\label{ExamplesSymbols}
    \begin{itemize}\item[]
        \item[(i)]Let $a\in C^{\infty}(\mathcal{U}\times\mathbb{R}^{d})$ be positively homogeneous of degree $m$ in the region $\vert\xi\vert\geq 1$, i.e.~
        \begin{align*}
            a(x,\lambda\xi)=\lambda^{m}a(x,\xi),\hspace*{1cm}\vert\xi\vert\geq 1
        \end{align*}
        for all $(x,\xi)\in\mathcal{U}$ and $\lambda\geq 1$. Then, $a\in\mathcal{S}^{m}_{1,0}(\mathcal{U})$. Symbols of this type are sometimes called \textit{homogeneous} and the set of all of them is denoted by $\mathcal{S}^{m}_{\mathrm{h}}(\mathcal{U})\subset\mathcal{S}^{m}(\mathcal{U})$. The principal symbol of a linear differential operator 
        \begin{align*}
            \sf{A}=\sum_{\alpha\in\mathbb{N}^{d},\,\vert\alpha\vert= k}a_{\alpha}\mathrm{D}^{\alpha}:C^{\infty}(\mathcal{U})\to C^{\infty}(\mathcal{U})
        \end{align*}
        with suitably regular coefficient functions $a_{\alpha}$ falls into this class.
        \item[(ii)]The symbol of $(\mathrm{id}-\Delta)^{-1}$ given by $a(x,\xi):=(1+\vert\xi\vert^{2})^{-1}$ is an element of $\mathcal{S}^{-2}_{1,0}(\mathcal{U})$.
        \item[(iii)]The exponential $a(x,\xi):=e^{i x\cdot\xi}$ is an element of $\mathcal{S}^{0}_{0,1}(\mathcal{U})$. 
        \item[(iv)]Every Schwartz function is a smoothing symbol, i.e.~$\mathcal{S}(\mathbb{R}^{d}\times\mathbb{R}^{d})\subset\mathcal{S}^{-\infty}(\mathbb{R}^{d}\times\mathbb{R}^{d})$. More generally, also $C_{\mathrm{bd}}^{\infty}(\mathbb{R}^{d},\mathcal{S}(\mathbb{R}^{d}))\subset\mathcal{S}^{-\infty}(\mathbb{R}^{d}\times\mathbb{R}^{d})$. In particular, any symbol of the form $(x,\xi)\mapsto\chi(\xi)$ for $\chi\in C^{\infty}_{\mathrm{c}}(\mathbb{R}^{d})$ is an element of $\mathcal{S}^{-\infty}(\mathbb{R}^{d}\times\mathbb{R}^{d})$.
    \end{itemize}
\end{examples}

The space of symbols $S^{m}_{\rho,\sigma}(\mathcal{U})$ comes equipped with a natural Fréchet space structure, namely the one induced by the family of semi-norms
\begin{align*}
    p_{\sf{K},\alpha,\beta}(a):=\sup_{(x,\xi)\in \sf{K}\times\mathbb{R}^{d}}\frac{\vert\partial_{x}^{\alpha}\partial_{\xi}^{\beta} a(x,\xi)\vert}{(1+\vert\xi\vert)^{m-\rho\vert\beta\vert+\sigma\vert\alpha\vert}}\, ,
\end{align*}
which assign to every symbol $a\in\mathcal{S}^{m}_{\rho,\sigma}(\mathcal{U})$ the \emph{optimal} constant $C_{\sf{K},\alpha,\beta}$ in the defining inequality. Additionally, we equip the spaces $\mathcal{S}^{\infty}(\mathcal{U})$ and $\mathcal{S}^{-\infty}(\mathcal{U})$ with their corresponding inductive and projective limit topologies, respectively.

\begin{lemma}\label{LemmaP}
    Let $m,n\in\mathbb{R}$ and $\rho,\sigma\in [0,1]$. The point-wise product is well-defined and continuous as a map
    \begin{align*}
        \cdot:\mathcal{S}^{m}_{\rho,\sigma}(\mathcal{U})\times\mathcal{S}^{n}_{\rho,\sigma}(\mathcal{U})\to\mathcal{S}^{m+n}_{\rho,\sigma}(\mathcal{U})\, .
    \end{align*}
    Furthermore, for $\alpha,\beta\in\mathbb{N}$, the partial derivatives are well-defined and continuous as maps 
    \begin{align*}
        \partial_{x}^{\alpha}\partial_{\xi}^{\beta}:\mathcal{S}^{m}_{\rho,\sigma}(\mathcal{U})\to\mathcal{S}^{m-\vert\beta\vert\rho+\vert\alpha\vert\sigma}_{\rho,\sigma}(\mathcal{U})\, .
    \end{align*}
\end{lemma}

\begin{proof}
    The first statement follows from a straightforward computation using the Leibniz rule:
    \begin{align*}
        \vert\partial_{x}^{\alpha}\partial_{\xi}^{\beta}(ab)(x,\xi)\vert &=\bigg\vert\sum_{\substack{\alpha_{1}+\alpha_{2}=\alpha\\\beta_{1}+\beta_{2}=\beta}}\binom{\alpha}{\alpha_{1}}\binom{\beta}{\beta_{1}}(\partial_{x}^{\alpha_{1}}\partial_{\xi}^{\beta_{1}}a(x,\xi))(\partial_{x}^{\alpha_{2}}\partial_{\xi}^{\beta_{2}}b(x,\xi))\bigg\vert\leq\\&\leq \underbrace{\bigg(\sum_{\substack{\alpha_{1}+\alpha_{2}=\alpha\\\beta_{1}+\beta_{2}=\beta}}\binom{\alpha}{\alpha_{1}}\binom{\beta}{\beta_{1}}C^{a}_{\sf{K},\alpha_{1},\beta_{1}}C^{b}_{\sf{K},\alpha_{2},\beta_{2}}\bigg)}_{=:C_{\sf{K},\alpha,\beta}}(1+\vert\xi\vert)^{m+n-\rho\vert\beta\vert+\sigma\vert\alpha\vert}
    \end{align*}
    for all compact $\sf{K}\subset\mathcal{U}$ and multi-indices $\alpha,\beta\in\mathbb{N}^{d}$, where $C_{\sf{K},\alpha\beta}^{a,b}$ are the corresponding constants for $a$ and $b$, respectively. The second claim follows from a similar calculation. 
\end{proof}

Having defined a suitable notion of symbols, we shall next introduce a suitable class of phase functions. In the following, we will use the notation $\mathbb{R}^{d}_{\ast}:=\mathbb{R}^{d}\backslash\{0\}$.

\begin{definition}\label{Def:Phase} (Phase Functions)\newline
    A \textit{phase function} is a map $\Phi\in C^{\infty}(\mathcal{U}\times\mathbb{R}^{d}_{\ast})$ with the following properties:
    \begin{align*}
        \text{(i)}&\hspace*{1cm}\mathrm{ran}(\Phi)\subset\mathbb{R}\\
        \text{(ii)}&\hspace*{0.9cm}\Phi(x,\lambda\xi)=\lambda\Phi(x,\xi)\hspace*{1cm}\forall (x,\xi)\in \mathcal{U}\times\mathbb{R}^{d}_{\ast},\,\lambda>0\\
        \text{(iii)}&\hspace*{0.8cm}\mathrm{d}\Phi\neq 0\quad\text{, i.e.~$\Phi$ is non-critical}
    \end{align*}
    The set of all phase functions will be denoted by $\mathcal{P}(\mathcal{U})$.
\end{definition}

More generally, one may also consider phase functions which take values in $\bb{C}$ if one assumes in addition that $\mathrm{Im}(\Phi)\geq 0$, see~\cite{Melin,Sjorstrand}. The term $\mathrm{d}\Phi$ denotes the total derivative of $\Phi$, i.e.~the $1$-form $\d\Phi\in\Omega^{1}(\mathcal{U}\times\mathbb{R}^{d}_{\ast})$ defined by
\begin{align*}
	\mathrm{d}\Phi=(\partial_{x^{i}}\Phi)\,\d x^{i}+(\partial_{\xi_{j}}\Phi)\,\d\xi_{j}\, .
\end{align*}
Hence, condition (iii) means that at every point $(x,\xi)\in \mathcal{U}\times\mathbb{R}_{\ast}^{d}$ at least one element of $\{\partial_{x^{i}}\Phi,\partial_{\xi_{j}}\Phi\}$ is non-zero. To sum up, a phase function is a real-valued smooth function on $\mathcal{U}\times\mathbb{R}^{d}_{\ast}$, which is positively homogeneous of degree one in $\xi$ and which does not have critical points. 

\begin{example}\label{Phase}
    An important example is provided by $\Phi(x,\xi):=x\cdot\xi$, where ``$\cdot$'' denotes the standard inner product on $\mathbb{R}^{d}$, as usual.
\end{example}

Now, let us fix a symbol $a\in\mathcal{S}^{m}_{\rho,\sigma}(\mathcal{U})$ and a phase function $\Phi\in\mathcal{P}(\mathcal{U})$. As already mentioned at the beginning of this section, the goal in the following is to study integrals of the type
\begin{align*}
   \mathrm{I}_{\Phi,a}(x):=\int_{\mathbb{R}^{d}}\,e^{i\Phi(x,\xi)}a(x,\xi)\,\mathrm{d}^{d}\xi\, .
\end{align*}
In full generality, the integral $\mathrm{I}_{\Phi,a}$ will not converge in the usual sense. As an example, consider the phase function $\Phi(x,\xi):=x\cdot\xi$ from Example \ref{Phase} together with the constant function $a(x,\xi)=1$, which is trivially contained in $\mathcal{S}^{-\infty}(\mathcal{U})$. Then, $\mathrm{I}_{\Phi,a}$ is given by 
\begin{align*}
   \mathrm{I}_{\Phi,a}(x):=\int_{\mathbb{R}^{d}}\,e^{ix\cdot\xi}\,\mathrm{d}^{d}\xi
\end{align*}
for all $x\in\mathcal{U}$, which is the Fourier transform of the constant $1$ function. This integral clearly does not convergence in the usual sense, however, can be understood in the sense of (tempered) distributions, in fact, it is exactly the $\delta$-distribution. In other words, for general $a,\Phi$, there is no hope for $\mathrm{I}_{\Phi,a}$ to be convergent in general. However, they can always be defined in a suitable way in the sense of distributions, which we lead us to the concept of \emph{oscillatory integrals}.

To start with, let us observe that $\mathrm{I}_{\Phi,a}$ makes actually sense for any $a\in\mathcal{S}^{m}_{\rho,\sigma}(\mathcal{U})$ whenever the order $m\in\mathbb{R}$ of the symbol is negative and sufficiently small compared to the dimension $d$. Furthermore, the regularity of the obtained map $\mathrm{I}_{\Phi,a}\:\mathcal{U}\to\bb{C}$ is controlled by $\vert m\vert-d$.  

\begin{lemma}\label{PropositionAbsoluteConvergence}
    Let $a\in\mathcal{S}^{m}_{\rho,\sigma}(\mathcal{U})$ and $\Phi\in\mathcal{P}(\mathcal{U})$. If $m<-d$, then the integral
    \begin{align*}
       \mathrm{I}_{\Phi,a}(x):=\int_{\mathbb{R}^{d}}\,e^{i\Phi(x,\xi)}a(x,\xi)\,\mathrm{d}^{d}\xi
    \end{align*}
    converges absolutely for all $x\in \mathcal{U}$. Moreover, $\mathrm{I}_{\Phi,a}\in C^{k}(\mathcal{U})$ for $k\in\mathbb{N}$ such that $m+k<-d$.
\end{lemma}

\begin{proof}
    By assumption, there is for every compact $\sf{K}\subset\mathcal{U}$ a constant $C_{\sf{K}}>0$ such that
    \begin{align*}
        \vert a(x,\xi)\vert\leq C_{\sf{K}}(1+\vert\xi\vert)^{m}
    \end{align*}
    for all $(x,\xi)\in \sf{K}\times\mathbb{R}^{d}$. Let us fix $x\in \mathcal{U}$ and $\sf{K}\subset\mathcal{U}$ with $x\in\sf{K}$. Then, we obtain
    \begin{align*}
       \int_{\mathbb{R}^{d}}\,\vert a(x,\xi)\vert\,\mathrm{d}^{d}\xi\leq C_{\sf{K}}\int_{\mathbb{R}^{d}}(1+\vert\xi\vert)^{m}\,\mathrm{d}^{d}\xi=\frac{2\pi^{\frac{d}{2}}}{\Gamma\big(\frac{d}{2}\big)}\int_{\bb{R}}(1+r)^{m}r^{d-1}\,\d r\,,
    \end{align*}
    which is convergent for $m<-d$. The fact that $\mathrm{I}_{\Phi,a}\in C^{k}(\mathcal{U})$ for $k\in\mathbb{N}$ such that $m+k<-d$ follows from the dominant convergence theorem and similar estimates.
\end{proof}

\begin{theorem}\label{OscIntThm} \emph{(Oscillatory Integrals)}\newline
    Let $\rho\in (0,1]$ and $\sigma\in [0,1)$. Then there exists exactly one continuous linear map
    \begin{align*}
       \mathrm{I}_{\Phi,\bullet}:\mathcal{S}^{\infty}_{\rho,\sigma}(\mathcal{U})\to\mathcal{D}^{\prime}(\mathcal{U})
    \end{align*}
    which for $a\in \mathcal{S}^{m}_{\rho,\sigma}(\mathcal{U})$ with $m<-d$ coincides with the (absolutely convergent) integral 
    \begin{align*}
       \mathrm{I}_{\Phi,a}(x):=\int_{\mathbb{R}^{d}}\,e^{i\Phi(x,\xi)}a(x,\xi)\,\mathrm{d}^{d}\xi\,.
    \end{align*}
\end{theorem}

\begin{proof}
    Uniqueness is clear, since $\mathcal{S}^{-\infty}(\mathcal{U})$ is dense in $\mathcal{S}^{m}(\mathcal{U})$ when taking the topology of any $\mathcal{S}^{m^{\prime}}(\mathcal{U})$ with $m^{\prime}>m$ (see for instance \cite[Proposition 1.7]{Sjorstrand}). For existence, we follow the proof in \cite{HormanderFourier} (see also the monographs \cite{Shubin,Sjorstrand}). The first step consists of constructing a linear differential operator $\sf{L}$ of the type
    \begin{align*}
        \sf{L}=\sum_{i=1}^{d}a_{i}(x,\xi)\partial_{\xi_{i}}+\sum_{j=1}^{d}b_{j}(x,\xi)\partial_{x^{j}}+c(x,\xi)
    \end{align*}
    for symbols $a_{i}\in\mathcal{S}^{0}(\mathcal{U})$ and $b_{j},c\in\mathcal{S}^{-1}(\mathcal{U})$ such that $\sf{L}^{t}(e^{i\Phi})=e^{i\Phi}$, where $\sf{L}^{t}=-\sum_{i}\partial_{\xi_{i}}\circ a_{i}-\sum_{j}\partial_{x_{j}}\circ b_{j}+c$ denotes the \textit{transpose of $\sf{L}$}, which for all $f,g\in C^{\infty}_{c}(\mathcal{U})$ is defined by
    \begin{align*}
        \int_{\mathbb{R}^{d}}\int_{\mathcal{U}}\sf{L}(f)g\,\mathrm{d}^{d}x\,\mathrm{d}^{d}\xi=\int_{\mathbb{R}^{d}}\int_{\mathcal{U}}f\sf{L}(g)\,\mathrm{d}^{d}x\,\mathrm{d}^{d}\xi\, .
    \end{align*}
  	The operator $\sf{L}$ plays the role of a \textit{regularisation}. In order to construct $\sf{L}$, we first observe that 
    \begin{align*}
        \frac{1}{i}\bigg(\sum_{i=1}^{d}(\partial_{\xi_{i}}\Phi)\vert\xi\vert^{2}\partial_{\xi_{i}}+\sum_{i=1}^{d}(\partial_{x_{i}}\Phi)\vert\xi\vert^{2}\partial_{x^{i}}\bigg)e^{i\Phi}=\underbrace{\bigg(\sum_{i=1}^{d}\vert\partial_{\xi_{i}}\Phi\vert^{2}\vert\xi\vert^{2}+\sum_{i=1}^{d}\vert\partial_{x^{i}}\Phi\vert^{2}\vert\xi\vert^{2}\bigg)}_{:=\Psi(x,\xi)}e^{i\Phi}\, .
    \end{align*}
    The function $\Psi\in C^{\infty}(\mathcal{U}\times \mathbb{R}^{d}_{\ast})$ is positively homogeneous of degree $2$, i.e. $\Psi(x,\lambda\xi)=\lambda^{2}\Psi(x,\xi)$ for all $\lambda>0$ and by construction, it holds that
    \begin{align*}
        \frac{1}{i\Psi(x,\xi)}\bigg(\sum_{i=1}^{d}(\partial_{\xi_{i}}\Phi)\vert\xi\vert^{2}\partial_{\xi_{i}}+\sum_{i=1}^{d}(\partial_{x^{i}}\Phi)\vert\xi\vert^{2}\partial_{x_{i}}\bigg)e^{i\Phi}=e^{i\Phi}\, .
    \end{align*}
    for $(x,\xi)\in \mathcal{U}\times\bb{R}^{d}_{\ast}$. It remains to show that we can get rid of the singularity at $\xi\to 0$ of $1/\Psi$. For this, let us choose a test function $\chi\in C_{\mathrm{c}}^{\infty}(\mathbb{R}^{d})$ with $\chi(\xi)=1$ for $\vert\chi\vert< 1/4$ and $\chi(\xi)=0$ for $\vert\xi\vert>1/2$. Then, we set
    \begin{align*}
        \sf{L}^{t}:=\frac{1-\chi(\xi)}{i\Psi(x,\xi)}\bigg(\sum_{i=1}^{d}(\partial_{\xi_{i}}\Phi)\vert\xi\vert^{2}\partial_{\xi_{i}}+\sum_{i=1}^{d}(\partial_{x^{i}}\Phi)\vert\xi\vert^{2}\partial_{x^{i}}\bigg)+\chi(\xi)\, ,
    \end{align*}
	which is the required operator satisfying $\sf{L}^{t}(e^{i\Phi})=e^{i\Phi}$. Let us now us the regulator $\sf{L}$ to prove the theorem: let $a\in\mathcal{S}^{m}_{\rho,\sigma}(\mathcal{U})$ be such that $m<-d$ so that $\mathrm{I}_{\Phi,a}$ is well-defined by Lemma~\ref{PropositionAbsoluteConvergence}. Then, for all $\varphi\in C^{\infty}_{\mathrm{c}}(\mathcal{U})$, we obtain
    \begin{equation}\label{OscIntProof}
        \begin{aligned}
            \langle\mathrm{I}_{\Phi,a},\varphi\rangle_{\mathcal{U}}=&\int_{\mathbb{R}^{d}}\int_{\mathcal{U}}e^{i\Phi(x,\xi)}a(x,\xi)\varphi(x)\,\mathrm{d}^{d}x\,\mathrm{d}^{d}\xi=\int_{\mathbb{R}^{d}}\int_{\mathcal{U}}((\sf{L}^{t})^{k}e^{i\Phi(x,\xi)})a(x,\xi)\varphi(x)\,\mathrm{d}^{d}x\,\mathrm{d}^{d}\xi=\\=&\int_{\mathbb{R}^{d}}\int_{\mathcal{U}}e^{i\Phi(x,\xi)}\sf{L}^{k}(a(x,\xi)\varphi(x))\,\mathrm{d}^{d}x\,\mathrm{d}^{d}\xi\, ,
        \end{aligned}
    \end{equation}
    where we used integration by parts in the last step. Note that $\sf{L}^{k}(a\varphi)\in\mathcal{S}^{m-ks}_{\rho,\sigma}(\mathcal{U})$ by Lemma \ref{LemmaP}, where $s:=\mathrm{min}\{\rho,1-\sigma\}$. If $s>0$, which is the case when $\rho>0$ and $\sigma<1$, the formula above allows us to define $\langle\mathrm{I}_{\Phi,a},u\rangle$ for arbitrary $m$ by choosing $k\in\mathbb{N}$ such that $m-ks<-d$, since in this case, the integral becomes absolutely convergent by Lemma~\ref{PropositionAbsoluteConvergence}. In other words, we define $\mathrm{I}_{\Phi,a}$ for arbitrary $a$ by the right-hand side of equation \eqref{OscIntProof}. By a density argument, we see that the choice of $k$ with the property $m-ks<-d$ does not matter.
\end{proof}

\begin{remark}
	For a fixed symbol $a\in\mathcal{S}^{m}_{\rho,\sigma}(\mathcal{U})$ and phase function $\Phi\in\mathcal{P}(\mathcal{U})$, one usually denotes the distribution $\mathrm{I}_{\Phi,a}\in\mathcal{D}^{\prime}(\mathcal{U})$ using the formal integral notation
	\begin{align*}
		\bb{C}\ni \langle\mathrm{I}_{\Phi,a},\varphi\rangle_{\mathcal{U}}=:\int_{\mathbb{R}^{d}}\int_{\mathcal{U}}e^{i\Phi(x,\xi)}a(x,\xi)\varphi(x)\,\d^{d}x\,\d^{d}\xi
	\end{align*}
	for all $\varphi\in C^{\infty}_{\mathrm{c}}(\mathcal{U})$. Another justification of this formula is the following fact: let $\chi\in C^{\infty}_{c}(\mathbb{R}^{d})$ be such that $\chi(\xi)=1$ in a neighbourhood of $0$. Then
    \begin{align*}
        \langle\mathrm{I}_{\Phi,a},\varphi\rangle_{\mathcal{U}}=\lim_{\varepsilon\to 0}\int_{\mathbb{R}^{d}}\int_{\mathcal{U}}\,e^{i\Phi(x,\xi)}\chi(\xi\varepsilon)a(x,\xi)\varphi(x)\,\mathrm{d}^{d}x\,\mathrm{d}^{d}\xi\, ,
    \end{align*}
    where the limit is taken in $\mathcal{D}^{\prime}(\mathcal{U})$ (equipped with the weak or strong dual topology; the choice does not matter, since $\mathcal{D}^{\prime}(\mathcal{U})$ with the strong topology is a Montal space). This can be shown as follows: using the regularisation operator $\sf{L}$ as defined in the previous proof, we have
    \begin{align}\label{sdffd}
        \int_{\mathbb{R}^{d}}\int_{\mathcal{U}}\,e^{i\Phi(x,\xi)}\chi(\xi\varepsilon)a(x,\xi)\varphi(x)\,\mathrm{d}^{d}x\,\mathrm{d}^{d}\xi=\int_{\mathbb{R}^{d}}\int_{\mathcal{U}}\,e^{i\Phi(x,\xi)}\sf{L}^{k}(\chi(\xi\varepsilon)a(x,\xi)\varphi(x))\,\mathrm{d}^{d}x\,\mathrm{d}^{d}\xi\, ,
    \end{align}
    where we used integration by parts $k\in\mathbb{N}$ times. Note that $a_{\varepsilon}(x,\xi):=\chi(\varepsilon\xi)a(x,\xi)$ is an element of $\mathcal{S}^{-\infty}_{\rho,\sigma}(\mathcal{U})$ and $a_{\varepsilon}\to a$ in $\mathcal{S}^{n}_{\rho,\sigma}(\mathcal{U})$ as $\varepsilon\to 0$ for every $n>m$. Now, using the dominant convergence theorem, we may interchange the integration and the limit, which shows that the right-hand side of \eqref{sdffd} is the same as \eqref{OscIntProof}. In particular, note that this implies that the result is independent of the chosen $\chi$. In other words, the oscillatory integral is the distribution
    \begin{align*}
    		\mathrm{I}_{\Phi,a}(x)=\lim_{\varepsilon\to 0}\int_{\mathbb{R}^{d}}\,e^{i\Phi(x,\xi)}\chi(\xi\varepsilon)a(x,\xi)\,\mathrm{d}^{d}\xi\, ,
    \end{align*}
    where the limit is taken in the space of distributions $\mathcal{D}^{\prime}(\mathcal{U})$.
\end{remark}

Before moving our discussion to pseudodifferential and Fourier integral operators, let us briefly discuss under which assumption $\mathrm{I}_{\Phi,a}\in\mathcal{D}^{\prime}(\mathcal{U})$ is actually a \emph{regular distribution}, i.e.~for which choice of phase functions $\Phi\in\mathcal{P}(\mathcal{U})$ and symbols $a\in\mathcal{S}^{m}_{\rho,\sigma}(\mathcal{U})$, we actually have that $\mathrm{I}_{\Phi,a}\in C^{\infty}(\mathcal{U})$.\footnote{Let us stress that this does not necessarily mean that the integral \textit{converges}, just that the map $x\mapsto\mathrm{I}_{\Phi,a}(x)$ is smooth, where $\mathrm{I}_{\Phi,a}(x)$ is still defined as an oscillatory integral.} An easy example is the case of a smoothing symbol $a\in \mathcal{S}^{-\infty}_{\rho,\sigma}(\mathcal{U})$, as a consequence of Lemma~\ref{PropositionAbsoluteConvergence}. More generally, we define the set
\begin{align*}
    C_{\Phi}:=\{(x,\xi)\in \mathcal{U}\times\mathbb{R}^{d}_{\ast}\mid \mathrm{grad}_{\xi}\Phi(x,\xi)=0\}\, ,
\end{align*}
where $\mathrm{grad}_{\xi}\Phi(x,\xi)$ denotes the gradient of $\Phi$ with respect to $\xi$, i.e.~$\Phi^{\prime}_{\xi}=(\partial_{\xi_{1}}\Phi,\dots,\partial_{\xi_{d}}\Phi)^{\sf{T}}$. Note that $C_{\Phi}$ is a conical neighbourhood of $\mathcal{U}\times\mathbb{R}^{d}_{\ast}$.

\begin{proposition} \emph{(Oscillatory Integrals and Regularity)}\label{PropRegu}\newline
    Let $\Phi\in\mathcal{P}(\mathcal{U})$ and $a\in\mathcal{S}^{m}_{\rho,\sigma}(\mathcal{U})$ for $\rho\in (0,1]$ and $\sigma\in [0,1)$. Then
    \begin{align*}
        \mathrm{sing}\,\mathrm{supp}(\mathrm{I}_{\Phi,a})\subset\pi C_{\Phi}\, ,
    \end{align*}
    where $\pi:\mathcal{U}\times\mathbb{R}^{d}_{\ast}\to \mathcal{U}$ denotes the canonical projection. In particular, if the symbol $a$ vanishes on a conical neighbourhood of $C_{\Phi}$, then $\mathrm{I}_{\Phi,a}\in C^{\infty}(\mathcal{U})$.
\end{proposition}

\begin{proof}
    Let $x\in \mathcal{U}\backslash (\pi C_{\Phi})$ be fixed. Then, $\xi\mapsto\Phi(x,\xi)$ is a phase function in $\mathcal{P}(\mathbb{R}^{d})$, by definition of $C_{\Phi}$. In particular, this means that the integral
    \begin{align*}
        \int_{\mathbb{R}^{d}}e^{i\Phi(x,\xi)}a(x,\xi)\,\mathrm{d}^{d}\xi=\lim_{\varepsilon\to 0}\int_{\mathbb{R}^{d}}e^{i\Phi(x,\xi)}a(x,\xi)\chi(\varepsilon\xi)\,\mathrm{d}^{d}\xi
    \end{align*}
    exists as an oscillatory integral \textit{for fixed $x$}. Taking derivatives with respect to $x$ and using the dominant convergence theorem to interchange the limit with the integral, one obtains integrals of the same type and repeating this procedure inductively, we conclude that $\mathrm{I}_{\Phi,a}\in C^{\infty}(\mathcal{U}\backslash (\pi C_{\Phi}))$.
\end{proof}

\begin{example}
	If we choose the phase function $\Phi(x,\xi):=x\cdot\xi$ from Example~\ref{Phase} together with an arbitrary symbol $a\in\mathcal{S}^{m}_{\rho,\sigma}(\mathcal{U})$, we obtain $\mathrm{sing}\,\mathrm{supp}(\mathrm{I}_{\Phi,a})=\{0\}$, i.e.~the distribution $\mathrm{I}_{\Phi,a}$ is singular at most at the origin.
\end{example}

\subsection{Pseudodifferential and Fourier Integral Operators}
In the following, let $\mathcal{U}\subset\mathbb{R}^{d}$ be open. After having discussed suitable classes of symbols and phase function, we are now in the position to define Fourier integral operators, which are linear operators whose Schwartz kernel is given by an oscillatory integral.

\begin{definition} (Fourier Integral and Pseudodifferential Operators)\newline
	Let $a\in\mathcal{S}^{m}_{\rho,\sigma}(\mathcal{U}\times\mathcal{U})$ with $(\rho,\sigma)\in (0,1]\times [0,1)$ and $\Phi\in\mathcal{P}(\mathcal{U}\times \mathcal{U})$. A linear and continuous operator $\sf{A}\:C^{\infty}_{\mathrm{c}}(\mathcal{U})\to\mathcal{D}^{\prime}(\mathcal{U})$ whose Schwartz kernel is given by $k_{\sf{A}}:=\mathrm{I}_{\Phi,a}\in\mathcal{D}^{\prime}(\mathcal{U}\times\mathcal{U})$ is called a \emph{Fourier integral operator with symbol $a$ and phase function $\Phi$}. Formally,
    \begin{align*}
        (\sf{A}\psi)(x)=\int_{\mathbb{R}^{d}}\int_{\mathcal{U}}e^{i\Phi(x,y,\xi)}a(x,y,\xi)\psi(y)\,\mathrm{d}^d y\,\mathrm{d}^d\,\xi
    \end{align*}
    for all $\psi\in C^{\infty}_{\mathrm{c}}(\mathcal{U})$. In the special case $\Phi(x,y,\xi)=(x-y)\cdot\xi$, we call $\sf{A}$ \textit{pseudodifferential operator of order $m$ and type $(\rho,\sigma)$}. The space of all operators is denoted by $\Psi^{m}_{\rho,\sigma}(\mathcal{U})$.
\end{definition}

As for symbols, we introduce the notation $\Psi^{\pm\infty}(\mathcal{U})$ for the set of pseudodifferential operators with symbols in $\mathcal{S}^{\pm\infty}(\mathcal{U}\times\mathcal{U})$. By definition, we have 
\begin{align*}
    \Psi^{\infty}(\mathcal{U})=\bigcup_{m\in\mathbb{R}}\Psi^{m}_{\rho,\sigma}(\mathcal{U})\hspace*{1cm}\text{and}\hspace*{1cm}\Psi^{-\infty}(\mathcal{U})=\bigcap_{m\in\mathbb{R}}\Psi^{m}_{\rho,\sigma}(\mathcal{U})
\end{align*}
for some choice of $(\rho,\sigma)\in (0,1]\times [0,1)$. Furthermore, as for symbol classes, we will use the special notation $\Psi^{m}(\mathcal{U}):=\Psi_{1,0}^{m}(\mathcal{U})$ for pseudodifferential operators of type $(1,0)$.

\begin{remark}
    Note that neither a Fourier integral operator nor a pseudodifferential operator does in general \textit{uniquely} define its phase function $\Phi$ and symbol $a$. In other words, different pairs $(\Phi,a)$ can give rise to the same operator. Furthermore, even for a fixed phase function $\Phi$, the corresponding operator does in general not determine its symbol uniquely.
\end{remark}

\begin{example}
	Any linear partial differential operator  $\sf{A}=\sum_{\alpha\in\mathbb{N}^{d},\,\vert\alpha\vert\leq k}a_{\alpha}\mathrm{D}^{\alpha}$ with smooth coefficient functions $a_{\alpha}\in C^{\infty}(\mathcal{U})$ is an element of $\Psi^{m}(\mathcal{U})$. Indeed,
	\begin{align*}
    		\sf{A}\psi(x)=\frac{1}{(2\pi)^{d}}\int_{\mathbb{R}^{d}}e^{ix\cdot\xi}a(x,\xi)\mathcal{F}(\psi(\xi)\,\mathrm{d}^{d}\xi=\frac{1}{(2\pi)^{d}}\int_{\mathbb{R}^{d}}\int_{\mathcal{U}}e^{i(x-y)\xi}a(x,\xi)\psi(y)\,\mathrm{d}^{d}y\,\mathrm{d}^{d}\xi
\end{align*}
for all $\psi\in C^{\infty}_{\mathrm{c}}(\mathcal{U})$, where $a(x,\xi)=\sum_{\vert\alpha\vert\leq k}a_{\alpha}(x)\xi^{\alpha}$ denotes the \emph{total symbol}. The first integral is absolutely convergent, while the double integral makes sense as an iterated one.
\end{example}

Now, in full generality general, a Fourier integral operator $\sf{A}$ maps test functions $C^{\infty}_{\mathrm{c}}(\mathcal{U})$ into distributions $\mathcal{D}^{\prime}(\mathcal{U})$. However, it turns out that under rather simple assumptions on the phase function $\Phi$, we obtain operators whose range is contained in the subset $C^{\infty}(\mathcal{U})\subset\mathcal{D}^{\prime}(\mathcal{U})$, i.e.~operators whose Schwartz kernel are semiregular. More precisely, let $\Phi\in\mathcal{P}(\mathcal{U}\times\mathcal{U})$ be an arbitrary phase function. Now, in general, the two maps 
\begin{align*}\Phi(x,\cdot,\cdot),\Phi(\cdot,y,\cdot)\in C^{\infty}(\mathcal{U}\times\mathbb{R}^{d}_{\ast})\end{align*} 
for \emph{fixed} $x,y\in\mathcal{U}$ are \textit{not} again phase function, since they do not necessarily satisfy the non-criticality assumption, i.e.~property (iii) in Definition~\ref{Def:Phase}. More precisely, by definition, $\Phi$ has the property that at any point $(x,y,\xi)$, at least one element of $\{\partial_{x	^{i}}\Phi,\partial_{y^{i}}\Phi,\partial_{\xi_{i}}\Phi\}$ is non-zero, however, it might be that at some point all the derivatives with respect to $x$ (or $y$) vanish. If the maps $\Phi(x,\cdot,\cdot)$ and $\Phi(\cdot,y,\cdot)$, however, happen to be non-critical individually, then $\mathrm{I}_{\Phi,a}$ is actually semiregular and in fact, even biregular:

\begin{proposition}\label{OperatorPhaseFunction}  Let $\Phi\in\mathcal{P}(\mathcal{U}\times\mathcal{U})$ be a phase function such that $\Phi(x,\cdot,\cdot),\Phi(\cdot,y,\cdot)\in\mathcal{P}(\mathcal{U})$ for every $x,y\in\mathcal{U}$. Then, the kernel $\mathrm{I}_{\Phi,a}\in\mathcal{D}^{\prime}(\mathcal{U}\times \mathcal{U})$ is biregular for any $a\in\mathcal{S}^{m}_{\rho,\sigma}(\mathcal{U}\times \mathcal{U})$ with $(\rho,\sigma)\in (0,1]\times [0,1)$. In particular, the corresponding Fourier integral operator $\sf{A}\colon C^{\infty}_{\mathrm{c}}(\mathcal{U})\to\mathcal{D}^{\prime}(\mathcal{U})$ has the following properties:
    \begin{itemize}
        \item[\emph{(i)}]$\sf{A}\psi\in C^{\infty}(\mathcal{U})$ for all $\psi\in C_{\mathrm{c}}^{\infty}(\mathcal{U})$ and $\sf{A}$ is continuous as a map $\sf{A}\:C^{\infty}_{\mathrm{c}}(\mathcal{U})\to C^{\infty}(\mathcal{U})$.
        \item[\emph{(ii)}]$\sf{A}$ has a unique continuous extension $\sf{A}\:\mathcal{E}^{\prime}(\mathcal{U})\to\mathcal{D}^{\prime}(\mathcal{U})$.
    \end{itemize}
\end{proposition}

\begin{proof}
	To show that $\mathrm{I}_{\Phi,a}(x,y)$ is individually smooth in $x$ and $y$, we employ similar arguments as in the proof of Proposition~\ref{PropRegu}. Consider the oscillatory integral 
	\begin{align*}
		\mathrm{I}_{\Phi,a}(x,y)=\int_{\mathbb{R}^{d}}e^{i\Phi(x,y,\xi)}a(x,y,\xi)\,\d^{d}\xi\, .
	\end{align*}
	By assumption, we know that both $\mathrm{I}_{\Phi,a}(x,\cdot)$ and $\mathrm{I}_{\Phi,a}(\cdot,y)$ are well-defined oscillatory integrals for fixed $x,y\in\mathcal{U}$, since clearly $a(x,\cdot,\cdot),a(\cdot,y,\cdot)\in\mathcal{S}^{m}_{\rho,\sigma}(\mathcal{U})$ for fixed $x,y\in\mathcal{U}$. Introducing a regularisation and taking derivatives with respect to $x$ (or $y$), we obtain again oscillatory integrals of the same types, showing that $\mathrm{I}_{\Phi,a}(x,\cdot)$ and $\mathrm{I}_{\Phi,a}(\cdot,y)$ are smooth maps on $\mathcal{U}$ for fixed $x,y$. Claim (i) and (ii) then follow from Proposition~\ref{Prop:Kernels}(i) and (ii).
\end{proof}

An important special case is provided by choosing $\Phi(x,y,\xi):=(x-y)\cdot\xi$, i.e.~pseudodifferential operators. We summarise this discussion in the following corollary.

\begin{corollary} \emph{(Properties of Pseudodifferential Operators)}\newline
Let $\sf{A}\in\Psi^{m}_{\rho,\sigma}(\mathcal{U})$ be a pseudodifferential operator.
\begin{itemize}
    \item[\emph{(i)}]$\sf{A}$ is continuous as an operator $\sf{A}\:C^{\infty}_{\mathrm{c}}(\mathcal{U})\to C^{\infty}(\mathcal{U})$ and $\sf{A}\:\mathcal{E}^{\prime}(\mathcal{U})\to\mathcal{D}^{\prime}(\mathcal{U})$.
    \item[\emph{(ii)}]$\sf{A}$ is \emph{pseudolocal}, i.e.~$\mathrm{sing}\,\mathrm{supp}(\sf{A}\psi)\subset\mathrm{sing}\,\mathrm{supp}(\psi)$ for all $\psi\in\mathcal{E}^{\prime}(\mathcal{U})$.
\end{itemize}  
\end{corollary}

\begin{proof}
	Claim (i) is a special case of Proposition~\ref{OperatorPhaseFunction}. For (ii), we first note that $\mathrm{sing}\,\mathrm{supp}(k_{\sf{A}})\subset\Delta$, where $\Delta=\{(u,u)\mid u\in\mathcal{U}\}$ denotes the diagonal, as a consequence of Proposition~\ref{PropRegu}. Now, let $\psi\in\mathcal{E}^{\prime}(\mathcal{U})$ be such that $\psi$ is regular in an open neighbourhood $\mathcal{V}\subset\mathcal{U}$ of some $x_{0}\in\mathcal{U}$, i.e.~$\psi\vert_{\mathcal{V}}\in C^{\infty}(\mathcal{V})$. Then, by (i), $\sf{A}\psi$ is regular in (a possibly smaller) neighbourhood $\mathcal{V}^{\prime}$ of $x_{0}$. Choose a test function $\varphi\in C^{\infty}_{\mathrm{c}}(\mathcal{V})$ such that $\varphi=1$ on $\mathcal{V}^{\prime}$. Then, $\psi=\varphi\psi+(1-\varphi)\psi$ where $\psi_{1}:=\varphi\psi\in C^{\infty}_{\mathrm{c}}(\mathcal{U})$ and $\psi_{2}:=(1-\varphi)\psi\in\mathcal{E}^{\prime}(\mathcal{U})$ vanishes on $\mathcal{V}^{\prime}$. Since $k_{\sf{A}}$ is biregular, we have $\sf{A}\psi_{1}\in C^{\infty}_{\mathrm{c}}(\mathcal{U})$. Moreover, $\sf{A}\psi_{2}\in \mathcal{D}^{\prime}(\mathcal{U})$ is regular in $\mathcal{V}^{\prime}$. Hence, $\sf{A}\psi$ is regular in $\mathcal{V}^{\prime}$.
\end{proof}

\begin{remark}\label{Rem:WeylQuant} (Kohn-Nirenberg and Weyl Quantisation)\newline
	If $\sf{A}$ is a pseudodifferential operator defined by a symbol $a$ depending only on the $x$-variable, it is usually referred to as the \emph{Kohn–Nirenberg quantisation} of $a$. This gives rise to a map 
	\begin{align*}
	\mathrm{Op}\:\mathcal{S}^{m}_{\rho,\sigma}(\mathcal{U})\mapsto\Psi^{m}_{\rho,\sigma}(\mathcal{U})\,\qquad\mathrm{Op}(a)\psi(x)=\frac{1}{(2\pi)^{d}}\int_{\mathbb{R}^{d}}\int_{\mathcal{U}}e^{i(x-y)\xi}a(x,\xi)\psi(y)\,\mathrm{d}^{d}y\,\mathrm{d}^{d}\xi\, .
\end{align*}	 
Besides the Kohn–Nirenberg quantisation, other types of \textit{quantisation maps} have also been introduced and studied in the literature within the context of pseudodifferential calculus. An important example is provided by the \emph{(Wigner-)Weyl quantisation}
\begin{align*}
	\mathrm{Op}_{\mathrm{W}}(a)\psi(x)=\frac{1}{(2\pi)^{d}}\int_{\mathbb{R}^{d}}\int_{\mathcal{U}}e^{i(x-y)\xi}a\bigg(\frac{x+y}{2},\xi\bigg)\psi(y)\,\mathrm{d}^{d}y\,\mathrm{d}^{d}\xi\, .
\end{align*}	 
This quantisation map was first introduced by Weyl in his work on group theory and quantum mechanics \cite{Weyl,WeylBook}. At the same time, Wigner \cite{Wigner} defined what is now known as the \textit{Wigner function} in his analysis of quantum corrections to statistical mechanics. Although not originally framed as such, the Wigner function effectively serves as an inverse to Weyl quantization. The connection between Weyl's quantisation and Wigner's transformation was only fully recognised in the late 1940s through the work of Groenewold \cite{Groenewold} and Moyal \cite{Moyal}. From the point of view of pseudodifferential calculus, the Weyl quantisation for a general class of symbols has been extensively studied by Hörmander \cite{HormanderWeyl} (see also \cite[Sec.~18.5]{HormanderIII}) and Dencker \cite{DenckerWeyl}, following earlier works of Grossmann-Loupias-Stein \cite{Grossmann}, Berezin-Shubin \cite{BerezinShubin} and Voros \cite{Voros}.

More generally, one can also consider an operator $\mathrm{Op}_{\tau}$ in which $(x-y)/2$ is replaced by $\frac{1}{2}(x\tau+(1-\tau)y)$ for $\tau\in [0,1]$. The Weyl quantisation has some advantages which makes it preferable in some applications: first of all, it is invariant under symplectic transformations. Furthermore, $\mathrm{Op}_{\mathrm{W}}(a)$ is formally self-adjoint if and only if $a$ is real-valued. More generally, it holds that $\mathrm{Op}_{\tau}(a)^{\ast}=\mathrm{Op}_{1-\tau}(\overline{a})$. We refer to \cite[Chap.~IV, Sec.~23]{Shubin} for more details.
\end{remark}

Next, let introduce the following terminology: a linear continuous operator $\sf{A}\:C^{\infty}_{\mathrm{c}}(\mathcal{U})\to\mathcal{D}^{\prime}(\mathcal{U})$ is called \textit{smoothing operator}, if it extends uniquely to a linear and continuous operator
\begin{align*}
    \sf{A}\colon \mathcal{E}^{\prime}(\mathcal{U})\to C^{\infty}(\mathcal{U})\, .
\end{align*}
Using Proposition~\ref{Prop:Kernels}, it is easy to see that this is the case if and only if the kernel $k_{\sf{A}}$ is regular, i.e.~$k_{\sf{A}}\in C^{\infty}(\mathcal{U}\times\mathcal{U})$. As it turns out, the set of smoothing operators is exactly $\Psi^{-\infty}(\mathcal{U})$:

\begin{proposition} A linear continuous operator $\sf{A}\colon C^{\infty}_{\mathrm{c}}(\mathcal{U})\to\mathcal{D}^{\prime}(\mathcal{U})$ is a smoothing operator if and only if $\sf{A}\in\Psi^{-\infty}(\mathcal{U})$.
\end{proposition}

\begin{proof}
    If $\sf{A}\in\Psi^{-\infty}(\mathcal{U})$, then clearly $k_{A}\in C^{\infty}(\mathcal{U}\times\mathcal{U})$, cf.~Lemma~\ref{PropositionAbsoluteConvergence}. Conversely, if $\sf{A}\:C^{\infty}_{\mathrm{c}}(\mathcal{U})\to\mathcal{D}^{\prime}(\mathcal{U})$ is a smoothing operator, then $k_{\sf{A}}\in C^{\infty}(\mathcal{U}\times \mathcal{U})$ and 
    \begin{align*}
        \sf{A}\psi(x)=\int_{\mathcal{U}}k_{\sf{A}}(x,y)\psi(y)\,\mathrm{d}^{d}y=\int_{\mathbb{R}^{d}}\int_{\mathcal{U}}k_{\sf{A}}(x,y)\psi(y)\chi(\xi)\,\mathrm{d}^{d}y\,\mathrm{d}^{d}\xi\,,
    \end{align*}
    where $\chi\in C^{\infty}_{\mathrm{c}}(\mathcal{U})$ is a test function normalised such that $\int_{\mathbb{R}^{d}}\chi(\xi)\,\mathrm{d}^{d}\xi=(2\pi)^{d}$. The map $a(x,y,\xi):=k_{\sf{A}}(x,y)e^{-i(x-y)\cdot\xi}\chi(\xi)$ is clearly contained in $\mathcal{S}^{-\infty}(\mathcal{U})$, which concludes the proof.
\end{proof}

Now, given two $\Psi$DOs $\sf{A}$ and $\sf{B}$, we aim to make sense of their compositions $\sf{A}\circ\sf{B}$. However, this is not possible in full generality. The issue arises because, for $\psi\in C^{\infty}_{\mathrm{c}}(\mathcal{U})$, the function $\sf{B}\psi$ will in general be not compactly supported and hence, $\sf{A}(\sf{B}\psi)$ may not be well-defined. To address this problem, we must restrict our attention to properly-supported operators.

\begin{proposition} \emph{(Properties of Properly-Supported Pseudodifferential Operators)}\newline
Let $\sf{A}\colon C^{\infty}_{\mathrm{c}}(\mathcal{U})\to C^{\infty}(\mathcal{U})$ be a properly-supported pseudodifferential operator.
	\begin{itemize}
	\item[\emph{(i)}]$A$ is linear and continuous as an operator of the form
	\begin{align*}
		&\sf{A}\colon C_{\mathrm{c}}^{\infty}(\mathcal{U})\to C_{\mathrm{c}}^{\infty}(\mathcal{U}) \quad && \sf{A}\colon C^{\infty}(\mathcal{U})\to C^{\infty}(\mathcal{U})\\	
		&\sf{A}\colon \mathcal{E}^{\prime}(\mathcal{U})\to \mathcal{E}^{\prime}(\mathcal{U}) \quad && \sf{A}\colon \mathcal{D}^{\prime}(\mathcal{U})\to \mathcal{D}^{\prime}(\mathcal{U})
	\end{align*}
	\item[\emph{(ii)}]The symbol $a\in\mathcal{S}^{m}_{\rho,\sigma}(\mathcal{U}\times\mathcal{U})$ of $\sf{A}$ can be chosen such that $a$ is properly supported in the sense that $\mathrm{supp}(a(\cdot,\cdot,\xi))\subset \mathcal{U}\times \mathcal{U}$ is proper for every fixed $\xi\in\mathbb{R}^{d}$.
	\end{itemize}
\end{proposition}

\begin{proof}
	Claim (i) is a direct consequence of Proposition~\ref{Prop:Kernels}(iii) and Proposition~\ref{OperatorPhaseFunction}. For (ii), let $a\in\mathcal{S}^{m}_{\rho,\sigma}(\mathcal{U}\times \mathcal{U})$ be a symbol of $\sf{A}$. We choose a test function $\chi\in C^{\infty}(\mathcal{U}\times \mathcal{U})$ such that $\chi=1$ in a neighbourhood of $\mathrm{supp}(k_{\sf{A}})$ and such that $\mathrm{supp}(\chi)$ is proper.\footnote{The existence of $\chi$ is proven in \cite[p.29]{Sjorstrand}. The idea is to choose a locally finite partition of unity $\{\varphi_{i}\}_{i\in\mathrm{I}}\subset C^{\infty}_{\mathrm{c}}(\mathcal{U})$. Then, $\varphi_{i}\boxtimes\varphi_{j}$ defines a partition of unity of $\mathcal{U}\times\mathcal{U}$ and $\chi:=\sum_{i,j}\varphi_{i}\boxtimes\varphi_{j}$ has the required properties.} Then, substituting $a^{\prime}(x,y,\xi):=\chi(x,y)a(x,y\chi)$ in $k_{\sf{A}}$ does not change $k_{\sf{A}}$ and $a^{\prime}$ is proper.
\end{proof}

In other words, the kernel of a properly supported pseudodifferential operator is supported ``sufficiently'' close to the diagonal $\Delta:=\{(u,u)\in \mathcal{U}\times\mathcal{U}\}$, which is the region containing the singular support of its Schwartz kernel.

\begin{example}
    Any linear partial differential operator  $\sf{A}=\sum_{\alpha\in\mathbb{N}^{d},\,\vert\alpha\vert\leq k}a_{\alpha}\mathrm{D}^{\alpha}$ is a properly supported $\Psi$DO, since $\mathrm{supp}(k_{\sf{A}})=\Delta$ in this case.
\end{example}

If $\sf{A},\sf{B}\in\Psi^{m}_{\rho,\sigma}(\mathcal{U})$ are two properly supported pseudodifferential operators, then the compositions $\sf{A}\circ \sf{B}$ and $\sf{B}\circ \sf{A}$ are well-defined, linear and continuous as operators from $C^{\infty}_{\mathrm{c}}(\mathcal{U})$ to $C^{\infty}_{\mathrm{c}}(\mathcal{U})$. As we will see below, they are actually again pseudodifferential operators.

Being properly supported is not a very restrictive property from the microlocal point of view, since every pseudodifferential operator is properly supported up to a smoothing operator:

\begin{proposition}\label{Prop:DecomPSI} Let $\sf{A}\in\Psi^{m}_{\rho,\sigma}(\mathcal{U})$. Then, there exist a properly-supported $\sf{A}_{\mathrm{pr}}\in\Psi^{m}_{\rho,\sigma}(\mathcal{U})$ such that $\sf{A}=\sf{A}_{\mathrm{pr}}+\sf{S}$ for some $\sf{S}\in\Psi^{-\infty}(\mathcal{U})$.
\end{proposition}

\begin{proof}
    Let $\chi\in C^{\infty}(\mathcal{U}\times\mathcal{U})$ be such that $\chi(x,y)=1$ in a neighbourhood of the diagonal $\Delta$ and such that $\mathrm{supp}(\chi)$ is proper. Then, we decompose the symbol $a\in\mathcal{S}^{m}_{\rho,\sigma}(\mathcal{U}\times\mathcal{U})$ of $\sf{A}$ as $a=a^{\prime}+s$, where $a^{\prime}(x,\xi):=\chi(x,\xi)a(x,\xi)$ and $s(x,\xi):=(1-\chi(x,\xi))a(x,\xi)$. The pseudodifferential operator $\sf{A}_{\mathrm{pr}}$ with symbol $a^{\prime}$ is properly supported and the pseudo-differential operator $\sf{S}$ with symbol $s$ is contained in $\Psi^{-\infty}(\mathcal{U})$, since $s=(1-\chi)a$ vanishes outside the diagonal. By construction $\sf{A}=\sf{A}_{\mathrm{pr}}+\sf{S}$.
\end{proof}

As discussed above, the symbol is not uniquely determined by the pseudodifferential operator in question. In the special case of differential operators, however, there always exists a preferred choice, namely the \emph{total symbol}. As it turns out, there exists a suitable generalisation of the total symbol also for general pseudodifferential operators. To start with, let us establish the following terminology: Let $a\in C^{\infty}(\mathcal{U}\times\mathbb{R}^{d})$ be a smooth function and $a_{j}\in\mathcal{S}^{m_{j}}_{\rho,\sigma}(\mathcal{U})$ be a sequence of symbols labelled by $j\in\mathbb{N}$ with $m_{j}\to -\infty$ as $j\to\infty$. We write 
\begin{align*}
    a(x,\xi)\sim\sum_{j=1}^{\infty}a_{j}(x,\xi)
\end{align*}
and call $\sum_{j=1}^{\infty}a_{j}(x,\xi)$ \textit{asymptotic expansion of $a$} if for any $r\in\mathbb{N}$ with $r\geq 2$ it holds that 
\begin{align*}
    a(x,\xi)-\sum_{j=1}^{r-1}a_{j}(x,\xi)\in\mathcal{S}^{\max_{j\geq r}m_{j}}_{\rho,\sigma}(\mathcal{U}).
\end{align*}
By definition, $a\in\mathcal{S}^{\max_{j}m_{j}}_{\rho,\sigma}(\mathcal{U})$. Furthermore, one can show that for every $\sum_{j=1}^{\infty}a_{j}(x,\xi)$ there exists a corresponding $a\in\mathcal{S}^{\infty}(\mathcal{U})$, unique modulo $\mathcal{S}^{-\infty}(\mathcal{U})$, such that $\sum_{j=1}^{\infty}a_{j}(x,\xi)$ is the asymptotic expansion of $a$. We refer to \cite[Proposition 3.5]{Shubin}, \cite[Proposition 1.8]{Sjorstrand} for details.

\begin{theorem}\label{theorem:Complete} \emph{(Total Symbol \cite[Theorem 3.1]{Shubin},  \cite[Theorem 3.4]{Sjorstrand})}\newline
Let $\sf{A}\in\Psi^{m}_{\rho,\sigma}(\mathcal{U})$ with $\rho>\sigma$ be properly supported with symbol $a\in\mathcal{S}^{m}_{\rho,\sigma}(\mathcal{U}\times \mathcal{U})$. Then, $\tau_{\sf{A}}(x,\xi):=e^{-ix\cdot\xi}\sf{A}(e^{i \bullet\cdot\xi})$ is an element of $\mathcal{S}^{m}_{\rho,\sigma}(\mathcal{U})$ and $\sf{A}=\mathrm{Op}(\tau_{\sf{A}})$, i.e.
\begin{align*}
    \sf{A}\psi(x)=\frac{1}{(2\pi)^{d}}\int_{\mathbb{R}^{d}}\int_{\mathcal{U}}e^{i(x-y)\cdot\xi}\tau_{A}(x,\xi)\psi(y)\,\mathrm{d}^{d}y\,\mathrm{d}^{d}\xi
\end{align*}
    for all $\psi\in C_{c}^{\infty}(\mathcal{U})$. The symbol $\tau_{\sf{A}}$ is called the \textit{(total) symbol of $\sf{A}$} and satisfies
    \begin{align*}
        \tau_{\sf{A}}(x,\xi)\sim\sum_{\alpha\in\mathbb{N}^{d}}\frac{1}{\alpha !}(\partial_{\xi}^{\alpha}\mathrm{D}_{y}^{\alpha}a(x,y,\xi))\vert_{y=x}
    \end{align*}
\end{theorem}

By Proposition \ref{Prop:DecomPSI}, every pseudodifferential operator $\sf{A}\in\Psi^{m}_{\rho,\sigma}(\mathcal{U})$ can be written as $\sf{A}=\sf{A}_{\mathrm{pr}}+\sf{S}$ with $\sf{A}_{\mathrm{pr}}\in\Psi^{m}_{\rho,\sigma}(\mathcal{U})$ being properly supported and $\sf{S}\in\Psi^{-\infty}(\mathcal{U})$. Hence, it makes sense to define the \textit{(total) symbol} of a general pseudodifferential operator $\sf{A}\in\Psi^{m}_{\rho,\sigma}(\mathcal{U})$ by
\begin{align*}
    \tau_{\sf{A}}:=[\tau_{\sf{A}_{\mathrm{pr}}}]\in\mathcal{S}^{m}_{\rho,\sigma}(\mathcal{U})/\mathcal{S}^{-\infty}\,.
\end{align*}
Now, if we consider the \textit{Kohn-Nirenberg quantisation map} $\mathrm{Op}\:\mathcal{S}^{m}_{\rho,\sigma}(\mathcal{U})\to\Psi^{m}_{\rho,\sigma}(\mathcal{U})$, as defined in Remark~\ref{Rem:WeylQuant}, the previous discussion implies that the induced map $\mathrm{Op}\:\mathcal{S}^{m}_{\rho,\sigma}(\mathcal{U})/\mathcal{S}^{-\infty}\to\Psi^{m}_{\rho,\sigma}(\mathcal{U})/\Psi^{-\infty}$ is bijective with inverse $[\sf{A}]\mapsto\tau_{\sf{A}}$. 

Next, let $\sf{A}\in\Psi^{m}_{\rho,\sigma}(\mathcal{U})$ be a pseudodifferential operator. The unique operator $A^{\ast}\:C^{\infty}_{\mathrm{c}}(\mathcal{U})\to C^{\infty}(\mathcal{U})$ satisfying
$\langle \sf{A}\varphi,\psi\rangle_{\sf{L}^{2}}=\langle \varphi,\sf{A}^{\ast}\psi\rangle_{\sf{L}^{2}}$ for all $\varphi,\psi\in C^{\infty}_{\mathrm{c}}(\mathcal{U})$ is called the \textit{formal adjoint} of $\sf{A}$. It is easy to see that $\sf{A}^{\ast}$ always exists and is a linear and continuous operator with Schwartz kernel $k_{\sf{A}^{\ast}}(x,y)=\overline{k_{\sf{A}}(y,x)}$. The adjoint is, in fact, also a $\Psi$DO:

\begin{proposition}\label{properties} \emph{(Algebra of Pseudodifferential Operators)}\newline
Let $\sf{A}\in\Psi^{m}_{\rho,\sigma}(\mathcal{U})$ and $\sf{B}\in\Psi^{n}_{\rho,\sigma}(\mathcal{U})$ with $\rho>\sigma$ be pseudodifferential operators and $\sf{B}$ be properly supported. Then,
\begin{itemize}
\item[\emph{(i)}]$\sf{A}\circ\sf{B}\in\Psi^{m+n}_{\rho,\sigma}(\mathcal{U})$ and if also $A$ is properly supported, then so is $\sf{A}\circ\sf{B}$. Moreover,
\begin{align*}
	\tau_{\sf{A}\circ \sf{B}}(x,\xi)\sim\sum_{\alpha\in\mathbb{N}^{d}}\frac{1}{\alpha!}\partial_{\xi}^{\alpha}\tau_{\sf{A}}(x,\xi)\mathrm{D}_{x}^{\alpha}\tau_{\sf{B}}(x,\xi)\, .
\end{align*}
\item[\emph{(ii)}]$\sf{A}^{\ast}\in\Psi^{m}_{\rho,\sigma}(\mathcal{U})$ and $A^{\ast}$ is properly supported if and only $A$ is. Moreover,
\begin{align*}
	\tau_{\sf{A}^{\ast}}(x,\xi)\sim\sum_{\alpha\in\mathbb{N}^{d}}\partial_{\xi}^{\alpha}\mathrm{D}_{x}^{\alpha}\overline{\tau_{\sf{A}}(x,\xi)}\, .
\end{align*}
\end{itemize}
\end{proposition}

\begin{proof}[Proof (sketch).]
	For (i), we assume w.l.o.g.~that $\sf{A}$ is properly supported. Then, 
	\begin{align*}
        (\sf{A}\circ \sf{B})\psi(x)=\int_{\mathbb{R}^{d}}\int_{\mathcal{U}}e^{i\Phi(x,y,\xi)}\tau_{A}(x,\xi)\tau_{B}(y,\xi)\psi(y)\,\mathrm{d}^d y\mathrm{d}^d\,\xi\, .
    \end{align*}
    Hence, by Lemma~\ref{LemmaP}, it is clear that $\sf{A}\circ \sf{B}\in\Psi^{m+n}_{\rho,\sigma}(\mathcal{U})$. Moreover, using Theorem~\ref{theorem:Complete}, we obtain
    \begin{align*}
    		\tau_{\sf{A}\circ\sf{B}}(x,\xi)&\sim\sum_{\alpha\in\mathbb{N}^{d}}\partial_{\xi}^{\alpha}\mathrm{D}_{x}^{\alpha}(\tau_{\sf{A}}(x,\xi)\tau_{\sf{B}}(y,\xi))\vert_{y=x}\sim\dots\sim \sum_{\alpha\in\mathbb{N}^{d}}\frac{1}{\alpha!}\partial_{\xi}^{\alpha}\tau_{\sf{A}}(x,\xi)\mathrm{D}_{x}^{\alpha}\tau_{\sf{B}}(x,\xi)\, ,
\end{align*}     
see \cite[Theorem~3.4.]{Shubin} for details. For (ii), we assume again w.l.o.g.~that $\sf{A}$ is properly supported. Now, if $a(x,y,\xi)$ is a symbol for $\sf{A}$, then $\overline{a(y,x,\xi)}$ is a symbol of $\sf{A}^{\ast}$. Hence, choosing $a(x,y,\xi)=\tau_{A}(x,\xi)$, the claim follows from Theorem~\ref{theorem:Complete}.
\end{proof}

\begin{remark}
	Consider the map $\#\colon \mathcal{S}^{m}_{\rho,\sigma}/\mathcal{S}^{-\infty}\times \mathcal{S}^{n}_{\rho,\sigma}/\mathcal{S}^{-\infty}\to \mathcal{S}^{m+n}_{\rho,\sigma}/\mathcal{S}^{-\infty}$ defined by
	\begin{align*}
		(a\# b)(x,\xi)\sim\sum_{\alpha\in\mathbb{N}^{d}}\frac{1}{\alpha!}\partial_{\xi}^{\alpha}a(x,\xi)\mathrm{D}_{x}^{\alpha}b(x,\xi)\, .
	\end{align*}
	Then, by definition, it holds that $\mathrm{Op}(a\# b)=\mathrm{Op}(a)\circ\mathrm{Op}(b)$ modulo $\Psi^{-\infty}(\mathcal{U})$. The bilinear pairing $\#$ can easily be seen to be associative and is usually called the \textit{Moyal product}.
\end{remark}

So far we have considered a fairly general class of pseudodifferential operators. In applications, however, it is usually more useful to consider a more narrow class of symbols, which allow for a well-defined calculus. One of the most important classes are those of \textit{classical symbols}, which mimic the behaviour of the symbols of differential operators.

\begin{definition} (Classical Symbols)\newline
    A function $a\in C^{\infty}(\mathcal{U}\times\mathbb{R}^{d})$ is called a \textit{classical symbol of order $m\in\mathbb{R}$}, if 
    \begin{align*}
        a(x,\xi)\sim\sum_{j=0}^{\infty}\psi(\xi)a_{m-j}(x,\xi)\,,
    \end{align*}
    where $\psi\in C^{\infty}(\mathbb{R}^{d})$ is such that $\psi(\xi)=0$ for $\vert\xi\vert\leq 1/2$ and $\psi(\xi)=1$ for $\vert\xi\vert\geq 1$ and where $\{a_{m-j}\}_{j=0}^{\infty}\subset C^{\infty}(\mathcal{U}\times\mathbb{R}^{d})$ are positive homogeneous of degree $m-j$ in $\xi$. We denote the set of all such symbols by $\mathcal{S}^{m}_{\mathrm{cl}}(\mathcal{U})$. 
\end{definition}

By the previous discussion, it is clear that $\mathcal{S}_{\mathrm{cl}}^{m}(\mathcal{U})\subset\mathcal{S}^{m}(\mathcal{U})$. Classical symbols are exactly those that have an asymptotic expansion in terms of homogeneous symbols, i.e. $a\sim\sum_{j=0}^{\infty}a_{m-j}$ for $a_{m-j}\in\mathcal{S}^{m-j}_{\mathrm{h}}(\mathcal{U})$, where the space $\mathcal{S}_{\mathrm{h}}^{m-j}(\mathcal{U})$ has been  introduced in Example \ref{ExamplesSymbols}(a). For this reason, classical symbols are sometimes also referred to as \textit{polyhomogenous}.

\begin{example}
	Every linear partial differential operator  with smooth coefficient functions is a properly supported classical $\Psi$DO.
\end{example}

\begin{definition} (Classical Pseudodifferential Operators)\newline
	A pseudodifferential operator $\sf{A}\in\Psi^{m}(\mathcal{U})$ for which there exists a symbol in $\mathcal{S}^{m}_{\mathrm{cl}}(\mathcal{U}\times \mathcal{U})$ is called a \textit{classical pseudodifferential operator of order $m$}. The subset of all such operators will be denoted by $\Psi_{\mathrm{cl}}^m(\mathcal{U})\subset\Psi^{m}(\mathcal{U})$.
\end{definition}

As a direct consequence of the asymptotic expansions derived in Proposition~\ref{properties} as well as Theorem~\ref{theorem:Complete}, we obtain the following properties:

\begin{proposition} \emph{(The Algebra of Classical $\Psi$DOs)}
\begin{itemize}
\item[\emph{(i)}]If $\sf{A}\in\Psi^{m}_{\mathrm{cl}}(\mathcal{U})$ is properly supported, then $\tau_{\sf{A}}\in\mathcal{S}_{\mathrm{cl}}^{m}(\mathcal{U})$.
	\item[\emph{(ii)}]If $\sf{A}\in\Psi^{m}_{\mathrm{cl}}(\mathcal{U})$ and $\sf{B}\in\Psi^{n}_{\mathrm{cl}}(\mathcal{U})$ are properly supported, then $\sf{A}\sf{B},\sf{B}\sf{A}\in\Psi_{\mathrm{cl}}^{m+n}(\mathcal{U})$.
	\item[\emph{(iii)}]If $\sf{A}\in\Psi^{m}_{\mathrm{cl}}(\mathcal{U})$ is properly supported, then $\sf{A}^{\ast}\in\Psi_{\mathrm{cl}}^{m}(\mathcal{U})$.
\end{itemize}
\end{proposition}

Last but not least, recall that the \textit{principal symbol} of a linear partial differential operator  $\sf{A}=\sum_{\alpha\in\mathbb{N}^{d},\,\vert\alpha\vert\leq k}a_{\alpha}\mathrm{D}^{\alpha}$ with coefficients $a_{\alpha}\in C^{\infty}(\mathcal{U})$ is the homogeneous function 
\begin{align*}
	\sigma_{\sf{A}}(x,\xi)=\sum_{\vert\alpha\vert=k}a_{\alpha}\xi^{\alpha}\, .
\end{align*}
The concept of a principal symbol generalises to classical pseudodifferential operators as follows.

\begin{definition} (Principal Symbol of a Classical $\Psi$DO)\newline
    Let $\sf{A}\in\Psi^{m}_{\mathrm{cl}}(\mathcal{U})$ be  properly supported with $\tau_{\sf{A}}\sim\sum_{j}a_{m-j}$. Then, the \textit{principal symbol of $\sf{A}$} is defined to be the homogeneous symbol $\sigma_{\sf{A}}:=a_{m}\in\mathcal{S}^{m}_{\mathrm{h}}(\mathcal{U})$.
\end{definition}

\begin{remark}\label{Rem:Ellip}
Let $\sf{A}\in\Psi^{m}_{\mathrm{cl}}(\mathcal{U})$. Then, the \textit{characteristic set of $\sf{A}$} is the set defined by
\begin{align*}
        \mathrm{char}(\sf{A})=\{(x,\xi)\in \mathcal{U}\times \mathbb{R}^{d}_{\ast}\mid \sigma_{\sf{A}}(x,\xi)=0\}\subset \sf{T}^{\ast}\mathbb{R}^{d}\backslash\{\textbf{0}\}\, .
    \end{align*}
    Using this definition, we call $\sf{A}\in\Psi^{m}_{\mathrm{cl}}(\mathcal{U})$ \textit{elliptic} if $\mathrm{char}(\sf{A})=\emptyset$. In the motivation section of this notes, we have stated that every elliptic differential operator can be inverted up to an error. The precise statement is as follows: Let $\sf{A}\in\Psi^{m}_{\mathrm{cl}}(\mathcal{U})$ be a properly supported elliptic pseudodifferential operator. Then, there exists a properly supported operator $\sf{B}\in\Psi^{-m}_{\mathrm{cl}}(\mathcal{U})$ such that
	\begin{align*}
		\sf{B}\circ\sf{A}-\mathrm{id},\, \sf{B}\circ \sf{A}-\mathrm{id}\in\Psi^{-\infty}(\mathcal{U})\,.
	\end{align*}
	Furthermore, $\sf{B}$ is unique up to $\Psi^{-\infty}(\mathcal{U})$. The operator $\sf{B}$ is called \textit{parametrix of $\sf{A}$}. The proof of this theorem is actually relatively easy once one has understand the machinery of pseudodifferential calculus. Details can be found in \cite[Chapter 4]{Sjorstrand} and \cite[Section I.5]{Shubin}.
	\end{remark}
	
\begin{remark}\label{Rem:WFSPsu} (Wavefront Set and Pseudodifferential Operators)\newline
	Let $\mathcal{U}\subset\bb{R}^{d}$ be open and $u\in\mathcal{D}^{\prime}(\mathcal{U})$. Then, it is not too hard to see that $x\notin\mathrm{sing}\,\mathrm{supp}(u)$ if and only if we can find a test function $\varphi\in C^{\infty}_{\mathrm{c}}(\mathcal{U})$ with $\varphi(x)\neq 0$ such that $\varphi u\in C^{\infty}(\mathcal{U})$. In other words, the singular support of $u$ can be characterised as
	\begin{align*}
		\mathrm{sing}\,\mathrm{supp}(u)=\bigcap_{\substack{\varphi\in C_{\mathrm{c}}^{\infty}(\mathcal{U})\\ \varphi u\in C^{\infty}(\mathcal{U})}}\{x\in\mathcal{U}\mid\varphi(x)=0\}\, .
	\end{align*}
	Now, as discussed in Section~\ref{Sec:Wavefront}, the wavefront set $\mathrm{WF}(u)$ of $u$ is a refinement of $\mathrm{sing}\,\mathrm{supp}(u)$ that also contains informations about the singular directions in Fourier space. Hence, there exists a natural characterisation of the wavefront set via pseudodifferential operators, analogues to the characterisation of the singular support above: $(x_{0},\xi_{0})\notin\mathrm{WF}(u)$ if and only if there exists an operator $\sf{A}\in\Psi_{\mathrm{cl}}^{0}(\mathcal{U})$ with $(x_{0},\xi_{0})\notin\mathrm{char}(\sf{A})$ such that $\sf{A}u\in C^{\infty}(\mathcal{U})$. In other words, 
	\begin{align}\label{eq:WFPSI}
        \mathrm{WF}(u)=\bigcap_{\substack{\sf{A}\in\Psi_{\mathrm{cl}}^{0}(\mathcal{U})\text{ properly supported}\\ \sf{A}u\in C^{\infty}(\mathcal{U})}}\mathrm{char}(\sf{A})\, .
    \end{align}
    This characterisation has in fact been used in the original work of Hörmander, see~\cite{HormanderFourier}. 
    
    Following \cite[Sec.~A.1.1]{Shubin}, let us give a short sketch of the proof of Eq.~\eqref{eq:WFPSI}: first, assume that $(x_{0},\xi_{0})\notin\mathrm{WF}(u)$. Then, we choose $f\in C_{\mathrm{c}}^{\infty}(\mathcal{U})$ such that $f(x)=1$ in a neighbourhood of $x_{0}$ and a closed conic neighbourhood $\mathcal{V}\subset\mathbb{R}^{d}$ of $\xi_{0}$, such that for every $N\in\mathbb{N}$ there is a constant $C_{N}>0$ such that 
    \begin{align*}
    		\vert \mathcal{F}(fu)(\xi)\vert\leq C_{N}(1+\vert\xi\vert)^{-N}\, .
    	\end{align*}
    	Now, let $\chi\in C^{\infty}(\mathbb{R}^{d})$ be supported in the neighbourhood $\mathcal{V}$ such that $\psi(\lambda \xi)=\psi(\xi)$ for $\lambda\geq 1$ and $\vert\xi\vert\geq 1$. Then
    \begin{align*}
    		\vert\chi(\xi)\mathcal{F}(fu)(\chi)\vert\leq C_{N}(1+\vert\xi\vert)^{-N}
    	\end{align*}
    	and hence in particular $\chi(\mathrm{D})(fu)\in C^{\infty}(\mathcal{U})$, where $\chi(\mathrm{D})$ is the operator with symbol $\chi$. Let us choose a localisation $\psi\in C^{\infty}_{\mathrm{c}}(\mathcal{U})$ such that $\psi(x)=1$ on a neighbourhood of $x_{0}$. Then, the pseudodifferential operator $\sf{A}:=\psi(x)\chi(\mathrm{D})f(x)$ clearly has the desired properties. For the reverse, assume that there there exists a properly-supported $\sf{A}\in\Psi_{\mathrm{cl}}^{0}(\mathcal{U})$ with $(x_{0},\xi_{0})\notin\mathrm{char}(\sf{A})$ such that $\sf{A}u\in C^{\infty}(\mathcal{U})$. Then, the idea is to use the fact that $\sf{A}$ is \textit{elliptic} at $(x_{0},\xi_{0})$, which means that we can find a \emph{parametrix} (see Remark~\ref{Rem:Ellip} above) $\sf{B}\in\Psi^{0}_{\mathrm{cl}}(\mathcal{U})$ such that $\sf{A}\sf{B}-\mathrm{id},\sf{B}\sf{A}-\mathrm{id}\in\Psi^{-\infty}(\mathcal{U})$. Now, we may write $u=\sf{B}\sf{A}u+(\mathrm{id}-\sf{B}\sf{A})u$. Since clearly also $\sf{B}\sf{A}u\in C^{\infty}(\mathcal{U})$, we conclude that $u$ is smooth in a neighbourhood of $x_{0}$. In particular, $(x_{0},\xi_{0})\notin\mathrm{WF}(u)$.
        
        As a direct consequence of Eq.~\eqref{eq:WFPSI}, it is straightforward to show that
        \begin{align*}
        		\mathrm{WF}(\sf{A}u)\subset\mathrm{WF}(\sf{A})\cap\mathrm{WF}(u)\,,\qquad \mathrm{WF}(u)\subset\mathrm{WF}(\sf{A}u)\cup\mathrm{char}(\sf{A})
        \end{align*}
        for all $u\in\mathcal{E}^{\prime}(\mathcal{U})$ and $\sf{A}\in\Psi_{\mathrm{cl}}^{m}(\mathcal{U})$, where $\mathrm{WF}(\sf{A})$ denotes the \emph{operator wavefront set}, i.e.~the set $\mathrm{WF}(\sf{A})\subset\sf{T}^{\ast}\mathcal{U}\backslash\{\textbf{0}\}$ defined by $(x,\xi)\notin\mathrm{WF}(\sf{A})$ if and only if the total symbol $\tau_{\sf{A}}$ is of order $-\infty$ near $x$ in a conic neighbourhood of $\xi$.
\end{remark}

Last but not least, let us discuss how pseudodifferential operators can be defined on manifolds. To start with, let $\mathcal{U}_{1},\mathcal{U}_{2}\subset\mathbb{R}^{d}$ be two open sets and $\kappa\colon \mathcal{U}_{1}\to \mathcal{U}_{2}$ be a diffeomorpism. Then, for any pseudodifferential operator $\sf{A}_{1}\in\Psi^{m}_{\rho,\sigma}(\mathcal{U}_{1})$ we can define an operator $\sf{A}_{2}\colon C_{\mathrm{c}}^{\infty}(\mathcal{U}_{2})\to C^{\infty}(\mathcal{U}_{2})$ in such a way that the following diagram commutes:

\begin{equation*}\begin{tikzcd}
C_{\mathrm{c}}^{\infty}(\mathcal{U}_{1}) \arrow[r,"\sf{A}_{1}"]  &C^{\infty}(\mathcal{U}_{1}) \\
C_{\mathrm{c}}^{\infty}(\mathcal{U}_{2}) \arrow[u,"\kappa^{\ast}"]\arrow[r,"\sf{A}_{2}"] & C^{\infty}(\mathcal{U}_{2})\arrow[u,"\kappa^{\ast}"]
\end{tikzcd}\end{equation*}

\noindent where $\kappa^{\ast}:C^{\infty}(\mathcal{U}_{2})\to C^{\infty}(\mathcal{U}_{1})$ denotes the \textit{pull-back} defined by $u\mapsto u\circ\kappa$. Its inverse, the \textit{push-forward}, is given by $u\mapsto \kappa_{\ast}= u\circ \kappa^{-1}$. In other words, the operator $\sf{A}_{2}$ is given by
\begin{align*}
	\sf{A}_{2}\varphi=(\kappa_{\ast}\circ A_{1}\circ\kappa^{\ast})\varphi=[\sf{A}_{1}(\varphi\circ\kappa)]\circ\kappa^{-1}
\end{align*}
for all $\varphi\in C^{\infty}_{\mathrm{c}}(\mathcal{U}_{2})$. Now, the question we would like to analyse in the following is under which assumption $\sf{A}_{2}$ is again a pseudodifferential operator. Suppose that $a_{1}\in\mathcal{S}^{m}_{\rho,\sigma}(\mathcal{U}\times \mathcal{U})$ is a symbol of $\sf{A}_{1}$. Then, $\sf{A}_{2}$ can be written as
\begin{align*}
    \sf{A}_{2}\varphi(x)=\frac{1}{(2\pi)^{d}}\int_{\mathbb{R}^{d}}\int_{\mathcal{U}}e^{i(\kappa^{-1}(x)-y)\cdot\xi}a_{1}(\kappa^{-1}(x),y,\xi)\varphi(\kappa(y))\,\mathrm{d}^{d}y\,\mathrm{d}^{d}\xi
\end{align*}
Now, let us perform a change of variables by setting $y=:\kappa^{-1}(z)$, which yields
\begin{align*}
    \sf{A}_{2}\varphi(x)=\frac{1}{(2\pi)^{d}}\int_{\mathbb{R}^{d}}\int_{\mathcal{U}}e^{i(\kappa^{-1}(x)-\kappa^{-1}(z))\cdot\xi}\underbrace{a_{1}(\kappa^{-1}(x),\kappa^{-1}(z),\xi)}_{=:a_{2}(x,z,\xi)}\varphi(z)\bigg\vert\mathrm{det}\frac{dz}{dy}\bigg\vert\,\mathrm{d}^{d}z\,\mathrm{d}^{d}\xi
\end{align*}
It is easy to see that $a_{2}\in\mathcal{S}^{m}_{\rho,\sigma}(\mathcal{U}\times \mathcal{U})$. Furthermore, $\Phi(x,z,\xi):=(\kappa^{-1}(x)-\kappa^{-1}(z))\cdot\xi$ defined a phase function in $\mathcal{P}(\mathcal{U}\times \mathcal{U})$. Hence, we conclude that $\sf{A}_{2}$ is a Fourier integral operator. Now, for $\rho+\sigma=1$ and $\sigma<\rho$, it turns out that $\sf{A}_{2}$ is again a pseudodifferential operator, i.e.~$\sf{A}_{2}\in\Psi^{m}_{\rho,\sigma}(\mathcal{U})$, since in this case one can perform another change of variables in $\xi$ to obtain the standard phase function. The principal symbol of $\sf{A}_{2}$ transforms covariantly as
\begin{align*}
	\sigma_{\sf{A}_{2}}(x,\xi)=\sigma_{\sf{A}_{1}}(\kappa^{-1}(x),((\d\kappa^{-1}_{x})^{t})^{-1}(\xi))
\end{align*}
for $(x,\xi)\in \mathcal{U}\times\mathbb{R}^{d}$ in this case. In particular, the principal symbol can be viewed as a smooth function on the cotangent bundle $\sf{T}^{\ast}\mathcal{U}$. For more details, see \cite[pp.~34f.]{Sjorstrand} and \cite[I.4]{Shubin}.

Now, let $\mathsf{M}$ be a smooth $d$-dimensional manifold. The previous discussion allows us to define pseudodifferential operators on $\mathsf{M}$ as follows:

\begin{definition} (Classical $\Psi$DO's on Manifolds)\newline
    Let $A:C^{\infty}_{\mathrm{c}}(\sf{M})\to C^{\infty}(\sf{M})$ be a linear and continuous operator. It is called a \textit{(classical) pseudodifferential operator of order $m$}, if for every chart $\varphi\colon \mathcal{U}\to\mathcal{V}\subset\bb{R}^{d}$ the linear operator $\sf{A}^{\prime}:=\varphi_{\ast}\circ \sf{A}_{1}\circ\varphi^{\ast}\colon C^{\infty}_{c}(\mathcal{V})\to C^{\infty}(V)$ is a (classical) pseudodifferential operator in $\Psi^{m}(\mathcal{V})$ (resp.~$\Psi^{m}_{\mathrm{cl}}(\mathcal{V})$). The set of all such operators is denoted by $\Psi^{m}(\sf{M})$ resp.~$\Psi^{m}_{\mathrm{cl}}(\sf{M})$.
\end{definition}

\begin{example}
	By the previous discussion of coordinate transformations, it is clear that any pseudofifferential operator $\sf{A}\in\Psi^{m}(\mathcal{U})$ is a pseudodifferential operator on the open submanifold $\mathcal{U}\subset\mathbb{R}^{d}$. Note, however, that not any pseudodifferential operator on the manifold $\mathcal{U}$ is of this form. In other words, there is slight notational ambiguity for when viewing $\mathcal{U}$ as a manifold.
\end{example}

More generally, if $\sf{E}\xrightarrow{\pi}{\sf{M}}$ is a smooth (finite-rank) $\bb{K}$-vector bundle over $\sf{M}$, we can also define pseudodifferential operators of the form
\begin{align*}
	\sf{A}\colon \Gamma^{\infty}_{\mathrm{c}}(\sf{M},\sf{E})\to\Gamma^{\infty}(\sf{M},\sf{E})\, .
\end{align*}
The set of all such operators is denoted by $\Psi^{m}(\sf{M},\sf{E})$ and by $\Psi^{m}_{\mathrm{cl}}(\sf{M},\sf{E})$ in the classical case. For this, one needs first to generalise the local theory on $\mathcal{U}\subset\bb{R}^{d}$ developed above to matrix-valued symbols $a\in C^{\infty}(\mathcal{U}\times\bb{R}^{d},\bb{C}^{m\times m})$, $m=\mathrm{rank}_{\bb{K}}(\sf{E})$, however, this does not present any additional complications, since everything reduces to a componentwise analysis. Furthermore, most of the definitions we did previously can be generalised to manifolds and to the vector bundle-valued case, like the notion of \textit{properly-supported} pseudodifferential operators, smoothing operators, the decomposition into a properly supported part and a smoothing error, etc. We also remark that any operator $\sf{A}\in\Psi^{m}(\sf{M},\sf{E})$ uniquely extends to a linear and continuous operator of the type $\sf{A}\:\mathcal{E}^{\prime}(\sf{E})\to\mathcal{D}^{\prime}(\sf{E})$ that is pseudolocal, i.e.~$\mathrm{sing}\,\mathrm{supp}(A\psi)\subset\mathrm{sing}\,\mathrm{supp}(\psi)$.

Last but not least, we remark that the symbol classes $\mathcal{S}^{m}(\mathcal{U})$ transform covariantly under the change of coordinates. This has two consequences: on the one hand, we can define a symbol class $\mathcal{S}^{m}(\sf{M},\sf{E})$ on the manifold $\sf{M}$ as the set of all maps $a\in \Gamma^{\infty}(\sf{T}^{\ast}\sf{M},\pi^{\ast}\mathrm{End}(\sf{E}))$, where $\pi\:\sf{T}^{\ast}\sf{M}\to\sf{M}$ denotes the bundle projection, such that in any local trivialisation $\phi\:\sf{E}_{\mathcal{U}}\to\mathcal{U}\times\bb{R}^{m}$ and compatible coordinate chart $\psi\:\mathcal{U}\xrightarrow{\cong}\mathcal{V}\subset\bb{R}^{d}$, the corresponding \emph{local representative} $\mathfrak{a}\in C^{\infty}(\mathcal{V}\times\bb{R}^{d},\bb{K}^{m\times m})$ satisfies
\begin{align*}
	\chi(x)\mathfrak{a}(x,\xi)\in \mathcal{S}^{m}(\mathcal{U},\bb{K}^{m\times m})
\end{align*}
for all $\chi\in C^{\infty}_{\mathrm{c}}(\mathcal{V})$. With this definition, we can define the \emph{principal symbol} $\tau_{\sf{A}}\in\mathcal{S}(\sf{M},\sf{E})$ for any $\sf{A}\in\Psi^{m}(\sf{M},\sf{E})$ in a similar way as in the local case. On the other hand, we can define a \emph{quantisation map} $\mathrm{Op}\:\mathcal{S}^{m}(\sf{M},\sf{E})\to\Psi^{m}(\sf{M},\sf{E})$ by batching together local representatives using a partition of unity and charts. We refer to \cite[Chap.~5]{Hintz} for more details on pseudodifferential calculus on manifolds.

\begin{remark} (Global Pseudodifferential Calculus)\newline
	As discussed above, the quantisation map in $\bb{R}^{d}$ establishes a canonical linear isomorphism between symbol classes and the corresponding pseudodifferential operators (mod $\mathcal{S}^{-\infty})$. Moreover, there are explicit formulas both for constructing an operator from its symbol and for recovering the symbol of a given operator. A central feature of this calculus is its covariance under coordinate changes. This local covariance, in turn, allows one to define symbol classes on manifolds with values in vector bundles. Furthermore, one can define pseudodifferential operators on manifolds acting on the sections of vector bundles by patching together local constructions using a partition of unity, as mentioned above. While sufficient for many applications, this approach depends, at least up to a smoothing operator, on non-canonical choices, such as coordinate systems and trivialisations, and thus lacks a fully intrinsic character. In fact, only the \emph{principal symbol} of a given pseudodifferential operator behaves tensorially and is globally well defined. Consequently, there is strong motivation to develop a coordinate-independent, geometrically natural theory of symbols and quantisation on manifolds. Several such approaches have been explored in the literature. Early work concentrated on the Kohn–Nirenberg quantisation for scalar-valued operators, notably by Bokobza-Haggiag \cite{Bokobza1,Bokobza2}, Widom \cite{Widow1,Widow2}, and Drager \cite{Drager}. More recent developments have extended these ideas to vector bundle-valued settings, as explored by Pflaum \cite{PflaumNormal} and Sharafutdinov \cite{MR2191866,MR2186590}. An invariant $\tau$-quantization on closed Riemannian manifolds for operators acting in $1$-forms has been developed in \cite{CapoferriCurl}. Further advancements in Weyl and $\tau$-quantisations within the scalar framework have been contributed by  Dereziński-Latosiński-Siemssen \cite{dls_weyl}, Underhill \cite{Underhill}, Unterberger \cite{Unterberger}, Liu-Qian \cite{LiuQian}, Pflaum \cite{PflaumWeyl}, Safarov \cite{Safarov0,Safarov,SafarovBook}, Fulling \cite{fulling} (see also~\cite{Fulling0,Kennedy}) and, more recently, by Levy \cite{Levy} and Andersson-Moser-Oancea-Paganini-Schmid~\cite{AnderssonSchmid} in the vector-bundle valued case.

More generally, a line of research concerns the definition of global Fourier integral operators and the global construction of propagators for hyperbolic equations. In particular, Laptev-Safarov-Vassiliev \cite{Laptev} (see also~\cite{SafarovVassiliev}) demonstrated that the propagator for a broad class of hyperbolic operators on compact manifolds can be represented as a \textit{global} Fourier integral operator with complex-valued phase function. Building upon this work and some applications discussed in \cite{Jakobson}, a global symbol calculus for Fourier integral operators on closed manifolds has been developed by Safarov \cite{Safarov2}. Furthermore, a geometric approach to global parametrices for the scalar wave operator on closed Riemannian manifolds, including a global invariant definition of the full symbol of the wave propagator and an algorithm for explicitly calculating its homogeneous components, was recently proposed by Capoferri-Levitin-Vassiliev \cite{Capoferri2}. This result has been generalised to the Lorentzian setting by Capoferri-Dappiaggi-Drago \cite{Capoferri3} for the wave operator on globally hyperbolic Lorentzian manifolds with compact Cauchy surfaces. Generalisations to the propagator for the Dirac operator have been obtained by Capoferri-Vassiliev \cite{Capoferri1} in the Riemannian and by Capoferri-Murro \cite{CapoferriMurro} in the Lorentzian setting.
\end{remark}

\subsection{Pseudodifferential Calculus on Manifolds of Bounded Geometry}\label{Subsec:BoundGeom}
Let $\sf{M}$ be a smooth manifold. As discussed above, two properly supported pseudodifferential operators can be composed. In particular, the space of properly supported $\Psi$DOs forms a well-defined algebra of operators. However, for more advanced applications, this calculus turns out to be too small. For instance, consider a complete Riemannian manifold $(\sf{M}, \sf{g})$. Its Laplace–Beltrami operator $\Delta$ is a classical and properly supported $\Psi$DO of order two, but its resolvent $(-\Delta+i)^{-1}$ is not properly supported. Now, if we work with compact manifolds, then every $\Psi$DO is properly supported, yielding a well-defined calculus, namely, the \emph{uniform pseudodifferential calculus on compact manifolds}. More generally, however, we need to identify a class of $\Psi$DOs that is small enough to allow for well-defined composition (and hence to form an algebra), but large enough to be closed under other operations of interest. Such a class is provided by the bounded pseudodifferential operators on Riemannian manifolds of bounded geometry, which we shall discuss now.

Throughout the following, let $(\sf{M},\sf{g})$ be a smooth and connected Riemannian manifold with Levi-Civita connection $\nabla$. We recall the following basic terminology of Riemannian geometry:
\begin{itemize}
	\item[(i)]By standard ODE theory, there exists a geodesic $\gamma_{p,v}$ such that $\gamma(0)=p$ and $\dot{\gamma}(0)=v$ for every $p\in\sf{M}$ and $v\in\sf{T}_{p}\sf{M}$. Furthermore, all such geodesics agree on their common domain and we shall denote the maximal possible open domain by $\mathrm{I}_{p,v}\subset\mathbb{R}$. Now, let $\mathcal{D}_{p}\sf{M}$ be the set of vectors $v\in\sf{T}_{p}\sf{M}$ such that $[0,1]\subset\mathrm{I}_{v}$. Then, the \emph{exponential map} $\mathrm{exp}_{p}\:\mathcal{D}_{p}\sf{M}\to\sf{M}$ at $p\in\sf{M}$ is the map defined by $\mathrm{exp}_{p}(v):=\gamma_{p,v}(1)$. In the case $\mathrm{I}_{p,v}=\bb{R}$ and hence $\mathcal{D}_{p}\sf{M}=\sf{T}_{p}\sf{M}$, the manifold $(\sf{M},\sf{g})$ is called \emph{(geodesically) complete} and by the the celebrated \textit{Theorem of Hopf-Rinow}, this is equivalent to require that $\sf{M}$ is complete as a metric space with respect to the metric induced by Riemannian distance.
	\item[(ii)]The exponential map $\mathrm{exp}_{p}$ is not a diffeomorphism in general, even in the case in which $(\sf{M},\sf{g})$ is complete, however, by the inverse function theorem, it is always a \textit{local} diffeomorphism. The largest radius $\mathrm{R}>0$ for which $\mathrm{exp}_{p}\:\mathcal{B}_{\mathrm{R}}^{\sf{g}_{p}}(0)\to\sf{M}$ is a diffeomorphism onto its image, is called the \textit{injectivity radius at $p$} and denoted by $r_{\sf{g}}(p)$, where $\mathcal{B}_{\mathrm{R}}^{\sf{g}_{p}}(0)$ denotes the open ball of radius $\mathrm{R}$ around the origin of the inner product space $(\sf{T}_{p}\sf{M},\sf{g}_{p})$. Accordingly, the \textit{injectivity radius of $\sf{M}$} is defined by 
	\begin{align*}
        r_{\sf{g}}:=\inf_{p\in\sf{M}}r_{\sf{g}}(p)=\inf_{p\in\sf{M}}\bigg\{\sup_{\mathrm{R}>0}\{\mathrm{exp}_{p}\vert_{\mathcal{B}^{\sf{g}_{p}}_{\mathrm{R}}(0)}\text{ is diffeomorphism onto its image}\}\bigg\}\in [0,\infty]
    \end{align*}
    The injectivity radius of the Euclidean space $(\bb{R}^{d},\delta)$ is $r_{\delta}=\infty$ and the injectivity radius of an arbitrary Riemannian manifold can be viewed as the largest radius of a ball around each point within which geodesics behave similarly to that of Euclidean space.
    \item[(iii)]For a given $p\in\sf{M}$, let us denote by $\Vert\cdot\Vert_{p}$ the obvious norm on $\sf{T}_{p}^{r,s}\sf{M}=\sf{T}_{p}\sf{M}^{\otimes r}\otimes\sf{T}^{\ast}_{p}\sf{M}^{\otimes s}$ defined by $\sf{g}_{p}$ and $\sf{g}_{p}^{-1}$. With this notation, we call a rank $(r,s)$-tensor field $\sf{T}\in\Gamma^{\infty}(\sf{T}^{r,s}\sf{M})$ \emph{uniformly bounded}, if $\Vert \sf{T}\Vert:=\sup_{p\in\sf{M}}\Vert \sf{T}_{p}\Vert_{p}<\infty$.
    \end{itemize}	
    
\begin{definition}\label{Definition:BoundGeom} A Riemannian manifold $(\sf{M},\sf{g})$ is said to have \emph{bounded geometry}, if $r_{\sf{g}}>0$ and if $\nabla^{k}\mathrm{Riem}(\sf{g})\in\Gamma^{\infty}(\sf{T}^{1,3+k}\sf{M})$ is uniformly bounded for all $k\in\bb{N}_{0}$, where $\mathrm{Riem}(\sf{g})\in\Gamma^{\infty}(\sf{T}^{1,3}\sf{M})$ denotes the Riemannian curvature tensor of $(\sf{M},\sf{g})$, as usual.
\end{definition}

In general, every Riemannian manifold with $r_{\sf{g}}>0$ can easily be seen to be complete and hence, every Riemannian manifold of bounded geometry is complete as well. Besides $(\bb{R}^{d},\delta)$, every compact Riemannian manifold has bounded geometry and every Lie group $\sf{G}$ equipped with a bi-invariant Riemannian metric has bounded geometry as well. We also note that ``bounded geometry'' is a purely geometric concept, since on every manifold there exists a Riemannian metric of bounded geometry, see \cite{Greene}. In other words, there are no topological obstructions to the existence of Riemannian metrics of bounded geometry.

In what follows, we introduce several definitions and constructions on Riemannian manifolds of bounded geometry, which will be used to develop the pseudodifferential calculus in this setting. The following exposition follows the discussion in \cite{GerardOulghaziWrochna} and \cite[Chap.~10]{GerardBook} (see also~\cite[App.~1]{ShubinPSI}). To start with, we provide an equivalent characterisation of bounded geometry, which will be very useful in applications. Let $\mathcal{U}\subset\mathbb{R}^{d}$ be open and let $\delta$ denote the Euclidean metric. Then, we denote by $\mathrm{BT}^{r,s}(\mathcal{U},\delta)$ the space of all rank $(r,s)$-tensor fields on $\mathcal{U}$, which are uniformly bounded with all their derivatives, equipped with its natural Fréchet space topology. For the case $r=s=0$, we also write $C^{\infty}_{\mathrm{bd}}(\mathcal{U}):=\mathrm{BT}^{0,0}(\mathcal{U},\delta)$. Then, by \cite[Thm.~2.2]{GerardOulghaziWrochna}, a Riemannian manifold $(\sf{M},\sf{g})$ is of bounded geometry if and only if for all $p\in\sf{M}$ there exists an open neighbourhood $\mathcal{U}_{p}$ of $p$ and a coordinate chart $\varphi_{p}\:\mathcal{U}_{p}\xrightarrow{\cong}\mathcal{B}_{1}(0)$, where $\mathcal{B}_{1}(0)\subset\mathbb{R}^{d}$ denotes the open ball of radius $1$ around the origin, such that 
\begin{align*}
	\text{(i)}&\quad (\sf{g}_{p}:=(\varphi_{p}^{-1})^{\ast}\sf{g})_{p\in\sf{M}}\text{ is a bounded family in }\mathrm{BT}^{0,2}(\mathcal{B}_{1}(0),\delta)\\
	\text{(ii)}&\quad \exists C>0:\quad C^{-1}\delta\leq g_{p}\leq C\delta\quad\forall p\in\sf{M}
\end{align*}

A family of charts $(\varphi_{i})_{i\in\mathrm{I}}$ with the above two properties is usually referred to as a \emph{family of bounded charts}. Given such a family, we can define a ``bounded version'' of various different objects of Riemannian geometry as follows:

\begin{itemize}
	\item[(1)]By \cite[Appendix 1, Lemma~1.2]{ShubinPSI}, there exists a countable good family of bounded charts $(\mathcal{U}_{n},\varphi_{n})_{n\in \mathbb{N}}$, covering $\sf{M}$, that is \textit{uniformly finite}, i.e.~there exists an $N\in\mathbb{N}$ such that for every $\mathrm{J}\subset \mathbb{N}$ with $\vert \mathrm{J}\vert\geq N$ we have that $\bigcap_{j\in \mathrm{J}}\mathcal{U}_{j}=\emptyset$. Such an atlas will be called \textit{bounded atlas}.
	\item[(2)]If $(\mathcal{U}_{n},\varphi_{n})_{n\in\mathbb{N}}$ is a bounded atlas, we call a smooth partition of unity $(\chi_{n})_{n\in\mathbb{N}}$ subordinate to $(\mathcal{U}_{n})_{n\in\bb{N}}$ \textit{bounded}, if the sequence $((\varphi_{n})_{\ast}\chi_{n})_{n\in\mathbb{N}}$ is a bounded family in $C_{\mathrm{bd}}^{\infty}(\mathcal{B}_{1}(0))$. One can show that such a partition of unity always exists. 
	\item[(3)]A tensor field $\sf{T}\in\Gamma^{\infty}(\sf{T}^{r,s}\sf{M})$ is called \emph{bounded} if the family $(\varphi_{p})_{\ast}T)_{p\in\sf{M}}$ is bounded in $\mathrm{BT}^{r,s}(\mathcal{B}_{1}(0),\delta)$ for some bounded family of charts $(\mathcal{U}_{p},\varphi_{p})$. The set of all such tensors is denoted by $\mathrm{BT}^{r,s}(\sf{M},g)$ and equipped with its natural Fréchet topology.
	\item[(4)]A finite-rank $\bb{K}$-vector bundle $\sf{E}\to\sf{M}$ is said to have \emph{bounded geometry}, if there exists a covering by bounded charts $(\mathcal{U}_{i},\varphi_{i})_{i\in\mathrm{I}}$, which does also provide a local trivialsation of $\sf{E}$ such that the family of transition functions is abounded family of matrix-valued functions. Examples include the trivial vector bundle, $\sf{T}^{r,s}\sf{M}$, $\bigwedge^{k}\sf{T}^{\ast}\sf{M}$ and the Clifford and spinor bundles in the case in which $(\sf{M},\sf{g})$ is also a \emph{spin} manifold. We define the space of all \emph{bounded sections} to be the set $\Gamma^{\infty}_{\mathrm{bd}}(\sf{M},\sf{E})$ consisting of sections $\psi\in\Gamma^{\infty}(\sf{E})$ for which $((\varphi_{i})_{\ast}\psi)_{i\in\mathrm{I}}$ defines a bounded family in $C_{\mathrm{bd}}^{\infty}(\bb{R}^{d},\bb{K}^{m})$. 
\end{itemize}

Let us now turn to pseudodifferential calculus on bounded geometry developed by Shubin \cite{ShubinPSI} and Kordyukov \cite{Kordyukov}. Many details can be found in \cite[Sec.~10.4]{GerardBook} in the scalar-valued case and in \cite[Chap.~5]{GerardStoskopf} in the vector bundle-valued case. Let us start with a good notion of ``bounded symbols'' on Riemannian manifolds of bounded geometry.

\begin{definition}\label{Definition:BoundSym}
	Let $(\sf{M},\sf{g})$ be a Riemannian manifold of bounded geometry and $\sf{E}$ a $\bb{K}$-vector bundle of bounded geometry over $\sf{M}$ with $m:=\mathrm{rank}_{\bb{K}}(\sf{E})$. We denote by $\mathcal{S}_{\mathrm{bd}}^{m}(\sf{M},\sf{E})$ the set of $a\in\Gamma^{\infty}(\sf{T}^{\ast}\sf{M},\pi^{\ast}\mathrm{End}(\sf{E}))$, such that for every $p\in\sf{M}$ it holds that $(\varphi_{p})_{\ast}a\in\mathcal{S}_{\mathrm{cl}}^{m}(\mathcal{B}_{1}(0),\mathbb{K}^{m\times m})$, and such that the family $((\varphi_{p})_{\ast}a)_{p\in\sf{M}}$ is bounded in $\mathcal{S}_{\mathrm{cl}}^{m}(\mathcal{B}_{1}(0),\mathbb{K}^{m\times m})$.
\end{definition}

An element $a\in\mathcal{S}_{\mathrm{bd}}^{m}(\sf{M},\sf{E})$ is called \emph{(classical) bounded symbol of order $m$}. Clearly $\mathcal{S}_{\mathrm{bd}}^{m}(\sf{M},\sf{E})\subset\mathcal{S}_{\mathrm{cl}}(\sf{M},\sf{E})$ and, as usual, we equip $\mathcal{S}_{\mathrm{bd}}^{m}(\sf{M},\sf{E})$ with its natural Fréchet space topology and define $\mathcal{S}_{\mathrm{bd}}^{\infty}(\sf{M},\sf{E})$ as the union/intersection over all $m\in\bb{R}$. 

Now, let us denote by $e\:\mathcal{S}_{\mathrm{cl}}^{m}(\mathcal{B}_{1}(0),\mathbb{C}^{m\times m})\to\mathcal{S}_{\mathrm{cl}}^{m}(\bb{R}^{d},\mathbb{C}^{k\times k})$ a continuous extension. Furthermore, let us choose a bounded atlas $(\mathcal{U}_{n},\varphi_{n})_{n\in\bb{N}}$ of $(\sf{M},\sf{g})$ and denote by $\sf{T}_{i}$ and $\widetilde{\sf{T}}_{i}$ the push-forwards of $\Gamma^{\infty}(\mathcal{U}_{n},\sf{E})$ and $\Gamma^{\infty}(\sf{T}^{\ast}\mathcal{U}_{n},\pi^{\ast}\mathrm{End}(\sf{E}))$, respectively, under $\varphi_{n}$ and the corresponding local trivialisation $\psi_{n}$ of $\sf{E}$. Now, if $a\in \mathcal{S}_{\mathrm{bd}}(\sf{M},\sf{E})$ is any bounded symbol, then we define its \emph{quantisation} by
\begin{align*}
	\mathrm{Op}_{\mathrm{bd}}(a):=\sum_{n\in\mathbb{N}}(\chi_{n}\sf{T}_{n}^{-1})\circ\mathrm{Op}(e\widetilde{\sf{T}}_{n}a)\circ (\chi_{n}\sf{T}_{n})\:\Gamma_{\mathrm{c}}^{\infty}(\sf{M},\sf{E})\to\Gamma^{\infty}(\sf{M},\sf{E})\, ,
\end{align*}
where $\mathrm{Op}(e\widetilde{\sf{T}}_{n}a)$ denotes the usual Kohn-Nirenberg quantisation of $e\widetilde{\sf{T}}_{n}a\in\mathcal{S}_{\mathrm{cl}}^{m}(\mathbb{R}^{d},\mathbb{C}^{m\times m})$. In general, the quantisation map depends on the various choices of $e,\chi_{n},\varphi_{n},\psi_{n}$, however, one can show that for any other quantisation map $\mathrm{Op}_{\mathrm{bd}}^{\prime}$ one has that
\begin{align*}
	\mathrm{Op}_{\mathrm{bd}}(a)-\mathrm{Op}^{\prime}_{\mathrm{bd}}(a)\in \mathcal{W}^{-\infty}(\sf{M},\sf{E})\, ,
\end{align*}
where $\mathcal{W}^{-\infty}(\sf{M},\sf{E})\subset\Psi^{-\infty}(\sf{M},\sf{E})$ denotes an ideal of smoothing operators defined by
\begin{align}\label{eq:BoundedSmoothing}
	\mathcal{W}^{-\infty}(\sf{M},\sf{E})\defeq\bigcap_{n\in\mathbb{N}}\mathcal{L}(\sf{H}^{-m}(\sf{M},\sf{E}),\sf{H}^{m}(\sf{M},\sf{E}))\,,
\end{align}
where $\mathcal{L}(\cdot,\cdot)$ denotes the space of bounded linear operators and $\sf{H}^{m}(\sf{M},\sf{E})$ the Sobolev spaces of order $m\in\mathbb{N}$. This gives rise to the following definition:

\begin{definition}\label{Definition:PSIDOBoun} (Pseudodifferential Calculus of Bounded Geometry)\newline
	Let $(\sf{M},\sf{g})$ be a Riemannian manifold of bounded geometry and $\sf{E}$ a $\bb{K}$-vector bundle of bounded geometry over $\sf{M}$. The space of \emph{bounded pseudodifferential operators} of order $m$ is defined by
	\begin{align*}
		\Psi^{m}_{\mathrm{bd}}(\sf{M},\sf{E}):=\mathrm{Op}_{\mathrm{bd}}(\mathcal{S}_{\mathrm{bd}}^{m}(\sf{M},\sf{E}))+\mathcal{W}^{-\infty}(\sf{M},\sf{E})\,.
	\end{align*}
	Furthermore, we set $\Psi^{\infty}_{\mathrm{bd}}(\sf{M},\sf{E}):=\bigcup_{m\in\mathbb{N}}\Psi^{m}_{\mathrm{bd}}(\sf{M},\sf{E})$.
\end{definition}

Clearly, $\Psi_{\mathrm{bd}}^{m}(\sf{M},\sf{E})\subset\Psi_{\mathrm{cl}}^{m}(\sf{M},\sf{E})$. Now, for general $a\in\mathcal{S}_{\mathrm{bd}}^{m}(\sf{M},\sf{E})$, the support of the kernel of $\mathrm{Op}_{\mathrm{bd}}(a)$ is contained in the set $\{(x,y)\in\sf{M}\times\sf{M}\mid \mathrm{d}(x,y)\leq C\}$ for some constant $C>0$, where $\mathrm{d}$ denotes the geodesic distance of $(\sf{M},\sf{g})$. In particular, this implies that $\mathrm{Op}_{\mathrm{bd}}(a)$ is properly-supported. As a consequence, operators of the form $\mathrm{Op}_{\mathrm{bd}}(a)$ can be composed. However, it is important to stress that for $a,b\in\mathcal{S}^{\infty}_{\mathrm{bd}}(\sf{M},\sf{E})$, the operator $\mathrm{Op}_{\mathrm{bd}}(a)\circ\mathrm{Op}_{\mathrm{bd}}(b)$ is in general \emph{not} again contained in $\mathrm{Op}_{\mathrm{bd}}(\mathcal{S}^{\infty}_{\mathrm{bd}}(\sf{M},\sf{E}))$. The reason for this is that the constant $C$ appearing in the set above is independent of $a$ and just depends on the chosen quantisation map. Furthermore, the support of $\mathrm{Op}_{\mathrm{bd}}(a)\circ\mathrm{Op}_{\mathrm{bd}}(b)$ is usually contained in a slightly larger set and hence $\mathrm{Op}_{\mathrm{bd}}(a)\circ\mathrm{Op}_{\mathrm{bd}}(b)=\mathrm{Op}_{\mathrm{bd}}(c)+\sf{R}$ for some $c\in\mathcal{S}^{\infty}_{\mathrm{bd}}(\sf{M},\sf{E})$ and $\sf{R}\in \mathcal{W}^{-\infty}(\sf{M},\sf{E})$. To sum up, the composition of operators in $\Psi^{m}_{\mathrm{bd}}(\mathcal{M})$ is well-defined. Similarly, one can show that the space $\Psi_{\mathrm{bd}}^{\infty}(\sf{M})$ closed under other important operations, such as taking the adjoint. 

Last but not least, we mention that \emph{time-dependent bounded pseudodifferential operators} on $(\sf{M},\sf{g})$ can be defined by 
\begin{align}\label{eq:Boundedsdfgde}
	C^{\infty}_{\mathrm{bd}}(\mathrm{I},\Psi^{m}_{\mathrm{bd}}(\sf{M})):=\mathrm{Op}_{\mathrm{bd}}(C^{\infty}_{\mathrm{bd}}(\mathrm{I},\mathcal{S}_{\mathrm{bd}}(\sf{M},\sf{E})))+C^{\infty}_{\mathrm{bd}}(\mathrm{I},\mathcal{W}^{-\infty}(\sf{M}))\,,
\end{align}
where $\mathrm{I}\subset\bb{R}$ is an open interval and $C^{\infty}_{\mathrm{bd}}(\mathrm{I},\mathcal{F})$ for some Fréchet space $\mathcal{F}$ with topology induced by a family of seminorms $(\Vert\cdot\Vert_{n})_{n\in\bb{N}}$ is the set of all smooth functions such that $\sup_{t\in\mathrm{I}}\Vert\partial_{t}^{p}f(t)\Vert_{n}<\infty$ for all $n,p\in\mathbb{N}$.

\subsection{Propagation of Singularities}\label{Subsec:PropSing}
In the last section of this exposition on microlocal analysis, we briefly discuss the \textit{propagation of singularities theorem}, which, roughly speaking, states that the singularities of solutions to partial and pseudodifferential equations governed by so-called ``real principal type operators'' travel along the \emph{bicharacteristic} flow determined by the associated \emph{Hamiltonian vector field.} 

To start with, we recall some terminology from \emph{Hamiltonian systems}, which clarifies the geometric picture underlying the forthcoming discussion. We refer to \cite{RudolphSchmidt1} for details.

\begin{itemize}
	\item[(i)]Let $(\sf{M},\omega)$ be a symplectic manifold, i.e.~a smooth $n$-manifold with non-degenerate and closed $\omega\in\Omega^{2}(\sf{M})$. Since every skew-symmetric matrix in odd dimensions is necessarily degenerate, we have $n=2d$ for some $d\in\bb{N}$. Now, for a given function $f\in C^{\infty}(\sf{M})$, we call $\sf{X}_{f}:=-(\d f)^{\sharp}$, where the isomorphism $\sharp\:\sf{T}^{\ast}\sf{M}\to\sf{T}\sf{M}$ is defined by $\omega$ in analogy to Riemannian geometry, the \textit{Hamiltonian vector field associated to $f$}. In other words, $\sf{X}_{f}\lrcorner\omega=-\d f$. Now, it is well known that around every point $p\in\sf{M}$, we can find a local coordinate chart $(x^{1},\dots,x^{d},\xi_{1},\dots,\xi_{d})$ such that $\omega=\d \xi_{i}\wedge \d x^{i}$ (cf.~\emph{Darbeaux Theorem}). With respect to this local coordinate system, the vector field $\sf{X}_{f}$ is given by
\begin{align*}
	\sf{X}_{f}=(\partial_{\xi_{i}}f)\partial_{x^{i}}-(\partial_{x^{i}}f)\partial_{\xi_{i}}\, .
\end{align*} 
The flow $\Phi^f\:\mathrm{I}\times\sf{M}\to\sf{M}$ of the vector field $\sf{X}_{f}$, where $\mathrm{I}\subset\bb{R}$ is some open time interval, is called the \textit{Hamiltonian flow of $f$}. By definition,
\begin{align*}
	\Phi^{f}_{0}(p)=p,\qquad \frac{\d}{\d t}\bigg\vert_{t=0}\Phi^{f}_{t}(p)=\sf{X}_{f}(p)\, .
\end{align*}
The corresponding integral curves are called \textit{trajectories} or \textit{orbits}. In other words, these are the curves $\gamma(t)=(x^{i}(t),\xi_{i}(t))$ satisfying the \emph{Hamiltonian equations}
\begin{align*}
\begin{cases}
	\dot{x}^{i}(t)&=\partial_{\xi_{i}}\gamma(x^{i}(t),\xi_{i}(t))\\
	\dot{\xi}_{i}(t)&=-\partial_{x^{i}}\gamma(x^{i}(t),\xi_{i}(t))
\end{cases}\, .
\end{align*}
\item[(ii)]The cotangent bundle $\sf{T}^{\ast}\sf{M}$ of any smooth $d$-manifold $\sf{M}$ can naturally be understood as a $2d$-dimensional symplectic manifold whose symplectic form is given as follows: consider a local chart $(\mathcal{U},\varphi=(x^{1},\dots,x^{d}))$ of $\sf{M}$ around some point $p\in\sf{M}$ and let $\pi\:\sf{T}^{\ast}\sf{M}\to\sf{M}$ denote the bundle projection. Then, $(\pi^{-1}(\mathcal{U}),\psi:=(x^{i},\xi_{i}))$ defined by
\begin{align*}
	\psi((p,\xi)):=(x^{1}(p),\dots,x^{d}(p),\xi_{1},\dots,\xi_{d})\in\bb{R}^{2d}\, ,
\end{align*}
where $\xi\in\sf{T}^{\ast}_{p}\sf{M}$ is locally written as $\xi=\xi_{i}\d x^{i}\vert_{p}$, is a local chart of $\sf{T}^{\ast}\sf{M}$ (see~\cite[Prop.~11.9]{LeeSmooth}). The $1$-form $\theta\in\Omega^{1}(\sf{T}^{\ast}\sf{M})$ defined by $\theta_{\xi}(\sf{X}):=\xi(\d\pi_{\xi}(\sf{X}))$ for $\xi\in\sf{T}^{\ast}\sf{M}$ and $\sf{X}\in\sf{T}_{\xi}(\sf{T}^{\ast}\sf{M})$, i.e.~locally $\theta:=\xi_{i}\d x^{i}$, is called the \textit{tautological $1$-form}, or \textit{Liouville-Poincaré form}, and $\omega:=\d\theta=\d \xi_{i}\wedge \d x^{i}$ defines the canonical symplectic form on $\sf{T}^{\ast}\sf{M}$.
\end{itemize}

Now, consider a pseudodifferential operator $\sf{A}\in\Psi^{m}(\sf{M})$ of order $m$ and denote its principal symbol by $a(x,\xi):=\sigma_{\sf{A}}(x,\xi)$. By definition, $a\in C^{\infty}(\sf{T}^{\ast}\sf{M})$, and we denote by $\sf{H}_{a}\in\mathfrak{X}(\sf{T}^{\ast}\sf{M})$ the Hamiltonian vector field associated to $a$. In local coordinates $(x^{i},\xi_{i})$ of $\sf{T}^{\ast}\sf{M}$, we have
\begin{align*}
	\sf{H}_{a}=(\partial_{\xi_{i}}a)\partial_{x^{i}}-(\partial_{x^{i}}a)\partial_{\xi_{i}}\, .
\end{align*}
Now, we call the integral curves of the Hamiltonian vector field associated to the pseudodifferential operator $\sf{A}$ the \textit{bicharacteristics} of $\sf{A}$. If a bicharacterstic $\gamma(t)=(x^{i}(t),\xi_{i}(t))$ satisfies $a(x^{i}(t),\xi_{i}(t))=0$ for all $t\in \mathrm{I}$, i.e.~if it is contained in $\mathrm{char}(\sf{A})$, it is called a \textit{null bicharacteristic.}

As a next step, let us state and sketch a proof of the propagation of singularities theorem for a simple class of operators. More precisely, we consider the Cauchy problem
\begin{align}\label{DiffEq}
	\begin{cases}
		(\partial_{t}-i\sf{A}(t))\psi&=\phi\\
		\psi\vert_{t=0}&=\mathfrak{f}
	\end{cases}\,,
\end{align}
where $\sf{A}\in C^{\infty}(\mathbb{R},\Psi^{1}_{\mathrm{cl}}(\bb{R}^{d}))$ is a time-dependent first-order pseudodifferential operator with (time-dependent) principal symbol $a\in C^{\infty}(\mathbb{R},\mathcal{S}^{1}_{\mathrm{h}}(\bb{R}^{d}))$. Now, one can show that the Cauchy problem for Eq.~\eqref{DiffEq} is well-posed, i.e.~there exists a unique solution $\psi\in C^{0}(\bb{R},\sf{H}^{s}(\bb{R}^{d}))\cap C^{1}(\bb{R},\sf{H}^{s-1}(\bb{R}^{d}))$ for a given source $\phi\in C^{0}(\bb{R},\sf{H}^{s}(\bb{R}^{d}))$ and initial datum $\mathfrak{f}\in\sf{H}^{s}(\bb{R}^{d})$. We refer to \cite[Chap.~7]{Hintz} and \cite[Sec.~7.7]{TaylorII} for a proof. Now, consider the corresponding \emph{evolution operator} $\mathcal{U}(s,t)\:\psi(t)\mapsto\psi(s)$. Furthermore, let us associate to \eqref{DiffEq} the Hamiltonian flow $\Phi_{t}:\sf{T}^{\ast}\mathbb{R}^{d}\to \sf{T}^{\ast}\mathbb{R}^{d}$, defined by $\Phi_{t}(x_{0},\xi_{0}):=\gamma(t)$ with $\gamma(t)$ such that $\gamma(0)=(x_{0},\xi_{0})$ and $\gamma^{\prime}(t)=\sf{H}_{a(t)}\vert_{\gamma(t)}$, where $\sf{H}_{a(t)}$ denotes the (time-dependent) Hamiltonian vector field. 

The following theorem is due to Egorov \cite{Egorov}, see also \cite[Sec.~7.2]{Hintz} or \cite[Sec.~7.8]{TaylorII}.

\begin{theorem}\label{Thm:Egorov} \emph{(Egorov Theorem)}\newline
	Let $\sf{P}_{0}\in\Psi^{m}(\mathbb{R}^{d})$ with symbol $p_{0}\in\mathcal{S}^{m}(\bb{R}^{d})$ and define $\sf{P}(t):=\mathcal{U}(t,0)\circ \sf{P}_{0}\circ \mathcal{U}(0,t)$. Then, there exists a (time-dependent) smoothing operator $\sf{R}$ such that $\sf{P}(t)-\sf{R}(t)\in\Psi^{m}(\mathbb{R}^{d})$ Moreover, the principal symbol of $\sf{P}(t)$ is given by $\sigma_{\sf{P}(t)}(\Phi_{t}(x_{0},\xi_{0}))=p_{0}(x_{0},\xi_{0})$.
\end{theorem}

\begin{proof}[Proof (sketch).]
	To start with, let us take the time derivative of the defining relation $\sf{P}(t):=\mathcal{U}(t,0)\circ \sf{P}_{0}\circ \mathcal{U}(0,t)$, which yields $\partial_{t}\sf{P}(t)=i[\sf{A}(t),\sf{P}(t)]$. The idea of the proof is then to construct an approximate solution $\sf{Q}(t)$ of this equation, i.e.~an operator $\sf{Q}(t)$ such that 
	\begin{align*}
		\partial_{t}\sf{Q}(t)=[\sf{A}(t),\sf{Q}(t)]+\sf{R}(t)
	\end{align*}
	with $\sf{Q}(0)=\sf{P}_{0}$, where $\sf{R}(t)$ is a one-parameter family of smoothing operators. This is done by writing its symbol as $q(t)\sim\sum_{j=0}^{\infty}q_{j}(t)$ for $q_{j}\in\mathcal{S}^{m-j}$, which yields a family of recursive \emph{transport equations} for  $q_{j}$. We refer to \cite[Thm.~7.3]{Hintz} and \cite[Thm.~8.1]{TaylorII} for details.
\end{proof}

As a direct consequence of the Egorov theorem, we obtain the following \emph{propagation of singularities} theorem for operators of the form $\partial_{t}-i\sf{A}(t)$.

\begin{corollary} \emph{(Propagation of Singularities for Operators of the Type $\partial_{t}-i\sf{A}(t)$)}\newline
	Consider the hyperbolic equation~\eqref{DiffEq} for $\sf{A}\in C^{\infty}(\bb{R},\Psi_{\mathrm{cl}}^{1}(\bb{R}^{d}))$ with source $\phi=0$ and initial datum $\mathfrak{f}\in\sf{H}^{-\mathrm{N}}(\bb{R}^{d})$ for some $\mathrm{N}\in\bb{N}$. Then, the unique solution $\psi$ satisfies
	\begin{align*}
		\mathrm{WF}(\psi(t))=\Phi_{t}\mathrm{WF}(\mathfrak{f}).
	\end{align*}
\end{corollary}

\begin{proof}
	Let $(x_{0},\xi_{0})\notin\mathrm{WF}(\mathfrak{f})$. Then, using the characterisation of the wavefront set in Remark~\ref{Rem:WFSPsu}, there is an operator $\sf{P}_{0}\in\Psi^{0}_{\mathrm{cl}}(\mathbb{R}^{d})$ such that $\sf{P}_{0}$ is elliptic at $(x_{0},\xi_{0})$, i.e.~$(x_{0},\xi_{0})\notin\mathrm{char}(\sf{P}_{0})$, and such that $\sf{P}_{0}\mathfrak{f}\in C^{\infty}(\mathbb{R}^{d})$. Then, by Theorem~\ref{Thm:Egorov}, $\sf{P}_{0}\mathfrak{f}=\mathcal{U}(0,t)\sf{P}(t)\mathcal{U}(t,0)\mathfrak{f}=\mathcal{U}(0,t)\sf{P}(t)\psi(t)$ and hence $\sf{P}(t)\psi(t)\in C^{\infty}(\mathbb{R}^{d})$. Now, since also $\Phi_{t}(x_{0},\xi_{0})\notin\mathrm{char}(\sf{P}(t))$, we conclude that $\Phi_{t}(x_{0},\xi_{0})\notin\mathrm{WF}(\psi(t))$. This shows $\mathrm{WF}(\psi(t))\subset\Phi_{t}\mathrm{WF}(\mathfrak{f})$. The reverse inclusion follows from switching the time direction.
\end{proof}

Now, let us briefly discuss the general situation. Let $\sf{M}$ be a smooth manifold. We say that a properly supported pseudodifferential operator $\sf{P}\in\Psi^{m}(\sf{M})$ is of \textit{real principal type}, if its principal symbol $p(x,\xi):=\sigma_{\sf{P}}(x,\xi)$ is real-valued and (positively) homogeneous of degree $m$, i.e.~$p(x,\lambda\xi)=\lambda^{m}p(x,\xi)$ for any $\lambda>0$ and $\xi\neq 0$, and if $\d p\neq 0$ on $\mathrm{char}(\sf{P})$. In particular, if $\sf{H}_{p}\in\mathfrak{X}(\sf{T}^{\ast}\sf{M})$ is the corresponding Hamiltonian vector field, then $\sf{H}_{p}\neq 0$ on $\mathrm{char}(\sf{P})$.

\begin{remark} 
	The equation $\sf{P}u=f$ for operators of real principal type is locally solvable, i.e.~for every $f\in C^{\infty}(\sf{M})$ there exists a local solution $u\in\mathcal{D}^{\prime}(\sf{M})$ on some compact set $\sf{K}\subset\sf{M}$ (for differential operators, this result is due to Hörmander \cite{Hormander55}). However, for complex coefficients, this is not true in general and a famous and surprisingly simple counterexample was provided by Lewy \cite{Lewy}. It was conjectured that a necessary and sufficient condition for local solvability is the so-called ``condition ($\Psi$)'' (cf.~\textit{Nirenberg-Treves conjecture} \cite{NirenbergTreves1,NirenbergTreves2}). The conjecture was proven to be true by Dencker \cite{Dencker}.
\end{remark}

\begin{theorem} \emph{(Propagation of Singularities for Operators of Real Principal Type)}\newline
	Let $\sf{P}\in\Psi^{m}(\sf{M})$ be a properly-supported pseudodifferential operator of real principal type and $u\in\mathcal{D}^{\prime}(\sf{M})$. Then $\mathrm{WF}(u)\backslash\mathrm{WF}(\sf{P}u)\subset\mathrm{char}(\sf{P})$ and $\mathrm{WF}(u)\backslash\mathrm{WF}(\sf{P}u)$ is invariant under the Hamiltonian flow. 
\end{theorem}

This theorem is due to Duistermaat-Hörmander \cite[Thm.~6.1.1]{DuistermaatHormander} (see also \cite[Thm.~26.1.1]{HormanderIV}). The idea of the proof is to use Fourier integral operators to conjugate the operator $\sf{P}$ in a microlocal sense to the operator $-i\partial_{t}$ for which the claim is obvious. An alternative proof, based on \emph{positive commutator arguments}, can, for instance, be found in \cite[Chap.~8]{Hintz}.

\begin{example}
As an example, consider the Beltrami d'Alembertian $\square=\sf{g}^{\alpha\beta}\nabla_\alpha\nabla_\beta$ on a Lorentzian manifold $(\sf{M},\sf{g})$. Its principal symbol is given by $\sigma_\square(x,\xi)=\sf{g}_{x}^{\alpha\beta}\xi_\alpha\xi_\beta$. The corresponding characteristic set is therefore the dual light cone, i.e.
\begin{align*}
	 \mathrm{char}(\square)=\{(x,\xi)\in\sf{T}^*\sf{M}\backslash\{\textbf{0}\}\mid \sf{g}_{x}^{\alpha\beta}\,\xi_\alpha\xi_\beta=0\}\,.
\end{align*}
The operator $\square$ is clearly of real principal type. The Hamiltonian equations associated with the principal symbol $\sigma_\square$ reduce to the \emph{geodesic equations} $\ddot{x}^{\gamma}+\Gamma^{\gamma}_{\alpha\beta}\dot{x}^{\alpha}\dot{x}^{\beta}=0$ with velocity vector $\xi_{\alpha}=\dot{x}_{\alpha}$. Hence, the wavefront set of any solution to the wave equation is a union of cotangent vectors along lightlike geodesics.
\end{example}

\section{Unbounded Operators and Spectral Theory}\label{App:FuncAna}
While the previous two parts of the appendix focused on microlocal analysis, we now turn to a different topic. More precisely, we provide brief definitions of some important concepts from functional and spectral analysis that were used in our discussion of the Hodge decomposition in Section~\ref{Sec:HodgeDecomp}. This section does not aim to be exhaustive, rather its purpose is to present the definitions employed, thereby ensuring that the thesis remains self-contained.

Throughout this section, let $(\mathcal{H}_{1},\langle\cdot,\cdot\rangle_{\mathcal{H}_{1}})$ and $(\mathcal{H}_{2},\langle\cdot,\cdot\rangle_{\mathcal{H}_{2}})$ be two (complex) Hilbert spaces. A \emph{(linear, densely-defined) operator} is a $\bb{C}$-linear map $\sf{A}\:\mathrm{dom}(\sf{A})\to\mathcal{H}_{2}$ whose \emph{domain} $\mathrm{dom}(\sf{A})\subset\mathcal{H}_{1}$ is a dense linear subspace. We recall the following definitions and basic facts:
\begin{itemize}
\item[(i)]For two densely-defined operators $\sf{A}\:\mathrm{dom}(\sf{A})\to\mathcal{H}_{2}$ and $\sf{B}\:\mathrm{dom}(\sf{B})\to\mathcal{H}_{2}$, we write $\sf{A}\subset\sf{B}$ if $\mathrm{dom}(\sf{A})\subset\mathrm{dom}(\sf{B})$ and $\sf{B}\vert_{\mathrm{dom}(\sf{A})}=\sf{A}$. Furthermore, if not explicitly stated otherwise, we equip the operators $\sf{A}+\sf{B}$ and $\sf{A}\sf{B}$ with the domains $\mathrm{dom}(\sf{A}+\sf{B})=\mathrm{dom}(\sf{A})\cap\mathrm{dom}(\sf{B})$ and $\mathrm{dom}(\sf{A}\sf{B})=\{\psi\in\mathrm{dom}(\sf{B})\mid \sf{B}\psi\in\mathrm{dom}(\sf{A})\}$. We stress that neither $\mathrm{dom}(\sf{A}+\sf{B})$ nor $\mathrm{dom}(\sf{A}\sf{B})$ is dense in general and they might as well be empty.
\item[(ii)]The \emph{adjoint} of a densely-defined operator $\sf{A}$ is the operator $\sf{A}^{\dagger}\:\mathrm{dom}(\sf{A}^{\dagger})\to\mathcal{H}_{1}$ with 
\begin{align*}
	\mathrm{dom}(\sf{A}^{\dagger}):=\{\psi\in\mathcal{H}_{2}\mid\exists\eta\in\mathcal{H}_{1}:\langle \psi,\sf{A}\varphi\rangle_{\mathcal{H}_{2}}=\langle \eta,\varphi\rangle_{\mathcal{H}_{1}}\text{ for all }\varphi\in\mathrm{dom}(\sf{A})\}\subset\mathcal{H}_{2}\,,
\end{align*}
in which case we set $\sf{A}^{\dagger}\psi:=\eta$. Note that $\mathrm{dom}(\sf{A}^{\dagger})$ is not necessarily dense and might in fact even be trivial. If $\mathcal{H}:=\mathcal{H}_{1}=\mathcal{H}_{2}$, we call $\sf{A}$ \emph{symmetric} if $\langle\sf{A}\psi,\varphi\rangle_{\mathcal{H}}=\langle\psi,\sf{A}\varphi\rangle_{\mathcal{H}}$ for all $\psi,\varphi\in\mathrm{dom}(\sf{A})$. In this case, $\sf{A}\subset\sf{A}^{\ast}$. If $\sf{A}=\sf{A}^{\dagger}$, i.e.~if  in addition $\mathrm{dom}(\sf{A}^{\dagger})=\mathrm{dom}(\sf{A})$, we call $\sf{A}$ \emph{self-adjoint}.
\item[(iii)]For a densely-defined operator $\sf{A}$, we denote by $\mathcal{G}(\sf{A}):=\{(\psi,\sf{A}\psi)\mid\psi\in\mathrm{dom}(\sf{A})\}\subset\mathcal{H}_{1}\oplus\mathcal{H}_{2}$ its \emph{graph}. If there exists an operators whose graph is given by the closure $\overline{\mathcal{G}(\sf{A})}$, we call $\sf{A}$ \emph{closable}. In this case, there exists exactly one operator $\overline{\sf{A}}\:\mathrm{dom}(\overline{\sf{A}})\to\mathcal{H}_{2}$ with $\mathcal{G}(\overline{\sf{A}})=\overline{\mathcal{G}(\sf{A})}$, which is called the \emph{closure of $\sf{A}$}. The domain $\mathrm{dom}(\overline{\sf{A}})$ can be characterised as the completion of $\mathrm{dom}(\sf{A})$ with respect to the \emph{graph norm} $\Vert\psi\Vert_{\mathcal{G}(\sf{A})}:=(\Vert\psi\Vert_{\mathcal{H}_{1}}^{2}+\Vert\sf{A}\psi\Vert_{\mathcal{H}_{2}}^{2})^{\frac{1}{2}}$ defined for $\psi\in\mathrm{dom}(\sf{A})$. One can show that $\sf{A}$ is closable if and only if its adjoint is densely-defined in which case it holds that $\overline{\sf{A}}=(\sf{A}^{\dagger})^{\dagger}$. There is yet another useful characterisation of the closure: $\sf{A}$ is closable if and only if for every $\omega\in\mathcal{H}_{1}$ and every pair of sequences $(\psi_{n})_{n\in\bb{N}}$ and $(\varphi_{n})_{n\in\bb{N}}$ in $\mathrm{dom}(\sf{A})$ with $\psi_{n}\to \omega$ and $\varphi_{n}\to\omega$ as $n\to\infty$ and with the property that $(\sf{A}\psi_{n})_{n\in\bb{N}}$ and $(\sf{A}\varphi_{n})_{n\in\bb{N}}$ are convergent in $\mathcal{H}_{2}$ it holds that $\lim_{n\to\infty}\sf{A}\psi_{n}=\lim_{n\to\infty}\sf{A}\varphi_{n}$. As a consequence, $\mathrm{dom}(\overline{\sf{A}})$ can also be characterised as
\begin{align*}
	\mathrm{dom}(\overline{\sf{A}})=\{\psi\in\mathcal{H}_{1}\mid \exists (\psi_{n})_{n\in\bb{N}}\in\mathrm{dom}(\sf{A})^{\mathbb{N}}\text{ with }\lim_{n\to\infty}\psi_{n}=\psi: (\sf{A}\psi_{n})_{n\in\bb{N}}\text{ is convergent}\}
\end{align*}
and the closure is given by $\overline{\sf{A}}\psi:=\lim_{n\to\infty}\sf{A}\psi_{n}$ for all $\psi\in\mathrm{dom}(\overline{\sf{A}})$ with $(\psi_{n})_{n\in\bb{N}}$ as above.
\item[(iv)]A densely-defined and closable operator $\sf{A}$ is called \emph{essentially self-adjoint}, if its closure $\overline{\sf{A}}$ is self-adjoint. It is not too hard to see that the adjoint $\sf{A}^{\dagger}$ of any densely-defined operator $\sf{A}$ is closed. Furthermore, it holds that
\begin{align*}
	\overline{\sf{A}}^{\dagger}=((\sf{A}^{\dagger})^{\dagger})^{\dagger}=\overline{\sf{A}^{\dagger}}=\sf{A}^{\dagger}\, .
\end{align*}
Hence, essentially self-adjointness is equivalent to the condition $\sf{A}^{\dagger}=\overline{\sf{A}}$. If $\sf{A}$ is essentially self-adjoint, one can show that $\overline{\sf{A}}$ is the \emph{unique} self-adjoint extension of $\sf{A}$. 
\end{itemize}

Having introduced the most important terminology for (not necessarily bounded) operators, let us briefly discuss the \emph{spectral theorem}. Let $\sf{A}\:\mathrm{dom}(\sf{A})\to\mathcal{H}$ with $\mathrm{dom}(\sf{A})\subset\mathcal{H}$ be a densely-defined operator on some $\bb{C}$-Hilbert space $(\mathcal{H},\langle\cdot,\cdot\rangle_{\mathcal{H}})$. We denote the \emph{resolvent set} of $\sf{A}$ by
\begin{align*}
	\rho(\sf{A}):=\{\lambda\in\bb{C}\mid (\sf{A}-\lambda\,\mathrm{id})\:\mathrm{dom}(\sf{A})\to\mathcal{H}\text{ is bijective with bounded inverse }\}\, .
\end{align*}
The \emph{spectrum} of $\sf{A}$ is then defined to be the complement in $\bb{C}$, i.e.~$\sigma(\sf{A}):=\bb{C}\backslash\rho(\sf{A})$. In the finite-dimensional case, in which $\sf{A}$ can be identified with a Hermitian matrix, $\sigma(\sf{A})$ is exactly the set of all \emph{eigenvalues} of $\sf{A}$. More generally, $\sigma(\sf{A})$ has both a discrete and continuous part. Moreover, we recall that $\sigma(\sf{A})$ is always a closed subset of $\bb{C}$. If $\sf{A}$ is bounded, then $\sigma(\sf{A})$ is in fact even a compact set. Last but not least, it is well known that a closed and symmetric densely-defined operator $\sf{A}$ is self-adjoint if and only if $\sigma(\sf{A})\subset\bb{R}$.

\begin{remark}
	If $\sf{A}$ is \emph{not} closed, then it is easy to show that $\sigma(\sf{A})=\bb{C}$. Hence, it is only useful to study the spectrum for \emph{closed} operators. Now, if $\sf{A}$ is closed, the \emph{closed graph theorem} implies that the inverse of $(\sf{A}-\lambda\,\mathrm{id})$ (if it exists) is automatically bounded and hence, the addition ``with bounded inverse'' is redundant in the above definition of $\rho(\sf{A})$ in this case. 
\end{remark}

Now, let $\sf{X}$ be any topology space and denote its \emph{Borel $\sigma$-algebra} by $\mathcal{B}(\sf{X})$. A \emph{projection-valued measure} (PVM), or \emph{spectral measure}, is a map $\mu\:\mathcal{B}(\sf{X})\to\mathcal{P}(\mathcal{H})$, where $\mathcal{P}(\mathcal{H})$ denotes the subset of $\mathcal{L}(\mathcal{H}):=\{\sf{T}\:\mathcal{H}\to\mathcal{H}\mid \text{linear and bounded}\}$ consisting of orthogonal projectors, with the following two properties:
\begin{align*}
	\text{(i)}&\quad \mu(\emptyset)=0\quad\text{and}\quad \mu(\sf{X})=\mathrm{id}_{\mathcal{H}}\\
	\text{(ii)}&\quad \forall (\sf{M}_{n})_{n\in\bb{N}}\in\mathcal{B}(\sf{X})^{\bb{N}}\text{ pairwise disjoint }:~\sum_{n=1}^{\infty}\mu(\sf{M}_{n})\psi=\mu\bigg(\bigcup_{n=1}^{\infty}\sf{M}_{n}\bigg)\psi\quad\forall\psi\in\mathcal{H}
\end{align*}
Note that these are similar conditions to the usual definitions of (scalar-valued) measures. In fact, for two given vectors $\psi,\varphi\in\mathcal{H}$, we obtain a \emph{complex measure} $\mu_{\psi,\varphi}\:\mathcal{B}(\mathcal{H})\to\mathbb{C}$ defined by $\mu_{\psi,\varphi}(\sf{M}):=\langle\mu(\sf{M})\psi,\varphi\rangle_{\mathcal{H}}$. In the case $\psi=\varphi$, we simply write $\mu_{\psi}:=\mu_{\psi,\psi}$. Now, if $f\:\sf{X}\to\bb{C}$ is any bounded and Borel-measurable function, we can define a \emph{spectral integral}
\begin{align*}
	\int_{\sf{X}}f(\lambda)\,\d\mu(\lambda)\in\mathcal{L}(\mathcal{H})
\end{align*}
in a similar way as one does in the case of the \emph{Lebesgue-Bochner integral}, i.e.~by defining the integral first on the level of \emph{simple step functions} and then extending it by a suitable limiting procedure. More generally, if $f$ is an \emph{unbounded} Borel-measurable function, one can define a densely-defined closed operator
\begin{align*}
	\int_{\sf{X}}f(\lambda)\,\d\mu(\lambda)\:\mathrm{dom}_{f}\to\mathcal{H}\,,\qquad \mathrm{dom}_{f}:=\bigg\{\psi\in\mathcal{H}\,\bigg\vert\, \int_{\sf{X}}\vert f(\lambda)\vert^{2}\,\d\mu_{\psi}(\lambda)<\infty\bigg\}
\end{align*}
by approximating $f$ by a sequence of \emph{bounded} functions $(f_{n})_{n\in\mathbb{N}}$ in $\sf{L}^{2}(\sf{X},\mu_{\psi})$ and by setting $\int_{\sf{X}}f(\lambda)\d\mu(\lambda)$ acting on $\psi$ as the corresponding limit of $\int_{\sf{X}}f_{n}(\lambda)\d\mu_{\psi}(\lambda)$.

\begin{theorem}\label{Thm:SpecThm} \emph{(Spectral Theorem for Unbounded Operators)}\newline
	Let $\sf{A}\:\mathrm{dom}(\sf{A})\to\mathcal{H}$ be a self-adjoint operator. Then, there exists a unique spectral measure $\mu^{\sf{A}}\:\mathcal{B}(\bb{R})\to\mathcal{P}(\mathcal{H})$ such that
	\begin{align*}
		\sf{A}=\int_{\sigma(\sf{A})}\lambda\,\d\mu^{\sf{A}}(\lambda)\,,\qquad \langle\sf{A}\psi,\varphi\rangle_{\mathcal{H}}=\int_{\sigma(\sf{A})}\lambda\,\d\mu_{\psi,\varphi}^{\sf{A}}(\lambda)\, .
	\end{align*}
	Moreover, $\mathrm{supp}(\mu^{\sf{A}})=\sigma(\sf{A})$ and, following the above discussion, we define a densely-defined closed operator $f(\sf{A})\:\mathrm{dom}(f(\sf{A}))\to\mathcal{H}$ for every Borel-measurable $f\:\sigma(\sf{A})\to\bb{C}$ by
	\begin{align*}
		f(\sf{A})=\int_{\sigma(\sf{A})}f(\lambda)\,\d\mu^{\sf{A}}(\lambda)\,,\,\,\, \mathrm{dom}(f(\sf{A}))=\bigg\{\psi\in\mathcal{H}\,\bigg\vert\, \Vert f(\sf{A})\psi\Vert_{\mathcal{H}}^{2}=\int_{\sigma(\sf{A})}\vert f(\lambda)\vert^{2}\,\d\mu_{\psi}^{\sf{A}}(\lambda)<\infty\bigg\}\,.
	\end{align*}
\end{theorem}

The assignment $f\mapsto f(\sf{A})$ is usually called the \emph{spectral calculus} of $\sf{A}$. For $f,g$ Borel-measurable, it has the following properties:
\begin{itemize}
	\item[(i)]$f(\sf{A})$ is bounded if and only if $f\in\sf{L}^{\infty}(\sigma(\sf{A}),\mu^{\sf{A}})$. In this case, $\Vert f(\sf{A})\Vert_{\mathrm{op}}=\Vert f\Vert_{\infty}$.
	\item[(ii)]$f(\sf{A})^{\ast}=\overline{f}(\sf{A})$. In particular, $f(\sf{A})$ is self-adjoint if and only if $f$ is real-valued.
	\item[(iii)]$(f+g)(\sf{A})=\overline{f(\sf{A})+g(\sf{A})}$ and $\mathrm{dom}(f(\sf{A})+g(\sf{A}))=\mathrm{dom}(f(\sf{A}))\cap\mathrm{dom}(g(\sf{A}))$.
	\item[(iv)]$(fg)(\sf{A})=\overline{f(\sf{A})g(\sf{A})}$ and $\mathrm{dom}(f(\sf{A})g(\sf{A}))=\mathrm{dom}(g(\sf{A}))\cap\mathrm{dom}((fg)(\sf{A}))$.
	\item[(v)]If $f\neq 0$ (almost everywhere), then $f(\sf{A})$ is invertible and $f(\sf{A})^{-1}=(1/f)(\sf{A})$.
	\item[(vi)]If $f(\lambda)\geq 0$ (almost everywhere), then $f(\sf{A})\geq 0$.
	\item[(vii)]If $f$ is continuous, then $\sigma(f(\sf{A}))=\overline{f(\sigma(\sf{A}))}$ where $f(\sigma(\sf{A})):=\{f(\lambda)\mid\lambda\in\sigma(\sf{A})\}$.
\end{itemize}
If \textbf{1}$\:1\mapsto 1$ denotes the constant $1$-function, then \textbf{1}$(\sf{A})=\mathrm{id}_{\mathcal{H}}$. Furthermore, $\mathrm{id}_{\sigma(\sf{A})}(\sf{A})=\sf{A}$ and for a polynomial $p(\lambda)=\sum_{i=0}^{n}\alpha_{n}\lambda^{n}$, we obtain $p(\sf{A})=\sum_{i=0}^{n}\alpha_{n}\sf{A}^{n}$. As a consequence of the above properties, we observe that the assignment $\Phi_{\sf{A}}\:C^{0}(\sigma(\sf{A}),\bb{C})\to\mathcal{L}(\mathcal{H}), f\mapsto f(\sf{A})$ is a continuous unital $\ast$-algebra homomorphism. For more details on the spectral theorem and proofs of the aforementioned statements, we refer to \cite{Schmudgen2,Werner,MorettiSpectral}. Last but not least, we note that the spectral theorem can be generalised to (unbounded) \emph{normal} operators.

\section{Linear Maps between Quotient Spaces}
\label{App:Quot}
In this section, we characterise linear maps and isomorphisms between quotient spaces of vector spaces,, which is particularly useful for establishing isomorphisms between the various phase spaces introduced in Section~\ref{Sec:HackSchenkel}.

\begin{itemize}
	\item[$\bullet$]Throughout the following, let $\sf{V}_{1},\sf{V}_{2}$ be vector spaces over the same field $\mathbb{F}$ and $\sf{W}_{1}\subset\sf{V}_{1},\sf{W}_{2}\subset\sf{V}_{2}$ be arbitrary linear subspaces thereof.
\end{itemize}

Now, proving whether a linear map $f\:\sf{V}_{1}\to\sf{V}_{2}$ induces a well-defined isomorphism between the quotient spaces $\sf{V}_{1}/\sf{W}_{1}$ and $\sf{V}_{2}/\sf{W}_{2}$ amounts to check the following conditions.

\begin{proposition}
    Let $f\:\sf{V}_{1}\to\sf{V}_{2}$ be a $\mathbb{F}$-linear map. If $f(\sf{W}_{1})\subset \sf{W}_{2}$, then $f$ induces a well-defined $\bb{F}$-linear map
    \begin{align*}
        [f]:\frac{\sf{V}_{1}}{\sf{W}_{1}}\to\frac{\sf{V}_{2}}{\sf{W}_{2}}
    \end{align*}
    defined by $[f]([v]):=[f(v)]$ for all $v\in\sf{V}_{1}$. Furthermore, the following holds:
    \begin{itemize}
        \item[\emph{(i)}]The map $[f]$ is injective if and only if $f^{-1}(\sf{W}_{2})\subset\sf{W}_{1}$.
        \item[\emph{(ii)}]The map $[f]$ is surjective if and only if $\sf{V}_{2}=f(\sf{V}_{1})+\sf{W}_{2}$.
    \end{itemize}
\end{proposition}

\begin{proof}
	We first show that $[f]$ is well-defined. If $v_{1}$ and $v_{2}$ are two representatives of the same equivalence class in $\sf{V}_{1}/\sf{W}_{1}$, then $v_{1}-v_{2}\in\sf{W}_{1}$. By assumption, $f(\sf{W}_{1})\subset \sf{W}_{2}$, which implies that $\sf{W}_{2}\ni f(v_{1}-v_{2})=f(v_{1})-f(v_{2})$, i.e.~$[f(v_{1})]=[f(v_{2})]$. We now turn to injectivity and surjectivity.
	\begin{itemize}
		\item[(i)]By definition, $[f]$ is injective if and only if $[f(v)]=0$ implies $[v]=0$ for all $v\in\sf{V}_{1}$. This is equivalent to saying that $f(v)\in\sf{W}_{2}$ implies $v\in\sf{W}_{1}$. This, in turn, is clearly equivalent to the inclusion $f^{-1}(\sf{W}_{2})\subset\sf{W}_{1}$, recalling that $f^{-1}(\sf{W}_{2}):=\{v\in\sf{V}_{1}\mid f(v)\in\sf{W}_{2}\}$. 
		\item[(ii)]The map $[f]$ is surjective if and only if for every $v_{2}\in\sf{V}_{2}$, there exists a $v_{1}\in\sf{V}_{1}$ such that $[f(v_{1})]=[v_{2}]$. In other words, surjectivity is equivalent to saying that for every $v_{2}\in\sf{V}_{2}$, there exists a $v_{1}\in\sf{V}_{1}$ and a $w\in\sf{W}_{2}$ such that $v_{2}-f(v_{1})=w$. Hence, $[f]$ is surjective if and only if we can write every $v_{2}\in\sf{V}_{2}$ as $v_{2}=f(v_{1})+w$ for some $v_{1}\in\sf{V}_{1}$ and $w\in\sf{W}_{2}$.
	\end{itemize}
\end{proof}

\begin{remark}
	Note that $f(\sf{W}_{1})\subset \sf{W}_{2}$ implies $\sf{W}_{1}\subset f^{-1}(\sf{W}_{2})$. Hence, if the quotient map $[f]$ is injective, we in fact have $f^{-1}(\sf{W}_{2})=\sf{W}_{1}$.
\end{remark}

%
%
%
%
%
%
%
%
%
%
%
%
%
%
%
\newpage 
\addtocontents{toc}{\vspace*{2ex}}
\addcontentsline{toc}{section}{\hspace{-16pt}\textbf{Bibliography}}
\markboth{Bibliography}{Bibliography}
\fancyhead[RE]{\nouppercase{Bibliography}}
\bibliographystyle{alpha}

\end{document}